\documentclass[10pt,english]{article}
\usepackage[T1]{fontenc}
\usepackage[utf8]{inputenc}
\usepackage[english]{babel}
\addto\captionsenglish{}
\usepackage[a4paper,margin=1.05in]{geometry}
\usepackage{amsmath,amsfonts,amssymb,amsthm,amscd}
\usepackage{lmodern}
 
\usepackage[mathcal]{euscript}
\usepackage{mathrsfs}
\usepackage{graphicx}
\usepackage{tikz}
\usepackage{subfig}
\usepackage{rotating}
\usepackage{array}
\usepackage{verbatim}
\usepackage{slashed}
\usepackage{tensor}
\usepackage[most]{tcolorbox}
\usepackage{microtype}
\usepackage{xcolor}
\usepackage{float}
\usepackage{needspace}
\usepackage{aliascnt}
\newtheorem{theorem}{Theorem}[section]
\newaliascnt{proposition}{theorem}
\newtheorem{proposition}[proposition]{Proposition}
\aliascntresetthe{proposition}
\newaliascnt{lemma}{theorem}
\newtheorem{lemma}[lemma]{Lemma}
\aliascntresetthe{lemma}
\newaliascnt{corollary}{theorem}
\newtheorem{corollary}[corollary]{Corollary}
\aliascntresetthe{corollary}
\newaliascnt{definition}{theorem}
\newtheorem{definition}[definition]{Definition}
\aliascntresetthe{definition}
\newaliascnt{example}{theorem}

\aliascntresetthe{example}
\newaliascnt{propositiondefinition}{theorem}

\aliascntresetthe{propositiondefinition}
\newaliascnt{remark}{theorem}

\aliascntresetthe{remark}
\newaliascnt{terminology}{theorem}

\aliascntresetthe{terminology}
\numberwithin{equation}{section} 
\usepackage[hidelinks]{hyperref}
\DeclareFontFamily{U}{BOONDOX-calo}{\skewchar \font=45}
\DeclareFontShape{U}{BOONDOX-calo}{m}{n}{<-> s*[1.05] BOONDOX-r-calo}{}
\DeclareFontShape{U}{BOONDOX-calo}{b}{n}{<-> s*[1.05] BOONDOX-b-calo}{}
\DeclareMathAlphabet{\mathcalboondox}{U}{BOONDOX-calo}{m}{n}
\SetMathAlphabet{\mathcalboondox}{bold}{U}{BOONDOX-calo}{b}{n}
\DeclareMathAlphabet{\mathbcalboondox}{U}{BOONDOX-calo}{b}{n}
\numberwithin{table}{section}
\numberwithin{figure}{section}
\newcommand \epss {\eps_\star}

\newcommand \gd g  
\newcommand \gconf g

\newcommand \givenCurv {\widetilde S} 
\newcommand \givennabla {\widetilde \nabla} 
\newcommand \givenBox {\Box_{\giveng}}
\newcommand \givenVol {\widetilde{\textrm{Vol}}}
\newcommand \givenT {\widetilde{T}}
\newcommand \givenG {\widetilde{\text{G}}}
\newcommand \givenRic {\widetilde{\textrm{Ric}}}
\newcommand \givenScal {\widetilde{R}}
\newcommand \givenR {\widetilde{R}}
\newcommand \be {\begin{equation}}
\newcommand \ee {\end{equation}} 
\newcommand \bel {\be\label} 
\newcommand \fTime {\mathbf T}
\newcommand \Minsk \eta
\newcommand \rhoH {{\mathbf r}^{\mathcal H}}
\newcommand \rhoE {{\mathbf r}^{\mathcal E}}
    
\newcommand \Fenergy      {\mathbf F}              
\newcommand \Eenergy      {\mathbf E}           
\newcommand \expeta \sigma 
 
\newcommand \ME {\mathcal{EM}} 
 
\renewcommand \H {\mathcal H}

\newcommand \MME 		{\Mcal^{\mathcal{EM}}}

\newcommand{\RR}{\mathbb{R}}
\newcommand{\del}{\partial}
\newcommand{\Mcal}{\mathcal{M}}
\newcommand{\Tr}{\text{Tr}}

\newcommand{\Gammaj}{\widetilde{\Gamma}}

\newcommand{\hfrak}{\mathsf{h}}
\newcommand{\ffrak}{\mathsf{f}}
\newcommand \Ufrak {\mathsf U}

\newcommand{\delb}{\overline{\del}}
\newcommand \sourterm {\mathcal F}
\newcommand \coeff    \kappa
\newcommand \fieldphi    {\phi}
\newcommand \fieldPhi    {\phi^{(2)}{}}
\newcommand \fieldvarphi {\phi^{(1)}{}}
\newcommand \coef \alpha 
\newcommand \bse {\begin{subequations}}
\newcommand \ese {\end{subequations}}
\newcommand \giveng {{\widetilde{g}}} 

\newcommand \Sch {\emph{Sch}}

\newcommand \eff {\mathrm{eff}}
\newcommand \Kcal {\mathcal K}
\newcommand \Lscr{\mathscr{L}}

\newcommand \rhon {\rho^\natural}

\newcommand \sourtermn {\sourterm^\natural}

\newcommand \la \langle
\newcommand \ra \rangle
\newcommand \bei  {\begin{itemize}}
\newcommand \eei {\end{itemize}}

\newcommand \vb {\overline v}
\newcommand \rhob {\overline \rho}

\newcommand \BoxChapeau {\widehat \Box}
\newcommand \BoxChapeaud {\BoxChapeau} 
\newcommand \Lcal {\mathcal L}
\newcommand \newperp  {\underline{\del}_\perp}

\newcommand \Hcal {\mathcal H}

\newcommand {\eps} \epsilon

\newcommand \Gammad  \Gamma    
\newcommand \Ocal {\mathcal O}

\newcommand \Mint {\Mcal^{\mathcal H}}
\newcommand \Mtran {\Mcal^{\mathcal M}}
\newcommand \Mext {\Mcal^{\mathcal E}}

\newcommand \crochet {\mathbf X}
\newcommand \delH {\del^{\mathcal H}} 
\newcommand \delsH {\slashed \del^{\mathcal H}}
\newcommand \PhiH {{\Phi^{\mathcal H}}{}}
\newcommand \PsiH {{\Psi^{\mathcal H}}{}}  

\newcommand \delN {\del^{\mathcal{N}}}
\newcommand \delsN {\slashed \del^{\mathcal{N}}}
\newcommand \delsEH {\slashed {\del}^{\mathcal{EH}}}
\newcommand \delsME {\slashed {\del}^{\mathcal{EM}}} 
\newcommand \Fbb {\mathbb F}
\newcommand \Gbb {\mathbb G}
\newcommand \Mbb {\mathbb M}
\newcommand \Pbb {\mathbb P}
\newcommand \Qbb {\mathbb Q}

\newcommand \dive {\text{div} \hskip.05cm}
\newcommand \delEH {\del^{\mathcal{E \hskip-.06cm H}}}

\newcommand \ord {\mathbf{ord} \hskip.05cm}

\newcommand \rank {\mathbf{rank} \hskip.05cm}
\newcommand \gMink {g_\emph{Mink}}

\newcommand \Sta {\textnormal{Sta}}
\newcommand \calE {\mathcal E}

\newcommand \underL {\underline L}
\newcommand \Ds {\mathbf D_s}
\newcommand \Bulk {\mathsf{Bulk}}
\newcommand \bulkdens {\mathsf{bulk}}

\begin{document}
\title{The General-Relativistic Limit of $f(R)$ Gravity}

\author{\large Philippe G. LeFloch$^1$ and Yue Ma$^2$}

\date{\today}

\maketitle
\footnotetext[1]{Laboratoire Jacques-Louis Lions and Centre National de la Recherche Scientifique, Sorbonne Universit\'e, 4~Place Jussieu, 75252 Paris, France. Email: {\sl contact@philippelefloch.org}.}
\footnotetext[2]{School of Mathematics and Statistics, Xi'an Jiaotong University, Xi'an, 710049 Shaanxi, People's Republic of China. Email: {\sl yuemath@mail.xjtu.edu.cn}.}
\setcounter{footnote}{2}

\begin{abstract}
We establish the general-relativistic limit for the \emph{\(f(R)\) theory of gravity} coupled to a massless scalar field, with small, asymptotically Euclidean, well-prepared Cauchy data in the near-Minkowski regime. First, we prove global existence and future causal geodesic completeness, and derive \emph{uniform estimates} for the metric, matter, and rescaled effective field for a broad class of modified Lagrangians \(f(R)\).
Second, we prove that as  \(f(R)\)
 approaches the Einstein--Hilbert Lagrangian \(R\), the physical and conformal metrics together with the matter field \emph{converge} to the Einstein--massless-scalar development of the limiting initial data set, uniformly on the entire future spacetime for finitely many space and time derivatives. The \emph{effective field}, defined by a nonlinear reparametrization of the physical scalar curvature, converges to the scalar curvature of the limiting Einstein development. The main analytical difficulty is that the mass of the effective field \emph{tends to infinity} when \(f(R)\) tends to \(R\), while its second-order evolution equation reduces to an algebraic trace relation in the limit. Building upon the \emph{Euclidean--hyperboloidal foliation method} from our earlier work (P.G. LeFloch and Y. Ma, Nonlinear stability of self-gravitating massive fields, Ann. PDE 10 (2024)), we analyze the field equations in suitable wave coordinates adapted to this problem, as a quasilinear system of coupled wave and Klein-Gordon equations whose nonlinearities do not satisfy the classical null condition. Small effective-field behavior may coexist with high-frequency oscillations and comparatively large derivatives, whose quadratic contributions to the metric equation must be controlled uniformly. We combine oscillatory estimates, suitably weighted interior and exterior energy estimates, and propagation estimates for the discrepancy from the Einstein trace relation. Quantitative convergence on compact regions, combined with spacetime decay estimates uniform along the admissible family of Lagrangians \(f(R)\), yields uniform convergence of the whole future development.
\end{abstract}

{\small

\setcounter{secnumdepth}{2}
\setcounter{tocdepth}{2}

\tableofcontents
}


\section{Introduction}
\label{section-1}
\providecommand{\coloneqq}{\mathrel{:=}}


\subsection{\texorpdfstring{The general-relativistic limit}{The general-relativistic limit}}
\label{section-1-1}

In \(f(R)\) gravity, the scalar curvature in the Einstein--Hilbert action is replaced by a nonlinear function~\cite{Buchdahl-1970,Sotiriou-2010}. We denote by \(\coeff>0\) the strength of its quadratic correction. The basic example is the Starobinsky model~\cite{Starobinsky-1979}, namely 
\be
\label{equa-STA}
f_\coeff(r)=r+ \frac{\coeff}{2}r^2, \qquad \coeff>0.
\ee
Here \(r\) is the scalar-curvature argument of the Einstein--Hilbert integrand. In the limit \(f_\coeff(r)\to r\) as \(\coeff\to0\), the global evolution problem from an initial hypersurface changes order: the modified metric equations are \emph{fourth order}, whereas Einstein's equations are \emph{second order}. The additional gravitational degree of freedom must therefore cease to evolve independently in the limit. In this context, we prove \emph{global nonlinear stability} and study the \emph{general-relativistic limit} for small data near Minkowski spacetime. Importantly, our estimates are \emph{uniform in the parameter} $\coeff$: the unknown metric and matter fields and finitely many of their spacetime derivatives converge uniformly on the entire future spacetime. This result completes a decade-long research program initiated with the authors’ memoir~\cite{LeFloch-Ma-2017} on the local-in-time theory. 

We use the so-called Jordan coupling, with a massless scalar field \(\psi_\coeff\) minimally coupled to the physical metric \(\widetilde g_\coeff\). Its stress-energy tensor generally has a nonzero trace, so the additional gravitational field remains coupled to matter. Writing \(\widetilde R_\coeff\) for the physical scalar curvature, we introduce the effective field \(\fieldphi_\coeff\) and the conformal metric \(g_\coeff\) by
\be
\label{equa:intro-conformal-fields}
\fieldphi_\coeff\coloneqq\coeff^{-1}\log f_\coeff'(\widetilde R_\coeff),\qquad g_\coeff\coloneqq e^{\coeff\fieldphi_\coeff}\widetilde g_\coeff.
\ee
In wave gauge for \(g_\coeff\), the equations form a second-order wave--Klein--Gordon system whose effective-field mass tends to infinity. Its natural rescalings are
\be
\label{equa:intro-rescaled-effective-fields}
\fieldPhi_\coeff\coloneqq\coeff\fieldphi_\coeff,\qquad \fieldvarphi_\coeff\coloneqq\coeff^{1/2}\fieldphi_\coeff.
\ee
The Einstein limit requires both disappearance of their contribution to the metric equation and recovery of the algebraic relation between scalar curvature and matter trace.

We build upon a series of paper concerned with the nonlinear stability of self-gravitating massive fields, in the near-Minkowski regime, in which we developed the \emph{Euclidean--hyperboloidal foliation method} for coupled systems of nonlinear wave and Klein-Gordon equations \cite{LeFloch-Ma-2014,LeFloch-Ma-2024b}. Independently, 
Ionescu and Pausader established global regularity for the Einstein--Klein--Gordon system with small, smooth, suitably localized initial data~\cite{Ionescu-2019,Ionescu-2022}. 
Both methods concern a \emph{fixed positive Klein--Gordon mass}. The novel issue addressed in the present memoir is to obtain metric, matter, and rescaled effective-field estimates that are \emph{uniform} as the effective-field mass $1/\coeff$ \emph{tends to infinity}, and to prove convergence to the Einstein--massless-scalar development. Our analysis in wave coordinates owes much to the pioneering work of Lindblad and Rodnianski, who introduced the weak null condition~\cite{Lindblad-Rodnianski-2003} and established the global nonlinear stability of (vacuum) Minkowski spacetime in this gauge~\cite{Lindblad-Rodnianski-2005,Lindblad-Rodnianski-2010}.


\subsection{\texorpdfstring{Global stability and convergence}{Global stability and convergence}}
\label{section-1-2}

\paragraph{Initial hypotheses and preparation}
We describe the hypotheses before stating the theorem. All initial quantities are prescribed on a fixed asymptotically Euclidean hypersurface.
\bei
\item \emph{The family of theories.} An admissible scaled theory has the form \(f_\coeff(r)=\coeff^{-1}f_*(\coeff r)\), with a fixed sufficiently regular profile satisfying \(f_*(0)=0\), \(f_*'(0)=f_*''(0)=1\), and \(f_*'>0\). On a fixed interval, \(\log f_*'\) is a diffeomorphism of the required regularity onto an interval containing a fixed neighborhood of zero.
\item \emph{Constrained initial data.} The conformal data prescribe the metric, second fundamental form, and values and derivatives along the conformal unit normal of both scalar fields. They satisfy the Hamiltonian and momentum constraints and the initial wave-gauge conditions. The same initial spatial coordinates, lapse, and shift are used for the family and its Einstein limit.
\item \emph{Smallness uniform in the parameter.} Weighted initial energies control the metric perturbation, matter, and rescaled effective field after commutation with translations and Lorentz boosts. The reduced equations determine the higher normal derivatives from the Cauchy data. The total size \(A_0\), including the reference geometry, is small independently of \(\coeff\). The reference metric is asymptotically Schwarzschild, with a small nonnegative mass \(m\) independent of \(\coeff\), and
\be
\label{equa:intro-relative-mass-hypothesis}
A_0\leq\varepsilon_0,\qquad 0\leq m\leq C_{\mathrm{ref}}A_0.
\ee
The constants are independent of \(\coeff\). We choose an auxiliary interface that expands slightly faster than the Minkowski light cones. The resulting energy flux is positive while the initial hypersurface and its data remain unchanged.
\item \emph{Compatibility and preparation.} Einstein compatibility means that the metric--matter data converge to constrained Einstein--massless-scalar data and that the rescaled effective-field data vanish in the weighted Sobolev norms. Well-preparedness additionally requires smallness of the discrepancy from the Einstein trace relation,
\be
\label{equa:intro-trace-residual}
\Theta_\coeff\coloneqq\fieldphi_\coeff-8\pi e^{-\fieldPhi_\coeff}g_\coeff(\nabla\psi_\coeff,\nabla\psi_\coeff),
\ee
where the gradient and contraction are taken with \(g_\coeff\). The preparation amplitude measures the commuted values of \(\Theta_\coeff\) and their \(\coeff^{1/2}\)-weighted first derivatives on the entire initial hypersurface, with spatial weights; this amplitude tends to zero.
\eei
The admissible theories are defined in Definition~\ref{def:admissible-f-family}, and compatibility and preparation in Definition~\ref{definition:Einstein-compatible-commuted-preparation}. The initial energy assumptions and the reference-mass conditions are collected in Theorem~\ref{theorem:final-global-stability-and-limit}. These assumptions concern Cauchy data, including the derivatives determined by the equations.

\paragraph{\texorpdfstring{The global limit theorem}{The global limit theorem}}
We compare the developments in wave coordinates \((t,x)\) on a fixed future domain \(\mathcal D^+\), using a foliation independent of \(\coeff\). The local energy topology of order \(q\) controls the values of the metric and matter in \(H^q\) on the foliation hypersurfaces and their first coordinate derivatives in \(H^{q-1}\), uniformly on compact regions. For the effective field, it controls \(\fieldPhi_\coeff\) in the same topology and \(\fieldvarphi_\coeff\) in \(H^q\).

The integer \(N\geq19\) denotes the number of controlled commutations. The spatial weights in the initial energies measure decay at spatial infinity. For the metric, the Schwarzschild contribution is first subtracted, and the differentiated perturbation is weighted by \((1+r)^\lambda\), where \(r=|x|\) and \(3/4<\lambda<1\). The lower threshold ensures integrability of quadratic source terms in the proof. A small loss exponent \(\delta>0\), chosen in terms of the spatial weights and \(N\), measures the polynomial growth in the uniform energy estimates. The resulting decay exponent for field values on the whole spacetime is
\be
\label{equa:intro-spacetime-decay-exponent}
\beta_\infty\coloneqq1-2\delta>0.
\ee
In the theorem, multiplication by \((1+t+r)^b\) expresses convergence with decay of order \(b\): the choice \(b=0\) gives ordinary uniform convergence, while \(0<b<\beta_\infty\) gives a stronger weighted conclusion. All these exponents are fixed independently of \(\coeff\).

\begin{theorem}[Global stability and the general-relativistic limit]
\label{theorem:main-global-limit}
Let \(N\geq19\) be an integer, let \(\coeff_0>0\), and consider an admissible scaled theory of regularity at least \(N+3\). For \(0<\coeff\leq\coeff_0\), let a smooth Einstein-compatible, well-prepared family of conformal Cauchy data for massless scalar matter satisfy the constraint, smallness, coordinate, and reference-mass conditions described above. Compatibility and preparation are required through order \(N-6\). Assume that the order-\(N\) energy is finite for each fixed \(\coeff>0\); its interior part need not be uniformly small. For \(A_0\) sufficiently small, with a threshold independent of \(\coeff\), the following conclusions hold.
\bei
\item \emph{Global existence.} Each data set determines a unique future global development in wave gauge. The physical metric \(\widetilde g_\coeff=e^{-\fieldPhi_\coeff}g_\coeff\) solves the original modified-gravity equations and is future causally geodesically complete.
\item \emph{Uniform estimates.} The interior metric and matter energies are uniform through order \(N-5\). The interior effective-field energy is uniform through order \(N-9\), and through order \(N-5\) after multiplication of its square-root norm by \(s^{-1/2}\). In the exterior, the metric and matter energies are uniform through order \(N\), and the effective-field energy through order \(N-5\); its order-\(N\) square-root norm carries the factor \(s^{-1}\). All these quantities have controlled polynomial growth with constants independent of \(\coeff\). The higher-order unscaled effective-field energy estimates keep their explicit parameter dependence.
\item \emph{Convergence on the entire future spacetime.} Let \((g_0,\psi_0)\) be the global Einstein--massless-scalar development of the limiting data. For every \(0\leq b<\beta_\infty\), one has, as \(\coeff\to0\), with componentwise absolute values,
\be
\label{equa:intro-global-uniform-limit}
\sup_{\mathcal D^+}(1+t+|x|)^b\bigl(|g_\coeff-g_0|+|\widetilde g_\coeff-g_0|+|\psi_\coeff-\psi_0|+|\fieldPhi_\coeff|+|\fieldvarphi_\coeff|\bigr)\longrightarrow0.
\ee
\item \emph{Uniform convergence of derivatives.} For every \(0\leq b<\beta_\infty\), the Cartesian spacetime derivatives of the physical and conformal metrics, matter, and rescaled effective fields converge uniformly on \(\mathcal D^+\) through order \(N-10\), with the same weight \((1+t+|x|)^b\). The unscaled field \(\fieldphi_\coeff\) converges uniformly to \(R[g_0]\), together with its derivatives through order \(N-12\), with this weight as well.
\item \emph{Local energy convergence and rates.} For \(1\leq q\leq N-7\), the same family converges without extraction in the local energy topologies described above, including at the initial hypersurface. Locally, \(\Theta_\coeff\to0\) in \(H^q\) and \(\coeff^{1/2}\nabla_{t,x}\Theta_\coeff\to0\) in \(H^{q-1}\), uniformly in the foliation parameter on compact intervals. If the compatibility and preparation moduli are \(\Ocal(\coeff^\sigma)\), with \(\sigma>0\), these local convergences have rate \(\Ocal(\coeff^{\min\{\sigma,1/2\}})\). For vanishing matter, the limit is the global vacuum Einstein development.
\eei
\end{theorem}

\paragraph{Spacetime decay and admissible weights}

The exponent \(\beta_\infty\) measures the spacetime decay kept in convergence, rather than a rate in \(\coeff\). It approaches one as the allowed energy-growth exponent \(\delta\) tends to zero; this choice precedes that of the smallness threshold. The interior field estimate has decay \(t^{-1+2\delta}\), whereas radial integration gives exterior decay closer to \(r^{-1}\). Thus the interior estimate determines the range \(b<1-2\delta\) stated above. The derivatives admit a larger range of weights, specified in Theorem~\ref{theorem:final-global-stability-and-limit}. The strict inequality in the weight makes the error outside a compact region tend to zero. Convergence at the limiting weight and its optimality remain separate questions. Writing \(r=|x|\), decay close to \(r^{-1}\) is consistent with the Schwarzschild contribution \(g^\star-\gMink=\Ocal(m/r)\) in the asymptotic end. Its stationary derivatives decay faster, but the time derivatives of an outgoing wave profile \(r^{-1}F(t-r)\), with \(F\) smooth, still have order \(r^{-1}\) near the light cone. Decay exponents close to one are therefore natural also for radiative derivatives.


\subsection{\texorpdfstring{Difficulties and strategy of proof}{Difficulties and strategy of proof}}
\label{section-1-3}

\paragraph{\texorpdfstring{Uniform control of the coupled evolution}{Uniform control of the coupled evolution}}
The change from a differential equation for the effective field to an algebraic trace relation is the main analytical difficulty. A field of small amplitude can have rapidly oscillating derivatives, whose quadratic contributions to the metric equation need not be small. We must control these contributions while the metric itself determines the wave operators and the propagation of matter. The proof addresses three coupled estimates.
\bei
\item \emph{A massive field with uniform constants.} Along hyperboloidal rays, the effective-field equation becomes a forced oscillator. We prove an estimate uniform in its frequency and apply it to the rescaled fields. We keep the parameter-dependent estimates for \(\fieldphi_\coeff\), needed for continuation at each fixed parameter, together with the estimates that tend to zero.
\item \emph{Nonlinear coupling without loss of derivatives.} The massless matter and massive effective field have different decay rates, and the matter trace forces the effective-field equation. We combine pointwise bounds for factors with fewer derivatives and energy bounds for factors with more derivatives. The wave-gauge identities improve the metric coefficients near the light cone, and induction on the number of Lorentz boosts controls the commutators.
\item \emph{Estimates across the interior and exterior regions.} Hyperboloidal hypersurfaces give decay in timelike directions; their Euclidean ends keep the spatial weights required by noncompact data. We combine the energy identities with their signed fluxes, which cancel at the interface. Radial, angular, and characteristic estimates control the values needed when rays enter the interior region.
\eei

\paragraph{\texorpdfstring{From compact regions to the whole future spacetime}{From compact regions to the whole future spacetime}}
Preparation controls the initial discrepancy from the Einstein trace relation. An estimate for its evolution, followed by a local energy estimate for the metric--matter difference, gives strong convergence on compact regions. The constants in this comparison may depend on the region. Uniform decay of the fields and their Cartesian derivatives gives the remaining step: outside a sufficiently large compact region, their weighted magnitudes are small independently of \(\coeff\). We first choose this region using decay, then pass to the limit inside it. This proves global weighted convergence of the fields and their derivatives; the algebraic rates in \(\coeff\) remain local.


\subsection{\texorpdfstring{Earlier results and organization}{Earlier results and organization}}
\label{section-1-4}

\paragraph{\texorpdfstring{The global convergence problem}{The global convergence problem}}
The local Cauchy problem and quantitative comparisons with Einstein gravity were studied in~\cite[Sections~6--9]{LeFloch-Ma-2017} for the quadratic modified Lagrangian and parameter-weighted small data. The global stability analysis in~\cite{LeFloch-Ma-2024b} gives estimates uniform in the modified theory when the dynamic perturbation is sufficiently small relative to the prescribed reference part, but leaves the convergence problem open; see its Section~2.3. Here we combine a propagation estimate for the discrepancy from the Einstein trace relation with a comparison of the metric and matter equations. The resulting limit holds on the entire future spacetime, with decay weights and finitely many space and time derivatives. The expanding interface allows the global estimates to use a small reference mass without requiring it to dominate the dynamic perturbation. The scalar--tensor formulation is related to~\cite{Brans-1961,Magnano-1994,Salgado-2007,Bhattacharyya-2023}; other modified-gravity models are considered in~\cite{Capozziello-2009,Briscese-2007,GarciaSalcedo-2010}.

Our estimates build on the approach pioneered by Klainerman~\cite{Klainerman-1985a} and H\"ormander~\cite{Hormander-1997} based on hyperboloids in Minkowski spacetime to deal with Klein-Gordon equation. The Euclidean-hyperboloidal method was developed for \emph{coupled} wave--Klein--Gordon systems on a \emph{curved spacetime}~\cite{LeFloch-Ma-2014,LeFloch-Ma-2016b,LeFloch-Ma-2023}, and especially to establish the nonlinear stability of self-gravitating massive fields~\cite{LeFloch-Ma-2016a,LeFloch-Ma-2018,LeFloch-Ma-2022,LeFloch-Ma-2024a}. Alternative approaches are due to Ionescu and Pausader~\cite{Ionescu-2014}, as well as, for compact perturbations only, by Wang~\cite{Wang-2020}. Related topics and results are treated  in~\cite{Katayama-2012a,Katayama-2012b,Klainerman-2020,Fang-2021,Dong-2021,Kauffman-2023}.

\paragraph{\texorpdfstring{Organization}{Organization}}
\bei
\item Part~I gives the physical and conformal equations, the initial data, the energy identities, and the oscillator estimate uniform in the mass.
\item Part~II establishes the interior pointwise and energy estimates, with the derivative counts needed for the nonlinear evolution and comparison.
\item Part~III gives the exterior estimates and the fluxes and traces required by the interior analysis. An expanding interface yields global stability under small initial norms and a small reference mass. The final section proves convergence on compact regions and then on the whole future spacetime.
\eei
For a first reading, Part~I and Sections~\ref{section-8}, \ref{section-12}, and~\ref{section-16} present the hypotheses, estimates, and limiting argument. Table~\ref{table:notation-proof} summarizes the notation.


\part{Formulation of the problem and technical estimates}
\providecommand{\coloneqq}{\mathrel{:=}}


\section{\texorpdfstring{The \(f(R)\) field equations and the Einstein limit}{The f(R) field equations and the Einstein limit}}
\label{section-2}


\subsection{Jordan coupling and the physical metric}
\label{section-2-1}
\paragraph{Purpose}
We use the Jordan coupling: matter is minimally coupled to the physical metric \(\giveng\), also called the Jordan metric. This choice of coupling is distinct from both the metric representation and the coordinate gauge. We first write the equations with \(\giveng\), then introduce the conformally related Einstein metric \(\gconf\), and finally impose wave coordinates for \(\gconf\). A suitably rescaled scalar curvature is treated as an additional unknown, the  effective field. The Jordan-coupled theory is thereby represented as a second-order Einstein--Klein--Gordon-type system in the conformal metric. Throughout this memoir, ``Jordan coupling'' specifies which metric enters the matter action, ``physical'' and ``conformal'' specify the metric variables, and ``wave gauge'' specifies the coordinates.

\paragraph{Einstein gravity}

We consider a four-dimensional spacetime \((\Mcal, \giveng) \), with \(\Mcal\simeq [0,+ \infty) \times\RR^3\), endowed with a Lorentzian metric \(\giveng\) of signature \((-,+,+,+) \). We denote by \(\givennabla \), \(\givenRic= (\givenR_{\alpha \beta}) \), and \(\givenScal\) its Levi-Civita connection, Ricci tensor, and scalar curvature. Greek indices range over \(0,1,2,3\), are raised and lowered with \(\giveng\), and repeated indices are summed.

The Einstein--Hilbert action of Einstein's theory of gravity is then
\be
\label{equa:action1}
\int_\Mcal \Big( \givenScal + 16 \pi \, L[\Psi, \giveng] \Big) \, d\givenVol,
\ee
where \(d\givenVol\) is the volume form of \((\Mcal, \giveng) \). Its critical points satisfy
\be
\label{equa-1:einstein-massif}
\givenG_{\alpha \beta} = 8\pi \, \givenT_{\alpha \beta}, \qquad \givenG_{\alpha \beta} \coloneqq \givenR_{\alpha \beta} - \frac{1}{2} \givenScal\, \giveng_{\alpha \beta}.
\ee
Here \(L[\Psi, \giveng] \) is the matter Lagrangian, possibly depending on a collection of fields \(\Psi\); the corresponding stress-energy tensor is
\be
\givenT_{\alpha \beta} \coloneqq
 - 2 \, \frac{\delta L}{\delta \giveng^{\alpha \beta}}[\Psi, \giveng] + \giveng_{\alpha \beta} \, L[\Psi, \giveng].
\ee
The twice-contracted Bianchi identities $\givennabla^\beta \givenG_{\alpha \beta} = 0$ yield the divergence equations 
\be
\label{equa-evolution-matter}
\givennabla^\beta \givenT_{\alpha \beta} = 0.
\ee

\paragraph{Modified gravity}

We replace the Einstein--Hilbert action by
\be
\label{equa:action}
\int_\Mcal \Big( f(\givenScal) + 16 \pi \, L[\Psi, \giveng] \Big) \, d\givenVol,
\ee
where \(f:\RR\to\RR\) is sufficiently regular and satisfies the small-curvature normalization \(f(r)=r+\Ocal(r^2) \). The critical points of \eqref{equa:action} satisfy
\be
\label{equa--FieldEquations}
\givenCurv_{\alpha \beta} = 8 \pi \, \givenT_{\alpha \beta}[\Psi, \giveng],
\ee
where the modified curvature tensor is defined by
\be
\label{equa--ModifiedCurv}
\givenCurv_{\alpha \beta} \coloneqq f'(\givenScal) \, \givenG_{\alpha \beta} - \frac{1}{2} \big( f(\givenScal) - \givenScal f'(\givenScal) \big) \giveng_{\alpha \beta} + \big( \giveng_{\alpha \beta} \, \givenBox - \givennabla_\alpha \givennabla_\beta \big) f'(\givenScal).
\ee
Here \(\givenBox \coloneqq \givennabla_\alpha \givennabla^\alpha \)\footnote{The reduced operator is denoted by \(\BoxChapeau_g\) (Section~\ref{section-4-1}) to distinguish it from the geometric operator \(\givenBox\). It corresponds to \(\widetilde\Box_g\) in~\cite[Section~4.1]{LeFloch-Ma-2017}.}, and \(\givenG\) is defined in \eqref{equa-1:einstein-massif}. Throughout this section, the covariant derivative \(\givennabla \) and the geometric wave operator \(\givenBox\) are formed with the physical Jordan metric \(\giveng\); after the conformal reduction, the untilded symbols \(\nabla \) and \(\Box_g\) refer instead to the conformal Einstein metric. Unlike \(\givenG\), which involves second derivatives of the metric, \(\givenCurv\) involves derivatives up to fourth order. A direct computation based on the twice-contracted Bianchi identities gives~\cite[Eq.~(1.8)]{LeFloch-Ma-2017}
\be
\label{equa:1112}
\givennabla^\alpha \givenCurv_{\alpha \beta} = 0.
\ee
Hence the matter fields satisfy the same divergence laws \eqref{equa-evolution-matter}. Thus the modified tensor \(\givenCurv\) plays, for the variational \(f(R) \) theory, the same divergence-free role as the Einstein tensor in general relativity, even though it contains higher derivatives of the metric.

\paragraph{Evolution of the scalar curvature}

Since \(\givenCurv\) contains the Hessian of \(f'(\givenR) \), it is natural to isolate an evolution equation for the scalar curvature. Taking the trace of \eqref{equa--ModifiedCurv} gives
\be
\label{equa--trace-gravity}
\Tr_{\giveng}(\givenCurv) = 3 \, \givenBox f'(\givenR) - 2 f(\givenR) + \givenR f'(\givenR).
\ee
Taking the trace of the field equations \eqref{equa--FieldEquations}, we obtain
\be
\label{Eq1-15}
3 \, \givenBox f'(\givenR) - 2 f(\givenR) + \givenR f'(\givenR) = 8 \pi \, \giveng^{\alpha \beta} \givenT_{\alpha \beta} \quad \text{ in } (\Mcal, \giveng).
\ee
On a curvature range where \(f''\) does not vanish, this is a Klein--Gordon equation for the scalar curvature \(\givenR\), and it plays a central role in the analysis.

\paragraph{Evolution of the matter field}

We consider scalar matter fields \(\psi:\Mcal\to\RR\) with energy-momentum tensor
\be
\label{equa:Talphabeta}
\givenT_{\alpha \beta} \coloneqq \givennabla_\alpha \psi \givennabla_\beta \psi - \Bigl( \frac{1}{2} \, \giveng^{\gamma \delta} \givennabla_\gamma \psi \givennabla_\delta \psi + V(\psi) \Bigr) \giveng_{\alpha \beta}.
\ee
Here \(V=V(\psi) \) is a prescribed potential. We assume \(V(0)=V'(0)=0\), so that the zero matter field is compatible with the Minkowski background. The evolution equation for the scalar matter field is
\be
\label{eq-KGG}
\givenBox \psi - V'(\psi) = 0 \quad \text{ in } (\Mcal, \giveng),
\ee
which is a nonlinear Klein--Gordon equation, or a wave equation when \(V\equiv0\). For simplicity, the main results below are stated for a massless field, corresponding to \(V\equiv0\). This evolution equation is consistent with \eqref{equa-evolution-matter}, since \(\givennabla^\beta \givenT_{\alpha \beta} = (\givenBox\psi-V'(\psi)) \givennabla_\alpha \psi\). The trace equation introduces the additional scalar degree of freedom into the Cauchy problem. After conformal rescaling, it becomes a Klein--Gordon equation whose effective mass and self-interaction are determined by the chosen function \(f\).


\subsection{Lagrangians of modified gravity}
\label{section-2-2}

\paragraph{Local assumptions on the Lagrangian}
The conformal parametrization replaces the fourth-order system for the physical metric \(\giveng\) by a second-order hyperbolic system. This formulation also makes explicit the additional scalar degree of freedom carried by the scalar curvature. It is therefore the natural formulation both for the global stability argument and for the study of the singular limit toward general relativity.

We first introduce the conformal parametrization for a sufficiently regular function \(f:\RR\to\RR\) satisfying, near the zero-curvature state,
\be
\label{Equa--condi-coeffi}
f(r) = r + { \frac{\coeff}{2}} r^2 + \Ocal(r^3), \quad r \to0,
\ee
for some coefficient \(\coeff>0\). Equivalently, we have
\be
f(0)=0, \quad f'(0)=1, \quad f''(0)= \coeff>0.
\ee
The normalization \(f'(0)=1\) recovers the Einstein--Hilbert integrand at first order, while \(f''(0)>0\) gives a positive mass to the additional scalar mode. On a sufficiently small curvature interval, \(f'(r)>0\) and \(r \mapsto f'(r) \) is locally invertible. These are  the local facts needed for the conformal change of variable below. The uniform hypotheses on the one-parameter family used in the global estimates are stated at the end of this section.

\paragraph{Logarithmic curvature variable and self-interaction potential}

We choose an open interval \(I_f\) containing zero such that \(f'>0\) and the logarithmic map \(r\mapsto\coeff^{-1} \log f'(r) \) is a diffeomorphism onto an open interval \(J_f\) containing zero. Its inverse is the curvature parametrization
\be
\hfrak(z) \coloneqq\bigl(\coeff^{-1} \log f'\bigr)^{-1}(z), \qquad z\in J_f.
\ee
Thus \(z\) denotes the scalar argument of this parametrization, and
\be
f'(\hfrak(z))=e^{\coeff z}.
\ee
Associated with \(f\), we introduce the \emph{self-interaction potential}
\be
U(r) \coloneqq \frac{r f'(r)-f(r)}{\coeff\,(f'(r))^2}, \quad  r\in I_f.
\ee
After the change of variable \(r= \hfrak(z) \), we write
\be
\ffrak \coloneqq f\circ\hfrak, \qquad \Ufrak \coloneqq U\circ\hfrak,
\ee
and the function \(\Ufrak\) is referred to as the self-interaction potential in the conformal formulation.

\paragraph{The quadratic model}

For instance, the \emph{Starobinsky model} is
\be
f_\Sta(r)\coloneqq r+ \frac{\coeff}{2}r^2, \qquad \coeff>0.
\ee
From this formula, we obtain
\be
f_\Sta'(r)=1+ \coeff r, \quad U_\Sta(r)= \frac{r^2}{2(1+ \coeff r)^2}.
\ee
Consequently, we have
\bel{eq1-09-Mars-2023}
\hfrak_\Sta(z) = \frac{1}{\coeff} \big(e^{\coeff z}-1\big), \quad \Ufrak_\Sta(z) = \frac{1}{2\coeff^2 e^{2\coeff z}} \big(e^{\coeff z}-1\big)^2
\ee
and, in particular, 
\be
\hfrak_\Sta(z) = z+ \frac{\coeff}{2}z^2 +\Ocal(\coeff^2z^3), \quad \Ufrak_\Sta(z) = \frac{1}{2}z^2 +\Ocal(\coeff z^3) \quad \text{as } z\to0.
\ee
Thus the scalar curvature variable and the logarithmic variable agree at leading order, while the self-interaction potential is quadratic to leading order,  with coefficient \(1/2\), uniformly in the singular regime.


\subsection{Conformal representation of the Jordan coupling}
\label{section-2-3}

\paragraph{Conformal metric and effective field}

The logarithmic variable agrees with the physical scalar curvature to first order near Minkowski spacetime. For an admissible scaling family, their difference is \(\Ocal(\coeff\givenR^2) \) on a fixed bounded curvature range. The conformal transformation makes this variable an independent unknown of a second-order system.

\begin{definition}
The \textbf{conformal Einstein metric} $g$ associated with a solution \((\Mcal, \giveng) \) of the \(f(R) \) field equations \eqref{equa--FieldEquations} is defined by
\be
\label{equa:conformgdef}
g_{\alpha \beta} \coloneqq e^{\coeff\fieldphi} \, \giveng_{\alpha \beta}, \quad \fieldphi \coloneqq \hfrak^{-1}(\givenR) =\frac{1}{\coeff} \log \big(f'(\givenR) \big), \qquad e^{\coeff\fieldphi} =f'(\givenR).
\ee
The scalar field \(\fieldphi:\Mcal\to\RR\), expressed in terms of the scalar curvature of the physical metric \(\giveng\), is called the \textbf{effective field}.
\end{definition}

With \(R_{\alpha\beta}\), \(\nabla\), and \(\Box_g\) denoting the Ricci tensor, connection, and wave operator of \(g\), the conformal identities proved in Section~\ref{section-7} give
\be
\label{eq2-10-mars-2023-a}
\givenCurv_{\alpha \beta} -{1\over2} \Tr_{\giveng}(\givenCurv) \, \giveng_{\alpha \beta} = e^{\coeff\fieldphi}R_{\alpha \beta} -{3\over2} \coeff^2 e^{\coeff\fieldphi} \nabla_\alpha \fieldphi\nabla_\beta \fieldphi -{\coeff\over2}e^{\coeff\fieldphi} \Ufrak(\fieldphi) \,g_{\alpha \beta},
\ee
while its trace is
\be
\label{eq2-10-mars-2023}
\Tr_{\giveng}(\givenCurv) = 3\coeff\,e^{2\coeff\fieldphi} \Box_g\fieldphi -\ffrak(\fieldphi) + \coeff e^{2\coeff\fieldphi} \Ufrak(\fieldphi) = e^{2\coeff\fieldphi} \Bigl( 3\coeff\, \Box_g\fieldphi -\Ufrak'(\fieldphi) \Bigr).
\ee
The traces satisfy \(\Tr_g(\givenCurv)=e^{-\coeff\fieldphi}\Tr_{\giveng}(\givenCurv)\). Thus these identities agree with \cite[Eqs.~(2.18), (2.19), and (2.26)]{LeFloch-Ma-2024b}, where the right-hand side of the effective-field equation is expressed using \(\Tr_g\).

The effective stress-energy tensor of the  effective field is
\be
\label{eq3-26-mars-2023}
T^{\mathrm{eff}}[\fieldphi,g]_{\alpha \beta} \coloneqq {3\over2} \coeff^2\nabla_\alpha \fieldphi\nabla_\beta \fieldphi -{3\over4} \coeff^2 g(\nabla \fieldphi, \nabla \fieldphi)g_{\alpha \beta} - {\coeff\over2} \Ufrak(\fieldphi) \,g_{\alpha \beta}.
\ee
Consequently, the modified gravity equations \eqref{equa--FieldEquations} and the trace-reversed identity \eqref{eq2-10-mars-2023-a} are equivalent to
\be
\label{eq4-26-mars-2023}
G_{\alpha \beta} = 8\pi e^{-\coeff\fieldphi} \givenT_{\alpha \beta}[\Psi, \giveng] + T^{\mathrm{eff}}[\fieldphi,g]_{\alpha \beta} \quad \text{in }(\Mcal,g),
\ee
where \(G_{\alpha \beta} \) is the Einstein tensor of \(g\). Thus, after the introduction of the  effective field \(\fieldphi\), the \(f(R) \) equations take the form of an Einstein system coupled to a Klein--Gordon-type scalar field and conformally transformed matter sources.

\paragraph{Evolution of the effective field}

From \eqref{eq2-10-mars-2023}, the  effective field satisfies
\be
\label{eq5-26-mars-2023}
3 \coeff \, \Box_g \fieldphi - \Ufrak'(\fieldphi) = e^{-2\coeff\fieldphi} \Tr_{\giveng}(\givenCurv).
\ee
Combining this identity with \(\Tr_{\giveng}(\givenCurv)=8\pi\, \Tr_\giveng(\givenT) \), we obtain
\be
\label{equa--effective-phi-equation}
3 \coeff \, \Box_g \fieldphi - \Ufrak'(\fieldphi) = 8\pi e^{-2\coeff\fieldphi} \Tr_\giveng(\givenT).
\ee
This is the Klein--Gordon-type equation satisfied by the  effective field.

\paragraph{Matter evolution in the conformal metric}

The matter equation \eqref{eq-KGG} satisfied by the scalar field \(\psi\) can likewise be rewritten in terms of the conformal metric:
\be
\label{equa-matter-conform}
\Box_g \psi - e^{-\coeff \fieldphi} \, V'(\psi) = \coeff \, g(\nabla \fieldphi, \nabla \psi).
\ee
In particular, for a massless scalar field, \(V\equiv0\), the equation reduces to
\be
\Box_g \psi = \coeff \, g(\nabla \fieldphi, \nabla \psi).
\ee

\paragraph{Summary of the conformal reduction}

The conformal Einstein equation and the effective-field equation imply the relation between \(\fieldphi\) and the scalar curvature of \(\giveng=e^{-\coeff\fieldphi}g\). Proposition~\ref{pro:40S} derives this relation directly. The wave-gauge constraint is propagated when a development is constructed from the reduced equations.

\begin{definition}
The \textbf{conformal formulation of \(f(R) \) gravity with self-gravitating matter} consists of the following equations imposed on a spacetime $(\Mcal,g)$.
\bei

\item The conformal Einstein metric \(g\) satisfies
\be
\label{eq4-26-mars-2023-rep}
G_{\alpha \beta} = 8\pi e^{-\coeff\fieldphi} \givenT_{\alpha \beta}[\Psi, \giveng] + T^{\mathrm{eff}}[\fieldphi,g]_{\alpha \beta}.
\ee

\item The effective field \(\fieldphi\) satisfies the nonlinear Klein--Gordon equation
\be
\label{equa--effective-phi-equation-rep}
3 \coeff \, \Box_g \fieldphi - \Ufrak'(\fieldphi) = 8\pi e^{-2\coeff\fieldphi} \Tr_\giveng(\givenT).
\ee

\item The massive or massless matter field \(\psi\) satisfies the nonlinear Klein--Gordon or wave equation
\be
\label{equa-matter-conform-rep}
\Box_g \psi - e^{-\coeff \fieldphi} \, V'(\psi) = \coeff \, g(\nabla \fieldphi, \nabla \psi),
\ee

\eei
\end{definition}

We regard \(\fieldphi\) as an independent unknown in this second-order system. The physical curvature is recovered from the conformal equations, as in \cite[Proposition~2.2]{LeFloch-Ma-2024b}.

\begin{proposition}[Recovery of the physical curvature]
\label{pro:40S}
Let \((\Mcal,g, \fieldphi, \psi) \) be a sufficiently regular solution of the conformal Einstein, effective-field, and matter equations \eqref{eq4-26-mars-2023-rep}--\eqref{equa-matter-conform-rep}, with \(\fieldphi\) in the domain of \(\hfrak\). The scalar curvature \(\givenR\) of \(\giveng=e^{-\coeff\fieldphi}g\) satisfies
\be
\label{equa:rhocondi}
\givenR= \hfrak(\fieldphi), \qquad f'(\givenR)=e^{\coeff\fieldphi}.
\ee
Consequently, \((\giveng, \psi) \) solves the original Jordan-coupled \(f(R) \) equations. For a solution constructed from the reduced wave-gauge equations, the Hamiltonian, momentum, and initial wave-gauge constraints imply the same conclusion.
\end{proposition}

The second conclusion combines propagation of the wave gauge with an algebraic recovery of the physical scalar curvature. These are the two steps in \cite[Proposition~5.4.1 and Lemmas~5.4.4--5.4.5]{LeFloch-Ma-2017}.

\begin{proof}
\bse
Taking the trace of the conformal Einstein equation gives
\be
\label{equa:physical-curvature-conformal-trace}
R_g=-8\pi e^{-2\coeff\fieldphi} \Tr_{\giveng}(\givenT) + \frac32\coeff^2g(\nabla \fieldphi, \nabla \fieldphi) +2\coeff\Ufrak(\fieldphi).
\ee
The scalar-curvature transformation and the effective-field equation therefore yield
\be
\label{equa:physical-curvature-recovery}
\givenR=e^{\coeff\fieldphi} \Bigl(R_g+3\coeff \, \Box_g\fieldphi-\frac32\coeff^2g(\nabla \fieldphi, \nabla \fieldphi) \Bigr)=e^{\coeff\fieldphi} \bigl(\Ufrak'(\fieldphi) +2\coeff\Ufrak(\fieldphi) \bigr)= \hfrak(\fieldphi).
\ee
The last equality follows by differentiating \(\Ufrak(\fieldphi)= (\hfrak(\fieldphi)e^{\coeff\fieldphi}-f(\hfrak(\fieldphi)))/(\coeff e^{2\coeff\fieldphi}) \) and using \(f'(\hfrak(\fieldphi))=e^{\coeff\fieldphi} \). Substitution into \eqref{eq2-10-mars-2023-a} gives the physical field equations.

For a reduced development, the Bianchi identity and the two scalar equations give a homogeneous subsidiary wave system for the contracted Christoffel symbols. The Hamiltonian and momentum constraints, together with the initial wave-gauge condition, give zero Cauchy data for this system. Uniqueness on each finite globally hyperbolic slab implies the wave gauge, as proved in \cite[Lemmas~5.4.4--5.4.5]{LeFloch-Ma-2017}. The preceding trace calculation then applies.
\ese
\end{proof}

The Jordan equations are fourth order in \(\giveng\), whereas the conformal equations are second order in \((g, \fieldphi, \psi) \). The additional pair of scalar Cauchy data represents the physical scalar curvature and its normal derivative, through \eqref{equa:rhocondi}.

\paragraph{Normalization of the potential}
\label{par:potential-normalizations}

The potential \(\Ufrak\) has the normalization used in \cite[Eqs.~(2.14), (2.18), and (2.26)]{LeFloch-Ma-2024b}. In~\cite{LeFloch-Ma-2017}, the variables are denoted by
\bse
\be
\label{eq:smf-notation-dictionary}
\rho_{\mathrm{SMF}} \coloneqq \frac{\coeff}{2} \fieldphi, \qquad W_{2, \mathrm{SMF}}(\rho_{\mathrm{SMF}}) \coloneqq -\coeff e^{\coeff\fieldphi} \Ufrak(\fieldphi).
\ee
The kinetic and potential terms therefore satisfy
\be
\label{eq:smf-kinetic-potential-dictionary}
6\nabla_\alpha \rho_{\mathrm{SMF}} \nabla_\beta \rho_{\mathrm{SMF}} = \frac32\coeff^2\nabla_\alpha \fieldphi\nabla_\beta \fieldphi, \qquad \frac12W_{2, \mathrm{SMF}}(\rho_{\mathrm{SMF}})g_{\alpha \beta} =-\frac{\coeff}{2}e^{\coeff\fieldphi} \Ufrak(\fieldphi)g_{\alpha \beta}.
\ee
\ese


\subsection{The general-relativistic limit}
\label{section-2-4}

\paragraph{Limiting regime}

For \(f(r)=r\), the field equations reduce to Einstein's equations. The admissible families defined below satisfy \(f_\coeff(r)\to r\) on every fixed bounded curvature interval as \(\coeff\to0\). At the level of the evolution equations, however, the limit is singular: the fourth-order Jordan equations become second order, or, in the conformal formulation, the mass of the effective field tends to infinity.

For small, well-prepared data, Theorem~\ref{theorem:final-global-stability-and-limit} compares the global \(f(R)\) development with the Einstein development of the limiting data. Sections~\ref{section-16-2} and~\ref{section:final-global-uniform-convergence} prove local energy convergence and uniform weighted convergence of the fields and their Cartesian derivatives on the future spacetime, including the initial hypersurface. The limiting system is the vacuum Einstein system when matter vanishes.

\paragraph{The infinite-mass transition}

We indicate the origin of the infinite-mass phenomenon. From \eqref{Equa--condi-coeffi}, we have, for small curvature,
\be
 -2f(\givenR) + \givenR f'(\givenR)=-\givenR+\Ocal(\givenR^3), \qquad \givenR\to0.
\ee
Consequently, the trace equation \eqref{Eq1-15} has, at leading order, the form
\be
3\coeff\, \givenBox \givenR-\givenR \simeq 8\pi\, \Tr_\giveng(\givenT).
\ee
Dividing this relation by \(3\coeff\) gives
\be
\label{equa:130}
\givenBox \givenR - {1\over 3\coeff} \givenR \simeq {8\pi\over 3\coeff} \Tr_\giveng(\givenT).
\ee
Thus the scalar curvature behaves, to leading order, as a Klein--Gordon field whose mass squared is of order \(1/\coeff\). The passage from \(f(R) \) gravity to Einstein gravity is therefore an infinite-mass limit for the effective field.

This observation is one of the main analytical features of the problem. Uniform estimates as \(\coeff\to0\) cannot be obtained by treating the effective field as a standard perturbative scalar field with fixed mass. Instead, the analysis keeps the three scales introduced above: fixed-parameter estimates with explicit singular dependence on \(\coeff\), rescaled estimates uniform in \(\coeff\), and vanishing estimates for the error in the Einstein trace relation.

In the limit, the effective-field equation becomes the Einstein trace relation \(\givenR=-8\pi\Tr_g(\givenT) \). The effective field is thereby determined by the limiting metric and matter field, rather than by an independent evolution equation.

\paragraph{Matter coupling and metric representation}

We finally distinguish two natural ways of coupling matter to the gravitational variables. In the Jordan coupling, the physical metric is \(\giveng\), and the matter fields are minimally coupled to \(\giveng\). This is the formulation used in the \(f(R) \) theory considered in this memoir:
\be
\label{eq1-26-mars-2023}
\givenCurv_{\alpha \beta} = 8\pi\, \givenT_{\alpha \beta}[\Psi, \giveng].
\ee
By contrast, in a distinct Einstein coupling, matter would be minimally coupled directly to the conformal metric \(g\). Its Einstein equation would instead take the form
\be
G_{\alpha \beta} = 8\pi\,T_{\alpha \beta}[\Psi,g] + T^{\mathrm{eff}}[\fieldphi,g]_{\alpha \beta}.
\ee
In particular, the Einstein-coupled model is not obtained by replacing the source in the Jordan equation \eqref{eq1-26-mars-2023}; it has a different matter equation and a different coupling to the  effective field.

This distinction is relevant both for the Cauchy problem and for the limiting process. The Jordan formulation is the natural physical formulation for the matter models considered here: the energy-momentum tensor is conserved with respect to the physical metric \(\giveng\), and the compatibility condition relating the  effective field to the scalar curvature of \(\giveng\) is propagated by the evolution. This is the setting in which the local Cauchy theory was developed in~\cite{LeFloch-Ma-2017}, and it is the setting adopted throughout this memoir.

\paragraph{Admissible families and hypotheses for the analysis}

We consider a family of Lagrangians \(f_\coeff\), with \(\coeff>0\), normalized by
\be
\coeff=f_\coeff''(0).
\ee
Thus \(\coeff\) measures the strength of the quadratic correction. A convenient model class is obtained from a fixed function \(f_*\), independent of \(\coeff\). To keep the curvature variable and its dimensionless rescaling distinct, we set \(y \coloneqq \coeff r\).  The variable \(r\) is the scalar-curvature argument of \(f_\coeff\), whereas \(y\) is the argument of the fixed profile \(f_*\).  We then set
\be
f_\coeff(r)\coloneqq{1\over \coeff}f_*(y),
\ee
where \(f_*\) satisfies
\be
f_*(0)=0, \quad f_*'(0)=1, \quad f_*''(0)=1.
\ee
Taylor expansion then gives
\be
f_\coeff(r)=r+ {\coeff\over2}r^2 + \coeff^2r^3\rho_*(y), \quad y= \coeff r\to0,
\ee
The remainder \(\rho_*\) is bounded near zero. The estimates apply to any fixed profile satisfying the following regularity and invertibility assumptions; their constants may depend on that profile, but not on \(\coeff\).

\begin{definition}
\label{def:admissible-f-family}
Let \(M\geq3\), and let \(r_f,s_f>0\).  A family is called an \textbf{admissible scaling family of modified Einstein--Hilbert integrands of order \(M\)} if there exists \(f_*\in C^M([-r_f,r_f]) \), independent of \(\coeff\), such that
\be
f_*(0)=0, \qquad f_*'(0)=1, \qquad f_*''(0)=1, \qquad f_*'>0,
\ee
the map \(y\mapsto\log f_*'(y) \) is a \(C^{M-1} \) diffeomorphism onto an interval containing \([-s_f,s_f] \), and the scaled family is defined by
\be
\label{Equa--condi-coeffi-ineq}
f_\coeff(r) \coloneqq {1\over\coeff}f_*(y), \qquad y= \coeff r, \qquad |y| \leq r_f .
\ee
\end{definition}

With \(\hfrak_*\coloneqq(\log f_*')^{-1}\), the curvature parametrization satisfies
\be
\label{equa:scaled-curvature-reconstruction}
\hfrak_\coeff(z)={1\over\coeff}\hfrak_*(\coeff z),\qquad |\coeff z|\leq s_f.
\ee
Thus a uniform bound for \(\coeff\fieldphi\) keeps the reconstructed Jordan curvature inside the domain of \(f_\coeff\), even though \(\fieldphi\) and the scalar curvature may be of order \(\coeff^{-1} \). We write simply \(f\) when a single modified theory is considered and \(f_\coeff\) when the singular parameter is relevant.  For a fixed \(\coeff>0\), the local conformal reduction uses only the corresponding smoothness and positivity assumptions.  Our parameter-uniform estimates use Definition~\ref{def:admissible-f-family}. The proof distinguishes fixed-parameter estimates, estimates for the rescaled effective field, and estimates tending to zero with \(\coeff\). The rescaled variables are introduced in Section~\ref{section-3-3}. 


\section{Cauchy data and Euclidean--hyperboloidal foliation}
\label{section-3}


\subsection{Geometric formulation of the initial value problem}
\label{section-3-1}

\paragraph{The \(f(R) \)-initial constraints}
We formulate the conformal \(f(R)\) Cauchy problem and define the Euclidean--hyperboloidal foliation, the reference metric, and the perturbation variables. Section~\ref{section-3-analysis} introduces the differential operators and energy functionals used in the global estimates. Table~\ref{table:notation-proof} summarizes the geometric notation. We distinguish throughout this section the two equivalent, but analytically different, formulations.
\bei
\item In the \emph{physical formulation}, the unknown metric is the Jordan metric \(\giveng\), matter is minimally coupled to \(\giveng\), and the metric equation is fourth order.
\item In the \emph{conformal formulation}, the unknowns are \((g, \fieldphi, \psi) \), with \(g=e^{\coeff\fieldphi} \giveng\); the equations form a second-order wave--Klein--Gordon system.  This is the formulation used for the estimates and the evolution argument.
\eei
Proposition~\ref{pro:40S} recovers the physical scalar curvature from every sufficiently regular conformal development. Since the effective field \(\fieldphi\) satisfies a second-order evolution equation, the initial value problem requires two additional data, namely the restriction of \(\fieldphi\) to the initial hypersurface and its normal derivative. Equivalently, we prescribe the spacetime scalar curvature, expressed through the compatibility condition, together with its normal derivative. We also allow for a scalar matter field \(\psi\), governed by a potential \(V=V(\psi) \). We refer to~\cite{ChoquetBruhat-2009} for the standard geometric formulation of the Einstein constraint equations.

The constraint equations below are obtained by projecting the conformal Einstein system onto the initial hypersurface. Equivalently, they are the Gauss--Codazzi relations for the conformal metric \(g\), with the effective field treated as an additional matter component.

\begin{definition}
\label{def36-22}
An \textbf{initial data set for the conformal system} consists of
\be
\mathbf d \coloneqq (\Sigma,g_0,k_0, \fieldphi_0, \fieldphi_1, \psi_0, \psi_1),
\ee
where \(\Sigma \) is a three-dimensional manifold, \(g_0\) is a Riemannian metric, \(k_0\) is a symmetric \(2\)-covariant tensor, and \((\fieldphi_0, \fieldphi_1) \), \((\psi_0, \psi_1) \) are the Cauchy data for the effective and matter fields. The second fundamental form has the convention \(k=-\tfrac12\mathcal L_n g|_{T\Sigma \times T\Sigma} \), where \(n\) is the future-directed conformal unit normal. These data satisfy
\be
\label{consttransHan}
R_{g_0} + \big(\Tr_{g_0}k_0\big)^2-|k_0|_{g_0}^2= \mu_0^{\mathrm{eff}} + \mu_0^{\mathrm{mat}}, \qquad \dive_{g_0}k_0-d\big(\Tr_{g_0}k_0\big)=J_0^{\mathrm{eff}} +J_0^{\mathrm{mat}},
\ee
where the effective-field densities are defined by
\be
\label{equa-initial-effective-field-densities}
\mu_0^{\mathrm{eff}} \coloneqq {3\over2} \coeff^2\Bigl(| \fieldphi_1|^2 + | \nabla_{g_0} \fieldphi_0|_{g_0}^2 + { {2\over3\coeff}} \Ufrak(\fieldphi_0) \Bigr), \qquad J_0^{\mathrm{eff}} \coloneqq -{3\over2} \coeff^2\fieldphi_1\, d\fieldphi_0,
\ee
and the matter densities are
\be
\label{equa-initial-matter-densities}
\mu_0^{\mathrm{mat}} \coloneqq 8\pi e^{-\coeff\fieldphi_0} \Bigl(| \psi_1|^2 + | \nabla_{g_0} \psi_0|_{g_0}^2 + { 2e^{-\coeff\fieldphi_0}V(\psi_0)} \Bigr), \qquad J_0^{\mathrm{mat}} \coloneqq -8\pi e^{-\coeff\fieldphi_0} \psi_1\, d\psi_0.
\ee
Here \(R_{g_0} \), \(\Tr_{g_0} \), and \(\dive_{g_0} \) are computed with respect to \(g_0\). The momentum signs correspond to this convention for \(k\), with \(\fieldphi_1=n\fieldphi\) and \(\psi_1=n\psi\). The Hamiltonian densities are the projections \(2T^{\mathrm{eff}}[\fieldphi,g](n,n) \) and \(16\pi e^{-\coeff\fieldphi_0} \givenT(n,n) \), respectively; this accounts for the factors of \(\coeff\) and of \(e^{-\coeff\fieldphi_0} \) in the two potential terms.
\end{definition}

\paragraph{Reconstruction of the physical Cauchy data}
\label{definition:physical-fR-compatible-data}
The physical metric \(\giveng=e^{-\coeff\fieldphi}g\) has future unit normal \(\widetilde n=e^{\coeff\fieldphi/2}n\). With the same sign convention for its second fundamental form, its initial geometry is
\be
\label{equa:physical-initial-geometry}
\widetilde g_0=e^{-\coeff\fieldphi_0}g_0, \qquad \widetilde k_0=e^{-\coeff\fieldphi_0/2} \Bigl(k_0 + \frac{\coeff}{2} \fieldphi_1g_0\Bigr).
\ee
We denote the physical scalar curvature on \(\Sigma \) by \(\mathcal R_J[\mathbf d] \), and its derivative along the \emph{conformal} unit normal by \(\mathcal R_{J,1}[\mathbf d] \). Proposition~\ref{pro:40S} gives
\be
\label{equa-initial-fR-compatibility}
\mathcal R_J[\mathbf d]= \hfrak_\coeff(\fieldphi_0), \qquad \mathcal R_{J,1}[\mathbf d]= \hfrak_\coeff'(\fieldphi_0) \fieldphi_1.
\ee
These identities determine the curvature data from the conformal scalar data. Derivatives along the physical normal acquire the factor \(e^{\coeff\fieldphi_0/2} \):
\be
\label{equa:physical-initial-normal-data}
\widetilde n\givenR|_\Sigma=e^{\coeff\fieldphi_0/2} \mathcal R_{J,1}[\mathbf d], \qquad \widetilde n\psi|_\Sigma=e^{\coeff\fieldphi_0/2} \psi_1.
\ee
Thus the conformal Hamiltonian and momentum constraints, together with a choice of initial wave coordinates, give the physical \(f(R) \) Cauchy data. The value and normal-derivative identities in \eqref{equa-initial-fR-compatibility} follow from the conformal equations.

\paragraph{Cauchy developments of $f(R)$ gravity}

\begin{definition}
\label{definition:conformal-Cauchy-data}
For an initial data set as in Definition~\ref{def36-22}, the \textbf{initial value problem for conformal \(f(R) \)-gravity} consists of finding a time-oriented Lorentzian manifold \((\Mcal,g) \), together with scalar fields \(\fieldphi, \psi\) on \(\Mcal\), such that the following conditions hold.
\bei

\item The conformal \(f(R) \)-gravity system \eqref{eq4-26-mars-2023}, \eqref{equa--effective-phi-equation}, and \eqref{equa-matter-conform} is satisfied.

\item There exists an embedding \(i:\Sigma \hookrightarrow\Mcal\) whose image \(\Sigma_0=i(\Sigma) \) is a spacelike Cauchy hypersurface, such that
\be
i^\star g=g_0,
\ee
and whose second fundamental form is \(k_0\).

\item The scalar fields induce the prescribed Cauchy data:
\be
\fieldphi|_{\Sigma_0} = \fieldphi_0, \quad \mathcal L_n\fieldphi|_{\Sigma_0} = \fieldphi_1, \quad \psi|_{\Sigma_0} = \psi_0, \quad \mathcal L_n\psi|_{\Sigma_0} = \psi_1,
\ee
where \(n\) denotes the future-oriented unit normal to \(\Sigma_0\) in \((\Mcal,g) \).

\eei
\end{definition}

A solution satisfying Definition~\ref{definition:conformal-Cauchy-data} is a \textbf{conformal Cauchy development}. Proposition~\ref{pro:40S} and the reconstruction above give the corresponding \textbf{modified-gravity Cauchy development} in the physical metric. The notion of maximal globally hyperbolic development is the standard one, adapted from the Einstein theory; see, for instance,~\cite{ChoquetBruhat-2009}. Compared with the vacuum Einstein equations, the conformal \(f(R) \) system contains the two additional Cauchy data \((\fieldphi_0, \fieldphi_1) \), associated with the effective field. These data cannot be prescribed freely: they enter the Hamiltonian and momentum constraints through \((\mu_0^{\mathrm{eff}},J_0^{\mathrm{eff}}) \), as in \eqref{consttransHan}.

Two simple observations will be useful.
\bei

\item If \(V(0)=V'(0)=0\), as in the massless case \(V\equiv0\), and \(\fieldphi_0= \fieldphi_1= \psi_0= \psi_1\equiv0\), the evolution reduces to the vacuum Einstein initial value problem.

\item If the matter field vanishes but \((\fieldphi_0,\fieldphi_1)\) does not, the effective field still contributes to the conformal Einstein equation. Such developments need not be vacuum Einstein developments.

\eei

The trivial solution has Minkowski conformal metric and vanishing scalar fields. The stability problem is therefore formulated as a small-data problem around this state, whereas the general-relativistic limit requires the smallness assumptions and estimates to be organized uniformly with respect to the parameter measuring the deviation from \(f(r)=r \).


\subsection{The Euclidean--hyperboloidal foliation}
\label{section-3-2}

\paragraph{Wave and massive regimes}
The global argument is carried out with the Euclidean--hyperboloidal foliation introduced for coupled wave and Klein--Gordon systems. Its purpose is to combine two geometries which are separately adapted to different parts of the evolution. In the interior of the future light cone, the hypersurfaces are hyperboloidal and therefore reflect the stronger decay of massive fields in timelike directions. Near null and spacelike infinity, the hypersurfaces become Euclidean and preserve the weighted control naturally associated with asymptotically Euclidean initial data. A merging domain connects these two parts smoothly.

This foliation supports both the wave estimates for the metric and matter field and the Klein--Gordon estimates for an effective field of mass \((3\coeff)^{-1/2}\).

In the wave coordinates \((t,x)\), we write \(r\coloneqq|x|\). The interior cone is \(\Kcal\coloneqq\{t-r>1\}\), and its boundary \(\Lscr\coloneqq\{t-r=1\}\) is the interior--exterior interface.

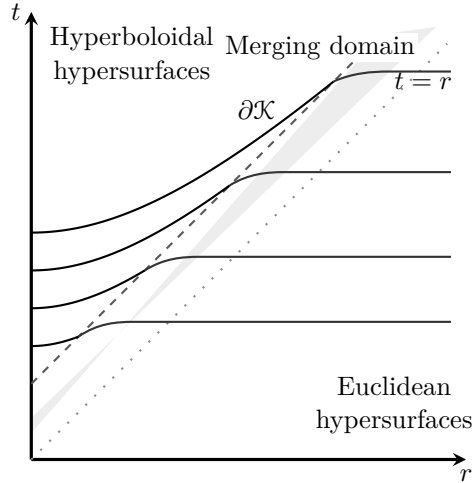
\begin{figure}[htbp]
\centering
\begin{tikzpicture}[
  x=1cm,y=1cm,
  font=\normalsize,
  every node/.style={text=black},
  axis/.style={very thick,->,>=stealth},
  hyperleaf/.style={draw=black,thick},
  euclleaf/.style={draw=black!80,thick},
  boundary/.style={draw=black!65,dashed,thick},
  lightcone/.style={draw=black!45,loosely dotted,thick},
  transition/.style={fill=black!7,draw=none}
]
  \path[transition] (0,.35) -- (4.62,5.62) -- (5.35,5.72) -- (0,.55) -- cycle;
  \draw[axis] (0,0) -- (5.75,0) node[below] {$r$};
  \draw[axis] (0,0) -- (0,5.92) node[left] {$t$};
  \draw[lightcone] (0,0) -- (5.50,5.50);
  \draw[boundary] (0,1) -- (4.62,5.62);
  \node[above left] at (3.35,4.38) {$\partial\Kcal$};
  \node[above right] at (4.68,4.78) {$t=r$};
  \foreach \ss/\xx/\yy in
    {1.50/.625/1.82,2.00/1.50/2.68,
     2.50/2.625/3.80,3.00/4.00/5.13} {
    \draw[hyperleaf] plot[domain=0:\xx,samples=80]
      (\x,{sqrt(\x*\x+ \ss*\ss)});
    \draw[euclleaf] (\xx,{sqrt(\xx*\xx+ \ss*\ss)})
.. controls ({\xx+.25},{\yy-.02}) and ({\xx+.50}, \yy)
.. ({\xx+.72}, \yy) -- (5.55, \yy);
  }
  \node[align=center] at (1.33,5.34)
    {Hyperboloidal\\hypersurfaces};
  \node[align=center,fill=white,inner sep=1pt] at (3.82,5.45) {Merging domain};
  \node[align=center] at (4.80,.72)
    {Euclidean\\hypersurfaces};
\end{tikzpicture}
\caption{Euclidean--hyperboloidal foliation in the \((r,t) \)-half-plane.
The hyperboloidal hypersurfaces are smoothly deformed across the merging domain into Euclidean hypersurfaces.  The interior--exterior interface is \(\Lscr= \partial\Kcal= \{r=t-1\} \).}
\label{fig:euclidean-hyperboloidal-foliation}
\end{figure}

\paragraph{The hyperboloidal time function}

We fix \(s_0\geq2\) and use the global coordinates
\be
(t,x)=(t,x^1,x^2,x^3),\qquad r=|x|,
\ee
introduced by the conformal wave gauge. In the interior domain, the hyperboloidal time is
\be
\label{equa-hyperboloidal-time-section3}
s \coloneqq \sqrt{t^2-r^2}.
\ee
For every \(s>0\), the corresponding Minkowski hyperboloid is
\be
\label{equa-hyperboloid-section3}
\Hcal_s  \coloneqq  \big\{(t,x):t>0, \ t^2-r^2=s^2\big\}.
\ee
When no ambiguity is possible, we identify \(\Hcal_s\) with its spatial parametrization
\be
x\longmapsto \big(\sqrt{s^2 + |x|^2},x\big).
\ee

We specify the global time function used in the exterior analysis. We take a smooth nondecreasing function \(\chi_{\mathrm{fol}}:\mathbb R\to[0,1] \) which vanishes on \((-\infty,0] \) and is equal to one on \([1,+ \infty) \), and define the hyperboloidal and Euclidean radii by
\be
\label{equa:foliation-radii-section3}
\rhoH(s) \coloneqq {s^2-1\over2}, \qquad \rhoE(s) \coloneqq {s^2 +1\over2}.
\ee
The foliation coefficient and the Euclidean--hyperboloidal time function are defined by
\bse
\be
\label{equa:foliation-coefficient-section3}
\xi(s,r) \coloneqq 1-\chi_{\mathrm{fol}} \bigl(r-\rhoH(s) \bigr),
\ee
\be
\label{equa:global-foliation-time-section3}
\del_r\fTime(s,r)={r\xi(s,r) \over\sqrt{s^2 +r^2}}, \qquad \fTime(s,0)=s.
\ee
\ese
Thus \(\xi =1\) for \(r\leq\rhoH(s) \), \(\xi =0\) for \(r\geq\rhoE(s) \), and \(\fTime(s,r)= \sqrt{s^2 +r^2} \) throughout the hyperboloidal domain. The global hypersurface and the associated coordinate Jacobian are
\be
\label{equa:global-foliation-leaf-jacobian-section3}
\Mcal_s \coloneqq \bigl\{(\fTime(s,|x|),x):x\in\mathbb R^3\bigr\}, \qquad J(s,r) \coloneqq \del_s\fTime(s,r)>0.
\ee
\paragraph{Interior, merging, and exterior domains}

The spacetime domain covered by the foliation is decomposed into
\be
\Mint, \quad \Mtran, \quad \Mext.
\ee
On each hypersurface, these regions are specified by
\be
\label{equa:foliation-regions-section3}
\Mint_s= \{r\leq\rhoH(s) \}, \qquad \Mtran_s= \{\rhoH(s) \leq r\leq\rhoE(s) \}, \qquad \Mext_s= \{r\geq\rhoE(s) \}.
\ee
The domain \(\Mint\) is the hyperboloidal interior described above. In \(\Mtran\), the time function interpolates smoothly between the hyperboloidal time and an asymptotically Euclidean time. In \(\Mext\), its level sets are Euclidean near spatial infinity. We denote the full hypersurface of the Euclidean--hyperboloidal foliation at time \(s\) by
\be
\Mcal_s = \Mint_s\cup\Mtran_s\cup\Mext_s.
\ee
We write \(\MME\coloneqq\Mtran\cup\Mext\) for the Euclidean--merging region. The interface \(\Lscr\) is the outer boundary of \(\Mint\) and the inner boundary of \(\MME\). The two traces at the interface are restrictions of the same Cauchy development.

Finally, the energy coefficient used below is
\be
\label{equa:global-foliation-zeta-section3}
\zeta(s,r)^2  \coloneqq 1-{r^2\xi(s,r)^2\over s^2 +r^2} ={s^2 +(1-\xi(s,r)^2)r^2\over s^2 +r^2}.
\ee
The coefficient \(\zeta \) agrees with \(s/t\) on the hyperboloidal part and with one on the Euclidean part. In particular, \(0\leq\del_r\fTime<1\), so every \(\Mcal_s\) is spacelike for the Minkowski metric; its spacelike character for \(g\) follows from the geometric bounds used in the energy estimates. We also fix a smooth nondecreasing function \(\aleph:\mathbb R\to[0,+ \infty) \) such that \(\aleph(q)=0\) for \(q\leq-1\) and \(\aleph(q)=q+1\) for \(q\geq0\), and define the exterior energy weight by
\be
\label{equa:global-foliation-exterior-weight-section3}
\crochet(t,r) \coloneqq 1+ \aleph(r-t).
\ee
In particular, \(\crochet=1\) on the interface \(\Lscr= \{r-t=-1\} \), so the weighted exterior flux and the unweighted interior flux cancel exactly when the two balance laws are added with opposite orientations. The Jacobian satisfies the quantitative bounds
\bse
\be
\label{equa:global-foliation-jacobian-interior-merging-section3}
J={s\over t} \quad\hbox{in } \Mint_s, \qquad \xi{s\over\sqrt{s^2 +r^2}} +(1-\xi){3s\over5} \leq J\leq \xi{s\over\sqrt{s^2 +r^2}} +(1-\xi)2s \quad\hbox{in } \Mtran_s,
\ee
\be
\label{equa:global-foliation-jacobian-exterior-section3}
{3s\over5} \leq J\leq2s \quad\hbox{in } \Mext_s.
\ee
\ese
Moreover, in the Euclidean--merging region we have
\be
\label{equa:global-foliation-zeta-bound-section3}
{1+ |r-t| \over r} \lesssim\zeta^2\leq\zeta \leq1.
\ee
These definitions are the specialization of the Euclidean--hyperboloidal foliation used in our preceding stability analysis~\cite[Section~3.1]{LeFloch-Ma-2024a}.

\paragraph{Degeneration near the light cone and the interface}

On \(\Mint_s\), the ratio between hyperboloidal and Cartesian time satisfies
\be
\label{equa-s-over-t-section3}
{s\over t} = \sqrt{1-{r^2\over t^2}}
\ee
and measures the position relative to the outgoing light cone. It is comparable to \(1\) in timelike directions and tends to \(0\) when \(r/t\to1\).

The interior portion of the foliation between two levels \(s_0\leq s_1\) is denoted by
\be
\label{equa-interior-slab-section3}
\Mint_{[s_0,s_1]}  \coloneqq  \bigcup_{s\in[s_0,s_1]} \Mint_s,
\ee
where \(\Mint_s\) denotes the truncated part of \(\Hcal_s\) contained in the interior domain of the Euclidean--hyperboloidal foliation. We also write
\be
\Mint \coloneqq \bigcup_{s\geq s_0} \Mint_s.
\ee
The truncation is determined by the merging with the Euclidean hypersurfaces. The hyperboloidal interior and the Euclidean--merging side meet along the conical hypersurface
\be
\label{equa-wave-cone-section3}
\Lscr=\del\Kcal=\{(t,x):r=t-1\},
\ee
already introduced as the interior--exterior interface. Its section on a hypersurface and its portion between two hypersurfaces are denoted by
\be
\Lscr_s \coloneqq \Lscr\cap\Mcal_s, \qquad \Lscr_{[s_0,s_1]} \coloneqq \Lscr\cap \bigcup_{s_0\leq s\leq s_1} \Mcal_s.
\ee
Although \(\Lscr\) is an interface in the spacetime, it is a lateral boundary of each truncated interior or exterior spacetime slab.

\paragraph{Normal-ray parametrization}

The Minkowski-normal ray derivative is defined by
\be
\label{equa-normal-ray-field-section3}
\newperp  \coloneqq  \del_t+ {x^a \over t} \del_a
\ee
and is proportional to the Minkowski scaling vector field:
\be
t\, \newperp = t\del_t+x^a \del_a.
\ee
It is orthogonal to the hyperboloids \(\Hcal_s\) for the Minkowski metric. Its integral curves are radial straight lines.  In detail, the curve through \((t_1,x_1) \) is $t \mapsto \Bigl(t,{t\over t_1}x_1\Bigr)$,  and both \(r/t\) and \(s/t\) remain constant along it. These curves will be used in Section~\ref{section-6} to reduce the singular Klein--Gordon equation to an ordinary differential equation. Followed backwards, such a curve meets either the initial hyperboloid or the interface \(\Lscr\). The corresponding lower ray parameter is
\be
\label{equa:lower-ray-parameter-section3}
\lambda_{\mathrm{in}}(s,x) \coloneqq \max\left\{s_0,{s\over t-r} \right\}.
\ee
Interface values on \(\Lscr\) are not independent data: they are provided by the estimates in the merging and exterior domains.


\subsection{Reference metric and perturbation variables}
\label{section-3-3}

\paragraph{Merging the Minkowski and Schwarzschild metrics}

We prescribe a reference metric, independent of \(\coeff\), which agrees with Schwarzschild near spatial infinity and with Minkowski in the interior.

The reference geometry is defined from the asymptotically Schwarzschild construction in~\cite[Section~1.2]{LeFloch-Ma-2022}; its Schwarzschild perturbation has the \(r^{-1}\) decay considered in~\cite{LeFloch-Ma-2024a}. In wave coordinates, the Schwarzschild metric \(g_{\Sch} \) is written as
\be
g_{\Sch,00} = - {r-m\over r+m}, \quad g_{\Sch,0a} =0, \qquad g_{\Sch,ab} = {r+m\over r-m} \, \omega_a \omega_b+ {(r+m)^2\over r^2} \big(\delta_{ab}-\omega_a \omega_b\big),
\ee
where \(\omega_a \coloneqq x_a/r \). We merge the Minkowski and Schwarzschild metrics as follows.  We take \(\chi^\star= \chi^\star(r) \) to be a regular cut-off function which vanishes for \(r \leq1/2\) and is identically \(1\) for \(r \geq3/4\). We choose \(0\leq m<1/2\) and restrict to \(t\geq2\). The Schwarzschild expression is then smooth on the support of the cut-off. At \(m=0\), the reference metric is Minkowski. We define
\be
\label{equa-defineMS-new}
g^\star = \gMink + \chi^\star(r) \, \chi^\star \!\Bigl({r \over t-1} \Bigr) \big(g_{\Sch}-\gMink\big).
\ee
Thus \(g^\star\) coincides with \(\gMink\) in the region
\be
{r \over t-1}< {1\over2},
\ee
and with \(g_{\Sch} \) in the exterior region
\be
{r \over t-1} \geq {3\over4},
\ee
which contains the outgoing light cone. This reference metric is designed to capture the leading asymptotic Schwarzschild behavior while leaving the interior geometry close to Minkowski space.

\paragraph{Schwarzschild exterior and reference transition}

We denote the product of the two reference cut-offs and its transition set by
\be
\label{equa:reference-transition-definition-part-I}
\vartheta^\star\coloneqq\chi^\star(r)\chi^\star\bigl(r/(t-1)\bigr),\qquad \mathcal A^\star\coloneqq\operatorname{supp}\bigl(\vartheta^\star(1-\vartheta^\star)\bigr)\cup\operatorname{supp}(d\vartheta^\star).
\ee
The reference construction is that of~\cite[Section~1.2, Eqs.~(1.14)--(1.16)]{LeFloch-Ma-2022}. Its Schwarzschild part has the light-bending sign; the outgoing coefficient of the full solution will be estimated in Section~\ref{section:eight-outgoing-characteristics}. We fix the mass \(m\) sufficiently small, independently of \(\coeff\). On the Euclidean--merging side, \(r\geq t-1\) and \(r\geq\rhoH(s_0)\geq3/2\), so both cut-offs equal one. Consequently,
\be
\label{equa:reference-Schwarzschild-on-exterior}
g^\star=g_{\Sch} \quad\hbox{on } \MME, \qquad \mathcal A^\star\cap\MME= \varnothing.
\ee
The consequences of \eqref{equa:reference-Schwarzschild-on-exterior} are as follows.
\bei
\item \emph{Exterior characteristic estimates.} The Schwarzschild form of the reference metric in the exterior permits the use of its tortoise coordinate in the characteristic estimate for the massless matter field. The energies remain defined on the Euclidean--hyperboloidal hypersurfaces.
\item \emph{Uniformity in the parameter.} The reference metric and foliation are independent of \(\coeff\). The estimates must control effective fields with mass of order \(\coeff^{-1/2}\) on this fixed geometry.
\item \emph{Distinct transition regions.} The set \(\mathcal A^\star\), where the reference metric changes, is disjoint from the foliation merging domain \(\Mtran\).
\eei

On the initial hypersurface, we compare the metric and second fundamental form with those induced by \(g^\star\). For \(a,b=1,2,3\), we write
\be
\label{equa-decomp-data-3}
g_{0ab} = g^\star_{0ab} +u_{0ab} = \delta_{ab} +h^\star_{0ab} +u_{0ab}, \quad k_{0ab} = k^\star_{0ab} +l_{0ab}.
\ee
Here \(g^\star_0\) is the metric induced by \(g^\star\) on the initial hypersurface, and \(k^\star_0\) is its second fundamental form with the sign convention of Definition~\ref{def36-22}. The pair \((u_0,l_0) \) is the perturbation of these geometric data.

We use the following terminology. The tensor \(h^\star_0\) is the {\it initial reference perturbation}; it is controlled in weighted high-order pointwise norms. The pair \((u_0,l_0) \) is the {\it dynamical perturbation}; it is controlled in weighted high-order energy norms and is propagated by the evolution. In the Euclidean--hyperboloidal method of~\cite{LeFloch-Ma-2024a}, these two components play different roles: the reference part is attached to a prescribed spacetime reference metric, while the dynamical perturbation is estimated through the nonlinear evolution.

Our  estimates distinguish between stability at fixed parameter, uniform control of the rescaled fields, and convergence for well-prepared data.

\paragraph{Spacetime perturbation variables}

The reference metric \(g^\star\) is prescribed once the mass parameter \(m\) and the cut-off function \(\chi^\star \) have been fixed. We decompose the conformal metric as
\be
\label{equa-spacetime-metric-decomposition-section3}
g_{\alpha \beta} = g^\star_{\alpha \beta} + u_{\alpha \beta}.
\ee
Equivalently, we write
\be
g_{\alpha \beta} = (g_{\mathrm{Mink}})_{\alpha \beta} + h_{\alpha \beta}^\star + u_{\alpha \beta}, \quad h_{\alpha \beta}^\star  \coloneqq  g^\star_{\alpha \beta}-(g_{\mathrm{Mink}})_{\alpha \beta}.
\ee
The tensor \(h^\star \) is a fixed long-range reference contribution, while \(u\) is the dynamical metric perturbation. This distinction will be preserved in all high-order norms: \(h^\star \) is estimated by explicit weighted pointwise bounds, whereas \(u\) is controlled through the nonlinear energy estimates.

We distinguish the covariant and contravariant perturbations by setting
\be
\label{equa:inverse-metric-perturbations}
(H^\star)^{\alpha \beta} \coloneqq (g^\star)^{\alpha \beta}-(\gMink)^{\alpha \beta}, \qquad \widehat H^{\alpha \beta} \coloneqq g^{\alpha \beta}-(g^\star)^{\alpha \beta}, \qquad H^{\alpha \beta} \coloneqq g^{\alpha \beta}-(\gMink)^{\alpha \beta}.
\ee
Thus \(H=H^\star+ \widehat H\), and the inverse-matrix identity is
\be
\label{equa:inverse-metric-dynamic-identity}
\widehat H^{\alpha \beta} =-g^{\alpha \mu}u_{\mu\nu}(g^\star)^{\nu\beta}.
\ee
The inverse perturbation \(\widehat H^{\alpha \beta} \) is the tensor denoted by \(u^{\alpha \beta} \) in our earlier work; it is not obtained from \(u_{\alpha \beta} \) by raising indices.

The remaining unknowns are the effective field \(\fieldphi\) and the matter field \(\psi\). We use the conformal rescaling
\be
\label{equa-rescaled-effective-field-variables-section3}
{\fieldPhi} \coloneqq \coeff\fieldphi,
\ee
and the energy-normalized rescaling
\be
\label{equa:energy-normalized-effective-field-section3}
{\fieldvarphi} \coloneqq \coeff^{1/2} \fieldphi.
\ee
The unscaled field \(\fieldphi\) enters the singular equation and the physical curvature relation. The field \(\fieldPhi\) enters the conformal metric and the differentiated couplings:
\be
e^{\coeff\fieldphi} =e^{\fieldPhi}, \quad \coeff\, \del\fieldphi = \del{\fieldPhi}.
\ee
The  energy-normalized effective field \(\fieldvarphi\) is the variable naturally seen by the undifferentiated massive part of the uniform energy.  The conversion identities are
\be
\label{equa-rescaled-effective-field-identities-section3}
\coeff^2| \del\fieldphi|^2=| \del{\fieldPhi}|^2, \quad \coeff\fieldphi^2= (\fieldvarphi)^2.
\ee
Thus the parameter-uniform pointwise and energy variables are naturally \(\del\fieldPhi\) and \(\fieldvarphi\), whereas \(\fieldphi\) is kept where its explicitly singular estimates are needed for existence and for the final trace-constraint argument.

The basic unknowns used in the reduced system are
\be
(u_{\alpha \beta}, \fieldphi, \psi),
\ee
while \(({\fieldPhi},{\fieldvarphi}) \) are used systematically in the conformal, derivative, energy, and pointwise estimates described above. The differential operators and high-order seminorms are defined below.


\section{Differential operators and energy hierarchy}
\label{section-3-analysis}


\subsection{Frames and differential operators}
\label{section-3-4}

\paragraph{Purpose}
We introduce the differential operators, pointwise seminorms, and energy functionals used in the singular limit. The frame conventions distinguish derivatives tangent to the foliation, Minkowski-normal ray derivatives, and future-normal derivatives for the evolving metric. The resulting energy hierarchy separates fixed-parameter, uniform, and vanishing estimates.

\paragraph{Cartesian frame}
The Cartesian frame is \((\del_\alpha)=(\del_t,\del_1,\del_2,\del_3)\). Greek indices range over \(0,1,2,3\), and lower-case Latin indices over \(1,2,3\). Spatial Cartesian indices are raised and lowered with \(\delta_{ab}\). Recalling \(r=|x|\), we use, for \(r>0\),
\be
\omega^a={x^a\over r}.
\ee

\paragraph{Semi-hyperboloidal frame}

In the interior domain, we introduce the frame
\be
\delH_0 \coloneqq \del_t, \quad \delH_a \coloneqq \delsH_a \coloneqq {x_a \over t} \del_t+ \del_a.
\ee
The vector fields \(\delsH_a \) are tangent to the Minkowski hyperboloids \(\Hcal_s\).  Their radial combination is
\be
\delsH_r \coloneqq \omega^a \delsH_a= \del_r+ {r\over t} \del_t.
\ee
The transversal derivative introduced in \eqref{equa-normal-ray-field-section3} satisfies
\be
\newperp = \del_t+ {x^a \over t} \del_a = {1\over t} \bigl(t\del_t+ r \del_r \bigr).
\ee
Thus \(\newperp\) is proportional to the scaling vector field and is normal to the hyperboloids with respect to the Minkowski metric. Tangential derivatives will be used to exploit the improved decay available along the hypersurfaces of the Euclidean--hyperboloidal foliation, whereas \(\newperp\) enters naturally in the energy identities and in the ODE reduction of the singular Klein--Gordon equation in Section~\ref{section-6}.

\paragraph{Semi-null and Euclidean--hyperboloidal derivatives}

In the exterior argument, two further families of derivatives must be distinguished. We complete the semi-null derivatives to a frame by setting
\be
\delN_0 \coloneqq \del_t, \qquad \delN_a \coloneqq \delsN_a \coloneqq \del_a+ \omega_a \del_t.
\ee
For an inverse two-tensor \(T\), its incoming semi-null component is
\be
T^{\mathcal N\,00}  \coloneqq T^{00}-2\omega_aT^{0a} + \omega_a \omega_bT^{ab}.
\ee
For the inverse-metric perturbations introduced in \eqref{equa:inverse-metric-perturbations}, we use the abbreviations
\be
H_N \coloneqq H^{\mathcal N\,00}, \qquad (H_N)^\star \coloneqq (H^\star)^{\mathcal N\,00}, \qquad \widehat H_N \coloneqq \widehat H^{\mathcal N\,00},
\ee
so that \(H_N= (H_N)^\star+ \widehat H_N\). We also write
\be
L \coloneqq \del_t+ \del_r, \qquad \underL \coloneqq \del_t-\del_r
\ee
for the outgoing and incoming Minkowski-null derivatives. The unindexed symbol \(L\) is distinct from the Lorentz boosts \(L_a\), defined below. The Euclidean--hyperboloidal frame adapted to the global hypersurfaces is
\be
\delEH_0 \coloneqq \del_t, \qquad \delEH_a \coloneqq \delsEH_a \coloneqq \del_a+ \omega_a \del_r\fTime(s,r) \del_t, \qquad \delsEH_r \coloneqq \omega^a \delsEH_a.
\ee
Its spatial vectors \(\delsEH_a \) are tangent to \(\Mcal_s\). On \(\MME= \Mtran\cup\Mext\), we denote its restriction by \(\delsME_a \coloneqq \delsEH_a \) and set
\be
\delsME_r \coloneqq \omega^a \delsME_a = \del_r+ \del_r\fTime(s,r) \del_t.
\ee
Along the graph \(t= \fTime(s,r) \), the operator \(\delsME_r\) is differentiation with respect to \(r\) at fixed \(s\). Thus \(\delsME_a \) is tangent to \(\Mcal_s\), while \(\delsN_a \) is tangent to the outgoing Minkowski cones. They satisfy
\be
\delsN_a-\delsME_a = \omega_a \bigl(1-\del_r\fTime(s,r) \bigr) \del_t.
\ee
On the hyperboloidal part, the corresponding relation with \(\delsH_a= \del_a+(x^a/t) \del_t\) is
\be
\delsH_a= \delsN_a-\omega_a{t-r\over t} \del_t.
\ee
The distinction is important: radial Hardy and Sobolev inequalities on a hypersurface use \(\delsME\), whereas the favorable bulk term in the exterior weighted current uses \(\delsN\). These derivative families are therefore not interchangeable: \(\delsH\) is tangent to the interior hyperboloids, \(\delsEH\) is adapted to the full Euclidean--hyperboloidal foliation, \(\delsME\) is its restriction to the merging and exterior regions, and \(\delsN\) is adapted to the outgoing null cones.  The future-normal derivative is denoted separately by \(n_s\), while an overbar is reserved for the \((s,x) \) coordinate frame.

\paragraph{Frame transformations}

We write
\be
\delH_\alpha = (\PhiH)_\alpha{}^\beta \del_\beta, \quad \del_\alpha = (\PsiH)_\alpha{}^\beta \delH_\beta, \qquad \PsiH= (\PhiH)^{-1},
\ee
for the corresponding change-of-frame relations. The coefficients of these matrices depend smoothly on \(x/t\) and remain uniformly bounded in the interior domain. We attach the superscript \(\mathcal H\) to components in the semi-hyperboloidal frame and use an overbar exclusively for the hyperboloidal coordinate frame \((\del_s, \del_a|_s) \) introduced in Section~\ref{section-6-2}. For instance, if \(T=T^{\alpha \beta} \del_\alpha \otimes\del_\beta \), then
\be
T^{\mathcal H\, \alpha \beta} = (\PsiH)_{\alpha'}{}^\alpha (\PsiH)_{\beta'}{}^\beta T^{\alpha'\beta'}.
\ee
The coordinate, ray, and unit-normal derivatives are related by
\be
\label{equa:frame-convention-crosswalk}
\delH_0= \del_t, \quad \delb_0 \coloneqq {s\over t} \del_t, \quad \delb_a \coloneqq \del_a \big|_s= \delsH_a, \quad \newperp= \del_t+ {x^a \over t} \del_a, \quad n_{\Hcal_s}^{\mathrm{Mink}} ={t\over s} \newperp= \delb_0 + {x^a \over s} \delb_a.
\ee
Thus \(\delH_0\), \(\delb_0\), and \(\newperp\) are distinct.  The symbol \(n_s\) used in the nonlinear energies denotes instead the future-directed \(g\)-unit normal to the whole hypersurface \(\Mcal_s\). The inverse-metric component \(H^{\mathcal H\,00} \) gives the normalized ray coefficient \(a=1-(t/s)^2H^{\mathcal H\,00} \) in Section~\ref{section-6}; this coefficient plays a distinguished role in the pointwise estimates. The matrix \(\PhiH\) gives the direct Cartesian-to-semi-hyperboloidal change of frame, \(\PsiH= \PhiH^{-1} \) is its inverse, and barred derivatives denote the hyperboloidal coordinate derivatives.

\paragraph{Commuting vector fields}

As usual, the nonlinear analysis uses a hierarchy of differential operators adapted to the Euclidean--hyperboloidal foliation. We introduce them here once and for all. Translations are used to measure ordinary derivatives, while rotations and Lorentz boosts preserve the principal part of the Minkowski wave operator and capture the geometry of the foliation. We denote by
\be
\del_\alpha, \quad \alpha=0,1,2,3,
\ee
the Cartesian translations, with \(\del_0= \del_t\). The spatial rotations and Lorentz boosts are
\be
\label{equa-commutation-vector-fields-section3}
\Omega_{ab}  \coloneqq  x_a \del_b-x_b\del_a, \quad L_a  \coloneqq  x_a \del_t+ t\del_a, \quad 1\leq a,b\leq3.
\ee
The boosts are tangent to the Minkowski hyperboloids, since $ L_a(s)=0. $ We observe that
\be
L_a=t\, \delsH_a.
\ee
Thus derivatives in the boost directions are weighted tangential derivatives along the interior hypersurfaces.

We shall use the collection
\be
\label{equa-Z-family-section3}
\mathscr Z  \coloneqq  \{\del_\alpha, \Omega_{ab},L_a \}.
\ee
The scaling vector field is not included in \(\mathscr Z\), since it does not commute with the Klein--Gordon operator. This is one of the reasons for using the Euclidean--hyperboloidal method rather than a conformal vector-field argument alone.

For the flat wave operator, it is well known that
\be
[\Box, \del_\alpha]=0, \quad [\Box, \Omega_{ab}]=0, \quad [\Box,L_a]=0.
\ee
For the reduced operator \(\BoxChapeau_g\), the corresponding commutators contain derivatives of the metric coefficients. Their structure will be analyzed in the high-order estimates below.

\paragraph{Two-parameter scale of pointwise norms}

We take \(I= (i_0,i_1,i_2,i_3) \) to be a Cartesian multi-index and \(J\) a finite ordered family of Lorentz boosts. For \(Z= \del^I L^J\), we define \(\ord Z\coloneqq|I|+ |J| \) and \(\rank Z\coloneqq|J| \). We write
\be
\del^I = \del_0^{i_0} \del_1^{i_1} \del_2^{i_2} \del_3^{i_3}, \qquad L^J = L_{j_1} \cdots L_{j_{|J|}}.
\ee
Rotations satisfy \(\Omega_{ab}=(x_a/t)L_b-(x_b/t)L_a\). In the interior, their coefficients and commuted derivatives are bounded, so rotations can be estimated through boosts. On unbounded exterior hypersurfaces, the factors \(x/t\) must instead be kept. We define the following pointwise seminorms using translations and boosts only:
\be
\label{equa-pointwise-semi-norm-pk-section3}
|u|_{p,k}  \coloneqq  \sum_{\substack{|I|+ |J| \leq p\\
|J| \leq k}} | \del^I L^J u|, \quad |T|_{p,k}  \coloneqq  \sum_{\substack{|I|+ |J| \leq p\\
|J| \leq k}} | \del^I L^J T|,
\ee
together with
\be
\label{equa-pointwise-derivative-semi-norms-section3}
| \del u|_{p,k}  \coloneqq  \sum_{\substack{|I|+ |J| \leq p\\
|J| \leq k}} | \del\del^I L^J u|, \quad | \delsH u|_{p,k}  \coloneqq  \sum_{\substack{|I|+ |J| \leq p\\
|J| \leq k}} | \delsH\del^I L^J u|.
\ee
The same convention is used for \(\newperp u\). The index \(p\) measures the \emph{total order}, while \(k\) counts the \emph{number of boosts}. This distinction is crucial in our proof, where boosts and translations carry different weights. Whenever no separate boost count is needed, we abbreviate
\be
|u|_p \coloneqq |u|_{p,p}, \quad | \del u|_p \coloneqq | \del u|_{p,p}.
\ee
Pointwise absolute values are Euclidean component norms in the specified frame. Tensor seminorms are sums over the components.

\begin{table}[H]
\centering
\renewcommand{\arraystretch}{1.18}
\begin{tabular}{>{\raggedright\arraybackslash}p{.25\textwidth}>{\raggedright\arraybackslash}p{.65\textwidth}}
\hline
\textbf{Notation} & \textbf{Meaning} \\
\hline
\(\Mcal_s\) &
Hypersurface of the Euclidean--hyperboloidal foliation \\

\(\Mint\), \(\Mtran\), \(\Mext\) &
Hyperboloidal, merging, and asymptotically Euclidean domains \\

\(\MME\), \(\Lscr\) &
Euclidean--merging side and its inner interface \\

\(\delH\), \(\delb\) &
Frames with \(\delH_0= \del_t\) and \(\delb_0= (s/t) \del_t\) \\

\(\delN\), \(\delsME\) &
Semi-null frame and derivatives tangent to the global hypersurfaces \\

\(\newperp\) &
Minkowski-normal ray derivative \\

\(Z= \del^I L^J\) &
Commutator built from translations and Lorentz boosts \\

\(\ord Z\), \(\rank Z\) &
Total differential order and boost rank \\

\hline
\end{tabular}
\caption{Geometric and differential notation used in the proof}
\label{table:notation-proof}
\end{table}


\subsection{Effective-field variables and energy scales} \label{section-3-5}

\paragraph{Rescalings and pointwise seminorms}
The variables \(\fieldPhi=\coeff\fieldphi\) and \(\fieldvarphi=\coeff^{1/2}\fieldphi\), defined in Section~\ref{section-3-3}, distinguish the differentiated couplings from the massive part of the energy. For the seminorms of \eqref{equa-pointwise-semi-norm-pk-section3}, their scaling is
\be
|\fieldPhi|_{p,k}=\coeff|\fieldphi|_{p,k},\qquad |\fieldvarphi|_{p,k}=\coeff^{1/2}|\fieldphi|_{p,k}.
\ee
The same relations hold after differentiation. We keep three classes of estimates: bounds for the unscaled field at fixed \(\coeff>0\), uniform bounds for the rescaled fields, and bounds tending to zero with \(\coeff\). The last class is needed to identify the Einstein limit. For a singular equation whose limiting relation is \(\rho=\sourterm\), the discrepancy is
\be
\label{equa-constraint-discrepancy-section3}
\rhon\coloneqq\rho-\sourterm.
\ee
Section~\ref{section-6} estimates this discrepancy along hyperboloidal rays, after the rescaling \(w=s^{3/2}\rho\) that removes the leading first-order term.

\paragraph{Flat energies and their parameter dependence}
For a scalar field \(v\), the flat hyperboloidal wave energy and the Klein--Gordon energy of mass \(c>0\) are
\be
\label{equa-wave-energy-schematic-section3}
\calE_{\mathrm w}(s,v)\coloneqq\int_{\Hcal_s}\bigl((s/t)^2|\del_tv|^2+|\delsH v|^2\bigr)\,dx,\qquad
\calE_{\mathrm{KG},c}(s,v)\coloneqq\calE_{\mathrm w}(s,v)+c^2\|v\|_{L^2(\Hcal_s)}^2.
\ee
The corresponding order-\(N\) energy is obtained by summing over translations and boosts:
\be
\label{equa-high-order-energy-section3}
\calE_{\mathrm w}^{N}(s,v)\coloneqq\sum_{|I|+|J|\leq N}\calE_{\mathrm w}(s,\del^I L^Jv),
\ee
and similarly for \(\calE_{\mathrm{KG},c}^{N}\). The linear part \(3\coeff\Box_g\fieldphi-\fieldphi\) determines the two effective-field energy scales
\be
\label{equa-degenerate-effective-field-energy-section3}
\calE_{\mathrm{eff}}^{\mathrm{sing}}(s,\fieldphi)\simeq
\int_{\Hcal_s}\bigl(3\coeff(s/t)^2|\del_t\fieldphi|^2+3\coeff|\delsH\fieldphi|^2+|\fieldphi|^2\bigr)\,dx,
\ee
\be
\label{equa-uniform-effective-field-energy-section3}
\calE_{\mathrm{eff}}^{\mathrm{unif}}(s,\fieldphi)\coloneqq\coeff\calE_{\mathrm{eff}}^{\mathrm{sing}}(s,\fieldphi)
\simeq\int_{\Hcal_s}\bigl(3(s/t)^2|\del_t\fieldPhi|^2+3|\delsH\fieldPhi|^2+|\fieldvarphi|^2\bigr)\,dx.
\ee
At order \(N\), we use
\be
\label{equa-high-order-effective-field-energies-section3}
\calE_{\mathrm{eff}}^{\mathrm{sing},N}(s,\fieldphi)\coloneqq
\sum_{|I|+|J|\leq N}\calE_{\mathrm{eff}}^{\mathrm{sing}}(s,\del^I L^J\fieldphi),\qquad
\calE_{\mathrm{eff}}^{\mathrm{unif},N}\coloneqq\coeff\calE_{\mathrm{eff}}^{\mathrm{sing},N}.
\ee
The singular energy controls \(\fieldphi\) and \(\coeff^{1/2}\del\fieldphi\); its rescaled counterpart controls the combinations entering the metric and matter equations. Section~\ref{section-5} defines the curved currents, including their weights and boundary fluxes.


\subsection{\texorpdfstring{Strategy of proof}{Strategy of proof}} \label{section-3-6}

The global stability analysis in~\cite{LeFloch-Ma-2024b} already provides estimates uniform in the modified theory, under its light-bending hypotheses on the initial geometry. The convergence argument requires, in addition, estimates that make the discrepancy from the Einstein trace relation tend to zero and a comparison with the limiting Einstein development. The estimates below serve the following purposes.
\bei
\item \emph{Reduced equations and energies.} Section~\ref{section-4} identifies every coupling coefficient in wave gauge, and Section~\ref{section-5} derives the singular and rescaled energy identities. The fixed-parameter estimates remain available for continuation at the highest order.
\item \emph{Pointwise control.} Section~\ref{section-6} gives the oscillator estimate uniform in the mass. The interior analysis combines it with the wave-gauge identities and estimates for commutators; the exterior analysis gives the values at rays entering through \(\Lscr\).
\item \emph{Global stability and convergence.} Parts~II and~III close the energy estimates simultaneously in the interior and exterior, keeping the signed boundary fluxes. Preparation then controls the error in the Einstein trace relation. Local comparison and uniform spacetime decay give the global convergence conclusion.
\eei


\section{Field equations in conformal wave gauge}
\label{section-4}


\subsection{Reduced conformal system}
\label{section-4-1}

\paragraph{Purpose}
We derive the wave-gauge equations for the conformal metric, effective field, and matter. Their explicit coupling coefficients distinguish the uniform source terms from the singular massive equation and identify the formal Einstein limit.

\paragraph{Conformal wave-gauge reduction}

From this point on, all equations are written in the conformal Einstein metric $g$. The physical metric is recovered from
\be
\giveng_{\alpha \beta} =e^{-\coeff\fieldphi}g_{\alpha \beta},
\ee
where \(\fieldphi\) is the  effective field introduced in Section~\ref{section-2}. We work in a global coordinate system
\be
(x^\alpha)= (t,x^a), \quad a=1,2,3,
\ee
and impose the \emph{conformal wave gauge condition}
\be
\label{equa:wcoo}
\Box_g x^\alpha=0, \quad \alpha=0,1,2,3
\ee
Equivalently, we require
\be
\label{equa:gamnul-0}
\Gamma^\alpha \coloneqq g^{\lambda \mu} \Gamma^\alpha_{\lambda \mu} =0.
\ee
We denote by
\be
\BoxChapeaud_g \coloneqq g^{\mu\nu} \del_\mu\del_\nu
\ee
the reduced wave operator. Under the gauge condition \eqref{equa:gamnul-0}, this operator coincides with the geometric wave operator on scalar functions. For the metric components, the Ricci tensor has the reduced form
\be
\label{equa:ricci-wave-gauge}
2R_{\alpha \beta} = -\BoxChapeaud_g g_{\alpha \beta} + \Fbb_{\alpha \beta}(g,g;\del g, \del g),
\ee
where \(\Fbb_{\alpha \beta} \) is quadratic in \(\del g\), with coefficients depending smoothly on \(g\). To specify these coefficients, we write \(\Gamma_{\alpha\beta\gamma}\coloneqq g_{\alpha\nu}\Gamma^\nu_{\beta\gamma}=(\del_\beta g_{\alpha\gamma}+\del_\gamma g_{\alpha\beta}-\del_\alpha g_{\beta\gamma})/2\). The expression in wave gauge is
\be
\label{equa:audit-exact-reduced-metric-source}
\Fbb_{\alpha\beta}=-(\del_\alpha g^{\lambda\mu})\Gamma_{\beta\lambda\mu}-(\del_\beta g^{\lambda\mu})\Gamma_{\alpha\lambda\mu} -(\del_\beta g^{\lambda\mu})\del_\alpha g_{\lambda\mu}-2\Gamma^\lambda_{\beta\mu}\Gamma^\mu_{\alpha\lambda}.
\ee
Indeed, differentiating \(g^{\lambda\mu}\Gamma_{\alpha\lambda\mu}=0\) eliminates the remaining second derivatives in the coordinate expression for Ricci, while \(\del_\lambda g^{\lambda\mu}=-g^{\lambda\mu}\Gamma^\nu_{\nu\lambda}\) cancels its two contracted-connection terms. The identity \(\del_\alpha g^{\lambda\mu}=-g^{\lambda\rho}g^{\mu\sigma}\del_\alpha g_{\rho\sigma}\) then fixes every inverse-metric coefficient before commutation. Its null and quasi-null decomposition is the standard wave-gauge decomposition used in \cite[Section~5.2]{LeFloch-Ma-2024b}.

\paragraph{Source terms in conformal gauge}

We recall the potential from Section~\ref{section-2-2}
\be
U(r)={r f'(r)-f(r)\over\coeff\,(f'(r))^2},\qquad \Ufrak=U\circ\hfrak,
\ee
and the conformal Einstein equations derived in Section~\ref{section-2} take the form
\be
\label{equa:Einstein-conformal-section4}
G_{\alpha \beta} = 8\pi e^{-\coeff\fieldphi} \givenT_{\alpha \beta}[\psi, \giveng] + T^{\mathrm{eff}}[\fieldphi,g]_{\alpha \beta},
\ee
where the stress-energy tensor of the  effective field is
\be
\label{equa:Tphi-section4}
T^{\mathrm{eff}}[\fieldphi,g]_{\alpha \beta} = {3\over2} \coeff^2\nabla_\alpha \fieldphi\nabla_\beta \fieldphi -{3\over4} \coeff^2g(\nabla \fieldphi, \nabla \fieldphi)g_{\alpha \beta} -{\coeff\over2} \Ufrak(\fieldphi)g_{\alpha \beta}.
\ee
The sign of the last term is important: the potential \(\Ufrak\) is positive near the origin, and therefore appears in \(T^{\mathrm{eff}}[\fieldphi,g] \) with the sign dictated by the Einstein equation.

For later use, we define the total conformal source tensor by
\be
S_{\alpha \beta}  \coloneqq  8\pi e^{-\coeff\fieldphi} \givenT_{\alpha \beta}[\psi, \giveng] + T^{\mathrm{eff}}[\fieldphi,g]_{\alpha \beta}.
\ee
Since \(G_{\alpha \beta} =S_{\alpha \beta} \), the Ricci curvature reads
\be
R_{\alpha \beta} = S_{\alpha \beta} - {1\over2}(\Tr_gS)g_{\alpha \beta}.
\ee
Combining this identity with \eqref{equa:ricci-wave-gauge} gives
\be
\BoxChapeaud_g g_{\alpha \beta} = \Fbb_{\alpha \beta}(g,g;\del g, \del g) -2S_{\alpha \beta} + (\Tr_gS)g_{\alpha \beta}.
\ee
For later use, we isolate the effective-field and matter contributions:
\be
\label{equa:ABs}
\aligned \Gbb_{\coeff}[\fieldphi,g]_{\alpha \beta} & \coloneqq  -3\coeff^2\nabla_\alpha \fieldphi\nabla_\beta \fieldphi -\coeff\Ufrak(\fieldphi)g_{\alpha \beta}, \\
\Mbb_{\coeff}[\fieldphi, \psi,g]_{\alpha \beta} & \coloneqq  -16\pi e^{-\coeff\fieldphi} \Bigl( \nabla_\alpha \psi\nabla_\beta \psi + e^{-\coeff\fieldphi}V(\psi)g_{\alpha \beta} \Bigr). \endaligned
\ee
These formulae hold for the scalar matter model considered here; their derivation is included in Section~\ref{section-7}.  That section also specifies the sign convention for the positive effective-field potential and gives the identities used in the singular limit. The blackboard symbol \(\Gbb_\coeff\) is reserved for this effective-field contribution.  It distinguishes the term from the Einstein tensor \(G\) and from the scalar growth functions introduced in the bootstrap argument.

The trace equation for the  effective field is
\be
\label{equa:phi-trace-section4}
3\coeff\, \Box_g\fieldphi - \Ufrak'(\fieldphi) = 8\pi e^{-2\coeff\fieldphi} \Tr_{\giveng} \big(\givenT[\psi, \giveng] \big).
\ee
For the scalar matter field, the tensorial trace is
\be
\label{equa:trace-source-explicit}
8\pi e^{-2\coeff\fieldphi} \Tr_{\giveng} \big(\givenT[\psi, \giveng] \big) = -8\pi e^{-\coeff\fieldphi}g(\nabla \psi, \nabla \psi) -32\pi e^{-2\coeff\fieldphi}V(\psi).
\ee
Finally, the matter equation becomes
\be
\label{equa:matter-conformal-section4}
\Box_g\psi-e^{-\coeff\fieldphi}V'(\psi) = \coeff\,g(\nabla \fieldphi, \nabla \psi).
\ee

\paragraph{Reduced conformal wave--Klein--Gordon system}
The three reduced equations form the following coupled system.

\begin{proposition}[Conformal \(f(R) \) equations in wave gauge]
\label{propo41}
In the conformal wave gauge \eqref{equa:wcoo}--\eqref{equa:gamnul-0}, the conformal formulation of \(f(R) \) gravity with scalar matter is the following wave--Klein--Gordon system:
\be
\label{MainPDE}
\aligned \BoxChapeaud_g g_{\alpha \beta}&= \Fbb_{\alpha \beta}(g,g;\del g, \del g) + \Gbb_{\coeff}[\fieldphi,g]_{\alpha \beta} + \Mbb_{\coeff}[\fieldphi, \psi,g]_{\alpha \beta}, \\
3\coeff\, \Box_g\fieldphi-\Ufrak'(\fieldphi)&= -8\pi e^{-\coeff\fieldphi}g(\nabla \psi, \nabla \psi)-32\pi e^{-2\coeff\fieldphi}V(\psi), \\
\Box_g\psi-e^{-\coeff\fieldphi}V'(\psi)&= \coeff\,g(\nabla \fieldphi, \nabla \psi). \endaligned
\ee
The initial data satisfy the Hamiltonian, momentum, and wave-gauge constraints. The physical scalar curvature is then recovered by \eqref{equa:rhocondi}.
\end{proposition}

Proposition~\ref{pro:40S} proves the equivalence with the Jordan-coupled \(f(R) \) equations on every sufficiently regular Cauchy development.

\paragraph{Initialization of the wave coordinates on the prescribed graph}

We construct the reduced initial jets directly on the prescribed graph. We set \(T_0(y)\coloneqq\fTime(s_0,|y|)\) and use the fixed auxiliary coordinates \(\tau\coloneqq t-T_0(x)\), \(y^a\coloneqq x^a\), so that \(\Sigma_0=\{\tau=0\}\). We denote the metric components in this chart by \(g_{\mathrm{aux},AB}\), with \(A,B\in\{\tau,1,2,3\}\), and set \(j_{\mathrm{aux}}\coloneqq\sqrt{|\det g_{\mathrm{aux}}|}\). On \(\Sigma_0\), a prescribed positive lapse \(\alpha\) and shift \(\beta\) determine
\bse
\be
\label{equa:audit-initial-auxiliary-metric}
g_{\mathrm{aux},\tau\tau}=-\alpha^2+(g_0)_{ab}\beta^a\beta^b,\qquad g_{\mathrm{aux},\tau a}=(g_0)_{ab}\beta^b,\qquad g_{\mathrm{aux},ab}=(g_0)_{ab}.
\ee
The second fundamental form and the scalar normal data give the initial derivatives
\be
\label{equa:audit-initial-auxiliary-velocities}
\del_\tau g_{\mathrm{aux},ab}=-2\alpha(k_0)_{ab}+(\mathcal L_\beta g_0)_{ab},\qquad \del_\tau v=\alpha v_1+\beta^a\del_{y^a}v_0\quad(v\in\{\fieldphi,\psi\}).
\ee
The original wave-coordinate conditions are \(\Box_gt=\Box_gx^a=0\). Since \(\Box_g\tau=-g_{\mathrm{aux}}^{ab}\del_{ab}T_0\), the auxiliary contracted Christoffel symbols and their density equations satisfy
\be
\label{equa:audit-initial-auxiliary-gauge}
\Gamma_{\mathrm{aux}}^\tau=g_{\mathrm{aux}}^{ab}\del_{ab}T_0,\qquad \Gamma_{\mathrm{aux}}^a=0,\qquad \del_A(j_{\mathrm{aux}}g_{\mathrm{aux}}^{AB})=-j_{\mathrm{aux}}\Gamma_{\mathrm{aux}}^B.
\ee
The identities
\be
\label{equa:audit-initial-lapse-shift-density}
j_{\mathrm{aux}}g_{\mathrm{aux}}^{\tau\tau}=-\frac{\sqrt{\det g_0}}\alpha,\qquad j_{\mathrm{aux}}g_{\mathrm{aux}}^{\tau a}=\frac{\sqrt{\det g_0}}\alpha\beta^a
\ee
\ese
show that these four gauge equations form an invertible triangular linear system for \(\del_\tau\alpha\) and \(\del_\tau\beta\). Their other coefficients depend only on the prescribed values, their spatial derivatives, and the fixed Hessian of \(T_0\). This determines the remaining first metric jets. The fixed coordinate transformation then gives the reduced initial data in \((t,x)\), without changing their differential order.


\subsection{Structure of the nonlinearities}
\label{section-4-2}

\paragraph{Metric nonlinearities and quasi-null structure}

We decompose the conformal metric as
\be
g_{\alpha \beta} = (\gMink)_{\alpha \beta} +h_{\alpha \beta}, \qquad h_{\alpha \beta} \coloneqq h^\star_{\alpha \beta} +u_{\alpha \beta},
\ee
where \(g_{\mathrm{Mink}} \) is the Minkowski metric. The metric equation in \eqref{MainPDE} has the schematic form
\be
\label{equa:metric-structure}
\Box h_{\alpha \beta} =-H^{\mu\nu} \del_\mu\del_\nu h_{\alpha \beta} + \Qbb_{\alpha \beta}(\del h, \del h) + \Pbb_{\alpha \beta}(\del h, \del h) + \Gbb_\coeff[\fieldphi,g]_{\alpha \beta} + \Mbb_\coeff[\fieldphi, \psi,g]_{\alpha \beta} +\Ocal(h\, \del h\, \del h).
\ee
Here \(\Box= (\gMink)^{\mu\nu} \del_\mu\del_\nu\), and \(H^{\mu\nu} \) is the inverse-metric perturbation in \eqref{equa:inverse-metric-perturbations}. The null and quasi-null expressions are the quadratic terms obtained by evaluating the coefficients of \(\Fbb\) at the Minkowski metric. Here \(\Qbb\) denotes combinations of classical null forms, while \(\Pbb\) denotes the quasi-null part of the reduced Einstein equations. The terms \(\Pbb\) are not null forms in the classical sense, but they satisfy the hierarchy used in the wave-coordinate stability theory. This is the same structure exploited in the Euclidean--hyperboloidal method for the Einstein--massive field systems.

The effective-field source in the metric equation is especially favorable:
\be
\Gbb_{\coeff}[\fieldphi,g]_{\alpha \beta} = -3\coeff^2\nabla_\alpha \fieldphi\nabla_\beta \fieldphi -\coeff\Ufrak(\fieldphi)g_{\alpha \beta}.
\ee
For the class of theories considered in this memoir, we have
\be
\label{equa:Ufrak-expansion-section4}
\Ufrak(\fieldphi) = {1\over2} \fieldphi^2 +\Ocal(\coeff\fieldphi^3), \quad \Ufrak'(\fieldphi) = \fieldphi+\Ocal(\coeff\fieldphi^2).
\ee
Thus the metric equation only contains the combinations
\be
\coeff^2(\del\fieldphi)^2, \quad \coeff\Ufrak(\fieldphi) \simeq\coeff\fieldphi^2,
\ee
which are compatible with estimates uniform in the general-relativistic limit.

The matter source \(\Mbb_{\coeff} \) is the usual scalar-field source, up to the conformal weights \(e^{-\coeff\fieldphi} \) and \(e^{-2\coeff\fieldphi} \). These weights remain harmless as long as the  conformal effective field \(\fieldPhi\) stays uniformly small.

\paragraph{Effective-field equation}

The singular nature of the limit is concentrated in the \(\fieldphi\)-equation. We define
\be
\label{equa:Wcoeff-def}
W_\coeff(\fieldphi)  \coloneqq  \Ufrak'(\fieldphi)-\fieldphi.
\ee
By \eqref{equa:Ufrak-expansion-section4}, we obtain
\be
\label{equa:Wcoeff-expansion}
W_\coeff(\fieldphi)=\Ocal(\coeff\fieldphi^2) \quad\text{near } \fieldphi =0.
\ee
Then \eqref{equa:phi-trace-section4} can be written as
\be
\label{equa:KG-effective-field-reduced}
3\coeff\, \Box_g\fieldphi-\fieldphi = \mathcal S[\psi, \fieldphi,g]+W_\coeff(\fieldphi),
\ee
where we define
\be
\label{equa:source-trace-def}
\mathcal S[\psi, \fieldphi,g]  \coloneqq  8\pi e^{-2\coeff\fieldphi} \Tr_{\giveng} \big(\givenT[\psi, \giveng] \big).
\ee
Equivalently, we have
\be
\label{equa:KG-massive}
\Box_g\fieldphi - {1\over3\coeff} \fieldphi = {1\over3\coeff} \mathcal S[\psi, \fieldphi,g] + {1\over3\coeff}W_\coeff(\fieldphi).
\ee
The  effective field therefore behaves as a Klein--Gordon field with mass squared \(1/(3\coeff) \). The general-relativistic limit is an infinite-mass limit.

The rescaled field \(\fieldPhi\) controls the differentiated coupling, while \(\fieldvarphi\) controls the massive part of the energy:
\be
\coeff^2| \del\fieldphi|^2=| \del{\fieldPhi}|^2, \quad \coeff\fieldphi^2= (\fieldvarphi)^2.
\ee
In these variables, we have
\be
\coeff\Ufrak(\fieldphi)=\Ocal((\fieldvarphi)^2), \quad W_\coeff(\fieldphi)=\Ocal((\fieldvarphi)^2),
\ee
at the schematic level relevant for the nonlinear estimates.

\paragraph{Matter equation}

The conformal matter equation is
\be
\Box_g\psi-e^{-\coeff\fieldphi}V'(\psi) = \coeff\,g(\nabla \fieldphi, \nabla \psi).
\ee
In the massless case \(V\equiv0\), this reduces to
\be
\label{equa:matter-massless}
\Box_g\psi = g(\nabla{\fieldPhi}, \nabla \psi).
\ee
Thus the matter field couples to the  effective field through the uniformly controlled derivative $\nabla{\fieldPhi}$, rather than through the possibly large derivative \(\nabla \fieldphi\).


\subsection{Formal general-relativistic limit}
\label{section-4-3}

\paragraph{Limiting equations}

We consider the formal limit \(\coeff\to0\). Since the conformal factor is \(e^{\coeff\fieldphi} =e^{\fieldPhi} \), uniform control and convergence of \({\fieldPhi} \) to zero imply
\be
g_{\alpha \beta}-\giveng_{\alpha \beta} = \big(e^{\coeff\fieldphi}-1\big) \giveng_{\alpha \beta} \longrightarrow0.
\ee
Thus the conformal and physical metrics coincide in the limiting theory.

The  effective-field equation
\be
3\coeff\, \Box_g\fieldphi-\fieldphi = \mathcal S[\psi, \fieldphi,g]+W_\coeff(\fieldphi)
\ee
formally degenerates into the constraint
\be
\label{equa:phi-limit-constraint}
\fieldphi = -\mathcal S[\psi,0,g].
\ee
This is the trace relation for the limiting Einstein-matter system, written in the variables of the conformal formulation. In the vacuum case, \(\mathcal S=0\), and the limiting constraint gives \(\fieldphi =0\). Hence the additional scalar degree of freedom of \(f(R) \) gravity is removed in the general-relativistic limit.

At the level of the reduced metric equation, we obtain the limiting Einstein system in wave gauge:
\be
\label{MainPDE-limit}
\BoxChapeaud_g g_{\alpha \beta} = \Fbb_{\alpha \beta}(g,g;\del g, \del g) + \mathcal M^{\rm Ein}_{\alpha \beta}[\psi,g],
\ee
where \(\mathcal M^{\rm Ein} \) is the matter source obtained in the limiting Einstein equations. For a scalar field, this limiting source is obtained from \(\Mbb_{\coeff} \) by setting \({\fieldPhi} =0\):
\be
\mathcal M^{\rm Ein}_{\alpha \beta}[\psi,g] = -16\pi \Bigl( \nabla_\alpha \psi\nabla_\beta \psi + V(\psi)g_{\alpha \beta} \Bigr).
\ee
The matter equation converges to
\be
\Box_g\psi-V'(\psi)=0.
\ee
In the massless case \(V\equiv0\), the limiting system is the reduced Einstein--massless-scalar field system; in the vacuum case, it is the reduced vacuum Einstein system.

\paragraph{Flat singular wave--Klein--Gordon model}
The singular structure of the full equations is already visible in the flat model
\be
\label{equa-WKGmodel}
\aligned \Box u&= Q(\del u, \del u) + \coeff^2(\del\fieldphi)^2 + \coeff\fieldphi^2 +(\del\psi)^2, \\
3\coeff\, \Box\fieldphi-\fieldphi&= (\del\psi)^2 + \coeff\fieldphi^2, \\
\Box\psi&= \coeff\, \del\fieldphi\, \del\psi. \endaligned
\ee
Here \(u\) represents the metric perturbation, \(\fieldphi\) the effective field, and \(\psi\) the matter field. As \(\coeff\to0\), the second equation becomes a constraint for \(\fieldphi\), while the first and third equations reduce to a coupled wave system. The massive field \(\fieldphi\) therefore \emph{does not remain an independent dynamical unknown} in the limiting theory.


\subsection{\texorpdfstring{Preparation for the Einstein limit}{Preparation for the Einstein limit}}
\label{section-4-4}

\paragraph{Effective-field energy for the convergence theory}

\begin{definition}
\label{definition:Einstein-trace-residual}
The \textbf{Einstein trace residual} is the scalar field defined by
\be
\label{equa:Einstein-trace-residual}
\Theta \coloneqq \fieldphi+ \mathcal S[\psi, \fieldphi,g].
\ee
Here \(\mathcal S\) is the matter trace source in the conformal effective-field equation \eqref{equa:source-trace-def}; setting \(\fieldPhi =0\) gives the trace source of the limiting Einstein system. Thus \(\Theta \) measures the discrepancy from the algebraic trace relation, whereas an interface trace is the restriction of a field to a hypersurface.
\end{definition}

Formally, the limit \(\coeff\to0\) should enforce \(\Theta \to0\). Applying \(3\coeff \, \Box_g-1\) to \(\Theta \), we obtain the identity
\be
\label{equa:trace-residual-equation}
3\coeff\, \Box_g\Theta-\Theta = W_\coeff(\fieldphi) +3\coeff\, \Box_g\mathcal S[\psi, \fieldphi,g].
\ee
The massive energy estimate applies to \(\Theta\) at its unscaled level. On a spacelike coordinate slice \(\Sigma_t\coloneqq\{t\}\times\mathbb R^3\), with a positive weight \(w\), its quadratic energy has the form
\be
\label{equa:trace-residual-energy}
\Eenergy^\eff(t, \Theta) \sim \int_{\Sigma_t} \Bigl( 3\coeff| \del\Theta|^2 + \Theta^2 \Bigr)w\, dx.
\ee
This energy controls the trace residual at its unscaled level. A vanishing initial energy, together with vanishing accumulated forcing, recovers the Einstein trace constraint. The matter equation and the Bochner identity express \(\Box_g\mathcal S[\psi, \fieldphi,g] \) using at most two derivatives of \(\psi\) and \(\fieldPhi\), as shown in \eqref{equa:eight-trace-source-Bochner}. The convergence estimate closes two orders below the existence hierarchy.

\paragraph{Asymptotically Euclidean norms and preparation}

We denote by \(q_\star\geq2\) the maximal order of the initial-data comparison. On the initial hypersurface \(\Sigma_0\simeq\mathbb R^3\), with coordinates \(x= (x^1,x^2,x^3) \), we fix a smooth Riemannian metric \(h_{\mathrm{aux}} \) equal to the Euclidean metric outside a compact set and a smooth positive weight \(w_0\) equal to \(1+ |x| \) there. We write \(|T|_{\mathrm{aux}} \) for the tensor norm and \(\nabla_{\mathrm{aux}} \) for the connection induced by \(h_{\mathrm{aux}} \), and set \(dx=dx^1dx^2dx^3\). The weighted asymptotically Euclidean Sobolev norm of order \(q\) is
\be
\label{equa:weighted-AE-Sobolev-topology}
\|v\|_{H^q_{\mathrm{AE}}}^2 \coloneqq \sum_{j=0}^q\int_{\Sigma_0}w_0^2| \nabla_{\mathrm{aux}}^jv|_{\mathrm{aux}}^2\, dx, \qquad 0\leq q\leq q_\star.
\ee
We use the componentwise sum for tuples.  Thus \(H^q_{\mathrm{AE}} \) is an ordinary weighted spatial Sobolev norm; it does not contain normal derivatives.  Any two choices of \(h_{\mathrm{aux}} \) and \(w_0\) with these properties yield equivalent norms, and the choice is fixed independently of \(\coeff\).

Denoting by \(n_0\) the future conformal unit normal to \(\Sigma_0\) and \(Z= \del^I L^J\), we define the \textbf{preparation amplitude} at order \(1\leq q\leq q_\star\) by
\be
\label{equa:commuted-preparation-amplitude}
\bigl(\mathrm{Prep}^{\mathrm{com}}_{\coeff,q}[\Theta] \bigr)^2 \coloneqq \sum_{\ord Z\leq q-1} \int_{\Sigma_0}w_0^2\Bigl(|Z\Theta|^2 + \coeff|n_0Z\Theta|^2 + \coeff| \nabla_{\mathrm{aux}}Z\Theta|^2\Bigr) \, dx.
\ee
We denote the initial-data size by \(\varepsilon>0\). The scalar initial rescalings are \(\fieldPhi_{0,\coeff}\coloneqq\coeff\fieldphi_{0,\coeff}\) and \(\fieldvarphi_{0,\coeff}\coloneqq\coeff^{1/2}\fieldphi_{0,\coeff}\). The notation \(n_0\fieldPhi_{0,\coeff}\) means the initial normal derivative of the spacetime field, namely \((n_0\fieldPhi_\coeff)|_{\Sigma_0}=\coeff\fieldphi_{1,\coeff}\).

\begin{definition}
\label{definition:Einstein-compatible-commuted-preparation}
Let \(\mathbf d_0\) be an Einstein--massless-scalar initial data set, and let  \(\mathbf d_\coeff^{\mathrm{gm}} \) denote the metric--matter component of the \(f_\coeff(R) \) data.  The metric--matter tuple is defined as 
\be
\label{equa:metric-matter-initial-tuple}
\mathbf d_\coeff^{\mathrm{gm}} \coloneqq(g_{0, \coeff},k_{0, \coeff}, \psi_{0, \coeff}, \psi_{1, \coeff}).
\ee
The family is \textbf{Einstein-compatible at order \(q_\star\)} relative to \(\mathbf d_0\) if there is a modulus \(\omega_\coeff\to0\) such that
\be
\label{equa:commuted-Einstein-compatibility}
\| \mathbf d_\coeff^{\mathrm{gm}}-\mathbf d_0\|_{H^{q_\star}_{\mathrm{AE}}} + \| \fieldvarphi_{0, \coeff} \|_{H^{q_\star}_{\mathrm{AE}}} + \| \fieldPhi_{0, \coeff} \|_{H^{q_\star}_{\mathrm{AE}}} + \|n_0\fieldPhi_{0, \coeff} \|_{H^{q_\star-1}_{\mathrm{AE}}} \leq\omega_\coeff\varepsilon.
\ee
An Einstein-compatible family consists of \textbf{well-prepared initial data sets at order \(q_\star\)} if \(\mathrm{Prep}^{\mathrm{com}}_{\coeff,q_\star}[\Theta] \to0\). It has \textbf{preparation rate \(\sigma>0\)} if \(\mathrm{Prep}^{\mathrm{com}}_{\coeff,q_\star}[\Theta] \leq C\coeff^\sigma \varepsilon\).
\end{definition}

The square of the preparation amplitude controls the initial singular trace-residual energy used in Section~\ref{section-15}: the weight \(w_0\simeq1+r\) dominates the energy weights, and the fixed initial geometry compares the derivatives. The operators \(Z=\del^I L^J\) include time derivatives, whose initial values are determined recursively by the reduced equations. Preparation therefore requires the corresponding higher normal jets, as well as the value and first normal derivative, to approach the Einstein relation. This is an initial-data condition.

The strength of this condition can be seen in the homogeneous oscillator \(A_\coeff\cos(t/\sqrt{3\coeff})\): each time differentiation costs a factor of order \(\coeff^{-1/2}\). At order \(q\), its commuted preparation has size comparable to \(|A_\coeff|\coeff^{-(q-1)/2}\), with constants depending on \(q\). High-order preparation suppresses this rapid mode strongly enough to yield convergence of unscaled time derivatives. The model illustrates the derivative count; the nonlinear preparation condition remains \eqref{equa:commuted-preparation-amplitude}.


\subsection{\texorpdfstring{Constrained families with a vacuum limit}{Constrained families with a vacuum limit}}
\label{section:prepared-vacuum-examples}

The constraint and preparation conditions admit geometrically nontrivial examples. The construction below also permits a nonzero effective field, with a mass fixed throughout the parameter-dependent family.

\begin{proposition}[Nontrivial well-prepared families]
\label{prop:nontrivial-small-mass-prepared-data}
Let \(N\geq19\), \(q_\star=N-6\), and \(0<\coeff_0\leq1\). On the fixed initial hypersurface, the constrained-data class of Theorem~\ref{theorem:final-global-stability-and-limit} contains geometrically nontrivial families with arbitrarily small positive mass \(m\), independent of \(\coeff\).
\begin{itemize}
\item For every admissible family of theories, such data can be chosen independent of \(\coeff\), with both scalar fields zero. Einstein compatibility and preparation then hold with zero discrepancy.
\item If the admissible profile \(f_*\) is smooth, such families can also be chosen with a nonzero effective-field Cauchy pair for every \(\coeff>0\), and with zero matter field. One can arrange \(\omega_\coeff\leq\coeff\) and \(\mathrm{Prep}^{\mathrm{com}}_{\coeff,q_\star}[\Theta]\leq\coeff\varepsilon\).
\end{itemize}
The second assertion applies, in particular, to the Starobinsky family.
\end{proposition}

\begin{proof}

\noindent{\bf 1.} \emph{Small vacuum data.} We first work on a Euclidean slice with coordinates \(y\in\mathbb R^3\). We choose a nonzero, smooth, compactly supported, even symmetric tensor \(\sigma\) satisfying \(\del^i\sigma_{ij}=0\) and \(\delta^{ij}\sigma_{ij}=0\). Such tensors can be constructed explicitly: with \(P_{ij}\coloneqq\delta_{ij}\Delta-\del_i\del_j\), a constant symmetric matrix \(B\), and an even compactly supported smooth function \(b_0\), we set
\bse
\be
\label{equa:constrained-TT-tensor}
\sigma_{ij}\coloneqq P_{ia}P_{jb}(B_{ab}b_0)-\tfrac12P_{ij}P_{ab}(B_{ab}b_0).
\ee
The identities \(\del^iP_{ij}=0\), \(P_{ia}P_{ib}=\Delta P_{ab}\), and \(P_{ii}=2\Delta\) give the two required constraints. A choice of \(B\) and \(b_0\) for which the polynomial symbol is nonzero gives \(\sigma\not\equiv0\). For a small \(a>0\), we solve
\be
\label{equa:constrained-vacuum-conformal-data}
\gamma_a=\chi_a^4\delta,\qquad K_a=a\chi_a^{-2}\sigma,\qquad \Delta\chi_a=-\tfrac18a^2|\sigma|^2\chi_a^{-7},\qquad\chi_a\longrightarrow1.
\ee
Indeed, the Newton potential map is
\be
\label{equa:constrained-Newton-map}
\chi(y)\longmapsto1+\frac{a^2}{32\pi}\int_{\mathbb R^3}\frac{|\sigma(z)|^2\chi(z)^{-7}}{|y-z|}\,dz.
\ee
For small \(a\), this is a contraction on \(1\leq\chi\leq1+C a^2\); elliptic regularity gives a smooth positive fixed point, depending smoothly on \(a\). Direct substitution gives \(\operatorname{tr}_{\gamma_a}K_a=0\), \(\operatorname{div}_{\gamma_a}K_a=0\), and \(R_{\gamma_a}=|K_a|_{\gamma_a}^2\), so the vacuum constraints hold. The ADM momentum is zero, and the mass is
\be
\label{equa:constrained-vacuum-mass}
m(a)=\frac{a^2}{16\pi}\int_{\mathbb R^3}|\sigma|^2\chi_a^{-7}\,dy=c_\sigma a^2+\Ocal(a^4),\qquad c_\sigma\coloneqq\frac1{16\pi}\int_{\mathbb R^3}|\sigma|^2\,dy>0.
\ee
\ese
These data are geometrically nontrivial. Their Codazzi tensor has leading term \(a(\del_i\sigma_{jk}-\del_j\sigma_{ik})\), which is not identically zero: if it vanished, the divergence-free condition would make every compactly supported component of \(\sigma\) harmonic and hence zero. Thus the vacuum development has nonzero curvature for small \(a>0\).

\noindent{\bf 2.} \emph{Asymptotics, the initial hypersurface, and the gauge.} We write \(R=|y|\). The compact even source gives \(\chi_a=1+m(a)/(2R)+\Ocal(a^2R^{-3})\), with the corresponding differentiated bounds. Outside a fixed ball, the radial change \(r=R+m(a)^2/(4R)\), smoothly extended inside, converts the Schwarzschild monopole to the harmonic radial coordinate used in the reference metric. Consequently, \(\gamma_a-\gamma_{\mathrm{Sch},m(a)}=\Ocal(a^2r^{-2})\), with one additional inverse power for each spatial derivative. We choose the lapse equal to the Schwarzschild lapse near infinity and the shift zero there, and impose the initial wave-gauge conditions. The resulting metric remainder has order \(r^{-2}\), its first jets have order \(r^{-3}\), and the vacuum reduced equations give the same differentiated decay at every finite order used here. Multiplication by the spatial factors in Lorentz boosts preserves the order needed for the initial weighted energies; their radial integrals are bounded by a constant times \(\int_1^\infty r^{2\lambda-4}\,dr\), which is finite for the weights of this memoir.

The prescribed initial hypersurface is a fixed compact deformation of a Euclidean time slice. For sufficiently small \(a\), the local vacuum Cauchy theory gives a development on a fixed bounded time interval containing this graph; its spacelikeness persists by continuity. Restriction to the graph gives small constrained data there. This uses only local existence and continuous dependence, as in~\cite[Chapter~VI]{ChoquetBruhat-2009}. We then assign the prescribed initial lapse and shift on this graph and reconstruct the reduced jets in wave gauge. All these operations are independent of \(\coeff\) for the vacuum family. Taking both scalar fields zero gives a family of constrained data for every admissible \(f_\coeff\), since \(f_\coeff(0)=0\) and \(f_\coeff'(0)=1\). Every commuted component of \(\Theta\) is zero. The initial energies are \(\Ocal(a)\), whereas \(m(a)\) is positive and comparable to \(a^2\). The reference amplitude can be chosen proportional to \(m(a)\), so the reference-mass condition in the theorem also holds.

\noindent{\bf 3.} \emph{Nonzero effective data with the same mass.} We now assume that \(f_*\) is smooth and fix a small \(a_0>0\), with \(m=m(a_0)\). We choose an even compactly supported smooth function \(b\not\equiv0\). For each fixed \(\coeff>0\), we prescribe on the preliminary Euclidean slice \(\fieldphi_0=\xi b\), \(\fieldphi_1=\psi_0=\psi_1=0\), and keep \(\gamma=\chi^4\delta\), \(K=a\chi^{-2}\sigma\). The momentum constraint still holds. The Hamiltonian constraint~\eqref{consttransHan} becomes
\bse
\be
\label{equa:constrained-effective-conformal-equation}
\Delta\chi=-\tfrac18a^2|\sigma|^2\chi^{-7}-\tfrac3{16}\coeff^2\xi^2\chi|\nabla_\delta b|^2-\tfrac18\coeff\chi^5\Ufrak_\coeff(\xi b).
\ee
At fixed \(\coeff\), the Newton potential construction gives a smooth branch near \(\xi=0\). Its mass satisfies \(m_\coeff(a,\xi)=m(a)+O_\coeff(\xi^2)\), because \(\Ufrak_\coeff(0)=\Ufrak_\coeff'(0)=0\). By~\eqref{equa:constrained-vacuum-mass}, \(m'(a_0)>0\). The implicit function theorem therefore gives
\be
\label{equa:constrained-fixed-mass-branch}
a=a_\coeff(\xi)=a_0+O_\coeff(\xi^2),\qquad m_\coeff(a_\coeff(\xi),\xi)=m.
\ee
\ese
The conformal factors are even. Equality of the masses cancels their \(r^{-1}\) difference, and parity removes the dipole term. Thus their difference is \(O_\coeff(\xi^2r^{-3})\), with all differentiated bounds. In particular, the metric difference belongs to the space \(H^{q_\star}_{\mathrm{AE}}\) of the equation~\eqref{equa:weighted-AE-Sobolev-topology}, including its zeroth-order weight \(w_0\simeq1+r\).

\noindent{\bf 4.} \emph{Preparation of all required initial jets.} At each fixed \(\coeff>0\), local continuous dependence around the vacuum development gives the same bounded interval needed to reach the prescribed graph, provided \(\xi\) is sufficiently small; the fixed-parameter local theory and its weighted difference estimate are given in Lemma~\ref{lem:seven-local-wave-gauge-propagation}. The effective field has compact causal support on this interval. Outside that support the equations are vacuum equations. The finite-time difference estimate with the weight \(w_0\), whose logarithmic derivatives are bounded, preserves the metric comparison in \(H^{q_\star}_{\mathrm{AE}}\) and the differentiated weighted norms, with constants depending on the fixed parameter and interval. We pull back the geometric data to the fixed graph, assign the same initial lapse and shift throughout the family, and determine their remaining reduced jets by the wave-gauge conditions. These finite-order operations are continuous at fixed \(\coeff\).

We denote by \(\mathcal N_\coeff(\xi)\) the sum of the left-hand side of~\eqref{equa:commuted-Einstein-compatibility}, the preparation amplitude at order \(q_\star\), and the seminorms of the differences from the vacuum data in the finitely many initial spaces required by the stability theorem. Here the limiting metric--matter data are the vacuum data just constructed. The \(r^{-3}\) difference controls the weighted metric--matter comparison, and the compact scalar support controls all scalar jets, including the normal derivatives determined by the reduced equations. Hence \(\mathcal N_\coeff(\xi)\to0\) as \(\xi\to0\), for each fixed \(\coeff>0\). We choose \(\xi_\coeff>0\) small enough that
\be
\label{equa:constrained-diagonal-preparation}
\mathcal N_\coeff(\xi_\coeff)\leq\coeff\varepsilon.
\ee
The vacuum data can first be chosen strictly inside all initial smallness bounds, so this selection preserves those bounds uniformly in \(\coeff\). It gives the required compatibility and preparation estimates with the same fixed positive mass \(m\). The effective Cauchy pair on the graph is not identically zero: otherwise backward uniqueness would force \(\xi_\coeff b=0\) on the preliminary slice. Since the effective field vanishes near spatial infinity, the physical and conformal ADM masses coincide. This proves both assertions.

\end{proof}


\subsection{\texorpdfstring{Constrained families with a nonzero matter limit}{Constrained families with a nonzero matter limit}}
\label{section:prepared-matter-examples}

The following construction keeps matter in the limiting Einstein system. Radial symmetry permits an exact adjustment of the mass, while a finite expansion of the initial jets enforces the commuted preparation condition.

\begin{proposition}[Well-prepared families with a nonzero matter limit]
\label{prop:nonzero-matter-prepared-families}
Suppose the admissible profile \(f_*\) is smooth, and fix \(N\geq19\) and \(q_\star=N-6\). After decreasing \(\coeff_0\) if necessary, the constrained-data class contains well-prepared families with a fixed, arbitrarily small positive mass and a nonzero limiting massless scalar field. The scalar curvature of the limiting Einstein spacetime is nonzero somewhere on the initial hypersurface. The families can be chosen so that
\be
\label{equa:nonzero-matter-preparation-rate}
\omega_\coeff\leq C\sqrt\coeff,\qquad \mathrm{Prep}^{\mathrm{com}}_{\coeff,q_\star}[\Theta]\leq C\coeff\varepsilon.
\ee
The metric differences from the limiting data have compact support on the initial hypersurface. All the initial norms required by the stability theorem are uniformly small, with constants uniform as the prescribed mass tends to zero.
\end{proposition}

The mass is fixed with respect to \(\coeff\). In the construction, the matter amplitude \(a_*>0\) satisfies \(m\simeq a_*^2\), while the initial-data size can be chosen as \(\varepsilon\simeq a_*\). Thus \(\varepsilon^2/m\) stays bounded above and below by positive constants; its smallness is not a hypothesis of the construction or of the evolution argument in Section~\ref{section:small-mass-moving-interface}.

\begin{proof}

\noindent{\bf 1.} \emph{Radial constraints and a prescribed mass.} We first construct time-symmetric data on a Euclidean slice. We choose a nonconstant smooth radial function \(b\), compactly supported in an annulus \(R_1<|y|<R_2\) outside the compact deformation of the prescribed initial graph. With \(a>0\), we prescribe \(\psi_0=ab\), \(\psi_1=\fieldphi_1=0\), \(k_0=0\), and take the initial spatial metric to be \(\chi^4\delta\). The effective-field value \(v=\fieldphi_0\) will be a smooth radial function supported in the same annulus. The momentum constraint vanishes, and the Hamiltonian constraint is
\bse
\be
\label{equa:nonzero-matter-radial-constraint}
\Delta\chi=-\pi a^2e^{-\coeff v}\chi|\nabla b|^2-\frac3{16}\coeff^2\chi|\nabla v|^2-\frac\coeff8\chi^5\Ufrak_\coeff(v),\qquad \chi\longrightarrow1.
\ee
The Newton potential is a contraction near \(\chi=1\) for small \(a\) and \(v\), in each fixed finite regularity class used below. The solution and its mass depend smoothly on their parameters, including \(\coeff=0\). This follows also by differentiating the contraction equation: its derivative in \(\chi\) has norm smaller than one. The smooth extension at \(\coeff=0\) follows from \(\Ufrak_\coeff(v)=v^2/2+\Ocal(\coeff v^3)\). Radial symmetry and compact support give
\be
\label{equa:nonzero-matter-radial-asymptotics}
\chi=c\quad(|y|<R_1),\qquad \chi=1+\frac{m}{2|y|}\quad(|y|>R_2).
\ee
At \(\coeff=0\), the effective field disappears from the constraint, and we have
\be
\label{equa:nonzero-matter-mass-derivative}
m_0(a)=\frac{a^2}{2}\int_{\mathbb R^3}\chi_0(a)|\nabla b|^2\,dy=c_ba^2+\Ocal(a^4),\qquad c_b\coloneqq\frac12\int_{\mathbb R^3}|\nabla b|^2\,dy>0.
\ee
\ese
We fix a small \(a_*>0\) and set \(m=m_0(a_*)\). We apply the implicit function theorem with the squared amplitude \(a^2\) as parameter; the derivative of the mass with respect to this parameter at zero is \(c_b>0\). Thus the correction introduces no inverse power of the small amplitude. For the effective values of size \(\Ocal(a_*^2)\) constructed below, the effective terms and the change of the matter coupling in the constraint have size \(\Ocal(\coeff a_*^4)\). The fixed-mass branch therefore satisfies \(a_\coeff^2=a_*^2+\Ocal(\coeff a_*^4)\), or \(a_\coeff=a_*+\Ocal(\coeff a_*^3)\). The conformal factor is identical to \(1+m/(2|y|)\) outside the annulus.

\noindent{\bf 2.} \emph{Finite expansion of the initial jets.} We describe a finite construction, with \(L=2N+4\). The smooth admissible scaling gives \(W_\coeff(z)=\coeff z^2\mathcal W_\coeff(\coeff z)\). In the reduced equations, the metric and matter equations at coefficient \(\coeff^j\) therefore depend on effective-field coefficients only through order \(j-1\). The coefficient of order \(j\) in the effective-field equation instead determines that effective-field coefficient algebraically.

More explicitly, we construct formal coefficients \(g_j,\psi_j,v_j\), for \(0\leq j\leq L\), on a neighborhood of the preliminary slice. We choose a fixed radial initial lapse, equal to one in the inner flat region and to the Schwarzschild lapse in the exterior, and zero initial shift. The initial wave-gauge conditions are imposed in coordinates invariant under time reversal: scalar fields, spatial metric components, and the time--time component are even, while the mixed time--space components are odd. At order zero, \((g_0,\psi_0)\) is the Einstein--scalar development of the radial time-symmetric data above, and
\bse
\be
\label{equa:nonzero-matter-leading-effective-field}
v_0=8\pi g_0^{\alpha\beta}\del_\alpha\psi_0\del_\beta\psi_0,\qquad v_0|_{\Sigma}=8\pi\chi_0^{-4}a_*^2|\nabla b|^2.
\ee
Suppose the coefficients below order \(j\) have been constructed. The coefficient of order \(j\) in the radial constraint and the fixed-mass equation determines the initial metric coefficient and the amplitude coefficient. These two equations use only \(v_0,\ldots,v_{j-1}\), because every occurrence of \(v\) in the constraint has a factor \(\coeff\). The coefficient equations for \((g_j,\psi_j)\) are the linearized reduced Einstein--scalar equations, with known smooth sources and these Cauchy data; the initial wave-gauge relations determine their coordinate jets. Local linear hyperbolic existence, or its finite Cauchy-jet recursion, determines all the finite jets needed here. Writing \([\coeff^j]\) for extraction of the coefficient of \(\coeff^j\), we define the effective coefficient by
\be
\label{equa:nonzero-matter-effective-recursion}
v_j=\bigl[\coeff^j\bigr]\left(3\coeff\Box_g\fieldphi-\mathcal S-W_\coeff(\fieldphi)\right),
\ee
\ese
where the expression on the right uses the just constructed metric--matter coefficients and the previously constructed effective coefficients. There is no \(v_j\) on that right-hand side: its occurrences in \(\mathcal S=-8\pi e^{-\coeff\fieldphi}g(d\psi,d\psi)\) and \(W_\coeff\) carry a positive power of \(\coeff\). This proves the triangular character of the recursion. All scalar coefficients are even under time reversal, so their initial normal derivatives vanish. Their initial values and finite time jets are supported in the annulus. Indeed, all matter jets vanish off that annulus, and the effective recursion then vanishes there by induction.

We set \(v^{[L]}_\coeff\coloneqq\sum_{j=0}^L\coeff^jv_j|_\Sigma\), prescribe this effective value and zero effective normal derivative, and solve the exact radial constraint with mass \(m\). Its exact conformal factor and matter amplitude agree with the formal ones through order \(L\). In every fixed smooth norm needed for the initial jets, their errors are therefore \(\Ocal(\coeff^{L+1})\). The implicit function theorem in the squared amplitude and the finite recursion preserve the amplitude bounds: the matter coefficients are \(\Ocal(a_*)\), and the effective coefficients are \(\Ocal(a_*^2)\), with no inverse amplitude factors. The mass derivative with respect to \(a^2\) is at least \(c_b/2\), and the finitely many coefficient norms are bounded uniformly on a small amplitude interval. Thus a single interval \(0\leq\coeff\leq\coeff_0\) works for every \(0\leq a_*\leq a_0\), after fixing \(\coeff_0,a_0\) sufficiently small.

\noindent{\bf 3.} \emph{Commuted preparation and derivative losses.} The formal solution through order \(L\) satisfies the three reduced equations up to \(\Ocal(\coeff^{L+1})\), with every fixed finite derivative needed below. The scalar equation has the form
\bse
\be
\label{equa:nonzero-matter-normal-recursion}
\del_t^2\fieldphi=-\frac{2g^{0a}\del_{ta}\fieldphi+g^{ab}\del_{ab}\fieldphi}{g^{00}}+\frac{\fieldphi+\mathcal S+W_\coeff(\fieldphi)}{3\coeff g^{00}},
\ee
where \(-g^{00}\) is bounded away from zero on the fixed initial coefficient range. The metric and matter equations contain no negative power of \(\coeff\). We denote the degree-\(L\) formal truncations by \(g_\coeff^{[L]},\psi_\coeff^{[L]},\fieldphi_\coeff^{[L]}\), and use \(g_\coeff,\psi_\coeff,\fieldphi_\coeff\) for the exact jets determined by the reduced equations from the constrained data. For every fixed spatial order \(r\) required in the initial norms and \(0\leq\ell\leq N+2\), induction gives
\be
\label{equa:nonzero-matter-amplitude-jet-errors}
\begin{aligned}
\|\del_t^\ell(g_\coeff-g_\coeff^{[L]})\|_{H^r(\Sigma)}+\|\del_t^\ell(\fieldphi_\coeff-\fieldphi_\coeff^{[L]})\|_{H^r(\Sigma)}&\leq C_r\coeff^{L+1-\ell}a_*^2,\\
\|\del_t^\ell(\psi_\coeff-\psi_\coeff^{[L]})\|_{H^r(\Sigma)}&\leq C_r\coeff^{L+1-\ell}a_*.
\end{aligned}
\ee
The constants are uniform for small \(a_*\) and \(\coeff\). The differences of jets have compact support: outside a fixed ball the exact and formal metric data, mass, lapse, and shift coincide, the scalar fields vanish, and the local recursion of the vacuum equations gives identical jets. This justifies the global spatial norms in~\eqref{equa:nonzero-matter-amplitude-jet-errors}. At a scalar step the division costs at most one power of \(\coeff\), while two additional time derivatives are determined; the estimate allows the larger loss of one power for each time derivative. The initial errors have order \(\coeff^{L+1}\), and the formal scalar equation error after division has order \(\coeff^L\). The metric corrections and effective-field coefficients are even in \(a_*\) and vanish to order two, whereas the matter coefficients are odd and vanish to order one. The equations and the squared-amplitude mass adjustment preserve these factors at each induction step. This proves the estimates with their amplitude dependence. Since \(L+1-(N+2)=N+3\), all the initial jets used by the order-\(N\) energies remain uniformly bounded.

The leading coefficient of \(\Theta=\fieldphi+\mathcal S\) is identically zero, not merely zero on the slice, by the definition of \(v_0\). Its formal initial derivatives are consequently \(\Ocal(\coeff a_*^2)\). Differentiating \(\Theta\) through order \(q_\star\) requires matter derivatives through order \(q_\star+1=N-5\), because \(\mathcal S\) is quadratic in first derivatives of \(\psi\). These orders lie within the jet comparison through \(N+2\). Since \(L=2N+4\), the exact-jet error is smaller at every derivative order used in the preparation amplitude. The coefficients of translations and boosts, and all spatial weights, are bounded on the annulus. Consequently, we obtain
\be
\label{equa:nonzero-matter-full-preparation}
\sum_{\operatorname{ord}Z\leq q_\star-1}\left(\|Z\Theta\|_{H^1(\Sigma)}+\|nZ\Theta\|_{L^2(\Sigma)}\right)\leq C\coeff a_*^2.
\ee
\ese
All these quantities vanish outside a fixed annulus, so the same estimate holds with the preparation weight \(w_0\). In the metric--matter comparison, the fixed mass makes the initial differences compactly supported. The effective fields satisfy \(\fieldPhi=\Ocal(\coeff a_*^2)\) and \(\fieldvarphi=\Ocal(\sqrt\coeff a_*^2)\). These bounds give the two rates in the statement, with \(\varepsilon\) comparable to \(a_*\).

\noindent{\bf 4.} \emph{The prescribed initial graph.} We choose the annulus far enough beyond the compact set on which that graph differs from a Euclidean time slice. In its interior the constructed data are \((c_\coeff^4\delta,0,0,0,0,0)\). They are induced by the explicit flat metric \(-d\tau^2+c_\coeff^4dy^2\). We replace the preliminary slice by the prescribed graph in this flat region and keep it unchanged near the annulus and in the exterior. The induced metric and second fundamental form satisfy the vacuum constraints there, and they agree with the preliminary data on the overlap where the graph is flat. Thus the graph gives an exact global constrained data set. Its spacelikeness follows from \(c_\coeff\to1\) as the data size tends to zero and the fixed strict spacelikeness of the graph. A fixed radial coordinate change, determined by the prescribed mass outside and equal to the identity near the inner graph, gives the harmonic Schwarzschild coordinate in the exterior. We assign the same initial lapse and shift throughout the family and reconstruct the reduced jets from the wave gauge. The scalar fields and all their jets vanish on the deformed inner portion; on their annular support the preliminary slice and the final graph coincide. The preparation estimate is therefore unchanged, up to fixed comparison constants.

The radial coordinate change takes place away from the Schwarzschild radius; its derivatives through the required orders are bounded uniformly for \(0\leq m\leq m_0\), with \(m_0\) fixed and small. The initial wave-gauge reconstruction is a smooth function of the bounded metric jets and the chosen lapse and shift. These transformations therefore preserve the preparation and compatibility bounds with constants uniform in \(\coeff\) and \(a_*\). All initial differences have compact support, and every order-\(N\) initial norm is small after reducing \(a_*\). The limiting matter field is nonzero, and \(v_0|_\Sigma>0\) wherever \(\nabla b\neq0\), so the Einstein scalar curvature is nonzero there. The mass is positive, fixed in \(\coeff\), and tends to zero with \(a_*\). This proves the proposition on a sufficiently small parameter interval, as required for the limit.

\end{proof}

Preparation supplements the uniform estimates for the coupled evolution. The metric and matter equations contain the effective field only through the controlled combinations \(\del\fieldPhi\), \(\fieldvarphi\), and the conformal factor. The singular estimate for \(\Theta\) identifies the limiting curvature. We derive the energy identities next and then the pointwise estimate along rays.


\section{Effective-field energy identities and \texorpdfstring{$\coeff$}{kappa}-dependent estimates}
\label{section-5}


\subsection{Energy scales and controlled quantities}
\label{section-5-1}

We derive the effective-field energy identities at order zero and after commutation, keeping the source, deformation, and boundary terms. Table~\ref{table-energy-hierarchy} gives the controlled quantities and their parameter dependence. The singular and rescaled energies express the same balance law at two normalizations. Their comparison with full normal and tangential derivatives is uniform on each fixed time interval; the global estimates keep the weights of the time-current energy.

A positive power of \(\coeff\) in an equation does not by itself imply that its contribution vanishes: for example, \(\coeff\fieldphi^2=\fieldvarphi^2\) can remain of order one. Convergence requires the additional preparation estimate, while the uniform bounds in the table control the nonlinear evolution.

\begin{table}[H]
\centering
\normalsize\renewcommand{\arraystretch}{1.25}
\begin{tabular}{
>{\raggedright\arraybackslash}p{.22\textwidth}
>{\raggedright\arraybackslash}p{.24\textwidth}
>{\raggedright\arraybackslash}p{.40\textwidth}}
\hline
Field & Controlled \(L^2\) quantities & \leavevmode Parameter conversion and role \\
\hline
Metric perturbation \(u\) & wave energy & Uniform with \(h^\star\) estimated separately \\
Matter field \(\psi\) & wave energy & Uniform matter control \\
 Effective field \(\fieldphi\) & \(\coeff| \nabla \fieldphi|^2 + | \fieldphi|^2\) & \(\del\fieldphi = \coeff^{-1} \del\fieldPhi = \coeff^{-1/2} \del\fieldvarphi\), fixed \(\coeff\) \\
Rescaled fields \(\fieldPhi, \fieldvarphi\) & \(| \nabla{\fieldPhi}|^2 + |{\fieldvarphi}|^2\) & \(\fieldPhi = \coeff\fieldphi\), \(\fieldvarphi = \coeff^{1/2} \fieldphi\), uniform \\
\hline
\end{tabular}
\caption{Energy scales and \(L^2\) conversion rules}
\label{table-energy-hierarchy}
\end{table}


\subsection{Basic energy identities}
\label{section-5-2}

\paragraph{Wave energy identity}

We consider a scalar field \(u\) and take \(X\) to be a future-directed timelike multiplier adapted to the Euclidean--hyperboloidal foliation. The standard \emph{wave stress-energy tensor} is
\be
\label{equa-wave-energy-tensor}
Q^{\mathrm{wave}}_{\alpha \beta}[u]  \coloneqq  \nabla_\alpha u\nabla_\beta u - {1\over2}g_{\alpha \beta}g(\nabla u, \nabla u).
\ee
If \(u\) satisfies \(\Box_g u=F\), then we obtain
\be
\label{equa-wave-current-divergence}
\nabla^\alpha \big( Q^{\mathrm{wave}}_{\alpha \beta}[u]X^\beta \big) = F\,Xu + {1\over2} Q^{\mathrm{wave} \, \alpha \beta}[u] \, \pi^{(X)}_{\alpha \beta},
\ee
where the deformation tensor of \(X\) is defined by
\be
\pi^{(X)}_{\alpha \beta}  \coloneqq  \nabla_\alpha X_\beta+ \nabla_\beta X_\alpha
\ee
These identities are \emph{uniform in the coefficient $\coeff$} and apply componentwise to the reduced metric equations and, with the appropriate mass term, to the matter equation.

\paragraph{Singular effective-field energy identity}

For the  effective field, the equation takes the form
\be
\label{equa-effective-field-energy-equation}
3\coeff\, \Box_g\fieldphi - \Ufrak'(\fieldphi) = F_\fieldphi,
\ee
where we define
\be
\label{equa-effective-field-energy-source}
F_\fieldphi  \coloneqq  8\pi e^{-2\coeff\fieldphi} \Tr_{\giveng}(\givenT).
\ee
For the scalar energy-momentum tensor \eqref{equa:Talphabeta}, this becomes
\be
F_\fieldphi = -8\pi e^{-\coeff\fieldphi}g(\nabla \psi, \nabla \psi) -32\pi e^{-2\coeff\fieldphi}V(\psi).
\ee
Thus \(F_\fieldphi =-8\pi e^{-\coeff\fieldphi}g(\nabla \psi, \nabla \psi) \) in the massless case $V \equiv 0$. The argument uses \eqref{equa-effective-field-energy-equation} and applies equally to more general sources satisfying the same integrability bounds.

The \emph{singular Klein--Gordon energy tensor} naturally associated with \eqref{equa-effective-field-energy-equation} is
\be
\label{equa-effective-field-energy-tensor}
Q^{\mathrm{eff}}_{\alpha \beta}[\fieldphi]  \coloneqq  3\coeff \Bigl( \nabla_\alpha \fieldphi\nabla_\beta \fieldphi - {1\over2}g_{\alpha \beta}g(\nabla \fieldphi, \nabla \fieldphi) \Bigr) - \Ufrak(\fieldphi)g_{\alpha \beta}.
\ee
The sign in the last term is consistent with the positive-potential convention fixed in Section~\ref{section-2}. In the small-field range used below, \(\Ufrak(\fieldphi) \geq0\), and the energy density of \eqref{equa-effective-field-energy-tensor} is positive. A direct calculation gives the divergence identity
\be
\label{equa-effective-field-energy-divergence}
\nabla^\alpha \bigl( Q^{\mathrm{eff}}_{\alpha \beta}[\fieldphi] \bigr) = \Bigl( 3\coeff\, \Box_g\fieldphi - \Ufrak'(\fieldphi) \Bigr) \nabla_\beta \fieldphi = F_\fieldphi\nabla_\beta \fieldphi.
\ee
Thus the coefficient $\coeff$ and the potential contribution are forced by the effective-field equation itself.

We denote by \(n_s\) the future-directed unit normal to a hypersurface of the Euclidean--hyperboloidal foliation and take \(X\) to be the multiplier introduced above. In the fixed spatial parametrization \(x\mapsto(T(s,x),x) \), let \(\gamma_s\) be the induced Riemannian metric and set
\be
\label{equa-KDL486}
j_s \coloneqq \sqrt{\det\gamma_s}, \qquad N_s \coloneqq j_sn_s.
\ee
Thus \(N_s\) is the future densitized normal relative to \(dx\), and every hypersurface flux is written below with the coordinate measure \(dx\).  Unless another measure is explicitly specified, all \(L^p(\Mcal_s) \), \(L^p(\Mint_s) \), and \(L^p(\MME_s) \) norms are taken with respect to \(dx\). If \((e_1,e_2,e_3) \) is any \(g\)-orthonormal frame tangent to \(\Mcal_s\), we define the full first-order energy derivative by
\be
\label{equa:foliation-derivative-norm}
| \Ds v|^2  \coloneqq  |n_sv|^2 + \sum_{A=1}^3|e_Av|^2.
\ee
Thus \(\Ds\) contains the future-normal derivative as well as the three intrinsic tangential derivatives; it is not the intrinsic gradient of \(\Mcal_s\).  Its definition does not depend on the chosen orthonormal tangent frame. We define the \emph{zero-order effective-field energy} by
\be
\label{equa-effective-field-energy-zero}
\calE_{\mathrm{eff}}^{\mathrm{sing}}(s, \fieldphi;X)  \coloneqq  \int_{\Mcal_s} Q^{\mathrm{eff}}_{\alpha \beta}[\fieldphi] \, \,N_s^\alpha X^\beta \, dx.
\ee
Here \(\Mcal_s\) denotes the whole hypersurface of the Euclidean--hyperboloidal foliation. Its interior part agrees with a hyperboloid, while its merging and exterior parts are asymptotically Euclidean.

We take \(X= \del_t\), as in the main estimates below. On each fixed interval \(s_0\leq s\leq {s_1}\), the foliation and metric bounds compare this multiplier and the densitized normal with the orthonormal frame. As long as \(| \coeff\fieldphi| \) remains sufficiently small, the energy density is positive definite and we obtain
\be
\label{equa-effective-field-energy-coercivity}
\calE_{\mathrm{eff}}^{\mathrm{sing}}(s, \fieldphi;X) \simeq_{s_1} \int_{\Mcal_s} \Bigl( 3\coeff\,| \Ds\fieldphi|^2 + 2\Ufrak(\fieldphi) \Bigr) \, dx \simeq \int_{\Mcal_s} \Bigl( \coeff| \Ds\fieldphi|^2 + | \fieldphi|^2 \Bigr) \, dx.
\ee
The first comparison has constants independent of \(\coeff\), but dependent on \({s_1}\); the subscript \({s_1}\) denotes this dependence throughout the energy comparisons below. In the flat hyperboloidal region, \(n_s= (t/s) \del_t+(x^a/s) \del_a \) and \(j_s=s/t\), so the comparison with the full normal-frame norm is not uniform as \({s_1}\to\infty\). The global estimates keep the weighted derivatives of the time-current energy. The second comparison is uniform in the small-field range and follows from
\be
\label{equa-effective-field-potential-quadratic}
\Ufrak(\fieldphi) = {1\over2} \fieldphi^2 + \Ocal(\coeff| \fieldphi|^3), \quad | \coeff\fieldphi| \ll1.
\ee
For each fixed \(\coeff>0\), \eqref{equa-effective-field-energy-coercivity} gives fixed-\(\coeff\) control of \(\fieldphi\) and \(\coeff^{1/2} \Ds\fieldphi\), as required for the nonlinear evolution. The estimate will be propagated and kept in our proof even though its derivative part degenerates in the general-relativistic limit.


\subsection{Zero-order singular and rescaled estimates}
\label{section-5-3}

\paragraph{Non-uniform effective-field energy}

We make explicit the energy estimate for $\fieldphi$ and, later on, all effective-field estimates.  We denote by
\be
\mathcal D_{[s_0,s]}  \coloneqq  \bigcup_{\tau\in[s_0,s]} \Mcal_\tau
\ee
the spacetime slab bounded by \(\Mcal_{s_0} \), \(\Mcal_s\), and, when the slab is truncated,  by the interior--exterior interface \(\Lscr_{[s_0,s]} \).  We define the energy current
\be
\label{equa-effective-field-energy-current}
J_\alpha^{(X)}[\fieldphi]  \coloneqq  Q^{\mathrm{eff}}_{\alpha \beta}[\fieldphi]X^\beta.
\ee
Combining \eqref{equa-effective-field-energy-divergence} with the deformation tensor identity gives
\be
\label{equa-effective-field-current-divergence}
\nabla^\alpha J_\alpha^{(X)}[\fieldphi] = F_\fieldphi\,X\fieldphi + {1\over2} Q^{\mathrm{eff} \, \alpha \beta}[\fieldphi] \, \pi^{(X)}_{\alpha \beta},
\ee
which we integrate over the spacetime slab. Here \(d\mu_g\coloneqq j_g\, dt\, dx\), with \(j_g\coloneqq\sqrt{| \det g|} \), is the spacetime volume measure. On a truncated slab, \(\mathsf F_{\mathrm{eff}}^{\Lscr} \) is the outward lateral flux of the vector current \(g^{\alpha \gamma}J_\gamma^{(X)} \); it is zero on a full slab. The exterior region has the opposite interface orientation.

\begin{proposition}[Zero-order effective-field energy identity]
\label{proposition-zero-order-effective-field-energy}
Let \(\fieldphi\) solve the equation~\eqref{equa-effective-field-energy-equation} on \(\mathcal D_{[s_0,s]}\), with sufficient regularity, finite energy, and integrable source and deformation terms. On a noncompact slab, assume that the boundary flux at spatial infinity vanishes along an exhaustion. Then one has
\be
\label{equa-zero-order-effective-field-balance}
\begin{aligned}
\calE_{\mathrm{eff}}^{\mathrm{sing}}(s,\fieldphi;X)-\mathsf F_{\mathrm{eff}}^{\Lscr}([s_0,s],\fieldphi;X) &=\calE_{\mathrm{eff}}^{\mathrm{sing}}(s_0,\fieldphi;X)\\
&\quad -\int_{\mathcal D_{[s_0,s]}}F_\fieldphi X\fieldphi\,d\mu_g-\frac12\int_{\mathcal D_{[s_0,s]}}Q^{\mathrm{eff}\,\alpha\beta}[\fieldphi]\pi^{(X)}_{\alpha\beta}\,d\mu_g.
\end{aligned}
\ee
Here \(\mathsf F_{\mathrm{eff}}^{\Lscr} \) is measured with the fixed interface orientation used in \eqref{equa:seven-exact-lateral-fluxes}.  The positive hypersurface energy is the negative of the final-hypersurface boundary flux of the geometric current \(Q^{\alpha}{}_{\beta}X^\beta \) for the signature \((-+++) \); Stokes' formula therefore gives the signs in \eqref{equa-zero-order-effective-field-balance}.  Equivalently, this identity is the effective-field analogue of the current identity \eqref{equa:eight-c-current-interior-identity}, divided by two.  The interface flux cancels with the exterior contribution when the two identities are written in that current normalization.
\end{proposition}

\begin{proof}
We first integrate the identity~\eqref{equa-effective-field-current-divergence} over a slab truncated at finite radius. The divergence theorem gives the final and initial energies, the oriented interface flux, and the outer boundary flux. We pass to the exhaustion in the hypotheses; absolute integrability gives convergence of the volume terms and finite energy gives convergence of the hypersurface integrals. The outer flux tends to zero, yielding the identity~\eqref{equa-zero-order-effective-field-balance}. In the nonlinear application, Section~\ref{section-15} justifies this exhaustion from the weighted energy bounds.
\end{proof}

The following unweighted estimate isolates the dependence on \(\coeff\). We assume here that \(X\) is uniformly timelike and that the lapse and the geometric energy-comparison factors are uniformly bounded on the slab under consideration. We also control the deformation tensor on every hypersurface. We apply the identity either to the full slab or to the sum of the interior and exterior slabs, so that the two oriented interface fluxes cancel. Using Cauchy--Schwarz, we obtain
\be
\label{equa-zero-order-effective-field-inequality}
\aligned \calE_{\mathrm{eff}}^{\mathrm{sing}}(s, \fieldphi;X)^{1/2} & \lesssim \calE_{\mathrm{eff}}^{\mathrm{sing}}(s_0, \fieldphi;X)^{1/2} + {1\over \coeff^{1/2}} \int_{s_0}^{s} \|F_\fieldphi(\tau) \|_{L^2(\Mcal_\tau)} \, d\tau \\
& \quad + \int_{s_0}^{s} \| \pi^{(X)}(\tau) \|_{L^\infty(\Mcal_\tau)} \calE_{\mathrm{eff}}^{\mathrm{sing}} (\tau, \fieldphi;X)^{1/2} \, d\tau, \endaligned
\ee
with these geometric factors included in the comparison constant. On the hyperboloidal portion of the global foliation, the flat lapse is \(t/s\) and grows near the interface. The global argument therefore uses the weighted inequalities in Sections~\ref{section-9-7} and~\ref{section-13-3}. The unweighted inequalities below display the parameter scale under the same bounded-factor convention. The identity \eqref{equa-zero-order-effective-field-balance} itself keeps all geometric factors and applies on the full slab. Commutation adds source terms without changing the singular effective-field normalization.

The energy identity displays explicitly \emph{what is lost as} \(\coeff\to0\). From \eqref{equa-effective-field-energy-coercivity}, we have
\be
\calE_{\mathrm{eff}}^{\mathrm{sing}} \simeq_{s_1} \| \fieldphi\|_{L^2(\Mcal_s)}^2 + \coeff\| \Ds\fieldphi\|_{L^2(\Mcal_s)}^2.
\ee
Thus the unscaled field remains controlled in \(L^2\), while the estimate for its derivatives deteriorates according to
\be
\| \Ds\fieldphi\|_{L^2(\Mcal_s)} \lesssim_{s_1} \coeff^{-1/2} \calE_{\mathrm{eff}}^{\mathrm{sing}}(s, \fieldphi;X)^{1/2}.
\ee
This \emph{derivative loss} reflects the degeneration of the equation into an algebraic constraint. The singular energy controls the unscaled field; the rescaled energy controls its contribution to the metric and matter equations.

\paragraph{Uniform rescaled effective-field energy}

Multiplying \eqref{equa-effective-field-energy-coercivity} by $\coeff$ gives
\be
\label{equa-uniform-rescaled-energy}
\calE_{\mathrm{eff}}^{\mathrm{unif}}(s, \fieldphi;X)  \coloneqq  \coeff\, \calE_{\mathrm{eff}}^{\mathrm{sing}}(s, \fieldphi;X) \simeq_{s_1} \int_{\Mcal_s} \Bigl( | \Ds{\fieldPhi}|^2 + |{\fieldvarphi}|^2 \Bigr) \, dx,
\ee
where ${\fieldPhi} = \coeff\fieldphi$ and ${\fieldvarphi} = \coeff^{1/2} \fieldphi$.  In particular, this expression satisfies
\be
\label{equa-uniform-rescaled-controls}
\| \Ds{\fieldPhi} \|_{L^2(\Mcal_s)} + \|{\fieldvarphi} \|_{L^2(\Mcal_s)} \lesssim_{s_1} \calE_{\mathrm{eff}}^{\mathrm{unif}} (s, \fieldphi;X)^{1/2},
\ee
with constants independent of \(\coeff\in(0, \coeff_0] \). Multiplying \eqref{equa-zero-order-effective-field-balance} by $\coeff$ yields the corresponding uniform balance law:
\be
\label{equa-uniform-rescaled-balance}
\begin{aligned}
&\calE_{\mathrm{eff}}^{\mathrm{unif}}(s,\fieldphi;X)-\coeff\mathsf F_{\mathrm{eff}}^{\Lscr}([s_0,s],\fieldphi;X)\\
&=\calE_{\mathrm{eff}}^{\mathrm{unif}}(s_0,\fieldphi;X)-\int_{\mathcal D_{[s_0,s]}}F_\fieldphi X\fieldPhi\,d\mu_g-\frac\coeff2\int_{\mathcal D_{[s_0,s]}}Q^{\mathrm{eff}\,\alpha\beta}[\fieldphi]\pi^{(X)}_{\alpha\beta}\,d\mu_g.
\end{aligned}
\ee
The source now couples to \(X{\fieldPhi} \), which is the uniformly controlled derivative variable. This is the form of the estimate that enters the uniform bootstrap argument. It is a rescaled version of the same balance law. It does not provide parameter-uniform control of the unscaled derivative \(\nabla \fieldphi\).

We propagate both normalizations: the singular energy controls \(\fieldphi\), while \eqref{equa-uniform-rescaled-energy} controls the rescaled fields in the metric and matter equations.


\subsection{High-order effective-field energies}
\label{section-5-4}

\paragraph{Commuted effective-field equation}
At positive differential order, we isolate the linear massive operator \(3\coeff \, \Box_g-1\) and treat the nonlinear remainder of the potential as a source. This makes every power of $\coeff$ explicit and avoids commuting the full nonlinear potential energy.  For multi-indices \(I,J\), we set
\be
\label{equa-commuted-effective-field}
\fieldphi_{I,J} \coloneqq \del^I L^J\fieldphi, \quad {\fieldPhi}_{I,J} \coloneqq \del^I L^J{\fieldPhi} = \coeff\fieldphi_{I,J}, \quad {\fieldvarphi}_{I,J} \coloneqq \del^I L^J{\fieldvarphi} = \coeff^{1/2} \fieldphi_{I,J}.
\ee
Since $\coeff$ is a constant parameter, no derivative of $\coeff$ is created by commutation. We use
\be
[Z, \Box_g] \coloneqq Z\Box_g-\Box_g Z,
\ee
and obtain immediately the following equation.

\begin{lemma}[\texorpdfstring{Commuted effective-field equation}{Commuted effective-field equation}]
For every \(Z= \del^I L^J\), the  commuted effective field satisfies the following singular Klein--Gordon equation:
\be
\label{equa-commuted-effective-field-equation}
3\coeff\, \Box_g\fieldphi_{I,J}-\fieldphi_{I,J} = \mathcal N_{I,J}^{\mathrm{eff}},
\ee
where the commuted source is defined by
\be
\label{equa-commuted-effective-field-source}
\mathcal N_{I,J}^{\mathrm{eff}}  \coloneqq  \del^I L^J\mathcal S[\psi, \fieldphi,g] + \del^I L^J W_\coeff(\fieldphi) - 3\coeff[\del^I L^J, \Box_g] \fieldphi.
\ee
\end{lemma}

The commutator term carries the explicit factor \(3\coeff\), and we have
\be
\label{equa-commutator-Phi-rewriting}
3\coeff[\del^I L^J, \Box_g] \fieldphi = 3[\del^I L^J, \Box_g]{\fieldPhi}.
\ee

In wave coordinates, \(\Box_g= \BoxChapeau_g\). We write \(g^{\alpha \beta} = (\gMink)^{\alpha \beta} +H^{\alpha \beta} \). For a translation or Lorentz boost \(X\), the commutator is
\be
\label{equa-curved-wave-commutator-structure}
[X, \Box_g] \fieldphi = (\mathcal L_XH)^{\alpha \beta} \del_\alpha \del_\beta \fieldphi,
\ee
where \(\mathcal L_XH\) is the Lie derivative of the contravariant tensor \(H\). For an ordered product \(Z=X_1\cdots X_d\), we denote by \(Z_A\) and \(\mathcal L_A\) the products over a subset \(A\subset\{1, \ldots,d\} \), in their original order. Repeated application of this identity yields
\be
\label{equa-curved-wave-commutator-coeff}
3\coeff[Z,\Box_g]\fieldphi=3\sum_{\varnothing\ne A\subset\{1,\ldots,d\}}(\mathcal L_AH)^{\alpha\beta}\del_\alpha\del_\beta Z_{A^c}\fieldPhi.
\ee
Each labelled nonempty subset occurs once; equal terms are combined with their Leibniz multiplicities when some operators coincide. The tensor Lie derivatives include the constant-matrix multiples of \(H\) arising from the boosts. Every term therefore involves a ``shorter'' operator on \(\fieldPhi\), including terms in which no Cartesian derivative falls on \(H\). This commutator has no negative power of \(\coeff\); the ordered identity is also used in Section~\ref{section-6-5}.

For the admissible scaling family of Definition~\ref{def:admissible-f-family}, with \(M\geq N+3\), the potential remainder has the factorization
\be
W_\coeff(\fieldphi)= \coeff\, \fieldphi^2\, \mathcal W_\coeff(\fieldPhi), \qquad \fieldPhi = \coeff\fieldphi,
\ee
where \(\mathcal W_\coeff\) is uniformly bounded in \(C^N([-c_0,c_0]) \). We denote by \(\operatorname{Part}(A)\) the set of partitions of the labelled subset \(A\), with the empty partition included when \(A=\varnothing\). For a scalar coefficient \(c\in C^d\) on the range of \(\fieldPhi\), repeated differentiation gives the composition formula
\be
\label{equa:audit-ordered-scalar-composition}
Z_A\bigl(c(\fieldPhi)\bigr)=\sum_{\pi\in\operatorname{Part}(A)}c^{(|\pi|)}(\fieldPhi)\prod_{B\in\pi}Z_B\fieldPhi.
\ee
The operators within every block keep their original order. For \(c(z)=e^{-az}\) with \(a>0\), the coefficient is \((-a)^{|\pi|}e^{-a\fieldPhi}\); for the potential remainder it is \(\mathcal W_\coeff^{(|\pi|)}(\fieldPhi)\). Applying the ordered product rule to its two additional field factors fixes every multiplicity in the following bound. For \(n=|I|+ |J| \leq N\), the chain rule and the ordered Leibniz formula give
\be
\label{equa-Wcoeff-commuted-estimate}
\big| \del^I L^J W_\coeff(\fieldphi) \big| \lesssim \sum_{\substack{2\leq q\leq n+2\\
(I_1,J_1) + \cdots+(I_q,J_q)= (I,J)}} \coeff^{q-1} C_{I_\ell,J_\ell} \prod_{\ell=1}^{q} \big| \del^{I_\ell}L^{J_\ell} \fieldphi\big|,
\ee
The finite sum runs over the distributions of translations and ordered boost operators among the factors; undifferentiated factors are allowed. The constants are uniform for \(\coeff\in(0, \coeff_0] \) and \(| \fieldPhi| \leq c_0\). Likewise, for every \(a>0\) and every positive differential order, we find
\be
\label{equa-conformal-weight-commuted}
\del^I L^J(e^{-a \coeff\fieldphi}) = e^{-a \coeff\fieldphi} \sum_{q=1}^{|I|+ |J|} \coeff^q \sum C_{a,I_\ell,J_\ell} \prod_{\ell=1}^{q} \del^{I_\ell}L^{J_\ell} \fieldphi.
\ee
Each differentiated effective-field factor in this expansion carries one power of \(\coeff\).

\paragraph{High-order degenerate estimate}

We use the energy tensor associated with the linear massive operator,
\be
\label{equa-linear-effective-field-energy-tensor}
Q^{\mathrm{lin}}_{\alpha \beta}[u]\coloneqq3\coeff\Bigl(\nabla_\alpha u\nabla_\beta u-\frac12g_{\alpha\beta}g(\nabla u,\nabla u)\Bigr)-\frac12u^2g_{\alpha\beta},
\ee
and define the commuted energy by
\be
\label{equa-high-order-degenerate-effective-field-energy}
\calE_{\mathrm{eff},N}^{\mathrm{sing}}(s,\fieldphi;X)\coloneqq\sum_{|I|+|J|\leq N}\int_{\Mcal_s}Q^{\mathrm{lin}}_{\alpha\beta}[\fieldphi_{I,J}]N_s^\alpha X^\beta\,dx.
\ee
At order zero, this energy replaces \(\Ufrak(\fieldphi)\) by \(\fieldphi^2/2\). It is equivalent to the nonlinear energy in the small-field range; the difference in the evolution equation is kept in the source \(W_\coeff(\fieldphi)\). With \(X= \del_t\) and \(s_0\leq s\leq {s_1}\), the same geometric comparison gives
\be
\label{equa-high-order-degenerate-coercivity}
\calE_{\mathrm{eff},N}^{\mathrm{sing}} \simeq_{s_1} \sum_{|I|+ |J| \leq N} \Bigl( \| \fieldphi_{I,J} \|_{L^2(\Mcal_s)}^2 + \coeff\| \Ds\fieldphi_{I,J} \|_{L^2(\Mcal_s)}^2 \Bigr).
\ee
We can extend Proposition~\ref{proposition-zero-order-effective-field-energy} as follows.

\begin{proposition}[High-order commuted effective-field identity]
\label{proposition-high-order-effective-field-energy}
Let \(\fieldphi\) solve the equation~\eqref{equa-effective-field-energy-equation} on \(\mathcal D_{[s_0,s]}\). Assume the regularity, integrability, and vanishing outer-flux hypotheses of Proposition~\ref{proposition-zero-order-effective-field-energy} for every commuted field of order at most \(N\). Then one has
\be
\label{equa-high-order-effective-field-balance}
\aligned & \calE_{\mathrm{eff},N}^{\mathrm{sing}}(s, \fieldphi;X) - \mathsf F_{\mathrm{eff},N}^{\Lscr} ([s_0,s], \fieldphi;X) \\
&= \calE_{\mathrm{eff},N}^{\mathrm{sing}}(s_0, \fieldphi;X) - \sum_{|I|+ |J| \leq N} \int_{\mathcal D_{[s_0,s]}} \mathcal N_{I,J}^{\mathrm{eff}}X\fieldphi_{I,J} \, d\mu_g - {1\over2} \sum_{|I|+ |J| \leq N} \int_{\mathcal D_{[s_0,s]}} Q^{\mathrm{lin} \, \alpha \beta}[\fieldphi_{I,J}] \pi^{(X)}_{\alpha \beta} \, d\mu_g \endaligned
\ee
Here \(\mathsf F_{\mathrm{eff},N}^{\Lscr} \) is the sum of the oriented commuted effective-field fluxes.  The energy estimate keeps it until its cancellation with the exterior flux. Consequently, one obtains
\be
\label{equa-high-order-effective-field-inequality}
\aligned \calE_{\mathrm{eff},N}^{\mathrm{sing}}(s, \fieldphi;X)^{1/2} & \lesssim \calE_{\mathrm{eff},N}^{\mathrm{sing}}(s_0, \fieldphi;X)^{1/2} + \coeff^{-1/2} \int_{s_0}^{s} \sum_{|I|+ |J| \leq N} \| \mathcal N_{I,J}^{\mathrm{eff}}(\tau) \|_{L^2(\Mcal_\tau)} \, d\tau \\
& \quad + \int_{s_0}^{s} \| \pi^{(X)}(\tau) \|_{L^\infty} \calE_{\mathrm{eff},N}^{\mathrm{sing}}(\tau, \fieldphi;X)^{1/2} \, d\tau. \endaligned
\ee
The inequality is applied after cancellation of the two interface fluxes; its constant depends on the geometric bounds above, \emph{ but not on \(\coeff\).} The factor \(\coeff^{-1/2} \) expresses the derivative scale of the unscaled energy.
\end{proposition}

\begin{proof}
\bse
The divergence of \(Q^{\mathrm{lin}}[\fieldphi_{I,J}] \) is
\be
\nabla^\alpha Q^{\mathrm{lin}}_{\alpha \beta}[\fieldphi_{I,J}] = \mathcal N_{I,J}^{\mathrm{eff}} \nabla_\beta \fieldphi_{I,J}.
\ee
Contracting with \(X\), integrating over \(\mathcal D_{[s_0,s]} \), and applying the divergence theorem yields \eqref{equa-high-order-effective-field-balance}. Moreover, we obtain
\be
\coeff^{1/2} \|X\fieldphi_{I,J} \|_{L^2(\Mcal_\tau)} \lesssim \calE_{\mathrm{eff},N}^{\mathrm{sing}}(\tau, \fieldphi;X)^{1/2},
\ee
which gives \eqref{equa-high-order-effective-field-inequality}.
\ese
\end{proof}

\paragraph{High-order uniform rescaled energy}

We obtain uniform energies by multiplying the singular energies by $\coeff$. We set
\be
\label{equa-high-order-uniform-effective-field-energy}
\calE_{\mathrm{eff},N}^{\mathrm{unif}}  \coloneqq  \coeff\, \calE_{\mathrm{eff},N}^{\mathrm{sing}}.
\ee
Consequently, we have
\be
\label{equa-high-order-uniform-coercivity}
\calE_{\mathrm{eff},N}^{\mathrm{unif}} \simeq_{s_1} \sum_{|I|+ |J| \leq N} \Bigl( \| \Ds{\fieldPhi}_{I,J} \|_{L^2(\Mcal_s)}^2 + \|{\fieldvarphi}_{I,J} \|_{L^2(\Mcal_s)}^2 \Bigr),
\ee
The comparison is uniform for \(\coeff\in(0, \coeff_0] \) on the fixed interval \([s_0,s_1] \). Multiplying \eqref{equa-high-order-effective-field-balance} by $\coeff$ gives
\be
\label{equa-high-order-uniform-balance}
\aligned & \calE_{\mathrm{eff},N}^{\mathrm{unif}}(s, \fieldphi;X) - \coeff\mathsf F_{\mathrm{eff},N}^{\Lscr} ([s_0,s], \fieldphi;X) \\
&= \calE_{\mathrm{eff},N}^{\mathrm{unif}}(s_0, \fieldphi;X) - \sum_{|I|+ |J| \leq N} \int_{\mathcal D_{[s_0,s]}} \mathcal N_{I,J}^{\mathrm{eff}}X{\fieldPhi}_{I,J} \, d\mu_g - {\coeff\over2} \sum_{|I|+ |J| \leq N} \int_{\mathcal D_{[s_0,s]}} Q^{\mathrm{lin} \, \alpha \beta}[\fieldphi_{I,J}] \pi^{(X)}_{\alpha \beta} \, d\mu_g. \endaligned
\ee
Hence we obtain
\be
\label{equa-high-order-uniform-inequality}
\aligned \calE_{\mathrm{eff},N}^{\mathrm{unif}}(s, \fieldphi;X)^{1/2} & \lesssim \calE_{\mathrm{eff},N}^{\mathrm{unif}}(s_0, \fieldphi;X)^{1/2} + \int_{s_0}^{s} \sum_{|I|+ |J| \leq N} \| \mathcal N_{I,J}^{\mathrm{eff}}(\tau) \|_{L^2(\Mcal_\tau)} \, d\tau \\
& \quad + \int_{s_0}^{s} \| \pi^{(X)}(\tau) \|_{L^\infty} \calE_{\mathrm{eff},N}^{\mathrm{unif}}(\tau, \fieldphi;X)^{1/2} \, d\tau. \endaligned
\ee
\emph{No negative power of $\coeff$ occurs} in this rescaled inequality.

\paragraph{Connection with the bootstrap hierarchy}

At every differential order we have the following structure:
\bei
\item
\(3\coeff[\del^I L^J, \Box_g] \fieldphi =3[\del^I L^J, \Box_g]{\fieldPhi} \);
\item
the commuted potential remainder carries a positive factor $\coeff$;
\item
each derivative of \(e^{-a \coeff\fieldphi} \) produces a positive power of $\coeff$;
\item
the unscaled energy inequality loses \(\coeff^{-1/2} \) on the source;
\item
the rescaled high-order inequality is uniform in $\coeff$.
\eei

The estimates above provide the \emph{fixed-parameter and rescaled effective-field normalizations} used later. The degenerate hierarchy controls \(\fieldphi\) and \(\coeff^{1/2} \nabla \fieldphi\), at the price of the explicit \(\coeff^{-1/2} \) loss on the source. The rescaled hierarchy controls ${\fieldvarphi}$ and $\nabla{\fieldPhi}$ without negative powers of $\coeff$. Section~\ref{section-8} combines these bounds with the metric and matter energies and with the pointwise estimates of Section~\ref{section-6}. The bootstrap norm will combine the uniform metric and matter energies, the degenerate effective-field energies, and the uniform rescaled effective-field energies. The assumptions and their improvement will be formulated after the commuted energy identities have been established.

\paragraph{Semiclassical interpretation}
Although \(\hslash\) is not Planck's constant in the present problem, the singular parameter admits a natural semiclassical interpretation.  Writing \(\coeff= \hslash^2\), the effective-field equation takes the singularly perturbed form \(3\hslash^2\Box_g\fieldphi-\fieldphi = \mathcal S[\psi, \fieldphi,g]+W_\coeff(\fieldphi) \), while the corresponding mass \(b_\coeff \coloneqq (3\coeff)^{-1/2} = (\sqrt{3} \, \hslash)^{-1} \) tends to infinity as \(\hslash\to0\).  The degenerate effective-field energy \(\| \fieldphi\|_{L^2}^2 + \hslash^2\| \nabla \fieldphi\|_{L^2}^2\) is therefore of semiclassical \(H^1_\hslash\)-type.  The rescaled variables \({\fieldvarphi} = \hslash\fieldphi\) and \({\fieldPhi} = \hslash^2\fieldphi\) yield the parameter-uniform and positive-power estimates required for the limit, whereas the unscaled energy is kept for the fixed-\(\hslash\) evolution. The hyperboloidal and Euclidean weights, as well as the boost hierarchy, are independent of \(\hslash\) and are thus compatible with this semiclassical energy scaling.


\section{Large-mass Klein--Gordon estimates}
\label{section-6}


\subsection{Singular equation and ray data}
\label{section-6-1}

\paragraph{Purpose}
This section derives raywise pointwise estimates for Klein--Gordon equations whose mass tends to infinity. These estimates are tailored to the effective field in \(f(R) \) gravity and are uniform in the singular parameter.

We consider equations of the form
\be
\label{equa:KG-singular-model}
-b^{-2} \BoxChapeau_g\rho+ \rho= \sourterm, \quad b^2={1\over3\coeff} \to + \infty.
\ee
Thus the general-relativistic limit \(\coeff\to0\) corresponds to \(b\to+ \infty\). Formally, the part $\BoxChapeau_g$ disappears in this limit and we expect to derive the algebraic relation
\be
\rho= \sourterm \quad \text{ when } \coeff \to 0.
\ee
The relevant unknown variable is taken to be the \emph{residual variable}
\be
\rhon \coloneqq \rho-\sourterm.
\ee
The purpose of the section is to obtain pointwise estimates for \(\rhon\), and for its normal derivative, that remain uniform as \(\coeff\to0\).

The argument separates three contributions:
\bei
\item the discrepancy from the limiting relation at the initial point of each ray;
\item the accumulated variation of the source along the ray;
\item the variation of the metric coefficient in the oscillator equation.
\eei
A ray starts either on the initial hyperboloid or on the interface \(\Lscr\), where its values are provided by the exterior estimates.

The proof proceeds by reduction along rays, a uniform oscillator estimate, recovery of the residual and its normal derivative, and commutation. Table~\ref{table-rescaled-variables} gives the variables used in this reduction. The traces on the initial hypersurface and at the interface remain explicit in the pointwise estimate.

\begin{table}[H]
\centering
\normalsize\renewcommand{\arraystretch}{1.18}
\begin{tabular}{>{\raggedright\arraybackslash}p{.24\textwidth}>{\raggedright\arraybackslash}p{.25\textwidth}>{\raggedright\arraybackslash}p{.41\textwidth}}
\hline
Representation & Scaling & Role in the estimates along rays \\
\hline
\(\rho\) & Klein--Gordon field & Unknown in the large-mass ODE \\
\(\rhon \coloneqq \rho-\sourterm\) & Discrepancy from \(\rho= \sourterm\) & Error in the limiting constraint \\
\(b^2={1\over3\coeff} \) & \(b\to+ \infty\) as \(\coeff\to0\) & Parameter in the uniform ODE bound \\
\hline
\end{tabular}
\caption{Variables specific to the raywise singular Klein--Gordon analysis}
\label{table-rescaled-variables}
\end{table}

\paragraph{Rays starting on the initial hyperboloid or at the interface} \label{sec:lower-ray-parameter}

We work in the interior domain \(\Mint\), foliated by the hyperboloids $ s^2=t^2-r^2$ with $ r=|x|. $ The relevant vector field is the Minkowski-normal ray derivative \(\newperp\) defined in \eqref{equa-normal-ray-field-section3}. It is orthogonal to the hyperboloids of the interior foliation and proportional to the scaling vector field. The decomposition below follows the integral curves of \(\newperp\).

We fix \((t_1,x_1) \in\Mint_{[s_0,s_1]} \), with $ x_1=r_1\omega$ and $\omega \in\mathbb S^2. $ The integral curve of \(\newperp\) through this point is
\be
t\mapsto y(t)=r(t) \omega, \quad r'(t)={r(t) \over t}, \quad r(t_1)=r_1,
\ee
and therefore we obtain
\be
r(t)={t\over t_1}r_1
\ee
and, by definition, the ratio \(r/t\) is constant along the curve.

In the Euclidean--hyperboloidal foliation, this curve is followed backward until it reaches the past boundary of the interior domain. Depending on the position of \((t_1,x_1) \),  the backward ray terminates \emph{either on the initial hyperboloid or on the interface \(\Lscr\)}. The latter case is essential here: the interior pointwise estimate must include \emph{interface values} (which are later controlled by the exterior analysis).

We denote this past terminal point by
\be
(t_{\mathrm{in}},x_{\mathrm{in}})= (t_{\mathrm{in}},r_{\mathrm{in}} \omega), \quad \lambda_{\mathrm{in}}^2=t_{\mathrm{in}}^2-r_{\mathrm{in}}^2.
\ee
If the curve reaches the interface \(\Lscr= \{r=t-1\} \), then we have
\be
r_{\mathrm{in}} =t_{\mathrm{in}}-1, \quad r_{\mathrm{in}} ={t_{\mathrm{in}} \over t_1}r_1,
\ee
and therefore we obtain
\be
\label{equa:tstar-sstar}
t_{\mathrm{in}} ={1\over1-r_1/t_1}, \quad \lambda_{\mathrm{in}}^2={1+ r_1/t_1\over1-r_1/t_1}.
\ee
Here \(\lambda_{\mathrm{in}} \) stands for \(\lambda_{\mathrm{in}}(\sqrt{t_1^2-r_1^2},x_1) \), with the function on the right defined in \eqref{equa:lower-ray-parameter-section3}. In the estimates below, all initial-hyperboloid or interface-trace contributions are measured at \(\lambda= \lambda_{\mathrm{in}} \).


\subsection{Reduction to an ODE along rays}
\label{section-6-2}

\paragraph{Hyperboloidal coordinate frame}

We use an overbar exclusively for the coordinates \((s,x) \), where \(t= \sqrt{s^2 +r^2} \), and define
\be
\overline v(s,x) \coloneqq v\bigl(\sqrt{s^2 +r^2},x\bigr),
\ee
so that the hyperboloidal coordinate derivatives defined in \eqref{equa:frame-convention-crosswalk} satisfy
\be
\label{equa:hyperboloidal-coordinate-frame-section6}
\delb_0= \del_s\big|_x={s\over t} \del_t, \qquad \delb_a= \del_a \big|_s= \delsH_a.
\ee
The semi-hyperboloidal derivatives themselves keep the notation \(\delH_0= \del_t\) and \(\delH_a= \delsH_a \) introduced in Section~\ref{section-3-4}. In particular, the barred derivative \(\delb_0\) is not the Cartesian derivative \(\del_t\).

\paragraph{Ray reduction}

Given a function \(v=v(t,x) \), we write \(\vb= \vb(s,x) \) when it is expressed in hyperboloidal variables.  For fixed \((s,x) \), we define
\be
\label{equa:rho-sx}
\rho_{s,x}(\lambda)  \coloneqq  \rhob\Bigl(\lambda,{\lambda \over s}x\Bigr) = \rho\Bigl( {\lambda \over s} \sqrt{s^2 + r^2}, {\lambda \over s}x \Bigr), \quad \lambda \in[\lambda_{\mathrm{in}},s].
\ee
Differentiating this identity, we obtain
\be
\label{equa:rho-sx-derivative}
\rho_{s,x}'(\lambda) = \Bigl(\delb_0 + {x_a \over s} \delb_a \Bigr) \rhob\Bigl(\lambda,{\lambda \over s}x\Bigr) = {1\over s} \sqrt{s^2 + r^2} \, (\newperp\rho) \Bigl( {\lambda \over s} \sqrt{s^2 +r^2},{\lambda \over s}x \Bigr).
\ee
We introduce the \emph{rescaled restriction}
\be
\label{equa:w-sx-def}
w_{s,x}(\lambda)  \coloneqq  \lambda^{3/2} \rhob\Bigl(\lambda,{\lambda \over s}x\Bigr).
\ee
The power \(\lambda^{3/2} \) is dictated by the Klein--Gordon scaling on hyperboloids. It cancels the leading first-order contribution generated by the radial decomposition and gives the \(s^{-3/2} \) decay.

We use the local hyperboloidal decomposition of~\cite[Lemma~3.4]{LeFloch-Ma-2016b}, keeping the mass explicit and keeping the values at the entry point of each ray. We write
\be
g^{\alpha \beta} = (\gMink)^{\alpha \beta} +H^{\alpha \beta},
\ee
and denote by \(H^{\mathcal H\, \alpha \beta} \) the components of the inverse metric perturbation in the semi-hyperboloidal frame of Section~\ref{section-3-4}.  We introduce the ray derivative
\be
\Lcal  \coloneqq  {t\over s} \del_t+ {x^a \over s} \del_a = \delb_0 + {x^a \over s} \delb_a.
\ee
A direct change-of-frame computation gives
\be
\label{equa:hyperboloidal-wave-decomposition-section6}
g^{\alpha \beta} \del_\alpha \del_\beta \rho = s^{-3/2} \Bigl({t\over s} \Bigr)^2 g^{\mathcal H\,00} \, \Lcal^2(s^{3/2} \rho) -\mathcal R_g[\rho],
\ee
where we define the remainder by
\be
\label{equa:R-flat-R2-R3}
\begin{aligned}
\mathcal R_g[\rho] &= -{3\over4}s^{-2} \rho -3s^{-1}{x^a \over s} \delsH_a \rho -{x^ax^b\over s^2} \delsH_a \delsH_b\rho -\sum_a \delsH_a \delsH_a \rho \\
& \quad + \Bigl({t\over s} \Bigr)^2H^{\mathcal H\,00} \Bigl( {3\over4}s^{-2} \rho + \bigl(3+(r/t)^2\bigr)t^{-1} \del_t\rho +3s^{-1}{x^a \over s} \delsH_a \rho \Bigr) \\
& \quad + \Bigl({t\over s} \Bigr)^2H^{\mathcal H\,00} \Bigl( {2x^a \over t} \del_t\delsH_a \rho + {x^ax^b\over s^2} \delsH_a \delsH_b\rho \Bigr) \\
& \quad -H^{\mathcal H\,a0} \delsH_a \del_t\rho -H^{\mathcal H\,0a} \del_t\delsH_a \rho -H^{\mathcal H\,ab} \delsH_a \delsH_b\rho -H^{\alpha \beta} \del_\alpha((\PsiH)_\beta{}^{\beta'}) \delH_{\beta'} \rho.
\end{aligned}
\ee
The signs follow from \(g^{-1} =g_{\mathrm{Mink}}^{-1} +H\): the radial principal term and the remainder cancel for a constant scalar field. Here \((\PsiH)_\beta{}^{\beta'} \) denotes the inverse semi-hyperboloidal change-of-frame matrix defined in Section~\ref{section-3-4}. The inverse metric satisfies
\be
g^{\mathcal H\,00} = -\Bigl({s\over t} \Bigr)^2 +H^{\mathcal H\,00},
\ee
and hence the coefficient of the raywise second derivative is
\be
\label{equa:raywise-metric-coefficient-section6}
a(t,x)  \coloneqq  1-\Bigl({t\over s} \Bigr)^2H^{\mathcal H\,00}(t,x).
\ee
The hyperboloidal coordinate frame is the one used in \cite[Section~2.2]{LeFloch-Ma-2016b}. In the ODE lemma of \cite[Section~3.4]{LeFloch-Ma-2016b}, however, the inverse-metric coefficient is introduced through \(g^{-1} =g_{\mathrm{Mink}}^{-1}-h_{\mathrm{LM}} \).  Thus, in the notation of that lemma, \(H=-h_{\mathrm{LM}} \), and the hyperboloidal component used there is \(\overline h_{\mathrm{LM}}^{00} = (t/s)^2h_{\mathrm{LM}}^{\mathcal H\,00} \); consequently, the factor denoted by \(a \) here is written \(1+ \overline h_{\mathrm{LM}}^{00} \) in that paper. For a source term \(\sourterm\), we set
\be
\label{equa:8333}
G_{s,x}(\lambda)  \coloneqq  \Big( \lambda^{3/2} \sourterm -b^{-2} \lambda^{3/2} \mathcal R_g[\rho] \Big) \Bigl({\lambda \over s}t,{\lambda \over s}x\Bigr).
\ee
This ODE reduction is uniform in the infinite-mass limit: formally, as \(b\to+ \infty\), the second derivative term in \eqref{equa:ODEKG} disappears and the equation becomes \(w_{s,x} =G_{s,x} \), which is the raywise form of the formal limiting relation \(\rho= \sourterm\).

\paragraph{Exact raywise ODE}

\begin{lemma}[ODE reduction of the singular Klein--Gordon equation]
\label{lem:KG-0}
Let \(\rho\) be a solution to
\be
\label{equa:KG-b}
-b^{-2} \BoxChapeau_g\rho+ \rho= \sourterm.
\ee
Then the rescaled restriction \(w_{s,x} \) satisfies the ordinary differential equation
\be
\label{equa:ODEKG}
b^{-2}a_{s,x}(\lambda) {d^2\over d\lambda^2}w_{s,x}(\lambda) + w_{s,x}(\lambda) = G_{s,x}(\lambda),
\ee
where the coefficient along the ray is defined by
\be
a_{s,x}(\lambda)  \coloneqq  a \Bigl( {\lambda \over s} \sqrt{s^2 +r^2},{\lambda \over s}x \Bigr).
\ee
The coefficient \(a_{s,x}\) remains in the principal part.
\end{lemma}

Keeping \(a_{s,x}\) permits cancellation of the large-frequency terms in the oscillator energy below. The ODE estimate will be applied with \(k_0=a_{s,x}-1\) and \(k_1=G_{s,x}\).

\begin{proof}
We multiply \eqref{equa:KG-b} by \(s^{3/2} \) and use \eqref{equa:hyperboloidal-wave-decomposition-section6}.  The identity \(g^{\mathcal H\,00} =-(s/t)^2 +H^{\mathcal H\,00} \) gives the coefficient \(b^{-2}a \) in front of \(\Lcal^2(s^{3/2} \rho) \).  Along the curve \(\lambda \mapsto(\lambda t/s, \lambda x/s) \), we have
\be
\Lcal(s^{3/2} \rho) = {d\over d\lambda}w_{s,x}(\lambda).
\ee
The remaining terms are those in \(G_{s,x} \), and hence we obtain \eqref{equa:ODEKG}.
\end{proof}


\subsection{Oscillator estimate uniform in the mass}
\label{section-6-3}

\paragraph{Energy estimate for a forced oscillator}

The following elementary lemma gives the uniform pointwise estimate. It quantifies the convergence of the singular ODE
\be
b^{-2}z''+z=k_1
\ee
toward the algebraic relation \(z=k_1\), while allowing a variable coefficient in front of \(z''\).

The estimate controls the initial discrepancy in position, the initial discrepancy in velocity divided by the mass, and the integral of \(|k_1''|\). The variation of \(k_0\) determines the amplification factor.

\begin{lemma}[\texorpdfstring{Oscillator estimate uniform in the mass}{Oscillator estimate uniform in the mass}]
\label{lem:K-G-ODE}
Let \(b>0\) and let \(k_0\in C^1([s_0,s_1])\) satisfy
\be
\sup_{[s_0,s_1]}|k_0| \leq {1\over2},
\ee
and let \(k_1\in C^2([s_0,s_1]) \). Let \(z\) be the solution of
\be
\label{equa:2233}
b^{-2} \big(1+k_0(\lambda) \big)z''(\lambda) +z(\lambda)=k_1(\lambda), \qquad z(s_0)=z_0, \qquad z'(s_0)=z_1.
\ee
Then, for all \(s\in[s_0,s_1] \), one has
\be
\label{equa:uniform-ODE-pointwise-bound}
\aligned &|z(s)-k_1(s)| + {1\over b}|z'(s)-k_1'(s)| \\
& \lesssim \exp\Bigl(C\int_{s_0}^s |k_0'(\lambda)| \, d\lambda \Bigr) \Bigl( |z_0-k_1(s_0)| + {1\over b}|z_1-k_1'(s_0)| + {1\over b} \int_{s_0}^s |k_1''(\lambda)| \, d\lambda \Bigr). \endaligned
\ee
The implicit constant is independent of \(b>0\).
\end{lemma}

\begin{proof}
\bse
We set \(u \coloneqq z-k_1\), \(v \coloneqq b^{-1}(z'-k_1') \), and \(a_* \coloneqq (1+k_0)^{-1} \).  The equation in \eqref{equa:2233} gives
\be
\label{equa:ODE-weighted-system}
u'=bv, \qquad v'=-ba_* u-b^{-1}k_1''.
\ee
Since \(|k_0| \leq1/2\), we use the weighted energy
\be
\label{equa:ODE-weighted-energy}
E^2 \coloneqq a_*|u|^2 + |v|^2, \qquad E\geq0,
\ee
whose square root \(E\) is uniformly equivalent to \(|u|+ |v| \). A differentiation gives the cancellation identity
\be
\label{equa:ODE-energy-cancellation}
(E^2)'=a_*'|u|^2-2b^{-1}v\,k_1'', \qquad |a_*'| \leq4|k_0'|.
\ee
The terms of size \(buv\) cancel. Consequently, we have
\be
\label{equa:ODE-weighted-inequality}
{d\over ds}E(s) \lesssim |k_0'(s)|E(s) +b^{-1}|k_1''(s)|.
\ee
Gronwall's inequality and the uniform equivalence of \(E\) with \(|u|+ |v| \) yield \eqref{equa:uniform-ODE-pointwise-bound}. At a zero of \(E\), we justify the differential inequality by first using \(\sqrt{E^2 + \epsilon^2} \) and then letting \(\epsilon\) decrease to zero.
\ese
\end{proof}

\paragraph{Initial oscillations and preparation}
Uniform convergence to \(k_1\) up to \(s=s_0\) requires \(z_0-k_1(s_0)\to0\). Lemma~\ref{lem:K-G-ODE} also controls the initial velocity discrepancy at scale \(b^{-1}\); convergence of unscaled derivatives requires stronger bounds. For \(k_0=0\), the discrepancy satisfies \(b^{-2}(z-k_1)''+(z-k_1)=-b^{-2}k_1''\), whose homogeneous solution is
\be
A\cos(b(\lambda-s_0)) +B\sin(b(\lambda-s_0)), \qquad A \coloneqq z_0-k_1(s_0), \qquad B \coloneqq {1\over b}(z_1-k_1'(s_0)).
\ee
Hence, in this example, the amplitude of the oscillatory component is  controlled by
\be
|z_0-k_1(s_0)|+ {1\over b}|z_1-k_1'(s_0)|.
\ee
In this constant-coefficient model, when this quantity tends to zero and \(b^{-1} \|k_1''\|_{L^1([s_0,s_1])} \to0\), the solution converges uniformly to \(k_1\) on the fixed interval. Otherwise a rapidly oscillating component may persist.


\subsection{\texorpdfstring{Estimates for commuted fields along rays}{Estimates for commuted fields along rays}}
\label{section-6-4}
\label{section-6-5}

\paragraph{Discrepancy and commuted equations}
We consider the singular equation
\be
\label{equa:KG-101}
-3\coeff\,g^{\alpha\beta}\del_\alpha\del_\beta\rho+\rho=\sourterm-\coeff f,\qquad
\rho|_{\Mint_{s_0}}=\rho_0,\qquad \delb_0\overline\rho|_{\Mint_{s_0}}=\rho_1.
\ee
Here \(\rho_1\) is the hyperboloidal-coordinate velocity. On the flat initial hyperboloid the geometric normal is \(n_{\Hcal_{s_0}}^{\mathrm{Mink}}=\delb_0+(x^a/s_0)\delb_a\), so its velocity is recovered by adding the tangential derivative of \(\rho_0\). The discrepancy from the limiting relation and its equation are
\be
\label{equa-cd55}
\rhon\coloneqq\rho-\sourterm,\qquad \sourtermn\coloneqq-\coeff f+3\coeff\,g^{\alpha\beta}\del_\alpha\del_\beta\sourterm,
\ee
\be
\label{equa:KG-101-sigma}
-3\coeff\,\BoxChapeau_g\rhon+\rhon=\sourtermn,\qquad
\rhon|_{\Mint_{s_0}}=\rho_0-\sourterm|_{\Mint_{s_0}},\qquad
\delb_0\overline{\rhon}|_{\Mint_{s_0}}=\rho_1-\delb_0\overline{\sourterm}|_{\Mint_{s_0}}.
\ee
We include the identity among the admissible operators \(Z=\del^I L^J\). For every such operator, we set \(\rhon_Z\coloneqq Z\rhon\) and obtain
\be
\label{equa:KG-101-sigma-commuted}
-3\coeff\,\BoxChapeau_g\rhon_Z+\rhon_Z=\mathcal R_Z,\qquad
\mathcal R_Z\coloneqq Z\sourtermn+3\coeff[Z,\BoxChapeau_g]\rhon.
\ee
If \(Z=X_1\cdots X_d\), the affine generators \(X_i\) are translations or boosts. For \(I\subset\{1,\ldots,d\}\), we denote by \(Z_I\) the scalar operators and by \(\mathcal L_I\) the tensor Lie derivatives in their original order. Since these generators are Minkowski Killing fields, the commutator is
\be
\label{equa:KG-commutator-high-order}
3\coeff[Z,\BoxChapeau_g]\rhon
=3\sum_{\varnothing\ne I\subset\{1,\ldots,d\}}(\mathcal L_IH)^{\alpha\beta}
\del_\alpha\del_\beta Z_{I^c}\fieldPhi^\natural,\qquad
\fieldPhi^\natural\coloneqq\coeff\rhon.
\ee
Indeed, \([X,H^{\alpha\beta}\del_{\alpha\beta}]=(\mathcal L_XH)^{\alpha\beta}\del_{\alpha\beta}\), and the ordered product rule gives the formula. Every Hessian component is kept, with no negative power of \(\coeff\). For an operator transverse to the initial hypersurface, \((Z\rhon)|_{\Sigma_0}\) is the restriction of a spacetime derivative; tangential differentiation of \(\rhon|_{\Sigma_0}\) alone does not determine this trace. The initial values of \(Z\rhon\) include the normal derivatives determined by the reduced equation; its values at the interface are provided by the exterior analysis.

\paragraph{Pointwise estimate along rays}
Along the ray \(\lambda\mapsto(\lambda t/s,\lambda x/s)\), with lower parameter \(\lambda_{\mathrm{in}}\), we define
\be
\label{equa:commuted-raywise-restriction}
\mathcal W_Z(\lambda)\coloneqq\lambda^{3/2}Z\rhon(\lambda t/s,\lambda x/s),
\ee
\be
\label{equa:commuted-raywise-source}
\mathcal G_Z(\lambda)\coloneqq
\bigl(\lambda^{3/2}\mathcal R_Z-3\coeff\lambda^{3/2}\mathcal R_g[Z\rhon]\bigr)
(\lambda t/s,\lambda x/s).
\ee
The boundary discrepancy and accumulated source variation are
\be
\label{equa:commuted-raywise-boundary-contribution}
B_Z^{\mathrm{ray}}(s,x)\coloneqq
\bigl(|\mathcal W_Z-\mathcal G_Z|+\sqrt{3\coeff}\,|\mathcal W_Z'-\mathcal G_Z'|\bigr)
\big|_{\lambda=\lambda_{\mathrm{in}}},
\ee
\be
\label{equa:commuted-raywise-source-contribution}
S_Z^{\mathrm{ray}}(s,x)\coloneqq
\int_{\lambda_{\mathrm{in}}}^s|\mathcal G_Z''(\lambda)|\,d\lambda.
\ee
We incorporate the variation of the coefficient \(a_{s,x}\) into
\be
\label{equa:C-high-order-natural}
C_Z^{\mathrm{ray}}(s,x)\coloneqq
\exp\Bigl(C\int_{\lambda_{\mathrm{in}}}^s|a_{s,x}'(\lambda)|\,d\lambda\Bigr)
\bigl(B_Z^{\mathrm{ray}}(s,x)+\sqrt{3\coeff}\,S_Z^{\mathrm{ray}}(s,x)\bigr),
\ee
\be
\label{equa:F-high-order-natural}
Q_Z^{\mathrm{ray}}(s,x)\coloneqq
|\mathcal G_Z(s)|+\sqrt{3\coeff}\,|\mathcal G_Z'(s)|+C_Z^{\mathrm{ray}}(s,x).
\ee
When a ray starts at the interface, \(B_Z^{\mathrm{ray}}\) contains the trace of the same Cauchy development, estimated from the exterior.

\begin{proposition}[Pointwise estimate for the commuted singular Klein--Gordon equation]
\label{proposition:large-mass-pointwise}
\label{prop:KG-high-order-pointwise}
Let \(\rho\) solve \eqref{equa:KG-101}, and let \(Z\) be an admissible operator, including the identity. Suppose that \(a_{s,x}\in C^1\), \(\mathcal G_Z\in C^2\), and \(\sup|a_{s,x}-1|\leq1/3\) on each closed ray interval. Then one has
\be
\label{equa:KG-high-order-ray-discrepancy}
|\mathcal W_Z(s)-\mathcal G_Z(s)|
+\sqrt{3\coeff}\,|\mathcal W_Z'(s)-\mathcal G_Z'(s)|
\lesssim C_Z^{\mathrm{ray}}(s,x),
\ee
\be
\label{equa:KG-high-order-total-pointwise}
s^{3/2}|Z\rhon(t,x)|+\coeff^{1/2}\frac ts\,s^{3/2}|\newperp Z\rhon(t,x)|
\lesssim Q_Z^{\mathrm{ray}}(s,x).
\ee
The constants are independent of \(0<\coeff\leq\coeff_0\), for fixed \(\coeff_0>0\) and \(s_0\geq1\). The statements apply whether the ray starts on the initial hyperboloid or at the interface.
\end{proposition}

The decay factor \(s^{-3/2}\) comes from the rescaling \(\mathcal W_Z=\lambda^{3/2}Z\rhon\), while the derivative factor \(\coeff^{1/2}\) reflects the mass scaling of \(\coeff|\del\rhon|^2+|\rhon|^2\). Applying the oscillator energy to the discrepancy \(\mathcal W_Z-\mathcal G_Z\) gives the source contribution \(\sqrt{3\coeff}\int_{\lambda_{\mathrm{in}}}^s|\mathcal G_Z''(\lambda)|\,d\lambda\). The two ray derivatives of the source thus yield a positive power of \(\coeff\). The amplification factor depends on the variation of \(a_{s,x}\), independently of the large oscillation frequency.

\begin{proof}
\bse
Lemma~\ref{lem:KG-0}, applied to \eqref{equa:KG-101-sigma-commuted}, gives
\be
\label{equa:ODEKG-nat}
3\coeff\,a_{s,x}(\lambda)\mathcal W_Z''(\lambda)+\mathcal W_Z(\lambda)
=\mathcal G_Z(\lambda).
\ee
Lemma~\ref{lem:K-G-ODE} with \(b=(3\coeff)^{-1/2}\), \(k_0=a_{s,x}-1\), and \(k_1=\mathcal G_Z\) proves the discrepancy estimate. Adding \(|\mathcal G_Z(s)|+\sqrt{3\coeff}\,|\mathcal G_Z'(s)|\) controls \(|\mathcal W_Z(s)|+\sqrt{3\coeff}\,|\mathcal W_Z'(s)|\) by \(Q_Z^{\mathrm{ray}}\). At the observation point, the chain rule gives
\be
\label{equa:commuted-raywise-derivative-recovery}
\mathcal W_Z'(s)=\tfrac32s^{1/2}Z\rhon(t,x)
+\frac ts\,s^{3/2}\newperp Z\rhon(t,x).
\ee
The first term is controlled by \(|\mathcal W_Z(s)|\), with a factor bounded by \(C\sqrt{\coeff_0}/s_0\). This proves the pointwise estimate, uniformly in \(\coeff\).
\ese
\end{proof}

\paragraph{Initial traces and the nonlinear application}
For comparison with the preparation amplitude, the spatial initial norm at order \(q\geq3\) is
\be
\label{equa:raywise-trace-amplitude}
\mathrm{Ray}_{\coeff,q}[\Theta]\coloneqq
\|\Theta|_{\Sigma_0}\|_{H^q_{\mathrm{AE}}}
+\coeff^{1/2}\|\mathcal L_n\Theta|_{\Sigma_0}\|_{H^{q-1}_{\mathrm{AE}}}.
\ee
On the fixed truncated initial hypersurface, Sobolev embedding gives
\be
\label{eqKS-zo-2}
|\rhon_0|+\coeff^{1/2}|\rhon_1|
\lesssim_{s_0}(1+r)^{-3/2}
\bigl(\|\rhon_0\|_{H^2(\Mint_{s_0})}+\coeff^{1/2}\|\rhon_1\|_{H^2(\Mint_{s_0})}\bigr).
\ee
Here \(\rhon_1=\delb_0\overline{\rhon}|_{\Mint_{s_0}}\), and \(r\leq\rhoH(s_0)\), so the radial weight is bounded. The relation between \(\delb_0\), the normal derivative, and tangential derivatives shows that \(\mathrm{Ray}_{\coeff,q}\) controls these initial values for \(\Theta\). The nonlinear convergence theorem uses the stronger commuted preparation amplitude \eqref{equa:commuted-preparation-amplitude}; the source terms in \(B_Z^{\mathrm{ray}}\) are estimated separately.

The estimate is an a priori inequality: \(\mathcal G_Z\) contains derivatives of \(Z\rhon\) through \(\mathcal R_g[Z\rhon]\), and the integral of \(\mathcal G_Z''\) requires two additional ray derivatives. In the nonlinear argument, mixed derivatives are instead written as a ray derivative plus an integrable source. The shifted oscillator energy \eqref{equa:seven-strong-ray-shifted-energy} in Section~\ref{section-coupled-pointwise} then keeps the same cancellation of the large frequency while avoiding differentiation of that source twice. Parts~II and~III provide the required derivative and boundary bounds. The unscaled energy estimate separately keeps the factor \(\coeff^{-1/2}\) in \eqref{equa-high-order-effective-field-inequality}.

\paragraph{Fixed positive mass}
The same estimate includes the fixed-mass equation of~\cite[Sections~3.3--3.4]{LeFloch-Ma-2016b},
\be
\label{equa:KG-standard}
-\BoxChapeau_gv+c^2v=f,\qquad c>0.
\ee

\begin{corollary}[Pointwise estimate for a fixed-mass Klein--Gordon field]
\label{prop:KG-standard-pointwise}
Let \(v\) solve \eqref{equa:KG-standard}. Suppose that its ray coefficient and source satisfy the regularity and smallness hypotheses of Proposition~\ref{proposition:large-mass-pointwise}. Then one has
\be
\label{equa:KG-standard-pointwise}
s^{3/2}|v(t,x)|+c^{-1}\frac ts\,s^{3/2}|\newperp v(t,x)|
\lesssim Q_{\mathrm{Id}}^{\mathrm{ray}}(s,x),
\ee
where the functional is evaluated with \(\rho=v\), \(\sourterm=0\), \(\coeff=(3c^2)^{-1}\), and singular source \(f_{\mathrm{sing}}=-3f\). Its boundary and source terms are those defined above. The constant is independent of \(c\geq c_0>0\).
\end{corollary}

\begin{proof}
Multiplication of \eqref{equa:KG-standard} by \(c^{-2}\) gives \eqref{equa:KG-101} with the substitutions in the statement. Proposition~\ref{proposition:large-mass-pointwise} applies with \(Z=\mathrm{Id}\) and \(\coeff_0=(3c_0^2)^{-1}\). Its constant therefore depends on \(c_0\), but not on \(c\geq c_0\).
\end{proof}


\section{Derivation of the conformal field equations}
\label{section-7}


\subsection{Conformal geometry} \label{section-7-1}

\paragraph{Purpose}
We verify the conformal identities underlying the physical curvature reconstruction and the wave-gauge sources. The potential normalization and transformed divergence identity are kept explicit.

\paragraph{Conformal metric identities}
We use the potential \(U(r)=(rf'(r)-f(r))/(\coeff(f'(r))^2)\) and \(\Ufrak=U\circ\hfrak\) from Section~\ref{section-2-2}. The conformal metric and its inverse satisfy
\be
g_{\alpha\beta}=e^{\coeff\fieldphi}\giveng_{\alpha\beta},\qquad g^{\alpha\beta}=e^{-\coeff\fieldphi}\giveng^{\alpha\beta}.
\ee
\bse
The Christoffel symbols of \(g\) and \(\giveng\) are related by
\be
\label{appendix-Christoffel}
\Gamma^\gamma_{\alpha \beta} = \Gammaj^\gamma_{\alpha \beta} + {\coeff\over2} \Bigl( \delta_\alpha^\gamma \del_\beta \fieldphi + \delta_\beta^\gamma \del_\alpha \fieldphi - g_{\alpha \beta}g^{\gamma \delta} \del_\delta \fieldphi \Bigr),
\ee
where \(\Gammaj\) denotes the Christoffel symbols of the physical metric \(\giveng\). In dimension \(4\), the Ricci tensors satisfy
\be
\label{appendix-Ricci-conformal}
R_{\alpha \beta} = \givenR_{\alpha \beta} -\coeff\, \givennabla_\alpha \givennabla_\beta \fieldphi + {\coeff^2\over2} \givennabla_\alpha \fieldphi\givennabla_\beta \fieldphi - \Bigl( {\coeff\over2} \givenBox\fieldphi + {\coeff^2\over2} \giveng(\givennabla \fieldphi, \givennabla \fieldphi) \Bigr) \giveng_{\alpha \beta}.
\ee
\ese
\bse
Equivalently, the scalar wave operators acting on a scalar field \(\psi\) are related by
\be
\label{appendix-wave-conformal}
\givenBox\psi = e^{\coeff\fieldphi} \Bigl( \Box_g\psi - \coeff\,g(\nabla \fieldphi, \nabla \psi) \Bigr).
\ee
The physical scalar equation is
\be
\givenBox\psi-V'(\psi)=0
\ee
and its conformal form is
\be
\label{appendix-matter-equation}
\Box_g\psi-e^{-\coeff\fieldphi}V'(\psi) = \coeff\,g(\nabla \fieldphi, \nabla \psi).
\ee
\ese

\paragraph{Transformed modified-gravity tensor}
By definition of the  effective field, we have
\be
\label{appendix-effective-field}
f'(\givenR)=e^{\coeff\fieldphi}, \quad \givenR= \hfrak(\fieldphi).
\ee
We set
\be
\ffrak \coloneqq f\circ\hfrak.
\ee
\bse
The modified curvature tensor is
\be
\givenCurv_{\alpha \beta} = f'(\givenR) \givenG_{\alpha \beta} -{1\over2} \big(f(\givenR)-\givenR f'(\givenR) \big) \giveng_{\alpha \beta} + \big(\giveng_{\alpha \beta} \givenBox-\givennabla_\alpha \givennabla_\beta \big) f'(\givenR),
\ee
and \eqref{appendix-Ricci-conformal} gives
\be
\label{appendix-conformal-curv}
\givenCurv_{\alpha \beta} -{1\over2} \Tr_{\giveng}(\givenCurv) \, \giveng_{\alpha \beta} = e^{\coeff\fieldphi}R_{\alpha \beta} -{3\over2} \coeff^2 e^{\coeff\fieldphi} \nabla_\alpha \fieldphi\nabla_\beta \fieldphi -{\coeff\over2}e^{\coeff\fieldphi} \Ufrak(\fieldphi) \,g_{\alpha \beta}.
\ee
Taking the trace gives
\be
\label{appendix-conformal-trace}
\Tr_{\giveng}(\givenCurv) = 3\coeff\,e^{2\coeff\fieldphi} \Box_g\fieldphi -\ffrak(\fieldphi) + \coeff e^{2\coeff\fieldphi} \Ufrak(\fieldphi).
\ee
\ese
The last two terms can be expressed through \(\Ufrak'\). Indeed, we have
\be
\label{appendix-U-identity}
\Ufrak'(\fieldphi) = e^{-2\coeff\fieldphi} \ffrak(\fieldphi)-\coeff\Ufrak(\fieldphi).
\ee
To verify this identity, we write \(r=\hfrak(\fieldphi)\), so that \(f'(r)=e^{\coeff\fieldphi}\), and obtain
\bse
\be
\Ufrak(\fieldphi)={r f'(r)-f(r)\over\coeff\,(f'(r))^2}.
\ee
Differentiating with respect to \(\fieldphi\) gives
\be
\Ufrak'(\fieldphi)={2f(r)-r f'(r)\over(f'(r))^2},
\ee
\ese
 which proves \eqref{appendix-U-identity}. Combining \eqref{appendix-conformal-trace} and \eqref{appendix-U-identity} yields the compact trace formula
\be
\label{appendix-conformal-trace-compact}
\Tr_{\giveng}(\givenCurv) = e^{2\coeff\fieldphi} \Bigl( 3\coeff\, \Box_g\fieldphi - \Ufrak'(\fieldphi) \Bigr).
\ee


\subsection{\texorpdfstring{Conformal field equations and potential}{Conformal field equations and potential}} \label{section-7-2}

\paragraph{Conformal Einstein system}
The physical field equations are $ \givenCurv_{\alpha \beta} = 8\pi\givenT_{\alpha \beta}[\psi, \giveng]. $ Using \eqref{appendix-conformal-curv}, their trace-reversed form is equivalent to
\be
R_{\alpha \beta} = 8\pi e^{-\coeff\fieldphi} \Bigl( \givenT_{\alpha \beta} -{1\over2} \Tr_{\giveng}(\givenT) \giveng_{\alpha \beta} \Bigr) + {3\over2} \coeff^2\nabla_\alpha \fieldphi\nabla_\beta \fieldphi + {\coeff\over2} \Ufrak(\fieldphi)g_{\alpha \beta}.
\ee
Taking the trace with respect to \(g\) and subtracting one half of the trace, we obtain
\be
\label{appendix-Einstein-conformal}
G_{\alpha \beta} = 8\pi e^{-\coeff\fieldphi} \givenT_{\alpha \beta}[\psi, \giveng] + T^{\mathrm{eff}}[\fieldphi,g]_{\alpha \beta},
\ee
where the effective-field tensor is
\be
\label{appendix-Tphi}
T^{\mathrm{eff}}[\fieldphi,g]_{\alpha \beta} = {3\over2} \coeff^2\nabla_\alpha \fieldphi\nabla_\beta \fieldphi -{3\over4} \coeff^2g(\nabla \fieldphi, \nabla \fieldphi)g_{\alpha \beta} -{\coeff\over2} \Ufrak(\fieldphi)g_{\alpha \beta}.
\ee
This is the stress-energy tensor used in Section~\ref{section-4}.

\paragraph{Wave-gauge reduction}
In wave coordinates, we have $ \Gamma^\alpha=g^{\lambda \mu} \Gamma^\alpha_{\lambda \mu} =0, $ and the Ricci tensor has the reduced form
\be
2R_{\alpha \beta} = -\BoxChapeaud_g g_{\alpha \beta} + \Fbb_{\alpha \beta}(g,g;\del g, \del g).
\ee
Combining this identity with \eqref{appendix-Einstein-conformal} gives
\be
\BoxChapeaud_g g_{\alpha \beta} = \Fbb_{\alpha \beta}(g,g;\del g, \del g) + \Gbb_{\coeff}[\fieldphi,g]_{\alpha \beta} + \Mbb_{\coeff}[\fieldphi, \psi,g]_{\alpha \beta},
\ee
where we define the effective-field contribution by
\be
\Gbb_{\coeff}[\fieldphi,g]_{\alpha \beta} = -3\coeff^2\nabla_\alpha \fieldphi\nabla_\beta \fieldphi -\coeff\Ufrak(\fieldphi)g_{\alpha \beta}.
\ee
For the scalar matter field, the physical energy-momentum tensor is
\be
\givenT_{\alpha \beta} = \givennabla_\alpha \psi\givennabla_\beta \psi - \Bigl( {1\over2} \giveng^{\mu\nu} \givennabla_\mu\psi\givennabla_\nu\psi + V(\psi) \Bigr) \giveng_{\alpha \beta}.
\ee
Its contribution to the reduced metric equation is
\be
\Mbb_{\coeff}[\fieldphi, \psi,g]_{\alpha \beta} = -16\pi e^{-\coeff\fieldphi} \Bigl( \nabla_\alpha \psi\nabla_\beta \psi + e^{-\coeff\fieldphi}V(\psi)g_{\alpha \beta} \Bigr).
\ee
Moreover, we find
\be
\Tr_{\giveng}(\givenT) = -e^{\coeff\fieldphi}g(\nabla \psi, \nabla \psi)-4V(\psi),
\ee
and therefore we obtain
\be
8\pi e^{-2\coeff\fieldphi} \Tr_{\giveng}(\givenT) = -8\pi e^{-\coeff\fieldphi}g(\nabla \psi, \nabla \psi) -32\pi e^{-2\coeff\fieldphi}V(\psi).
\ee
Together with \eqref{appendix-conformal-trace-compact}, this gives the Klein--Gordon equation for \(\fieldphi\) stated in \eqref{MainPDE}.

\paragraph{Expansion of the effective-field potential}
\label{section-7-3}
The admissible scaling family satisfies
\be
f_\coeff(r)={1\over \coeff}f_*(y), \qquad y \coloneqq \coeff r, \quad f_*(0)=0, \quad f_*'(0)=1, \quad f_*''(0)=1,
\ee
on a fixed sufficiently small interval of \(|y| \). Its Taylor expansions are
\be
f_\coeff(r) = r+ {\coeff\over2}r^2 +\Ocal(\coeff^2r^3), \quad f_\coeff'(r) = 1+ \coeff r+\Ocal(\coeff^2r^2).
\ee
Since \(f'(\hfrak(\fieldphi))=e^{\coeff\fieldphi} \), it follows that
\bse
\be
\hfrak(\fieldphi)= \fieldphi+\Ocal(\coeff\fieldphi^2)
\ee
Consequently, we obtain
\be
\Ufrak(\fieldphi) = {1\over2} \fieldphi^2 +\Ocal(\coeff\fieldphi^3), \quad \Ufrak'(\fieldphi) = \fieldphi+\Ocal(\coeff\fieldphi^2).
\ee
In particular, we set
\be
W_\coeff(\fieldphi) \coloneqq \Ufrak'(\fieldphi)-\fieldphi.
\ee
Then we find
\be
W_\coeff(\fieldphi)=\Ocal(\coeff\fieldphi^2).
\ee
\ese
For an admissible family of order \(M\), the coefficient functions in the rescaled quadratic and cubic remainders are \(C^{M-3} \), uniformly in \(\coeff\).~\autoref{lem:seven-admissible-potential-factorization} gives their factorization. For the Starobinsky model \(f_\Sta(r)=r+(\coeff/2)r^2\), we have explicitly
\be
\hfrak_\Sta(\fieldphi)={1\over \coeff}(e^{\coeff\fieldphi}-1), \quad \Ufrak_\Sta(\fieldphi) = {1\over 2\coeff^2 e^{2\coeff\fieldphi}} (e^{\coeff\fieldphi}-1)^2.
\ee

\paragraph{Divergence identity}
\bse
The tensor \(\givenCurv\) is divergence-free with respect to the physical Jordan metric:
\be
\givennabla^\alpha \givenCurv_{\alpha \beta} =0
\ee
When the same covariant two-tensor is differentiated and traced with respect to the conformal Einstein metric, this identity becomes
\be
\label{eq2-26-mars-2023}
\nabla^\alpha \givenCurv_{\alpha \beta} = \coeff\, \nabla^\mu\fieldphi\, \givenCurv_{\mu\beta} - {\coeff\over2} \Tr_g(\givenCurv) \nabla_\beta \fieldphi.
\ee
\ese
Indeed, for any symmetric two-tensor \(T\), we have
\be
\nabla^\alpha T_{\alpha \beta} = e^{-\coeff\fieldphi} \givennabla^\alpha T_{\alpha \beta} + \coeff\,g^{\mu\nu}T_{\mu\beta} \del_\nu\fieldphi - {\coeff\over2} \Tr_g(T) \del_\beta \fieldphi.
\ee
Applying this identity to \(T_{\alpha \beta} = \givenCurv_{\alpha \beta} \) gives \eqref{eq2-26-mars-2023}.

\part{Bootstrap in the asymptotically hyperboloidal domain}


\section{Interior norms and bootstrap assumptions} 
\label{section-8}


\subsection{Strategy of proof in the interior}
\label{section-8-1}

\paragraph{Rescaled fields and the interior hierarchy}

We consider the massless matter model, \(V\equiv0\), on the hyperboloidal part of the Cauchy development. Our arguments must control the metric, the matter field, and the effective field. The rescaled fields \(\fieldvarphi = \coeff^{1/2} \fieldphi\) and \(\fieldPhi = \coeff\fieldphi\) must have estimates uniform in \(\coeff\), while recovering \(\fieldphi\) involves negative powers of \(\coeff\). The well-preparation issue will arise only in the convergence argument.  The interior proof proceeds in three steps.
\bei
\item \emph{Pointwise estimates.} Schematically, the wave gauge improves the metric component estimates, the large-mass ray argument controls the effective field, and a characteristic integration controls the matter field near the light cone. These bounds place the low-order factor of each quadratic source in \(L^\infty\).

\item \emph{Energy estimates.} As usual, the higher-order factor in each quadratic source is measured in the energy norm. A boost on its coefficient \emph{lowers the rank} of the remaining factor, permitting induction on the number of boosts. A radial identity controls the high-order metric factor containing boosts only.

\item \emph{Continuation.} The geometric and high-order energy bounds give us a control of the interior part; however, we also need the exterior argument (cf. Part III) in order to control the solution along the boundary (or interface) between the two domains.
\eei

\paragraph{Small growth exponents}

For a nonnegative energy norm \(E\), a constant \(a \geq0\), and an integrable source \(F\) on finite intervals, Gronwall's inequality gives
\be
\label{eq:small-growth-gronwall}
E'(s) \leq a s^{-1}E(s) +F(s) \quad\Longrightarrow\quad E(s) \leq (s/s_0)^aE(s_0) + \int_{s_0}^s(s/\tau)^aF(\tau) \, d\tau.
\ee
A small coefficient of \(s^{-1} \) produces a \emph{slow polynomial growth}, whereas \(s^{-1-\gamma} \), with \(\gamma>0\), is integrable. At positive rank, the exponents also account for the boosts falling on some coefficients. Their gaps control the resulting time integrals. At the higher effective-field level, the rescaled energy may have an additional factor \(s^{1/2} \): after this factor is removed, the source coefficient \(s^{-3/2} \) becomes borderline. (All growth exponents below refer to energy norms, not their squares.)

\paragraph{Uniform and fixed-parameter interior hierarchies}

The hyperboloidal energy and Hardy arguments follow~\cite[Sections~2.3 and~2.5]{LeFloch-Ma-2016b}; the local wave-gauge calculation follows~\cite[Section~4.7]{LeFloch-Ma-2018}. Every integration below keeps its interface terms. The differentiation levels \(p_{\mathbf{high}}=N-5\) and \(p_{\mathbf{low}}=N-9\) leave four additional derivatives for the sphere Sobolev and characteristic estimates. A \emph{boost-only} operator contains Lorentz boosts and no Cartesian translation. The lower rescaled energies have slow growth uniformly in \(\coeff\); at the higher level the effective-field energy allows an additional factor \(s^{1/2}\). Exact rescaling then recovers the norms needed for continuation at fixed \(\coeff>0\).

\paragraph{Interior domain and traces at the interface}
\label{paragraph-7-scope}

The interior bootstrap argument concerns the portion of the solution contained in the hyperboloidal domain
\be
\Mint_{[s_0,s_1]} = \bigcup_{s_0\leq s\leq s_1} \Mint_s.
\ee
Throughout Part~II, we fix \(s_0\geq2\).  The compact interval preceding \(s_0\) is treated as usual by a local well-posedness argument.  This choice also ensures that the interior--exterior interface lies in the exact Schwarzschild portion of the reference metric from its first hypersurface onward. Consequently, \(R_s= (s^2-1)/2\) is positive on every hypersurface under consideration, and the interface trace seminorms and moving-boundary identities below are nondegenerate. We recall that the \emph{interior--exterior interface} is \(\Lscr_{[s_0,s_1]} \), and all of the functions under consideration \emph{need not vanish} on this hypersurface: every Hardy, characteristic, ray, and energy estimate keeps the corresponding \emph{interface terms} on \(\Lscr_{[s_0,s_1]} \). The fluxes and the distinct trace norms are estimated in the exterior domain and must be closed simultaneously. The \emph{exterior estimates used in the interior argument} are separated as follows.
\begin{itemize}
\item The coarse exterior energies control the source portions of the wave equations.

\item The good metric spacetime integral controls the integrated boost-only trace.

\item The improved interface bounds (cf.~\eqref{equa:seven-interface-data-bounds}) give the final interior bootstrap improvement.
\end{itemize}

The issue to keep in mind throughout is that the Euclidean--hyperboloidal estimates of \cite[Sections~3--6 and~14]{LeFloch-Ma-2024a} and \cite[Sections~2 and~4--6]{LeFloch-Ma-2024b} must here be \emph{uniform in the mass}
\be
b_\coeff= (3\coeff)^{-1/2} \to + \infty
\qquad \text{ as \(\coeff\to0\).}
\ee
 The two metric perturbations of interest are
\be
g=g^\star+u, \qquad g^{\alpha \beta} = (\gMink)^{\alpha \beta} +H^{\alpha \beta},
\ee
so that \(u\) is the \emph{covariant perturbation of the reference metric}, whereas \(H\) is the \emph{contravariant perturbation of the Minkowski metric.} We use\footnote{In the interior domain, the rotations need not be added as independent generators, since $\Omega_{ab}{\coloneqq}x_a \del_b-x_b\del_a ={x_a \over t}L_b-{x_b\over t}L_a$.} the hyperboloidal commutator family~\cite{LeFloch-Ma-2016b,LeFloch-Ma-2018}:
\be
Z= \del^I L^J, \quad \ord Z=|I|+ |J|, \quad \rank Z=|J|.
\ee

\paragraph{Uniform potential factorization and gauge propagation}

We rely on Definition~\ref{def:admissible-f-family} and analyze the asymptotic properties of the potential functions associated with the function $f(R)$. 

\begin{lemma}[\texorpdfstring{Uniform factorization of the potentials}{Uniform factorization of the potentials}]
\label{lem:seven-admissible-potential-factorization}
Let \(f_\coeff\) be an admissible scaling family of order \(M\geq N+3\). For some fixed \(c_0>0\), whenever \(| \fieldPhi| \leq c_0\), one has
\bse
\be
\label{equa:seven-uniform-potential-hypothesis}
\coeff\Ufrak_\coeff(\fieldphi)= \frac12(\fieldvarphi)^2 +(\fieldvarphi)^2\fieldPhi\, \mathcal U_{1, \coeff}(\fieldPhi), \qquad W_\coeff(\fieldphi)= (\fieldvarphi)^2\mathcal W_\coeff(\fieldPhi), 
\ee
where the coefficients satisfy the uniform bounds 
\be
\label{equa:seven-uniform-potential-CN}
\sup_{0<\coeff\leq\coeff_0} \Bigl(\| \mathcal U_{1, \coeff} \|_{C^N([-c_0,c_0])} + \| \mathcal W_\coeff\|_{C^N([-c_0,c_0])} \Bigr)< + \infty.
\ee
\ese
\end{lemma}
\begin{proof}
\bse
We set \(\hfrak_*= (\log f_*')^{-1} \) and define the rescaled potential by
\be
\label{equa:admissible-rescaled-potential}
\mathcal U_*(z)\coloneqq \bigl(\hfrak_*(z)e^z-f_*(\hfrak_*(z)) \bigr)e^{-2z}, \qquad \Ufrak_\coeff(\fieldphi)= \coeff^{-2} \mathcal U_*(\coeff\fieldphi).
\ee
The inverse map belongs to \(C^{M-1} \), and differentiating gives
\be
\label{equa:admissible-rescaled-potential-derivative}
\mathcal U_*'(z)= \hfrak_*(z)e^{-z}-2\mathcal U_*(z), \qquad \mathcal U_*(0)= \mathcal U_*'(0)=0, \qquad \mathcal U_*''(0)=1.
\ee
Hence \(\mathcal U_*\in C^M\). Taylor's formula defines two \(C^{M-3} \) functions, including at \(z=0\), by
\be
\label{equa:admissible-rescaled-remainders}
\mathcal U_{1, \coeff}(z)\coloneqq \frac12\int_0^1(1-\tau)^2\mathcal U_*'''(\tau z) \, d\tau, \qquad \mathcal W_\coeff(z)\coloneqq \int_0^1(1-\tau) \mathcal U_*'''(\tau z) \, d\tau.
\ee
They satisfy \(\mathcal U_*(z)=z^2/2 +z^3\mathcal U_{1, \coeff}(z) \) and \(\mathcal U_*'(z)=z+z^2\mathcal W_\coeff(z) \). Substitution of \(z= \fieldPhi = \coeff\fieldphi\), with \((\fieldvarphi)^2= \coeff\fieldphi^2\), proves the factorization and its uniform \(C^N\) bound.
\ese
\end{proof}

Local evolution is used at each fixed positive parameter. Its existence time may depend on that parameter and on the finite initial norms.

\begin{lemma}[Local evolution and propagation of the wave gauge]
\label{lem:seven-local-wave-gauge-propagation}
Fix \(\coeff>0\), \(N\geq19\), an admissible profile of order \(M\geq N+3\), and \(V=0\). Assume that the initial hypersurface is uniformly spacelike and that \(\fieldPhi=\coeff\fieldphi\) takes values in a compact subinterval of its admissible range. Assume that the weighted order-\(N\) energies of the first derivatives of \(u\), the scalar first derivatives, and the effective-field mass terms are finite, with the asymptotically Euclidean normalization of the wave fields. The reduced conformal system has a unique local solution and depends continuously on the Cauchy data in the corresponding norms at one lower order. Its order-\(N\) energies remain finite. The solution extends past a finite foliation time if these energies stay bounded, the strict hyperbolicity and spacelikeness bounds persist, and \(\fieldPhi\) stays in a compact subinterval of its admissible range. The local times and constants may depend on \(\coeff\) and on the finite initial energies. If the Hamiltonian, momentum, and wave-gauge constraints hold initially, then \(\Gamma^\mu=g^{\alpha\beta}\Gamma^\mu_{\alpha\beta}\) vanishes throughout the development.
\end{lemma}
\begin{proof}
\noindent{\bf 1.} \emph{Local construction on the fixed initial graph.} We use \(\tau=t-T_0(x)\), \(y=x\), with the prescribed \(T_0(x)=\fTime(s_0,|x|)\), to flatten \(\Sigma_0\). For the construction, we replace both scalar wave operators in \eqref{MainPDE} by \(\BoxChapeau_g\) in the original coordinates and then express the reduced system in this fixed chart. The equations keep a diagonal strictly hyperbolic principal operator, whose coefficients depend on the fields rather than their derivatives. The coordinate change is smooth, is a time translation outside a compact set, and preserves the differential order. The initialization in the equations~\eqref{equa:audit-initial-auxiliary-metric}--\eqref{equa:audit-initial-lapse-shift-density} gives the reduced jets. We apply the local iteration to \(u=g-g^\star\) and the two scalar fields. On every bounded time interval, \(g^\star\) is an exact stationary vacuum metric outside a fixed ball, so its independent forcing has bounded support and all its differentiated coefficients are bounded. The inverse metric and scalar sources have the finite composition regularity needed for the iteration: \(\mathcal U_*\in C^M\) gives \(\Ufrak_\coeff'\in C^{M-1}\), hence at least \(C^{N+2}\) regularity. Thus the finite-order proof of~\cite[Appendix~III, Theorem~4.6 and Corollary~4.8]{ChoquetBruhat-2009} applies. We use uniformly local \(H^{N+1}\times H^N\) Cauchy bounds, obtained from the initial energies, normalization, and Sobolev inequality on unit balls. The local existence time is independent of the centers of the balls, by~\cite[Appendix~III, Theorem~4.14]{ChoquetBruhat-2009}; uniqueness and finite propagation combine these solutions, as in Theorem~4.13 there. These constants may depend on the fixed \(\coeff\).

\noindent{\bf 2.} \emph{Weighted norms and continuation.} We keep \(g^\star\) and regularize the dynamic Cauchy data. The commuted energy identities and the product rule bound the weighted energies on a short interval depending on their initial size and on \(\coeff\). The Hardy inequality controls the undifferentiated dynamic metric; the Schwarzschild term remains in \(g^\star\). On a finite slab, derivatives of the logarithms of the weights are bounded, so derivatives of a weight contribute bounded multiples of the same energies. The difference estimate at one lower order gives continuous dependence and propagates the additional weighted Sobolev comparisons used here, including the \(H^{q_\star}_{\mathrm{AE}}\) comparison of Section~\ref{section-4-4}. The regularized iteration converges in this lower norm, while the energy bounds preserve order \(N\). Bounded energies and strict coefficient margins give a lower bound for the time of a restarted local solution; restarting near a finite terminal time proves continuation, as in~\cite[Appendix~III, Section~5]{ChoquetBruhat-2009}. The smallness threshold independent of \(\coeff\) in the global theorem will follow from the coupled estimates below.

\noindent{\bf 3.} \emph{Propagation of the wave gauge.} 
\bse
The total source tensor in the conformal Einstein equation is
\be
\label{equa:seven-total-source-subsidiary}
\mathcal T_{\alpha \beta}{\coloneqq}8\pi e^{-\fieldPhi} \givenT_{\alpha \beta} + \frac32\del_\alpha \fieldPhi\del_\beta \fieldPhi-\Bigl(\frac34g^{\mu\nu} \del_\mu\fieldPhi\del_\nu\fieldPhi+ \frac\coeff2\Ufrak(\fieldphi) \Bigr)g_{\alpha \beta}.
\ee
The reduced scalar equations give the divergence formula
\be
\label{equa:seven-total-source-gauge-divergence}
\nabla^\alpha \mathcal T_{\alpha \beta} =-\Gamma^\mu\Bigl(8\pi e^{-\fieldPhi} \del_\mu\psi\del_\beta \psi+ \frac32\del_\mu\fieldPhi\del_\beta \fieldPhi\Bigr).
\ee
To derive it, we differentiate the matter factor and the effective-field stress separately. The matter-trace terms cancel by the effective-field equation. The remaining terms are the geometric scalar-equation residuals, equal to \(-\Gamma^\mu\del_\mu\psi\) and \(-3\coeff\Gamma^\mu\del_\mu\fieldphi\). Taking the divergence of the reduced Einstein equation and using the Bianchi identity therefore gives the homogeneous subsidiary system
\be
\label{equa:seven-subsidiary-wave}
\Box_g\Gamma^\mu +A^\mu{}_{\nu}{}^\alpha(g, \del g) \del_\alpha \Gamma^\nu +B^\mu{}_{\nu}(g, \del g, \del^2g,{\fieldPhi}, \del\fieldPhi, \psi, \del\psi) \Gamma^\nu =0.
\ee
The initial constraints imply that \(\Gamma^\mu\) and its future normal derivative vanish initially. Uniqueness for this linear system on each finite domain of dependence gives \(\Gamma=0\).  
\ese
\end{proof}


\subsection{\texorpdfstring{Energy norms and bootstrap assumptions}{Energy norms and bootstrap assumptions}}
\label{section-8-2}
\label{section-8-3}

\paragraph{Metric and matter energies}

We take \(X_\H{\coloneqq} \del_t\), future timelike in the range under consideration, and use the future unit normal \(n_s\) defined by the factorisation $ N_s=j_sn_s$ (cf.~\eqref{equa-KDL486}). All hypersurface integrals are written with the measure \(dx\). For a scalar component \(q\), the wave energy is
\bse
\be
\label{equa:seven-exact-wave-energy}
\calE_{\mathrm w}^{\H}(s,q) \coloneqq \int_{\Mint_s} Q^{\mathrm{wave}}_{\alpha \beta}[q] \,N_s^\alpha X_\H^\beta \, dx,
\ee
where \(Q^{\mathrm{wave}} \) is defined in \eqref{equa-wave-energy-tensor}.  Metric components are summed componentwise.  For the massless matter equation, it is more convenient to use instead its weighted divergence-form energy
\be
\label{equa:seven-exact-matter-energy}
\calE_{\psi}^{\H}(s,q) \coloneqq \int_{\Mint_s} e^{-{\fieldPhi}}Q^{\mathrm{wave}}_{\alpha \beta}[q] \,N_s^\alpha X_\H^\beta \, dx.
\ee
Since \(|{\fieldPhi}| \leq c_0\), this energy is \emph{uniformly equivalent} to \(\calE_{\mathrm w}^{\H}(s,q) \).  For the commuted linear effective-field equation, we introduce
\be
\label{equa:seven-exact-effective-field-tensors}
Q^{\mathrm{sing}}_{\alpha \beta}[q] {\coloneqq} 3\coeff \Bigl( \nabla_\alpha q\nabla_\beta q -{1\over2}g_{\alpha \beta}g(\nabla q, \nabla q) \Bigr) -{1\over2}q^2g_{\alpha \beta}, \qquad Q^{\mathrm{unif}}_{\alpha \beta}[q] {\coloneqq} \coeff Q^{\mathrm{sing}}_{\alpha \beta}[q],
\ee
and we define (for both singular or uniform quantities) 
\be
\label{equa:seven-exact-effective-field-energies}
\calE_{\mathrm{eff}}^{\mathrm{a}, \H}(s,q) \coloneqq \int_{\Mint_s} Q^{\mathrm{a}}_{\alpha \beta}[q] \,N_s^\alpha X_\H^\beta \, dx, \quad \mathrm a \in\{\mathrm{sing}, \mathrm{unif} \}.
\ee
\ese
The flat density is
\be
e_\H[q] {\coloneqq} \Big({s\over t} \del_tq\Big)^2 + \sum_a| \delsH_aq|^2.
\ee
There are constants \(0<c_\H<C_\H<+ \infty\), independent of \(0<\coeff\leq\coeff_0\), such that, pointwise on every interior hypersurface,
\be
\label{equa:seven-adapted-coercivity-hypothesis}
c_\H e_\H[q] \leq Q^{\mathrm{wave}}_{\alpha \beta}[q] \,N_s^\alpha X_\H^\beta \leq C_\H e_\H[q].
\ee
The same inequalities hold after multiplication by \(e^{-{\fieldPhi}} \), and, after adding the corresponding positive zeroth-order density, for \(Q^{\mathrm{sing}} \) and \(Q^{\mathrm{unif}} \).  Condition \eqref{equa:seven-adapted-coercivity-hypothesis} is also propagated to all times: the quadratic energy form remains uniformly comparable to \(e_\H\).  It holds initially for the admissible data, and the rank-zero metric and reference estimates below keep its constants in \([c_\H/2,2C_\H] \).

We denote by \(d\Sigma_\alpha^{\Lscr, \mathrm{int}} \) the directed surface element on \(\Lscr_{[s_0,s]} \), with the orientation induced by the interior slab. We define the corresponding \textbf{oriented fluxes along the boundary} by
\be
\label{equa:seven-exact-lateral-fluxes}
\begin{aligned}
\mathsf F_{\mathrm w}^{\Lscr}(s,q;s_0) &\coloneqq \int_{\Lscr_{[s_0,s]}} Q^{\mathrm{wave} \, \alpha}{}_{\beta}[q]X_\H^\beta \, d\Sigma_\alpha^{\Lscr, \mathrm{int}}, \\
\mathsf F_{\psi}^{\Lscr}(s,q;s_0) &\coloneqq \int_{\Lscr_{[s_0,s]}} e^{-{\fieldPhi}}Q^{\mathrm{wave} \, \alpha}{}_{\beta}[q]X_\H^\beta \, d\Sigma_\alpha^{\Lscr, \mathrm{int}}, \\
\mathsf F_{\mathrm{eff}}^{\mathrm a, \Lscr}(s,q;s_0) &\coloneqq \int_{\Lscr_{[s_0,s]}} Q^{\mathrm a \, \alpha}{}_{\beta}[q]X_\H^\beta \, d\Sigma_\alpha^{\Lscr, \mathrm{int}}.
\end{aligned}
\ee
In particular, we have \(\mathsf F_{\mathrm{eff}}^{\mathrm{unif}, \Lscr} = \coeff\mathsf F_{\mathrm{eff}}^{\mathrm{sing}, \Lscr} \).  For the metric perturbation $u= (u_{\alpha \beta})$, we sum over all metric components and set
\be
\mathsf F_{\mathrm w}^{\Lscr}(s,Zu;s_0) \coloneqq \sum_{0\leq\alpha \leq\beta \leq3} \mathsf F_{\mathrm w}^{\Lscr} (s,Zu_{\alpha \beta};s_0),
\ee
so every metric flux below is the componentwise sum with this convention. 

The interface formulas in Part~II are written for the fixed boundary $t-r=1$. The expanding interface of Section~\ref{section:small-mass-moving-interface} uses the actual directed surface element and the opposite induced orientations; the energy and flux normalizations are given in \eqref{equa:audit-moving-interface-energy-definition} and \eqref{equa:eight-mobile-flux-opposite-normalizations}. The effective-field rescaling is unchanged.

Finally, we introduce the \textbf{boost-counting scale of energies} 
\be
\label{equa:seven-wave-matter-norms}
\mathbf W_{p,k}(s)^2 \coloneqq \sum_{Z\preccurlyeq(p,k)} \calE_{\mathrm w}^{\H}(s,Zu), \quad \mathbf P_{p,k}(s)^2 \coloneqq \sum_{Z\preccurlyeq(p,k)} \calE_{\psi}^{\H}(s,Z\psi).
\ee 
As usual, the notation $Z\preccurlyeq(p,k)$ stands for the collection of $Z$ with order at most $p$ and rank at most $k$.

\paragraph{Effective-field energy functionals}

For the  effective field, we define the \emph{fixed-parameter} \textbf{high-order energy} and its robust rescaling by
\be
\label{equa:seven-DU-definition}
D_{p,k}(s)^2\coloneqq \sum_{Z\preccurlyeq(p,k)} \calE_{\mathrm{eff}}^{\mathrm{sing}, \H}(s,Z\fieldphi), \qquad U_{p,k}(s)^2\coloneqq \coeff D_{p,k}(s)^2= \sum_{Z\preccurlyeq(p,k)} \calE_{\mathrm{eff}}^{\mathrm{unif}, \H}(s,Z\fieldphi).
\ee
Obviously, \(D_{p,k} \) and \(U_{p,k} \) are \emph{two normalizations} of the same effective-field energy. Both are used because they control different combinations of effective-field variables and satisfy different source inequalities. The statements of Section~\ref{section-5} give the following property.

\begin{lemma}[\texorpdfstring{Comparison of the effective-field energies}{Comparison of the effective-field energies}]
Under the comparison \eqref{equa:seven-adapted-coercivity-hypothesis}, the singular energies satisfy
\be
\label{equa:seven-D-coercivity}
D_{p,k}(s)^2 \simeq \sum_{Z\preccurlyeq(p,k)} \int_{\Mint_s} \bigg( |Z\fieldphi|^2 + \coeff\Big({s\over t} \Big)^2| \del Z\fieldphi|^2 + \coeff| \delsH Z\fieldphi|^2 \bigg) \, dx
\ee
whereas their re-scalings satisfy 
\be
\label{equa:seven-U-coercivity}
U_{p,k}(s)^2 \simeq \sum_{Z\preccurlyeq(p,k)} \int_{\Mint_s} \bigg( |Z{\fieldvarphi}|^2 + \Big({s\over t} \Big)^2| \del Z{\fieldPhi}|^2 + | \delsH Z{\fieldPhi}|^2 \bigg) \, dx,
\ee
where ${\fieldPhi} = \coeff\fieldphi$ and ${\fieldvarphi} = \coeff^{1/2} \fieldphi$.
\end{lemma}

All implied constants in \eqref{equa:seven-D-coercivity}--\eqref{equa:seven-U-coercivity} are \emph{independent of \(0<\coeff\leq\coeff_0\).}   
 In our interior analysis, here and below, norms without an indicated domain \emph{are always taken on \(\Mint_s\)}. We find 
\be
\label{equa:seven-coeff-conversion-L2}
\begin{aligned}
\| \,| \fieldphi|_{p,k} \|_{L^2} + \coeff^{-1} \| \,|{\fieldPhi}|_{p,k} \|_{L^2} & \lesssim D_{p,k} = \coeff^{-1/2}U_{p,k}, \\
\Bigl\| \Big({s\over t} \Big)| \del\fieldphi|_{p,k} \Bigr\|_{L^2} + \| \,| \delsH\fieldphi|_{p,k} \|_{L^2} & \lesssim \coeff^{-1/2}D_{p,k} = \coeff^{-1}U_{p,k}, \\
\| \,|{\fieldvarphi}|_{p,k} \|_{L^2} + \Bigl\| \Big({s\over t} \Big)| \del{\fieldPhi}|_{p,k} \Bigr\|_{L^2} + \| \,| \delsH{\fieldPhi}|_{p,k} \|_{L^2} & \lesssim U_{p,k}.
\end{aligned}
\ee
We use the following convention throughout this section:
\be
C \quad\hbox{is independent of } \coeff\in(0, \coeff_0], \quad C_\coeff \quad\hbox{may diverge as } \coeff\to0.
\ee
Indeed, estimates containing the symbol \(C_\coeff\) are typically \emph{non-uniform}.

\paragraph{Interior bootstrap assumptions}

We introduce the \emph{high differentiation order} $p_{\mathbf{high}}$ and the \emph{low differentiation order} $p_{\mathbf{low}}$, together with constants arising in our estimates, by
\be
\label{equa:seven-differentiation-levels}
p_{\mathbf{high}}{\coloneqq}N-5, \quad p_{\mathbf{low}}{\coloneqq}N-9, \quad A_0\coloneqq \epss+ C_0\eps, \quad A\coloneqq C_1A_0.
\ee
We take \(N\geq19\), so that \(p_{\mathbf{low}} \geq10\). With \(p_{\mathrm w}{\coloneqq} \lfloor p_{\mathbf{high}}/2\rfloor\), the derivative counts used in the interior estimates are
\be
\label{equa:seven-derivative-counts}
p_{\mathrm w} +3\leq p_{\mathbf{low}}, \qquad \left\lfloor\frac{p_{\mathbf{low}} +2}{2} \right\rfloor\leq p_{\mathbf{low}}-3, \qquad p_{\mathbf{low}} +2 +2=p_{\mathbf{high}}.
\ee
The first inequality is relevant to deal with the low factor and three characteristic or angular commutations; the other two are relevant to deal with the low--high products after taking angular and Sobolev derivatives. If more than \(p_{\mathbf{low}} \) derivatives fall on a high coefficient, at most three commutations remain on the complementary Hessian. Its additional translation and three pointwise derivatives use at most seven low-order commutations. The fixed constant \(C_{\mathrm{rk}} \geq1\) is chosen once and for all, larger than every rank-loss constant occurring in the commutator, trace, and source estimates. Motivated by our earlier studies~\cite{LeFloch-Ma-2016b,LeFloch-Ma-2018}, we set
\be
\label{equa:seven-exponents}
\rho{\coloneqq}{\delta \over64}, \quad 0<\theta \ll\delta<{1\over8}, \quad C_{\mathrm{rk}}(N+4) \theta<\rho, \quad {3\over4}<\mu<1, \quad 3C_{\mathrm{rk}} \theta<2\mu-1, \quad 0<\zeta_\star\leq {1\over2}.
\ee
We fix the exterior metric weight \(\lambda \) and the growth exponent \(\delta \) before starting the bootstrap, with
\be
\label{equa:seven-global-exponent-restrictions}
\frac34<\lambda<\mu<1, \qquad 0<\delta \leq\frac{\mu-\lambda}{2}, \qquad \delta \bigl(26(p_{\mathbf{low}} +2) +115\bigr)<32.
\ee
These restrictions are choices of fixed exponents, \underline{independent of \(\coeff\) and of the bootstrap terminal time.} For \(0\leq k\leq N\), the growth exponents are
\be
\label{equa:seven-wave-rank-exponents}
\begin{aligned}
\chi_k&\coloneqq \rho+C_{\mathrm{rk}}k\theta, & \alpha_k&\coloneqq3\rho+C_{\mathrm{rk}}k\theta, & \beta_k&\coloneqq5\rho+C_{\mathrm{rk}}k\theta, \\
\widetilde\mu_k&\coloneqq10\rho+C_{\mathrm{rk}}(k+2) \theta, & \nu_k&\coloneqq12\rho+C_{\mathrm{rk}}(k+2) \theta.
\end{aligned}
\ee
The two differentiation levels are \(p_{\mathbf{low}} \) and \(p_{\mathbf{high}} \); at either level, \(D_{p,k} \) and \(U_{p,k} \) denote the fixed-parameter and rescaled norms.

For the \emph{sharp low-order pointwise hierarchy} and for the two energy levels, we also set
\be
\label{equa:seven-level-dependent-exponents}
\eta_k\coloneqq C_{\mathrm{rk}}k\theta, \quad \eta_{\Lscr}{\coloneqq}C_{\mathrm{rk}}N\theta, \quad\qquad a_{p,k}{\coloneqq} \begin{cases}
\alpha_k,&p=p_{\mathbf{low}}, \\
\nu_k,&p=p_{\mathbf{high}},
\end{cases} \quad b_{p,k}{\coloneqq} \begin{cases}
\beta_k,&p=p_{\mathbf{low}}, \\
\nu_k,&p=p_{\mathbf{high}}.
\end{cases}
\ee
Thus we have
\be
0<\chi_k<\alpha_k<\beta_k<\widetilde\mu_k<\nu_k <{13\delta \over64}< {\delta \over4}.
\ee
Table~\ref{table:interior-rank-growth-exponents} summarizes the role of each exponent within our bootstrap estimates. 
\bei

\item At the lower level the effective-field, matter, and metric energies are closed successively.   This hierarchy remains uniform in the limit \(\coeff\to0\).

\item At the higher level, the uniform estimate for \(s^{-1/2}U\) also gives the fixed-\(\coeff\) bound for \(D= \coeff^{-1/2}U\). Thus the \emph{high-order continuation estimate keeps an explicit singular factor},  while remaining finite on every finite interval.

\eei

\begin{table}[H]
\centering
\normalsize

\renewcommand{\arraystretch}{1.16}
\begin{tabular}{>{\raggedright\arraybackslash}p{.18\textwidth}>{\raggedright\arraybackslash}p{.70\textwidth}}
\hline
Exponent & Controlled quantity \\
\hline
\(\eta_k\) & Pointwise and interface estimates \\
\(\chi_k\) & Improved lower matter energy \\
\(\alpha_k\) & Lower rescaled effective-field energy \\
\(\beta_k\) & Lower matter bootstrap energy \\
\(\widetilde\mu_k\) & Integrated metric source \\
\(\nu_k\) & Metric and normalized high energies \\
\hline
\end{tabular}
\caption{Powers of \(s\) in the two-level rank-growth hierarchy}
\label{table:interior-rank-growth-exponents}
\end{table}

\paragraph{Energy bootstrap hierarchy}

For \(p\in\{p_{\mathbf{high}},p_{\mathbf{low}} \} \), we set
\be
\omega_{p_{\mathbf{high}}}(s)\coloneqq s^{1/2}, \quad \omega_{p_{\mathbf{low}}}(s)\coloneqq1.
\ee
To write both differentiation levels together, we set
\be
\widehat U_{p,k}(s)\coloneqq \omega_p(s)^{-1}U_{p,k}(s), \quad \widehat P_{p,k}(s)\coloneqq \mathbf P_{p,k}(s), \quad \widehat W_{p,k}(s)\coloneqq \mathbf W_{p,k}(s).
\ee
Finally, to the component $H^{\H00}$ we also associate 
\be
\label{equa:seven-high-H00-radial-profile}
\bigl( \mathbf H_{p,k}(s) \bigr)^2 \coloneqq \sum_{\substack{p_{\mathbf{low}}<|J| \leq\min\{p,k\}}} \|s^{-1}L^JH^{\H00} \|_{L^2(\Mint_s)}^2.
\ee
This norm controls the \emph{boost-only metric coefficient} $H^{\H00}$ beyond the lower differentiation level. Its boundary term is kept in the radial Hardy estimate.

\begin{definition}
The \textbf{bootstrap assumptions} on a bootstrap interval \([s_0,s_1] \) are defined as follows.

\bei

\item For \(p\in\{p_{\mathbf{low}},p_{\mathbf{high}} \} \) and every \(0\leq k\leq p\), the uniform energy bounds are
\bse
\label{equa:seven-uniform-bootstrap}
\be
\label{equa:seven-uniform-bootstrap-effective-field}
\omega_p(s)^{-1}U_{p,k}(s) \leq A \, s^{a_{p,k}} \qquad \text{ (effective-field),}
\ee
\be
\label{equa:seven-uniform-bootstrap-matter}
\mathbf P_{p,k}(s) \leq A \, s^{b_{p,k}} \qquad \text{ (matter),}
\ee
\be
\label{equa:seven-uniform-bootstrap-metric}
\mathbf W_{p,k}(s) \leq A \, s^{\nu_k} \qquad \text{ (metric),}
\ee
\ese

\item
The high-order \(H^{\H00} \) coefficient estimate is
\be
\label{equa:seven-uniform-high-radial-bootstrap}
\mathbf H_{p_{\mathbf{high}},k}(s) \leq A \, s^{\nu_k}, \quad 0\leq k\leq p_{\mathbf{high}}.
\ee
\eei
\end{definition}

\paragraph{Robust and fixed-parameter levels}

Since \(\nu_k<\delta/4\), these assumptions imply
\be
\label{equa:seven-uniform-bootstrap-coarse}
\mathbf W_{p_{\mathbf{low}},p_{\mathbf{low}}}(s) + \mathbf P_{p_{\mathbf{low}},p_{\mathbf{low}}}(s) +U_{p_{\mathbf{low}},p_{\mathbf{low}}}(s) \leq 3A s^\delta.
\ee
The derivative counts in the equation~\eqref{equa:seven-derivative-counts} leave the Sobolev and characteristic orders needed by the low--high products. Throughout the improvement step, we reduce the data size so that
\be
\label{equa:seven-smallness-versus-rho}
CA<{\rho\over4}.
\ee
In parallel, for every fixed \(0<\coeff\leq\coeff_0\), we assume
\be
\label{equa:seven-degenerate-bootstrap}
D_{p_{\mathbf{low}},p_{\mathbf{low}}}(s) \leq B_\coeff s^\delta.
\ee
The constant \(B_\coeff\) is finite for each fixed $\coeff$ and is chosen explicitly below; it is permitted, and in general expected, to contain \underline{at least the loss} \(\coeff^{-1/2} \).

We define the fixed-parameter continuation norm by
\be
\label{equa:seven-fixed-high-energy}
\mathsf E_{k}^{\mathrm{tame}}(s) \coloneqq \mathbf W_{p_{\mathbf{high}},k}(s) + \mathbf P_{p_{\mathbf{high}},k}(s) +D_{p_{\mathbf{high}},k}(s) + \mathbf H_{p_{\mathbf{high}},k}(s), \quad 0\leq k\leq p_{\mathbf{high}}.
\ee
The superscript \(\mathrm{tame}\) indicates that the nonlinear source estimates are linear in the factor carrying the largest number of derivatives. The fixed-parameter norm is recovered from the uniform high hierarchy by rescaling in \eqref{equa:seven-fixed-high-rescaling}; its growth is given by \eqref{equa:seven-tame-high-final}.

\paragraph{Summary of the growth rates}

Table~\ref{table:interior-bootstrap-growth} summarizes the interior energy bounds.

\begin{table}[H]
\centering
\normalsize

\renewcommand{\arraystretch}{1.16}
\begin{tabular}{>{\raggedright\arraybackslash}p{.30\textwidth}>{\raggedright\arraybackslash}p{.32\textwidth}>{\raggedright\arraybackslash}p{.28\textwidth}}
\hline
\textbf{Quantity} & \textbf{Bound in \(s\)} & \leavevmode{\textbf{Dependence on \(\coeff\)}} \\
\hline
\(U_{p_{\mathbf{low}},k} \) & \(A s^{\alpha_k} \) & Uniform in $\coeff$ \\
\(\mathbf P_{p_{\mathbf{low}},k} \) & \(A s^{\beta_k} \) & Uniform in $\coeff$ \\
\(\mathbf W_{p_{\mathbf{low}},k} \) & \(A s^{\nu_k} \) & Uniform in $\coeff$ \\
\(U_{p_{\mathbf{high}},k} \) & \(A s^{1/2 + \nu_k} \) & Uniform in $\coeff$ \\
\(\mathbf P_{p_{\mathbf{high}},k} \), \(\mathbf W_{p_{\mathbf{high}},k} \), \(\mathbf H_{p_{\mathbf{high}},k} \) & \(A s^{\nu_k} \) & Uniform in $\coeff$ \\
\(D_{p_{\mathbf{low}},p_{\mathbf{low}}} \) & \(B_\coeff s^\delta \) & \leavevmode{For each fixed \(\coeff>0\)} \\
\(\mathsf E_k^{\mathrm{tame}} \) & \(C(1+ \coeff^{-1/2})A s^{1/2 + \nu_k} \) & \leavevmode{For each fixed \(\coeff>0\)} \\
\hline
\end{tabular}
\caption{Growth bounds for the interior energies}
\label{table:interior-bootstrap-growth}
\end{table}

We also introduce analogous notation for norms along the interior--exterior interface, as follows. We use 
\be
\mathsf F_{\mathrm w}^{\Lscr,p,k}, \quad
\mathsf F_{\psi}^{\Lscr,p,k},
\quad \mathsf F_{\mathrm{eff}}^{\mathrm{unif}, \Lscr,p,k},
\quad \mathsf F_{\mathrm{eff}}^{\mathrm{sing}, \Lscr,p,k}
\ee
 to denote the sums over \(Z\preccurlyeq(p,k) \) of the oriented fluxes in \eqref{equa:seven-exact-lateral-fluxes}: the metric uses \(\mathsf F_{\mathrm w}^{\Lscr}(s,Zu;s_0) \), the matter field uses the weighted flux \(\mathsf F_{\psi}^{\Lscr}(s,Z\psi;s_0) \), and the two effective-field quantities use \(q=Z\fieldphi\).  By definition we have
\be
\label{equa:seven-flux-rescaling}
\mathsf F_{\mathrm{eff}}^{\mathrm{unif}, \Lscr,p,k} = \coeff\, \mathsf F_{\mathrm{eff}}^{\mathrm{sing}, \Lscr,p,k}.
\ee
For \(\mathrm a \in\{\mathrm{unif}, \mathrm{sing} \} \), we abbreviate the \emph{three oriented interface fluxes} by
\be
\mathcal F_{p,k}^{\Lscr, \mathrm a}(\tau;s_0) \coloneqq \big| \mathsf F_{\mathrm w}^{\Lscr,p,k}(\tau;s_0) \big|^{1/2} + \big| \mathsf F_{\psi}^{\Lscr,p,k}(\tau;s_0) \big|^{1/2} + \big| \mathsf F_{\mathrm{eff}}^{\mathrm a, \Lscr,p,k}(\tau;s_0) \big|^{1/2}.
\ee
\paragraph{\texorpdfstring{Bounds at the interface from the exterior estimates}{Bounds at the interface from the exterior estimates}}

The interior argument requires the following collection of estimates along the interface.  The exterior analysis will establish them in Proposition~\ref{proposition:eight-exterior-interface-outputs}.

\begin{definition}
The \textbf{interface bounds required by the interior argument} are the following estimates, for \(p=p_{\mathbf{low}} \) and \(0\leq k\leq p_{\mathbf{low}} \):
\bse
\label{equa:seven-lateral-bootstrap}
\be
\label{equa:seven-lateral-uniform}
\mathcal F_{p,k}^{\Lscr, \mathrm{unif}}(s;s_0) \leq A \, s^{C_{\mathrm{rk}}N\theta},
\ee
\be
\label{equa:seven-lateral-degenerate}
\omega_p(s)^{-1} \big| \mathsf F_{\mathrm{eff}}^{\mathrm{sing}, \Lscr, p,k} \big|^{1/2} \leq B_\coeff s^{C_{\mathrm{rk}}N\theta}.
\ee
\ese
\end{definition}

We take absolute values since the sign of the oriented interior flux depends on the chosen normal and multiplier.  The argument uses the control of the same geometric flux from the Euclidean side.  In particular, the matter term is the weighted flux containing \(e^{-{\fieldPhi}} \).

We set
\be
R_s\coloneqq{s^2-1\over2}, \quad \mathbb S_s\coloneqq \Mint_s\cap\Lscr = \{t= (s^2 +1)/2, \ r=R_s\},
\ee
and define the \emph{\(L^2\) interface trace seminorm}
\be
\label{equa:seven-L2-trace-definition}
\bigl( \mathsf T_{2;p,k}[v](s) \bigr)^2 \coloneqq R_s^{-1} \sum_{Z\preccurlyeq(p,k)} \|Zv\|_{L^2(\mathbb S_s)}^2.
\ee
The exterior bootstrap gives, independently of the spacetime flux,
\bse
\label{equa:seven-trace-bootstrap}
\be
\label{equa:seven-trace-bootstrap-uniform}
\mathsf T_{2;p,k}[u](s) + \mathsf T_{2;p,k}[\psi](s) + \omega_p(s)^{-1} \mathsf T_{2;p,k}[{\fieldvarphi}](s) \leq A \, s^{C_{\mathrm{rk}}N\theta},
\ee
\be
\label{equa:seven-trace-bootstrap-degenerate}
\omega_p(s)^{-1} \mathsf T_{2;p,k}[\fieldphi](s) \leq B_\coeff s^{C_{\mathrm{rk}}N\theta}.
\ee
\ese
Here \(p=p_{\mathbf{low}} \) and \(0\leq k\leq p_{\mathbf{low}} \). The metric and massless traces, like their energies and fluxes, are left unnormalized.  The singular unscaled effective-field trace is kept separately. Pointwise traces used by the characteristic and ray arguments are introduced separately below and require two additional angular derivatives.  


\subsection{\texorpdfstring{Sobolev, Hardy, and source estimates}{Sobolev, Hardy, and source estimates}}
\label{section-8-4}
\label{section-8-5}

\paragraph{Low-order energy consequences}

We set \(\gamma_{p_{\mathbf{low}}} = \delta \). Equations \eqref{equa:seven-uniform-bootstrap}--\eqref{equa:seven-D-coercivity} imply, for \(p=p_{\mathbf{low}} \) and \(v\in\{u, \psi\} \),
\be
\label{equa:seven-direct-L2}
\Bigl\| \Big({s\over t} \Big)| \del v|_p\Bigr\|_{L^2} + \| \,| \delsH v|_p\|_{L^2} \lesssim A s^{\gamma_p}, \quad \Bigl\| \Big({s\over t} \Big)| \del{\fieldPhi}|_p\Bigr\|_{L^2} + \| \,| \delsH{\fieldPhi}|_p\|_{L^2} + \| \,|{\fieldvarphi}|_p\|_{L^2} \lesssim A s^{\gamma_p}.
\ee
whereas the bootstrap estimates give
\be
\label{equa:seven-direct-L2-singular}
\| \,| \fieldphi|_p\|_{L^2} \lesssim B_\coeff s^{\gamma_p}, \qquad \Bigl\| \Big({s\over t} \Big)| \del\fieldphi|_p\Bigr\|_{L^2} + \| \,| \delsH\fieldphi|_p\|_{L^2} \lesssim \coeff^{-1/2}B_\coeff s^{\gamma_p}.
\ee
The stronger estimate
\be
\Bigl\| \Big({s\over t} \Big)| \del\fieldphi|_p\Bigr\|_{L^2} + \| \,| \delsH\fieldphi|_p\|_{L^2} \lesssim\coeff^{-1}A s^{\gamma_p}
\ee
also follows from the uniform estimates; however, it is \emph{more singular} and will be used only when a term is naturally written in the rescaled field \(\del{\fieldPhi} \).

\paragraph{Direct pointwise consequences}

For convenience, we state the hyperboloidal Sobolev inequality in the form used on the truncated interior domain.

\begin{proposition}[\texorpdfstring{Hyperboloidal Sobolev inequality}{Hyperboloidal Sobolev inequality}]
\label{lem:seven-hyperboloidal-Sobolev}
For every sufficiently regular scalar function \(v\) on \(\Mint_s\), one has
\be
\label{equa:seven-hyperboloidal-Sobolev}
\sup_{\Mint_s}t^{3/2}|v| \lesssim \sum_{|J| \leq2} \|L^Jv\|_{L^2(\Mint_s)}.
\ee
The implicit constant is independent of \(s\), \(v\), and \(\coeff\).
\end{proposition}
\begin{proof}
On the restriction \(v_s(x)\coloneqq v(\sqrt{s^2 + |x|^2},x) \), the boosts satisfy \(L_a v=t\del_av_s\). The standard Sobolev inequality on rescaled cubes contained in the truncated hyperboloidal domain gives the factor \(t^{-3/2} \). This is the estimate of~\cite[Proposition~2.3]{LeFloch-Ma-2022}, on the hyperboloidal domain with measure \(dx\); its hypotheses allow a nonzero trace on the boundary sphere.
\end{proof}

Applying \eqref{equa:seven-hyperboloidal-Sobolev} after two additional commutations yields the following full-rank bounds. At a rank-resolved level, the two commutations cost two orders and two boost ranks; for \(v\in\{u, \psi\} \),
\bse
\be
\label{equa:seven-direct-pointwise-uniform}
{s\over t}| \del v|_{p-2} + | \delsH v|_{p-2} \lesssim A t^{-3/2}s^{\gamma_p}, \qquad {s\over t}| \del{\fieldPhi}|_{p-2} + | \delsH{\fieldPhi}|_{p-2} + |{\fieldvarphi}|_{p-2} \lesssim A t^{-3/2}s^{\gamma_p}.
\ee
\paragraph{Auxiliary high-order pointwise consequences}

At the auxiliary high level, the same argument must account for its rank loss.  The robust high bootstrap gives, for \(0\leq k\leq p_{\mathbf{high}}-2\) and \(v\in\{u, \psi\} \),
\be
\label{equa:seven-direct-pointwise-robust-high}
\begin{aligned}
{s\over t}| \del v|_{p_{\mathbf{high}}-2,k} + | \delsH v|_{p_{\mathbf{high}}-2,k} & \lesssim C_NA t^{-3/2}s^{\nu_{k+2}}, \\
{s\over t}| \del{\fieldPhi}|_{p_{\mathbf{high}}-2,k} + | \delsH{\fieldPhi}|_{p_{\mathbf{high}}-2,k} + |{\fieldvarphi}|_{p_{\mathbf{high}}-2,k} & \lesssim C_NA t^{-3/2}s^{1/2 + \nu_{k+2}}.
\end{aligned}
\ee
The factor \(s^{1/2} \) occurs only in the effective-field line and is the normalization in \eqref{equa:seven-uniform-bootstrap-effective-field}.

For the  unscaled effective field, the corresponding statements are
\be
\label{equa:seven-direct-pointwise-singular}
| \fieldphi|_{p-2} \lesssim B_\coeff t^{-3/2}s^{\gamma_p}, \qquad {s\over t}| \del\fieldphi|_{p-2} + | \delsH\fieldphi|_{p-2} \lesssim \coeff^{-1/2}B_\coeff t^{-3/2}s^{\gamma_p},
\ee
The robust effective-field energy also gives
\be
{s\over t}| \del\fieldphi|_{p-2} + | \delsH\fieldphi|_{p-2} \lesssim \coeff^{-1}A t^{-3/2}s^{\gamma_p}.
\ee
\ese
These negative powers are the \emph{explicit cost of recovering} the unscaled derivatives from either effective-field energy.

Since \(\psi\) is massless, its energy does not control \(\| \psi\|_{L^2} \).  Whenever an undifferentiated matter field occurs, we use the Hardy estimate below, which controls \(t^{-1}Z\psi\), or the wave characteristic estimate.

\begin{table}[H]
\centering
\normalsize\renewcommand{\arraystretch}{1.18}
\begin{tabular}{>{\raggedright\arraybackslash}p{.29\textwidth}>{\raggedright\arraybackslash}p{.61\textwidth}}
\hline
\textbf{Functional} & \leavevmode{\textbf{Controlled quantities and use}} \\
\hline
\(\calE_{\mathrm w}^{\H}(s,q) \), \(\calE_{\psi}^{\H}(s,q) \) & Metric and weighted massless-matter energies on \(\Mint_s\) \\
\(\calE_{\mathrm{eff}}^{\mathrm{sing}, \H}(s,q) \) & Fixed-\(\coeff\) effective-field energy for the unscaled field \\
\(\calE_{\mathrm{eff}}^{\mathrm{unif}, \H}(s,q) \) & Uniform effective-field energy for \((\del\fieldPhi, \fieldvarphi) \) \\
\(\mathbf W_{p,k}(s) \), \(\mathbf P_{p,k}(s) \) & Metric and matter energies of order \(p\) and rank \(k\) \\
\(D_{p,k}(s) \), \(U_{p,k}(s) \) & Effective-field normalizations with \(U_{p,k} = \coeff^{1/2}D_{p,k} \) \\
\(\widehat W_{p,k} \), \(\widehat P_{p,k} \), \(\widehat U_{p,k} \) & Energies on the growth hierarchy \\
\(\mathsf F_{\mathrm w}^{\Lscr} \), \(\mathsf F_{\psi}^{\Lscr} \), \(\mathsf F_{\mathrm{eff}}^{\mathrm a, \Lscr} \) & Oriented fluxes cancelling their exterior counterparts \\
\(\mathsf T_{2;p,k}[v] \), \(\mathbf H_{p,k} \) & Sphere trace and high boost-only \(H^{\H00} \) profile \\
\(\mathsf E_k^{\mathrm{tame}} \) & Fixed-\(\coeff\) continuation energy with linear top order \\
\hline
\end{tabular}
\caption{Relevant functionals for the interior bootstrap argument}
\label{table:interior-functionals}
\end{table}

\paragraph{Hardy inequality with the interface trace}

On a truncated hyperboloid, Hardy's inequality uses the tangential derivatives and the sphere trace defined in \eqref{equa:seven-L2-trace-definition}.  

\begin{lemma}[Interior Hardy inequality with a sphere trace]
For every sufficiently regular \(v\), one has
\be
\label{equa:seven-Hardy-with-boundary}
\|t^{-1}|v|_{p,k} \|_{L^2(\Mint_s)} \lesssim \bigl\|| \delsH v|_{p,k} \bigr\|_{L^2(\Mint_s)} + \mathsf T_{2;p,k}[v](s).
\ee
Here the right-hand side is controlled by the corresponding wave energy and its sphere trace.
\end{lemma}

\begin{proof}
\bse
We fix a commuted field and restrict it to the graph: \(f(x)=Zv(\sqrt{s^2 + |x|^2},x) \), so \(\del_af= \delsH_aZv\). The vector field \(x/r^2\) has divergence \(r^{-2} \) in three dimensions. Integration on \(B_{R_s} \), with a vanishing inner boundary obtained by regularization, gives
\be
\label{equa:seven-Hardy-tangential-identity}
\int_{B_{R_s}} \frac{f^2}{r^2} \, dx =R_s^{-1} \int_{\mathbb S_s}f^2\, dS -2\int_{B_{R_s}}f\, \frac{x}{r^2} \cdot\nabla f\, dx.
\ee
Cauchy--Schwarz and \(t\geq r\) therefore imply
\be
\label{equa:seven-Hardy-tangential-bound}
\|f/t\|_2\leq\|f/r\|_2 \leq2\| \nabla f\|_2 +R_s^{-1/2} \|f\|_{L^2(\mathbb S_s)}.
\ee
Summing over the finite commutator family proves the claim, with a constant depending only on the order.
\ese
\end{proof}

\paragraph{Radial \(L^2\) estimate with a moving sphere}

For the high-coefficient commutator we need a second, very different, estimate.  It is important here to differentiate the moving radial boundary \(R_s= (s^2-1)/2\), so that \(R_s'=s\).  If \(v_s(x)\coloneqq v(\sqrt{s^2 + |x|^2},x) \), direct differentiation of \( s^{-2} \int_{|x| \leq R_s}|v_s(x)|^2dx \) gives the following moving-boundary estimate.

\begin{proposition}[\texorpdfstring{Radial integration with the moving boundary}{Radial integration with the moving boundary}] One has
\be
\label{equa:seven-radial-L2-moving-boundary}
\|s^{-1}v\|_{L^2(\Mint_s)} \leq \|s_0^{-1}v\|_{L^2(\Mint_{s_0})} + \int_{s_0}^{s} \tau^{-1} \Bigl\|{\tau\over t} \del_tv\Bigr\|_{L^2(\Mint_\tau)}d\tau + \left\| \tau^{-1/2} \|v\|_{L^2(\mathbb S_\tau)} \right\|_{L^2((s_0,s);d\tau)}.
\ee
\end{proposition}

Indeed, if \(E_v(s)\coloneqq s^{-1} \|v_s\|_{L^2(B_{R_s})} \), then
\be
\label{equa:seven-radial-L2-differentiation}
{d\over ds}E_v(s)^2 \leq 2E_v(s)s^{-1} \Bigl\|{s\over t} \del_tv\Bigr\|_{L^2(\Mint_s)} +s^{-1} \|v\|_{L^2(\mathbb S_s)}^2.
\ee
The elementary inequality \(E^2{}'\leq2Ef+b\Rightarrow E(s) \leq E(s_0) + \int f+ \sqrt{\int b} \) proves \eqref{equa:seven-radial-L2-moving-boundary}.

For an expanding interior ball, we denote its radius by $R_b(s)$ and its boundary by $\mathbb S_s^b$. The sphere term in \eqref{equa:seven-radial-L2-differentiation} becomes $s^{-2}R_b'(s)\|v\|_{L^2(\mathbb S_s^b)}^2$, and the integrated trace in \eqref{equa:seven-radial-L2-moving-boundary} becomes the $L^2(d\tau)$ norm of $\tau^{-1}\sqrt{R_b'(\tau)}\|v\|_{L^2(\mathbb S_\tau^b)}$. The identity~\eqref{equa:audit-moving-interface-radial-profile} and $R_b'(s)\simeq s$ preserve the ensuing powers and commutation ranks.

Because the boundary derivative of the formally stronger expression with \(s^{-1}(s/t)^{-1/4}v\) contains the factor \(R_s'=s\), we use \eqref{equa:seven-radial-L2-moving-boundary} with its full integrated trace. For example, if
\be
\label{equa:seven-radial-pointwise-trace}
\|v\|_{L^\infty(\mathbb S_\tau)} \leq K\tau^{-2 + C_{\mathrm{rk}}k\theta}, \quad k\geq1,
\ee
then \(R_\tau\simeq\tau^2\), and we obtain
\be
\label{equa:seven-radial-trace-integration}
\int_{s_0}^{s} \tau^{-1} \|v\|_{L^2(\mathbb S_\tau)}^2d\tau \lesssim {K^2\over C_{\mathrm{rk}}k\theta} s^{2C_{\mathrm{rk}}k\theta}.
\ee
Thus the boundary term in \eqref{equa:seven-radial-L2-moving-boundary} is the interface trace, while the other two terms are interior quantities.

\paragraph{Parameter dependence and effective-field source}

The source estimates distinguish the unscaled and rescaled effective-field variables. In the small-field range under consideration, smooth coefficients depending on \(g\) and \({\fieldPhi} \) are uniformly bounded together with all derivatives required below.  The effective-field contribution in the metric equation is
\bse
\be
\Gbb_{\coeff, \alpha \beta} = -3\, \del_\alpha{\fieldPhi} \, \del_\beta{\fieldPhi} -\coeff\Ufrak(\fieldphi)g_{\alpha \beta},
\ee
where the potential term satisfies
\be
\label{equa:seven-Ufrak-rescaled}
\coeff\Ufrak(\fieldphi) = {1\over2}(\fieldvarphi)^2 + {(\fieldvarphi)^2{\fieldPhi} \, \mathcal U_{1, \coeff}({\fieldPhi})}.
\ee
Consequently, after commutation we have
\be
\label{equa:seven-G-source-structure}
\begin{aligned}
|Z\Gbb_\coeff| & \lesssim \sum_{Z_1Z_2\preccurlyeq Z} | \del Z_1{\fieldPhi}| \,| \del Z_2{\fieldPhi}| + \sum_{Z_1Z_2\preccurlyeq Z} |Z_1{\fieldvarphi}| \,|Z_2{\fieldvarphi}| \\
& \quad + \sum_{\substack{q\geq3\\
Z_1\cdots Z_q\preccurlyeq Z}} \Bigl( \prod_{\ell=1}^{2}|Z_\ell{\fieldvarphi}| \Bigr) \Bigl( \prod_{\ell=3}^{q}|Z_\ell{\fieldPhi}| \Bigr) + \mathcal C_Z(u,{\fieldPhi},{\fieldvarphi}),
\end{aligned}
\ee
Here the last term has the explicit pointwise bound
\be
\label{equa:seven-G-source-metric-product}
\mathcal C_Z(u,{\fieldPhi},{\fieldvarphi}) \lesssim \sum_{\substack{q\geq2, \ Z_0Z_1\cdots Z_q\preccurlyeq Z\\
\ord Z_0\geq1}} |Z_0(u+h^\star)| \, |Z_1{\fieldvarphi}| \,|Z_2{\fieldvarphi}| \prod_{\ell=3}^{q}|Z_\ell{\fieldPhi}|.
\ee
\ese
Indeed, in coordinate components the derivative product \(\del{\fieldPhi} \, \del{\fieldPhi} \) has no metric coefficient.  The only additional differentiated metric factor comes from \(\coeff g_{\alpha \beta} \Ufrak(\fieldphi) \), and \eqref{equa:seven-Ufrak-rescaled} gives \eqref{equa:seven-G-source-metric-product}.  Thus every effective-field derivative in \(\mathcal C_Z\) is a derivative of \({\fieldPhi} \) or ${\fieldvarphi}$, and every undifferentiated quadratic effective-field factor is a factor ${\fieldvarphi}$. Formula \eqref{equa:seven-G-source-structure}, rather than a statement about unscaled \(\del\fieldphi\), is the robust structure used in the metric estimates.

\paragraph{Matter stress and trace terms}

Since we have assumed \(V\equiv0\), the matter contribution is
\bse
\bel{equa-matterc}
\Mbb_{\coeff, \alpha \beta} = -16\pi e^{-{\fieldPhi}} \del_\alpha \psi\, \del_\beta \psi.
\ee
Thus its leading term is a matter--matter product.  Derivatives of the conformal weight generate derivatives of \({\fieldPhi} \):
\be
\label{equa:seven-M-source-structure}
|Z\Mbb_\coeff| \lesssim \sum_{\substack{q\geq2\\
Z_1\cdots Z_q\preccurlyeq Z}} \Bigl( \prod_{\ell=1}^{q-2}|Z_\ell{\fieldPhi}| \Bigr) | \del Z_{q-1} \psi| \,| \del Z_q\psi|.
\ee
There is no derivative of \(u\) in \eqref{equa:seven-M-source-structure}: the formula is written in coordinate components, and the only differentiated coefficient is the conformal weight \(e^{-{\fieldPhi}} \). 

In contrast, metric derivatives occur instead in the contracted source \(\mathcal S\) in the trace equation: 
\be
\label{equa:seven-S-exact}
\mathcal S = -8\pi e^{-{\fieldPhi}}g(\nabla \psi, \nabla \psi) \quad (V\equiv0).
\ee
\ese
It has no global factor $\coeff$.  This fact is responsible for the singular source norm in the \(D\)-energy inequality below.  The nonlinear potential remainder is
\bse
\be
\label{equa:seven-W-exact}
W_\coeff(\fieldphi) = {\coeff\fieldphi^2\mathcal W_\coeff({\fieldPhi})} = {(\fieldvarphi)^2\mathcal W_\coeff({\fieldPhi})}.
\ee
Hence we obtain
\be
\label{equa:seven-W-product}
|ZW_\coeff(\fieldphi)| \lesssim \sum_{\substack{q\geq2\\
Z_1\cdots Z_q\preccurlyeq Z}} |Z_1{\fieldvarphi}| \,|Z_2{\fieldvarphi}| \prod_{\ell=3}^{q}|Z_\ell{\fieldPhi}|.
\ee
\ese
We observe that there is no additional factor $\coeff$ in front of \(ZW_\coeff\), and its leading term is quadratic.

\paragraph{Matter equation}

Finally, the massless matter equation is $ \Box_g\psi = g(\nabla{\fieldPhi}, \nabla \psi) $ or, equivalently,
\be
\label{equa:seven-matter-divergence-form}
\nabla_\alpha \bigl(e^{-{\fieldPhi}} \nabla^\alpha \psi\bigr)=0.
\ee
This divergence form is \emph{convenient at top order}: it is uniform at the level of the variables propagated in \eqref{equa:seven-uniform-bootstrap}, and it prevents the factor \(\del{\fieldPhi} \, \del\psi\) with all derivatives on \({\fieldPhi} \) from being estimated as an unrelated inhomogeneous wave source.

The usual low--high decomposition places the factor of order at most \(p_{\mathbf{low}}-2\) in \(L^\infty\) and the other factor in the appropriate energy norm.  We must also distinguish the massless stress tensor from the Klein--Gordon-type sources.  Indeed, for the matter contribution \eqref{equa-matterc} we have
\be
\label{equa:seven-massless-direct-source}
\| \,| \Mbb_\coeff|_{p_{\mathbf{low}}} \|_{L^2(\Mint_s)} \lesssim A^2s^{-1+2\delta},
\ee
and in general this cannot be improved to \(s^{-3/2 +2\delta} \).  On the boundary \(r=t-1\), the two factors \(t/s\) needed to recover Cartesian derivatives from the wave energy leave exactly the borderline factor \(s^{-1} \).

For our interior estimates, for all \(0\leq j\leq p_{\mathbf{low}}-2\) we define the sup-norm
\be
\mathcal Q_j(s) \coloneqq \Bigl\|{t\over s}| \del\psi|_{p_{\mathbf{low}}-2,j} \Bigr\|_{L^\infty(\Mint_s)}
\ee
and Leibniz' rule gives 
\be
\label{equa:seven-massless-source-rank}
\| \,| \Mbb_\coeff|_{p,k} \|_{L^2} \lesssim C\sum_{j=0}^{\min\{k,p_{\mathbf{low}}-2\}} \mathcal Q_j(s) \mathbf P_{p,k-j}(s) + \mathcal R^{\fieldPhi}_{p,k}(s),
\ee
where a derivative falling on \(e^{-{\fieldPhi}} \) is put in \(\mathcal R^{\fieldPhi}_{p,k} \). In detail, we find
\be
\label{equa:seven-massless-conformal-weight-remainder}
\mathcal R^{\fieldPhi}_{p,k}(s) \lesssim \sum_{\substack{q\geq3, \ \ord(Z_1\cdots Z_q) \leq p\\
\rank(Z_1\cdots Z_q) \leq k\\
\ord(Z_1\cdots Z_{q-2}) \geq1}} \Bigl\| \Bigl(\prod_{\ell=1}^{q-2}|Z_\ell{\fieldPhi}| \Bigr) | \del Z_{q-1} \psi| \,| \del Z_q\psi| \Bigr\|_{L^2(\Mint_s)}.
\ee
The condition on the total order of the first factors expresses the fact that at least one commutation has fallen on the conformal weight; the undifferentiated weight is included in the bounded coefficient of the principal quadratic term.  The usual low--high split in \eqref{equa:seven-massless-conformal-weight-remainder}, together with \eqref{equa:seven-coeff-conversion-L2}, never divides by $\coeff$. Hyperboloidal Sobolev, the fact that \(t\leq Cs^2\) on \(\Mint_s\), and \eqref{equa:seven-uniform-bootstrap-matter} imply the bound obtained directly from the energy hierarchy: 
\be
\mathcal Q_j(s) \lesssim s^{-1} \mathbf P_{p_{\mathbf{low}},j+2}(s) \lesssim A \, s^{-1+ \beta_{j+2}}.
\ee
Consequently, we obtain
\be
\label{equa:seven-massless-source-inhomogeneous}
\| \,| \Mbb_\coeff|_{p,k} \|_{L^2} \lesssim CA^2s^{-1+ \widetilde\mu_k} + CA^2s^{-3/2 + C_{\mathrm{rk}}k\theta}.
\ee
This conservative form is used only at \(p=p_{\mathbf{low}} \). Its first term is borderline but quadratic, and we have
\be
\int_{s_0}^{s} \tau^{-1+ \widetilde\mu_k} \, d\tau \leq \widetilde\mu_k^{-1}s^{\widetilde\mu_k}.
\ee
Since \(\widetilde\mu_k<\nu_k\), it lies strictly below the metric energy growth.  The characteristic hierarchy below gives a sharper form, but the conservative estimate and its explicit factor \(\widetilde\mu_k^{-1} \) will also be kept.

\paragraph{Effective-field source}

The contracted source \(\mathcal S\) on the right-hand side of \eqref{equa:KG-effective-field-reduced} requires a separate estimate. In the semi-hyperboloidal frame, we have
\be
\label{equa:seven-null-contraction}
\begin{aligned}
|g(\nabla v, \nabla w)| & \lesssim | \delsH v| \,| \del w| + | \del v| \,| \delsH w| + \Bigl( \Big({s\over t} \Big)^2 + |H^{\H00}| \Bigr)| \del v| \,| \del w| \\
& \quad + |H| \Bigl( | \delsH v| \,| \del w| + | \del v| \,| \delsH w| \Bigr).
\end{aligned}
\ee
Thus only the refined component \(H^{\H00} \) multiplies a \emph{bad--bad product.} The other metric components multiply a product containing \emph{at least one good derivative.}  Together with \eqref{equa:seven-G-source-structure} and \eqref{equa:seven-W-product}, the direct robust bootstrap estimates give, for \(p=p_{\mathbf{low}} \),
\be
\label{equa:seven-structured-source-rank}
\begin{aligned}
& 
\| \,| \Gbb_\coeff|_{p,k} \|_{L^2} + \| \,| \mathcal S|_{p,k} \|_{L^2} + \| \,|W_\coeff|_{p,k} \|_{L^2} 
\\
& \lesssim CA s^{-3/2 + \beta_{k+2}} \big(U_{p,k} + \mathbf P_{p,k} \big) 
+ CA\sum_{j=1}^{k} s^{-3/2 + \beta_{j+2}} \big(U_{p,k-j} + \mathbf P_{p,k-j} \big) + CA^2s^{-3/2 + \beta_{k+2}}.
\end{aligned}
\ee
For the product \(\del\fieldPhi\,\del\fieldPhi\), the lower-order derivative is estimated from the mass term at one additional translation order. With \(d\leq\lfloor p_{\mathbf{low}}/2\rfloor\) and \(j\leq d\), the two Sobolev boosts require \(d+3\leq p_{\mathbf{low}}\), and give
\be
\label{equa:reaudit-effective-derivative-mass-control}
|\del\fieldPhi|_{d,j}\lesssim\sqrt\coeff\,A t^{-3/2}s^{\alpha_{j+2}},\qquad \frac ts|\del\fieldPhi|_{d,j}\lesssim\sqrt\coeff\,A s^{-3/2+\alpha_{j+2}}.
\ee
The second inequality uses \(t\geq s\). Pairing it with the weighted derivative in the higher-order energy gives the effective-field derivative contribution in the estimate~\eqref{equa:seven-structured-source-rank}, since \(\alpha_{j+2}<\beta_{j+2}\) and \(\coeff\leq\coeff_0\). The estimate \eqref{equa:seven-structured-source-rank} for \(\mathcal S\) uses \eqref{equa:seven-null-contraction}; it is not deduced from \eqref{equa:seven-massless-direct-source}.  At the first stage, before the refined metric estimate is available, the terms containing \(H\) are kept in low--high form and are absorbed by the metric bootstrap.  The weight \(s^{-3/2 + \beta_{j+2}} \) is integrable when it multiplies an unnormalized wave energy.  At top effective-field order, however, \(s^{-3/2}U_{p_{\mathbf{high}},k} =s^{-1}(s^{-1/2}U_{p_{\mathbf{high}},k}) \); this is a triangular borderline term, not an ``integrable remainder.''  The separate energy hierarchy below keeps this distinction explicit.

We compare the contribution of the trace source to the two effective-field energies.  In the unscaled identity, the source is paired with \(\del Z\fieldphi\), while \(D_{p,k} \) controls only \(\coeff^{1/2} \del Z\fieldphi\).  Cauchy--Schwarz therefore produces the factor \(\coeff^{-1/2} \|Z\mathcal S\|_{L^2} \).  After multiplication by \(\coeff\), the derivative becomes \(\del Z{\fieldPhi} \), which is controlled by \(U_{p,k} \), and the same source enters with the uniform size \(\|Z\mathcal S\|_{L^2} \).  The two expressions below are therefore the two normalizations of the same source pairing:
\be
\label{equa:seven-source-normalizations-preview}
\text{degenerate scale: } \coeff^{-1/2} \|Z\mathcal S\|_{L^2}, \qquad \text{robust scale: } \|Z\mathcal S\|_{L^2}.
\ee
The first expression may blow up.  It will be kept in the fixed-$\coeff$ continuation estimate.



\section{Interior pointwise estimates} 
\label{section-9}


\subsection{Wave gauge, characteristics, and the rank hierarchy}
\label{section-9-1}

\paragraph{Gauge bounds and ray reduction}

We establish wave-gauge bounds and a first metric estimate, then derive the effective-field ray equation and the ordered commutators. Sections~\ref{section-coupled-pointwise} and~\ref{section-characteristic-pointwise} will combine these identities with source estimates. The traces at the interface and the boost-ranks are specified explicitly. Tables~\ref{table:interior-rank-growth-exponents}--\ref{table:interior-bootstrap-growth} summarize the list of growth exponents, and Figure~\ref{fig:interior-decay-allocation} shows the relevant regions.

\begin{figure}[H]
\centering

\begin{tikzpicture}[x=.62cm,y=.62cm,font=\normalsize,every node/.style={text=black}]
  \fill[black!3] (0,0) -- plot[domain=2.1:8.8,samples=80]
    ({\x-1.25*pow(\x,.3333)},{\x}) -- (0,8.8) -- cycle;
  \fill[black!5] plot[domain=2.1:8.8,samples=80]
    ({\x-1.25*pow(\x,.3333)},{\x}) -- (7.8,8.8) -- (0,1) -- cycle;
  \draw[->,thick] (0,0) -- (8.8,0) node[below] {\(x\)};
  \draw[->,thick] (0,0) -- (0,9.2) node[left] {\(t\)};
  \draw[dotted] (0,0) -- (8.7,8.7) node[below right] {\(x=t\)};
  \draw[thick,dashed] (0,1) -- (7.8,8.8);
  \node[rotate=45,anchor=south,fill=white,inner sep=1pt] at (6.8,7.8) {\(\Lscr:\ x=t-1\)};
  \draw[thick,dash dot] plot[domain=2.1:8.8,samples=80]
    ({\x-1.25*pow(\x,.3333)},{\x});
  \node[fill=white,inner sep=1pt] at (1.75,5.45) {\(q=t-x\simeq t^{1/3} \)};
  \draw[black,thick] plot[domain=0:1.50,samples=60]
    ({\x},{sqrt(4+ \x*\x)});
  \draw[black,thick] plot[domain=0:3.145,samples=60]
    ({\x},{sqrt(7.29+ \x*\x)});
  \draw[black,thick] plot[domain=0:5.28,samples=60]
    ({\x},{sqrt(11.56+ \x*\x)});
  \draw[black,thick] plot[domain=0:7.50,samples=60]
    ({\x},{sqrt(16+ \x*\x)});
  \node[anchor=east] at (2.25,3.45) {\(\H_{s_1} \)};
  \node[anchor=east] at (4.10,4.95) {\(\H_{s_2} \)};
  \node[align=center] at (1.45,7.25) {\emph{central} \\\emph{region}};
  \node[align=center] at (5.15,7.35) {\emph{lightlike} \\\emph{region}};
  \node[align=center,rotate=45] at (7.55,5.55) {Euclidean--merging\\domain};
  \draw[black!55] (11.15,.3) -- (11.15,8.8);
  \node[anchor=west,align=left,text width=4.25cm] at (11.45,7.15)
    {\emph{Hyperboloidal Sobolev} \\[-1mm]
     estimate in the central region\\[-1mm]
     \(\Longrightarrow\ t^{-3/2}s^{\gamma} \) pointwise decay};
  \node[anchor=west,align=left,text width=4.25cm] at (11.45,3.55)
    {\emph{Sobolev estimate on the} \\[-1mm]
     \emph{spheres} \\[-1mm]
     and characteristic integration\\[-1mm]
     \(\Longrightarrow\ t^{-1}s^{\eta_k} \) for the bad wave derivative};
\end{tikzpicture}
\caption{Principal decay estimates in the half-plane \((t,x) \), with \(x\geq0\) and \(q=t-x\).  The dash-dotted curve \(q\simeq t^{1/3} \) marks the transition scale between the central and lightlike estimates.  The hyperboloidal Sobolev estimate gives the direct \(t^{-3/2} \) decay in the central region.  In the lightlike region, the Sobolev estimate on the spheres is combined with characteristic integration to control the bad wave derivative with \(t^{-1} \) decay.  The dashed line \(\Lscr\) is the inner boundary of the Euclidean--merging domain.}
\label{fig:interior-decay-allocation}
\end{figure}

\paragraph{Wave-gauge control and traces at the interface}

We use \(\widehat H^{\H00} \) for the \(00\)-component of the inverse-metric perturbation in the semi-hyperboloidal frame.  It is algebraically determined by the covariant perturbation \(u=g-g^\star\) through \eqref{equa:inverse-metric-dynamic-identity}.  The wave-coordinate condition yields
\bse
\be
\label{equa:seven-wave-gauge}
\begin{aligned}
| \del\widehat H^{\H00}|_{p,k} & \lesssim | \delsH u|_{p,k} + \Big({s\over t} \Big)^2| \del u|_{p,k} + t^{-1}|u|_{p,k} + | \del h^\star|_{p,k} + t^{-1}|h^\star|_{p,k} \\
& \quad + \sum_{\substack{p_1+p_2\leq p\\
k_1+k_2\leq k}} \Bigl(\begin{aligned}
&|u|_{p_1,k_1}| \del u|_{p_2,k_2} + |u|_{p_1,k_1}| \del h^\star|_{p_2,k_2} \\
& \quad + |h^\star|_{p_1,k_1}| \del u|_{p_2,k_2} + |h^\star|_{p_1,k_1}| \del h^\star|_{p_2,k_2}
\end{aligned} \Bigr).
\end{aligned}
\ee
Together with the Hardy estimate, this gives the following explicit bound at the lower level
\be
\label{equa:seven-dLuH00}
\sum_{a=1}^3\| \,| \del L_a \widehat H^{\H00}|_{N-6{,}p_{\mathbf{low}}-1} \|_{L^2(\Mint_s)} \lesssim A s^{\delta} +A^2s^{2\delta} + \mathsf T_{2;N-5,N-5}[u](s) + \mathsf T_{2;p_{\mathbf{low}},p_{\mathbf{low}}}[u](s) +A_0, 
\ee
which involves a trace contribution. 
\ese
(We observe that the high-order interface estimate derived in \eqref{equa:seven-e-normalized-trace-bound}, below, together with the metric bootstrap bounds, gives the coarser bound \(CA s^{2\delta} \)).

We introduce the three pointwise interface quantities used below. We denote by \(\mathscr G(u) \) the semi-hyperboloidal components containing \emph{at least one tangential direction:}
\be
\mathscr G(u) \coloneqq \bigl\{u(\delH_\alpha, \delH_\beta):0\leq\alpha, \beta \leq3, (\alpha, \beta) \neq(0,0) \bigr\}.
\ee
We then set 
\be
\label{equa:seven-metric-pointwise-traces}
\begin{aligned}
\bigl( \mathsf B_{00;k}(s) \bigr)^2 &\coloneqq \max_{0\leq j\leq k} \ \sup_{s_0\leq\tau\leq s} \tau^{2 (2-\eta_j)}R_\tau^{-2} \sum_{\substack{|J|=j\leq p_{\mathbf{low}} \\
|K| \leq2}} \| \Omega^K L^JH^{\H00} \|_{L^2(\mathbb S_\tau)}^2, \\
\mathsf B_{\mathrm{char};p,k}(s) &\coloneqq \max_{w\in\mathscr G(u)} \sup_{\Lscr_{[s_0,s]}} t^{-\eta_k/2} | (\del_t-\del_r) (r w)|_{p,k}, \\
\mathsf B_{\psi;k}(s) &\coloneqq \sup_{\Lscr_{[s_0,s]}} t^{-\eta_k/2} |(\del_t-\del_r) (r \psi)|_{p_{\mathbf{low}}-3,k},
\end{aligned}
\ee
which are associated with the $(00)$-component, the interface metric component, and the interface matter component, respectively.

The two rotations in the first seminorm $\mathsf B_{00;k}(s)$ are the two additional angular commutations used when applying the Sobolev inequality on the spheres.  Since \(p_{\mathbf{high}}-p_{\mathbf{low}} =4\), they remain \emph{inside the high family} even when \(|J|=p_{\mathbf{low}} \).  The normalization \(\eta_j\) measures the target number of boosts. Whenever these rotations are rewritten in an ordinary interior energy norm, the ranks \(j+1\) and \(j+2\) are charged explicitly. The boundary traces in the characteristic and ray estimates can require one further angular commutation.  On \(\Lscr\), where \(r \simeq t\simeq s^2\), the exterior gain \(r^{\mu-1/2} \) absorbs this three-rank loss since
\be
\eta_{j+3}-\eta_j=3C_{\mathrm{rk}} \theta<2\mu-1.
\ee
The first quantity in \eqref{equa:seven-metric-pointwise-traces} is cumulative in the boost rank; note the passage from the covariant perturbation \(u\) to the contravariant component \(H^{\H00} \).  The factor \(R_\tau^{-1} \), together with the two rotations, is the Sobolev factor on \(\mathbb S_\tau\); hence we obtain the pointwise Sobolev bound
\be
\|L^JH^{\H00} \|_{L^\infty(\mathbb S_\tau)}^2 \leq (C_N)^2 R_\tau^{-2} \sum_{|K| \leq2} \| \Omega^KL^JH^{\H00} \|_{L^2(\mathbb S_\tau)}^2.
\ee 
These trace norms are the required outputs of the exterior estimates in Section~\ref{section:exterior-characteristics}, not consequences of \eqref{equa:seven-lateral-bootstrap} alone.  The analogous initial traces on \(\Mint_{s_0} \) are controlled by initial Sobolev and are already included in \(A_0\); they are not part of the lateral seminorms above.

\paragraph{Direct wave-gauge estimates}

The commuted wave-gauge estimate \eqref{equa:seven-wave-gauge} first gives a derivative bound using only the coarse metric energy and Hardy estimates. For every ordered translation--boost operator \(Z\) with \(\ord Z\leq p_{\mathbf{low}} \), we have
\be
\label{equa:seven-direct-H00-gradient}
| \del ZH^{\H00}| \lesssim A t^{-3/2}s^\delta +A^2t^{-1}s^{-1+2\delta} +A_0t^{-2}.
\ee
The linear terms are \(\delsH Zu\), \((s/t)^2\del Zu\), and \(t^{-1}Zu\); the quadratic term keeps both factors in \(u\del u\). This is the local commuted gauge calculation of~\cite[Section~4.7]{LeFloch-Ma-2018}. The good derivatives also control the Cartesian gradient through \(\del_a=\delsH_a-(x_a/t)\del_t\). It uses at most \(\ord Z+3\leq p_{\mathbf{high}} \) energy orders. 

For a boost operator \(L^J\), radial integration at fixed physical time from \(r\) to \(t-1\) gives
\be
\label{equa:seven-direct-H00-value}
|L^JH^{\H00}| \lesssim A\Bigl(\frac st\Bigr)^2t^{-1/2}s^\delta +A^2\Bigl(\frac st\Bigr)^2s^{-1+2\delta} + \bigl(A_0 + \mathsf B_{00;|J|}(s) \bigr)t^{-1+ \eta_{|J|}/2}, \qquad |J| \leq p_{\mathbf{low}}.
\ee
Indeed, the interval has length at most \(s^2/t\), and the quadratic integral is bounded by
\be
\label{equa:seven-direct-H00-radial-integral}
\int_r^{t-1}(t^2-\rho^2)^{-1/2 + \delta} \, d\rho \lesssim t^{-1}s^{1+2\delta}.
\ee
The boundary has parameter \(\sqrt{2t-1} \leq s\), so its value is controlled by the cumulative trace. If this boundary lies below the initial hyperboloid, the integration stops on that hyperboloid; this bounded-time contribution is included in \(A_0\). The same estimates apply to tensor Lie derivatives: boosts preserve \(dt-(x^a/t)dx^a \) up to a bounded scalar factor, while translations produce \(t^{-1} \) frame terms.

At rank zero, \eqref{equa:seven-direct-H00-value} controls the ray coefficient directly. Writing \(\chi =s/t\), we obtain
\be
\label{equa:seven-direct-ray-coefficient-value}
| \chi^{-2}H^{\H00}| \lesssim A\chi^{1/2}s^{-1/2 + \delta} +A^2s^{-1+2\delta} + \bigl(A_0 + \mathsf B_{00;0}(s) \bigr) \chi^{-1}s^{-1}.
\ee
Since \(\chi s\geq1\) in the interior, this gives the required smallness of the principal coefficient from the coarse bootstrap and the interface output.

\paragraph{Energy on the bounded initial interval}

We fix the height of the Euclidean end of the initial hypersurface by
\be
\label{equa:seven-cone-initial-time-ceiling}
T_0\coloneqq\fTime(s_0,\rhoE(s_0)),\qquad s_0\leq\fTime(s_0,r)\leq T_0.
\ee
The bounded initial range can be estimated on the available foliation hypersurfaces themselves. We use \(p_{\mathbf{high}}=N-5\geq14\), write \(\langle r\rangle\coloneqq\sqrt{1+r^2}\), and use the following norm on the full hypersurface; the domains and normal densities are adapted when the moving foliation is used:
\bse
\be
\label{equa:audit-bounded-initial-energy}
\begin{aligned}
\mathcal E_{\mathrm{bd}}(s)^2&\coloneqq\sum_{\ord Z\leq p_{\mathbf{high}}}\Bigl(\|\langle r\rangle^\lambda\del Zu\|_{L^2(\Mcal_s)}^2+\|\langle r\rangle^{\lambda-1}Zu\|_{L^2(\Mcal_s)}^2\\
&\quad+\|\langle r\rangle^\mu(\del Z\psi,\del Z\fieldPhi,Z\fieldvarphi)\|_{L^2(\Mcal_s)}^2\Bigr).
\end{aligned}
\ee
For \(s\leq T_0\), the Cartesian times on \(\Mcal_s\) are bounded by a constant fixed by the initial geometry, and the geometric energy coefficients have positive lower bounds. We apply the current identities with the spatial weights \(\langle r\rangle^{2\lambda}\) and \(\langle r\rangle^{2\mu}\) on the full hypersurface. The two orientations cancel at the interface because each field has the same weight on both sides. The weighted Hardy inequality controls the dynamic metric term. A principal product with a high coefficient is estimated by this Hardy norm and a lower Hessian. We use the coarse high-order bounds for the lower factor: including its Hessian and the exterior Sobolev derivatives requires at most \(\lfloor p_{\mathbf{high}}/2\rfloor+7\leq p_{\mathbf{high}}\) commutations. The scalar decay also absorbs the difference between the weights: a high metric coefficient in a scalar equation leaves the bounded multiplier \(\langle r\rangle^{\mu+1-\lambda}|\del^2Z_{\mathrm{low}}\psi|\lesssim A\langle r\rangle^{-\lambda}\), and the same bound holds for \(\fieldPhi\). A high scalar factor in the metric stress leaves \(\langle r\rangle^{\lambda-\mu}|\del Z_{\mathrm{low}}\psi|\lesssim A\langle r\rangle^{\lambda-2\mu-1}\). Every derivative of a weight is bounded by a fixed multiple of that weight. In the effective-field equation we keep the constant mass on the left: its energy controls \(Z\fieldvarphi\), and the nonlinear potential factors through the rescaled fields. These terms and the metric commutators therefore introduce no negative power of \(\coeff\). The independent reference forcing is supported in a fixed bounded region on this time interval and has norm at most \(CA_0\). The product estimates and the current identities give
\be
\label{equa:audit-bounded-initial-Gronwall}
\mathcal E_{\mathrm{bd}}(s)^2\leq CA_0^2+C\int_{s_0}^s(1+A)\mathcal E_{\mathrm{bd}}(\sigma)^2\,d\sigma,\qquad \sup_{s_0\leq s\leq\min\{s_1,T_0\}}\mathcal E_{\mathrm{bd}}(s)\leq C A_0.
\ee
\ese
The constants depend on the fixed geometry, exponents, profile bounds, and \(\coeff_0\), but not on \(\coeff\), \(C_1\), or the available terminal time \(s_1\), since \(A\leq1\). This argument is performed only on the existing slab. For an interior target with \(s<T_0\), the inequality \(t-r\geq1\) gives \(t\leq T_0^2\); on the adapted domains, \(t-r\geq3/4\) gives the same bounded-time conclusion with a fixed constant. The direct Sobolev estimates therefore prove the required bound in this range. For \(s\geq T_0\), the inequality \(\fTime(s,r)\geq s\) places the entire region between \(\Sigma_0\) and \(t=T_0\) in the existing slab. The terminal graph is \(1\)-Lipschitz, so \(\fTime(s,|y|)\geq t-|x-y|\); the flat causal past of the target down to this Cartesian slice also stays in the slab. The same properties hold for the adapted foliation. All subsequent representations based on \(t=T_0\) use this distinction, with the bounded initial range controlled by the equation~\eqref{equa:audit-bounded-initial-Gronwall}.

\paragraph{A first pointwise estimate for the wave fields}

\begin{proposition}[First wave field estimate]
\label{lem:seven-first-full-metric-refinement}
Under the coarse energy and spacetime bounds, the propagated wave gauge, and the prescribed reference and initial data estimates, the full Cartesian perturbation \(h=g-\gMink\) and the matter field satisfy
\be
\label{equa:seven-first-full-metric-refinement}
|h|+ |H|+ | \psi| \leq C_{N, \delta, \lambda}A t^{-1}s^{2\delta}, \qquad t-r\geq1.
\ee
The constant is independent of \(0<\coeff\leq\coeff_0\). This estimate concerns the undifferentiated Cartesian components.
\end{proposition}

\begin{proof}

\noindent{\bf 1.} \emph{Interior flat source.} The lower direct estimates control the nonprincipal metric source by \(CA^2t^{-1}s^{-2 +2\delta} \). This includes both matter stress tensors: the effective-field factors are \(\del\fieldPhi\) and the mass-controlled value \(\fieldvarphi\), since \(\coeff U(\fieldphi) \lesssim(\fieldvarphi)^2\). The matter equation has the smaller nonprincipal source
\bse
\be
\label{equa:seven-first-wave-matter-source}
|g(\del\fieldPhi, \del\psi)| \lesssim A^2t^{-2}s^{-1+2\delta} \leq A^2t^{-1}s^{-2 +2\delta}.
\ee
The smallness of \((t/s)^2H^{\H00} \), provided by \eqref{equa:seven-direct-ray-coefficient-value} and~\autoref{lem:seven-pointwise-metric-interface-bulk}, permits inversion of the semi-hyperboloidal Hessian identity. It gives
\be
\label{equa:seven-first-full-metric-Hessian}
| \del^2h|+ | \del^2\psi| \lesssim A t^{1/2}s^{-3+ \delta} +A^2t s^{-4+2\delta}.
\ee
Both summands are kept. The three coefficient terms in \eqref{equa:seven-direct-H00-value}, multiplied by these two summands and divided by \(A^2t^{-1}s^{-2 +2\delta} \), give respectively
\be
\label{equa:seven-first-full-metric-six-products}
\frac st, \quad A t^{-1/2}s^\delta, \quad A t^{-1/2}s^\delta, \quad A^2s^{-1+2\delta}, \quad t^{1/2}s^{-1-\delta}, \quad \frac{At}{s^2}.
\ee
They are bounded because \(s\leq t\lesssim s^2\), \(\delta<1/2\), and \(A\leq1\). The other principal products contain a good Hessian or a frame factor \(t^{-1} \), and are bounded by \(CA^2t^{-2}s^{-1+2\delta} \). Consequently, we obtain
\be
\label{equa:seven-first-full-metric-flat-source}
| \Box h|+ | \Box\psi| \lesssim A^2t^{-1}s^{-2 +2\delta} \lesssim A^2t^{-2 + \delta}(t-r)^{-1+ \delta}.
\ee
\ese
The positive Kirchhoff representation estimates this interior contribution directly, following~\cite[Sections~3.3 and~11.2]{LeFloch-Ma-2018}. For a target \((T,x)\), we set \(R\coloneqq|x|\), \(P\coloneqq T+R\), and \(Q\coloneqq T-R\geq1\), and denote the contribution by \(\mathcal I_{\mathrm{int}}\). Its source coordinates \(p'\coloneqq\tau+r'\), \(q'\coloneqq\tau-r'\) satisfy \(Q\leq p'\leq P\), \(0<q'\leq Q\), and \(\tau\simeq p'\). The source bound gives
\be
\label{equa:audit-interior-Kirchhoff-kernel}
|\mathcal I_{\mathrm{int}}(T,x)|\lesssim\frac{A^2}{R}\int_Q^P(p')^{-1+\delta}\,dp'\int_0^Q(q')^{-1+\delta}\,dq'\lesssim_\delta A^2T^{-1+\delta}Q^\delta.
\ee
Since \(s^2=PQ\), this is at most \(C_\delta A^2T^{-1}s^{2\delta}\). At \(R=0\), the quotient is interpreted by continuity. The same positive majorant applies when the interior source is restricted by \(q'\geq q_b(\tau)\geq3/4\), as in the moving-domain argument.

\noindent{\bf 2.} \emph{Dynamic and reference-metric contributions in the exterior.} We set \(w=2 +r-t\) in the exterior. The radial estimate \eqref{equa:eight-massive-coefficient-flat-value}, extended over the bounded merging strip, and the direct second-derivative estimate give
\bse
\be
\label{equa:seven-first-full-metric-exterior-source}
|S_{\mathrm{ext,dyn}}| \lesssim A^2r^{-2}t^\delta w^{-a_{\mathrm{ext}}} +A_0^2r^{-4}, \qquad a_{\mathrm{ext}}{\coloneqq}2\lambda-\tfrac12>1.
\ee
For the matter equation, the product of the direct effective-field and matter jets has weight \(w^{-\mu-\lambda} \), which is stronger than \(w^{-a_{\mathrm{ext}}} \). The dynamic principal term has the same coefficient and Hessian bounds as the metric term. The terms \(H^\star\del^2u\) and \(H^\star\del^2\psi\) are treated separately below. With the same target and source coordinates, the positive Kirchhoff kernel is bounded by
\be
\label{equa:seven-first-full-metric-exterior-kernel}
\frac{CA^2}{R} \int_Q^P (p')^{-1+ \delta} \int_{\min\{2T_0-p',1\}}^1(2-q')^{-a_{\mathrm{ext}}} \, dq'\, dp' \lesssim A^2T^{-1+ \delta} \lesssim A^2T^{-1}s^{2\delta}.
\ee
\ese
The inner integral is zero when \(p'<2T_0-1\), since the corresponding causal section is empty. The estimate follows from \(a_{\mathrm{ext}}>1\) and \((P-Q)^{-1} \int_Q^P p^{-1+ \delta}dp\lesssim_\delta P^{-1+ \delta} \); at \(R=0\) the latter expression has its continuous limiting value.

For \(H^\star\del^2v\), with \(v\in\{u, \psi\} \), we divide at \(w=4t^b\), where \(b= \delta/\lambda \). In the inner region the direct bound is \(CmA r^{-2}t^{\delta/2}w^{-\lambda} \). Its \(q'\)-integral adds at most \((p')^{b(1-\lambda)} \), and \(\delta/2 +b(1-\lambda) \leq\delta \). In the outer region, the exterior lightlike Hessian reduction adds \(w^{-1} \), with small coefficient \((t/w)|H-H^\star| \lesssim At^{-\delta/2} \); the reference coefficient is \(\Ocal(m/w) \). The far spacelike reduction gives the same extra factor. The resulting weight \(w^{-1-\lambda} \) is integrable, so \eqref{equa:seven-first-full-metric-exterior-kernel} also controls this term. Only coarse fixed-order scalar jets enter this reduction. For the stationary reference source \(A_0^2r^{-4}\), the inner \(q'\)-integral is \(\Ocal((p')^{-2})\), as in~\cite[Section~3.3]{LeFloch-Ma-2018}; its contribution is therefore at most \(CA_0^2/(PQ)\leq CA_0^2/T\).

\noindent{\bf 3.} \emph{Initial data and parameter dependence.} The reference data give \(CA_0/T\). For the dynamic free wave, we restrict the existing solution to \(t=T_0\). The coarse estimates on the strip identified above, with weights comparable to their initial values, give the following bounds with amplitude \(C_{T_0}A\):
\bse
\be
\label{equa:seven-first-full-metric-data}
|f(y)| \lesssim A\langle y\rangle^{-\lambda-1/2}, \qquad |g(y)|+ | \nabla f(y)| \lesssim A\langle y\rangle^{-\lambda-3/2}.
\ee
For completeness, these are radial Sobolev estimates on annuli of radius \(r\), followed by two angular derivatives. The required weighted derivatives \(r\nabla g\) and \(r\nabla^2f\) follow from the initial traces of \(\del_bL_a v\) and \(\del_tL_a v\), for \(v\in\{u, \psi\} \):
\be
\label{equa:seven-first-full-metric-data-identities}
\del_b(L_av)|_{T_0} = \delta_{ab}g+x_a \del_bg+T_0\del_{ab}f, \qquad x_a \Delta f= (\del_tL_av)|_{T_0}-\del_af-T_0\del_ag +x_a \Box v|_{T_0}.
\ee
\ese
The polar Laplacian controls the radial second derivative; its angular terms are controlled by the sphere derivatives. The initial quadratic source uses only \(\del\fieldPhi\), \(\fieldvarphi\), and the wave derivatives. At most six fixed-order commutations are needed, within the lower families. Kirchhoff's data formula uses \(\int_0^\infty\rho\langle\rho\rangle^{-\lambda-3/2}d\rho< + \infty\). For \(R\leq T/2\), the initial sphere instead lies in \(|y| \simeq T\), giving the stronger bound \(CAT^{-\lambda-1/2} \). Thus the free wave is bounded by \(CA/T\). Combining the three contributions proves the claim; smooth inversion transfers it from \(h\) to \(H\). All massive factors were measured in the uniform lower energy, so the constants introduce \emph{no negative power of \(\coeff\).}

\end{proof}

\paragraph{Interpolation and commutator for wave components}

\begin{lemma}[Boost interpolation]
\label{lem:seven-boost-v-interpolation}
Under the hypotheses of~\autoref{lem:seven-first-full-metric-refinement}, put \(\ell=p_{\mathbf{low}} +2\) and \(b=13\delta/64\). For \(v\in\{h, \psi,H,K\} \), where \(K=j_gg^{-1}-\gMink^{-1} \), one has
\be
\label{equa:seven-wave-v-interpolation}
|L^Jv| \leq C_N A t^{-1+j/(2\ell)}s^{2\delta(1-j/\ell) +bj/\ell}, \qquad |J|=j\leq\ell.
\ee
Let \(\mathcal R(z,h, \psi) \) be smooth with uniformly bounded derivatives for \(|z| \leq1\) and small \((h, \psi) \), and vanish to second order in \((h, \psi) \). Then one has
\be
\label{equa:seven-wave-quadratic-v-remainder}
\frac{s^2}{t}|L^J\mathcal R(x/t,h, \psi)| \leq C_NA^2s^{-1/2 +3\delta} \leq C_NA^2, \qquad |J| \leq\ell.
\ee
\end{lemma}

\begin{proof}
\bse

\noindent{\bf 1.} \emph{Bounds at the two interpolation orders.} We apply hyperboloidal Sobolev to \(t^{-1}L^Jv\), first for \(v=h, \psi\), and use \eqref{equa:seven-Hardy-with-boundary}. The two additional boosts use \(\ell+2=p_{\mathbf{high}} \) orders. The normalized sphere trace is bounded by \(C_NAs^{-1+ \delta} \), and the contribution of the reference metric is estimated separately. Thus the two endpoints are
\be
\label{equa:seven-wave-interpolation-endpoints}
F_0\coloneqq C_NAt^{-1}s^{2\delta}, \qquad F_\ell{\coloneqq}C_NAt^{-1/2}s^b, \qquad |v| \leq F_0, \quad |L^Jv| \leq F_\ell\quad(|J| \leq\ell).
\ee
A ball of radius \(ct\) in the spatial coordinates of the hyperboloid, intersected with its interior part, is uniformly regular after rescaling, including at the interface. The Sobolev constant is therefore uniform without extending across the boundary. The trace is used only in the Hardy inequality.

\noindent{\bf 2.} \emph{One-sided boost curves.} For \(r/t\geq1/2\), we choose short boost curves with initial direction \(\xi\cdot x/r\leq-1/2\). Along them \(L_\xi(t-r)=-(t-r) \xi\cdot x/r\), so \(t-r\geq1\) is preserved. They remain on the same hyperboloid and have comparable \(t\). When \(r/t<1/2\), the strict margin \(s_0/\sqrt3>1\) permits a fixed short interval in every direction. Endpoint interpolation on these intervals gives
\be
\label{equa:seven-wave-directional-interpolation}
|L_\xi^jv| \leq C_\ell F_0^{1-j/\ell}(F_\ell+F_0)^{j/\ell}, \qquad 0\leq j\leq\ell.
\ee
Indeed, one-sided finite differences with step \(a \) give the bound \(C(a^{-j}F_0 +a^{\ell-j}F_\ell) \); we choose \(a \) proportional to \(\min\{1,(F_0/F_\ell)^{1/\ell} \} \). A finite collection of inward directions, rotated with \(x/r\), recovers the symmetrized derivatives with a uniform constant; the direction stays fixed along each individual curve. The identities \([L_a,L_b]=z_aL_b-z_bL_a \) and \(L_az_b= \delta_{ab}-z_az_b\) then recover ordered operators by induction, with lower-order bounded coefficients. Since \(F_0\lesssim F_\ell\), this proves \eqref{equa:seven-wave-v-interpolation} for \(h, \psi\).

\noindent{\bf 3.} \emph{Smooth compositions.} Each differentiated term of degree \(n\) and total wave order \(d\leq j\) is bounded by \(C_NF_0^{n-d/\ell}F_\ell^{d/\ell} \). Smooth inversion and the determinant formula therefore give the same interpolation for \(H,K\). Tensorial Lie operators add constant matrices acting on lower scalar operators. For \(\mathcal R\), every term has \(n\geq2\), so its bound is at most \(C_NF_0F_\ell\). Derivatives of \(z=x/t\) have bounded coefficients. Finally, we find
\be
\label{equa:seven-wave-quadratic-interpolation-weight}
\frac{s^2}{t}F_0F_\ell\lesssim A^2t^{-5/2}s^{2 +2\delta+b} \leq A^2s^{-1/2 +3\delta},
\ee
in which \(\delta<1/8\). This completes the proof. \qedhere 
\ese
\end{proof}

\paragraph{Coefficient and characteristic estimates}

The coupled pointwise argument uses the following coefficient estimate, proved in Section~\ref{section:seven-boosted-coefficients}:
\be
\label{equa:seven-preliminary-H00}
|H^{\H00}|_{p_{\mathbf{low}}-2,k} \leq C_NA t^{-1}s^{\eta_k}.
\ee
The boost-only terms will follow from Proposition~\ref{prop:seven-polynomial-coefficient-reduction} and the rank estimate~\eqref{equa:seven-coupled-coefficient-ray-bound}, proved in Section~\ref{section:seven-boosted-coefficients}. An operator containing a translation uses \eqref{equa:seven-direct-H00-gradient}; its additional time decay gives the same bound. The scalar component and tensor Lie conventions differ only by lower operators with bounded frame coefficients. The rank-zero ray argument itself uses the direct gauge estimate, so the proof starts without this improved coefficient as a premise.

For \(w\in\mathscr G(u) \) and \(v=Zw\), we set \(W_{Z,w}{\coloneqq} \underL(rv) \).  The same zero-lightlike-weight calculation as for the matter field below gives the identity
\bse
\be
\label{equa:seven-characteristic}
\mathbb P_H^N W_{Z,w} = {4r \over4-H_N} \Bigl( -\Box_gv + \sum_{a<b}(r^{-1} \Omega_{ab})^2v + r^{-1}H_NX^N[v] + \slashed H^N[v] \Bigr),
\ee
where \(\mathbb P_H^N\), \(X^N\), and \(\slashed H^N\) are defined in \eqref{equa:seven-characteristic-operator-and-errors}.  We denote by \(\gamma_{t,x} \) the backward integral curve of \(\mathbb P_H^N\) through \((t,x) \), parametrized by its time coordinate and stopped at its first intersection with \(\Mint_{s_0} \), \(\Lscr\), or \(\Sigma_{\mathrm c} \).  The sign-free geometric estimates \eqref{equa:seven-characteristic-small-rH}-- \eqref{equa:seven-characteristic-dtds} apply verbatim. The full characteristic source is defined in \eqref{equa:seven-j-characteristic-source-definition}. Combining its integrated bound \eqref{equa:seven-j-ray-allocation-bound} with \eqref{equa:seven-preliminary-H00}, reconstruction using \(r| \underL v| \leq|W_{Z,w}|+ |v| \) and the wave gauge gives the following estimate for \(0\leq k\leq p_{\mathbf{low}}-3\):
\be
\label{equa:seven-sharp-metric}
| \del w|_{p_{\mathbf{low}}-3,k} + |H^{\H00}|_{p_{\mathbf{low}}-2,k} \leq C_NA t^{-1}s^{\eta_k}.
\ee
\ese

\paragraph{Boost-only coefficient bound}

The high-order commutators require the boost-only version
\be
\label{equa:seven-sharp-boost-only-H00}
|L^JH^{\H00}(t,x)| \leq C_NA t^{-1}s^{\eta_{|J|}}, \qquad |J| \leq p_{\mathbf{low}}.
\ee
The proof in Section~\ref{section:seven-boosted-coefficients} uses the available counts
\be
\label{equa:seven-boost-only-available-orders}
|J|+4\leq p_{\mathbf{high}}, \qquad \left\lfloor\frac{p_{\mathbf{low}} +2}{2} \right\rfloor\leq p_{\mathbf{low}}-3 \quad(N\geq19).
\ee
The first count allows two angular differentiations and two Sobolev differentiations on the high wave factor. The second places the lower effective-field factor in the strong ray family. The polynomial projection keeps the moving-frame terms, while the effective-field amplitudes remain unknown until the rank induction.

\paragraph{High-rank radial profile}

The quasi-linear commutator has a triangular component.  If a Cartesian translation falls on the coefficient, its derivative is controlled directly by the metric energy and the complementary Hessian is a low factor.  If no translation falls on the coefficient, it has the boost-only form \(L^JH^{\H00} \).  Estimate \eqref{equa:seven-sharp-boost-only-H00} puts it in \(L^\infty\) whenever \(|J| \leq p_{\mathbf{low}} \).  Higher boost-only coefficients occur at the high energy level and must be kept; they are treated by the radial \(L^2\) profile below.

\paragraph{Radial \(L^2\) control of the metric coefficient}

The high boost-only coefficient cannot be discarded.  For \(v\in\{u, \psi,{\fieldPhi} \} \), we define its contribution by

\bse
\be
\label{equa:seven-critical-radial-remainder}
\mathbf R_{v;p,k}^{\mathrm{crit}}(s) = \Bigl\| \sum_{\substack{|J_1|>p_{\mathbf{low}}, \ |J_1|+ |I_2|+ |J_2| \leq p\\
|J_1|+ |J_2| \leq k}} |L^{J_1}H^{\H00}| \,| \del^2\del^{I_2}L^{J_2}v| \Bigr \|_{L^2(\Mint_s)}.
\ee
It is empty when \(p=p_{\mathbf{low}} \), but not when \(p=p_{\mathbf{high}} \).  At the high level, the corresponding members of the standard commutator family remain and must be estimated.

To estimate this term without losing a derivative, we use the profile \(\mathbf H_{p,k} \) defined in \eqref{equa:seven-high-H00-radial-profile} and introduce the corresponding interface trace norm
\be
\label{equa:seven-high-H00-L2-boundary}
\bigl( \mathsf B_{2;p,k}^{H}(s) \bigr)^2 = \sum_{\substack{p_{\mathbf{low}}<|J| \leq\min\{p,k\}}} \|L^JH^{\H00} \|_{L^2(\mathbb S_s)}^2.
\ee
\ese
Applying \eqref{equa:seven-radial-L2-moving-boundary} componentwise and using the algebraic equivalence between derivatives of \(H\) and of \(u\) gives
\bse
\be
\label{equa:seven-high-H00-radial-derived}
\begin{aligned}
\mathbf H_{p,k}(s) & \leq C_N\mathbf H_{p,k}(s_0) + C_N\int_{s_0}^{s} \tau^{-1} \mathbf W_{p,k}(\tau) \, d\tau \\
& \quad + C_N\int_{s_0}^{s} \kappa_u^\star(\tau) \mathsf T_{2;p,k}[u](\tau) \, d\tau + C_NA_0 + C_N \left\| \tau^{-1/2} \mathsf B_{2;p,k}^{H}(\tau) \right\|_{L^2((s_0,s);d\tau)}.
\end{aligned}
\ee
Here \(\kappa_u^\star \) is the integrable metric kernel defined in \eqref{equa:seven-metric-integrable-kernel}. Differentiating the inverse-metric identity~\eqref{equa:inverse-metric-dynamic-identity} and separating the terms according to the number of derivatives on each factor gives
\be
\label{equa:reaudit-inverse-metric-time-derivative}
\Bigl\|{s\over t} \del_tL^JH\Bigr\|_{L^2(\Mint_s)} \leq C_N\mathbf W_{p,k}(s) + C_NA s^{-1/2 + \delta} \mathsf T_{2;p,k}[u](s) + C_NA_0s^{-1/2}.
\ee
\ese
The commutators \([\del_t,L] \) have lower boost rank. The resulting products are controlled by \(\mathbf W_{p,k}\) when the differentiated factor has high order, and by the Hardy norm and its sphere trace when the undifferentiated factor has high order. After the factor \(s^{-1} \) in \eqref{equa:seven-radial-L2-moving-boundary}, the latter contribution is bounded by \(\kappa_u^\star(s) \mathsf T_{2;p,k}[u](s) \). The reference contribution is supported where \(r\gtrsim s\), with \(|(s/t)\del_tL^Jh^\star|\lesssim C_Nm s r^{-3}\). Its squared spatial integral is bounded by \(C_Nm^2s^2\int_{cs}^{\infty}r^{-4}\,dr\lesssim C_Nm^2s^{-1}\). The additional factor \(s^{-1}\) in the radial inequality therefore gives the integrable term \(C_NA_0s^{-3/2}\), whose integral is bounded by \(C_NA_0\). Thus the radial profile is an interior consequence of the metric energy and of an \(L^2(\mathbb S_s) \) trace; it is not an exterior volume norm. In particular, if
\bse
\be
\label{equa:seven-high-H00-boundary-input}
\mathcal T^H_{p,k}(s)= \sup_{s_0\leq\tau\leq s} \tau^{-\nu_k} \mathsf B_{2;p,k}^{H}(\tau),
\ee
we then obtain
\be
\label{equa:seven-high-H00-radial-growth}
\begin{aligned}
\mathbf H_{p,k}(s) & \leq C_N\mathbf H_{p,k}(s_0) + C_N\int_{s_0}^{s} \tau^{-1} \mathbf W_{p,k}(\tau) \, d\tau \\
& \quad + C_N\int_{s_0}^{s} \kappa_u^\star(\tau) \mathsf T_{2;p,k}[u](\tau) \, d\tau + C_NA_0 + {C_N\over \sqrt{\nu_k}} \mathcal T^H_{p,k}(s)s^{\nu_k}.
\end{aligned}
\ee
\ese


\subsection{\texorpdfstring{The unscaled effective field along rays}{The unscaled effective field along rays}}
\label{section-9-2}

\paragraph{Ray equation and coefficient variation}

We apply the fixed-mass interior argument with its dependence on $\coeff$ explicit.  We use the inverse metric perturbation \(H^{\alpha \beta} =g^{\alpha \beta}-(\gMink)^{\alpha \beta} \) and the normalized ray coefficient \(a \) defined in \eqref{equa:raywise-metric-coefficient-section6}. We parameterize the ray by
\bse
\be
\gamma_{s,x}(\lambda) = \Bigl({\lambda t\over s},{\lambda x\over s} \Bigr), \quad \lambda \in[\lambda_{\mathrm{in}},s],
\ee
where \(\lambda_{\mathrm{in}} \) is defined in \eqref{equa:lower-ray-parameter-section3}. The intersection with \(r=t-1\) occurs at \(\lambda=s/(t-r) \); if this value lies below \(s_0\), the ray reaches \(\Mint_{s_0} \) first. In either case \(s_0\leq\lambda_{\mathrm{in}} \leq s\). For the expanding interior, the ray instead meets the initial hypersurface or the moving interface, with entry parameter given by \eqref{equa:audit-moving-interface-ray-entry}. Its differential equation is unchanged, and its entry time remains comparable to the square of that parameter. Interior characteristics likewise stop at the moving boundary; their geometric estimates use an absolute bound for $rH_N$ and remain valid for $t-r\geq3/4$. We define the renormalized ray variable by
\be
Y_Z(\lambda) \coloneqq \lambda^{3/2}Z\fieldphi\bigl(\gamma_{s,x}(\lambda) \bigr).
\ee
\ese
The commuted trace equation and its raywise reduction yield 
\bse
\be
\label{equa:seven-commuted-trace}
3\coeff \, \Box_gZ\fieldphi-Z\fieldphi = Z\mathcal S+ZW_\coeff -3\coeff[Z, \Box_g] \fieldphi,
\ee
which, after performing the hyperboloidal decomposition, takes the ODE form
\be
\label{equa:seven-raw-ODE}
3\coeff a_{s,x}(\lambda)Y_Z''(\lambda) +Y_Z(\lambda) =F_Z(\lambda), \qquad a_{s,x}(\lambda) =a \bigl(\gamma_{s,x}(\lambda) \bigr).
\ee
\ese

\paragraph{Exact remainder in the ray equation}

The following notation distinguishes the radial ray derivative from the semi-hyperboloidal derivatives:
\be
\label{equa:seven-ray-remainder-frame}
\chi{\coloneqq}s/t, \quad z^a\coloneqq x^a/t, \quad y^a\coloneqq x^a/s, \quad E_a\coloneqq \delsH_a, \quad D\coloneqq \Lcal, \quad h\coloneqq \chi^{-2}H^{\H00}, \quad a\coloneqq1-h.
\ee
The scalar \(\chi\) and the vectors \(z,y\) are constant along each ray. The remaining metric coefficients needed below are
\be
\label{equa:seven-ray-remainder-coefficients}
B_H^a\coloneqq H^{\H0a}, \qquad b_H\coloneqq t^{-1} \Bigl( 2H^{0a}z_a-H^{ab}z_az_b-\sum_a H^{aa} \Bigr).
\ee
In wave coordinates, a direct expansion gives, for every scalar function \(v\),
\be
\label{equa:seven-ray-remainder-exact}
\Box_gv=-a s^{-3/2}D^2(s^{3/2}v)-R_g[v],
\ee
where the remainder is
\be
\label{equa:seven-ray-remainder-formula}
\begin{aligned}
R_g[v]&=-\frac{3a}{4s^2}v-\frac{3a}{s}y^aE_av -(\delta^{ab} +H^{ab} +a y^ay^b)E_aE_bv\\
& \quad -2(B_H^a-hz^a) \del_tE_av -\Bigl(b_H-\frac{h(3+ |z|^2)}t\Bigr) \del_tv.
\end{aligned}
\ee
The spatial components \(H^{ab} \) in this formula are the Cartesian components, which coincide with their semi-hyperboloidal spatial components.

\begin{proof}
\bse
The frame relation \(\del_a=E_a-z_a \del_t\) yields
\be
\label{equa:seven-ray-remainder-cartesian-expansion}
\Box_gv= (-\chi^2 +H^{\H00}) \del_t^2v+2(-z^a+B_H^a) \del_tE_av +(\delta^{ab} +H^{ab})E_aE_bv + \Bigl(-\frac{3+ |z|^2}t+b_H\Bigr) \del_tv.
\ee
Since \(D= \chi\del_t+y^aE_a \), \(D\chi =Dy^a=0\), and \([E_a, \del_t]=z_a t^{-1} \del_t\), we have
\be
\label{equa:seven-ray-remainder-second-derivative}
\chi^2\del_t^2v=D^2v-2z^a \del_tE_av-y^ay^bE_aE_bv-\frac{|z|^2}t\del_tv.
\ee
A substitution, followed by
\be
\label{equa:seven-ray-remainder-renormalization}
s^{-3/2}D^2(s^{3/2}v)=D^2v+ \frac3sDv+ \frac{3}{4s^2}v,
\ee
proves the identity. In particular, \(a=1-\chi^{-2}H^{\H00} \) is the principal coefficient with the sign in \eqref{equa:seven-ray-remainder-exact}.
\ese
\end{proof}

The full commuted source in the singular equation is consequently
\be
\label{equa:seven-ray-remainder-commuted-source}
F_Z(\lambda)= \lambda^{3/2} \Bigl(-Z\mathcal S-ZW_\coeff+3[Z, \Box_g] \fieldPhi-3R_g[Z\fieldPhi] \Bigr) (\gamma_{s,x}(\lambda)).
\ee
For an ordered translation--boost operator, the commutator is the sum
\be
\label{equa:seven-ray-remainder-Lie-commutator}
[Z, \Box_g] \fieldPhi = \sum_{\varnothing\ne I\subset\{1, \ldots, \ord Z\}} (\mathcal L_I g^{-1})^{\alpha \beta} \del_{\alpha \beta}Z_{I^c} \fieldPhi.
\ee 

The direct gauge estimate \eqref{equa:seven-direct-ray-coefficient-value} gives the ellipticity needed by the oscillator. For each fixed ray, we also keep its full variation: 
\be
\label{equa:seven-ray-coefficient-ellipticity}
\frac23\leq a_{s,x} \leq\frac43, \qquad 
\Lambda_{s,x}{\coloneqq} \int_{\lambda_{\mathrm{in}}}^{s}|a_{s,x}'(\lambda)| \, d\lambda.
\ee
The commuted gauge and the coarse metric bounds give
\be
\label{equa:seven-ray-coefficient-derivative}
|a_{s,x}'(\lambda)| \lesssim A\chi^{-3/2} \lambda^{-3/2 + \delta} +A^2\chi^{-2} \lambda^{-2 +2\delta} +A_0\chi^{-1} \lambda^{-2}.
\ee
The factor \(\chi =s/t\) is constant on the ray. Uniform control of the variation is obtained on the restarted interval \([\lambda_*,s] \) in \eqref{equa:seven-strong-ray-variation-tail-integral}; the algebraic estimate treats the preceding interval. The full-ray formula below keeps its actual exponential \(e^{C\Lambda_{s,x}} \).

For the subsequent pointwise bounds, we keep the trace amplitude
\be
\label{equa:seven-trace-amplification}
C_{\mathrm{tr}}(s)\coloneqq \exp\Big(C\big(A_0 +A^2 + \mathsf B_{00;1}(s) + \mathsf B_{\mathrm{char};p_{\mathbf{low}}-3,1}(s) \big) \Big) \geq1.
\ee
This quantity measures the specified traces at the interface. The uniform shifted-energy estimate is proved separately on the oscillator tail.

\paragraph{Uniform raywise energy estimate}

We group the raywise energy, its differential inequality, and its integrated estimate as follows. We introduce 
\bse
\be
\mathcal O_Z^2\coloneqq3\coeff^2a_{s,x}|Y_Z'|^2 + \coeff|Y_Z|^2
\ee
and we multiply \eqref{equa:seven-raw-ODE} by \(\coeff Y_Z'\) and obtain
\be
\mathcal O_Z'\lesssim|a_{s,x}'| \mathcal O_Z+ |F_Z|.
\ee
Therefore, we have
\be
\label{equa:seven-raw-ODE-estimate}
\coeff|Y_Z'(s)|+ \coeff^{1/2}|Y_Z(s)| \lesssim e^{C\Lambda_{s,x}} \Bigl( \coeff|Y_Z'(\lambda_{\mathrm{in}})| + \coeff^{1/2}|Y_Z(\lambda_{\mathrm{in}})| + \int_{\lambda_{\mathrm{in}}}^{s}|F_Z(\lambda)| \, d\lambda \Bigr).
\ee
\ese
In particular, the source integral in \eqref{equa:seven-raw-ODE-estimate} \emph{has no factor like \(\coeff^{1/2} \).}

\paragraph{Normalization by the factor \((t/s)^2\)}

The lightlike factor in the pointwise norm is \((t/s)^2\). Multiplication of the oscillator estimate by this factor, which is constant on each ray, gives
\be
\label{equa:seven-raw-pointwise}
\begin{aligned}
\Bigl(\frac ts\Bigr)^2s^{3/2} \Bigl(|Z\fieldvarphi|+ \frac st| \del Z\fieldPhi| \Bigr) & \leq Ce^{C\Lambda_{s,x}} \Bigl(\mathcal B_Z^\fieldphi(s,x) + \Bigl(\frac ts\Bigr)^2s^{1/2}| \fieldPhi|_{\ord Z+1, \rank Z+1} \\
& \quad + \Bigl(\frac ts\Bigr)^2\int_{\lambda_{\mathrm{in}}}^{s}|F_Z(\lambda)| \, d\lambda \Bigr).
\end{aligned}
\ee
The initial expression, in the same normalization, is
\be
\label{equa:seven-raw-boundary}
\mathcal B_Z^\fieldphi(s,x)\coloneqq \Bigl(\frac ts\Bigr)^2 \left[\lambda^{3/2} \bigl(|Z\fieldvarphi|+ | \del_\lambda Z\fieldPhi| \bigr) + \lambda^{1/2}| \fieldPhi|_{\ord Z+1, \rank Z+1} \right]_{\lambda= \lambda_{\mathrm{in}}}.
\ee
The associated \textbf{interface amplitude} is then defined as 
\be
\label{equa:seven-ray-trace-definition}
\mathsf B_{\mathrm{ray};p,k}^{\fieldphi}(s_1)\coloneqq \sup_{Z\preccurlyeq(p,k)} \sup_{\substack{(t,x) \in\Mint_{[s_0,s_1]} \\
\lambda_{\mathrm{in}} =s/(t-r) \geq s_0}} \lambda_{\mathrm{in}}^{-\eta_{\rank Z}} \mathcal B_Z^\fieldphi(s,x).
\ee
Proposition~\ref{lem:seven-strong-massive-ray-en} below will bound this amplitude by \(C_NA\) for \(p\leq p_{\mathbf{low}}-3\). The initial-hyperboloid contribution is controlled by the prescribed initial norms.

\paragraph{Algebraic region and oscillator tail}

The massive equation gives an additional lightlike gain before the oscillator estimate is used. The relevant threshold is
\be
\label{equa:seven-exact-light-like-compensation}
b_{\mathrm{ray}}{\coloneqq}3\rho+2C_{\mathrm{rk}} \theta, \qquad \lambda_*\coloneqq \max\left\{\lambda_{\mathrm{in}}, \chi^{-1/(2b_{\mathrm{ray}})} \right\}.
\ee
Along a fixed ray, \(\chi\) is constant. The algebraic region \(\chi\leq\lambda^{-2b_{\mathrm{ray}}} \) is its initial portion. If the ray reaches \(\lambda_*\), the strong algebraic estimate gives the oscillator data there. Otherwise the algebraic estimate covers the entire ray in the current slab. The remaining source integrals are taken only over \([\lambda_*,s] \); their mixed second derivative is treated by the shifted oscillator energy below.

\paragraph{The massive equation at the interface}

\begin{proposition}[\texorpdfstring{Effective-field data at the initial point of a ray}{Effective-field data at the initial point of a ray}]
\label{lem:seven-strong-massive-ray-en}
Assume the bootstrap estimates, the propagated wave gauge, and the effective-field equation are on a finite slab. Let \(p\leq p_{\mathbf{low}}-3\) and \(0\leq k\leq p\). At an interface point with hyperboloidal parameter \(\lambda \), the expression
\be
\label{equa:seven-strong-massive-ray-e-definition}
\widehat{\mathcal B}_Z(\lambda)\coloneqq \Bigl(\frac t\lambda \Bigr)^2 \left[\lambda^{3/2} \bigl(|Z\fieldvarphi|+ | \Lcal Z\fieldPhi| \bigr) + \lambda^{1/2}| \fieldPhi|_{p+1,k+1} \right], \qquad Z\preccurlyeq(p,k),
\ee
satisfies
\be
\label{equa:seven-strong-massive-ray-e-bound}
\lambda^{-\eta_k} \widehat{\mathcal B}_Z(\lambda) \leq C_{N,s_0, \coeff_0}A\lambda^{-1/2 +2\delta-\eta_k} \leq C_{N,s_0, \coeff_0}A.
\ee
The same expression on the initial-hyperboloid portion is bounded by the prescribed initial data. The estimate uses at most \(p+3\leq p_{\mathbf{low}} \) effective-field energy orders.
\end{proposition}

\begin{proof}
\bse
On \(q=1\), we have \(t= (\lambda^2 +1)/2\), so \(t\simeq\lambda^2\).~\autoref{lem:eight-displaced-cone-coefficient} controls the bad null coefficient of every required tensor Lie derivative by \(C_N(A_0 +A)t^{-1} \). Replacing \(dt-dr\) by \(dt-(x^a/t)dx^a \) adds \(\Ocal(t^{-1}) \) times full metric components. Thus the same bound applies to the bad semi-hyperboloidal coefficient. All remaining Hessian terms have a good derivative.

The massive equation and its ordered commutations give
\be
\label{equa:seven-strong-massive-ray-e-identity}
Z\fieldvarphi =3\sqrt\coeff\bigl(Z(\Box_g\fieldPhi)-ZG_\Phi\bigr), \qquad G_\Phi{\coloneqq} \tfrac13(\mathcal S+W_\coeff).
\ee
The tensor product rule in \eqref{equa:seven-ray-remainder-Lie-commutator} and the preceding coefficient bound yield, on the interface,
\be
\label{equa:seven-strong-massive-ray-e-operator}
|Z(\Box_g\fieldPhi)| \leq C_Nt^{-1} \Bigl(| \del^2\fieldPhi|_{p,k} + | \del\fieldPhi|_{p+1,k+1} \Bigr) \leq C_NA t^{-2} \lambda^\delta.
\ee
A Cartesian Hessian costs one translation in the kinetic energy; a good Hessian costs one boost and the factor \(t^{-1} \). The two Sobolev boosts give the total \(p+3\). The coarse lower estimate \(A\lambda^\delta \) is sufficient here.

The matter trace is quadratic in first derivatives of \(\psi\). The potential factorization in \eqref{equa:seven-matter-product-rescaled-sources} expresses the other term in \(\sqrt\coeff\,ZG_\Phi\) as a uniformly smooth multiple of \(\sqrt\coeff\,Z((\fieldvarphi)^2) \). The direct lower estimates therefore give
\be
\label{equa:seven-strong-massive-ray-e-value}
\sqrt\coeff\,|ZG_\Phi| \leq C_{N, \coeff_0} \sqrt\coeff\,A^2t^{-2} \lambda^{2\delta}, \qquad |Z\fieldvarphi| \leq C_{N, \coeff_0} \sqrt\coeff\,A\lambda^{-4+2\delta}.
\ee
The differentiated smooth coefficients have additional decaying factors; the total commutation order and rank are unchanged.

For the ray derivative, the mass term of the energy is used at one additional translation or boost order. Since \(\fieldPhi = \sqrt\coeff\, \fieldvarphi\), it gives
\be
\label{equa:seven-strong-massive-ray-e-mass-derivative}
| \del Z\fieldPhi|+ \sum_a|L_aZ\fieldPhi|+ | \fieldPhi|_{p+1,k+1} \leq C_N\sqrt\coeff\,A t^{-3/2} \lambda^\delta.
\ee
This estimate again costs only \(p+3\) orders. The expression $ \Lcal= \frac\lambda t\del_t+ \frac{x^a}{\lambda t}L_a$ has coefficients bounded by \(C\lambda^{-1} \) on the interface. Hence \(| \Lcal Z\fieldPhi| \leq C_N\sqrt\coeff\,A\lambda^{-4+ \delta} \). A substitution in \eqref{equa:seven-strong-massive-ray-e-definition} proves \eqref{equa:seven-strong-massive-ray-e-bound}, because \(2\delta<1/2\). On \(\Mint_{s_0} \), both \(t/s_0\) and the coefficients in the last identity are bounded in terms of \(s_0\), so the prescribed initial norms give the remaining assertion.
\ese
\end{proof}


\subsection{\texorpdfstring{Commuted equations and the Einstein trace relation}{Commuted equations and the Einstein trace relation}}
\label{section-9-3}

\paragraph{Einstein trace-residual equation}

We now consider the \emph{Einstein trace residual} \(\Theta \) from Definition~\ref{definition:Einstein-trace-residual}.  The terminology refers to the tensorial trace in the limiting Einstein equations and is unrelated to the traces at the interface \(\Lscr\).  The formal limiting trace constraint is \(\fieldphi =-\mathcal S[\psi,0,g] \).  From the trace equation, we obtain
\bse
\be
\label{equa:seven-Theta-equation}
3\coeff \, \Box_g\Theta-\Theta = W_\coeff(\fieldphi) +3\coeff \, \Box_g\mathcal S.
\ee
Equivalently, we consider the unknown of Section~\ref{section-6}, say \(q_{\mathrm{KG}}{\coloneqq} \fieldphi\), and take
\be
\sourterm=-\mathcal S, \qquad f= \coeff^{-1}W_\coeff.
\ee
\ese
The equation~\eqref{equa:KG-101} is the negative of the trace equation. After the interior hierarchy closes, Proposition~\ref{proposition:seven-local-vanishing-from-robust} will give residual convergence on fixed interior slabs. The energy estimate on the full domain additionally propagates the full weighted preparation amplitude. We commute \eqref{equa:seven-Theta-equation}, substitute the field equations in \(\Box_g\mathcal S\), and apply the singular energy identity after cancellation of the two interface fluxes. The initial term tends to zero for well-prepared data, with its \(\coeff^{1/2} \)-weighted normal derivative. The final metric--matter comparison then gives convergence to the prescribed Einstein development.

\paragraph{Commuted source hierarchy}

Since \(\coeff\) is constant, we introduce the commuted effective-field source
\bse
\be
\mathcal N_Z^{\mathrm{eff}} {\coloneqq} Z\mathcal S+ZW_\coeff-3\coeff[Z, \Box_g] \fieldphi,
\ee
so that \eqref{equa:seven-commuted-trace} takes the form
\be
3\coeff \, \Box_gZ\fieldphi-Z\fieldphi = \mathcal N_Z^{\mathrm{eff}},
\ee
while the commutator satisfies
\be
\label{equa:seven-commutator-Phi}
3\coeff[Z, \Box_g] \fieldphi = 3[Z, \Box_g]{\fieldPhi}.
\ee
\ese
Thus the robust commutator contains no negative power of $\coeff$. For an ordered translation--boost operator \(Z\), let \(\mathcal L_{Z_1}H\) denote the corresponding ordered tensor Lie derivative, with \(Z_1\) nonempty in every proper commutator partition. In the semi-hyperboloidal frame the principal terms satisfy
\be
\label{equa:seven-effective-field-commutator-decomposition}
\begin{aligned}
|3\coeff[Z, \Box_g] \fieldphi| & \lesssim \sum_{\substack{Z_1Z_2\preccurlyeq Z\\\ord Z_1\geq1}} |(\mathcal L_{Z_1}H)^{\H00}| \,| \del_t^2Z_2\fieldPhi|
\\
& \quad + \sum_{\substack{Z_1Z_2\preccurlyeq Z\\\ord Z_1\geq1}} | \mathcal L_{Z_1}H| \Bigl(| \del\delsH Z_2\fieldPhi|+ t^{-1}| \del Z_2\fieldPhi| \Bigr).
\end{aligned}
\ee
Both sums keep the second derivatives of the effective field. The first includes translated metric coefficients as well as boost-only coefficients; the latter require the refined bad-component estimate. The second contains a tangential derivative, while differentiation of the frame contributes the \(t^{-1} \) term. These terms are estimated separately in the low--high argument below.

For the matter equation, we set
\bse
\be
\mathscr L_{\fieldPhi} v \coloneqq e^{\fieldPhi} \nabla_\alpha \bigl(e^{-{\fieldPhi}} \nabla^\alpha v\bigr) = \Box_gv-g(\nabla{\fieldPhi}, \nabla v).
\ee
Then the commuted divergence-form equation is
\be
\label{equa:seven-commuted-matter}
\mathscr L_{\fieldPhi} Z\psi = [\mathscr L_{\fieldPhi},Z] \psi =:\mathcal K_Z^\psi.
\ee
\ese
In ordinary wave form this is the identity \(\Box_gZ\psi =Z(g(\nabla{\fieldPhi}, \nabla \psi))-[Z, \Box_g] \psi\), whereas the divergence form is used for the top-order \(L^2\) estimate. No singular coefficient in $\coeff$ is present.  For the metric equation, we distinguish the prescribed localization residual of \(g^\star \) from its dynamic interaction with \(u\).  We define
\bse
\be
\label{equa:seven-reference-localization-residual}
\mathcal R^\star {\coloneqq} \Fbb(g^\star,g^\star;\del g^\star, \del g^\star) -\BoxChapeaud_{g^\star}g^\star
\ee
and the dynamic source by
\be
\label{equa:seven-reference-dynamic-source}
\mathcal N^\star[u]{\coloneqq} \Fbb(g,g;\del g, \del g)-\Fbb(g^\star,g^\star;\del g^\star, \del g^\star) +(\BoxChapeaud_{g^\star}-\BoxChapeaud_g)g^\star.
\ee
Then \(\mathcal N^\star[0]=0\), and the equation for \(u=g-g^\star \) is
\be
\label{equa:seven-exact-u-reference-split}
\BoxChapeaud_g u = \mathcal N^\star[u] + \Gbb_\coeff+ \Mbb_\coeff+ \mathcal R^\star.
\ee
Consequently, we have
\be
\BoxChapeaud_gZu = Z\bigl(\mathcal N^\star[u]+ \Gbb_\coeff + \Mbb_\coeff+ \mathcal R^\star \bigr) -[Z, \BoxChapeaud_g]u.
\ee
\ese
Only \(\mathcal R^\star \) is prescribed; the term \(\mathcal N^\star[u] \) is estimated as an interior metric nonlinearity. Terms containing a differentiated reference field or a prescribed reference error remain in the dynamic reference interaction and in \(\mathcal R^\star \), respectively; undifferentiated \(h^\star \) factors are absorbed in the bounded coefficients of \(\Fbb\). The low--high estimate for this dynamic source, including its Hardy interface-trace term and the differentiated reference coefficients, is proved in~\autoref{lem:seven-dynamic-metric-source-commutator}.

\paragraph{Commutator estimate}

We set
\bse
\be
\mathcal C_{{\fieldPhi};p,k}(s)^2 \coloneqq \sum_{Z\preccurlyeq(p,k)} \|[Z, \Box_g]{\fieldPhi} \|_{L^2(\Mint_s)}^2 .
\ee
The rank decomposition of \eqref{equa:seven-effective-field-commutator-decomposition}, with the high boost-only part isolated by \eqref{equa:seven-critical-radial-remainder}, yields
\be
\label{equa:seven-commutator-L2-triangular}
\begin{aligned}
\mathcal C_{{\fieldPhi};p,k}(s) & \lesssim CA s^{-1}U_{p,k} + CA\sum_{j=1}^{k} s^{-1+ C_{\mathrm{rk}}j\theta}U_{p,k-j} + C\mathcal B_3^{\mathrm{low}}(s)s^{-3/2} \mathbf W_{p,k} \\
& \quad + C\mathcal B_3^{\mathrm{low}}(s) \sum_{j=1}^{k}s^{-3/2 + \eta_j} \mathbf W_{p,k-j} + CA^2s^{-3/2 + C_{\mathrm{rk}}k\theta} + C\mathbf R_{{\fieldPhi};p,k}^{\mathrm{crit}}(s).
\end{aligned}
\ee
Here no derivative is lost.  Indeed, \(| \del^2{\fieldPhi}|_{p-1,k-1} \) is a first derivative of a commuted field of order at most \(p\), and therefore we have
\be
\Bigl\|{s\over t}| \del^2{\fieldPhi}|_{p-1,k-1} \Bigr\|_{L^2} \lesssim U_{p,k-1}.
\ee
For the first term of \eqref{equa:seven-effective-field-commutator-decomposition}, the rank-zero sharp bound \(|H^{\H00}| \lesssim A t^{-1} \) and \(t/s\) from the preceding display give \(CA s^{-1}U_{p,k-1} \).  If \(j\geq1\) boosts fall on \(H^{\H00} \), the effective-field factor has rank at most \(k-j\); the pointwise trace estimate contributes only \(s^{C_{\mathrm{rk}}j\theta} \).  The remaining terms are standard low--high products with a good derivative or an explicit factor \(t^{-1} \). If a translated high coefficient is paired with a low effective-field Hessian, we keep the metric energy rather than inserting its bootstrap bound:
\be
\| \,| \del Z_1H| \,| \del^2Z_2{\fieldPhi}| \|_2 \leq \Bigl\|{s\over t}| \del Z_1H| \Bigr\|_2 \Bigl\|{t\over s}| \del^2Z_2{\fieldPhi}| \Bigr\|_\infty \leq C_N\mathcal B_3^{\mathrm{low}}s^{-3/2 + \eta_{\rank Z_2}} \mathbf W_{p, \rank Z_1}.
\ee
These are the two \(\mathbf W\)-terms. The high-coefficient component is \(\mathbf R_{{\fieldPhi};p,k}^{\mathrm{crit}} \) and has not been discarded. The metric commutator is estimated together with the dynamic source \(\mathcal N^\star[u] \) in~\autoref{lem:seven-dynamic-metric-source-commutator}.  The corresponding matter commutator has \(\mathbf P\) and \(\mathbf R_{\psi;p,k}^{\mathrm{crit}} \) in place of the effective-field quantities.  Thus only the same-rank noncritical coefficient is \(CA s^{-1} \); each high-rank coefficient remains a separate remainder and is estimated inside the interior by \eqref{equa:seven-high-H00-radial-derived} and \eqref{equa:seven-critical-remainder-rank-bounds}.

If \eqref{equa:seven-commutator-L2-triangular} is rewritten for the unscaled field, we obtain
\be
\sum_{Z\preccurlyeq(p,k)} \|[Z, \Box_g] \fieldphi\|_{L^2(\Mint_s)}^2 = \coeff^{-2} \bigl( \mathcal C_{{\fieldPhi};p,k}(s) \bigr)^2.
\ee
\ese
This singular identity is valid and is kept whenever an unscaled commutator is required.  In the equation itself, however, it is always multiplied by \(3\coeff\), as in \eqref{equa:seven-commutator-Phi}.


\section{Effective-field estimates along rays}  
\label{section-coupled-pointwise}


\subsection{\texorpdfstring{Ray estimates and the coupled hierarchy}{Ray estimates and the coupled hierarchy}}
\label{section-9-4}

\paragraph{Coupling of rays and characteristics}

Our pointwise argument involves the massive field along hyperboloidal rays, the tangential metric components along wave characteristics, and the lightlike derivative \((t/s)| \del\psi| \) needed by the massless stress tensor. All three use the same boost-rank weights. The characteristic estimate uses the transported variable \(\underL(rZ\psi) \), its interface trace on \(\Lscr\), and the hyperboloidal Sobolev bound with exponent \(\beta_k\).

\paragraph{Weighted norms for rays and characteristics}

The differentiation level is \(p_{\mathbf{low}}-3\).~\autoref{lem:seven-strong-effective-field-ray} below concerns the massive ray estimate, and~\autoref{lem:seven-j-ray-allocation} states the remaining characteristic source estimate and its consequence. We collect the initial and interface amplitudes as follows:
\be
\mathcal A_{p,k}^{\mathrm{bd}}(s) \coloneqq A_0 +A+A^2 + \mathsf B_{00;k}(s) + \mathsf B_{\mathrm{char};p,k}(s) + \mathsf B_{\psi;k}(s) + \mathsf B_{\mathrm{ray};p,k}^{\fieldphi}(s).
\ee
The factor \(s^{-\eta_i} \) separates the boost-rank growth. With the incoming null derivative \(\underL\) from Section~\ref{section-3-4}, we define
\be
\label{equa:seven-j-ray-norm}
\begin{aligned}
\mathcal Y_{p,i}(\lambda) &\coloneqq \sup_{\Mint_\lambda} \Big({t\over \lambda} \Big)^2\lambda^{3/2} \Bigl( \coeff^{1/2}| \fieldphi|_{p,i} + \coeff{\lambda \over t}| \del\fieldphi|_{p,i} \Bigr), \\
\mathcal W_i^\psi(\lambda) &\coloneqq \sup_{\substack{Z\preccurlyeq(p_{\mathbf{low}}-3,i), \ \rank Z=i\\
(t,x) \in\Mint_\lambda, \ r/t\geq3/5}} | \underL(rZ\psi) (t,x)|, \\
\mathcal W_{p,i}^u(\lambda) &\coloneqq \max_{w\in\mathscr G(u)} \sup_{\substack{Z\preccurlyeq(p,i), \ \rank Z=i\\
(t,x) \in\Mint_\lambda, \ r/t\geq3/5}} | \underL(rZw) (t,x)|, \\
\mathcal V_{p,k}(s_1) &\coloneqq \max_{0\leq i\leq k} \ \sup_{s_0\leq\lambda \leq {s_1}} \lambda^{-\eta_i} \bigl( \mathcal Y_{p,i}(\lambda) + \mathcal W_i^\psi(\lambda) + \mathcal W_{p,i}^u(\lambda) \bigr).
\end{aligned}
\ee
We also set
\be
\label{equa:seven-pointwise-integrable-kernel}
\kappa_{\mathrm{pt}}(\lambda) \coloneqq \lambda^{-3/2 + \delta} + \lambda^{-2 +5\delta} + \lambda^{-1-\zeta_\star}.
\ee

We keep the commuted metric coefficients as unknown amplitudes on the finite slab. With \(p_s\coloneqq p_{\mathbf{low}}-3\), we define
\be
\label{equa:seven-pointwise-coefficient-amplitudes}
\mathcal K_i(s_1)\coloneqq\sup_{s_0\leq s\leq s_1}\ \sup_{\Mint_s}ts^{-\eta_i}\max_{|I|=i}|(\mathcal L_{L^I}H)^{\H00}|,\qquad 0\leq i\leq p_{\mathbf{low}},
\ee
\be
\label{equa:seven-strong-ray-coupled-amplitude}
\mathcal V^{(2)}_k(s_1)\coloneqq\sup_{s_0\leq s\leq s_1}s^{-\eta_k}\mathcal Y_{p_s,k}(s),\qquad0\leq k\leq p_s.
\ee
The following estimate keeps its dependence on \(\mathcal K_i\). Section~\ref{section:seven-j-coefficient-closure} will bound the metric and effective-field amplitudes together by induction on the boost rank.

\begin{proposition}[Effective-field estimate with the \((t/s)^2\) weight]
\label{lem:seven-strong-effective-field-ray}
Assume the coarse energy, Hardy, and exterior spacetime bounds, the propagated wave gauge, and the effective-field equation on a finite slab \([s_0,s_1]\), with the prescribed initial norms and potential. For sufficiently small \(A\), the amplitudes satisfy
\be
\label{equa:seven-strong-ray-coupled-inequality}
\mathcal V^{(2)}_0(s_1)\leq C_NA,\qquad \mathcal V^{(2)}_k(s_1)\leq C_NA\Bigl(1+\sum_{i=1}^k\mathcal K_i(s_1)\Bigr)+\frac{C_N}{\eta_k}\sum_{i=1}^k\mathcal K_i(s_1)\mathcal V^{(2)}_{k-i}(s_1),\qquad1\leq k\leq p_s.
\ee
In particular, the following pointwise estimates hold.
\begin{itemize}
\item At boost rank zero, the direct metric estimate \eqref{equa:seven-direct-ray-coefficient-value} and the interface estimate of~\autoref{lem:seven-pointwise-metric-interface-bulk} give
\be
\label{equa:seven-strong-effective-field-rank-zero}
\mathcal Y_{p,0}(s) \leq C_NA, \qquad p\leq p_{\mathbf{low}}-3, \quad s_0\leq s\leq {s_1}.
\ee
\item At positive boost ranks, one additionally assumes \eqref{equa:seven-preliminary-H00} for the commuted bad metric coefficients. If \(C_NA\leq C_{\mathrm{rk}} \theta \), with a sufficiently small fixed constant, then
\be
\label{equa:seven-strong-effective-field-ray-bound}
\mathcal Y_{p,k}(s) \leq C_NA s^{\eta_k}, \qquad p\leq p_{\mathbf{low}}-3, \quad 1\leq k\leq p, \quad s_0\leq s\leq {s_1}.
\ee
\end{itemize}
The constants may depend on the fixed exponents, \(N\), \(s_0\), and \(\coeff_0\), and are independent of \(s_1\) and \(0<\coeff\leq\coeff_0\).
\end{proposition}

\begin{proof}

We take \(p=p_s\) and use the source in the equation~\eqref{equa:seven-ray-remainder-commuted-source}. Along a fixed ray, we write \(c=C_{\mathrm{rk}} \theta \), \(\chi =s/t\), \(D= \Lcal\), \(E_a= \delsH_a \), \(h= \chi^{-2}H^{\H00} \), and \(a=1-h\). We also set \(B_0=A_0 + \mathsf B_{00;0}(s_1) \);~\autoref{lem:seven-pointwise-metric-interface-bulk} gives \(B_0\leq C_NA\). The direct coefficient estimate keeps three terms:
\bse
\be
\label{equa:seven-strong-ray-direct-h}
|h| \leq C_N\Bigl(A\chi^{1/2} \lambda^{-1/2 + \delta} +A^2\lambda^{-1+2\delta} +B_0\chi^{-1} \lambda^{-1} \Bigr).
\ee
Since \(\chi\lambda \geq1\), this estimate gives \(2/3\leq a \leq4/3\) for sufficiently small \(A\). We use the exponent
\be
\label{equa:seven-strong-ray-source-exponent}
e_{\mathrm{ray}}{\coloneqq}3\delta+ \nu_N, \qquad e_{\mathrm{ray}} +5b_{\mathrm{ray}}<\tfrac12.
\ee
Indeed, \(\rho= \delta/64\), \(c(N+4)<\rho\), \(\delta<1/8\), and \(N\geq19\).

\noindent{\bf 1.} \emph{Algebraic portion and starting data.} The identity \(Z\fieldPhi =3\coeff\bigl(Z(\Box_g\fieldPhi)-ZG_\Phi\bigr) \), the lower kinetic estimate for its Hessian, and the lower mass estimate for \(\del Z\fieldvarphi\) give
\be
\label{equa:seven-strong-ray-algebraic-bound}
\chi^{-2} \lambda^{3/2} \Bigl(| \fieldvarphi|_{p,k} + \chi| \del\fieldPhi|_{p,k} \Bigr) (\gamma_{s,x}(\lambda)) \leq C_NA\lambda^{\eta_k} \Bigl(\chi^{1/2} \lambda^{b_{\mathrm{ray}}} + \lambda^{-1/2 +e_{\mathrm{ray}}}\Bigl(1+\sum_{i=1}^k\mathcal K_i(s_1)\Bigr) \Bigr).
\ee
The leading term is the flat coefficient \(\chi^2\del_t^2Z\fieldPhi\). Its undifferentiated metric correction has the small multiplier \(h\). The first two terms of the equation~\eqref{equa:seven-strong-ray-direct-h} decay, and its trace term contributes at most \(C_NAB_0\lambda^{-1/2+\alpha_{k+2}}\), since \(\chi\lambda\geq1\). A translated coefficient uses the equation~\eqref{equa:seven-direct-H00-gradient}. For a nonempty boost-only coefficient, the lower kinetic estimate gives
\be
\label{equa:seven-strong-ray-coupled-algebraic}
\chi^{-2}\lambda^{3/2}|(\mathcal L_IH)^{\H00}\del_t^2Z_{I^c}\fieldPhi|\lesssim A\mathcal K_i(s_1)\chi^{-1/2}\lambda^{-1+\eta_k+b_{\mathrm{ray}}}\lesssim A\mathcal K_i(s_1)\lambda^{-1/2+\eta_k+b_{\mathrm{ray}}}.
\ee
\ese
A good Hessian is bounded by \(t^{-1}|\del\fieldPhi|_{p+1,k+1}\); the null matter trace and nonlinear potential have stronger decay. The largest kinetic count is one translation and two Sobolev boosts, so \(p+3\leq p_{\mathbf{low}}\).

The mass term at one additional translation gives \(|\del Z\fieldPhi|\leq C_N\sqrt\coeff A t^{-3/2}\lambda^{\alpha_{k+2}}\). On \(\chi\leq\lambda^{-2b_{\mathrm{ray}}}\), the normalized starting size is therefore at most \(C_NA(1+\sum_{i=1}^k\mathcal K_i(s_1))\). If this portion ends before the target, the oscillator starts at \(\lambda_*\) defined by the equation~\eqref{equa:seven-exact-light-like-compensation}. The identity \(D=(\lambda/t)\del_t+x^a(\lambda t)^{-1}L_a\), together with the mass norm at one additional boost order, controls its velocity. At an original starting point, Proposition~\ref{lem:seven-strong-massive-ray-en} and the initial norms give the same bound.

\noindent{\bf 2.} \emph{Variation on the oscillator interval.} The direct gauge estimate \eqref{equa:seven-direct-H00-gradient} and the coarse bound for the other metric components imply
\bse
\be
\label{equa:seven-strong-ray-coefficient-tail}
|Dh|+ |D\beta| \leq C_N\Bigl(A\chi^{-3/2} \lambda^{-3/2 + \delta} +A^2\chi^{-2} \lambda^{-2 +2\delta} +A_0\chi^{-1} \lambda^{-2} \Bigr), \qquad \beta^a\coloneqq B_H^a-hz^a.
\ee
Here \(\chi\) is constant on the ray, and \(|D H^{\H00}| \lesssim\chi^{-1}| \del H^{\H00}| \). We differentiate the metric itself, so the cumulative trace amplitude \(B_0\) is not differentiated. Since \(\chi^{-b} \leq\lambda_*^{2bb_{\mathrm{ray}}} \), the inequalities \(\delta+3b_{\mathrm{ray}}<1/2\) and \(2\delta+4b_{\mathrm{ray}}<1\) give
\be
\label{equa:seven-strong-ray-variation-tail-integral}
\int_{\lambda_*}^{s}|a'(\lambda)| \, d\lambda \leq C_N(A+A^2 +A_0) \leq C_NA.
\ee
\ese
The interval in this estimate begins at \(\lambda_*\); the preceding portion is controlled by the algebraic estimate.

\noindent{\bf 3.} \emph{Mixed derivative and shifted oscillator energy.} For \(V= \chi^{-2} \lambda^{3/2}Z\fieldphi\), the mixed term in the source satisfies
\bse
\be
\label{equa:seven-strong-ray-mixed-primitive}
\begin{aligned}
6\chi^{-2} \lambda^{3/2} \beta^a \del_tE_aZ\fieldPhi&=P'+J_{\mathrm{mix}}, \qquad P\coloneqq6\chi^{-3} \lambda^{3/2} \beta^aE_aZ\fieldPhi, \\
J_{\mathrm{mix}}&\coloneqq-6\chi^{-3}(\lambda^{3/2} \beta^a)'E_aZ\fieldPhi-6\chi^{-3} \lambda^{3/2} \beta^ay^bE_bE_aZ\fieldPhi.
\end{aligned}
\ee
This follows from \(\del_t= \chi^{-1}(D-y^bE_b) \), since \(\chi\) and \(y\) are constant along the ray. We write the equation as \(3\coeff aV''+V=P'+F_0\). The \textbf{shifted oscillator energy} is
\be
\label{equa:seven-strong-ray-shifted-energy}
U\coloneqq \sqrt\coeff V, \qquad Q\coloneqq \coeff V'-\frac{P}{3a}, \qquad \mathcal E\coloneqq U^2 +3aQ^2.
\ee
Its derivative and the resulting inequality are
\be
\label{equa:seven-strong-ray-shifted-energy-identity}
\mathcal E'=3a'Q^2 +2QF_0 + \frac{2Pa'}aQ+ \frac{2PU}{3a \sqrt\coeff}, \qquad (\sqrt{\mathcal E})'\leq C|a'| \sqrt{\mathcal E} +C\Bigl(|F_0|+ |a'||P|+ \frac{|P|}{\sqrt\coeff} \Bigr).
\ee
\ese
The factor \(\coeff^{-1/2} \) is kept and compensated by \(\fieldPhi = \sqrt\coeff\fieldvarphi\) inside \(P\).

\noindent{\bf 4.} \emph{Integrable source terms.} For one good derivative, the mass estimate uses at most \(p+3\) lower orders. For two good derivatives, it uses \(p+4\leq p_{\mathbf{high}} \), keeping the additional high factor \(\lambda^{1/2} \). We obtain
\bse
\be
\label{equa:seven-strong-ray-mass-tangential-jets}
|E Z\fieldPhi| \leq C_N\sqrt\coeff A\chi^{5/2} \lambda^{-5/2 + \alpha_{k+3}}, \qquad |E^2Z\fieldPhi| \leq C_N\sqrt\coeff A\chi^{7/2} \lambda^{-3+ \nu_{k+4}}.
\ee
The bound \(|B_H| \leq C_NA\chi^{1/2} \lambda^{-1/2 + \delta} \) and the three terms of \eqref{equa:seven-strong-ray-direct-h} control \(\beta \). Its quadratic part contributes \(C_NA^3\chi^{-1/2} \lambda^{-2 + \alpha_{k+3} +2\delta} \) to \(|P|/\sqrt\coeff\), and \(C_NA^3\chi^{-1/2} \lambda^{-5/2 + \nu_{k+4} +2\delta} \) to the term containing \(\beta y E^2Z\fieldPhi\). Both have stronger decay than the bounds below. Its trace part has amplitude \(AB_0\lesssim A^2\).

We denote by \(F_{\mathrm{reg}} \) the flat and unmixed metric remainders, \(J_{\mathrm{mix}} \), the null matter trace, the nonlinear potential, and the good-Hessian commutators. The pointwise estimates give
\be
\label{equa:seven-strong-ray-regular-source-bounds}
\begin{aligned}
\frac{|P|}{\sqrt\coeff}& \leq C_NA^2\lambda^{\eta_k} \Bigl(\lambda^{-3/2 +e_{\mathrm{ray}}} + \chi^{-3/2} \lambda^{-2 +e_{\mathrm{ray}}} \Bigr), \\
|F_{\mathrm{reg}}|& \leq C_NA\lambda^{\eta_k} \Bigl(\chi^{-1/2} \lambda^{-3/2 +e_{\mathrm{ray}}} + \chi^{-2} \lambda^{-5/2 +e_{\mathrm{ray}}} + \chi^{-5/2} \lambda^{-3+e_{\mathrm{ray}}} \Bigr).
\end{aligned}
\ee
The first source power comes from the flat remainder with the high mass loss. The other two come from the linear and quadratic parts of \(D\beta \). In a good-Hessian commutator, a nonempty coefficient operator leaves at most \(p-1\) field operators, so its mass estimate costs at most \(p+3\) lower orders. After extracting \(\lambda^{\eta_k} \leq s^{\eta_k} \), each kernel has a bounded integral on \([\lambda_*,s] \), by \eqref{equa:seven-strong-ray-source-exponent}. The \(|a'||P| \) term uses the bounded primitive and \eqref{equa:seven-strong-ray-variation-tail-integral}.

\noindent{\bf 5.} \emph{Bad commutators and boost rank.} For \(\ell_{\H} =dt-z_a dx^a= \chi\, ds\), we have
\be
\label{equa:seven-strong-ray-bad-Lie-coefficient}
\mathcal L_{L_a} \ell_{\H} =-z_a \ell_{\H}, \qquad (\mathcal L_{L_a}H) (\ell_{\H}, \ell_{\H})=L_a(H^{\H00}) +2z_aH^{\H00}.
\ee
An operator containing a translation can be reordered with a translation first, since a translation--boost commutator is a translation. Differentiating \(\ell_{\H} \) by a translation contributes \(\Ocal(t^{-1}) \) times full metric components. The direct gauge estimate and these frame terms give the same-rank kernel
\be
\label{equa:seven-strong-ray-integrable-commutator-kernel}
K(\lambda)\coloneqq C_N\Bigl(A\lambda^{-3/2 + \delta} +A^2\lambda^{-2 +2\delta} +A_0\lambda^{-2} \Bigr), \qquad \|K\|_{L^1(s_0, \infty)} \leq C_NA.
\ee
The Hessian of the complementary field is a first derivative of an  operator with one additional translation and total order at most \(p\). At rank zero every nonempty coefficient operator has this form.

At positive ranks, a coefficient operator containing \(i\geq1\) boosts and no translations has size \(\mathcal K_i(s_1)t^{-1}\lambda^{\eta_i}\). Its complementary field has order at most \(p-i\), and hence
\be
\label{equa:seven-strong-ray-coupled-bad-source}
\chi^{-2}\lambda^{3/2}|(\mathcal L_IH)^{\H00}\del_t^2Z_{I^c}\fieldPhi|\lesssim\mathcal K_i(s_1)\lambda^{-1+\eta_i}\mathcal Y_{p,k-i}(\lambda).
\ee
\ese
The extra translation representing the Hessian leaves its order at most \(p-i+1\leq p\), and preserves the rank bound \(k-i\).

\noindent{\bf 6.} \emph{Recovery and rank induction.} The shifted energy controls \(\sqrt\coeff V\) and \(\coeff V'\), since the primitive is bounded. The identity \(\chi\del_t\fieldPhi =D\fieldPhi-y^aE_a \fieldPhi\) recovers the Cartesian derivative. The extra tangential and zeroth-order terms have strong normalization at most \(C_N\sqrt\coeff A\chi^{-1/2} \lambda^{-1+ \alpha_{k+3}} \), which is bounded by \(C_NA\lambda^{-1/2 + \alpha_{k+3}} \) because \(\chi\lambda \geq1\). Spatial derivatives follow from \(\del_a=E_a-z_a \del_t\).

At rank zero, the starting data and regular sources give \(C_NA\), and the only unknown-dependent kernel is the integrable \(K\). Gronwall proves the equation~\eqref{equa:seven-strong-effective-field-rank-zero}, independently of \(\mathcal K_0\). At positive ranks, the preceding commutator estimate and \(\eta_i+\eta_{k-i}=\eta_k\) give
\be
\label{equa:seven-strong-ray-coupled-integral}
s^{-\eta_k}\int_{s_0}^s\lambda^{-1+\eta_i}\mathcal Y_{p_s,k-i}(\lambda)\,d\lambda\leq\frac{\mathcal V^{(2)}_{k-i}(s_1)}{\eta_k}.
\ee
Together with the starting bound, this proves the equation~\eqref{equa:seven-strong-ray-coupled-inequality}. Under the additional coefficient hypothesis, \(\mathcal K_i\leq C_NA\); finite rank induction with \(C_NA\leq c\) then proves the equation~\eqref{equa:seven-strong-effective-field-ray-bound}. Restriction of the operator families gives every smaller order \(p\leq p_s\).\qedhere

\end{proof}

\paragraph{Characteristic source integrals}

The estimate of the effective-field component is provided by~\autoref{lem:seven-strong-effective-field-ray}. For the two wave components, their right-hand side, for a commuted scalar component \(v\), is denoted by
\be
\label{equa:seven-j-characteristic-source-definition}
\mathcal R_{\mathrm{ch}}[v] {\coloneqq} \frac{4r}{4-H_N} \Bigl(-\Box_gv+ \sum_{a<b}(r^{-1} \Omega_{ab})^2v +r^{-1}H_NX^N[v]+ \slashed H^N[v] \Bigr).
\ee
Thus \(\mathbb P_H^N\underL(rv)= \mathcal R_{\mathrm{ch}}[v] \). This expression includes the commuted equation, angular derivatives, frame terms, and all metric and matter couplings.

\begin{proposition}[Conditional reduction to characteristic source estimates]
\label{lem:seven-j-ray-allocation}
Let \(p=p_{\mathbf{low}}-3\), and consider a finite slab \([s_0,s_1]\) satisfying the coarse hypotheses of Proposition~\ref{lem:seven-strong-effective-field-ray}, the commuted metric coefficient bound~\eqref{equa:seven-preliminary-H00}, and the characteristic geometry~\eqref{equa:seven-characteristic-small-rH}--\eqref{equa:seven-characteristic-endpoint-alternative}. The initial and interface values and the values on the central boundary are measured by \(\mathcal A_{p,j}^{\mathrm{bd}}\) and the direct coarse bounds.

Suppose that, for each rank \(0\leq j\leq p\), the following integrated estimate has been established for every \(v=Z\psi\) and \(v=Zw\), \(w\in\mathscr G(u) \), occurring in \(\mathcal W_j^\psi\) and \(\mathcal W_{p,j}^u\). If \(\gamma \) is its backward characteristic ending at hyperboloidal time \(\sigma \leq {s_1}\), stopped at the first intersection with the initial, interface, or central boundary, and \(\tau\) is physical time along \(\gamma \), then
\be
\label{equa:seven-j-ray-allocation-bound}
\begin{aligned}
\sigma^{-\eta_j} \int_\gamma| \mathcal R_{\mathrm{ch}}[v]| \, d\tau & \leq C_N(A+A_0) +C_NA\int_{s_0}^{\sigma} \kappa_{\mathrm{pt}}(\lambda) \Bigl(\frac{\lambda}{\sigma} \Bigr)^{\eta_j} \mathcal V_{p,j}(\lambda) \, d\lambda \\
& \quad +C_NA\sum_{i =1}^{j} \sigma^{-\eta_j} \int_{s_0}^{\sigma} \lambda^{-1+ \eta_i+ \eta_{j-i}} \mathcal V_{p,j-i}(\lambda) \, d\lambda.
\end{aligned}
\ee
The sum is empty when \(j=0\). Under the smallness condition \(C_NA\leq C_{\mathrm{rk}} \theta \), this estimate implies the coupled bound \eqref{equa:seven-j-ray-norm-bound} below.  
\end{proposition}

\begin{proof}
\bse

\noindent{\bf 1.} \emph{Effective-field contribution.} We apply the effective-field estimate proved in Proposition~\ref{lem:seven-strong-effective-field-ray} and obtain
\be
\label{equa:seven-j-effective-field-contribution}
\max_{0\leq j\leq k} \sup_{s_0\leq\lambda \leq {s_1}} \lambda^{-\eta_j} \mathcal Y_{p,j}(\lambda) \leq C_NA.
\ee
Its mixed source is controlled by the shifted oscillator energy. The coupled reduction uses this proved output, with the coefficient hypotheses of that lemma, rather than an absolute integral of the original effective-field forcing.

\noindent{\bf 2.} \emph{Characteristic integration and endpoints.} The two transport identities give
\be
\label{equa:seven-j-characteristic-integrated-identity}
\underL(rv) (\gamma(\tau_1)) = \underL(rv) (\gamma(\tau_0)) + \int_{\tau_0}^{\tau_1} \mathcal R_{\mathrm{ch}}[v](\gamma(\tau)) \, d\tau.
\ee
The value on the initial hypersurface is controlled by \(A_0\). On \(\Lscr\), \(t\simeq s^2\), so the characteristic trace weights agree with the powers \(s^{-\eta_j} \). Commuting \(r\) and the radial frame adds lower-order characteristic terms and good derivatives, controlled by the same cumulative trace amplitudes and the coarse estimates. At \(\Sigma_{\mathrm c} \), we have \(t\simeq r\simeq s\); the direct Sobolev and Hardy bounds give \(C_NA s^{-1/2 + \delta} \) for both wave fields. Thus every normalized boundary value is bounded by \(C_N\mathcal A_{p,j}^{\mathrm{bd}}(s_1) \).

\noindent{\bf 3.} \emph{Induction over the boost rank.} Combining the preceding two steps with \eqref{equa:seven-j-ray-allocation-bound}, then taking the suprema in \eqref{equa:seven-j-ray-norm}, gives the following inequality:
\be
\label{equa:seven-ray-Volterra}
\begin{aligned}
\mathcal V_{p,k}(s_1) & \leq C_N\max_{0\leq j\leq k} \mathcal A_{p,j}^{\mathrm{bd}}(s_1) +C_NA\int_{s_0}^{s_1} \kappa_{\mathrm{pt}}(\lambda) \mathcal V_{p,k}(\lambda) \, d\lambda \\
& \quad +C_NA\max_{1\leq j\leq k} \sum_{i =1}^{j} \sup_{s_0\leq\sigma \leq {s_1}} \sigma^{-\eta_j} \int_{s_0}^{\sigma} \lambda^{-1+ \eta_i+ \eta_{j-i}} \mathcal V_{p,j-i}(\lambda) \, d\lambda.
\end{aligned}
\ee
The last line is zero at rank zero. The maximum over \(j\) keeps the normalization of each rank in the cumulative norm. Since \(\eta_i+ \eta_{j-i} = \eta_j\), each lower-rank term satisfies
\be
\label{equa:seven-j-triangular-operator-bound}
\sup_{s_0\leq\sigma \leq {s_1}} \sigma^{-\eta_j} \int_{s_0}^{\sigma} \lambda^{-1+ \eta_i+ \eta_{j-i}} \mathcal V_{p,j-i}(\lambda) \, d\lambda \leq\frac{\mathcal V_{p,j-i}(s_1)}{\eta_j}, \qquad 1\leq i\leq j.
\ee
The same-rank kernel \(\kappa_{\mathrm{pt}} \) is integrable. Rank-zero Gronwall, followed by finite induction with \(C_NA/(C_{\mathrm{rk}} \theta) \) sufficiently small, proves
\be
\label{equa:seven-j-ray-norm-bound}
\mathcal V_{p,k}(s_1) \leq C_N\max_{0\leq j\leq k} \mathcal A_{p,j}^{\mathrm{bd}}(s_1), \qquad p=p_{\mathbf{low}}-3, \quad0\leq k\leq p.
\ee
The constants are independent of the terminal time and of \(0<\coeff\leq\coeff_0\). Restricting the commutator families gives the corresponding smaller-order conclusions.
\ese
\end{proof}

Propositions~\ref{lem:seven-matter-characteristic-source-integral} and \ref{lem:seven-good-metric-characteristic-source} below will prove both characteristic source estimates. Their rank-zero conclusions use the coarse bounds and direct gauge geometry; their positive-rank conclusions require the commuted metric coefficient hypothesis.

\paragraph{Pointwise bounds for the metric and scalar fields}

Under the coarse hypotheses of Proposition~\ref{lem:seven-strong-effective-field-ray}, the coefficient bound~\eqref{equa:seven-preliminary-H00}, and the characteristic source estimate~\eqref{equa:seven-j-ray-allocation-bound}, we obtain, for \(0\leq k\leq p_{\mathbf{low}}-3\),
\bse
\label{equa:seven-refined-pointwise-system}
\be
\label{equa:seven-refined-metric}
|H^{\H00}|_{p_{\mathbf{low}}-2,k} + \max_{w\in\mathscr G(u)}| \del w|_{p_{\mathbf{low}}-3,k} \lesssim C\theta^{-1} \mathcal A_{p_{\mathbf{low}}-3,k}^{\mathrm{bd}}(s) t^{-1}s^{\eta_k},
\ee
\be
\label{equa:seven-refined-effective-field}
\coeff^{1/2}| \fieldphi|_{p_{\mathbf{low}}-3,k} + \coeff{s\over t}| \del\fieldphi|_{p_{\mathbf{low}}-3,k} \lesssim C\theta^{-1}C_{\mathrm{tr}}(s) \mathcal A_{p_{\mathbf{low}}-3,k}^{\mathrm{bd}}(s) \Big({s\over t} \Big)^2 s^{-3/2 + \eta_k},
\ee
\be
\label{equa:seven-characteristic-matter}
{t\over s}| \del\psi|_{p_{\mathbf{low}}-3,k} \lesssim C\theta^{-1}C_{\mathrm{tr}}(s) \mathcal A_{p_{\mathbf{low}}-3,k}^{\mathrm{bd}}(s) s^{-1+ \eta_k}.
\ee
The effective-field and massless-wave bounds follow from the separate ray and characteristic lemmas. The good-metric estimate additionally requires its characteristic source integral. For \(0\leq k\leq p_{\mathbf{low}}-2\), the direct hyperboloidal Sobolev estimate at the next differential order gives
\be
\label{equa:seven-refined-matter}
{s\over t}| \del\psi|_{p_{\mathbf{low}}-2,k} + | \delsH\psi|_{p_{\mathbf{low}}-2,k} \lesssim CA\,t^{-3/2}s^{\beta_{k+2}}.
\ee
\ese
The last estimate is kept because of its stronger explicit \(t\)-weight. The characteristic estimate gives the lightlike derivative needed for the non-null stress tensor, while the direct line is used only in terms which acquire the outer high-order factor \(s^{-1/2} \).   The exponents follow from  Gronwall estimates and induction on the rank, since \(CA<\rho/4\) and \(C_{\mathrm{rk}}(N+4) \theta<\rho\).  When the exterior trace quantities satisfy the prescribed interface bounds below, \(\mathcal A_{p,k}^{\mathrm{bd}} \lesssim A_0 +A+A^2\) and \(C_{\mathrm{tr}} \lesssim C\), with a constant independent of \(s\) and $\coeff$.

For later use, we define
\be
\label{equa:seven-Dpsi-definition}
\mathcal D_{\psi,j}(s) \coloneqq \Bigl\|t^{1/2}s\,| \del\psi|_{p_{\mathbf{low}}-2,j} \Bigr\|_{L^\infty(\Mint_s)} + \Bigl\|t^{3/2}| \delsH\psi|_{p_{\mathbf{low}}-2,j} \Bigr\|_{L^\infty(\Mint_s)}.
\ee
The hyperboloidal Sobolev estimate gives
\be
\label{equa:seven-Dpsi-Sobolev}
\mathcal D_{\psi,j}(s) \lesssim \mathbf P_{p_{\mathbf{low}},j+2}(s), \quad 0\leq j\leq p_{\mathbf{low}}-2.
\ee
Here the two Sobolev commutations on the spheres are rotations; they cost the two additional differential orders and, after conversion to the standard family, the two boost ranks on the right-hand side. The two weights in \eqref{equa:seven-Dpsi-definition} are the two parts of \(t^{3/2}e_\psi\), where \(e_\psi = (s/t)| \del\psi|+ | \delsH\psi| \).

The direct Sobolev bound \eqref{equa:seven-Dpsi-Sobolev} is used in the lower-order energy argument and keeps the growth \(s^{\beta_{j+2}} \). At high order, the required coefficient is instead the spatially weighted bound \eqref{equa:seven-sufficient-massless-gain}, with exponent \(\eta_j\). The characteristic geometry remains relevant: in the exact Schwarzschild zone,
\be
(g_{\Sch})^{tt} =-{r+m\over r-m}, \quad (g_{\Sch})^{rr} ={r-m\over r+m},
\ee
and therefore the semi-null coefficient satisfies
\be
\label{equa:seven-Schwarzschild-null-tail}
(H_N)^\star =-{4mr \over r^2-m^2} =-{4m\over r} +\Ocal(m^3r^{-3}).
\ee
Along an outgoing characteristic, \(d(t-r)/dt=2m/r+\Ocal(m^2r^{-2}) \), so the Minkowskian lightlike distance acquires a logarithmic Schwarzschild correction.~\autoref{lem:seven-massless-core-gain} uses the wave representation on the full domain and direct Sobolev to establish the sufficient bound on the whole interior. Its intermediate order satisfies
\be
\label{equa:seven-wave-intermediate-order}
p_{\mathrm w} = \left\lfloor{p_{\mathbf{high}} \over2} \right\rfloor, \quad p_{\mathrm w} +3\leq p_{\mathbf{low}} \quad(N\geq19),
\ee
Direct Sobolev suffices when \(t\geq s^{1+2b} \), with \(b=5\rho+2C_{\mathrm{rk}} \theta \); the cone representation treats the remaining targets. Thus the wave and effective-field estimates use different decay arguments while preserving the same commutation ranks.


\subsection{\texorpdfstring{Higher-order estimates for the effective field}{Higher-order estimates for the effective field}}
\label{section:seven-higher-effective-values}

The metric equation requires estimates through \(p=p_{\mathbf{low}}\), three orders above the estimate with the \((t/s)^2\) weight. We keep \(c=C_{\mathrm{rk}}\theta\), \(\eta_k=ck\), and the coefficient amplitudes from the equation~\eqref{equa:seven-pointwise-coefficient-amplitudes}, and define
\be
\label{equa:seven-weak-ray-amplitudes}
\mathcal Y^{(0)}_{p,k}(s)\coloneqq\sup_{\Mint_s}s^{3/2}\Bigl(|\fieldvarphi|_{p,k}+\frac st|\del\fieldPhi|_{p,k}\Bigr),\qquad\mathcal V^{(0)}_k(s_1)\coloneqq\sup_{s_0\leq s\leq s_1}s^{-\eta_k}\mathcal Y^{(0)}_{p,k}(s).
\ee

\begin{proposition}[Effective-field estimate without $(t/s)^2$ weight]
\label{lem:seven-weak-effective-field-ray}
Assume the coarse energy, Hardy, and exterior spacetime bounds on \([s_0,s_1] \), with the propagated wave gauge, prescribed Cauchy norms, and effective-field equation. For sufficiently small \(A\), one has
\be
\label{equa:seven-weak-effective-field-ray-bound}
\mathcal V^{(0)}_0(s_1) \leq C_NA, \qquad \mathcal V^{(0)}_k(s_1) \leq C_NA\bigl(1+ \sum_{i =1}^k\mathcal K_i(s_1) \Bigr) + \frac{C_N}{\eta_k} \sum_{i =1}^k\mathcal K_i(s_1) \mathcal V^{(0)}_{k-i}(s_1), \qquad 1\leq k\leq p.
\ee
The constants may depend on the fixed exponents, \(N\), \(s_0\), and \(\coeff_0\), and are independent of \(s_1\) and \(0<\coeff\leq\coeff_0\). The rank-zero estimate is independent of \(\mathcal K_0(s_1) \).
\end{proposition}

\begin{proof}

We use \(\chi =s/t\), \(z=x/t\), \(y=x/s\), \(D= \Lcal\), \(E_a= \delsH_a \), and \(a=1-\chi^{-2}H^{\H00} \) along each ray. These three ratios are constant there. The direct gauge estimate and the interface bound give \(2/3\leq a \leq4/3\). We set \(\nu_*=13\rho\); the choice \(c(N+4)<\rho= \delta/64\) bounds every fixed rank excess below by \(\nu_*\).

\noindent{\bf 1.} \emph{Algebraic estimate and oscillator initial data.} We set $\lambda_*\coloneqq \max\{\lambda_{\mathrm{in}}, \chi^{-4} \}$. If \(\lambda_*\) exceeds the target parameter, we use only the algebraic estimate. The high kinetic and mass bounds, with their factor \(\lambda^{1/2} \) kept, give
\bse
\be
\label{equa:seven-weak-ray-direct-jets}
| \del^2Z\fieldPhi| \lesssim A t^{-1/2} \lambda^{-1/2 + \nu_{k+3}}, \qquad | \del Z\fieldPhi| \lesssim\sqrt\coeff A t^{-3/2} \lambda^{1/2 + \nu_{k+3}}.
\ee
The mass estimate in the second inequality is applied to a translation  operator. In the identity \(Z\fieldPhi =3\coeff\bigl(Z(\Box_g\fieldPhi)-ZG_\Phi\bigr) \), the flat bad Hessian gives \(CA\chi^{5/2} \lambda^{1/2 + \nu_*} \) after the weak normalization and division by \(\lambda^{\eta_k} \). Its undifferentiated metric correction has the small multiplier \(\chi^{-2}H^{\H00} \). A nonempty boost-only coefficient gives
\be
\label{equa:seven-weak-ray-algebraic-coefficient}
\lambda^{3/2} \mathcal K_i t^{-1} \lambda^{\eta_i}| \del^2Z_{I^c} \fieldPhi| \lesssim A\mathcal K_i\chi^{3/2} \lambda^{-1/2 + \eta_k+ \nu_*}.
\ee
Translated coefficients use the direct gauge, and good Hessians carry \(t^{-1} \). The potential and null matter trace have stronger powers. Thus, with \(e_a=3\delta+ \nu_*<1/2\), the normalized starting size is bounded by
\be
\label{equa:seven-weak-ray-algebraic-en}
CA\Bigl(\chi^{5/2} \lambda^{1/2 + \nu_*} + \lambda^{-1/2 +e_a} \Bigl(1+ \sum_{i =1}^k\mathcal K_i(s_1) \Bigr) \Bigr).
\ee
On the algebraic portion, \(\chi\leq\lambda^{-1/4} \), so the first term is at most \(\lambda^{-1/8+ \nu_*} \). The derivative part has the same bound. The identity \(D\fieldPhi = \chi\del_t\fieldPhi+ \lambda^{-1}z^aL_a \fieldPhi\) also controls the oscillator velocity. Its extra term, and \(\lambda^{-1} \fieldPhi\), have normalized size at most \(C\sqrt\coeff A\chi^{3/2} \lambda^{-1/2 + \nu_{k+3}} \).

At an original location on \(t-r=1\), we have \(\chi\simeq\lambda^{-1} \), and the direct high mass gives
\be
\label{equa:seven-weak-ray-interface-data}
\lambda^{3/2}|Z\fieldvarphi| \lesssim A\lambda^{-1+ \nu_{k+2}}, \qquad \lambda^{3/2}|DZ\fieldPhi|+ \lambda^{1/2}|Z\fieldPhi| \lesssim\sqrt\coeff A\lambda^{-2 + \nu_{k+3}}.
\ee
\ese
The bounded part of the initial hyperboloid is controlled by the prescribed Sobolev norms, with a constant depending on \(s_0\). The largest derivative count in this step is \(p+3\leq p_{\mathbf{high}} \).

\noindent{\bf 2.} \emph{Coefficient variation and mixed primitive.} The direct estimate \eqref{equa:seven-strong-ray-coefficient-tail} yields
\bse
\be
\label{equa:seven-weak-ray-variation}
\int_{\lambda_*}^s|a'| \, d\lambda \lesssim A\chi^{1/2-4\delta} +A^2\chi^{2-8\delta} +A_0\chi^3\lesssim A.
\ee
We use \(V= \lambda^{3/2}Z\fieldphi\). Multiplying the mixed primitive by \(\chi^2\) gives
\be
\label{equa:seven-weak-ray-primitive}
\begin{aligned}
P&\coloneqq6\chi^{-1} \lambda^{3/2} \beta^aE_aZ\fieldPhi, \qquad \beta^a\coloneqq B_H^a-\chi^{-2}H^{\H00}z^a, \\
J_{\mathrm{mix}}&\coloneqq-6\chi^{-1}(\lambda^{3/2} \beta^a)'E_aZ\fieldPhi-6\chi^{-1} \lambda^{3/2} \beta^ay^bE_bE_aZ\fieldPhi.
\end{aligned}
\ee
The oscillator has the form \(3\coeff aV''+V=P'+F_0\). Its shifted energy \(U= \sqrt\coeff V\), \(Q= \coeff V'-P/(3a) \), \(\mathcal E=U^2 +3aQ^2\) satisfies \eqref{equa:seven-strong-ray-shifted-energy-identity}.

Both good mass jets are estimated at the high level:
\be
\label{equa:seven-weak-ray-good-jets}
|EZ\fieldPhi| \lesssim\sqrt\coeff A\chi^{5/2} \lambda^{-2 + \nu_{k+3}}, \qquad |E^2Z\fieldPhi| \lesssim\sqrt\coeff A\chi^{7/2} \lambda^{-3+ \nu_{k+4}}.
\ee
Their largest count is \(p+4=p_{\mathbf{high}} \). The first full metric estimate \eqref{equa:seven-first-full-metric-refinement} gives \(| \beta| \lesssim A\chi^{-1} \lambda^{-1+2\delta} \), hence
\be
\label{equa:seven-weak-ray-primitive-bound}
\frac{|P|}{\sqrt\coeff} \lesssim A^2\chi^{1/2} \lambda^{-3/2 +2\delta+ \nu_*+ \eta_k}.
\ee
\ese
This kernel is integrable because \(2\delta+ \nu_*<1/2\). It controls the initial shift and, together with \eqref{equa:seven-weak-ray-variation}, the term \(|a'||P| \). The factor \(\coeff^{-1/2} \) is compensated by the mass factor \(\fieldPhi = \sqrt\coeff\fieldvarphi\).

\noindent{\bf 3.} \emph{Sources and tangential Hessians.} The flat remainder has size \(C\sqrt\coeff A\chi^{3/2} \lambda^{-3/2 + \nu_*+ \eta_k} \). For the term \(h t^{-1} \del_tZ\fieldPhi\), the coefficient bound is \(h= \chi^{-2}H^{\H00} \lesssim A\chi^{-1} \lambda^{-1+2\delta} \). Its source is therefore at most \(C\sqrt\coeff A^2\chi^{3/2} \lambda^{-3/2 +2\delta+ \nu_*+ \eta_k} \). The terms containing \(D\beta \) have the three bounds
\bse
\be
\label{equa:seven-weak-ray-mixed-errors}
C\sqrt\coeff\lambda^{\eta_k} \Bigl(A^2\lambda^{-2 + \delta+ \nu_*} +A^3\chi^{-1/2} \lambda^{-5/2 +2\delta+ \nu_*} +AA_0\chi^{1/2} \lambda^{-5/2 + \nu_*} \Bigr).
\ee
The middle term is integrable from \(\lambda_*\), since \(\chi^{-1/2} \lambda_*^{-3/2 +2\delta+ \nu_*} \leq\chi^{11/2-8\delta-4\nu_*} \leq1\). The remaining terms of \(J_{\mathrm{mix}} \) are stronger.

In a good-Hessian commutator, a coefficient operator of order at least four leaves a field operator whose mass estimate costs at most \(p\) orders. The other cases are boost-only coefficient operators of length \(i =1,2,3\).~\autoref{lem:seven-boost-v-interpolation}, weakened from order \(p+2\) to \(p\geq10\), gives
\be
\label{equa:seven-weak-ray-interpolated-coefficient}
| \mathcal L_{L^I}H| \lesssim A t^{-1+ \vartheta_i} \lambda^{e_i}, \qquad \vartheta_i\coloneqq \frac{i}{2p} \leq\frac3{20}, \qquad e_i\coloneqq2\delta \Bigl(1-\frac ip\Bigr) + \delta \frac ip\leq2\delta.
\ee
The bounds at the two interpolation orders are \(|H| \lesssim At^{-1} \lambda^{2\delta} \) and \(|H|_p\lesssim At^{-1/2} \lambda^\delta \), within the available derivative count. The good-Hessian product is consequently bounded by
\be
\label{equa:seven-weak-ray-high-good-source}
C\sqrt\coeff A^2\chi^{7/2-\vartheta_i} \lambda^{-3/2 + \vartheta_i+e_i+ \nu_*+ \eta_k}, \qquad \frac12-\vartheta_i-e_i-\nu_*>\frac{191}{2560}.
\ee
The frame terms have the same or stronger bounds. The null matter trace uses at most \(p+3\) wave orders and gives \(CA^2\lambda^{-3/2 +2\nu_*+ \eta_k} \).

For the nonlinear potential, its factor with fewer differentiations uses the lower mass estimate. With \(e_{\mathrm{pot}} =3\rho+2c\), the resulting term is
\be
\label{equa:seven-weak-ray-potential-kernel}
CA\lambda^{-3/2 +e_{\mathrm{pot}}} \sum_{i =0}^k\lambda^{\eta_i} \mathcal Y^{(0)}_{p,k-i}(\lambda) \leq C_NA\lambda^{-3/2 +e_{\mathrm{pot}} + \eta_p} \mathcal Y^{(0)}_{p,k}(\lambda).
\ee
\ese
The exponent satisfies \(e_{\mathrm{pot}} + \eta_p<1/2\). This keeps the larger factor in the unknown amplitude; it keeps the actual high energy loss in every direct estimate.

\noindent{\bf 4.} \emph{Commutators with two time derivatives.} A nonempty boost-only coefficient of rank \(i\) leaves a Hessian which is a first derivative of an operator with one additional translation, of order at most \(p-i+1\leq p\) and rank at most \(k-i\). Thus we obtain
\bse
\be
\label{equa:seven-weak-ray-bad-source}
\lambda^{3/2}|(\mathcal L_{L^I}H)^{\H00} \del_t^2Z_{I^c} \fieldPhi| \lesssim\mathcal K_i(s_1) \lambda^{-1+ \eta_i} \mathcal Y^{(0)}_{p,k-i}(\lambda), \qquad i\geq1.
\ee
A coefficient operator containing a translation uses the gauge estimate and its \(t^{-1} \) frame terms. It gives the integrable kernel \eqref{equa:seven-strong-ray-integrable-commutator-kernel}, at the same or a lower rank. In particular every nonempty coefficient operator at rank zero has an integrable kernel.

\noindent{\bf 5.} \emph{Recovery of all derivatives.} The shifted energy controls \(U= \lambda^{3/2}Z\fieldvarphi\) and \(\coeff V'\), since \(P\) is bounded. The identities
\be
\label{equa:seven-weak-ray-recovery}
\lambda^{3/2} \chi\del_tZ\fieldPhi = \coeff V'-\tfrac32\lambda^{1/2}Z\fieldPhi-\lambda^{3/2}y^aE_aZ\fieldPhi, \quad \del_aZ\fieldPhi =E_aZ\fieldPhi-z_a \del_tZ\fieldPhi
\ee
hold also for translation operators. The extra terms in the first line are bounded by 
\be
C\sqrt\coeff A\chi^{3/2} \lambda^{-1/2 + \nu_*+ \eta_k}.
\ee
 The additional good spatial term has an extra factor \(\chi^2\). These powers are bounded after rank normalization. Commuting a boost through a translation operator adds lower operators already present in \(\mathcal Y^{(0)}_{p,k} \).

The variation estimate and the integrable same-rank kernels give a uniform Gronwall factor. The remaining integral satisfies
\be
\label{equa:seven-weak-ray-rank-integral}
s^{-\eta_k} \int_{s_0}^s\lambda^{-1+ \eta_i} \mathcal Y^{(0)}_{p,k-i}(\lambda) \, d\lambda \leq\frac{\mathcal V^{(0)}_{k-i}(s_1)}{\eta_k}, \qquad k\geq1.
\ee
\ese
The initial data estimate and this identity prove \eqref{equa:seven-weak-effective-field-ray-bound}. At rank zero the first full metric estimate gives the uncommuted mixed coefficient, so \(\mathcal K_0\) is absent.
\qedhere

\end{proof}


\section{Characteristic estimates for the wave fields} 
\label{section-characteristic-pointwise}


\subsection{Characteristic estimates for the massless field}
\label{section-9-5}

\paragraph{Transport variable and wave equation}

As we now show, the characteristic equations give pointwise bounds for the massless field and the tangential metric components \emph{near the light cone.} We combine these bounds with the effective-field ray estimates, keep the initial and interface values, and eventually deduce the \emph{low-order Hessian and source estimates} needed by the energy argument.

To estimate the massless matter stress tensor at lower orders, we need a pointwise bound on  \((t/s)| \del\psi| \) near the light cone. We use \(L= \del_t+ \del_r\), \(\underL= \del_t-\del_r\), and the semi-null coefficient \(H_N\) from Section~\ref{section-3-4}, and set
\be
\label{equa:seven-matter-characteristic-variable}
q\coloneqq t-r, \qquad v\coloneqq Z\psi, \qquad W_Z\coloneqq \underL(rv).
\ee
Direct spherical decomposition, as in~\cite[Section~8.3, Eq.~(8.28)]{LeFloch-Ma-2024a}, gives the following identity whenever \(4-H_N>0\):
\bse
\be
\label{equa:seven-matter-characteristic-identity}
\mathbb P_H^N W_Z= \frac{4r}{4-H_N} \Bigl(-\Box_gv+r^{-2} \Delta_{\mathbb S^2}v+r^{-1}H_NX^N[v]+ \slashed H^N[v] \Bigr),
\ee
where the operator and remainders are defined by
\be
\label{equa:seven-characteristic-operator-and-errors}
\mathbb P_H^N\coloneqq \del_t+ \frac{4+H_N}{4-H_N} \del_r, \qquad X^N[v]{\coloneqq} \tfrac12\underL v,
\ee
\be
\label{equa:seven-matter-characteristic-exact-remainder}
\aligned
\slashed H^N[v]
& \coloneqq
H_N\Bigl(\tfrac12L\underL v+ \tfrac14L^2v\Bigr) +2(H^{0a}-H^{ab} \omega_b) \del_t\delsN_av 
\\
& \quad +H^{ab} \delsN_a \delsN_bv-r^{-1}(\delta_{ab}-\omega_a \omega_b)H^{ab} \del_tv.
\endaligned
\ee
\ese
The angular operator is \(\Delta_{\mathbb S^2} = \sum_{a<b} \Omega_{ab}^2\). 

Indeed, the semi-null expansion separates \(H_N\del_t^2v\), and \(r\underL^2v= \underL W_Z+2\underL v\). Every remaining Hessian term contains an outgoing or angular derivative, so
\be
\label{equa:seven-matter-characteristic-remainder-size}
| \slashed H^N[v]| \leq C|H| \Bigl(| \del\delsN v|+r^{-1}| \del v| \Bigr).
\ee

\paragraph{Near-light-cone geometry and characteristic transport}

The null and semi-hyperboloidal frame components satisfy 
\be
\label{equa:seven-null-hyperboloidal-component-comparison}
|H_N-H^{\H00}| \lesssim {q\over t}|H|.
\ee
The direct gauge bound \eqref{equa:seven-direct-H00-value}, the interface estimate of~\autoref{lem:seven-pointwise-metric-interface-bulk}, and \(q\geq1\) therefore give
\be
\aligned
r|H_N| \leq r|H^{\H00}|+Cq|H| 
& \leq C_N\bigl(Aqt^{-1/2}s^\delta+A^2qs^{-1+2\delta} +A_0 + \mathsf B_{00;0}(s) +q|H| \bigr) 
\\
& \leq C_N(A_0 +A)q.
\endaligned
\ee
Consequently, we have
\be
\label{equa:seven-characteristic-small-rH}
|H_N| \leq h_0<1, \quad |rH_N| \leq\varepsilon_{\mathrm{ch}}q, \quad \varepsilon_{\mathrm{ch}} \leq C_N(A_0 +A).
\ee
If \(\gamma(\tau)= (\tau,r(\tau), \omega) \) is an integral curve of \(\mathbb P_H^N\), then
\be
\label{equa:seven-characteristic-q-s-derivatives}
{dr \over d\tau} ={4+H_N\over4-H_N}, \quad {dq\over d\tau} =-{2H_N\over4-H_N}, \quad {ds\over d\tau} ={q-2rH_N/(4-H_N) \over s}.
\ee
For \(C\varepsilon_{\mathrm{ch}}<1/2\), the last quantity is positive and satisfies
\be
\label{equa:seven-characteristic-dtds}
c{s\over q} \leq {d\tau\over ds} \leq C{s\over q}.
\ee
Every backward characteristic, stopped at its first exit from \(\{q\geq1, \ r/t\geq3/5, \ s\geq s_0\} \), therefore reaches at least one of the following boundary pieces:
\be
\label{equa:seven-characteristic-endpoint-alternative}
\Mint_{s_0}, \quad \Lscr= \{q=1\}, \quad\hbox{or} \quad \Sigma_{\mathrm c}{\coloneqq} \{r/t=3/5\}.
\ee
At an intersection, either corresponding trace estimate applies. The third alternative is an interior terminal point and must not be counted as exterior data. On it, \(t=5s/4\), \(r=3s/4\), and \(q=s/2\). Hyperboloidal Sobolev and the Hardy inequality with its interface trace give, using \(\beta_{j+2}<1/2\),
\be
\label{equa:seven-central-characteristic-endpoint}
|W_Z| \leq r| \del Z\psi|+ |Z\psi| \leq C_NA s^{-1/2 + \beta_{j+2}} \leq C_NA \quad\hbox{on } \Sigma_{\mathrm c}.
\ee
The angular operator must be counted both in differential order and, after conversion to the standard family, in boost rank.  Since we have
\be
\label{equa:seven-angular-translation-reserve}
r^{-1} \Omega_{ab} = \omega_a \delsN_b-\omega_b\delsN_a, \quad r(r^{-1} \Omega_{ab})^2Z\psi = (r^{-1} \Omega_{ab}) (\Omega_{ab}Z\psi),
\ee
one angular derivative is a good derivative, while the other and the two Sobolev commutations on the spheres are rotations.  Thus, for \(Z\preccurlyeq(p_{\mathbf{low}}-3,j) \), one commuted rotation and the two Sobolev rotations reach exactly the full lower family of order \(p_{\mathbf{low}} \), with boost rank at most \(j+3\). Consequently, we have
\be
\label{equa:seven-angular-characteristic-integral}
\left|r \sum_{a<b}(r^{-1} \Omega_{ab})^2Z\psi\right| \leq C_NA t^{-3/2}s^{\beta_{j+3}} \leq C_NA t^{-3/2}s^{\nu_j}, \qquad \int_\gamma \left|r \sum_{a<b}(r^{-1} \Omega_{ab})^2Z\psi\right| \, d\tau \leq C_NA.
\ee
Indeed, \(\beta_{j+3}<\nu_j\), since \(\nu_j-\beta_{j+3} =7\rho-C_{\mathrm{rk}} \theta>0\). Moreover, \(d\tau/ds\lesssim t/s\) and \(t\geq cs\) in the near-cone region, so the last integral is bounded by \(C_NA\int_{s_0}^s\lambda^{-3/2 + \nu_j}d\lambda \). The three angular commutations are therefore charged explicitly to the lower family.  Their rank loss is absorbed by the exponent gap above, and the angular term is not inserted into a Gronwall kernel.

\paragraph{Rank induction for the characteristic norm}

For the rank induction, we define
\be
\label{equa:seven-Wpsi-characteristic-norm}
\mathcal Z_{\psi,j}(s_1) \coloneqq \max_{0\leq i\leq j} \ \sup_{\substack{Z\preccurlyeq(p_{\mathbf{low}}-3,i), \ \rank Z=i\\
(t,x) \in\Mint_{[s_0,s_1]}, \ r/t\geq3/5}} (t^2-r^2)^{-\eta_i/2}|W_Z(t,x)|.
\ee
This is the \(W\)-component of \(\mathcal V_{p_{\mathbf{low}}-3,j}(s_1) \).  On \(\Lscr\), we have \(t\simeq s^2\), so the lateral weight in \(\mathsf B_{\psi;j} \) is equivalent to the weight in this definition.
\paragraph{Integrated source estimate}

\begin{proposition}[Matter characteristic source integral]
\label{lem:seven-matter-characteristic-source-integral}
Let \(p=p_{\mathbf{low}}-3\). On a finite bootstrap slab, one assumes the coarse energy and Hardy bounds through \(p_{\mathbf{high}} \), the wave gauge, and \eqref{equa:seven-characteristic-small-rH}--\eqref{equa:seven-characteristic-dtds}. The coefficient \(1-(t/s)^2H^{\H00} \) is bounded above and below by positive fixed constants. The positive-rank coefficient and effective-field bounds are
\be
\label{equa:seven-matter-characteristic-pointwise-inputs}
|(\mathcal L_{L^J}H)^{\H00}| \leq C_NA t^{-1}s^{\eta_{|J|}}, \quad1\leq|J| \leq p, \qquad | \del\fieldPhi|_{p,j} \leq C_NA\frac st s^{-3/2 + \eta_j}, \quad0\leq j\leq p.
\ee
The first estimate is stated for the tensor Lie derivative; the cumulative scalar estimate \eqref{equa:seven-preliminary-H00}, including its lower-rank terms, implies it. The second estimate is provided by the strong ray lemma under its coefficient hypotheses. Then, for every \(v=Z\psi\) of order at most \(p\) and rank \(j\), and every characteristic ending at hyperboloidal parameter \(\sigma \leq {s_1}\),
\be
\label{equa:seven-matter-characteristic-integrated-source}
\sigma^{-\eta_j} \int_\gamma| \mathcal R_{\mathrm{ch}}[Z\psi]| \, d\tau \leq C_NA+C_NA\sum_{i =1}^{j} \sigma^{-\eta_j} \int_{s_0}^{\sigma} \lambda^{-1+ \eta_i+ \eta_{j-i}} \mathcal Z_{\psi,j-i}(\lambda) \, d\lambda.
\ee
Here \(\tau\) is physical time, and \(\mathcal Z_{\psi,j} \) is the normalized characteristic norm in \eqref{equa:seven-Wpsi-characteristic-norm}. In particular, this proves the matter instance of \eqref{equa:seven-j-ray-allocation-bound}. The positive-rank metric bound is a separate hypothesis. At rank zero, only the rank-zero effective-field estimate and the geometric bounds are used.
\end{proposition}

\begin{proof}

\noindent{\bf 1.} \emph{Commuted transport source.} The remainders are given by \eqref{equa:seven-matter-characteristic-identity}--\eqref{equa:seven-matter-characteristic-remainder-size}. For an ordered translation--boost operator, the equation and the coupling have the expansions
\bse
\be
\label{equa:seven-matter-characteristic-ordered-sources}
\begin{aligned}
\Box_gZ\psi&=ZF_\psi-\sum_{I\ne\varnothing}(\mathcal L_IH)^{\alpha \beta} \del_{\alpha \beta}Z_{I^c} \psi, \qquad F_\psi =g^{-1}(d\fieldPhi,d\psi), \\
ZF_\psi&= \sum_{I_0\sqcup I_1\sqcup I_2}(\mathcal L_{I_0}g^{-1}) (dZ_{I_1} \fieldPhi,dZ_{I_2} \psi).
\end{aligned}
\ee
Thus the source includes every principal and coupling commutator.

\noindent{\bf 2.} \emph{A stronger good-Hessian estimate.} The wave equation, the direct energies, and \eqref{equa:seven-matter-characteristic-pointwise-inputs} imply
\be
\label{equa:seven-matter-characteristic-light-like-jets}
| \del^2Z\psi| \leq C_NA t^{1/2}s^{-3+ \nu_{j+3}}, \qquad | \del\delsN Z\psi| \leq C_NA t^{-3/2}s^{-1+ \nu_{j+3}}.
\ee
For the first estimate, \(\delsH_a=t^{-1}L_a \) gives \(| \del\delsH Z\psi|+ t^{-1}| \del Z\psi| \leq C_NA t^{-3/2}s^{-1+ \beta_{j+3}} \). Solving the semi-hyperboloidal wave equation for \(\del_t^2Z\psi\) costs \((t/s)^2\). The null contraction and the strong effective-field derivative yield
\be
\label{equa:seven-matter-characteristic-coupling-bound}
|ZF_\psi| \leq C_NA^2t^{-5/2}s^{-1/2 + \nu_{j+3}}.
\ee
Here a bad effective-field derivative is paired with a good matter derivative. When the good derivative falls on the effective field, the mass term gives \(| \delsH\fieldPhi|_{d,i} \leq C_N\sqrt\coeff A t^{-5/2}s^{\alpha_{i+3}} \), at \(d\leq p\). The bad--bad coefficient is \(-(s/t)^2 +H^{\H00} \); its differentiated metric contributions use the coefficient bound in the statement or the wave gauge. The potential is absent from the matter equation.

The principal commutators are handled by induction on total operator order. A boost-only bad coefficient has size \(C_NA t^{-1}s^{\eta_i} \) and multiplies a lower-order Hessian; \(t/s^2\leq1\) and \(\eta_i+ \nu_{j-i+3} = \nu_{j+3} \) preserve the target weight. A translated bad coefficient is controlled by the gauge. Its additional factors, relative to the target Hessian weight, are bounded by \(C_NA s^{-1+12\rho+4c} \) and \(C_NA^2s^{-1+24\rho+8c} \), where \(c=C_{\mathrm{rk}} \theta \). Good-Hessian commutators have the stronger factor \(C_NA s^{-1/2 +12\rho+4c} \). This proves the first estimate by finite induction. The maximum kinetic count is one translation or boost followed by two Sobolev boosts, namely \(p+3\leq p_{\mathbf{low}} \).

For the second estimate, the identity
\be
\label{equa:seven-matter-characteristic-good-frame}
\delsN_a= \delsH_a+ \frac qt\omega_a \del_t, \qquad \left| \del\Bigl(\frac qt\omega_a \Bigr) \right| \leq Ct^{-1}, \qquad qt\leq s^2
\ee
\ese
gives the additional factor \(s^{-1} \) in the good Hessian. This factor makes the regular metric terms integrable.

\noindent{\bf 3.} \emph{Physical time and the angular term.} Along a characteristic, \eqref{equa:seven-characteristic-q-s-derivatives} gives \(d\tau/d\lambda= \lambda/(q-2rH_N/(4-H_N)) \simeq t/\lambda \). In the angular term, \eqref{equa:seven-angular-translation-reserve} keeps one rotation as a good derivative and converts the other to one boost. The two Sobolev commutations therefore cost \(p+3\leq p_{\mathbf{low}} \). Estimate~\eqref{equa:seven-angular-characteristic-integral} gives its integrable majorant \(C_NA\lambda^{-3/2 + \beta_{j+3}} \). The Hardy boundary term is bounded by the exterior radial trace \(C_NA\lambda^{-1+ \delta} \), which is absorbed into the direct lower-order growth.

\noindent{\bf 4.} \emph{Regular metric terms, coupling, and translated coefficients.} The coarse pointwise metric bound is \(|H| \leq C_NA t^{-1/2} \lambda^{\nu_2} \). Equations \eqref{equa:seven-matter-characteristic-light-like-jets} and \eqref{equa:seven-characteristic-dtds} give
\bse
\be
\label{equa:seven-matter-characteristic-regular-integrands}
\frac t\lambda \Bigl(r|H|| \del\delsN Z\psi|+ |H|| \del Z\psi| \Bigr) \leq C_NA^2\lambda^{-2 + \eta_j+24\rho+9c}, \qquad \frac t\lambda r|ZF_\psi| \leq C_NA^2\lambda^{-2 + \nu_{j+3}}.
\ee
The first bound includes the explicit \(H_N\underL v/2\) term. Good-Hessian principal commutators are expanded in the semi-hyperboloidal frame; their full coefficient is bounded by \(C_NA t^{-1/2} \lambda^{\nu_{i+2}} \), and their good Hessian satisfies the direct estimate in step 2. Their frame term carries \(t^{-1} \) and obeys the same majorant.

For a coefficient operator containing a translation, the gauge gives
\be
\label{equa:seven-matter-characteristic-translated-coefficient}
|(\mathcal L_IH)^{\H00}| \leq C_N\Bigl(A t^{-3/2} \lambda^{\nu_{i+2}} +A^2t^{-1} \lambda^{-1+ \eta_i+24\rho+8c} +A_0t^{-2} \Bigr),
\ee
where \(i\) is its boost rank. Multiplication by the Hessian in \eqref{equa:seven-matter-characteristic-light-like-jets}, by \(r\), and by \(d\tau/d\lambda \) gives terms bounded by \(C_N(A^2 +A^3) \lambda^{-2 + \eta_j+36\rho+13c} \), together with stronger reference terms. The offsets \(\beta_{j+3}-\eta_j\), \(24\rho+9c\), and \(36\rho+13c\) are less than \(\delta \). Thus all terms treated so far have normalized integral bounded by \(C_NA\).

\noindent{\bf 5.} \emph{Boost-only bad coefficients.} The identity \(\mathcal L_{L_a} \ell_{\H} =-(x_a/t) \ell_{\H} \) implies that tensor boost commutation preserves the bad contraction up to its lower-rank scalar contractions. If \(I\) contains \(i\geq1\) boosts and no translations, the first bound in \eqref{equa:seven-matter-characteristic-pointwise-inputs} applies directly to \((\mathcal L_IH)^{\H00} \). The Hessian of its complementary operator is a first derivative of an operator with one additional translation, of order at most \(p-i+1\leq p\) and rank at most \(j-i\). Characteristic reconstruction and the direct good derivative and pointwise estimates give
\be
\label{equa:seven-matter-characteristic-lower-rank-Hessian}
| \del^2Z_{I^c} \psi| \leq Ct^{-1} \lambda^{\eta_{j-i}} \mathcal Z_{\psi,j-i}(\lambda) +C_NA t^{-3/2} \lambda^{\nu_{j-i+2}}.
\ee
\ese
After multiplication by \(r\, d\tau/d\lambda \), the first term contributes $ C_NA\lambda^{-1+ \eta_i+ \eta_{j-i}} \mathcal Z_{\psi,j-i}(\lambda). $ The second contributes an integrable term bounded by \(C_NA^2\lambda^{-3/2 + \eta_j+ \delta} \). This accounts for every remaining principal commutator and proves \eqref{equa:seven-matter-characteristic-integrated-source}.

\end{proof}

The initial value, the interface amplitude \(\mathsf B_{\psi;k} \), and the direct central-boundary bound can be inserted in the transport identity. Finite rank induction with \(C_NA\leq C_{\mathrm{rk}} \theta \) gives
\be
\label{equa:seven-Wpsi-sharp}
\mathcal Z_{\psi,k}(s_1) \leq C_N\Bigl(A_0 +A+ \max_{0\leq j\leq k} \mathsf B_{\psi;j}(s_1) \Bigr).
\ee
The exterior low-order matter estimate bounds the last amplitude by \(C_NA\). The positive-rank conclusion requires the metric coefficient hypothesis used in the strong ray and characteristic estimates.
\paragraph{Characteristic reconstruction and singular consequences}

Finally, in the near-cone region, we have
\be
\label{equa:seven-characteristic-reconstruction}
| \del Z\psi| \lesssim r^{-1} \bigl(|W_Z|+ |Z\psi| \bigr) + | \delsN Z\psi|.
\ee
The definition of \(\mathcal Z_{\psi,j} \) and \eqref{equa:seven-Wpsi-sharp} give \(|W_Z| \leq C_N(A_0 +A+ \max_{i\leq j} \mathsf B_{\psi;i})s^{\eta_j} \). The pointwise bound for \(r^{-1}|Z\psi|\) follows from Hardy--Sobolev; the good derivative is controlled by the direct hyperboloidal estimate, and the bad derivative by the preceding bound for \(W_Z\). In \(r/t\leq3/5\), the direct estimate is already stronger. This proves the characteristic matter estimate at rank zero from the coarse bounds, and at positive ranks under the metric coefficient hypothesis of~\autoref{lem:seven-matter-characteristic-source-integral}. The norm \(\mathcal D_{\psi,0} \) keeps its direct Sobolev growth.

Dividing \eqref{equa:seven-refined-effective-field} by the powers of $\coeff$ also gives
\be
\label{equa:seven-refined-effective-field-singular}
\begin{aligned}
| \fieldphi|_{p_{\mathbf{low}}-3,k} & \lesssim C\theta^{-1} \coeff^{-1/2}C_{\mathrm{tr}}(s) \mathcal A_{p_{\mathbf{low}}-3,k}^{\mathrm{bd}}(s) \Big({s\over t} \Big)^2s^{-3/2 + \eta_k}, \\
{s\over t}| \del\fieldphi|_{p_{\mathbf{low}}-3,k} & \lesssim C\theta^{-1} \coeff^{-1}C_{\mathrm{tr}}(s) \mathcal A_{p_{\mathbf{low}}-3,k}^{\mathrm{bd}}(s) \Big({s\over t} \Big)^2s^{-3/2 + \eta_k}.
\end{aligned}
\ee
Alternatively, the degenerate pointwise bound \eqref{equa:seven-direct-pointwise-singular} loses only \(\coeff^{-1/2} \) on a derivative but has the weaker direct decay.  Both estimates are useful, and neither dominates the other in every nonlinear term.


\subsection{\texorpdfstring{Good metric components and low-order Hessians}{Good metric components and low-order Hessians}}
\label{section-good-metric-characteristics}

\paragraph{Tangential metric components}

The tangential projection of the reduced metric equation removes its leading plane-null quadratic tensor. The remaining potential is estimated by the lower mass energy. The normalized metric characteristic norm is
\be
\label{equa:seven-good-metric-characteristic-norm}
\mathcal Z_{u,k}(s_1)\coloneqq \max_{0\leq j\leq k} \sup_{s_0\leq\lambda \leq {s_1}} \lambda^{-\eta_j} \mathcal W^u_{p,j}(\lambda), \qquad p=p_{\mathbf{low}}-3.
\ee

\begin{proposition}[Good metric characteristic source integral]
\label{lem:seven-good-metric-characteristic-source}
On a finite bootstrap slab, the coarse energy and Hardy bounds, the propagated wave gauge, the prescribed reference estimates, and the characteristic geometry \eqref{equa:seven-characteristic-small-rH}--\eqref{equa:seven-characteristic-dtds} are imposed. The uncommuted semi-hyperboloidal principal coefficient is uniformly elliptic. At positive boost ranks, the additional coefficient hypothesis is
\be
\label{equa:seven-good-metric-characteristic-coefficient}
|(\mathcal L_{L^I}H)^{\H00}| \leq C_NA t^{-1}s^{\eta_{|I|}}, \qquad 1\leq |I| \leq p.
\ee
For every scalar operator \(v=Zw\), with \(w\in\mathscr G(u) \), \(\ord Z\leq p\), and \(\rank Z=j\), the backward characteristic ending at hyperboloidal time \(\sigma \leq {s_1}\) satisfies
\be
\label{equa:seven-good-metric-characteristic-integrated-source}
\sigma^{-\eta_j} \int_\gamma| \mathcal R_{\mathrm{ch}}[Zw]| \, d\tau \leq C_NA+C_NA\sum_{i =1}^j\sigma^{-\eta_j} \int_{s_0}^{\sigma} \lambda^{-1+ \eta_i+ \eta_{j-i}} \mathcal Z_{u,j-i}(\lambda) \, d\lambda.
\ee
The constants are independent of \(s_1\) and \(0<\coeff\leq\coeff_0\). The sum is empty at rank zero. This proves the metric instance of \eqref{equa:seven-j-ray-allocation-bound}, with the positive-rank coefficient hypothesis kept.
\end{proposition}

\begin{proof}

\noindent{\bf 1.} \emph{Tangential projection and frame terms.} We set \(z=x/t\), \(E_a= \delsH_a \), and
\bse
\be
\label{equa:seven-good-metric-projection-family}
\mathcal A_{\H}T= (T(E_a, \del_\beta),T(E_a,E_b))_{a,b, \beta}.
\ee
For \(T_Z= \mathcal L_Zu\), write \(V_Z= \mathcal A_{\H}T_Z\), component by component. Symmetry and the identity
\be
\label{equa:seven-good-metric-component-equivalence}
T(E_a, \del_b)=T(E_a,E_b)-z_bT(E_a, \del_t)
\ee
give both conversions between this family and \(\mathscr G(T) \), since \(\delH_0= \del_t\). All coefficients are bounded polynomials in \(z\). The frame identities are
\be
\label{equa:seven-good-metric-frame-identities}
\relax[L_i,E_a]=-z_aE_i, \quad [\del_t,E_a]=-t^{-1}z_a \del_t, \quad [\del_i,E_a]=t^{-1} \delta_{ia} \del_t, \quad [\Box,z_a]f=2t^{-1}E_af-2t^{-2}z_af.
\ee
Boosts preserve the good component family. Translations introduce \(t^{-1} \) times raw components of lower-order tensor operators. Product differentiation therefore gives
\be
\label{equa:seven-good-metric-projection-commutator}
|[\Box_g, \mathcal A_{\H}]T_Z| \lesssim t^{-1}|ET_Z|+ t^{-2}|T_Z| + |H| \bigl(t^{-1}| \del T_Z|+ t^{-2}|T_Z| \bigr).
\ee
\ese
The coefficients are bounded polynomials in \(z\).

\noindent{\bf 2.} \emph{Projected quadratic and massive sources.} The plane-null calculation \eqref{equa:eight-good-metric-plane-null-Christoffel}--\eqref{equa:eight-good-metric-projected-quadratic}, polarized and combined with the commuted wave gauge, applies to every ordered tensor Lie partition. Since \(E_a= \delsN_a-(q/t) \omega_a \del_t\) and \(q/t\leq(s/t)^2\), it yields
\bse
\be
\label{equa:seven-good-metric-projected-quadratic}
| \mathcal A_{\H} \mathbb F| \lesssim |Eh| \,| \del h|+(s/t)^2| \del h|^2 + |H| \,| \del h|^2.
\ee
The last term keeps the nonlinear remainder in the wave-gauge identity. The scalar derivative sources contain \(E\fieldPhi\, \del\fieldPhi\) and \(E\psi\, \del\psi\). The potential is a smooth bounded coefficient times \((\fieldvarphi)^2\). Its lower mass bound is \(|Z\fieldvarphi| \lesssim At^{-3/2}s^{\alpha_{j+2}} \). Thus the nonprincipal source, including the projection terms, satisfies
\be
\label{equa:seven-good-metric-nonprincipal-source}
|S_Z| \lesssim A t^{-5/2}s^{\nu_{j+2}} +A_0t^{-3} +A^2t^{-2}s^{-1+ \eta_j+24\rho+8c} +A^3t^{-3/2}s^{-2 + \eta_j+36\rho+12c}, \qquad c=C_{\mathrm{rk}} \theta.
\ee
The potential and the quadratic term weighted by \((s/t)^2\) have the stronger weight \(A^2t^{-3} \). Smooth coefficient derivatives add decaying metric components; their boost ranks remain in the sum \(\eta_j\). The mass estimate costs \(p+2\) orders, so every effective-field factor belongs to the lower family. No strong ray estimate is used in this step.

\noindent{\bf 3.} \emph{Hessians of the projected metric components.} The projected commuted principal terms are $ -(\mathcal L_IH)^{\alpha \beta} \del_{\alpha \beta}V_{Z\setminus I}$ with $I\ne\varnothing$, up to the frame terms above. Solving the semi-hyperboloidal wave equation and inducting on total operator order gives
\be
\label{equa:seven-good-metric-light-like-Hessians}
| \del^2V_Z| \lesssim A t^{1/2}s^{-3+ \nu_{j+3}}, \qquad | \del\delsN V_Z| \lesssim A t^{-3/2}s^{-1+ \nu_{j+3}}.
\ee
The direct good Hessian costs one additional commutation and two Sobolev boosts, hence \(p+3\leq p_{\mathbf{low}} \). After multiplication of \eqref{equa:seven-good-metric-nonprincipal-source} by \((t/s)^2\), its quadratic and cubic terms, relative to the first bound above, have factors at most \(A s^{-1/2 +12\rho+3c} \) and \(A^2s^{-1+24\rho+7c} \). A boost-only bad coefficient has the bounded factor \(C_NA(t/s^2)s^{\eta_i} \) on the lower-order Hessian. Translated bad coefficients use \eqref{equa:seven-matter-characteristic-translated-coefficient}; their additional factors are \(A s^{-1+12\rho+4c} \) and \(A^2s^{-1+24\rho+8c} \). Good principal coefficients multiply direct good Hessians. This proves the first estimate by finite induction. The identity \eqref{equa:seven-matter-characteristic-good-frame} proves the second.

Repeated use of \eqref{equa:seven-good-metric-frame-identities} expresses \(Zw\) as bounded polynomials in \(z\) times projected tensor operators, plus \(t^{-d}P(z) \mathcal L_Yu\), where \(d\geq1\) and \(\ord Y\leq\ord Z-d\). The inverse conversion has the same form. For these raw frame errors, the unprojected lightlike reduction keeps
\be
\label{equa:seven-good-metric-raw-frame-Hessian}
| \del^2\mathcal L_Yh| \lesssim A t^{1/2}s^{-3+ \nu_{\rank Y+3}} +A^2t s^{-4+ \eta_{\rank Y} +24\rho+8c} +A_0t^{-1}s^{-2}.
\ee
\ese
The extra \(t^{-1} \) gives integrable characteristic sources, including the reference term. Derivatives of a bounded polynomial in \(z\) have the good flat contraction in \eqref{equa:seven-good-metric-frame-identities}. Consequently the scalar and tensor formulations differ only by integrable source terms.

\noindent{\bf 4.} \emph{Physical time and integrable contributions.} The Jacobian is \(d\tau/d\lambda= \lambda/(q-2rH_N/(4-H_N)) \simeq t/\lambda \). For the angular term, one rotation is a good derivative and one is a bounded combination of boosts. The two Sobolev boosts therefore give
\bse
\be
\label{equa:seven-good-metric-angular-source}
|r^{-1} \Delta_{\mathbb S^2}(Zw)| \lesssim A t^{-3/2}s^{\nu_{j+3}}, \qquad p+3=p_{\mathbf{low}}.
\ee
After conversion to \(\lambda \), the angular and linear frame terms are bounded by \(C_NA\lambda^{-3/2 + \nu_{j+3}} \). The quadratic, cubic, translated-coefficient, and regular metric terms, including \(H_N\underL v/2\), are bounded by
\be
\label{equa:seven-good-metric-integrable-sources}
C_N(A^2 +A^3) \lambda^{-2 + \eta_j+36\rho+13c}.
\ee
Here \(t^{1/2} \lesssim\lambda \), and \(36\rho+13c<\delta<1/8\). Their normalized integrals are bounded by \(C_NA\).

\noindent{\bf 5.} \emph{Strictly lower boost rank.} For a nonempty boost-only coefficient of rank \(i\), the complementary Hessian is a first derivative of a translation operator of total order at most \(p\) and rank at most \(j-i\). Characteristic reconstruction and the direct good derivative and pointwise estimates give
\be
\label{equa:seven-good-metric-lower-rank-Hessian}
| \del^2V_{Z\setminus I}| \lesssim t^{-1} \lambda^{\eta_{j-i}} \mathcal Z_{u,j-i}(\lambda) +A t^{-3/2} \lambda^{\nu_{j-i+2}}.
\ee
\ese
Multiplication by \eqref{equa:seven-good-metric-characteristic-coefficient}, by \(r\), and by \(d\tau/d\lambda \) leaves \(C_NA\lambda^{-1+ \eta_i+ \eta_{j-i}} \mathcal Z_{u,j-i} \). The second term is integrable. This accounts for every remaining principal commutator and proves the source estimate.

\end{proof}

The initial and central endpoints satisfy the direct bounds. The interface boundary value is measured by \(\mathsf B_{\mathrm{char};p,k} \). Finite rank induction, with \(C_NA\leq C_{\mathrm{rk}} \theta \), gives
\be
\label{equa:seven-good-metric-characteristic-output}
\mathcal Z_{u,k}(s_1) \leq C_N\Bigl(A+A_0 + \max_{0\leq j\leq k} \mathsf B_{\mathrm{char};p,j}(s_1) \Bigr).
\ee
Proposition~\ref{lem:eight-good-metric-characteristic} below will bound the interface term by \(C_NA\). At rank zero this uses only the coarse bounds and the direct gauge geometry. Positive ranks keep \eqref{equa:seven-good-metric-characteristic-coefficient}.

\paragraph{Low-order Hessians and refined sources}

For an operator \(Z\) of order at most three and rank at most \(q\), the semi-hyperboloidal wave decomposition gives
\be
\label{equa:seven-low-wave-Hessian-identity}
s^2| \del^2Zv| \lesssim t\bigl(| \del LZv|+ | \del Zv| \bigr) + t^2| \Box_gZv|, \qquad v\in\{u, \psi\}.
\ee
Here \(| \del LZv| \) is summed over the three boosts. The rank-zero coefficient bound makes the coefficient of \(\del_t^2Zv\) comparable to \((s/t)^2\). The propagated wave gauge identifies \(\Box_g\) with the reduced operator on Cartesian metric components. On a finite slab ending at \(s_1\), we define
\be
\label{equa:seven-low-boundary-amplitude}
\mathcal B_q^{\mathrm{low}}(s)\coloneqq C_{\mathrm{tr}}(s) \max_{0\leq i\leq p_{\mathbf{low}}-3} \mathcal A_{p_{\mathbf{low}}-3,i}^{\mathrm{bd}}(s).
\ee
\be
\label{equa:seven-cone-refined-amplitude-dependence}
B_{s_1}{\coloneqq}1+ \theta^{-1} \mathcal B_3^{\mathrm{low}}(s_1), \qquad \varepsilon_N\coloneqq \delta \, \mathbf1_{\{N=19\}}.
\ee
Constants below may depend on \(B_{s_1}\) and \(\coeff_0\). The loss \(\varepsilon_N\) will be used only at the additional angular order when \(N=19\). For \(0\leq q\leq3\), the Hessian estimates distinguish the three fields.

The refined matter and effective-field estimates give
\be
\label{equa:seven-low-Hessian-bounds}
| \del^2\psi|_{3,q} \lesssim C\theta^{-1} \mathcal B_q^{\mathrm{low}}(s)s^{-2 + \eta_{q+1}}, \qquad | \del^2\fieldPhi|_{3,q} \lesssim C\theta^{-1} \mathcal B_q^{\mathrm{low}}(s) \frac st s^{-3/2 + \eta_q}.
\ee
The matter Hessian uses the refined first derivative at the additional boost rank \(q+1\). The effective-field Hessian uses its refined first derivative after one additional translation. The full metric Hessian keeps its direct growth. The structured lightlike estimate \eqref{equa:seven-cone-light-like-Hessian}, extended to the spatial origin, gives
\be
\label{equa:seven-low-full-metric-Hessian}
| \del^2u|_{3,q} + | \del^2H|_{3,q} \lesssim C_{N,B_{s_1}, \coeff_0}A s^{-2 + \nu_{q+3}}.
\ee
Its hypotheses are the direct metric energies, the refined good metric components, and the separate bad coefficient estimate. The same-rank weighted consequence of the refined matter estimate is
\be
\label{equa:seven-weighted-Hessian-same-rank}
\frac ts| \del^2\psi|_{3,q} \lesssim C\theta^{-1} \mathcal B_q^{\mathrm{low}}(s)s^{-1+ \eta_q}.
\ee
Products involving a full metric Hessian and a translated high coefficient are estimated by the commuted gauge below. Boost commutation preserves the bad contraction because
\be
\label{equa:seven-Hessian-boost-frame}
\mathcal L_{L_a} \ell_{\mathcal H} =-(x_a/t) \ell_{\mathcal H}, \qquad \ell_{\mathcal H} =dt-(x_b/t)dx^b.
\ee
Translation derivatives of this frame have size \(t^{-1} \). Induction on total order therefore controls the principal commutators with their separate bad-coefficient bounds. Each low Hessian estimate uses only orders below \(p_{\mathbf{low}} \).

\paragraph{Critical commutators and the radial metric profile}

The low Hessian bounds give a radial-profile estimate for the high boost-only coefficients. We distinguish it from the time-integrated estimate used in the energy argument.
\bse
\label{equa:seven-critical-remainder-rank-bounds}
\be
\label{equa:seven-critical-remainder-wave-bound}
\mathbf R_{v;p_{\mathbf{high}},k}^{\mathrm{crit}}(s) \lesssim C_{N,B_{s_1}, \coeff_0}A\sum_{j=0}^{\min\{3,k\}}s^{-1+ \nu_{j+3}} \mathbf H_{p_{\mathbf{high}},k-j}(s), \qquad v\in\{u, \psi\}.
\ee
This follows from the lightlike Hessian estimate \(| \del^2v|_{3,j} \lesssim C_{N,B_{s_1}, \coeff_0}A s^{-2 + \nu_{j+3}} \) and the identity between the coefficient norm and \(s\mathbf H\). The time-integrated estimate used at high order follows from the weighted variation and initial arguments in Section~\ref{section-weighted-Hessians}. For the effective field, \eqref{equa:seven-low-Hessian-bounds} instead gives
\be
\label{equa:seven-critical-remainder-effective-field-bound}
\mathbf R_{\fieldPhi;p_{\mathbf{high}},k}^{\mathrm{crit}}(s) \leq\frac{C_N}{\theta} \mathcal B_3^{\mathrm{low}}(s) \sum_{i =0}^{\min\{3,k\}}s^{-1/2 + \eta_i} \mathbf H_{p_{\mathbf{high}},k-i}(s).
\ee
\ese
In both estimates the coefficient identity is \(\|L^JH^{\H00} \|_2=s\|s^{-1}L^JH^{\H00} \|_2\). The weighted variation and initial estimates in Sections~\ref{section-weighted-Hessians}--\ref{section-e-bulk} give the quadratic forcing in the high-order energy inequality.

At the uniform lower level the critical remainder is empty.  At the fixed-$\coeff$ continuation level it is nonempty and is estimated by \eqref{equa:seven-high-H00-radial-derived}-- \eqref{equa:seven-critical-remainder-rank-bounds}.  This separation is what prevents a top-order existence estimate from being mistaken for a uniform convergence bound.

Using \eqref{equa:seven-refined-pointwise-system} in the low--high decompositions sharpens \eqref{equa:seven-structured-source-rank}, but the trace source must remain separate from the two effective-field nonlinearities. More explicitly, we obtain
\bse
\label{equa:seven-refined-genuine-sources}
\be
\label{equa:seven-refined-GW-source}
\| \,| \Gbb_\coeff|_{p,k} \|_{L^2} + \| \,|W_\coeff|_{p,k} \|_{L^2} \lesssim C\theta^{-1} \mathcal B_3^{\mathrm{low}}(s) s^{-3/2 + \eta_k}U_{p,k} + C\theta^{-1} \bigl(\mathcal B_3^{\mathrm{low}}(s) \bigr)^2 s^{-3/2 + \eta_k},
\ee
\be
\label{equa:seven-refined-S-source-before-low}
\| \,| \mathcal S|_{p,k} \|_{L^2} \lesssim C\theta^{-1} \mathcal B_3^{\mathrm{low}}(s) \sum_{j=0}^{\min\{k,p_{\mathbf{low}}-2\}} s^{-3/2 + \beta_{j+2}} \mathbf P_{p,k-j} + C\theta^{-1} \bigl(\mathcal B_3^{\mathrm{low}}(s) \bigr)^2 s^{-3/2 + \beta_{k+2}}.
\ee
\ese
Here and below an exponent whose index exceeds \(N\) is evaluated at \(N\). To prove \eqref{equa:seven-refined-S-source-before-low}, we set \(e_\psi = (s/t)| \del\psi|+ | \delsH\psi| \).  Formula \eqref{equa:seven-null-contraction} bounds every principal product by \((t/s)e_{\psi, \mathrm{low}}e_{\psi, \mathrm{high}} \), whereas \((t/s)t^{-3/2} =t^{-1/2}s^{-1} \leq s^{-3/2} \). The direct estimate \eqref{equa:seven-refined-matter}, with the two additional derivatives required by the Sobolev estimate, gives the exponent \(\beta_{j+2} \).  Since every \(\beta_{j+2}<1/2\) in the indicated range, this coefficient is integrable at the uniform lower level. At \(p=p_{\mathbf{high}} \), the same low--high decomposition is used in the normalized robust system of~\autoref{lem:seven-robust-high-triangular}. Once the uniform high-order estimate is established, rescaling in \eqref{equa:seven-fixed-high-rescaling} gives the unscaled high effective-field bound.

The characteristic estimate \eqref{equa:seven-characteristic-matter} also gives the sharper stress bound
\be
\label{equa:seven-massless-source-characteristic}
\| \,| \Mbb_\coeff|_{p,k} \|_{L^2} \lesssim CA s^{-1} \mathbf P_{p,k} + CA\sum_{j=1}^{\min\{k,p_{\mathbf{low}}-3\}} s^{-1+ \eta_j} \mathbf P_{p,k-j} + CA^2s^{-3/2 + \eta_k}.
\ee
At the lower level we also keep the fully internal estimate \eqref{equa:seven-massless-source-inhomogeneous}; it contributes the explicit growth \(A^2\widetilde\mu_k^{-1}s^{\widetilde\mu_k} \), which remains below the metric exponent \(\nu_k\).  Finally, \(g(\nabla{\fieldPhi}, \nabla \psi) \) is  absent from \eqref{equa:seven-refined-genuine-sources}: its commuted null contraction is estimated separately in the matter-energy argument below.


\section{Interior energy identities and low-order bounds}  
\label{section-10}


\subsection{Interior energy identities and dynamic reference terms}
\label{section-9-6}

\paragraph{Wave current and reference geometry}

We now derive the interior energy identities with their oriented interface fluxes, estimate the dynamic reference terms, and improve the lower energy bounds. The interface flux cancels its exterior counterpart only after the two identities have been added. The deformation tensor is defined by $\pi^{(X_\H)}_{\alpha \beta} {\coloneqq} \nabla_\alpha(X_\H)_\beta+ \nabla_\beta(X_\H)_\alpha$. For its semi-hyperboloidal components, we use
\be
\widehat\pi^{\mathcal H\, \alpha \beta} {\coloneqq}-\big(\mathcal L_{X_\H}g^{-1} \big)^{\mathcal H\, \alpha \beta}
\ee
We define the adapted deformation norm by
\be
\label{equa:seven-adapted-deformation-norm}
\Pi_\H^{\mathrm{ad}}(s) \coloneqq \Bigl\| {t\over s}| \widehat\pi^{\mathcal H\,00}| + \sum_a| \widehat\pi^{\mathcal H\,0a}| + {s\over t} \sum_{a,b}| \widehat\pi^{\mathcal H\,ab}| \Bigr\|_{L^\infty(\Mint_s)}.
\ee
The weights in \eqref{equa:seven-adapted-deformation-norm} cannot be omitted near the cone.  Coarea and \(2(s/t)| \del_tq|| \delsH q| \leq e_\H[q] \) give, for every one of the wave, matter, degenerate, and uniform currents,
\be
\label{equa:seven-adapted-deformation-pairing}
\left| {1\over2} \int_{\Mint_s} Q^{\alpha \beta}[q] \pi^{(X_\H)}_{\alpha \beta} {s\over t}j_g\, dx \right| \leq C\Pi_\H^{\mathrm{ad}}(s) \calE_Q^\H(s,q).
\ee
\paragraph{Localized Schwarzschild reference estimates}

The reference estimates follow from the localized Schwarzschild metric. We set

\bse
\be
\vartheta^\star(t,r) \coloneqq \chi^\star(r) \chi^\star \!\Bigl({r \over t-1} \Bigr), \quad h^\star=g^\star-\gMink = \vartheta^\star(g_{\Sch}-\gMink),
\ee
and assume
\be
0\leq m\leq m_0<{1\over4}, \quad m\leq C_{\mathrm{ref}} \epss\leq C_{\mathrm{ref}}A_0.
\ee
The first restriction keeps the explicit Schwarzschild coefficients regular on the support of the radial cut-off.  We define
\be
\label{equa:seven-reference-transition-set}
\mathcal A^\star {\coloneqq} \operatorname{supp} \bigl(\vartheta^\star(1-\vartheta^\star) \bigr) \cup\operatorname{supp}(d\vartheta^\star).
\ee
It is contained in
\be
\left\{{1\over2} \leq r \leq{3\over4}, \ {r \over t-1} \geq{1\over2} \right\} \cup \left\{r \geq{1\over2}, \ {1\over2} \leq {r \over t-1} \leq{3\over4} \right\}.
\ee
\ese
The first set lies in a fixed compact spacetime region.  On the second set, every truncated hyperboloid satisfies
\bse
\be
\label{equa:seven-reference-transition-geometry}
1+ t+ r \simeq1+s, \quad \operatorname{Vol}(\mathcal A^\star \cap\Mint_s) \leq C(1+s)^3.
\ee
Translations of \(r/(t-1) \) gain one power of \((1+ t+ r)^{-1} \), whereas Lorentz boosts preserve its size.  Direct differentiation of the Schwarzschild metric consequently gives, for \(|I|+ |J| \leq N+4\),
\be
\label{equa:seven-reference-pointwise-bounds}
| \del^IL^Jh^\star| \leq C_{N, \chi,m_0}m(1+ t+ r)^{-1-|I|}.
\ee
\ese

All constants here and below are independent of $\coeff$.

The reference localization residual \(\mathcal R^\star \) was defined in \eqref{equa:seven-reference-localization-residual}.  It vanishes where \(g^\star= \gMink\) and where \(g^\star=g_{\Sch} \), since the latter is Ricci flat and is written in wave coordinates.  Hence, we have \(\operatorname{supp} \mathcal R^\star \subset\mathcal A^\star \), and
\bse
\be
\label{equa:seven-reference-residual-pointwise}
| \del^IL^J\mathcal R^\star| \leq C_{N, \chi,m_0}m\, \mathbf 1_{\mathcal A^\star}(1+ t+ r)^{-3-|I|}, \quad |I|+ |J| \leq N.
\ee
Combining \eqref{equa:seven-reference-transition-geometry} and \eqref{equa:seven-reference-residual-pointwise} yields
\be
\label{equa:seven-reference-residual-L2}
\| \,| \mathcal R^\star|_{p,k} \|_{L^2(\Mint_s)} \leq C_Nm(1+s)^{-3/2}, \quad p\leq p_{\mathbf{high}}, \quad0\leq k\leq p.
\ee
We write \(\widehat\pi_\star^{\mathcal H\, \alpha \beta}{\coloneqq} -(\mathcal L_{X_\H}(g^\star)^{-1})^{\mathcal H\, \alpha \beta} \). The reference metric is stationary in the Minkowski and Schwarzschild regions. On the transition set, \eqref{equa:seven-reference-transition-geometry} gives
\be
\label{equa:seven-reference-deformation-direct}
\Pi_\star^{\mathrm{ad}}(s) \leq C_Nm(1+s)^{-2}.
\ee
\ese
Consequently, for the fixed \(\zeta_\star\) chosen above, we obtain the following bounds:
\bse
\label{equa:seven-reference-interior-bounds}
\be
\label{equa:seven-reference-adapted-deformation}
\Pi_\star^{\mathrm{ad}}(s) \leq C A_0s^{-1-\zeta_\star},
\ee
\be
\label{equa:seven-reference-source-bound}
\| \,| \mathcal R^\star|_{p,k} \|_{L^2(\Mint_s)} \leq C_N A_0s^{-1-\zeta_\star}, \quad p\leq p_{\mathbf{high}}, \quad 0\leq k\leq p.
\ee
\ese
These estimates for the explicit glued metric are established before the interior equations are solved; they are not interface information provided by the exterior analysis.  By contrast, \((\BoxChapeaud_{g^\star}-\BoxChapeaud_g)g^\star \) depends on \(u\) and remains in the dynamic source \(\mathcal N^\star[u] \).

\paragraph{Dynamic reference source and metric commutator}

For a tensor field, \(| \del v|_{\mathrm g;p,k} \) denotes the sum of the derivatives of the semi-hyperboloidal components containing at least one tangential index, and we define
\be
\mathbf e_{\H}[v] {\coloneqq}| \delsH v|+ \Big({s\over t} \Big)^2| \del v|, \quad \mathbf e_{\H}[v]_{p,k} {\coloneqq}| \delsH v|_{p,k} + \Big({s\over t} \Big)^2| \del v|_{p,k}.
\ee
The two quantities in the commuted metric equation are
\bse
\be
\label{equa:seven-dynamic-metric-source-norms}
\bigl(\mathcal N^\star_{p,k}(s) \bigr)^2\coloneqq \sum_{Z\preccurlyeq(p,k)} \|Z\mathcal N^\star[u] \|_{L^2(\Mint_s)}^2, \qquad \bigl(\mathcal C_{u;p,k}(s) \bigr)^2\coloneqq \sum_{Z\preccurlyeq(p,k)} \|[Z, \BoxChapeaud_g]u\|_{L^2(\Mint_s)}^2.
\ee
Since the metric wave energy has no zeroth-order term, the norm naturally associated with an undifferentiated metric factor is
\be
\label{equa:seven-metric-Hardy-energy}
\mathbf Q^u_{p,k}(s) \coloneqq \mathbf W_{p,k}(s) + \mathsf T_{2;p,k}[u](s).
\ee
Indeed, \eqref{equa:seven-Hardy-with-boundary} gives
\be
\label{equa:seven-metric-Hardy-energy-control}
\|t^{-1}|u|_{p,k} \|_{L^2(\Mint_s)} \lesssim\mathbf Q^u_{p,k}(s).
\ee
\ese

The coefficient profile of the explicit reference metric is
\bse
\be
\label{equa:seven-reference-coefficient-profile}
\mathbf q_N^\star(s) \coloneqq \max_{Z\preccurlyeq(N+2,N+2)} \Bigl\| {t\over s}| \del Zh^\star| + t| \del^2Zh^\star| + t^{-1}|Zh^\star| \Bigr\|_{L^\infty(\Mint_s)}.
\ee
Estimate \eqref{equa:seven-reference-pointwise-bounds} and the stationarity of the Schwarzschild zone imply
\be
\label{equa:seven-reference-coefficient-decay}
\mathbf q_N^\star(s) \leq C_NA_0s^{-2}.
\ee
We set
\be
\label{equa:seven-metric-integrable-kernel}
\kappa_u^\star(s)\coloneqq As^{-3/2 + \delta} +A^2s^{-3/2 +2\delta} + \mathbf q_N^\star(s).
\ee
\be
\label{equa:seven-metric-integrable-kernel-L1}
\int_{s_0}^{\infty} \kappa_u^\star(\tau) \, d\tau\leq C_N(A+A^2 +A_0) \leq C_N(A+A_0).
\ee
\ese

\paragraph{Dynamic source and metric commutator estimate}

\begin{proposition}[Dynamic metric source and metric commutator]
\label{lem:seven-dynamic-metric-source-commutator}
Let \(p\in\{p_{\mathbf{low}},p_{\mathbf{high}} \} \) and \(0\leq k\leq p\). Under the metric bootstrap, the bad-coefficient estimate \eqref{equa:seven-preliminary-H00}, and the good-component estimate \eqref{equa:seven-refined-metric}, one has
\be
\label{equa:seven-dynamic-metric-source-commutator-L2}
\begin{aligned}
\mathcal N^\star_{p,k}(s) + \mathcal C_{u;p,k}(s) & \lesssim C_N\mathcal B_3^{\mathrm{low}}(s)s^{-1} \mathbf W_{p,k}(s) + \frac{C_N}{\theta} \mathcal B_3^{\mathrm{low}}(s) \sum_{j=1}^{k}s^{-1+ \eta_j} \mathbf W_{p,k-j}(s) \\
& \quad +C_N\mathbf R_{u;p,k}^{\mathrm{crit}}(s) +C_N\kappa_u^\star(s) \mathbf Q^u_{p,k}(s) +C_NAA_0s^{-2 + \delta}.
\end{aligned}
\ee
The critical remainder is absent at \(p=p_{\mathbf{low}} \), and every constant is independent of \(0<\coeff\leq\coeff_0\).
\end{proposition}

\begin{proof}

We first estimate the differentiated coefficients, then the interactions with the reference metric, and finally the principal commutator. This separates coefficients controlled pointwise from the high boost-only coefficient that remains in the radial norm.

\medskip\noindent\emph{Low-order coefficient bounds.} The sharp coefficient and good-component bounds, together with the direct estimate for \(\mathbf e_\H[u] \), control the profile
\bse
\be
\label{equa:seven-low-metric-coefficient-definition}
\mathbf c_j^u(s)\coloneqq \left\| \frac ts\Bigl(|H^{\H00}|_{p_{\mathbf{low}}-2,j} + | \del u|_{\mathrm g;p_{\mathbf{low}}-3,j} + \mathbf e_\H[u]_{p_{\mathbf{low}}-2,j} \Bigr) \right\|_{L^\infty(\Mint_s)}.
\ee
The lower factor in a product of total order \(p\leq p_{\mathbf{high}} \) has order at most \(\lfloor p/2\rfloor\leq p_{\mathbf{low}}-3\). Thus this good-component range covers every low--high product in the metric source. The bounds in \eqref{equa:seven-low-metric-coefficient} hold for this profile. Consequently, we have
\be
\label{equa:seven-low-metric-coefficient}
\mathbf c_j^u(s) \leq \begin{cases}
C_N\mathcal B_3^{\mathrm{low}}(s)s^{-1},&j=0, \\
C_N\theta^{-1} \mathcal B_3^{\mathrm{low}}(s) s^{-1+ \eta_j},&1\leq j\leq p_{\mathbf{low}}-2.
\end{cases}
\ee
The physical-time and rank weights give the conversion
\be
\label{equa:seven-metric-coefficient-rank-conversion}
\frac ts\,t^{-1}s^{\eta_j} =s^{-1+ \eta_j}.
\ee
Thus the characteristic and ray estimates use the same hyperboloidal rank growth. Direct hyperboloidal Sobolev, applied after \eqref{equa:seven-Hardy-with-boundary}, also gives
\be
\label{equa:seven-low-undifferentiated-metric-coefficient}
\Bigl\|{t\over s}t^{-1}|u|_{p_{\mathbf{low}}-2,j} \Bigr\|_{L^\infty(\Mint_s)} \leq C_NAs^{-3/2 + \delta}.
\ee
\ese
Its interface trace term is \(\mathsf T_{2;p_{\mathbf{low}},j+2}[u] \), not an energy flux.

\medskip\noindent\emph{Dynamic reference interactions.} Derivatives of the smooth metric coefficients also produce cubic terms. We collect their absolute values in
\bse
\be
\label{equa:seven-dynamic-cubic-products}
\mathcal C_Z^u\coloneqq\sum_{\substack{3\leq q\leq\ord Z+3,\ Z_0\cdots Z_{q-1}\preccurlyeq Z}}|Z_0u|\,|\del Z_1u|\,|\del Z_2u|\prod_{j=3}^{q-1}|Z_ju|.
\ee
Only finitely many products, with bounded smooth coefficients, occur at each differentiation order. Polarization of the reduced Einstein source and the commuted wave-gauge identity give
\be
\label{equa:seven-dynamic-F-structure}
\begin{aligned}
|Z\{\Fbb(g,g;\del g, \del g) -\Fbb(g^\star,g^\star;\del g^\star, \del g^\star) \}| & \lesssim \sum_{Z_1Z_2\preccurlyeq Z} | \del Z_1u| \, \mathbf e_{\H}[Z_2u] + \sum_{Z_1Z_2\preccurlyeq Z} | \del Z_1u|_{\mathrm g}| \del Z_2u|_{\mathrm g} \\
& \quad + t^{-1} \sum_{Z_1Z_2\preccurlyeq Z} |Z_1u| \,| \del Z_2u| +\mathcal C_Z^u + \mathcal I_Z^\star[u].
\end{aligned}
\ee
The cubic sum includes derivatives of the inverse-metric coefficients and the nonlinear part of the commuted gauge relation. An undifferentiated reference factor satisfies \(|h^\star|\lesssim m/t\lesssim m(s/t)^2\) in the interior and is included in the first sum. If the factor with the largest number of commutations is differentiated, use its \((s/t)\)-weighted energy; otherwise use its Hardy norm, keeping the sphere trace. The direct bounds on the other factors give
\be
\label{equa:seven-dynamic-cubic-L2}
\sum_{Z\preccurlyeq(p,k)}\|\mathcal C_Z^u\|_2\lesssim C_NA^2s^{-2+2\delta}\mathbf Q^u_{p,k},\qquad \frac ts|u_{\mathrm{low}}|\,|\del u_{\mathrm{low}}|+t|\del u_{\mathrm{low}}|^2\lesssim A^2s^{-2+2\delta}.
\ee
Additional undifferentiated low-order factors are bounded and small. Since \(\delta<1/8\), the coefficient is integrable and belongs to \(C_N\kappa_u^\star\mathbf Q^u_{p,k}\). The terms where a derivative falls on the reference metric, together with \(Z((\BoxChapeaud_{g^\star}-\BoxChapeaud_g)g^\star) \), satisfy
\be
\label{equa:seven-dynamic-reference-interaction-L2}
\sum_{Z\preccurlyeq(p,k)} \| \mathcal I_Z^\star[u] \|_{L^2(\Mint_s)}^2 \leq C_N\bigl(\mathbf q_N^\star(s) \bigr)^2 \bigl(\mathbf Q^u_{p,k}(s) \bigr)^2.
\ee
Indeed, the two representative products satisfy
\be
\label{equa:seven-reference-interaction-products}
\| \del h^\star \del u\|_2 \leq \|(t/s) \del h^\star \|_\infty \|(s/t) \del u\|_2, \qquad \|u\del^2h^\star \|_2 \leq \|t\del^2h^\star \|_\infty\|t^{-1}u\|_2.
\ee
\ese

In the product \(| \del u| \mathbf e_\H[u] \), both placements are controlled directly: the low first derivative has norm at most \(C_NAs^{-3/2 + \delta} \), and the low good factor satisfies \(\|(t/s) \mathbf e_\H[u] \|_\infty\leq C_NAs^{-3/2 + \delta} \). The high factors use the good energy and the weighted Cartesian derivative, respectively. The term \(t^{-1}|u|| \del u| \) is estimated in the same manner, keeping the Hardy trace for a high undifferentiated factor. Only the product \(| \del u|_{\mathrm g}| \del u|_{\mathrm g} \) uses the refined good-component estimate. Differentiated frame coefficients contribute \(t^{-1}u\); their low factor is bounded directly and their high factor by Hardy. They belong to the integrable term \(C_N\kappa_u^\star\mathbf Q^u\).

\medskip\noindent\emph{Principal commutator terms.} For the commutator, we use \eqref{equa:seven-effective-field-commutator-decomposition} with \({\fieldPhi} \) replaced by \(u\).  Low coefficients are handled by \eqref{equa:seven-low-metric-coefficient}. For a high coefficient containing a translation, we first separate its bad semi-hyperboloidal contraction. After reordering translations, we denote the resulting tensor coefficient by \(P\) and set \(v=Z_2u\), with \(\ord Z_2\leq3\). The commuted gauge and Hardy inequality give
\bse
\be
\label{equa:seven-translated-high-bad-coefficient-L2}
\|P^{\H00} \|_2\lesssim C_N(1+As^\delta) \mathbf Q^u_{p,k} +C_NA_0s^{-1/2}.
\ee
The nonlinear placements have coefficients \((t/s)|H_{\mathrm{low}}| \) and \(t| \del H_{\mathrm{low}}| \), both bounded by \(C_NAs^\delta \) since \(t\lesssim s^2\). The undifferentiated high factor uses the Hardy norm with its trace. Translation derivatives of the frame contribute \(t^{-1} \) and obey the same bound. Direct Sobolev also gives
\be
\label{equa:seven-translated-high-direct-low-Hessians}
\| \del^2v\|_\infty\lesssim A s^{-3/2 + \delta}, \qquad \left\| \frac ts G_2[v] \right\|_\infty\lesssim A s^{-5/2 + \delta}.
\ee
Here \(G_2[v] \) denotes the good Hessians and their frame terms. They are linear combinations of \(t^{-1} \del L_av\), \(t^{-1} \del v\), and \(t^{-2}L_av\). These bounds use at most seven energy orders. We estimate \(P^{\H00} \del_t^2v\) with the unweighted gauge norm, and the remaining contractions with \(\|(s/t)P\|_2\). This gives
\be
\label{equa:seven-high-translation-coefficient}
\|P^{\alpha \beta} \del_{\alpha \beta}v\|_2 \lesssim C_N\bigl(As^{-3/2 + \delta} +A^2s^{-3/2 +2\delta} \bigr) \mathbf Q^u_{p,k} +C_NAA_0s^{-2 + \delta}.
\ee
The good-Hessian contribution has the stronger coefficient \(As^{-5/2 + \delta} \). All these powers are integrable. The cumulative rank \(k\) contains the lower ranks in the differentiated products. The independent reference forcing satisfies
\be
\label{equa:seven-translated-high-reference-forcing}
\int_{s_0}^{s_1} AA_0\tau^{-2 + \delta} \, d\tau\lesssim AA_0\leq A^2.
\ee
\ese
It therefore contributes to the quadratic forcing in the lower and high energy improvements. A high boost-only coefficient remains in \(\mathbf R_{u;p,k}^{\mathrm{crit}} \). The other frame terms are controlled by Hardy, with their traces at the interface, and contribute to \(\kappa_u^\star\mathbf Q^u\). This proves the result.

\end{proof}

The adapted deformation is controlled directly by the rank-zero wave gauge. Since \(X_\H= \del_t\), contraction with \(\ell_\H=dt-(x^a/t)dx^a \) gives
\be
\label{equa:seven-deformation-gauge-component}
\widehat\pi^{\H00} =-\del_t(H^{\H00}) +\Ocal(t^{-1}|H|).
\ee
The gauge estimate controls its weighted bad component, while the mixed and tangential components use the direct full-gradient estimate. Consequently, we obtain
\be
\label{equa:seven-deformation-bound}
\Pi_\H^{\mathrm{ad}}(s) \leq C_N\bigl(As^{-3/2 + \delta} +A^2s^{-2 +2\delta} +A_0s^{-2} \bigr) + \Pi_\star^{\mathrm{ad}}(s).
\ee
Every term is integrable. The explicit reference contribution is kept separately.


\subsection{Interior energy balance laws}
\label{section-9-7}

\paragraph{Wave energy identity}

We consider a commuted wave field \(v\) satisfying \(\Box_gv=F_v\).  The energy identity on the interior slab has the form
\bse
\be
\label{equa:seven-exact-wave-balance}
\calE_{\mathrm w}^{\H}(s,v) -\mathsf F_{\mathrm w}^{\Lscr}(s,v;s_0) = \calE_{\mathrm w}^{\H}(s_0,v) -\int_{\Mint_{[s_0,s]}}F_vX_\H v\, d\mu_g -{1\over2} \int_{\Mint_{[s_0,s]}} Q^{\mathrm{wave} \, \alpha \beta}[v] \pi^{(X_\H)}_{\alpha \beta} \, d\mu_g.
\ee
Consequently, we obtain
\be
\label{equa:seven-wave-energy-identity}
\calE_{\mathrm w}^{\H}(s,v)^{1/2} \leq \calE_{\mathrm w}^{\H}(s_0,v)^{1/2} + \sup_{s_0\leq\sigma \leq s} \big| \mathsf F_{\mathrm w}^{\Lscr}(\sigma,v;s_0) \big|^{1/2} +C\int_{s_0}^{s} \Bigl( \|F_v(\tau) \|_{L^2(\Mint_\tau)} + \Pi_\H^{\mathrm{ad}}(\tau) \calE_{\mathrm w}^{\H}(\tau,v)^{1/2} \Bigr) \, d\tau.
\ee
\ese
We observe that the boundary flux has no prescribed sign. The supremum in \eqref{equa:seven-wave-energy-identity} is controlled by the interface amplitudes used below. 

\paragraph{Weighted matter balance and exchange cancellation}

For the massless field, the top-order identity comes from \eqref{equa:seven-matter-divergence-form}.  For every scalar \(v\), we have the following identity:
\bse
\be
\label{equa:seven-matter-current-divergence}
\begin{aligned}
\nabla^\alpha \Bigl( e^{-{\fieldPhi}}Q^{\mathrm{wave}}_{\alpha \beta}[v]X_\H^\beta \Bigr) &= e^{-{\fieldPhi}}(\mathscr L_{\fieldPhi} v)X_\H v + {1\over2}e^{-{\fieldPhi}} Q^{\mathrm{wave} \, \alpha \beta}[v] \pi^{(X_\H)}_{\alpha \beta} \\
& \quad + {1\over2}e^{-{\fieldPhi}}(X_\H{\fieldPhi}) g(\nabla v, \nabla v).
\end{aligned}
\ee
Thus \(v=Z\psi\) satisfies the balance
\be
\label{equa:seven-exact-matter-balance}
\begin{aligned}
\calE_\psi^{\H}(s,Z\psi) -\mathsf F_\psi^{\Lscr}(s,Z\psi;s_0) &= \calE_\psi^{\H}(s_0,Z\psi) -\int_{\Mint_{[s_0,s]}} e^{-{\fieldPhi}} \mathcal K_Z^\psi X_\H Z\psi\, d\mu_g \\
& \quad -{1\over2} \int_{\Mint_{[s_0,s]}} e^{-{\fieldPhi}}Q^{\mathrm{wave} \, \alpha \beta}[Z\psi] \pi^{(X_\H)}_{\alpha \beta} \, d\mu_g \\
& \quad -{1\over2} \int_{\Mint_{[s_0,s]}} e^{-{\fieldPhi}}(X_\H{\fieldPhi}) g(\nabla Z\psi, \nabla Z\psi) \, d\mu_g.
\end{aligned}
\ee
\ese
At order zero, the effective-field and matter balance laws have an exact exchange cancellation.  Indeed, by \eqref{equa:seven-S-exact}, we have
\bse
\be
\label{equa:seven-zero-order-exchange-cancellation}
\int_{\Mint_{[s_0,s]}} \mathcal S X_\H{\fieldPhi} \, d\mu_g +8\pi\int_{\Mint_{[s_0,s]}}e^{-{\fieldPhi}}(X_\H{\fieldPhi}) g(\nabla \psi, \nabla \psi) \, d\mu_g =0.
\ee
Thus the sum of the uniform effective-field balance and \(16\pi\) times the matter balance contains no zeroth-order exchange term. At positive order, the contribution of the two top polarizations to this sum has the signed form
\be
\label{equa:seven-top-exchange-good-derivative}
\begin{aligned}
&{16\pi e^{-{\fieldPhi}}g^{\alpha \beta}\del_\beta\psi\bigl(\del_\alpha Z\psi\,\del_tZ\fieldPhi-\del_\alpha Z\fieldPhi\,\del_tZ\psi\bigr)}\\
&=16\pi e^{-{\fieldPhi}}g^{a\beta}\del_\beta\psi\bigl(\delsH_aZ\psi\,\del_tZ\fieldPhi-\delsH_aZ\fieldPhi\,\del_tZ\psi\bigr).
\end{aligned}
\ee
\ese
The \(\alpha=0\) contribution and the two bad--bad frame corrections cancel. This is an antisymmetric null contraction, which we use at its normalization.  At the lower energy level, we estimate the commutator directly with the full Sobolev norm \eqref{equa:seven-Dpsi-definition}; at the high level, the pairing is estimated by the two-region bounds of~\autoref{lem:seven-massless-two-region-gain}.

\paragraph{Commuted matter source}

With the convention \([A,B]=AB-BA\), the commuted equation is \(\mathscr L_{\fieldPhi} Z\psi = [\mathscr L_{\fieldPhi},Z] \psi\).  The uncommuted term \(g(\nabla{\fieldPhi}, \nabla Z\psi) \) stays on the left.  Every remaining derivative coupling is therefore a sum of terms

\bse
\be
(Z_0g^{-1}) (\nabla Z_1{\fieldPhi}, \nabla Z_2\psi), \quad Z_0Z_1Z_2\preccurlyeq Z, \quad \ord Z_0 + \ord Z_1\geq1,
\ee
together with the standard metric wave commutator and lower-order frame terms.  If \(Z_2\psi\) is the low factor, then \eqref{equa:seven-null-contraction} and \((t/s)t^{-3/2} =t^{-1/2}s^{-1} \leq s^{-3/2} \) give
\be
\label{equa:seven-matter-null-L2}
\|g(\nabla Z_1{\fieldPhi}, \nabla Z_2\psi) \|_{L^2(\Mint_s)} \lesssim s^{-3/2} \mathcal D_{\psi,j}(s)U_{p,k-j}(s).
\ee
If \(Z_1{\fieldPhi} \) is the low factor, the same decomposition puts its good or weighted bad derivative in \(L^\infty\) and the derivative of \(Z_2\psi\) in the matter energy.  For later use, we define the squared commuted-source norm by
\be
\bigl(\mathcal K_{p,k}^\psi(s) \bigr)^2 \coloneqq \sum_{Z\preccurlyeq(p,k)} \| \mathcal K_Z^\psi(s) \|_{L^2(\Mint_s)}^2.
\ee
Combining the two alternatives with the metric commutator estimate yields
\be
\label{equa:seven-matter-commutator-L2}
\begin{aligned}
\mathcal K_{p,k}^\psi(s) & \lesssim CA s^{-1} \mathbf P_{p,k}(s) + CA\sum_{j=1}^{k}s^{-1+ C_{\mathrm{rk}}j\theta} \mathbf P_{p,k-j}(s) \\
& \quad + C\sum_{j=0}^{\min\{k,p_{\mathbf{low}}-2\}} s^{-3/2} \mathcal D_{\psi,j}(s) \bigl(U_{p,k-j}(s) + \mathbf W_{p,k-j}(s) \bigr) + CA^2s^{-3/2 + \widetilde\mu_k} + C\mathbf R_{\psi;p,k}^{\mathrm{crit}}(s).
\end{aligned}
\ee
\ese
No derivative is lost: the largest new drift factor is \(\nabla Z_1{\fieldPhi} \), which is controlled by \(U_{p,k-j} \), while the high-metric--low-Hessian alternative is expressed through \(\mathbf W\) and \(\mathbf R_{\psi;p,k}^{\mathrm{crit}} \), rather than being suppressed.

\paragraph{Matter bulk term and triangular energy inequality}

The last bulk term in \eqref{equa:seven-exact-matter-balance} is integrable. Indeed, co-area and the semi-hyperboloidal decomposition give
\bse
\be
\label{equa:seven-matter-weight-bulk}
\left| {1\over2} \int_{\Mint_s}e^{-{\fieldPhi}}(X_\H{\fieldPhi}) g(\nabla v, \nabla v){s\over t}j_g\, dx \right| \lesssim \| \del{\fieldPhi} \|_{L^\infty(\Mint_s)} \calE_\psi^\H(s,v),
\ee
and \eqref{equa:seven-refined-effective-field} implies \(\| \del{\fieldPhi} \|_{L^\infty} \lesssim CA s^{-3/2 + \chi_0} \).  Hence the balance, Cauchy--Schwarz, and \eqref{equa:seven-matter-commutator-L2} give
\be
\label{equa:seven-matter-energy-triangular-preliminary}
\begin{aligned}
\mathbf P_{p,k}(s) & \leq \mathbf P_{p,k}(s_0) + \sup_{s_0\leq\sigma \leq s}| \mathsf F_\psi^{\Lscr,p,k}(\sigma;s_0)|^{1/2} + C\int_{s_0}^{s} \Pi_\H^{\mathrm{ad}}(\tau) \mathbf P_{p,k}(\tau) \, d\tau \\
& \quad + CA\int_{s_0}^{s} \tau^{-1} \mathbf P_{p,k}(\tau) \, d\tau + CA\sum_{j=1}^{k} \int_{s_0}^{s} \tau^{-1+ C_{\mathrm{rk}}j\theta} \mathbf P_{p,k-j}(\tau) \, d\tau \\
& \quad + C\sum_{j=0}^{\min\{k,p_{\mathbf{low}}-2\}} \int_{s_0}^{s} \tau^{-3/2} \mathcal D_{\psi,j}(\tau) \bigl(U_{p,k-j}(\tau) + \mathbf W_{p,k-j}(\tau) \bigr) \, d\tau \\
& \quad + C\int_{s_0}^{s} \mathbf R_{\psi;p,k}^{\mathrm{crit}}(\tau) \, d\tau + CA^2.
\end{aligned}
\ee
\ese
No term on the right of \eqref{equa:seven-matter-energy-triangular-preliminary} contains a negative power of $\coeff$.  The only lateral term is the weighted matter flux already present in \eqref{equa:seven-exact-matter-balance}; the direct commutator estimate creates no additional cone correction. Inequality \eqref{equa:seven-matter-energy-triangular-preliminary} applies at \(p=p_{\mathbf{low}} \). At \(p=p_{\mathbf{high}} \), the two-region matter estimate enters the coupled energy inequality of~\autoref{lem:seven-robust-high-triangular}.

\paragraph{Singular and robust effective-field balance laws}

For the trace equation, the divergence theorem gives the following two balance laws:
\bse
\label{equa:seven-exact-effective-field-balances}
\be
\label{equa:seven-exact-effective-field-balance-degenerate}
\begin{aligned}
D_{p,k}(s)^2 -\mathsf F_{\mathrm{eff}}^{\mathrm{sing}, \Lscr,p,k}(s;s_0) &= D_{p,k}(s_0)^2 -\sum_{Z\preccurlyeq(p,k)} \int_{\Mint_{[s_0,s]}} \mathcal N_Z^{\mathrm{eff}}X_\H(Z\fieldphi) \, d\mu_g \\
& \quad -{1\over2} \sum_{Z\preccurlyeq(p,k)} \int_{\Mint_{[s_0,s]}} Q^{\mathrm{sing} \, \alpha \beta}[Z\fieldphi] \pi^{(X_\H)}_{\alpha \beta} \, d\mu_g,
\end{aligned}
\ee
\be
\label{equa:seven-exact-effective-field-balance-uniform}
\begin{aligned}
U_{p,k}(s)^2 -\mathsf F_{\mathrm{eff}}^{\mathrm{unif}, \Lscr,p,k}(s;s_0) &= U_{p,k}(s_0)^2 -\sum_{Z\preccurlyeq(p,k)} \int_{\Mint_{[s_0,s]}} \mathcal N_Z^{\mathrm{eff}}X_\H(Z{\fieldPhi}) \, d\mu_g \\
& \quad -{1\over2} \sum_{Z\preccurlyeq(p,k)} \int_{\Mint_{[s_0,s]}} Q^{\mathrm{unif} \, \alpha \beta}[Z\fieldphi] \pi^{(X_\H)}_{\alpha \beta} \, d\mu_g.
\end{aligned}
\ee
\ese
Here the second equality is exactly $\coeff$ times the first one; in particular, the two fluxes are not independent.

\paragraph{Effective-field source pairings and normalized inequalities}

In the coordinates \((s,x) \), the co-area formula is
\be
\label{equa:seven-interior-co-area}
d\mu_g=j_g{s\over t} \, ds\, dx, \quad j_g\coloneqq| \det g|^{1/2} \simeq1.
\ee
Consequently, every source component \(Z\mathcal R\) satisfies the following inequalities:
\be
\label{equa:seven-exact-source-pairings}
\begin{aligned}
\left| \int_{\Mint_s}j_gZ\mathcal R\,{s\over t} \del_tZ\fieldphi\, dx \right| 
& \lesssim \coeff^{-1/2} \|Z\mathcal R\|_{L^2(\Mint_s)}D_{p,k}(s), \\
\left| \int_{\Mint_s}j_gZ\mathcal R\,{s\over t} \del_tZ{\fieldPhi} \, dx \right| & \lesssim \|Z\mathcal R\|_{L^2(\Mint_s)}U_{p,k}(s).
\end{aligned}
\ee
The two square-root inequalities obtained from \eqref{equa:seven-exact-effective-field-balances} are therefore
\bse
\label{equa:seven-effective-field-energy-normalizations}
\be
\label{equa:seven-D-energy-inequality}
\begin{aligned}
D_{p,k}(s)& \leq D_{p,k}(s_0) + \sup_{s_0\leq\sigma \leq s} \big| \mathsf F_{\mathrm{eff}}^{\mathrm{sing}, \Lscr,p,k}(\sigma;s_0) \big|^{1/2} \\
& \quad +C\coeff^{-1/2} \int_{s_0}^{s} \sum_{Z\preccurlyeq(p,k)} \| \mathcal N_Z^{\mathrm{eff}}(\tau) \|_{L^2(\Mint_\tau)} \, d\tau +C\int_{s_0}^{s} \Pi_\H^{\mathrm{ad}}(\tau)D_{p,k}(\tau) \, d\tau,
\end{aligned}
\ee
\be
\label{equa:seven-U-energy-inequality}
\begin{aligned}
U_{p,k}(s)& \leq U_{p,k}(s_0) + \sup_{s_0\leq\sigma \leq s} \big| \mathsf F_{\mathrm{eff}}^{\mathrm{unif}, \Lscr,p,k}(\sigma;s_0) \big|^{1/2} 
\\
& \quad +C\int_{s_0}^{s} \sum_{Z\preccurlyeq(p,k)} \| \mathcal N_Z^{\mathrm{eff}}(\tau) \|_{L^2(\Mint_\tau)} \, d\tau +C\int_{s_0}^{s} \Pi_\H^{\mathrm{ad}}(\tau)U_{p,k}(\tau) \, d\tau.
\end{aligned}
\ee
\ese
The factor \(\coeff^{-1/2} \) in \eqref{equa:seven-D-energy-inequality} is unavoidable.  In particular, the trace source \(\mathcal S\) does not cancel it.  Multiplying the energy identity by $\coeff$, or equivalently testing the equation with \(X_\H Z{\fieldPhi} \), gives \eqref{equa:seven-U-energy-inequality} without a negative power.
Taking \(\mathcal R= \mathcal S\) in \eqref{equa:seven-exact-source-pairings} gives the two trace estimates used below.  This is the distinction between the fixed-$\coeff$ existence hierarchy and the robust hierarchy.


\subsection{\texorpdfstring{Rank-weighted energy inequalities}{Rank-weighted energy inequalities}}
\label{section-10-1}

\paragraph{Normalized energies and interface terms}

We now normalize the energies and estimate the lower effective-field, matter, and metric energies successively. The high-order argument additionally requires the sufficient massless bound in Section~\ref{section-10-3}. Exact rescaling then recovers the fixed-parameter norm.
First, recalling \(\widehat U_{p,k} = \omega_p^{-1}U_{p,k} \), for \(p\in\{p_{\mathbf{high}},p_{\mathbf{low}} \} \), the remaining normalizations are
\bse
\be
\label{equa:seven-normalized-energies}
X_{p,k}(s) \coloneqq \widehat W_{p,k}(s) + \widehat P_{p,k}(s) + \widehat U_{p,k}(s), \qquad Y_{p,k}^{\mathrm{sing}}(s)\coloneqq \omega_p(s)^{-1}D_{p,k}(s).
\ee
Only the high-order effective-field norm is divided by \(s^{1/2} \); the metric and matter norms keep their original normalization.

\paragraph{Initial values and interface contributions}

We keep the initial and interface contributions visible.  For \(s_0\leq\tau\leq s\), we set
\be
\label{equa:seven-energy-inputs}
\begin{aligned}
\mathcal I^U_{p,k}(s) &\coloneqq s_0^{-a_{p,k}} \omega_p(s_0)^{-1}U_{p,k}(s_0) + \sup_{s_0\leq\tau\leq s} \tau^{-a_{p,k}} \omega_p(\tau)^{-1} | \mathsf F_{\mathrm{eff}}^{\mathrm{unif}, \Lscr,p,k} (\tau;s_0)|^{1/2}, \\
\mathcal I^P_{p,k}(s) &\coloneqq s_0^{-b_{p,k}} \mathbf P_{p,k}(s_0) + \sup_{s_0\leq\tau\leq s} \tau^{-b_{p,k}} | \mathsf F_{\psi}^{\Lscr,p,k} (\tau;s_0)|^{1/2}, \\
\mathcal I^{P, \mathrm l}_{k}(s) &\coloneqq s_0^{-\chi_k} \mathbf P_{p_{\mathbf{low}},k}(s_0) + \sup_{s_0\leq\tau\leq s} \tau^{-\chi_k}| \mathsf F_{\psi}^{\Lscr,p_{\mathbf{low}},k} (\tau;s_0)|^{1/2}, \\
\mathcal I^W_{p,k}(s) &\coloneqq s_0^{-\nu_k} \mathbf W_{p,k}(s_0) + \sup_{s_0\leq\tau\leq s} \tau^{-\nu_k} | \mathsf F_{\mathrm w}^{\Lscr,p,k} (\tau;s_0)|^{1/2}, \\
\mathcal I^{\mathrm{sing}}_{p,k}(s) &\coloneqq s_0^{-\delta} \omega_p(s_0)^{-1}D_{p,k}(s_0) + \sup_{s_0\leq\tau\leq s} \tau^{-\delta} \omega_p(\tau)^{-1} | \mathsf F_{\mathrm{eff}}^{\mathrm{sing}, \Lscr,p,k} (\tau;s_0)|^{1/2}.
\end{aligned}
\ee
These functions are nondecreasing and contain only the actual oriented fluxes.  The reference source is bounded directly by \eqref{equa:seven-reference-source-bound}. At \(p=p_{\mathbf{low}} \), the critical remainders are empty by their defining order restriction. Moreover, the two effective-field interface terms are not independent.  From \eqref{equa:seven-DU-definition}, \eqref{equa:seven-flux-rescaling}, \(s_0\geq1\), and \(a_{p,k}<\delta \), we have the comparison
\be
\label{equa:seven-input-rescaling}
\mathcal I^{\mathrm{sing}}_{p,k}(s) \leq \coeff^{-1/2} \mathcal I^U_{p,k}(s).
\ee
\ese

\paragraph{Hardy interface trace}

For the metric interface trace in the Hardy estimate, we set
\bse
\be
\label{equa:seven-normalized-metric-trace-input}
\mathcal T^u_{p,k}(s) \coloneqq \max_{0\leq q\leq k} \ \sup_{s_0\leq\tau\leq s} \tau^{-\eta_{\Lscr}} \mathsf T_{2;p,q}[u](\tau).
\ee
By \eqref{equa:seven-exponents}, \(\delta+ \eta_{\Lscr}<1/2\).  We therefore have
\be
\label{equa:seven-integrated-metric-Hardy-kernel}
\int_{s_0}^{s} \kappa_u^\star(\tau) \mathbf Q^u_{p,k}(\tau) \, d\tau \leq \int_{s_0}^{s} \kappa_u^\star(\tau) \mathbf W_{p,k}(\tau) \, d\tau +C_N(A+A_0) \mathcal T^u_{p,k}(s).
\ee
Indeed, we have \(\mathsf T_{2;p,k}[u](\tau) \leq\mathcal T^u_{p,k}(s) \tau^{\eta_{\Lscr}} \), while the trace kernel satisfies
\be
\label{equa:seven-integrated-metric-trace-kernel}
\sup_{s\geq s_0} \int_{s_0}^{s} \kappa_u^\star(\tau) \tau^{\eta_{\Lscr}} \, d\tau \leq C_N(A+A_0).
\ee
\ese
Thus the Hardy interface trace is an explicit interface quantity multiplied by an integrable coefficient; it is neither discarded nor identified with a lateral energy flux.

\paragraph{High-order interface amplitude}

The auxiliary robust high-order quantity contains the two metric traces in addition to the three normalized high energies and fluxes. To keep its definition readable, we first collect the integrated bulk and Hardy traces in \(\mathcal I^{\mathrm{tr}}_{p,k} \).  We define
\be
\label{equa:seven-robust-high-interface-input}
\begin{aligned}
\mathcal I^{\mathrm{tr}}_{p,k}(\sigma)&\coloneqq \left\| \tau^{-1/2} \mathsf B_{2;p,k}^{H}(\tau) \right\|_{L^2((s_0, \sigma);d\tau)} + \int_{s_0}^{\sigma} \kappa_u^\star(\tau) \mathsf T_{2;p,k}[u](\tau) \, d\tau+ \frac1\sigma \int_{s_0}^{\sigma} \mathsf T_{2;p,k}[u](\tau) \, d\tau, \\
\mathcal I_{\mathrm h,k}^{\mathrm{rob}}(s)&\coloneqq \frac{A_0}{\zeta_\star} + \max_{0\leq q\leq k} \Bigl\{\mathcal I^U_{p_{\mathbf{high}},q}(s) + \mathcal I^P_{p_{\mathbf{high}},q}(s) + \mathcal I^W_{p_{\mathbf{high}},q}(s) +s_0^{-\nu_q} \mathbf H_{p_{\mathbf{high}},q}(s_0) \\
& \quad + \sup_{s_0\leq\sigma \leq s} \sigma^{-\nu_q} \mathcal I^{\mathrm{tr}}_{p_{\mathbf{high}},q}(\sigma) \Bigr\}.
\end{aligned}
\ee
The last term of \(\mathcal I^{\mathrm{tr}} \) is the normalized Hardy-trace average used to recover the radial metric profile. The term \(A_0/\zeta_\star\) is the integrated prescribed reference source. The other terms are initial values, oriented interface fluxes, and sphere traces. The matter energy and its flux keep their original scale: their propagation uses the massless divergence form \eqref{equa:seven-matter-divergence-form}.  Only the uniform effective-field energy and flux carry the factor \(\omega_{p_{\mathbf{high}}}^{-1} =s^{-1/2} \).

The fixed-parameter interface quantity used in the argument is
\be
\label{equa:seven-fixed-high-interface-input}
\begin{aligned}
\mathcal I_k^{\mathrm{tame}}(s) &= \mathbf W_{p_{\mathbf{high}},k}(s_0) + \mathbf P_{p_{\mathbf{high}},k}(s_0) +D_{p_{\mathbf{high}},k}(s_0) + \mathbf H_{p_{\mathbf{high}},k}(s_0) \\
& \quad + \sup_{s_0\leq\tau\leq s} \mathcal F_{p_{\mathbf{high}},k}^{\Lscr, \mathrm{sing}}(\tau;s_0) + \mathcal I^{\mathrm{tr}}_{p_{\mathbf{high}},k}(s).
\end{aligned}
\ee
Exact rescaling of the effective-field energy and flux gives, on each finite interval,
\be
\label{equa:seven-fixed-interface-from-robust}
\mathcal I_k^{\mathrm{tame}}(s_1) \leq C_N(1+ \coeff^{-1/2}){s_1}^{1/2 + \nu_k} \mathcal I_{\mathrm h,k}^{\mathrm{rob}}(s_1).
\ee


\subsection{Low-order effective-field and massless energies}
\label{section-10-2}

\paragraph{Robust effective-field estimate}

At \(p=p_{\mathbf{low}} \), the critical sum in \eqref{equa:seven-critical-radial-remainder} is empty and \(\widehat U_{p_{\mathbf{low}},k} =U_{p_{\mathbf{low}},k} \).  Equations \eqref{equa:seven-commutator-L2-triangular}, \eqref{equa:seven-refined-genuine-sources}, and \eqref{equa:seven-U-energy-inequality} give
\bse
\be
\label{equa:seven-U-triangular}
\begin{aligned}
U_{p_{\mathbf{low}},k}(s) & \leq C\mathcal I^U_{p_{\mathbf{low}},k}(s)s^{\alpha_k} + C_N\theta^{-1} \bigl(\mathcal B_3^{\mathrm{low}}(s) \bigr)^2 + C\mathcal B_3^{\mathrm{low}}(s) \int_{s_0}^{s} \tau^{-1}U_{p_{\mathbf{low}},k}(\tau) \, d\tau \\
& \quad + C\mathcal B_3^{\mathrm{low}}(s) \sum_{j=1}^{k} \int_{s_0}^{s} \tau^{-1+ \eta_j}U_{p_{\mathbf{low}},k-j}(\tau) \, d\tau.
\end{aligned}
\ee
Rank-zero Gronwall and induction on \(k\) yield
\be
\label{equa:seven-U-improved-raw}
U_{p_{\mathbf{low}},k}(s) \leq C_N\mathcal U_k^{\mathbf{low}}(s)s^{\alpha_k}, \quad \mathcal U_k^{\mathbf{low}}(s) \coloneqq \max_{0\leq q\leq k} \left\{ \mathcal I^U_{p_{\mathbf{low}},q}(s) + {\bigl(\mathcal B_3^{\mathrm{low}}(s) \bigr)^2 \over \rho\theta} \right\}.
\ee
\ese

\paragraph{Massless energy estimate}

We use only the direct bound \eqref{equa:seven-Dpsi-Sobolev}.  Under the bootstrap it gives \(\mathcal D_{\psi,j} \lesssim As^{\beta_{j+2}} \).  The exponent gaps in \eqref{equa:seven-wave-rank-exponents} imply
\bse
\be
\beta_{j+2} + \alpha_{k-j}<\frac12, \quad \beta_{j+2} + \nu_{k-j}<\frac12.
\ee
Consequently, every mixed term in \eqref{equa:seven-matter-energy-triangular-preliminary} is integrable at the lower level.  The critical remainder is again empty.  Rank-zero Gronwall uses \(CA<\rho/4\), while positive rank is triangular, and hence we obtain
\be
\label{equa:seven-P-low-improved-raw}
\mathbf P_{p_{\mathbf{low}},k}(s) \leq C_N\mathcal L_k(s)s^{\chi_k},
\ee
where the matter amplitude \(\mathcal L_k(s) \) is defined by
\be
\label{equa:seven-low-matter-amplitude}
\mathcal L_k(s) \coloneqq \max_{0\leq q\leq k} \left\{ \mathcal I^{P, \mathrm l}_{q}(s) + {\bigl(\mathcal B_3^{\mathrm{low}}(s) \bigr)^2\over \rho\theta} + {\mathcal B_3^{\mathrm{low}}(s) \over \rho\theta} \mathcal U_q^{\mathbf{low}}(s) \right\}.
\ee
\ese
This is stronger than the low bootstrap target since \(\chi_k<\beta_k\).

\paragraph{Integrated metric source}

For insertion in the metric wave inequality, we define
\bse
\be
\mathbf S^u_{p,k}(s) \coloneqq \mathcal N^\star_{p,k}(s) + \mathcal C_{u;p,k}(s) + \| \,| \mathcal R^\star|_{p,k} \|_{L^2(\Mint_s)}.
\ee
At the lower level,~\autoref{lem:seven-dynamic-metric-source-commutator}, \eqref{equa:seven-integrated-metric-Hardy-kernel}, and \eqref{equa:seven-reference-source-bound} give
\be
\label{equa:seven-low-integrated-dynamic-metric-source}
\begin{aligned}
\int_{s_0}^{s} \mathbf S^u_{p_{\mathbf{low}},k}(\tau) \, d\tau & \lesssim C_N\mathcal B_3^{\mathrm{low}}(s) \int_{s_0}^{s} \tau^{-1} \mathbf W_{p_{\mathbf{low}},k}(\tau) \, d\tau + {C_N\over \theta} \mathcal B_3^{\mathrm{low}}(s) \sum_{j=1}^{k} \int_{s_0}^{s} \tau^{-1+ \eta_j} \mathbf W_{p_{\mathbf{low}},k-j}(\tau) \, d\tau \\
& \quad + C_N\int_{s_0}^{s} \kappa_u^\star(\tau) \mathbf W_{p_{\mathbf{low}},k}(\tau) \, d\tau + C_N\Bigl((A+A_0) \mathcal T^u_{p_{\mathbf{low}},k}(s) + {A_0\over \zeta_\star} \Bigr)s^{\nu_k}.
\end{aligned}
\ee
\ese
The same-rank integral carrying \(\kappa_u^\star \) has an \(L^1\) coefficient and is handled by ordinary Gronwall.  For the triangular term, \(\eta_j+ \nu_{k-j} = \nu_k\).

\paragraph{Metric energy estimate}

At \(p=p_{\mathbf{low}} \), we use the conservative stress estimate \eqref{equa:seven-massless-source-inhomogeneous}, or the sharper \eqref{equa:seven-massless-source-characteristic}.  The reference source is integrated directly by \eqref{equa:seven-reference-source-bound}.  The wave energy inequality gives
\bse
\be
\label{equa:seven-W-triangular-low}
\begin{aligned}
\mathbf W_{p_{\mathbf{low}},k}(s) & \leq C\mathcal I^W_{p_{\mathbf{low}},k}(s)s^{\nu_k} +  {C_N\bigl(\mathcal B_3^{\mathrm{low}}(s) \bigr)^2\over \widetilde\mu_k}s^{\widetilde\mu_k} + {C_NA_0\over \zeta_\star} + C_N(A+A_0) \mathcal T^u_{p_{\mathbf{low}},k}(s)s^{\nu_k} \\
& \quad + C_N\mathcal B_3^{\mathrm{low}}(s) \int_{s_0}^{s} \tau^{-1} \mathbf W_{p_{\mathbf{low}},k}(\tau) \, d\tau + {C_N\over \theta} \mathcal B_3^{\mathrm{low}}(s) \sum_{j=1}^{k} \int_{s_0}^{s} \tau^{-1+ \eta_j} \mathbf W_{p_{\mathbf{low}},k-j}(\tau) \, d\tau \\
& \quad + C_N\int_{s_0}^{s} \kappa_u^\star(\tau) \mathbf W_{p_{\mathbf{low}},k}(\tau) \, d\tau + C\mathcal B_3^{\mathrm{low}}(s) \sum_{j=0}^{k} \int_{s_0}^{s} \tau^{-1+ \eta_j} \bigl(\mathbf P_{p_{\mathbf{low}},k-j} +U_{p_{\mathbf{low}},k-j} \bigr) (\tau) \, d\tau.
\end{aligned}
\ee
Since \(\eta_j+ \chi_{k-j} = \chi_k<\nu_k\) and \(\eta_j+ \alpha_{k-j} = \alpha_k<\nu_k\), this yields
\be
\label{equa:seven-W-improved-raw}
\mathbf W_{p_{\mathbf{low}},k}(s) \leq C_N\mathcal W_k^{\mathbf{low}}(s)s^{\nu_k},
\ee
where we define
\be
\label{equa:seven-low-metric-amplitude}
\begin{aligned}
\mathcal M_q^W(s) &\coloneqq \mathcal I^W_{p_{\mathbf{low}},q}(s) + {A_0\over \zeta_\star} +(A+A_0) \mathcal T^u_{p_{\mathbf{low}},q}(s) + {\bigl(\mathcal B_3^{\mathrm{low}}(s) \bigr)^2\over \widetilde\mu_q} + {\mathcal B_3^{\mathrm{low}}(s) \over \rho\theta} \bigl(\mathcal U_q^{\mathbf{low}}(s) + \mathcal L_q(s) \bigr), \\
\mathcal W_k^{\mathbf{low}}(s) &\coloneqq \max_{0\leq q\leq k} \mathcal M_q^W(s).
\end{aligned}
\ee
\ese


\section{Wave representation and cone estimates}  
\label{section-central-wave}


\subsection{Central-region estimate for the massless wave}
\label{section-10-3}

\paragraph{Core and lightlike regions}
 
The high-order energy estimates require a pointwise bound for the massless field with the rank exponent \(\eta_j\). We reduce this bound to a signed wave representation, derive the wave-gauge identities, and estimate the differentiated cone operators. Section~\ref{section-central-sources} applies these estimates to the metric sources and separates the remaining contributions.

First of all, the massless unknown is a wave variable and does not enjoy the same massive pointwise estimate available for \({\fieldPhi} \). We group the definitions of the two regions with their elementary geometric consequence:
\bse
\be
\label{equa:seven-massless-two-regions}
q\coloneqq t-r, \quad \Mint_{s, \mathrm{core}} {\coloneqq} \Mint_s\cap\{q\geq2t^{1/3} \}, \quad \Mint_{s, \mathrm{opt}} {\coloneqq} \Mint_s\cap\{1\leq q\leq3t^{1/3} \}.
\ee
These regions cover \(\Mint_s\).  For \(s\geq3\), they have a nonempty overlap; the compact range \(s_0\leq s\leq3\) is controlled by the local estimates and absorbed into the initial term.  The lightlike-region geometry is
\be
\label{equa:seven-light-like-region-geometry}
(t,x) \in\Mint_{s, \mathrm{opt}} \quad\Longrightarrow\quad t\geq c s^{3/2},
\ee
\ese
since \(s^2=q(t+ r) \leq6t^{4/3} \). More generally, \(q\leq Ct^a \) implies \(t\gtrsim s^{2/(1+a)} \). Thus \(a=1/3\) is the largest splitting exponent for which \(t^{-1} \lesssim s^{-3/2} \) follows from this geometry alone. The following estimate treats the whole interior and gives the coefficient needed in the high-order energy inequality.

\begin{proposition}[Pointwise estimate for the massless wave]
\label{lem:seven-massless-core-gain}
Assume \(N\geq19\), \(3/4<\lambda<\mu<1\), \(0<\delta \leq(\mu-\lambda)/2\), and the bootstrap hypotheses in the interior and exterior regions on \([s_0,s_1] \):
\begin{itemize}
\item {the two interior energy levels of Section~\ref{section-8-3} satisfy their bootstrap bounds with amplitude \(A\); the exterior hierarchy controls the metric and matter energies through order \(N\), the uniformly rescaled effective-field energy through order \(p_{\mathbf{high}}\), and the order-\(N\) effective-field energy with its factor \(s^{-1}\), together with the metric good-derivative spacetime integral and Hardy norms; all these fields belong to the same solution to the Cauchy problem;}
\item the reference metric, the initial weighted norms, and the geometric bootstrap bounds are those fixed in the Cauchy problem.
\end{itemize}
If the amplitude satisfies
\bse
\be
\label{equa:seven-massless-core-smallness}
C_Np_{\mathrm w} \bigl(A+ \mathcal B_3^{\mathrm{low}}(s_1) \bigr) \leq\varepsilon_{\mathrm{cw}} \theta,
\ee
for a sufficiently small fixed \(\varepsilon_{\mathrm{cw}}>0\), then, for \(s_0\leq s\leq {s_1}\) and \(0\leq j\leq p_{\mathrm w} \), one has
\be
\label{equa:seven-massless-core-gain}
\sup_{\Mint_s} \Bigl(| \del\psi|_{p_{\mathrm w},j} + \frac ts| \delsH\psi|_{p_{\mathrm w},j} \Bigr) \leq C_{\mathrm{cw}}(A+A^2)s^{-3/2 + \eta_j}.
\ee
In particular, one obtains
\be
\label{equa:seven-sufficient-massless-gain}
\sup_{\Mint_s} \Bigl(| \del\psi|_{p_{\mathrm w},j} + \frac ts| \delsH\psi|_{p_{\mathrm w},j} \Bigr) \leq C_{\mathrm{cw}} \theta^{-1} \mathcal B_3^{\mathrm{low}}(s)s^{-3/2 + \eta_j}.
\ee
\ese
The constant depends on the fixed geometric parameters and exponent gaps, but is independent of \(s_1\) and \(0<\coeff\leq\coeff_0\).
\end{proposition}

\paragraph{Reduction to the central targets}

The proof separates two target regions. With \(b\coloneqq5\rho+2C_{\mathrm{rk}} \theta \), direct Sobolev already gives
\be
\label{equa:seven-direct-massless-central-reduction}
| \del\psi|_{p_{\mathrm w},j} + \frac ts| \delsH\psi|_{p_{\mathrm w},j} \lesssim A t^{-1/2}s^{-1+ \eta_j+b} \leq A s^{-3/2 + \eta_j} \quad\text{when }t\geq s^{1+2b}.
\ee
On the remaining targets \(t<s^{1+2b} \), the proof has three steps:
\begin{itemize}
\item the signed cone formula and the wave gauge control the metric sources, including their trace at the interface;
\item a mass-weighted product correction estimates the matter source, its cutoff terms, and its nonzero Cauchy data;
\item the homogeneous waves and all source contributions are combined in Section~\ref{section-central-homogeneous}.
\end{itemize}
The exponent \(\eta_j\) is essential: replacing it by \(\beta_{j+2} = \eta_j+b\) would give the nonintegrable coefficient \(s^{-1+b} \) after normalization of the high effective-field energy.


\subsection{\texorpdfstring{Signed cone integrals and boundary traces}{Signed cone integrals and boundary traces}}
\label{section-central-cone}

\paragraph{Signed wave representation in the central region}

We prove \eqref{equa:seven-sufficient-massless-gain} using the flat wave representation of the solution. The principal metric source keeps its sign until the bad derivatives have been integrated by parts. For a fixed target \((t,x) \), we parameterize its past cone by
\be
\label{equa:seven-central-past-cone}
r=|x|, \quad Q=t-r, \quad R=t-\tau, \quad y=x+R\omega, \quad r'=|y|, \quad q'= \tau-r', \quad \omega \in\mathbb S^2.
\ee
Here \(r',q'\) refer to the source point, whereas \(r,Q\) refer to the target. Along the cone, differentiation at fixed \(\omega \) is \(D= \del_\tau-\omega \cdot\nabla_y\). For \(r'>0\), we set \(e=y/r'\), \(c=1+ \omega \cdot e\), and \(\omega_T= \omega-(\omega \cdot e)e\). Direct differentiation and the cosine identity give
\be
\label{equa:seven-central-cone-transversality}
c= \frac{(Q-q') (t+r-q')}{2r'R}, \qquad Dq'=c, \qquad Dc=-\frac{c(2-c)}{r'}.
\ee
On \(|q'| \leq Q/2\), the source lies away from the axis \(r'=0\) and \(c\gtrsim Q/t\). With \(L_y= \del_\tau+ \del_{r'} \), \(\underL_y= \del_\tau-\del_{r'} \), and \(\nabla_{\tan} \) the projection of \(\nabla_y\) onto the source sphere, we obtain the decomposition
\be
\label{equa:seven-central-bad-derivative}
\underL_y= \frac2cD-\frac{2-c}{c}L_y+ \frac2c\omega_T\cdot\nabla_{\tan}.
\ee
We partition the source with a smooth factor \(\chi(q'/Q) \), equal to one for \(|q'| \leq Q/4\) and zero for \(|q'| \geq Q/2\). This leaves the unknown uncut. For a smooth coefficient \(a \), application of \eqref{equa:seven-central-bad-derivative} to \(\underL_yv\), followed by integration by parts, yields
\be
\label{equa:seven-central-signed-integration}
\begin{aligned}
\int_{t_0}^{t} \chi Ra \, \underL_y^2v\, d\tau &= \left[\frac{2\chi Ra}{c} \underL_yv\right]_{t_0}^{t} -\int_{t_0}^{t}D\Bigl(\frac{2\chi Ra}{c} \Bigr) \underL_yv\, d\tau\\
& \quad -\int_{t_0}^{t} \frac{\chi Ra(2-c)}c L_y\underL_yv\, d\tau + \int_{t_0}^{t} \frac{2\chi Ra}c\omega_T\cdot\nabla_{\tan} \underL_yv\, d\tau.
\end{aligned}
\ee
All fields in this formula are evaluated at \((\tau,y) \); \(t_0\) is the lower time of the representation. Its trace must be controlled from the prescribed Cauchy hypersurface. The upper trace vanishes because \(\chi\) vanishes near the cone vertex. The differentiated coefficient is explicitly
\be
\label{equa:seven-central-differentiated-coefficient}
D\Bigl(\frac{Ra}{c} \Bigr)= \frac RcDa-\frac ac+ \frac{Ra(2-c)}{r'c}, \qquad D\chi = \frac cQ\chi'(q'/Q).
\ee
The bad--bad part of \(H^{\alpha \beta} \del_{\alpha \beta}v\) is \((H_N/4) \underL_y^2v\); the other principal terms contain a good derivative, with additional lower-order terms from differentiating the spherical frame. Formula \eqref{equa:seven-central-signed-integration} applies with \(a=H_N/4\), and likewise to each commuted coefficient. It replaces the bad Hessian by a first derivative and good Hessians. At \(x=0\), we have \(c=2\), \(r'=R\), and \(D(Ra/c)=0\) for the coefficient \(a=-m/r'\), consistent with \eqref{equa:seven-Schwarzschild-null-tail}. Thus this leading contribution reduces to the initial trace and source-partition terms.

\paragraph{Two integrations in the incoming source coordinate}

A second integration removes the bad derivative of the unknown from the volume integral. For \(r>0\), we set \(Q_+=t+r\) and \(p'= \tau+r'\). The past cone is described by \(Q\leq p'\leq Q_+ \), \(q'\leq Q\), and an azimuth \(\vartheta \). The polar angle \(\lambda \) of \(y\) relative to \(x\) satisfies
\be
\label{equa:seven-cone-polar-angle}
\cos\lambda= \frac{(p'-t) (t-q') +r^2}{r(p'-q')}, \qquad \alpha_{\mathrm c}{\coloneqq} \del_{q'} \lambda=-\frac{\sqrt{r^2-(p'-t)^2}}{(p'-q') \sqrt{(t-q')^2-r^2}}.
\ee
On \(q'\leq Q/2\), direct differentiation gives \(| \alpha_{\mathrm c}| \leq Q^{-1} \) and \(| \del_{q'} \alpha_{\mathrm c}| \leq4Q^{-2} \). Both estimates are uniform as \(r\to0\). If \(v^\sharp\) denotes the restriction of \(v\) to this parameterized cone, then \(\del_{q'}v^\sharp= \frac12\underL_yv+ \alpha_{\mathrm c}v_\lambda \), where angular derivatives of \(v\) are taken at fixed \((\tau,r') \) before restriction. For \(C_a=r'a \), including the source partition in \(a \), two integrations on any interval \([q_-,q_+] \subset(-\infty,Q/2] \) yield
\be
\label{equa:seven-cone-double-integration}
\begin{aligned}
\int_{q_-}^{q_+}C_a \underL_y^2v\, dq' &=4\left[C_a \del_{q'}v^\sharp-(\del_{q'}C_a)v^\sharp-2C_a \alpha_{\mathrm c}v_\lambda \right]_{q_-}^{q_+} \\
& \quad +4\int_{q_-}^{q_+} \Bigl((\del_{q'}^2C_a)v^\sharp +(2\alpha_{\mathrm c} \del_{q'}C_a+C_a \del_{q'} \alpha_{\mathrm c})v_\lambda +C_a \alpha_{\mathrm c}^2v_{\lambda \lambda} \Bigr)dq'.
\end{aligned}
\ee
Here derivatives of \(C_a \) are total derivatives along the cone. The formula follows by differentiating the chain rule once more and integrating \(\del_{q'}^2v^\sharp-2\del_{q'}(\alpha_{\mathrm c}v_\lambda) \) by parts. The flat retarded measure is \((8\pi r)^{-1}r'\, dp'\, dq'\, d\vartheta \). Thus \eqref{equa:seven-cone-double-integration} applies directly to its bad--bad source. The upper trace vanishes for a source partition supported in \(q'<Q/2\). The lower trace is determined by the domain: for a constant initial time it lies at \(q'=2t_0-p'\); for the truncated hyperboloidal interior it lies at \(q'= \max\{1,s_0^2/p'\} \), and includes an interface contribution.

\begin{proposition}[Weighted cone integrals]
\label{lem:seven-weighted-cone-integrals}
For \(0<\gamma<1\), \(0<Q<Q_+ \), and \(r= (Q_+-Q)/2\), one has
\bse
\be
\label{equa:seven-cone-principal-kernel}
\frac1r\int_Q^{Q_+} \int_0^{Q/2} \frac{(p'q')^{-\gamma}}{p'} \, dq'\, dp' \leq\frac{2^\gamma}{1-\gamma}(QQ_+)^{-\gamma}.
\ee
\be
\label{equa:seven-cone-angular-kernel}
\frac1r\int_Q^{Q_+} \int_0^{Q/2} \frac{(p'q')^{-\gamma}}{Q^2} \, dq'\, dp' \leq\frac{2^\gamma}{(1-\gamma)^2Q}(QQ_+)^{-\gamma}.
\ee
\ese
The same estimates hold for the continuous limits as \(Q_+ \downarrow Q\), and for every measurable subset of the integration rectangle.
\end{proposition}

\begin{proof}
\bse
We integrate first in \(q'\), then in \(p'\). With \(z=Q_+/Q\geq1\), the two integrals, divided respectively by \((QQ_+)^{-\gamma} \) and \(Q^{-1}(QQ_+)^{-\gamma} \), equal
\be
\label{equa:seven-cone-kernel-ratios}
\frac{2^\gamma}{\gamma(1-\gamma)} \frac{z^\gamma-1}{z-1}, \qquad \frac{2^\gamma}{(1-\gamma)^2} \frac{z-z^\gamma}{z-1}.
\ee
Concavity of \(z\mapsto z^\gamma \) gives the first bound; the second follows from \(z^\gamma \geq1\). The ratios extend continuously at \(z=1\), with respective limits
\be
\label{equa:seven-cone-kernel-center-limits}
\frac{2^\gamma}{1-\gamma}, \qquad \frac{2^\gamma}{1-\gamma}.
\ee
Positivity gives the assertion for subsets.
\ese
\end{proof}

The first estimate gives the critical power without an additional growth exponent. Indeed, on the source region \(0<q'\leq Q/2\), \(Q\leq p'\leq Q_+ \), we have \(\sigma{\coloneqq} \sqrt{p'q'} \leq s\), where \(s^2=QQ_+ \). Consequently, for nonnegative \(\eta_i, \eta_j\) with \(\eta_i+ \eta_j= \eta_k\), we have
\be
\label{equa:seven-cone-rank-weighted-kernel}
\frac1r\int_Q^{Q_+} \int_0^{Q/2} \frac{\sigma^{-3/2 + \eta_i+ \eta_j}}{p'} \, dq'\, dp' \lesssim s^{-3/2 + \eta_k}.
\ee
In particular, the first volume term in \eqref{equa:seven-cone-double-integration} has the desired bound if the differentiated coefficient is bounded by \(B(p')^{-1} \sigma^{\eta_i} \) and the unknown by \(G\sigma^{-3/2 + \eta_j} \). This is an estimate of the integral operator; the coefficient bound must still be derived from the metric equations. Formula \eqref{equa:seven-cone-angular-kernel} gives an additional factor \(Q^{-1} \) for angular and source-partition terms with coefficient \(\Ocal(Q^{-2}) \). In the remaining target region \(t<s^{1+2b} \), we have \(Q\geq\frac12s^{1-2b} \), so this factor can absorb a positive growth exponent smaller than \(1-2b\).

\paragraph{The reference-metric volume contribution}

For the glued reference metric, the differentiated coefficients can be estimated directly. Its radial partition is \(\vartheta^\star= \chi^\star(r') \chi^\star(r'/(\tau-1)) \), and we set \(a^\star= (H_N)^\star/4\). Inverting the time and radial coefficients in \eqref{equa-defineMS-new} gives
\be
\label{equa:seven-cone-reference-coefficient}
(H_N)^\star=-\frac{4mr'\vartheta^\star}{(r')^2-m^2(1-2\vartheta^\star)^2}, \qquad C_{a^\star}{\coloneqq} \tfrac14r'(H_N)^\star.
\ee
On \(Q\leq p'\leq Q_+ \), \(0<q'\leq Q/2\), and \(Q\geq8\), we have \(r'\geq p'/4\) and \(\tau-1\geq p'/4\). The radial cutoff is constant there, and \(r'/(\tau-1)= (p'-q')/(p'+q'-2) \). Its first two \(q'\)-derivatives are \(\Ocal((p')^{-1}) \) and \(\Ocal((p')^{-2}) \). Since the rational denominator in \eqref{equa:seven-cone-reference-coefficient} stays away from zero,
\be
\label{equa:seven-cone-reference-derivatives}
|C_{a^\star}| \lesssim m, \qquad | \del_{q'}C_{a^\star}| \lesssim m(p')^{-1}, \qquad | \del_{q'}^2C_{a^\star}| \lesssim m(p')^{-2}.
\ee
These are total cone derivatives because the reference coefficient is radial. After multiplication by \(\chi(q'/Q) \), each volume coefficient in \eqref{equa:seven-cone-double-integration} is bounded by \(C m Q^{-2} \), including the angular coefficients. In the exact Schwarzschild region, \(C_{a^\star} =-m(r')^2/((r')^2-m^2) \); its leading constant \(-m\) disappears under differentiation.

We denote by \(\mathcal V_\star[v] \) the volume contribution of \(((H_N)^\star/4) \underL_y^2v\) after these integrations, restricted to the portion above the initial hypersurface with \(1\leq q'\leq Q/2\). For \(v= \del Z\psi\), \(\ord Z\leq p_{\mathrm w} \), and \(\rank Z\leq k\leq p_{\mathrm w} \), two angular derivatives and the two Sobolev commutations use at most \(p_{\mathrm w} +4\leq p_{\mathbf{high}} \) derivatives. The auxiliary high matter bootstrap therefore gives
\be
\label{equa:seven-cone-angular-Sobolev}
\sum_{a=0}^2| \del_\lambda^a v| \lesssim C_N A\tau^{-1/2} \sigma^{-1+ \nu_{k+4}} \lesssim C_N A\sigma^{-3/2 + \nu_{k+4}}.
\ee
We used \(\tau\geq\sigma \). Applying \eqref{equa:seven-cone-angular-kernel} with \(\gamma=3/4\), then \(Q\geq\frac12s^{1-2b} \) and \(\nu_{k+4} = \eta_k+12\rho+6C_{\mathrm{rk}} \theta \) in the remaining target region, yields
\be
\label{equa:seven-cone-reference-volume-gain}
| \mathcal V_\star[\del Z\psi](t,x)| \lesssim C_NmA Q^{-1}s^{-3/2 + \nu_{k+4}} \lesssim C_NmA s^{-3/2 + \eta_k-\gamma_\star}, \qquad \gamma_\star{\coloneqq}1-22\rho-10C_{\mathrm{rk}} \theta>0.
\ee
This estimates the volume integral from the reference metric, including its angular terms and gluing region. The lower trace remains separate, with its initial and interface pieces determined by \eqref{equa:seven-cone-double-integration}.

\paragraph{Cancellation of the interface trace}

We apply the signed cone identity to the same coefficient and unknown on both sides of \(q'=1\), with the same source partition. For fixed \((p', \vartheta) \), its boundary expression is
\be
\label{equa:seven-cone-trace-expression}
\mathcal J_a[v]{\coloneqq}C_a \del_{q'}v^\sharp-(\del_{q'}C_a)v^\sharp-2C_a \alpha_{\mathrm c}v_\lambda.
\ee
The interior integral contributes \(-4\mathcal J_a[v]|_{q'=1} \), and the exterior integral contributes \(+4\mathcal J_a[v]|_{q'=1} \). These cancel pointwise before integration in \((p', \vartheta) \), since the fields and their first derivatives have the same traces. This is the cancellation of the boundary terms in \eqref{equa:seven-cone-double-integration}, independently of the energy-flux identity.

We use the initial-time ceiling \(T_0\) fixed in the equation~\eqref{equa:seven-cone-initial-time-ceiling}. On the source partition, \(p'\geq Q\) and \(|q'| \leq Q/2\), so \(\tau\geq Q/4\). For \(Q>4T_0\), this region lies above the initial hypersurface; equivalently, at its intersection with the past cone, \(q'=2\tau-p'<-Q/2\). Both \(\chi(q'/Q) \) and its derivative vanish there, as they do at \(q'= \pm Q/2\). Thus the near-cone contribution has only volume terms.

In the remaining target region \(t<s^{1+2b} \), the bounded range \(Q\leq4T_0\) implies \(s\leq(8T_0)^{1/(1-2b)} \). The direct bound \eqref{equa:seven-direct-massless-central-reduction} therefore gives, for \(0\leq j\leq p_{\mathrm w} \),
\be
\label{equa:seven-cone-bounded-target-range}
| \del\psi|_{p_{\mathrm w},j} + \frac ts| \delsH\psi|_{p_{\mathrm w},j} \lesssim A(8T_0)^{b/(1-2b)}s^{-3/2 + \eta_j}.
\ee
This constant is independent of \(\coeff\). The remaining wave field estimates concern the differentiated dynamic coefficient, the complementary and exterior volumes, and the homogeneous Cauchy contribution. The commuted estimates must also count the extra boost in \((t/s) \delsH_aZ\psi =s^{-1}L_aZ\psi\).


\subsection{\texorpdfstring{Wave-gauge identities for the metric coefficient}{Wave-gauge identities for the metric coefficient}}
\label{section-central-coefficient}

\paragraph{Wave-gauge control of the incoming coefficient}

The coefficient derivatives in the signed representation can be related to good derivatives by the divergence form of the wave gauge. The identities below hold for \(r>0\), and their estimates are used on \(t/3\leq r<t\). We set \(\omega=x/r\) and
\be
\label{equa:seven-densitized-metric-definition}
j_g= \sqrt{| \det g|}, \qquad K^{\alpha \beta} =j_gg^{\alpha \beta}-(\gMink)^{\alpha \beta}, \qquad \Pi_{ab} = \delta_{ab}-\omega_a \omega_b.
\ee
Thus \(\del_\alpha K^{\alpha \beta} =0\). We use the radial components and tangential vector
\be
\label{equa:seven-densitized-radial-components}
K_{r0} = \omega_aK^{a0}, \quad K_{rr} = \omega_a \omega_bK^{ab}, \quad \operatorname{tr}_{\mathbb S^2}K= \Pi_{ab}K^{ab}, \quad K_T^a= \Pi_{ab}(K^{b0}-K^{bc} \omega_c).
\ee
The incoming component is \(K_N=K^{00}-2K_{r0} +K_{rr} =j_gH_N\). Contracting the divergence identity with \(\ell= (1,-\omega) \), and using \(\del_a \omega_b=r^{-1} \Pi_{ab} \), gives
\be
\label{equa:seven-densitized-null-gauge}
\underL K_N=-L(K^{00}-K_{rr})-\frac4r(K_{r0}-K_{rr})-\frac2r\operatorname{div}_{\mathbb S^2}K_T-\frac2r\operatorname{tr}_{\mathbb S^2}K.
\ee
Indeed, the vector \(J^\alpha=K^{\alpha \beta} \ell_\beta \) has components \(J^t=K^{00}-K_{r0} \), \(J^r=K_{r0}-K_{rr} \), and \(J^\top=K_T\), whereas \(\del_\alpha J^\alpha=-r^{-1} \operatorname{tr}_{\mathbb S^2}K\). The spherical divergence formula yields \eqref{equa:seven-densitized-null-gauge}. Since \([L, \underL]= [\Omega, \underL]=0\) and \(\underL(r^{-1})=r^{-2} \), another incoming derivative satisfies
\be
\label{equa:seven-densitized-second-null-bound}
| \underL^2K_N| \lesssim |L\underL K|+r^{-1}| \Omega \underL K|+r^{-1}| \underL K|+r^{-2}(| \Omega K|+ |K|) \lesssim t^{-1}| \del K|_{1,1} + \frac qt| \del^2K|+ t^{-2}|K|_{1,1}.
\ee
The norms on the right refer to Cartesian tensor components. The second inequality uses \(L=t^{-1} \omega^aL_a+(q/t) \del_t\) and the boost representation of rotations. The full Cartesian derivatives keep their direct growth. The source estimates below use the gauge identity with one incoming integration by parts and the lightlike Hessian bounds, rather than a refined bound for every metric component.

\paragraph{An exact decomposition of the second incoming derivative}

The metric equation can replace the mixed second derivative in the gauge identity. We first work in physical source coordinates \(p=t+r\), \(q=t-r\), with \(\del_q= \underL/2\) at fixed \((p, \omega) \), before restricting a field to a target cone. For a symmetric divergence-free Cartesian tensor \(P\), we set
\be
\label{equa:seven-dynamic-divergence-components}
F=P^{00}-P_{rr}, \quad G=P_{r0}-P_{rr}, \quad T= \Pi_{ab}P^{ab}, \quad P_N=F-2G, \quad P_T= \Pi(P^{\cdot0}-P^{\cdot b} \omega_b).
\ee
We denote by \(\Gamma= (\Box P)^{00}-\omega_a \omega_b(\Box P)^{ab} \) the radial contraction of its Cartesian wave operator. The decomposition is
\be
\label{equa:seven-dynamic-divergence-identity}
\del_q^2(rP_N)=C_P+ \operatorname{div}_{\mathbb S^2}V_P,
\ee
where the scalar and tangential vector on the right are
\be
\label{equa:seven-dynamic-divergence-coefficients}
\begin{aligned}
C_P&= \frac r4\Gamma+ \frac12\del_tF-\underL G-\frac12\underL T+ \frac{G+T-P_{rr}}r, \\
V_P&=-\frac12\underL P_T+ \frac{P_T}{2r}-\frac1{4r} \Bigl(\nabla_{\mathbb S^2}F+2P^{ab} \nabla_{\mathbb S^2}(\omega_a \omega_b) \Bigr).
\end{aligned}
\ee
To derive this formula, we differentiate \eqref{equa:seven-densitized-null-gauge} for \(P\), using \(\del_qr=-1/2\). The radial wave decomposition then gives
\be
\label{equa:seven-dynamic-cartesian-wave-projection}
\Gamma=-L\underL F+ \frac2r\del_rF+ \frac1{r^2} \Bigl(\Delta_{\mathbb S^2}P^{00}-\omega_a \omega_b\Delta_{\mathbb S^2}P^{ab} \Bigr).
\ee
Finally, \(\Delta_{\mathbb S^2}(\omega_a \omega_b)=2\delta_{ab}-6\omega_a \omega_b\) implies
\be
\label{equa:seven-dynamic-angular-divergence-conversion}
\Delta_{\mathbb S^2}P^{00}-\omega_a \omega_b\Delta_{\mathbb S^2}P^{ab} = \operatorname{div}_{\mathbb S^2} \Bigl(\nabla_{\mathbb S^2}F+2P^{ab} \nabla_{\mathbb S^2}(\omega_a \omega_b) \Bigr) -2T+4P_{rr}.
\ee
Substitution proves \eqref{equa:seven-dynamic-divergence-identity}. In particular, the second incoming derivative is expressed through the metric wave equation, first derivatives, and an angular divergence.

\paragraph{Tangential gauge and angular derivatives}

The angular divergence also benefits from the wave gauge. We write \(P_{T0} = \Pi P^{\cdot0} \), \(P_{Tr} = \Pi P\omega \), and \(P_{TT} = \Pi P\Pi\). Tangential projection of \(\del_\alpha P^{\alpha b} =0\) gives
\be
\label{equa:seven-dynamic-tangential-gauge}
\del_tP_{T0} + \del_rP_{Tr} + \frac3rP_{Tr} + \frac1r\operatorname{div}_{\mathbb S^2}P_{TT} =0, \qquad \underL P_T=-L(P_{T0} +P_{Tr})-\frac6rP_{Tr}-\frac2r\operatorname{div}_{\mathbb S^2}P_{TT}.
\ee
The factor \(3/r\) follows from the two cross terms in the decomposition of the spatial tensor: \(\omega \otimes P_{Tr} \) contributes \(\del_rP_{Tr} +2P_{Tr}/r\), and \(P_{Tr} \otimes\omega \) contributes \(P_{Tr}/r\). Thus \(V_P\) contains only outgoing derivatives and inverse-radius multiples of angular derivatives or tensor components.

Tensor Lie derivatives \(K^Z\) along translations and boosts remain divergence-free, since these vector fields are affine and have zero Cartesian divergence. On the source region \(1\leq q\leq Q/2\), \(Q\leq p\leq Q_+ \), we have \(r\geq p/4\), \(t\simeq p\), and \(\sigma= \sqrt{pq} \). The direct Sobolev and Hardy bounds imply, for an operator of order \(d\) and rank \(j\),
\be
\label{equa:seven-dynamic-direct-coefficient-bounds}
|K|_{d,j} \lesssim A p^{-1/2} \sigma^{\nu_{j+2}}, \qquad | \del K|_{d,j} \lesssim A p^{-1}q^{-1/2} \sigma^{\nu_{j+2}}, \qquad d+2\leq p_{\mathbf{high}}.
\ee
Since \(L=t^{-1} \omega^aL_a+(q/t) \del_t\), the tangential gauge equation and one angular derivative give
\be
\label{equa:seven-dynamic-angular-vector-gain}
|V_{K^Z}|+ | \nabla_{\mathbb S^2}V_{K^Z}| \lesssim C_N A p^{-3/2} \sigma^{\nu_{j+4}} \lesssim C_N A p^{-1} \sigma^{\eta_j}, \qquad d+4\leq p_{\mathbf{high}}.
\ee
The first inequality uses two commutations before the two Sobolev commutations. For the second, \(\sigma \leq p\) and \(\nu_{j+4}-\eta_j=12\rho+6C_{\mathrm{rk}} \theta<1/2\). The angular divergence therefore has the required coefficient decay at the original rank. The scalar term \(C_{K^Z} \) additionally requires the reduced metric equation.

\paragraph{Near-light-cone Hessians and outgoing derivatives}

A single incoming integration will require an outgoing derivative of the first gauge identity. For this purpose, the direct high growth rates can be kept in a stronger lightlike Hessian estimate. On \(1\leq q\leq p/2\), for \(Z\preccurlyeq(d,j) \), \(d\leq p_{\mathrm w} \), we have
\be
\label{equa:seven-cone-light-like-Hessian}
| \del^2K^Z|+ | \del^2Z\psi| \lesssim A p^{-1}q^{-3/2} \sigma^{\nu_{j+3}}.
\ee
The same estimate holds for commuted smooth metric compositions that vanish at the Minkowski metric. The constants keep the dependence on \(B_{s_1}\) from \eqref{equa:seven-cone-refined-amplitude-dependence}.

Constants in the lightlike and cone source estimates below may depend on \(N,B_{s_1}\), but not on \(\coeff\) or otherwise on the terminal slab parameter. We choose \(0<\delta \leq(\mu-\lambda)/2<1/8\). The exceptional loss \(\varepsilon_N\) concerns estimates using one additional boost or angular order. We prove \eqref{equa:seven-cone-light-like-Hessian} by induction on \(d\), using the semi-hyperboloidal decomposition
\be
\label{equa:seven-cone-light-like-Hessian-reduction}
\sigma^2| \del^2w| \lesssim p\bigl(| \del L_aw|+ | \del w| \bigr) +p^2| \Box_gw|.
\ee
The rank-zero metric estimate makes the time coefficient comparable to \((\sigma/p)^2\). The first term has the bound in \eqref{equa:seven-cone-light-like-Hessian} by \eqref{equa:seven-dynamic-direct-coefficient-bounds}, since \(p/\sigma^2=q^{-1} \).

For the metric derivative sources, the null block pairs a direct Cartesian derivative with a direct good derivative. Its weight is \(A^2p^{-5/2}q^{-1/2} \), with the direct rank powers; the extra factor \(p^{-1/2} \) absorbs the fixed lower-rank growth. The quasi-null block pairs a refined good metric component with a direct derivative. Its lower operator has order at most \(\lfloor d/2\rfloor\leq p_{\mathbf{low}}-3\), also at the additional angular order. Thus these terms have the required weight \(A p^{-2}q^{-1/2} \sigma^{\nu_{j+2}} \). The separate bad coefficient bound is used only in the principal commutators. For a smooth metric composition, its extra quadratic first derivatives are estimated algebraically after the metric Hessian has been obtained: their direct weight \(p^{-2}q^{-1} \) is bounded by \(p^{-1}q^{-3/2} \), since \(q\leq p\).

The matter and effective-field derivative sources pair a refined low factor with a direct high factor. For the latter, the weights \(p^{-1} \sigma^{-1/2} \) and \(p^{-1}q^{-1/2} \sigma^{1/2 + \nu_{j+2}} \) give the required source bound. The potential weights satisfy
\be
\label{equa:seven-cone-potential-Hessian-weight}
\frac pq\bigl(p^{-2} \sigma^{1/2} \bigr) \bigl(p^{-3/2} \sigma^{1/2 + \nu_{j+2}} \bigr) \leq p^{-1}q^{-3/2} \sigma^{\nu_{j+2}}.
\ee
We used \(q/p\leq1\). In the equation for \(Z\psi\), every effective-field derivative at order \(d\leq p_{\mathrm w} \) lies in the refined range \(p_{\mathbf{low}}-3\); pairing it with the direct matter derivative gives the same source bound. A highest-order derivative on a smooth coefficient is estimated by its direct pointwise bound and two lower differentiated factors. The reference residual is \(\Ocal(A_0p^{-3}) \). Each of these terms satisfies \eqref{equa:seven-cone-light-like-Hessian} after multiplication by \(p/q\).

In a principal commutator, the Hessian has smaller commutation order. Its bad coefficient is bounded by a refined amplitude times \(p^{-1} \sigma^{\eta_i} \). By the induction hypothesis, its contribution relative to \eqref{equa:seven-cone-light-like-Hessian} carries the bounded factor \(q^{-1} \), since \(\eta_i+ \nu_{j-i+3} = \nu_{j+3} \). Every other Hessian contains a semi-hyperboloidal derivative and is bounded by \(p^{-1}(| \del L_aw|+ | \del w|) \). The additional factor in the resulting product is \(A p^{-1/2} \sigma^{\nu_2} \), which is bounded because \(\sigma \leq p\) and \(\nu_2<1/2\). Translation derivatives of the frame have an additional factor \(p^{-1} \). This proves the induction and also covers the quadratic derivative terms in a smooth metric composition.

Applying \(L=t^{-1} \omega^aL_a+(q/t) \del_t\) to first derivatives, then differentiating this identity once more for the metric, gives
\be
\label{equa:seven-cone-outgoing-Hessian-bounds}
|L\del K^Z|+ |L\del Z\psi| \lesssim A p^{-2}q^{-1/2} \sigma^{\nu_{j+3}}, \qquad |L^2K^Z| \lesssim A p^{-5/2} \sigma^{\nu_{j+4}}.
\ee
The second estimate uses \(p^{-2}|L_aL_bK^Z| \), \(qp^{-2}| \del L_aK^Z| \), and \(q^2p^{-2}| \del^2K^Z| \). All have the required weight. The derivative count is \(p_{\mathrm w} +4\leq p_{\mathbf{high}} \), while refined coefficients use at most \(p_{\mathrm w} \leq p_{\mathbf{low}}-3\). Thus these bounds include \(N=19\) and require no outgoing derivative of a massive source.

\paragraph{The additional angular order}

One additional angular derivative is needed in the cone integral. The metric Hessian argument extends to \(d\leq p_{\mathrm w}+1\). At \(N=19\), we have \(p_{\mathbf{low}}=10\), \(p_{\mathbf{high}}=14\), and \(p_{\mathrm w}=7\). A coefficient of order eight and rank at most seven is covered by the general refined estimate. For the remaining boost-only coefficient, we first prove its boundary trace. On \(\mathbb S_\sigma\), where \(t=(\sigma^2+1)/2\) and \(r=R_\sigma=(\sigma^2-1)/2\),
\bse
\be
\label{equa:audit-exceptional-angular-conversion}
\Omega_{ab}=\frac{x_a}{t}L_b-\frac{x_b}{t}L_a,\qquad L_a(x_b/t)=\delta_{ab}-(x_a/t)(x_b/t).
\ee
The coefficients in two iterations of this identity are bounded on this sphere. We also expand the differentiated component \(H^{\H00}\) into Cartesian inverse-metric components there. Both conversions are made on the boundary, before radial integration, so no unbounded factor \(r/t\) enters the exterior integral. For a decaying Cartesian component \(v\), the fundamental theorem of calculus on the hypersurface and Cauchy--Schwarz give
\be
\label{equa:audit-exceptional-radial-trace}
\|v\|_{L^2(\mathbb S_\sigma)}^2\leq R_\sigma^2\!\int_{R_\sigma}^{\infty}r^{-2}\crochet^{-2\lambda}\,dr\,\|\crochet^\lambda\delsME_rv\|_{L^2(\MME_\sigma)}^2\leq\frac{3^{2\lambda}}{2\lambda-1}\|\crochet^\lambda\delsME_rv\|_{L^2(\MME_\sigma)}^2.
\ee
Indeed, \(\crochet\geq(1+r-R_\sigma)/3\) on this ray. The radial derivative is a first derivative already controlled by the wave energy. The two rotations therefore require at most ten commutations, within the exterior order-\(N\) bootstrap. Smooth inversion places the highest dynamic factor in this trace norm and the lower factors in their pointwise norms; the reference contribution has sphere norm \(\Ocal(m)\). Hence
\be
\label{equa:seven-cone-exceptional-trace}
\sum_{\substack{|J|\leq8\\|K|\leq2}}\|\Omega^KL^JH^{\H00}\|_{L^2(\mathbb S_\sigma)}^2\leq C_NA^2\sigma^{2\delta},\qquad \mathsf B_{00;8}(\sigma)\leq C_NA\sigma^\delta.
\ee
\ese
The second bound follows from the definition~\eqref{equa:seven-metric-pointwise-traces}, since \(\tau^2/R_\tau\) is bounded and \(\eta_j\geq0\) for \(s_0\leq\tau\leq\sigma\). The interior coefficient requires the separate coefficient--ray argument completed in~\eqref{equa:seven-coupled-coefficient-ray-bound}. In that argument, interpolation controls the quadratic remainders of the inverse-metric and determinant conversions, while the linear terms use the polynomial contraction of the densitized inverse metric and its trace. The conversion~\eqref{equa:seven-polynomial-commuted-conversion} and the simultaneous rank induction for the two massive amplitudes then prove~\eqref{equa:seven-sharp-boost-only-H00} through order \(p_{\mathbf{low}}\). This argument uses the coarse energy estimates, independently of the central wave gain being proved here. In particular,
\be
\label{equa:audit-exceptional-interior-coefficient}
|L^JH^{\H00}|\leq C_NA p^{-1}\sigma^{\eta_8}\leq C_NA p^{-1}\sigma^{\eta_8+\varepsilon_N},\qquad |J|=8,\quad\varepsilon_N=\delta\quad(N=19).
\ee
We keep the weaker last bound in the source estimates. The exceptional coefficient leaves an uncommuted Hessian in the differential-order induction. All other bad coefficients have lower order or contain a translation. The ordered Leibniz partitions below contain at most \(p_{\mathrm w}+5=12\) commutations, so two exceptional coefficients of order eight cannot occur in one product. Thus at most one loss \(\varepsilon_N\) is transmitted, including after the outgoing differentiation. For \(N\geq20\), the additional coefficient order belongs to the general refined range.

For matter at this last order, the term \(g(\del Z\fieldPhi, \del\psi) \) has a high effective-field derivative outside the refined range. We keep its null structure. With \(b= \beta_2\), direct low matter estimates and \(\fieldPhi = \coeff^{1/2} \fieldvarphi\) give
\be
\label{equa:seven-cone-last-order-null-factors}
| \delsH\psi| \lesssim A p^{-3/2} \sigma^b, \quad | \del\psi| \lesssim A p^{-1}q^{-1/2} \sigma^b, \quad | \delsH Z\fieldPhi| \lesssim\coeff^{1/2}A p^{-5/2} \sigma^{1/2 + \nu_{j+3}}.
\ee
The good effective-field derivative uses one boost on its value. Its bad derivative keeps the high factor \(\sigma^{1/2} \). In the two mixed products and the flat bad--bad product, the additional factors are bounded respectively by \(p^{-1/2} \sigma^{1/2} \), \(p^{-3/2} \sigma^{1/2} \), and \(p^{-1}q^{1/2} \sigma^{1/2} \), all at most one. The correction \(H^{\H00} \) gives the bounded factor \(p^{-1}q^{-1/2} \sigma^{1/2} \) times a refined rank-zero amplitude. Every differentiated metric coefficient leaves at most seven commutations on \(\fieldPhi\). Thus the induction proves, throughout \(d\leq p_{\mathrm w} +1\),
\be
\label{equa:seven-cone-one-angular-order}
| \del^2Z\psi| \lesssim A p^{-1}q^{-3/2} \sigma^{\nu_{j+3} +b+ \varepsilon_N}, \qquad |L\del Z\psi| \lesssim A p^{-2}q^{-1/2} \sigma^{\nu_{j+3} +b+ \varepsilon_N}.
\ee
For \(N\geq20\), this extra order already lies in the refined range; the additional factor \(\sigma^b\) is harmless. The conservative derivative count, including \(L^2K\) at this order, is \(p_{\mathrm w} +5\leq p_{\mathbf{high}} \), equal to \(12\leq14\) at \(N=19\). Rotations are converted to boosts in the source region \(r'\simeq t\); derivatives of the conversion coefficients have the corresponding inverse-radius factors.

\paragraph{Density conversion and differentiation on the target cone}

The cone coefficient is a component of \(H\), rather than of \(K\). With \(J=j_g^{-1} \) and \(C=rK_N\), its conversion is
\be
\label{equa:seven-dynamic-density-conversion}
\del_q^2(rH_N)=JC_K-\nabla_{\mathbb S^2}J\cdot V_K+2(\del_qJ) (\del_qC) +(\del_q^2J)C + \operatorname{div}_{\mathbb S^2}(JV_K).
\ee
Scalar and tensor commutations are related by the frame identities, with \(\ell_N=dt-\omega_a dx^a \),
\be
\label{equa:seven-dynamic-null-frame-commutations}
\mathcal L_{L_a} \ell_N=-\omega_a \ell_N-(q/r) \Pi_{ab}dx^b, \qquad \mathcal L_{\del_t} \ell_N=0, \qquad \mathcal L_{\del_a} \ell_N=-r^{-1} \Pi_{ab}dx^b.
\ee
Each \(q/r\) error has one fewer boost on the tensor, whereas a translation-frame error has one fewer differential order. These distinctions must be kept in estimating the differentiated products in \eqref{equa:seven-dynamic-density-conversion}.

A superscript \(\sharp\) denotes restriction to the fixed target cone. For any scalar coefficient \(C\), its total cone derivative differs from its physical incoming derivative by
\be
\label{equa:seven-dynamic-total-cone-second-derivative}
\del_q^2 C^\sharp= (\del_q^2C)^\sharp+2\alpha_{\mathrm c}(\del_qC)_\lambda^\sharp+(\del_q\alpha_{\mathrm c})C_\lambda^\sharp+ \alpha_{\mathrm c}^2C_{\lambda \lambda}^\sharp.
\ee
The first gauge identity, followed by one angular derivative, gives
\be
\label{equa:seven-dynamic-mixed-coefficient}
| \del_q(r(K^Z)_N)|+ | \del_\lambda \del_q(r(K^Z)_N)| \lesssim C_NA p^{-1/2} \sigma^{\nu_{j+4}}.
\ee
Hence its mixed cone term has the factor \(Q^{-1}p^{-1/2} \). To distinguish the target from the source coordinates, we write \(r_{\mathrm{tar}} = (Q_+-Q)/2\) and \(s_{\mathrm{tar}} = \sqrt{QQ_+} \) in the following integrals. The mixed integral satisfies
\be
\label{equa:seven-dynamic-mixed-cone-kernel}
\frac1{r_{\mathrm{tar}}} \int_Q^{Q_+} \int_0^{Q/2} \frac{(pq)^{-3/4}}{Q\sqrt p} \, dq\, dp\lesssim Q^{-1/2}s_{\mathrm{tar}}^{-3/2}.
\ee
Indeed, with \(z=Q_+/Q\) and \(w=z^{1/4} \), division by \(Q^{-1/2}s_{\mathrm{tar}}^{-3/2} \) gives the bounded expression \(2^{19/4}w^2/((w+1) (w^2 +1)) \). In the remaining target region, \(Q^{-1/2} \lesssim s_{\mathrm{tar}}^{-1/2 +b} \); this strict gain absorbs the additional Sobolev growth in these mixed terms.

\paragraph{The last differential order and the remaining source estimates}

In \(r\Gamma \), a direct high Hessian paired with a refined low coefficient produces the factor \(p^{-1}q^{-1/2} \sigma^e\), after the amplitudes and rank weights have been extracted, where \(e=12\rho+4C_{\mathrm{rk}} \theta<1/2\). For a wave factor with weight \(\sigma^{-3/2} \), this factor has a strict integral gain:
\be
\label{equa:seven-dynamic-last-order-kernel}
\begin{aligned}
\frac1{r_{\mathrm{tar}}} \int_Q^{Q_+} \int_1^{Q/2} \frac{(pq)^{-3/4+e/2}}{p\sqrt q} \, dq\, dp & \leq\frac{2}{1/4-e/2}Q^{-7/4+e/2} \\
& \lesssim s_{\mathrm{tar}}^{-3/2-\gamma_e}, \qquad \gamma_e\coloneqq (1-2b) (7/4-e/2)-3/2>0.
\end{aligned}
\ee
We integrate \(q^{-5/4+e/2} \) from \(1\), bound the average of \(p^{-7/4+e/2} \) by its value at \(Q\), and use \(Q\gtrsim s_{\mathrm{tar}}^{1-2b} \). This estimate can handle a direct high Hessian without increasing the threshold \(N\geq19\).

For the twice-integrated representation, the remaining source estimates are as follows. Section~\ref{section-central-target-derivatives} develops a single-integration argument for the interior bad--bad metric volume.
\begin{itemize}
\item The reduced metric source in \(C_{K^Z} \), together with the determinant products in \eqref{equa:seven-dynamic-density-conversion}, must satisfy the cone estimates at every required differential order.
\item In the translated wave equation, a differentiated metric coefficient can multiply an untranslated \(Z\psi\). Its direct pointwise bound has weight \(\sigma^{-1/2} \), not \(\sigma^{-3/2} \); the stronger weight in \eqref{equa:seven-dynamic-last-order-kernel} is therefore unavailable for this term.
\item The source portions outside the near-cone partition, including \(q'<1\), require their own estimates. The extra boost in \((t/s) \delsH_aZ\psi =s^{-1}L_aZ\psi\) must be included in the same derivative count.
\end{itemize}


\subsection{\texorpdfstring{Differentiated cone integrals}{Differentiated cone integrals}}
\label{section-central-target-derivatives}

\paragraph{Radial target derivatives of the signed cone integral}

Differentiating the cone representation itself provides an alternative to first translating the wave equation. For a scalar integrand independent of the target, angular integration by parts leaves only that integrand and its outgoing derivative. We fix the target direction \(n\), write \(Q=T-r_{\mathrm{tar}} \), \(Q_+=T+r_{\mathrm{tar}} \), and initially suppose \(0<Q<Q_+ \). Source coordinates are \(p= \tau+r'\), \(q= \tau-r'\). We normalize the source interval by
\be
\label{equa:seven-target-cone-normalized-coordinate}
p=Q+(Q_+-Q) \xi, \qquad 0\leq\xi\leq1, \qquad \frac{dp}{r_{\mathrm{tar}}} =2\, d\xi.
\ee
The source direction has polar angle \(\lambda \) relative to \(n\) and azimuth \(\vartheta \). The cosine formula is equivalently
\be
\label{equa:seven-target-cone-half-angle}
\tan^2\frac\lambda2= \frac{(1-\xi) (Q-q)}{\xi(Q_+-q)}.
\ee
For \(|q|<Q/2\), differentiation at fixed \((\xi,q, \vartheta,n) \), or at fixed \((Q,Q_+,q, \vartheta,n) \) for the last derivative, gives
\be
\label{equa:seven-target-cone-angle-derivatives}
\lambda_Q= \frac{\sin\lambda}{2(Q-q)}, \qquad \lambda_{Q_+} =-\frac{\sin\lambda}{2(Q_+-q)}, \qquad \lambda_\xi =-\frac{\sin\lambda}{2\xi(1-\xi)}.
\ee
These formulas follow by differentiating the logarithm of \eqref{equa:seven-target-cone-half-angle} and using \(d\lambda/d\log\tan(\lambda/2)= \sin\lambda \).

We take the scalar integrand \(W=W(p,q, \lambda, \vartheta;Q,Q_+) \) to be smooth in physical source coordinates, bounded at the polar axes, and sufficiently integrable for the following differentiations. A superscript \(\sharp\) denotes its restriction to the parameterized cone. We write \(W_Q^{\mathrm{exp}} \) and \(W_{Q_+}^{\mathrm{exp}} \) for derivatives of the explicit target dependence, holding \((p,q, \lambda, \vartheta) \) fixed; these derivatives vanish for a target-independent physical source. Also, \(W_p= \tfrac12L_yW\) at fixed \((q, \lambda, \vartheta,Q,Q_+) \). The chain rule and \(\del_\xi W^\sharp= (Q_+-Q)W_p^\sharp+ \lambda_\xi W_\lambda^\sharp\) give the following formulas after elimination of the angular derivatives:
\be
\label{equa:seven-target-cone-angle-elimination}
\begin{aligned}
\del_QW^\sharp&= (W_Q^{\mathrm{exp}})^\sharp+ \frac{(1-\xi) (p-q)}{Q-q}W_p^\sharp-\frac{\xi(1-\xi)}{Q-q} \del_\xi W^\sharp, \\
\del_{Q_+}W^\sharp&= (W_{Q_+}^{\mathrm{exp}})^\sharp+ \frac{\xi(p-q)}{Q_+-q}W_p^\sharp+ \frac{\xi(1-\xi)}{Q_+-q} \del_\xi W^\sharp.
\end{aligned}
\ee

We take a fixed source interval \(\mathcal J\), for example \(\mathbb R\) for the full source or \([1, \infty) \) for its interior volume, and a smooth partition \(\chi(q/Q) \) supported in \(|q|<Q/2\). We denote the product integration domain by \(\Xi = \mathcal J\times[0,1] \times[0,2\pi] \), with measure \(dq\, d\xi\, d\vartheta \), and set
\be
\label{equa:seven-target-cone-integral}
\mathcal I_W(Q,Q_+,n)\coloneqq2\int_\Xi\chi(q/Q)W^\sharp\, dq\, d\xi\, d\vartheta.
\ee
Integration by parts in \(\xi\) yields
\be
\label{equa:seven-target-cone-integral-derivatives}
\begin{aligned}
\del_Q\mathcal I_W&=2\int_\Xi\chi\Bigl((W_Q^{\mathrm{exp}})^\sharp+ \frac{(1-\xi) (p-q)}{Q-q}W_p^\sharp+ \frac{1-2\xi}{Q-q}W^\sharp\Bigr) -2\int_\Xi\frac q{Q^2} \chi'(q/Q)W^\sharp, \\
\del_{Q_+} \mathcal I_W&=2\int_\Xi\chi\Bigl((W_{Q_+}^{\mathrm{exp}})^\sharp+ \frac{\xi(p-q)}{Q_+-q}W_p^\sharp-\frac{1-2\xi}{Q_+-q}W^\sharp\Bigr).
\end{aligned}
\ee
The traces at \(\xi =0,1\) vanish because \(\xi(1-\xi)=0\) and \(W^\sharp\) is bounded. The formula applies to each volume term after the incoming integrations by parts. If all partition factors have already been included in \(W\), we take the external factor \(\chi\equiv1\); its derivative term disappears, while the actual partition derivatives remain in \(W_Q^{\mathrm{exp}} \). The factors \(\alpha_{\mathrm c} \), \(\del_q\alpha_{\mathrm c} \), and other explicit target coefficients must likewise be included in these explicit derivatives.

For the bounded initial time \(T_0\), the range \(Q\geq8T_0\) puts the partition above the initial hypersurface: \(\tau= (p+q)/2\geq Q/4\geq2T_0\). Thus the initial trace and its partition derivatives vanish. The signed traces at the interface from the incoming integrations remain separate from this volume formula and are estimated in \eqref{equa:seven-exterior-metric-retained-trace}. Outside this large-target range, moving initial boundaries must be kept and differentiated, and the fixed-domain formula alone does not account for them.

Equations \eqref{equa:seven-target-cone-angle-elimination}--\eqref{equa:seven-target-cone-integral-derivatives} concern time and radial derivatives with respect to the target with \(n\) fixed: \(\del_T= \del_Q+ \del_{Q_+} \) and \(\del_{r_{\mathrm{tar}}} =-\del_Q+ \del_{Q_+} \). Their normalized expressions have finite limits as \(Q_+ \downarrow Q\), under the smoothness and integrability conditions. The dependence on the target direction is treated next for a scalar physical source.

\paragraph{Angular target derivatives}

We first assume that \(W\) is a scalar function of the physical source point, with no explicit dependence on the target direction. We choose the target axis \(n=e_3\); a variation towards \(e_1\) rotates the source by \(\mathcal R= \cos\vartheta \, \del_\lambda-\cot\lambda \sin\vartheta \, \del_\vartheta \). Integration in the periodic azimuth gives
\be
\label{equa:seven-target-cone-azimuth-integration}
\int_0^{2\pi} \mathcal R W^\sharp\, d\vartheta= \int_0^{2\pi} \cos\vartheta \Bigl(W_\lambda^\sharp+ \cot\lambda \,W^\sharp\Bigr)d\vartheta.
\ee
We set \(a_-=Q-q\), \(a_+=Q_+-q\), \(D=Q_+-Q\), and \(h= \sqrt{\xi(1-\xi)} \). The identities for the angle imply
\be
\label{equa:seven-target-cone-angular-cancellation}
\beta{\coloneqq} \frac{(p-q)h}{\sqrt{a_-a_+}} =-\lambda_\xi^{-1}, \qquad \del_\xi\beta+ \cot\lambda= \frac{2Dh}{\sqrt{a_-a_+}}.
\ee
Indeed, \(\cot\lambda= ((a_-+a_+) \xi-a_-)/(2h\sqrt{a_-a_+}) \); differentiation of \(\beta= (a_-+D\xi)h/\sqrt{a_-a_+} \) proves the second identity. We substitute \(W_\lambda^\sharp=-\beta \del_\xi W^\sharp+D\beta W_p^\sharp\) into \eqref{equa:seven-target-cone-azimuth-integration} and integrate in \(\xi\). The traces vanish because \(\beta=0\) at \(\xi =0,1\). Since \(D/r_{\mathrm{tar}} =2\), the target tangential derivative is
\be
\label{equa:seven-target-cone-angular-target-derivative}
e_1\cdot\nabla_x\mathcal I_W =4\int_\Xi\frac{\chi(q/Q)h\cos\vartheta}{\sqrt{a_-a_+}} \Bigl(2W^\sharp+(p-q)W_p^\sharp\Bigr), \qquad x=r_{\mathrm{tar}}e_3.
\ee
A shift of the azimuth gives the other tangential direction. If \(W\) has explicit target-direction dependence, its corresponding term \(2r_{\mathrm{tar}}^{-1} \int_\Xi\chi\, \del_n^{\mathrm{exp}}W^\sharp\) must be added. Thus \eqref{equa:seven-target-cone-angular-target-derivative} applies directly to a scalar coefficient contracted only with the source radial frame, but not automatically to a component defined using a target-dependent polar frame.

\paragraph{Angular terms in the source frame}

The angular terms in the incoming integration can be expressed in the source frame. For \(\omega=y/|y| \) and \(\Pi_\omega=I-\omega \otimes\omega \), we set
\be
\label{equa:seven-cone-invariant-angular-coefficients}
D_{\mathrm c}{\coloneqq} (T-q)^2-|x|^2, \qquad \mathbf B\coloneqq D_{\mathrm c}^{-1} \Pi_\omega x, \qquad \mathbf B_1\coloneqq \frac{2(T-q)-x\cdot\omega}{D_{\mathrm c}} \mathbf B.
\ee
Source derivatives are taken at fixed \((p, \omega) \), before restriction to the cone. The total incoming derivative is \(\mathcal D_q= \del_q+ \mathbf B\cdot\nabla_{\mathbb S^2} \), since \(\alpha_{\mathrm c} \del_\lambda= \mathbf B\cdot\nabla_{\mathbb S^2} \). For the covariant derivative on the unit sphere, we have
\be
\label{equa:seven-cone-invariant-angular-derivative}
\del_q\mathbf B+ \nabla_{\mathbf B} \mathbf B= \mathbf B_1.
\ee
Indeed, \(\del_q\mathbf B=2(T-q) \mathbf B/D_{\mathrm c} \), while \(\Pi_\omega x= \nabla_{\mathbb S^2}(x\cdot\omega) \) and \(\nabla_{\mathbb S^2}^2(x\cdot\omega)=-(x\cdot\omega)g_{\mathbb S^2} \). For \(C=r'a \), including the partition, a single incoming integration therefore gives
\be
\label{equa:seven-cone-invariant-single-integration}
\int_{q_-}^{q_+}C\underL_y^2v\, dq =4[C\del_qv]_{q_-}^{q_+}-4\int_{q_-}^{q_+} \Bigl((\del_qC) \del_qv+ \mathbf B\cdot\nabla_{\mathbb S^2}(C\del_qv) \Bigr)dq.
\ee
This follows from \(\mathcal D_q(C\del_qv)=C\del_q^2v+(\del_qC) \del_qv+ \mathbf B\cdot\nabla_{\mathbb S^2}(C\del_qv) \). In particular, this form leaves only a first incoming derivative of the coefficient. If we integrate once more, the volume integrand in \eqref{equa:seven-cone-double-integration} becomes
\be
\label{equa:seven-cone-invariant-volume}
\mathcal V[C,v]= (\del_q^2C)v+2\mathbf B\cdot\nabla_{\mathbb S^2}((\del_qC)v) + \mathbf B_1\cdot\nabla_{\mathbb S^2}(Cv) +(\mathbf B\otimes\mathbf B):\nabla_{\mathbb S^2}^2(Cv).
\ee
The corresponding boundary expression is \(C\del_qv-(\del_qC)v-\mathbf B\cdot\nabla_{\mathbb S^2}(Cv) \). Expansion by the product rule recovers \eqref{equa:seven-cone-double-integration}. Every factor is restricted to the cone after the physical source derivatives have been taken.

On \(|q| \leq Q/2\), we have \(D_{\mathrm c} \geq QQ_+/4=s^2/4\), and direct differentiation gives
\be
\label{equa:seven-cone-invariant-target-bounds}
| \mathbf B| \lesssim Q^{-1}, \quad | \mathbf B_1| \lesssim Q^{-2}, \quad | \del_{T,x} \mathbf B| \lesssim Q^{-2}, \quad | \del_{T,x} \mathbf B_1| \lesssim Q^{-3}.
\ee
With the physical source point held fixed, the target semi-hyperboloidal derivative satisfies
\be
\label{equa:seven-cone-invariant-good-target-derivative}
\delsH_i\mathbf B= \frac{\Pi_\omega e_i}{D_{\mathrm c}} + \frac{2q x_i\Pi_\omega x}{T D_{\mathrm c}^2}, \qquad | \delsH\mathbf B| \lesssim s^{-2}, \qquad | \delsH\mathbf B_1| \lesssim s^{-2}Q^{-1}.
\ee
The coefficients \(\mathbf B\) and \(\mathbf B_1\) are smooth at \(x=0\), where both vanish. Thus their target derivatives introduce neither a polar-axis singularity nor an inverse power of the target radius. The partition keeps its explicit dependence on \(Q=T-|x| \).

\paragraph{A differentiated integral estimate}

For a target-independent scalar physical source \(W\), we take \(\mathcal J= [1, \infty) \), \(\eta \geq0\), and \(0\leq a<1\). If the source and its outgoing derivative satisfy
\be
\label{equa:seven-target-cone-source-condition}
|W|+p|W_p| \leq M p^{-3/2}q^{-a} \sigma^\eta, \qquad \sigma= \sqrt{pq}.
\ee
Then, for \(r_{\mathrm{tar}}>0\), the integral and its target derivatives satisfy
\be
\label{equa:seven-target-cone-differentiated-bound}
T^{1/2}| \mathcal I_W|+T^{1/2}s| \del\mathcal I_W|+T^{3/2}| \delsH\mathcal I_W| \lesssim M Q^{-a}s^\eta, \qquad s= \sqrt{QQ_+}.
\ee
The constants depend on \(a \) and the fixed partition. To prove these estimates, we extend the positive scalar integration bound from \(q\geq1\) to \(q\geq0\). Direct integration gives
\be
\label{equa:seven-target-cone-v-kernel}
I_a\coloneqq \frac1{r_{\mathrm{tar}}} \int_Q^{Q_+} \int_0^{Q/2}p^{-3/2}q^{-a} \, dq\, dp = \frac{2^aQ^{-a}}{1-a} \, \frac{2\sqrt Q}{\sqrt{Q_+}(\sqrt{Q_+} + \sqrt Q)}.
\ee
We use \(\sigma \leq s\), \(a_-\geq Q/2\), and \(a_+ \geq Q_+/2\) in the differentiated identities. The factor \(p-q\) multiplying \(W_p\) is bounded by \(p\), and is covered by \eqref{equa:seven-target-cone-source-condition}. The partition derivative is bounded by a constant times \(Q^{-1} \), with the same support. These identities therefore give
\be
\label{equa:seven-target-cone-component-bounds}
| \mathcal I_W|+Q| \del_Q\mathcal I_W|+Q_+ | \del_{Q_+} \mathcal I_W|+s| \nabla_{\tan,x} \mathcal I_W| \lesssim M I_as^\eta.
\ee
These bounds follow from differentiating the actual representation, not its scalar majorant. Since \(T= (Q+Q_+)/2\), the formula with \(a=0\) yields \(I_0\lesssim T^{-1/2} \), \(I_0/Q\lesssim T^{-1/2}s^{-1} \), and \(I_0/s\lesssim T^{-3/2} \); also \(I_a=2^a(1-a)^{-1}Q^{-a}I_0\). For the radial semihyperboloidal derivative, we use
\be
\label{equa:seven-target-cone-hyperboloidal-target}
\delsH_r\mathcal I_W=-\frac QT\del_Q\mathcal I_W+ \frac{Q_+}{T} \del_{Q_+} \mathcal I_W.
\ee
It is bounded by \(C M I_aT^{-1}s^\eta \); the tangential semihyperboloidal derivatives equal the ordinary tangential derivatives. This proves \eqref{equa:seven-target-cone-differentiated-bound}.

The factor \(q^{-a} \) can compensate for source growth. A bound \(M p^{-3/2}q^{-a} \sigma^{\eta_k+e} \), with \(e\geq0\), gives on \(Q\geq\frac12s^{1-2b} \)
\be
\label{equa:seven-target-cone-growth-compensation}
T^{1/2}s| \del\mathcal I_W|+T^{3/2}| \delsH\mathcal I_W| \lesssim M s^{\eta_k-\gamma}, \qquad \gamma=a(1-2b)-e.
\ee
Thus \(a=1/2\) permits every loss \(e<1/2-b\). This conclusion uses the same weighted bound on \(pW_p\); a bound on \(W\) alone controls only the value of the integral.

All these estimates are uniform as \(r_{\mathrm{tar}} \downarrow0\). Under the smoothness and integrability hypotheses, the integral has a direction-independent value at the center and a locally Lipschitz extension there. The partition depends on \(Q=T-|x| \), so this separate cone contribution need not be differentiable at the center. Condition \eqref{equa:seven-target-cone-source-condition} is an integral-operator hypothesis to be established for the actual signed source, including its outgoing derivative. The full post-integration-by-parts source also has explicit target-dependent factors; their derivatives require their own bounds through \eqref{equa:seven-target-cone-integral-derivatives}.


\section{Metric source estimates for the massless wave} 
\label{section-central-sources}


\subsection{\texorpdfstring{One integration in the incoming source coordinate}{One integration in the incoming source coordinate}}
\label{section-central-incoming-source}

\paragraph{Metric sources on the past light cone}

We now estimate the metric source on the past light cone. A single incoming integration, the wave gauge, and the radial boost control the near-cone interior terms. A Cartesian representation treats the complementary interior region. Three exterior source regions and the trace at the interface complete the metric calculation. All constants keep the dependence on \(B_{s_1}\) in \eqref{equa:seven-cone-refined-amplitude-dependence}.

\paragraph{Principal metric source in the cone integral}

We apply the differentiated integral estimate to the first volume term of the single-integration formula \eqref{equa:seven-cone-invariant-single-integration}. Here the physical source coefficient is independent of the target; the partition is kept outside it. For an operator \(Z\preccurlyeq(d,j) \), \(d\leq p_{\mathrm w} \), we write \(H^Z\) for its ordered tensor Lie derivative and set
\be
\label{equa:seven-cone-single-coefficient-definition}
C_Z^K\coloneqq r'(K^Z)_N, \qquad C_Z^H\coloneqq r'(H^Z)_N, \qquad r'= \tfrac12(p-q).
\ee
The first gauge identity expresses \((C_Z^K)_q\) through \(r'LK^Z\), angular derivatives of \(K^Z\), and \(K^Z\) itself. Its outgoing derivative is controlled by \eqref{equa:seven-cone-outgoing-Hessian-bounds}. Consequently, we obtain
\be
\label{equa:seven-cone-single-densitized-bounds}
|C_Z^K|+p|(C_Z^K)_p| \lesssim A p^{1/2} \sigma^{\nu_{j+3}}, \qquad |(C_Z^K)_q| \lesssim A p^{-1/2} \sigma^{\nu_{j+3}}, \qquad p|(C_Z^K)_{pq}| \lesssim A p^{-1/2} \sigma^{\nu_{j+4}}.
\ee
The passage to the inverse metric keeps the differentiated density. Since \(H=J(K+ \gMink^{-1})-\gMink^{-1} \), with \(J=j_g^{-1} \), tensor Leibniz differentiation and the null contraction give
\be
\label{equa:seven-cone-single-density-products}
C_Z^H= \sum_{Z_1Z_2\preccurlyeq Z}c_{Z_1Z_2}(Z_1J)C_{Z_2}^K.
\ee
The fixed constants \(c_{Z_1Z_2} \) are the Leibniz coefficients. The Minkowski terms vanish because the commutators are Minkowski Killing fields and \(\gMink^{-1}(\ell_N, \ell_N)=0\). For a typical product \(J_1C_2\), the only terms in its two required derivatives are
\be
\label{equa:seven-cone-single-density-derivatives}
(J_1C_2)_q=J_1(C_2)_q+(J_1)_qC_2, \qquad (J_1C_2)_{pq} = (J_1)_p(C_2)_q+J_1(C_2)_{pq} +(J_1)_{pq}C_2 +(J_1)_q(C_2)_p.
\ee
The direct composition estimates and \eqref{equa:seven-cone-outgoing-Hessian-bounds} control the derivatives of \(J_1\). The largest sum of growth exponents is \(\nu_{i+3} + \nu_{j-i+3} = \eta_j+24\rho+10C_{\mathrm{rk}} \theta \). Using \(q\geq1\) and \(A\leq1\), we obtain
\be
\label{equa:seven-cone-single-inverse-bounds}
|(C_Z^H)_q|+p|(C_Z^H)_{pq}| \lesssim A p^{-1/2} \sigma^{\eta_j+e_C}, \qquad e_C\coloneqq24\rho+10C_{\mathrm{rk}} \theta.
\ee
For \(\ord Z_1+ \ord Z_2\leq p_{\mathrm w} \) and total rank at most \(k\), the principal physical source is \(W= (C_{Z_1}^H)_q(Z_2\psi)_q\). The direct matter derivative and its outgoing bound in \eqref{equa:seven-cone-outgoing-Hessian-bounds} therefore prove
\be
\label{equa:seven-cone-single-actual-source}
|W|+p|W_p| \lesssim A^2p^{-3/2}q^{-1/2} \sigma^{\eta_k+e}, \qquad e\coloneqq36\rho+15C_{\mathrm{rk}} \theta.
\ee
The factor \(1/4\) in the bad--bad metric coefficient affects only the constant. If \(\mathcal V_{\mathrm{pr}} \) denotes this scalar volume contribution with its partition, \eqref{equa:seven-target-cone-differentiated-bound} with \(a=1/2\) gives, on the remaining target region,
\be
\label{equa:seven-cone-single-principal-gain}
T^{1/2}| \mathcal V_{\mathrm{pr}}|+T^{1/2}s| \del\mathcal V_{\mathrm{pr}}|+T^{3/2}| \delsH\mathcal V_{\mathrm{pr}}| \lesssim A^2s^{\eta_k-\gamma_{\mathrm{pr}}}, \qquad \gamma_{\mathrm{pr}}{\coloneqq} \tfrac12-41\rho-17C_{\mathrm{rk}} \theta>0.
\ee
This proves a strict gain for the scalar volume of the actual tensorially commuted metric coefficient, including its density. It uses the reduced equations only to estimate lightlike Hessians; the massive source is never differentiated in an outgoing direction.

\paragraph{Angular terms and derivatives of the cone partition}

We factor the partition as \(C= \chi(q/Q)C_0\) in \eqref{equa:seven-cone-invariant-single-integration}. The angular term is \(\chi\mathbf B\cdot\mathbf F\), where \(\mathbf F= \nabla_{\mathbb S^2}(C_0\del_qv) \) is independent of the target. More generally, for a physical source vector satisfying
\be
\label{equa:seven-angular-one-integration-source}
| \mathbf F|+p| \del_p\mathbf F| \leq M p^{-1/2}q^{-1/2} \sigma^\eta, \qquad \eta \geq0,
\ee
we prove the following estimate, with \(z=Q_+/Q\):
\be
\label{equa:seven-angular-one-integration-target}
T^{1/2}| \mathcal I_{\mathbf B\cdot\mathbf F}|+T^{1/2}s| \del\mathcal I_{\mathbf B\cdot\mathbf F}|+T^{3/2}| \delsH\mathcal I_{\mathbf B\cdot\mathbf F}| \lesssim M\sqrt z\,Q^{-1/2}s^\eta.
\ee
The relevant positive kernel has the value
\be
\label{equa:seven-angular-one-integration-kernel}
J_{1/2}{\coloneqq} \frac1{r_{\mathrm{tar}}} \int_Q^{Q_+} \int_0^{Q/2}p^{-1/2}q^{-1/2} \, dq\, dp=sI_{1/2}.
\ee
Indeed, the ratio of the two \(p\)-integrals is \(\sqrt{QQ_+} \). At fixed physical source point, we have \(\del_p\mathbf B=0\) and
\be
\label{equa:seven-angular-one-integration-explicit-derivatives}
\del_Q^{\mathrm{exp}} \mathbf B=-\frac{\Pi_\omega n}{2(Q-q)^2}, \quad \del_{Q_+}^{\mathrm{exp}} \mathbf B= \frac{\Pi_\omega n}{2(Q_+-q)^2}, \quad r_{\mathrm{tar}}^{-1} \del_n^{\mathrm{exp}} \mathbf B[e]= \frac{\Pi_\omega e}{D_{\mathrm c}}, \quad e\cdot n=0.
\ee
We put \(d=r_{\mathrm{tar}}/s\). The source-coordinate terms in the differentiated representation are bounded using \(| \mathbf B| \lesssim r_{\mathrm{tar}}s^{-2} \) and \(J_{1/2} =sI_{1/2} \); the explicit terms use \eqref{equa:seven-angular-one-integration-explicit-derivatives}. Together they give
\be
\label{equa:seven-angular-one-integration-component-bounds}
\begin{aligned}
| \mathcal I_{\mathbf B\cdot\mathbf F}|& \lesssim MdI_{1/2}s^\eta, \qquad Q| \del_Q\mathcal I_{\mathbf B\cdot\mathbf F}| \lesssim M(d+s/Q)I_{1/2}s^\eta, \\
Q_+ | \del_{Q_+} \mathcal I_{\mathbf B\cdot\mathbf F}|& \lesssim M(d+s/Q_+)I_{1/2}s^\eta, \qquad s| \nabla_{\tan,x} \mathcal I_{\mathbf B\cdot\mathbf F}| \lesssim M(1+d)I_{1/2}s^\eta.
\end{aligned}
\ee
For the radial semi-hyperboloidal derivative, the source-coordinate terms have size \(MdT^{-1}I_{1/2}s^\eta \), while the explicit coefficient term has size \(Ms^{-2}J_{1/2}s^\eta \) by \eqref{equa:seven-cone-invariant-good-target-derivative}. Since \(d/T\leq s^{-1} \), their sum is bounded by \(MI_{1/2}s^{-1+ \eta} \). The kernel bounds and \(1+d\lesssim\sqrt z\) now prove \eqref{equa:seven-angular-one-integration-target}, uniformly towards the center.

For the actual metric coefficient \(C_0=C_{Z_1}^H/4\) and \(v=Z_2\psi\), the direct value and outgoing bounds for smooth metric compositions give \(|C_0|+p|(C_0)_p| \lesssim A p^{1/2} \sigma^{\nu_{i+3}} \). One angular derivative costs one further boost. Combining these bounds with \eqref{equa:seven-cone-one-angular-order} proves \eqref{equa:seven-angular-one-integration-source} with
\be
\label{equa:seven-angular-one-integration-actual-source}
M\lesssim A^2, \qquad \eta= \eta_k+e_{\mathrm{ang}}, \qquad e_{\mathrm{ang}}{\coloneqq}24\rho+12C_{\mathrm{rk}} \theta+b+ \varepsilon_N.
\ee
The partition term is \(Q^{-1} \chi'(q/Q)C_0\del_qv\). The same differentiated formulas, now with the scalar kernel \(J_{1/2} \), give the bound in \eqref{equa:seven-angular-one-integration-target}: differentiating the explicit factor \(Q^{-1} \) costs \(Q^{-2} \), and \(J_{1/2}/Q= \sqrt z I_{1/2} \). The derivative \(\chi''\) is bounded on the same partition region. Its source needs no additional angular order and is covered by \(e_{\mathrm{ang}} \).

Finally, \(\sqrt z=Q_+/s\leq2s^{2b} \) and \(Q^{-1/2} \lesssim s^{-1/2 +b} \) on the remaining target region. Thus the angular and partition volumes have the strict exponent gap
\be
\label{equa:seven-angular-one-integration-gain}
\gamma_{\mathrm{ang}}{\coloneqq} \tfrac12-3b-e_{\mathrm{ang}} = \tfrac12-44\rho-20C_{\mathrm{rk}} \theta-\varepsilon_N>0.
\ee
Together with \eqref{equa:seven-cone-single-principal-gain}, this estimates the interior near-cone volume of the bad--bad metric source. Its signed interface trace is kept and estimated in \eqref{equa:seven-exterior-metric-retained-trace}.


\subsection{\texorpdfstring{Radial boost reduction of the metric source}{Radial boost reduction of the metric source}}
\label{section-central-other-metric}

\paragraph{Radial and angular decomposition}

The remaining metric terms contain a good derivative. We express their second radial derivatives through the radial boost, so that their outgoing source derivatives use only the lightlike Hessians already estimated. In source coordinates \((p,q, \omega) \), with \(r'= (p-q)/2\), we set
\be
\label{equa:seven-radial-boost-definition}
L_r\coloneqq \omega^aL_a=p\del_p-q\del_q, \qquad L_r\omega=0, \qquad [\del_p,L_r]= \del_p, \qquad [\del_q,L_r]=-\del_q.
\ee
Consequently, \(L_r^2v= \omega^a \omega^bL_aL_bv\), and direct expansion gives
\be
\label{equa:seven-radial-boost-Hessians}
v_{pq} =p^{-1} \bigl(\del_q(L_rv) +v_q+qv_{qq} \bigr), \qquad v_{pp} =p^{-2} \bigl((L_r^2-L_r)v+2q\del_q(L_rv) +q^2v_{qq} \bigr).
\ee
For a symmetric Cartesian tensor \(P\), we write \(P^{r0} = \omega_aP^{a0} \), \(P^{rr} = \omega_a \omega_bP^{ab} \), and use the source-sphere projection \(\Pi =I-\omega \otimes\omega \). Its relevant components are
\be
\label{equa:seven-radial-metric-components}
P_N=P^{00}-2P^{r0} +P^{rr}, \quad P_L=P^{00} +2P^{r0} +P^{rr}, \quad F_P=P^{00}-P^{rr}, \quad T_P= \Pi_{ab}P^{ab}.
\ee
We denote the projected vector components by \(P_{T0} = \Pi P^{\cdot0} \) and \(P_{Tr} = \Pi P\omega \), and the tangential spatial tensor by \(P_{TT} = \Pi P\Pi\). All sphere derivatives below are at fixed \((p,q) \). The angular and radial-frame terms in the Cartesian Hessian are
\be
\label{equa:seven-radial-metric-frame-terms}
\begin{aligned}
E_P[v]&= \frac2{r'} \bigl((P_{T0} +P_{Tr}) \cdot\nabla_{\mathbb S^2}v_p+(P_{T0}-P_{Tr}) \cdot\nabla_{\mathbb S^2}v_q\bigr) \\
& \quad + \frac{T_P}{r'}(v_p-v_q) + \frac1{r'^2} \bigl(P_{TT}:\nabla_{\mathbb S^2}^2v-2P_{Tr} \cdot\nabla_{\mathbb S^2}v\bigr).
\end{aligned}
\ee
The last term includes the derivative of the radial frame. Substitution of \eqref{equa:seven-radial-boost-Hessians} into \(P^{\alpha \beta} \del_{\alpha \beta}v=P_Nv_{qq} +2F_Pv_{pq} +P_Lv_{pp} +E_P[v] \) yields the decomposition
\be
\label{equa:seven-radial-metric-full-decomposition}
P^{\alpha \beta} \del_{\alpha \beta}v= (P_N+a_P)v_{qq} +R_P[v], \qquad a_P\coloneqq2\frac qpF_P+ \frac{q^2}{p^2}P_L,
\ee
where the regular part is
\be
\label{equa:seven-radial-metric-regular-part}
R_P[v]= \frac{P_L}{p^2}(L_r^2-L_r)v+ \frac2p\Bigl(F_P+ \frac qpP_L\Bigr) \del_q(L_rv) + \frac{2F_P}p v_q+E_P[v].
\ee
The combined bad coefficient also equals \((4t^2/p^2)P^{\H00} \). We estimate \(a_P\) separately: its explicit factor \(q/p\) gives the required incoming derivative bound.

\paragraph{The regular metric volume}

We take \(P=H^{Z_1} \), \(v=Z_2\psi\), with \(\ord Z_1+ \ord Z_2\leq p_{\mathrm w} \) and total rank at most \(k\). Direct composition estimates and \eqref{equa:seven-cone-outgoing-Hessian-bounds} give, when \(Z_1\) has rank \(i\),
\be
\label{equa:seven-radial-metric-coefficient-bounds}
|P|+p|P_p| \lesssim A p^{-1/2} \sigma^{\nu_{i+3}}, \qquad |P_q|+p|P_{pq}| \lesssim A p^{-1}q^{-1/2} \sigma^{\nu_{i+3}}.
\ee
The matter factors in \(R_P[v] \) have either one derivative and one further boost or rotation, or a value with two further boosts or rotations. The first type uses \eqref{equa:seven-cone-one-angular-order}; the second uses direct Hardy--Sobolev and \(L=t^{-1} \omega^aL_a+(q/t) \del_t\). Thus, for an operator \(Y\) of order at most \(p_{\mathrm w} +1\) and rank \(j\), and for \(0\leq\ell\leq2\), respectively,
\be
\label{equa:seven-radial-metric-matter-factors}
| \del Y\psi|+p| \del_p\del Y\psi| \lesssim A p^{-1}q^{-1/2} \sigma^{\nu_{j+3} +b+ \varepsilon_N}, \qquad | \Omega^\ell Z_2\psi|+p| \del_p\Omega^\ell Z_2\psi| \lesssim A p^{-1/2} \sigma^{\nu_{\rank Z_2 + \ell+3}}.
\ee
For \(W_{\mathrm{reg}} =r'R_P[v] \), these estimates prove
\be
\label{equa:seven-radial-metric-regular-source}
|W_{\mathrm{reg}}|+p|(W_{\mathrm{reg}})_p| \lesssim A^2p^{-3/2}q^{-1/2} \sigma^{\eta_k+e_{\mathrm{reg}}}, \qquad e_{\mathrm{reg}}{\coloneqq}24\rho+12C_{\mathrm{rk}} \theta+b+ \varepsilon_N.
\ee
Indeed, the terms containing \((L_r^2-L_r)v\) or two sphere derivatives of \(v\) have weight \(A^2p^{-2} \), which is at most \(A^2p^{-3/2}q^{-1/2} \). Their maximal growth exponent is \(\nu_{i+3} + \nu_{k-i+5} = \eta_k+24\rho+12C_{\mathrm{rk}} \theta \). The terms containing \(\del_q(L_rv) \), \(v_q\), or one sphere derivative of \(v_p,v_q\) have weight \(A^2p^{-3/2}q^{-1/2} \); the extra matter order contributes at most \(b+ \varepsilon_N\). Since \(\omega_p= \omega_q=0\), differentiating \(\del_q(L_rv) \) in \(p\) uses \(\omega^a \del_p\del_qL_av\), already covered by \eqref{equa:seven-cone-one-angular-order}. The same estimates apply after the outgoing derivative hits a radial coefficient or a projected tensor component.

The scalar cone estimate \eqref{equa:seven-target-cone-growth-compensation} therefore controls this volume, with the strict gap
\be
\label{equa:seven-radial-metric-regular-gap}
\gamma_{\mathrm{reg}} = \tfrac12-b-e_{\mathrm{reg}} = \tfrac12-34\rho-16C_{\mathrm{rk}} \theta-\varepsilon_N>0.
\ee

\paragraph{Additional coefficient of the second incoming derivative}

We put \(C_{\mathrm{ex}} =r'a_P=A_1F_P+A_2P_L\), where
\be
\label{equa:seven-radial-metric-extra-coefficients}
A_1=q-q^2/p, \quad A_2=q^2/(2p)-q^3/(2p^2), \quad |A_m| \lesssim q, \quad |(A_m)_q|+p|(A_m)_{pq}| \lesssim1, \quad |(A_m)_p| \lesssim q/p.
\ee
The projections defining \(F_P,P_L\) are independent of \(p,q\). The product rule and \eqref{equa:seven-radial-metric-coefficient-bounds} give
\be
\label{equa:seven-radial-metric-extra-bounds}
|C_{\mathrm{ex}}|+p|(C_{\mathrm{ex}})_p| \lesssim A p^{1/2} \sigma^{\nu_{i+3}}, \qquad |(C_{\mathrm{ex}})_q|+p|(C_{\mathrm{ex}})_{pq}| \lesssim A p^{-1/2} \sigma^{\nu_{i+3}}.
\ee
In particular, the mixed term \(pA_mP_{pq} \) has weight \(A p^{-1}q^{1/2} \), bounded by \(A p^{-1/2} \); the other differentiated products have the same bound. A single incoming integration applies to \(C_{\mathrm{ex}}v_{qq} \). Its scalar volume satisfies \eqref{equa:seven-target-cone-source-condition} with \(a=1/2\) and growth \(\eta_k+24\rho+10C_{\mathrm{rk}} \theta \). One angular derivative of \(C_{\mathrm{ex}}v_q\) has the bound \eqref{equa:seven-angular-one-integration-source}, with growth \(\eta_k+e_{\mathrm{reg}} \). Its angular and partition volumes therefore have the gap \(\gamma_{\mathrm{ang}} \) in \eqref{equa:seven-angular-one-integration-gain}.

Combining this calculation with the estimate for \(P_Nv_{qq} \), the interior near-cone volume \(\mathcal V_{\mathrm{met}} \) of the principal metric source satisfies
\be
\label{equa:seven-radial-metric-total-gain}
T^{1/2}| \mathcal V_{\mathrm{met}}|+T^{1/2}s| \del\mathcal V_{\mathrm{met}}|+T^{3/2}| \delsH\mathcal V_{\mathrm{met}}| \lesssim A^2s^{\eta_k-\gamma_{\mathrm{ang}}}.
\ee
The largest derivative count is \(p_{\mathrm w} +5\leq p_{\mathbf{high}} \), equal to \(12\leq14\) when \(N=19\). The calculation uses lightlike matter Hessians at order at most \(p_{\mathrm w} +1\). The incoming integration keeps its interface trace, which must be added to the exterior contribution.


\subsection{\texorpdfstring{Metric source integrals in the complementary interior region}{Metric source integrals in the complementary interior region}}
\label{section-central-complement}

\paragraph{Cartesian differentiation of the cone integral}

On the complementary source region, Cartesian differentiation avoids the factor \((Q-q)^{-1} \) in the lightlike coordinates. We put \(h=1-\chi\) for positive arguments, so that \(h(q/Q)=0\) for \(q\leq Q/4\) and \(h(q/Q)=1\) for \(q\geq Q/2\). For \(Q>4T_0\), a Cartesian source \(F\) contributes
\be
\label{equa:seven-complement-cartesian-integral}
\mathcal I_F^{\mathrm c}(T,x)= \frac1{4\pi} \int_{T_0}^T(T-\tau) \int_{\mathbb S^2}h(q/Q)F(\tau,y) \, d\omega \, d\tau, \qquad y=x+(T-\tau) \omega.
\ee
Here \(p= \tau+ |y| \), \(q= \tau-|y| \), and \(\sigma= \sqrt{pq} \). On the support of this integral, \(p\geq Q\), \(Q/4\leq q\leq Q\), \(p\leq Q_+ \), and \(\sigma \leq s\). Suppose that, for \(\eta,e\geq0\),
\be
\label{equa:seven-complement-cartesian-source}
|F| \leq M p^{-3/2}q^{-3/2} \sigma^{\eta+e}, \qquad | \nabla_yF| \leq M p^{-3/2}q^{-5/2} \sigma^{\eta+e}.
\ee
Differentiating \eqref{equa:seven-complement-cartesian-integral} at fixed \((\tau, \omega) \), a spatial target derivative acts on \(F\) as \(\nabla_y\), whereas the time derivative acts as \(\omega \cdot\nabla_y\). The latter also differentiates \(T-\tau\); its upper-limit term vanishes. Every derivative of the partition costs at most \(C/Q\). Since \(p\geq\tau\) and \(q\geq Q/4\), the required scalar integrals satisfy
\be
\label{equa:seven-complement-cartesian-kernels}
\int_{Q/4}^T(T-\tau) \tau^{-3/2}Q^{-3/2} \, d\tau\lesssim TQ^{-2}, \qquad \int_{Q/4}^T(T-\tau) \tau^{-3/2}Q^{-5/2} \, d\tau+ \int_{Q/4}^T\tau^{-3/2}Q^{-3/2} \, d\tau\lesssim TQ^{-3}.
\ee
Thus the actual representation and its derivatives obey
\be
\label{equa:seven-complement-cartesian-bounds}
| \mathcal I_F^{\mathrm c}| \lesssim MTQ^{-2}s^{\eta+e}, \qquad | \del\mathcal I_F^{\mathrm c}|+ | \delsH\mathcal I_F^{\mathrm c}| \lesssim MTQ^{-3}s^{\eta+e}.
\ee
The partition is constant near \(y=0\), since there \(q=p\geq Q\). Hence source-axis derivatives cause no singularity. At the target center, the separate contribution is locally Lipschitz, as in the near-cone integral; the partition derivatives cancel when the two contributions are added. The lower initial trace is absent because \(\tau\geq q\geq Q/4>T_0\). On \(T\leq s^{1+2b} \), the bound \(Q\geq\frac12s^{1-2b} \) gives
\be
\label{equa:seven-complement-cartesian-gain}
T^{1/2}| \mathcal I_F^{\mathrm c}|+T^{1/2}s| \del\mathcal I_F^{\mathrm c}|+T^{3/2}| \delsH\mathcal I_F^{\mathrm c}| \lesssim M s^{\eta-\gamma}, \qquad \gamma{\coloneqq} \tfrac12-11b-e.
\ee

\paragraph{The third matter derivative}

We use the smooth semi-hyperboloidal frame \(\delsH_a= \del_{y_a} +(y_a/\tau) \del_\tau\), including at \(y=0\). The lightlike Hessian proof in \eqref{equa:seven-cone-light-like-Hessian-reduction} extends to \(p/2\leq q\leq p\): in that region \(p,q, \sigma, \tau\) are comparable and the frame coefficients are smooth. We write \(c=C_{\mathrm{rk}} \theta \) in this calculation. If \(G_2[w] \) denotes any Hessian component with a semi-hyperboloidal derivative, its expression uses \(\tau^{-1} \del L_aw\), \(\tau^{-1} \del w\), and bounded smooth coefficients. Consequently, for \(\ord Z\leq p_{\mathrm w} \), \(\rank Z=j\), we have
\be
\label{equa:seven-complement-good-Hessian}
|G_2[Z\psi]| \lesssim A p^{-2}q^{-1/2} \sigma^{\eta_j+12\rho+5c}, \qquad | \del G_2[Z\psi]| \lesssim A p^{-2}q^{-3/2} \sigma^{\eta_j+17\rho+8c+ \varepsilon_N}.
\ee
The second bound uses \eqref{equa:seven-cone-one-angular-order} at one further boost. The third Cartesian derivative satisfies
\be
\label{equa:seven-complement-third-matter-derivative}
| \del^3Z\psi| \lesssim A p^{-1}q^{-5/2} \sigma^{\eta_j+36\rho+16c+ \varepsilon_N} +A^2p^{-1}q^{-2} \sigma^{\eta_j+8\rho+4c}.
\ee
To prove this estimate, we apply the lightlike Hessian reduction to \(\del Z\psi\) and induct on the commutation order. Its tangential term is a known Hessian at one further order and rank, multiplied by \(q^{-1} \), and has the first weight in \eqref{equa:seven-complement-third-matter-derivative}. In the differentiated matter source, the largest product is \(\del^2Z_1\fieldPhi\, \del Z_2\psi\). Both factors admit direct estimates from the lower energy level:
\be
\label{equa:seven-complement-lower-energy-product}
| \del^2Z_1\fieldPhi| \lesssim A p^{-1}q^{-1/2} \sigma^{\alpha_{i+2}}, \qquad | \del Z_2\psi| \lesssim A p^{-1}q^{-1/2} \sigma^{\beta_{j-i+2}}.
\ee
The first factor uses \(\ord Z_1+3\leq p_{\mathrm w} +3\leq p_{\mathbf{low}} \), equal to \(10\) at \(N=19\). Thus its estimate uses the lower effective-field energy, with no additional \(\sigma^{1/2} \). Multiplication by \(p/q\) from the Hessian reduction gives the second term in \eqref{equa:seven-complement-third-matter-derivative}, because \(\alpha_{i+2} + \beta_{j-i+2} = \eta_j+8\rho+4c\). When the translation falls on the matter factor instead, the refined effective-field derivative and \eqref{equa:seven-cone-one-angular-order} give the first weight or a stronger one.

A differentiated metric coefficient paired with a known matter Hessian gives \(A^2p^{-1}q^{-3} \), after the reduction, and is bounded by the first term. The remaining bad principal commutators have a lower matter commutation order; their coefficient is bounded by \(C_{N,B_{s_1}}p^{-1} \sigma^{\eta_i} \), and hence adds only \(C_{N,B_{s_1}}/q\) to the inductive estimate. The last coefficient order at \(N=19\) contains a translation and has rank at most \(p_{\mathrm w} \), within the refined coefficient range. The other commutators use \eqref{equa:seven-complement-good-Hessian}. Derivatives of smooth frame coefficients add inverse-time powers. Finite induction preserves both weights; the first growth loss \(36\rho+16c+ \varepsilon_N\) bounds their sums. The conservative derivative count is still \(p_{\mathrm w} +5\leq p_{\mathbf{high}} \).

A weaker third-derivative bound will avoid this exceptional loss in the Hessian product. We apply the same reduction to \(\del Z\psi\), but estimate the tangential Hessian \(\del^2L_aZ\psi\) directly from the high energy. Its contribution is \(A p^{-1}q^{-3/2} \sigma^{\nu_{j+3}} \). The differentiated matter source is controlled by \eqref{equa:seven-complement-lower-energy-product}; its growth \(\eta_j+8\rho+4c\) is smaller than \(\nu_{j+3} \). Every last-order metric coefficient contains a translation and has rank at most \(p_{\mathrm w} \). The finite induction therefore uses only \(B_{s_1}\), and gives
\be
\label{equa:seven-cone-weak-third-matter-jet}
| \del^3Z\psi| \lesssim A p^{-1}q^{-3/2} \sigma^{\nu_{j+3}}, \qquad |L\del^2Z\psi| \lesssim A p^{-2}q^{-1/2} \sigma^{\nu_{j+3}}.
\ee
For the second estimate, we use \(L=t^{-1} \omega^aL_a+(q/t) \del_t\) and direct high Sobolev on \(L_a \del^2Z\psi\). Both bounds cost at most \(p_{\mathrm w} +4\) high orders. Their weaker power of \(q\) is sufficient for the product estimate below.

\paragraph{Metric source and exponent gap}

For \(F_{\mathrm{met}} =H^{Z_1\, \alpha \beta} \del_{\alpha \beta}Z_2\psi\), with total order at most \(p_{\mathrm w} \) and total rank at most \(k\), we separate \((H^{Z_1})^{\H00} \del_t^2Z_2\psi\) from the good Hessians. The direct metric value bound gives the source estimate in \eqref{equa:seven-complement-cartesian-source}. In its spatial derivative, the four possible weights, apart from growth factors, are
\be
\label{equa:seven-complement-metric-gradient-weights}
A^2p^{-2}q^{-2}, \qquad A^2p^{-3/2}q^{-5/2}, \qquad C_{N,B_{s_1}}A^2p^{-2}q^{-2}, \qquad A^2p^{-5/2}q^{-3/2}.
\ee
They arise, respectively, when the derivative hits the coefficient, the first term of \eqref{equa:seven-complement-third-matter-derivative}, its second term, or a good Hessian. Only the third case needs the refined bad coefficient bound \(|(H^{Z_1})^{\H00}| \lesssim C_{N,B_{s_1}}p^{-1} \sigma^{\eta_i} \); the others keep the direct factor \(A\). This distinction preserves the overall amplitude \(A^2\) without comparing the trace amplitude with \(A\). Since \(p\geq q\geq1\), all four weights are bounded by \(C_{N,B_{s_1}}A^2p^{-3/2}q^{-5/2} \). The largest growth loss is
\be
\label{equa:seven-complement-metric-gap}
e_{\mathrm{met}}{\coloneqq}48\rho+20c+ \varepsilon_N, \qquad \gamma_{\mathrm{met}}{\coloneqq} \tfrac12-11b-e_{\mathrm{met}} = \tfrac12-103\rho-42c-\varepsilon_N>\tfrac16.
\ee
Therefore \eqref{equa:seven-complement-cartesian-gain} estimates the complementary interior metric volume with \(M=C_{N,B_{s_1}}A^2\), \(\eta= \eta_k\), and gap \(\gamma_{\mathrm{met}} \). The lower bound follows from \(\delta<1/8\), \(\rho= \delta/64\), and \(c<\rho/23\). Together with \eqref{equa:seven-radial-metric-total-gain}, this treats the principal metric volume throughout the interior source region. The separate matter source, the exterior volume, and the interface terms remain necessary for \eqref{equa:seven-sufficient-massless-gain}.


\subsection{\texorpdfstring{Exterior metric sources and traces at the interface}{Exterior metric sources and traces at the interface}}
\label{section-central-exterior-metric}

\paragraph{Three exterior source regions}

The exterior principal source can be estimated without an incoming integration. We use the closing exterior energy and the weighted Sobolev and Hardy bounds proved in Section~\ref{section-central-exterior-matter}. We choose \(0<\delta \leq(\mu-\lambda)/2\), within the prescribed strict gap between the exterior weights, and put \(a= \delta/\lambda \). Then \(\delta<1/8\) and \(a<1/6\). With \(w=1+(-q)_+ \), a smooth physical partition separates the following regions:
\begin{itemize}
\item The exterior collar \(w\leq4p^a \), with cutoff equal to one for \(w\leq2p^a \), where direct scalar derivatives suffice.
\item The intermediate region \(w\geq2p^a \), \(q\geq-p/3\), where \(r'\simeq\tau\), \(|q| \simeq w\), and the wave equation improves the incoming Hessian.
\item The strictly spacelike region \(q\leq-p/4\), where \(r'\geq5\tau/3\) and a radial--tangential reduction controls the Cartesian Hessian.
\end{itemize}
The partition can be chosen from smooth functions of \(w/p^a \) and \(q/p\), with a smooth equivalent of \(w\) in the first cutoff. Each cutoff has bounded \(p\del_p\)-derivative. It depends only on the source, and its three terms sum to one. We consider a commuted source \(P^{\alpha \beta} \del_{\alpha \beta}v\), where \(P=H^{Z_1} \), \(v=Z_2\psi\), and the total order is at most \(p_{\mathrm w} \). For brevity in this calculation, we write \(E=p^{\delta/2} \); the direct exterior estimates give
\be
\label{equa:seven-exterior-metric-direct-bounds}
| \del Z\psi|+ | \del Z\fieldPhi| \lesssim Ap^{-1}w^{-\mu}E, \quad |ZH|+p|LZH| \lesssim Ap^{-1}w^{1-\lambda}E, \quad | \del ZH| \lesssim Ap^{-1}w^{-\lambda}E.
\ee
The prescribed reference metric satisfies the same inequalities. The derivative orders used below are counted explicitly.

\paragraph{The direct collar estimate}

Applying \eqref{equa:seven-exterior-metric-direct-bounds} at one additional translation order and using \(L= (r')^{-1}(\omega^aL_a-q\del_{r'}) \) gives
\be
\label{equa:seven-exterior-collar-Hessian}
| \del^2Z\psi| \lesssim Ap^{-1}w^{-\mu}E, \qquad p|L\del^2Z\psi| \lesssim Ap^{-1}w^{1-\mu}E.
\ee
The second inequality uses at most \(p_{\mathrm w} +2 +5\) energy orders, including the exterior Sobolev step. Thus the physical radial source \(W_{\mathrm{col}} =r'P^{\alpha \beta} \del_{\alpha \beta}v\), multiplied by its collar cutoff, satisfies
\be
\label{equa:seven-exterior-collar-metric-source}
|W_{\mathrm{col}}|+p|(W_{\mathrm{col}})_p| \lesssim A^2p^{-1+ \delta}w^{2-\lambda-\mu}, \qquad \int\bigl(|W_{\mathrm{col}}|+p|(W_{\mathrm{col}})_p| \bigr) \, dq\lesssim A^2p^{-\alpha_{\mathrm{col}}},
\ee
where \(\alpha_{\mathrm{col}} =1-\delta-a(3-\lambda-\mu)>1-3\delta \). Indeed, \(0<2-\lambda-\mu<1/2\), so integration over \(w\leq4p^a \) contributes \(p^{a(3-\lambda-\mu)} \). For large \(p\), this collar has \(\tau\geq p/4\) and lies above the initial-time slice.

\paragraph{Hessians in the intermediate region}

Here \(| \tau^2-(r')^2| \simeq pw\). The semi-hyperboloidal Hessian calculation is algebraic and gives, with this absolute value,
\be
\label{equa:seven-exterior-intermediate-Hessian-reduction}
| \del^2f| \lesssim w^{-1} \bigl(| \del L_bf|+ | \del f| \bigr) +(p/w)| \Box_gf|.
\ee
The perturbation is absorbed because \((p/w)|H| \lesssim AEw^{-\lambda} \lesssim Ap^{-\delta/2} \). Applying the matter equation and inducting on the commutation order yields
\be
\label{equa:seven-exterior-intermediate-matter-jets}
\begin{aligned}
| \del^2Z\psi|& \lesssim Ap^{-1}w^{-1-\mu}E, && \ord Z\leq p_{\mathrm w} +1, \\
| \del^3Z\psi|& \lesssim Ap^{-1}w^{-2-\mu}E+A^2p^{-1}w^{-1-2\mu}E^2, && \ord Z\leq p_{\mathrm w}.
\end{aligned}
\ee
The first estimate uses the direct derivative of \(L_bZ\psi\). Its matter source has the additional bounded factor \(AEw^{-\mu} \). A metric commutator has a lower Hessian and the factor \(AEw^{-\lambda} \), so finite induction applies. For the second estimate, the tangential term is the first estimate at one additional boost order. The largest differentiated matter product is \(\del^2\fieldPhi\, \del\psi\), for which the effective-field Hessian is estimated directly. This produces the second weight, with no negative power of \(\coeff\). The largest count is \(p_{\mathrm w} +7\).

Using the outgoing identity again, the intermediate radial source obeys
\be
\label{equa:seven-exterior-intermediate-metric-source}
|W_{\mathrm{mid}}|+p|(W_{\mathrm{mid}})_p| \lesssim A^2p^{-1+ \delta}w^{-\lambda-\mu} +A^3p^{-1+3\delta/2}w^{1-\lambda-2\mu}.
\ee
Both incoming weights are integrable, since \(\lambda+ \mu>1\) and \(\lambda+2\mu>2\). Their integral is bounded by \(CA^2p^{-1+3\delta/2} \). The partition derivatives have the same bound, and \(\tau\geq p/3\) removes the initial-time intersection for large targets.

\paragraph{The Cartesian Hessian in the spacelike region}

The exterior hypersurface is Euclidean in this region, with \(\zeta=1\). Simultaneous \(H^2\) Sobolev on a unit radial cylinder times \(\mathbb S^2\) gives
\be
\label{equa:seven-exterior-spacelike-Sobolev}
r'w^\mu|f| \lesssim\sum_{b+ |J| \leq2} \|w^\mu\del_{r'}^b\Omega^Jf\|_{L^2(\mathcal A_{r'})}.
\ee
The cylinder may be enlarged by a fixed amount. In sphere charts this is the three-dimensional \(H^2\) inequality; its Fourier proof uses the integrability of \((1+ | \xi|^2)^{-2} \). Two rotations cost four boosts, a mixed radial--rotation derivative costs three, and two radial derivatives cost two. Thus this local estimate costs four commutations, whereas the merging region used five.

We use \(\Pi =I-\omega \otimes\omega \) and \(\Omega_{ca} =y_c\del_a-y_a \del_c\), with \(\omega_0=0\). Differentiation gives, for every Cartesian index \(\beta \),
\be
\label{equa:seven-exterior-spacelike-tangential-Hessian}
r'\Pi_a^{\ i} \del_{i\beta}f= \sum_c\omega_c\del_\beta(\Omega_{ca}f)-\omega_\beta \del_af+ \delta_{a \beta} \del_{r'}f.
\ee
Thus every Hessian component with one spatial tangential index is bounded by \((r')^{-1}(| \del\Omega f|+ | \del f|) \). The remaining three radial--time components are determined by \(L_rf_\tau=r'f_{\tau\tau} + \tau f_{\tau r'} \), \(L_rf_{r'} =r'f_{\tau r'} + \tau f_{r'r'} \), and the wave trace. Elimination yields
\be
\label{equa:seven-exterior-spacelike-radial-Hessian}
\Bigl(1-\frac{\tau^2}{(r')^2} \Bigr)f_{r'r'} = \Box f-\Pi^{ab} \del_{ab}f+(r')^{-1}L_rf_\tau-\tau(r')^{-2}L_rf_{r'}.
\ee
Since \(\tau/r'\leq3/5\), the radial coefficient is bounded away from zero. Commuting a boost past one derivative adds a first derivative. After absorption of the small metric perturbation, we obtain the full Cartesian inequality
\be
\label{equa:seven-exterior-spacelike-full-Hessian}
| \del^2f| \lesssim(r')^{-1} \bigl(| \del L_bf|+ | \del\Omega_{ab}f|+ | \del f| \bigr) + | \Box_gf|.
\ee
We apply it to \(Z\psi\) and then to \(\del Z\psi\), using the direct exterior scalar bounds and induction on the order for the metric commutators. The last angular term is \(\del\Omega \del Z\psi\), a first jet at operator order at most \(p_{\mathrm w} +3\). The four-order Sobolev estimate therefore requires \(p_{\mathrm w} +7\leq p_{\mathbf{high}} \), including at \(N=19\). The matter source and its derivative have weight \(A^2p^{-2-2\mu}E^2\), bounded by the linear term. Consequently, we obtain
\be
\label{equa:seven-exterior-spacelike-matter-jets}
| \del^2Z\psi|+ | \del^3Z\psi| \lesssim Ap^{-2-\mu}E, \qquad \ord Z\leq p_{\mathrm w}.
\ee
All coefficients in this argument involve \(y/r'\) and \(\tau/r'\), which are bounded here. Combining \eqref{equa:seven-exterior-spacelike-matter-jets} with the metric coefficient gives
\be
\label{equa:seven-exterior-spacelike-metric-source}
|W_{\mathrm{far}}|+p|(W_{\mathrm{far}})_p| \lesssim A^2p^{-\lambda-\mu+ \delta}, \qquad \int\bigl(|W_{\mathrm{far}}|+p|(W_{\mathrm{far}})_p| \bigr) \, dq\lesssim A^2p^{-\alpha_{\mathrm{far}}},
\ee
where \(\alpha_{\mathrm{far}} = \lambda+ \mu-1-\delta \). The incoming interval has length at most \(Cp\). The largest outgoing product is \(r'pH\del^3\psi\), so the second- and third-derivative weight in \eqref{equa:seven-exterior-spacelike-matter-jets} suffices.

\paragraph{The initial strip and the interface trace}

The time cutoff used in Section~\ref{section-central-exterior-matter} provides a smooth zero extension of the source below \(T_0\). Its outgoing derivative is supported on a bounded incoming interval, where the sharper source value gives \(|p\chi_0'W_{\mathrm{far}}| \lesssim A^2p^{-\lambda-\mu} \). The complementary time strip can be estimated directly by its Cartesian Kirchhoff integral. For large \(Q\), its source points have \(r'\geq Q/2\), and
\be
\label{equa:seven-exterior-metric-initial-source}
|P^{\alpha \beta} \del_{\alpha \beta}v|+ | \nabla_y(P^{\alpha \beta} \del_{\alpha \beta}v)| \lesssim C_{N,T_0}A^2(r')^{-2-\lambda-\mu}.
\ee
Differentiation of the actual integral at fixed source time and sphere direction gives a value bound \(CA^2TQ^{-2-\lambda-\mu} \) and the same bound for its Cartesian derivatives. In the normalized target norm, this has positive gap \(\lambda+ \mu-1/2-(9+2\lambda+2\mu)b\).

The three unbounded source regions share the exponent
\be
\label{equa:seven-exterior-metric-c-exponent}
\alpha_{\mathrm{met}}{\coloneqq} \min\{1-3\delta, \lambda+ \mu-1-\delta \}, \qquad \tfrac12 + \delta<\alpha_{\mathrm{met}}<1.
\ee
The differentiated negative-\(q\) cone identities in Section~\ref{section-central-exterior-matter} therefore give
\be
\label{equa:seven-exterior-metric-target-gain}
T^{1/2}| \mathcal I_{\mathrm{ext}}|+T^{1/2}s| \del\mathcal I_{\mathrm{ext}}|+T^{3/2}| \delsH\mathcal I_{\mathrm{ext}}| \lesssim A^2Q^{1/2-\alpha_{\mathrm{met}}}z^{1-\alpha_{\mathrm{met}}} \lesssim A^2s^{-\gamma_{\mathrm{ext,met}}},
\ee
where \(\gamma_{\mathrm{ext,met}} = \alpha_{\mathrm{met}}-1/2-(3-2\alpha_{\mathrm{met}})b>\delta-2b>0\). These estimates use only the physical source and its outgoing derivative.

The interior incoming integration keeps its signed trace at \(q=1\). We estimate that trace separately, with the coefficient \(C=r'(P_N+a_P) \) from the radial decomposition. The exterior boundary values and one outgoing derivative give
\be
\label{equa:seven-exterior-metric-retained-trace}
|C\del_qv|+p| \del_p(C\del_qv)| \lesssim A^2p^{-1+ \delta}, \qquad q=1.
\ee
The same target-derivative identities apply with evaluation at \(q=1\) in place of the incoming integral. The exponent \(1-\delta \) gives the positive normalized gap \(\gamma_{\mathrm{bd}} =1/2-\delta-(1+2\delta)b\). This keeps the sign and coefficient prescribed by the interior identity and bounds its actual contribution. The estimate uses at most \(p_{\mathrm w} +6\) orders. Thus the exterior volume, the finite initial-time contribution, and the trace at the interface of the original principal metric source are controlled.

On a moving interface, we write its incoming source coordinate as $q=a_b(p)$. The outgoing derivative of its signed trace is then $\del_p+a_b'(p)\del_q$. The bound~\eqref{equa:audit-moving-interface-trace-source-bound} keeps the same exponent and the order $p_{\mathrm w}+6$; the additional term containing $a_b'$ decays faster. Differentiation of the source domain also contributes the trace~\eqref{equa:audit-moving-interface-cone-target-trace}. The estimates in Section~\ref{section:mobile-interior-transfer} keep the gap $\gamma_{\mathrm{bd}}$.


\section{Matter sources and wave reconstruction}  


\subsection{\texorpdfstring{Mass-weighted treatment of the matter source}{Mass-weighted treatment of the matter source}}
\label{section-central-matter-correction}

\paragraph{A product correction from the massive equation}

We now estimate the matter source on the past light cone. A product correction uses the large mass in the interior, while the exterior coupling admits a direct weighted estimate. We control the terms produced by the cutoff between the interior and exterior regions, the local correction, and its Cauchy contribution. The homogeneous wave and lightlike-region products then give the corresponding terms in the pointwise and energy estimates.

First of all, the direct derivative of \(g(\nabla \fieldPhi, \nabla \psi) \) contains an effective-field Hessian. We use the mass in its equation before estimating this derivative. The two scalar equations give
\be
\label{equa:seven-matter-product-correction-definition}
F_\psi{\coloneqq}g(\nabla \fieldPhi, \nabla \psi)= \Box_g\psi, \qquad G_\Phi{\coloneqq} \Box_g\fieldPhi-(3\coeff)^{-1} \fieldPhi = \tfrac13(\mathcal S+W_\coeff).
\ee
Metric compatibility and \(\Box_g(df)=d\Box_gf+ \operatorname{Ric}_g(df, \cdot) \) imply the product identity
\be
\label{equa:seven-matter-product-correction-identity}
\begin{aligned}
\Box_gF_\psi&= (3\coeff)^{-1}F_\psi+R_\psi, \\
R_\psi&\coloneqq g(\nabla G_\Phi, \nabla \psi) +g(\nabla \fieldPhi, \nabla F_\psi) +2\nabla^2\fieldPhi:\nabla^2\psi+2\operatorname{Ric}_g(\nabla \fieldPhi, \nabla \psi).
\end{aligned}
\ee
The colon contracts both Hessian indices with \(g^{-1} \). The two Ricci terms, from the commuted gradients of the two scalars, have the same sign. Thus the corrected matter variable satisfies
\be
\label{equa:seven-matter-product-corrected-equation}
\Box_g(\psi-3\coeff F_\psi)=-3\coeff R_\psi.
\ee
In the flat wave representation, the same identity reads, for every translation--boost operator \(Z\),
\be
\label{equa:seven-matter-product-flat-equation}
\Box(Z\psi-3\coeff ZF_\psi)=-Z(H^{\alpha \beta} \del_{\alpha \beta} \psi) +3\coeff Z(H^{\alpha \beta} \del_{\alpha \beta}F_\psi)-3\coeff ZR_\psi.
\ee
The additional principal metric term, the local correction, and its two Cauchy traces are therefore part of the representation. We estimate the commuted identity as a whole; its individual Leibniz factors need not satisfy the uncommuted scalar equations.

\paragraph{Lower massive energy and third derivatives}

The rescaled potential factorization in \eqref{equa:seven-uniform-potential-hypothesis} and trace reversal of the Einstein equation give
\be
\label{equa:seven-matter-product-rescaled-sources}
\begin{aligned}
\coeff G_\Phi&=-\tfrac{8\pi\coeff}{3}e^{-\fieldPhi}g(\nabla \psi, \nabla \psi) + \tfrac13\fieldPhi^2\mathcal W_\coeff(\fieldPhi), \\
\coeff\operatorname{Ric}_{g, \alpha \beta}&= \tfrac{3\coeff}{2} \del_\alpha \fieldPhi\del_\beta \fieldPhi+8\pi\coeff e^{-\fieldPhi} \del_\alpha \psi\del_\beta \psi + \fieldPhi^2\bigl(\tfrac14+ \tfrac12\fieldPhi\mathcal U_{1, \coeff}(\fieldPhi) \bigr)g_{\alpha \beta}.
\end{aligned}
\ee
Their differentiated coefficients are uniform in the prescribed rescaled range. In particular, \(\coeff\del G_\Phi\) consists of \(\fieldPhi\del\fieldPhi\), \(\coeff\del^2\psi\del\psi\), and products with additional differentiated coefficients.

For \(\ord Z\leq p_{\mathrm w} \), \(\rank Z=j\), the lower direct energy, rather than the refined trace amplitude, controls the mass term after one translation. The resulting bounds are
\be
\label{equa:seven-matter-product-weighted-third-jet}
\coeff| \del^3Z\fieldPhi| \lesssim A q^{-3/2} \sigma^{\nu_{j+5}}, \qquad \coeff|L\del^2Z\fieldPhi| \lesssim A p^{-1}q^{-1/2} \sigma^{\nu_{j+5}}.
\ee
We prove the first bound by applying \eqref{equa:seven-cone-light-like-Hessian-reduction} to \(\del Z\fieldPhi\), multiplying by \(\coeff\), and inducting on the commutation order. The mass contribution is \((p/q)| \del Z\fieldPhi| \lesssim A q^{-3/2} \sigma^{\alpha_{j+2}} \), using the lower energy. The tangential term has weight \(\coeff A p^{-1}q^{-3/2} \sigma^{1/2 + \nu_{j+3}} \), bounded by this weight since \(\sigma \leq p\). The products in \eqref{equa:seven-matter-product-rescaled-sources}, estimated with the lightlike matter Hessian and the direct lower factors, have additional inverse powers of \(p\) or \(q\). The bad metric commutators contribute a bounded multiple of \(q^{-1} \) times a lower-order third derivative; the remaining commutators contain a good derivative. The coefficient of largest order at \(N=19\) contains a translation and lies in the refined coefficient range. Finite induction proves the estimate, with dependence on \(N,B_{s_1}, \coeff_0\) and no negative power of \(\coeff\).

The identity \(L=t^{-1} \omega^aL_a+(q/t) \del_t\) gives the second bound. All direct tangential terms use at most \(p_{\mathrm w} +4\) high energy orders. Independently, direct high Sobolev gives
\be
\label{equa:seven-matter-product-direct-outgoing-jet}
\coeff|L\del^2Z\fieldPhi| \lesssim A p^{-2}q^{1/2} \sigma^{1/2 + \nu_{j+3}}.
\ee
Both estimates will be used: their relative strength depends on \(q\).

\paragraph{The Hessian product in the near-cone volume}

For the physical radial source \(W_2=r'\coeff Z(\nabla^2\fieldPhi:\nabla^2\psi) \), with \(\ord Z\leq p_{\mathrm w} \), \(\rank Z=k\), the two outgoing estimates imply
\be
\label{equa:seven-matter-product-minimum-source}
|W_2|+p|(W_2)_p| \lesssim A^2\sigma^{\eta_k}(pq)^{e_2/2} \min\{p^{-3/4}q^{-3/4},q^{-2} \}, \qquad e_2\coloneqq29\rho+14c.
\ee
The largest term is the outgoing derivative of the effective-field Hessian. Multiplication of \eqref{equa:seven-matter-product-direct-outgoing-jet} by the matter Hessian and by \(r'p\) gives \(p^{-3/4}q^{-3/4} \); using \eqref{equa:seven-matter-product-weighted-third-jet} instead gives \(q^{-2} \). The rank exponents add, and \(\nu_{i+5} + \nu_{k-i+3} +b= \eta_k+e_2\). When the outgoing derivative hits the matter Hessian, we use \eqref{equa:seven-cone-weak-third-matter-jet} and the lower direct effective-field Hessian. Their product, multiplied by \(r'p\), has weight \(A^2p^{-1}q^{-1} \sigma^{\eta_k+15\rho+7c} \). Since \(p\geq q\geq1\), this weight is bounded by the minimum in \eqref{equa:seven-matter-product-minimum-source}, with no exceptional loss. The lower Hessian uses \(p_{\mathrm w} +3\leq p_{\mathbf{low}} \). The connection terms in the covariant Hessians and the differentiated inverse metrics have additional first-derivative factors and inverse-time decay; the same estimates bound them.

The fixed growth factor is kept inside the source integral. Splitting its positive majorant at \(q=p^{3/5} \) gives
\be
\label{equa:seven-matter-product-minimum-kernel}
\int_1^\infty(pq)^{e_2/2} \min\{p^{-3/4}q^{-3/4},q^{-2} \} \, dq \leq\Bigl(\frac1{1/4+e_2/2} + \frac1{1-e_2/2} \Bigr)p^{-a_2}, \qquad a_2\coloneqq \tfrac35-\tfrac45e_2.
\ee
Here \(0<e_2<1/8\), so \(1/2<a_2<3/5\). With \(z=Q_+/Q\), the remaining averaged integral is bounded by \(C Q^{-a_2}z^{-a_2} \), uniformly at \(z=1\). The target-derivative formulas control \(Q\del_Q\), \(Q_+ \del_{Q_+} \), and \(s\nabla_{\tan,x} \) by this same kernel. Hence the normalized value and derivatives of this volume are bounded by \(CA^2Q^{1/2-a_2}z^{1-a_2}s^{\eta_k} \). On the remaining target region, this proves a strict gain with
\be
\label{equa:seven-matter-product-Hessian-gap}
\gamma_2\coloneqq a_2-\tfrac12-(3-2a_2)b>0.
\ee
The bounds \(\rho<1/384\) and \(c<\rho/23\) give \(\gamma_2>1/80\). Thus the fixed Sobolev growth is absorbed without comparing \(A\) with the refined trace amplitude.

\paragraph{The other product sources}

We denote by \(R_\psi^{\mathrm{rem}} \) the sum of the other three terms in \eqref{equa:seven-matter-product-correction-identity}. Expanding \eqref{equa:seven-matter-product-rescaled-sources}, their differentiated radial source satisfies
\be
\label{equa:seven-matter-product-other-source}
|r'\coeff ZR_\psi^{\mathrm{rem}}|+p| \del_p(r'\coeff ZR_\psi^{\mathrm{rem}})| \lesssim A^2p^{-1}q^{-3/2} \sigma^{\eta_k+e_{\mathrm{rem}}}, \qquad e_{\mathrm{rem}}{\coloneqq}60\rho+30c+ \varepsilon_N.
\ee
We verify the outgoing derivative by listing the products. Apart from smooth coefficients depending on \(g\) and \(\fieldPhi\), whose derivatives are uniformly bounded in the prescribed range, \(\coeff R_\psi^{\mathrm{rem}}\) consists of
\be
\label{equa:audit-matter-three-factor-products}
\fieldPhi\del\fieldPhi\del\psi,\qquad \coeff\del^2\psi(\del\psi)^2,\qquad \coeff\del^2\fieldPhi\del\fieldPhi\del\psi,\qquad \coeff(\del\fieldPhi)^2\del^2\psi,
\ee
together with the products
\be
\label{equa:audit-matter-higher-products}
\coeff\del g(\del\fieldPhi)^2\del\psi,\quad \coeff\del g(\del\psi)^3,\quad \coeff(\del\fieldPhi)^3\del\psi,\quad \coeff\del\fieldPhi(\del\psi)^3,\quad \fieldPhi^2\del\fieldPhi\del\psi.
\ee
The factors \(\del g\) account for derivatives of inverse-metric coefficients. These lists follow from the equations~\eqref{equa:seven-matter-product-rescaled-sources} and \(F_\psi=g^{\alpha\beta}\del_\alpha\fieldPhi\del_\beta\psi\). To display the decay powers, first omit the rank-growth factors. The lower energy and the outgoing decomposition give
\be
\label{equa:audit-matter-lower-factors}
|\fieldPhi|\lesssim A p^{-3/2},\qquad |\del\fieldPhi|+|\del^2\fieldPhi|+|\del\psi|\lesssim A p^{-1}q^{-1/2},\qquad |L\fieldPhi|+|L\del\fieldPhi|\lesssim A p^{-2}q^{1/2}.
\ee
The constants may depend on \(\coeff_0\). The matter Hessian estimates and the equation~\eqref{equa:seven-matter-product-weighted-third-jet} give
\be
\label{equa:audit-matter-outgoing-factors}
|\del^2\psi|\lesssim A p^{-1}q^{-3/2},\qquad |L\del\psi|+|L\del^2\psi|\lesssim A p^{-2}q^{-1/2},\qquad \coeff|L\del^2\fieldPhi|\lesssim A p^{-1}q^{-1/2}.
\ee
When the outgoing derivative falls on the effective-field Hessian, its largest product is
\be
\label{equa:audit-matter-largest-product}
r'p\coeff|L\del^2\fieldPhi|\,|\del\fieldPhi|\,|\del\psi|\lesssim A^3p^{-1}q^{-3/2}.
\ee
When it falls on a first effective-field derivative, the resulting weight is at most \(A^3p^{-2}q^{-1/2}\), which is bounded by the preceding weight since \(q\leq p\). In the potential product, its placements on \(\fieldPhi\), \(\del\fieldPhi\), and \(\del\psi\) give respectively \(A^3p^{-2}q^{-1/2}\), \(A^3p^{-5/2}\), and \(A^3p^{-5/2}q^{-1}\); these satisfy the same bound. The differentiated matter Hessian gives \(A^3p^{-2}q^{-3/2}\). Every product with four factors gains one further undifferentiated field factor or a first derivative. Differentiating a smooth coefficient or \(r'\) is controlled by these same estimates.

We apply this calculation to each Leibniz partition of \(Z\). The largest scalar jet uses \(p_{\mathrm w}+5\leq p_{\mathbf{high}}\) orders, and the lower first derivatives use \(p_{\mathrm w}+3\leq p_{\mathbf{low}}\). Restoring the rank powers gives at most \(\eta_k+60\rho+30c+\varepsilon_N\). Since \(A^3+A^4\lesssim A^2\), this proves the equation~\eqref{equa:seven-matter-product-other-source}.
The corresponding positive kernel is
\be
\label{equa:seven-matter-product-other-kernel}
\frac1{r_{\mathrm{tar}}} \int_Q^{Q_+} \int_1^{Q/2}p^{-1}q^{-3/2} \, dq\, dp\lesssim Q^{-1}.
\ee
The lower limit is essential because this majorant is not integrable at \(q=0\). The differentiated cone identities give the same kernel for the normalized coordinate derivatives. On \(T\leq s^{1+2b} \), the normalized physical derivatives therefore have the positive gap
\be
\label{equa:seven-matter-product-other-gap}
\gamma_{\mathrm{rem}}{\coloneqq} \tfrac12-5b-e_{\mathrm{rem}} = \tfrac12-85\rho-40c-\varepsilon_N>0.
\ee

\paragraph{The additional principal metric source}

We apply the radial decomposition \eqref{equa:seven-radial-metric-full-decomposition} to \(v= \coeff Z_2F_\psi\), with \(\ord Z_1+ \ord Z_2\leq p_{\mathrm w} \). If \(\rank Z_2=j\), the lower direct Hessian estimates and \eqref{equa:seven-matter-product-weighted-third-jet} give
\be
\label{equa:seven-matter-product-metric-jets}
| \del v|+ |L\del v| \lesssim A^2p^{-2}q^{-1} \sigma^{\eta_j+e_v}, \qquad e_v\coloneqq29\rho+14c.
\ee
The outgoing derivative of the effective-field Hessian uses the mass-weighted third-derivative estimate. When it hits the matter Hessian, we use \eqref{equa:seven-cone-weak-third-matter-jet}; every term contains two direct scalar amplitudes. For one additional boost \(B\), including the angular derivatives needed in the cone identity, direct high Sobolev for the Hessian and the lower massive estimate for its outgoing derivative yield
\be
\label{equa:seven-matter-product-metric-boost-jets}
| \del Bv| \lesssim A^2p^{-7/4}q^{-3/4} \sigma^{\eta_j+e_v}, \qquad |L\del Bv| \lesssim A^2p^{-2}q^{-1} \sigma^{\eta_j+e_v+ \varepsilon_N}.
\ee
Here \(e_v\) includes the extra rank. The first bound keeps the factor \(\sigma^{1/2} \) of the high energy; the second uses the lower bound for the mass term. At \(N=19\), its lower Sobolev step costs \(p_{\mathrm w} +3=p_{\mathbf{low}} \), while the tangential high step costs at most \(p_{\mathrm w} +5\). To extend \eqref{equa:seven-matter-product-weighted-third-jet} by this extra boost, the largest coefficient, of order nine, is estimated by the direct high bound and paired with the uncommuted effective-field Hessian. After the factor \(p/q\) from the Hessian reduction, this contributes \(p^{-1/2}q^{-3/2} \), which has additional decay. The coefficients of order eight use the refined general estimate, except for the boost-only coefficient, which contributes \(\sigma^{\varepsilon_N} \) by \eqref{equa:seven-cone-exceptional-trace}. Two additional boost derivatives and their outgoing derivative, used in the regular radial terms, follow from direct high Sobolev.

After the single incoming integration, the scalar source has weight \(A^3p^{-3/2}q^{-1/2} \sigma^{\eta_k+53\rho+24c+ \varepsilon_N} \). The angular source has weight \(A^3p^{-1/2}q^{-1/2} \sigma^{\eta_k+41\rho+20c+ \varepsilon_N} \); the latter loss includes the extra angular derivative of the coefficient. The same bounds hold for their outgoing derivatives multiplied by \(p\). The regular radial source has the stronger weight \(A^3p^{-3/2}q^{-3/4} \sigma^{\eta_k+36\rho+17c+b+ \varepsilon_N} \). At most one factor uses the exceptional order, as follows from the Leibniz distribution before estimating each product. For example, the largest differentiated angular product is \(pC L\del Bv\), whose weight is \(A^3p^{-1/2}q^{-1} \), and hence is bounded by the angular weight. Applying the scalar, angular, and regular cone estimates gives the  positive gap
\be
\label{equa:seven-matter-product-metric-gap}
\gamma_F\coloneqq \tfrac12-58\rho-26c-\varepsilon_N>0.
\ee
The volume amplitude is \(A^3\leq A^2\). Thus every near-cone interior volume in \eqref{equa:seven-matter-product-flat-equation} is estimated. Its traces at the interface are added to the exterior contribution.

\paragraph{Complementary Hessian and product volumes}

We apply the Cartesian formula \eqref{equa:seven-complement-cartesian-integral} on \(q\geq Q/4\). For \(F_2= \coeff Z(\nabla^2\fieldPhi:\nabla^2\psi) \), the lower massive third derivative and the lightlike matter Hessian give
\be
\label{equa:seven-matter-complement-Hessian-source}
|F_2| \lesssim A^2p^{-2}q^{-2} \sigma^{\eta_k+e_2}, \qquad | \nabla_yF_2| \lesssim A^2p^{-1}q^{-3} \sigma^{\eta_k+e_2}.
\ee
The largest differentiated product is \(\coeff\del^3\fieldPhi\, \del^2\psi\). Its exponent is \(\nu_{i+5} + \nu_{k-i+3} +b= \eta_k+e_2\). When the derivative falls on the matter Hessian, the first term in \eqref{equa:seven-complement-third-matter-derivative} has an additional \(p^{-1} \), which absorbs its extra growth \(10\rho+4c+ \varepsilon_N<1/4\); its second term has stronger decay and a smaller exponent. The connection terms have additional differentiated factors. On this source region, \(p/q\leq4Q_+/Q\lesssim s^{4b} \). We therefore obtain
\be
\label{equa:seven-matter-complement-Hessian-weight}
p^{-1}q^{-3} = (p/q)^{1/2}p^{-3/2}q^{-5/2} \lesssim s^{2b}p^{-3/2}q^{-5/2}, \qquad \gamma_{2, \mathrm c}{\coloneqq} \tfrac12-13b-e_2= \tfrac12-94\rho-40c>0.
\ee
We first differentiate the actual integral and then apply this positive majorant; the factor \(s^{2b} \) belongs only to the estimate. Formula \eqref{equa:seven-complement-cartesian-gain} proves the complementary Hessian bound with amplitude \(C_{N,B_{s_1}, \coeff_0}A^2\) and gap \(\gamma_{2, \mathrm c} \). The other three product sources satisfy \eqref{equa:seven-complement-cartesian-source} with \(e=e_{\mathrm{rem}} \): their largest differentiated effective-field factor has weight \(\coeff\del^3\fieldPhi\, \del\fieldPhi\, \del\psi\lesssim A^3p^{-2}q^{-5/2} \), and the remaining terms decay faster. Their complementary gap is \(1/2-11b-e_{\mathrm{rem}}>0\).

\paragraph{Complementary principal volume and the local correction}

For the additional principal term we need one more massive derivative. The smooth Hessian reduction applied to \(\del^2Z\fieldPhi\), multiplied by \(\coeff\), gives
\be
\label{equa:seven-matter-complement-fourth-effective-jet}
\coeff| \del^4Z\fieldPhi| \lesssim C_{N,B_{s_1}, \coeff_0}Aq^{-3/2} \sigma^{\nu_{j+6}}, \qquad \ord Z\leq p_{\mathrm w}, \quad \rank Z=j.
\ee
Its mass contribution is \((p/q)| \del^2Z\fieldPhi| \), controlled by the lower energy with \(p_{\mathrm w} +3\) orders. We estimate the tangential term directly by \(A p^{-1}q^{-3/2} \sigma^{1/2 + \nu_{j+4}} \), using at most \(p_{\mathrm w} +5\) high orders; \(\sigma \leq p\) absorbs the additional growth. The twice differentiated potential and matter trace contain only \(\fieldPhi\del^2\fieldPhi\), \((\del\fieldPhi)^2\), \(\coeff\del^3\psi\del\psi\), and \(\coeff(\del^2\psi)^2\), with differentiated smooth coefficients. Their direct bounds have additional decay. The bad metric commutators are treated by induction on the order. At \(N=19\), a coefficient of order eight contains a translation and has rank at most seven. The largest coefficient, of order nine, contains two translations and is estimated directly against the uncommuted lower effective-field Hessian; its additional factor \(p^{-1/2} \) absorbs the fixed growth. Thus this fourth-derivative estimate uses only \(B_{s_1}\).

For \(v= \coeff Z_2F_\psi\), with \(\rank Z_2=j\), the product rule yields
\be
\label{equa:seven-matter-complement-correction-jets}
| \del^2v|+ | \del^3v| \lesssim A^2p^{-1}q^{-2} \sigma^{\eta_j+e_v}, \qquad | \del G_2[v]| \lesssim A^2\bigl(p^{-2}q^{-2} \sigma^{\varepsilon_N} +p^{-3}q^{-1} \sigma^{1/2} \bigr) \sigma^{\eta_j+e_v}.
\ee
Only \(\coeff\del^4\fieldPhi\, \del\psi\) needs \eqref{equa:seven-matter-complement-fourth-effective-jet}; direct high estimates suffice for \(\coeff\del\fieldPhi\, \del^4\psi\). The good Hessian uses \(\tau^{-1} \del L_av\). Its differentiated effective-field factor at the last boost order keeps \(\sigma^{1/2} \), as shown in \eqref{equa:seven-matter-complement-correction-jets}.

We split \(P^{\alpha \beta} \del_{\alpha \beta}v=P^{\H00} \del_t^2v+PG_2[v] \). The refined estimate \(|P^{\H00}| \lesssim C_{N,B_{s_1}}p^{-1} \sigma^{\eta_i} \) gives weight \(A^2p^{-2}q^{-2} \) when the gradient hits the bad Hessian. The good terms have spatial weights \(A^3p^{-5/2}q^{-2} \) and \(A^3p^{-7/2}q^{-1} \sigma^{1/2} \); the first also carries \(\sigma^{\varepsilon_N} \). Their spatial decay is bounded by \(A^2p^{-3/2}q^{-5/2} \), and their growth is included in \(e_{\mathrm{met}} \). The differentiated coefficient has stronger decay. Thus \eqref{equa:seven-complement-cartesian-source} holds with \(M=C_{N,B_{s_1}, \coeff_0}A^2\) and \(e=e_{\mathrm{met}} \). The complementary principal volume has the positive gap \(\gamma_{\mathrm{met}} \) of \eqref{equa:seven-complement-metric-gap}.

Finally, at the target the direct lower first and second scalar derivatives give
\be
\label{equa:seven-matter-local-correction}
| \coeff ZF_\psi|+ | \del(\coeff ZF_\psi)| \lesssim A^2T^{-1}s^{-2 + \eta_k+8\rho+4c}, \qquad \gamma_{\mathrm{loc}}{\coloneqq} \tfrac32-b-8\rho-4c>0.
\ee
Since \(| \delsH v| \leq2| \del v| \) and \(T\leq s^{1+2b} \), this controls the normalized local correction with amplitude \(A^2s^{\eta_k-\gamma_{\mathrm{loc}}} \). The near-cone and complementary calculations cover all interior source volumes of \eqref{equa:seven-matter-product-flat-equation}. Section~\ref{section-central-exterior-matter} localizes this correction and treats the exterior matter source. Its homogeneous Cauchy contribution is estimated in Section~\ref{section-central-homogeneous}.


\subsection{\texorpdfstring{Exterior matter and the interface cutoff}{Exterior matter and the interface cutoff}}
\label{section-central-exterior-matter}

\paragraph{Direct estimates from the exterior energy}

The matter coupling admits a direct estimate in the exterior source region. We use the closing exterior energy, which bounds the weighted first derivatives of both scalar fields by \(As_{\mathrm e}^{\delta} \), where \(s_{\mathrm e} \) is the exterior foliation parameter. The normalization of the effective field has weight one at this order, as specified in \eqref{equa:eight-exterior-order-weight}. We take \(3/4<\lambda<\mu<1\) and \(0<\delta \leq(\mu-\lambda)/2\), within the choice of exponents in \eqref{equa:eight-exterior-weight-gaps}. On an exterior hypersurface, \(\tau\simeq s_{\mathrm e}^2\). For \(q= \tau-r'\leq1\), the weight \(w=1+(-q)_+ \) is comparable to \(\crochet\).

We justify the pointwise estimate across the merging annulus, where the factor \(\zeta \) in the energy can be small. For a Cartesian component \(f= \del Zv\), \(v\in\{\fieldPhi, \psi\} \), the energy controls \(w^\mu\zeta f\) and the hypersurface-tangent derivative \(w^\mu\del_r^{\ME}f\) at one further translation order. Indeed, \(\del_a^{\ME} = \delsN_a-(1-T_r) \omega_a \del_t\) and \(1-T_r\leq\zeta \). In the Euclidean region \(r'\geq\rhoE(s_{\mathrm e}) \), we have \(\zeta=1\). The one-dimensional Sobolev inequality on \([r',r'+1] \), followed by two angular derivatives on the sphere, gives
\be
\label{equa:seven-exterior-local-sphere-bound}
r'w^\mu|f| \lesssim\sum_{|J| \leq2} \Bigl(\|w^\mu\Omega^Jf\|_{L^2(\mathcal A_{r'})} + \|w^\mu\del_r^{\ME} \Omega^Jf\|_{L^2(\mathcal A_{r'})} \Bigr), \qquad \mathcal A_{r'}{\coloneqq} \{r'\leq|y| \leq r'+1\}.
\ee
These norms use \(dy\) on the hypersurface. The same inequality at \(\rhoE(s_{\mathrm e}) \) gives the outer trace of the merging annulus. Integration of the hypersurface-radial derivative transports this trace across the annulus, whose width is one and on which \(w\simeq1\). Cauchy--Schwarz therefore preserves the estimate. Two rotations cost four boosts, and the radial Sobolev step costs one translation. Consequently, throughout the exterior region, we have
\be
\label{equa:seven-exterior-safe-scalar-bound}
r'w^\mu\bigl(| \del Z\fieldPhi|+ | \del Z\psi| \bigr) \lesssim A s_{\mathrm e}^{\delta}, \qquad \ord Z+5\leq p_{\mathbf{high}}.
\ee
The metric estimate follows directly from radial integration. On the flat part of the hypersurface, we set \(R= \rhoE(s_{\mathrm e}) \), \(\widehat w=1+r'-R\), and \(\beta=2\lambda-1>0\). The weight \(\widehat w\) is comparable to \(w\). For a commuted perturbation \(Zu\), the asymptotically Euclidean normalization gives \(Zu\to0\) at infinity. Integrating the derivative of \((r')^2\widehat w^\beta|Zu|^2\) and keeping its positive terms yields
\be
\label{equa:seven-exterior-metric-v-Hardy}
\begin{aligned}
R^2\|Zu(R, \cdot) \|_{L^2(\mathbb S^2)}^2 + \beta \| \widehat w^{\lambda-1}Zu\|_2^2 +2\int_{r'\geq R}(r')^{-1} \widehat w^\beta|Zu|^2\, dy &=-2\int_{r'\geq R} \widehat w^\beta Zu\, \del_{r'}Zu\, dy\\
& \leq2\| \widehat w^{\lambda-1}Zu\|_2\| \widehat w^\lambda \del_{r'}Zu\|_2.
\end{aligned}
\ee
The outer boundary vanishes by Cauchy--Schwarz in \(Zu(r', \omega)=-\int_{r'}^\infty\del_\rho Zu(\rho, \omega) \, d\rho\) and the tail of the weighted derivative norm. The identity controls both the weighted \(L^2\) norm of \(Zu\) and the trace at \(R\); the width-one transport again covers the merging annulus. Applied after rotations and the local Sobolev estimate, it gives \(r'w^{\lambda-1}|ZH|+r'w^\lambda| \del ZH| \lesssim As_{\mathrm e}^\delta \). The prescribed reference term is estimated separately by its explicit inverse-radius decay. The outgoing identity \(L= (r')^{-1}(\omega^aL_a-q\del_{r'}) \), applied to scalar derivatives, yields
\be
\label{equa:seven-exterior-outgoing-scalar-bound}
|L\del Z\fieldPhi|+ |L\del Z\psi| \lesssim A(r')^{-2}w^{1-\mu}s_{\mathrm e}^{\delta}, \qquad \ord Z\leq p_{\mathrm w}.
\ee
This uses at most \(p_{\mathrm w} +6\leq p_{\mathbf{high}} \) orders. On the source cone, \(p\geq Q\), \(q\leq1\), and \(\tau\geq T_0\), so \(r'\simeq p\), \(w\lesssim p\), and \(s_{\mathrm e}^2\lesssim p\), once \(Q\) is large. The metric estimate and its outgoing derivative are bounded by \(CAp^{-\lambda+ \delta/2} \). Product differentiation of \(F_\psi =g(\nabla \fieldPhi, \nabla \psi) \) therefore gives
\be
\label{equa:seven-exterior-direct-matter-source}
|W|+p|W_p| \lesssim A^2p^{-1+ \delta}w^{1-2\mu}, \qquad W=r'ZF_\psi, \quad \ord Z\leq p_{\mathrm w}.
\ee

\paragraph{Cone integration for negative incoming coordinate}

We separate the bounded initial-time interval by a smooth function \(\chi_0(\tau) \), zero for \(\tau\leq T_0\) and one for \(\tau\geq T_0 +1\). The source \(\chi_0ZF_\psi\) extends smoothly by zero below \(T_0\). Its outgoing derivative still satisfies \eqref{equa:seven-exterior-direct-matter-source}: where \(\chi_0'\neq0\), \(w\simeq p\), and the extra factor \(p\) is absorbed by the factor \(w\) in that estimate. We can thus use the fixed source interval \(-\infty<q\leq1\).

The angle and target-derivative identities of Section~\ref{section-central-target-derivatives} hold on this interval as well. In fact, \(Q-q\geq Q-1\), \(Q_+-q\geq Q_+-1\), and \(p-q\leq2p\) on the support of the source. With the external partition equal to one, \eqref{equa:seven-target-cone-integral-derivatives} and \eqref{equa:seven-target-cone-angular-target-derivative} bound \(Q\del_Q\mathcal I_W\), \(Q_+ \del_{Q_+} \mathcal I_W\), and \(s\nabla_{\tan,x} \mathcal I_W\) by the same positive kernel as \(\mathcal I_W\). Writing \(z=Q_+/Q\), we obtain
\be
\label{equa:seven-exterior-matter-cone-kernel}
\int_{-p+2T_0}^{1}w^{1-2\mu} \, dq\lesssim p^{2-2\mu}, \qquad \frac1{r_{\mathrm{tar}}} \int_Q^{Q_+}p^{-\alpha_{\mathrm{ext}}} \, dp\lesssim Q^{-\alpha_{\mathrm{ext}}}z^{-\alpha_{\mathrm{ext}}}, \quad \alpha_{\mathrm{ext}}{\coloneqq}2\mu-1-\delta.
\ee
Here \(1/2<\alpha_{\mathrm{ext}}<1\). The second estimate follows from the primitive of \(p^{-\alpha_{\mathrm{ext}}} \); after division by its right-hand weight, the remaining factor is \(2(z-z^{\alpha_{\mathrm{ext}}})/((1-\alpha_{\mathrm{ext}}) (z-1)) \), bounded also at \(z=1\). Thus the actual cone integral satisfies
\be
\label{equa:seven-exterior-matter-target-gain}
T^{1/2}| \mathcal I_W|+T^{1/2}s| \del\mathcal I_W|+T^{3/2}| \delsH\mathcal I_W| \lesssim A^2Q^{1/2-\alpha_{\mathrm{ext}}}z^{1-\alpha_{\mathrm{ext}}} \lesssim A^2s^{-\gamma_{\mathrm{ext}}},
\ee
where the last estimate is on \(T\leq s^{1+2b} \), and
\be
\label{equa:seven-exterior-matter-positive-gap}
\gamma_{\mathrm{ext}}{\coloneqq} \alpha_{\mathrm{ext}}-\tfrac12-(3-2\alpha_{\mathrm{ext}})b>0, \qquad b=5\rho+2c.
\ee
Indeed, \(\delta \leq\mu-\lambda \) and \(\lambda>3/4\) imply \(\alpha_{\mathrm{ext}}-1/2>\delta \), whereas \(3-2\alpha_{\mathrm{ext}}<2\) and \(b<\delta(5+2/23)/64\). The source integral and the derivative identities have finite limits at \(r_{\mathrm{tar}} =0\).

\paragraph{The finite initial-time contribution}

The source \((1-\chi_0)ZF_\psi\) acts only for \(T_0\leq\tau\leq T_0 +1\). At bounded time, \eqref{equa:seven-exterior-safe-scalar-bound} gives decay \(\Ocal(A^2r'^{-2-2\mu}) \) for this source and its Cartesian derivatives through order two. Their products with \(r'^2\) are square integrable since \(\mu>3/4\). Two additional boosts involve at most the same two spatial moments. The derivative count is \(p_{\mathrm w} +7\leq p_{\mathbf{high}} \), equal to \(14\leq14\) at \(N=19\).

We denote by \(w_{Z,0} \) the corresponding flat forced wave with zero Cauchy data at \(T_0\). Its two-boost energy on \(T_0 +1\) is bounded by \(CA^2\). In deriving this bound, the differentiated initial equation contributes the moments \(r'ZF_\psi\) and \(r'^2\del ZF_\psi\); the boosted initial jets need not vanish. After \(T_0 +1\), this wave is free. For \(s\geq T_0 +1\), the homogeneous estimate proved in Section~\ref{section-central-homogeneous} gives
\be
\label{equa:seven-exterior-matter-initial-strip}
| \del w_{Z,0}|+(t/s)| \delsH w_{Z,0}| \lesssim A^2t^{-1/2}s^{-1} \leq A^2s^{-3/2}.
\ee

\paragraph{Localization of the product correction}

To combine the direct exterior estimate with the interior massive equation, we choose \(a_{\mathrm m} \in C^\infty(\mathbb R) \), equal to zero on \(( -\infty,1] \) and one on \([2, \infty) \). For \(q= \tau-r'\), the flat wave operator satisfies
\be
\label{equa:seven-matter-localized-commutator}
[\Box,a_{\mathrm m}(q)]F_\psi =-2a_{\mathrm m}'(q) (LF_\psi+F_\psi/r').
\ee
The second derivative of \(a_{\mathrm m} \) cancels because \(dq\) is null. The localized flat equation is therefore
\be
\label{equa:seven-matter-localized-equation}
\Box(\psi-3\coeff a_{\mathrm m}F_\psi)=-H^{\alpha \beta} \del_{\alpha \beta} \psi+(1-a_{\mathrm m})F_\psi+3\coeff a_{\mathrm m}H^{\alpha \beta} \del_{\alpha \beta}F_\psi -3\coeff a_{\mathrm m}R_\psi+6\coeff a_{\mathrm m}'(LF_\psi+F_\psi/r').
\ee
Every commutation acts on the entire right-hand side. The term \(Z((1-a_{\mathrm m})F_\psi) \) is covered by the direct estimate: on the additional strip \(1\leq q\leq2\), the lower interior scalar estimates give \(p^{-1+ \delta} \), and all commuted derivatives of \(a_{\mathrm m} \) are bounded. The initial correction \(3\coeff a_{\mathrm m}F_\psi\) is spatially compact on \(\tau=T_0\), and its jets satisfy the estimates of Section~\ref{section-central-homogeneous}. We estimate the new cutoff terms on their fixed-width strip. The bounded target range follows from the direct estimates; below we take \(Q\geq\max\{16,8(T_0 +1) \} \).

\paragraph{Direct high derivatives on the cutoff strip}

On \(1\leq q\leq2\), we have \(r'\simeq\tau\simeq p\) and \(\sigma \simeq p^{1/2} \). The direct high bound for \(\del\fieldPhi\) has the factor \(p^{-1} \sigma^{1/2} \), whereas that for \(\del\psi\) has the factor \(p^{-1} \). Their product thus decays like \(p^{-7/4} \). For \(v_Z= \coeff ZF_\psi\), \(\ord Z\leq p_{\mathrm w} \), and \(\rank Z=j\), the product rule gives
\be
\label{equa:seven-matter-cutoff-direct-jets}
| \del^mB^\ell v_Z| \lesssim A^2p^{-7/4} \sigma^{\eta_j+e_0}, \qquad m+ \ell\leq3, \qquad e_0\coloneqq24\rho+12c.
\ee
Here \(B^\ell\) denotes an additional operator of \(\ell\) boosts. On this strip a rotation is a bounded combination of one boost, so the same bound covers the angular derivatives below. Each differentiated inverse-metric factor has an additional \(p^{-1/2} \); this decay absorbs its fixed growth. Using \(L=2p^{-1}(L_r+q\del_q) \), with \(L_r= \omega^aL_a \), we obtain
\be
\label{equa:seven-matter-cutoff-outgoing-jets}
|L\del^mB^\ell v_Z| \lesssim A^2p^{-11/4} \sigma^{\eta_j+e_0}, \qquad m+ \ell\leq2.
\ee
The second outgoing derivative follows from the identity
\be
\label{equa:seven-matter-cutoff-second-outgoing}
L^2v_Z=4p^{-2} \bigl((L_r^2-L_r)v_Z+2q\del_qL_rv_Z+q^2\del_q^2v_Z\bigr), \qquad |L^2v_Z| \lesssim A^2p^{-15/4} \sigma^{\eta_j+e_0}.
\ee
The largest scalar derivative is a fourth Cartesian derivative of \(\fieldPhi\), estimated directly at the high energy level. The count is at most \(p_{\mathrm w} +6\leq p_{\mathbf{high}} \). Every coefficient \(\Lambda=Z_0a_{\mathrm m} \) or \(Z_0a_{\mathrm m}'\) supported on the strip satisfies
\be
\label{equa:seven-matter-cutoff-coefficients}
| \Lambda|+ | \del_q\Lambda|+p| \del_p\Lambda|+p| \del_p\del_q\Lambda|+ | \nabla_{\mathbb S^2} \Lambda|+p| \del_p\nabla_{\mathbb S^2} \Lambda| \leq C_{N,a_{\mathrm m}}.
\ee
This follows from \(L_aq=-q\omega_a \), \(\del_pq=0\), and the bounded frame coefficients on the strip. The identities \([L_a,L]=-\omega_aL+(q/r') \Pi_{ab} \del_b\) and \([\del_a,L]= (r')^{-1} \Pi_{ab} \del_b\) show that the commuted flat-cutoff term in \eqref{equa:seven-matter-localized-equation} is a sum of \(cLv_Y\), \(p^{-1}c\del v_Y\), and \(p^{-1}cv_Y\), with supported bounded coefficients and \(|c|+p|c_p| \lesssim1\). Consequently its physical radial source obeys
\be
\label{equa:seven-matter-cutoff-flat-source}
|W_{\mathrm{flat}}|+p|(W_{\mathrm{flat}})_p| \lesssim A^2p^{-7/4} \sigma^{\eta_k+e_0}.
\ee

\paragraph{Incoming and angular cutoff terms}

Tensorial commutation of the added principal term gives \(\Lambda P^{\alpha \beta} \del_{\alpha \beta}v\), where \(v= \coeff Z_2F_\psi\) and the total order on \(\Lambda,P,v\) is at most \(p_{\mathrm w} \). If a derivative hits \(a_{\mathrm m} \), this entire source is supported on the strip. The direct metric bound \(|P|+p|P_p| \lesssim Ap^{-1/2} \sigma^{\nu_{i+3}} \), together with \eqref{equa:seven-matter-cutoff-outgoing-jets}, gives
\be
\label{equa:seven-matter-cutoff-principal-source}
|r'\Lambda P^{\alpha \beta} \del_{\alpha \beta}v|+p| \del_p(r'\Lambda P^{\alpha \beta} \del_{\alpha \beta}v)| \lesssim A^3p^{-5/4} \sigma^{\eta_k+e_1}, \qquad e_1\coloneqq48\rho+24c.
\ee
For the term with undifferentiated \(a_{\mathrm m} \), the radial decomposition and incoming integration used in the interior remain applicable. With \(C=r'P_N\) or \(C=r'a_P\), the signed identity is
\be
\label{equa:seven-matter-cutoff-signed-integration}
\begin{aligned}
\int\Lambda C\del_q^2v\, dq&= [\Lambda C\del_qv]-\int\bigl(\Lambda \del_qC\del_qv+(\del_q\Lambda)C\del_qv\bigr) \, dq\\
& \quad -\int\mathbf B\cdot\bigl(\Lambda \nabla_{\mathbb S^2}(C\del_qv) +(\nabla_{\mathbb S^2} \Lambda)C\del_qv\bigr) \, dq.
\end{aligned}
\ee
All physical derivatives are taken before restriction to the cone. The new scalar term is \((\del_q\Lambda)C\del_qv\). Its largest outgoing product is \(pC\del_p\del_qv\), which has weight \(A^3p^{1+1/2-11/4} =A^3p^{-5/4} \). The same bound holds for its value. The angular vectors on the strip have this weight as well, by one extra angular derivative in \eqref{equa:seven-matter-cutoff-direct-jets}--\eqref{equa:seven-matter-cutoff-outgoing-jets}. The regular radial terms have the stronger weight \(A^3p^{-9/4} \). These bounds all hold with loss \(e_1\). The term \(\Lambda \del_qC\del_qv\) away from the strip is its already estimated interior volume.

The target partition equals one on the strip. Where its derivative is supported, \(a_{\mathrm m} =1\) and all differentiated cutoff terms vanish. Multiplication of the other product sources by \(a_{\mathrm m} \), or by its commuted derivatives, preserves their physical source bounds because \(| \Lambda|+p| \Lambda_p| \lesssim1\). The complementary interior region has \(q\geq Q/4\geq4\), so its source estimates are unchanged.

For a scalar radial source supported in \(1\leq q\leq2\), the compact incoming interval improves the cone kernel. If \(|W|+p|W_p| \leq Mp^{-a} \sigma^{\eta+e} \), with \(a>1\), direct integration and the target-derivative identities give
\be
\label{equa:seven-matter-cutoff-compact-kernel}
\frac1{r_{\mathrm{tar}}} \int_Q^{Q_+} \int_1^2p^{-a} \, dq\, dp\lesssim Q^{-a}z^{-1}, \qquad T^{1/2}| \mathcal I_W|+T^{1/2}s| \del\mathcal I_W|+T^{3/2}| \delsH\mathcal I_W| \lesssim M Q^{1/2-a}s^{\eta+e}.
\ee
The angular vectors are covered by the invariant angular estimate, using \(p^{-5/4} \lesssim p^{-1/2}q^{-1/2} \) on their support. The scalar gaps \(5/4-5b/2-e_0\) and \(3/4-3b/2-e_1\) are larger than the  angular gap
\be
\label{equa:seven-matter-cutoff-c-gap}
\gamma_{\mathrm{cut}}{\coloneqq} \tfrac12-3b-e_1= \tfrac12-63\rho-30c>0.
\ee
Thus all new cutoff volumes have normalized amplitude at most \(CA^2s^{\eta_k-\gamma_{\mathrm{cut}}} \). The trace in \eqref{equa:seven-matter-cutoff-signed-integration} vanishes at \(q=1\), because \(a_{\mathrm m} \) and its derivatives vanish there. The integration for the undifferentiated cutoff crosses \(q=2\) without a boundary; every differentiated cutoff vanishes there. The localized Cauchy correction remains in the representation. At the targets under consideration, \(Z(a_{\mathrm m}F_\psi)=ZF_\psi\), so the local correction is the one already estimated in \eqref{equa:seven-matter-local-correction}. Together these estimates treat the matter contribution on the past light cone. Section~\ref{section-central-exterior-metric} treats the original exterior metric source and its trace at the interface.


\subsection{\texorpdfstring{Homogeneous waves and lightlike-region products}{Homogeneous waves and lightlike-region products}}
\label{section-central-homogeneous}

\paragraph{The homogeneous wave contribution}

We recall the time \(T_0\) from \eqref{equa:seven-cone-initial-time-ceiling}. For each \(Z\preccurlyeq(p_{\mathrm w},j) \), we denote by \(w_Z\) the solution of the flat wave equation whose value and first time derivative on \(t=T_0\) agree with those of \(v=Z\psi\). The region \(r\geq\rhoE(s_0) \) on this slice is part of the prescribed initial hypersurface. The remaining region is bounded and is controlled by the direct bootstrap estimates. We prove that
\be
\label{equa:seven-homogeneous-wave-gain}
| \del w_Z(t,x)|+ \frac ts| \delsH w_Z(t,x)| \lesssim C_{N,s_0} A t^{-1/2}s^{-1} \leq C_{N,s_0}A s^{-3/2}, \qquad s\geq T_0.
\ee
The range \(s<T_0\) in the remaining target region \(t<s^{1+2b} \) is bounded and follows from the direct estimate. The only issue in applying the free wave energy estimate is the conversion of the commuted Cauchy jets: second time derivatives of the flat and physical equations differ. We give that conversion explicitly.

All norms in the next calculation are taken on \(t=T_0\), and \(\lambda, \mu\) denote the exterior weight exponents, with \(3/4<\lambda<\mu<1\). We write \(\langle r\rangle= \sqrt{1+r^2} \), and \(\mathcal D\) for a bound on the physical norms \(\| \langle r\rangle^\mu\del Z'\psi\|_{L^2} \), with \(\ord Z'\leq p_{\mathrm w} +2\), together with the corresponding norms on the bounded region. Thus \(\mathcal D\lesssim C_{N,s_0}A\). The direct coefficient estimates, including the fixed reference metric, give for \(\ord Y\leq p_{\mathrm w} +1\)
\be
\label{equa:seven-homogeneous-slice-coefficients}
|YH| \lesssim (m+A) \langle r\rangle^{-\lambda}, \quad | \del YH| \lesssim(m+A) \langle r\rangle^{-1-\lambda}, \quad | \del Y\fieldPhi| \lesssim A\langle r\rangle^{-1-\mu}.
\ee
Only the rescaled effective field occurs here. At spatial infinity, we use \(\Omega_{ab} = [L_a,L_b] \): the two rotations in the sphere Sobolev estimate cost four boosts, and the radial estimate costs one additional derivative. The coefficient estimates therefore use at most \(p_{\mathrm w} +6\leq p_{\mathbf{high}} \) derivatives. The fixed time slice has \(\crochet\simeq\langle r\rangle\) outside a bounded set. Hence these estimates and the bound for \(\mathcal D\) follow from the existing interior and exterior bootstrap norms, with constants independent of \(\coeff\).

\paragraph{Weighted spatial derivatives of the Cauchy jets}

For \(v=Z\psi\), we write the commuted physical equation and the corresponding flat source as
\be
\label{equa:seven-homogeneous-commuted-source}
g^{\alpha \beta} \del_{\alpha \beta} v=F_Z, \qquad S_Z\coloneqq \Box v=-H^{\alpha \beta} \del_{\alpha \beta} v+F_Z.
\ee
Each metric term in \(F_Z\) has a commuted coefficient \(YH\) and a second derivative of a lower-order matter field. Its other terms contain \(\del Y\fieldPhi\) and a derivative of a matter field, with bounded metric factors. All coefficient factors can be placed in \(L^\infty\) at this differentiation level. The strict weight inequalities \(1-\lambda<\mu\), \(2-2\lambda<\mu\), and \(1-\lambda<2\mu\) then imply
\be
\label{equa:seven-homogeneous-physical-source-weights}
\| \langle r\rangle F_Z\|_{L^2} + \| \langle r\rangle^{2-\lambda} \del F_Z\|_{L^2} \lesssim C_N(m+A) \mathcal D.
\ee
A differentiated metric commutator contributes \((m+A) \| \langle r\rangle^{2-2\lambda} \del^3 Z_2\psi\|_{L^2} \), where \(\ord Z_2<\ord Z\). The effective-field contribution is bounded by \(A\| \langle r\rangle^{1-\lambda-\mu} \del^{\leq2}Z_2\psi\|_{L^2} \). Both are controlled by \(\mathcal D\).

On the fixed time slice, the boost identity is \(L_a v=T_0\del_a v+x_a \del_t v\). Its first derivatives control \(\langle r\rangle\del_t\del_\alpha v\). We apply the spatial elliptic equation for \(v\), with weight \(\langle r\rangle\), to control the remaining second derivatives. This yields
\be
\label{equa:seven-homogeneous-weighted-second-jets}
\| \langle r\rangle\del^2 v\|_{L^2} \lesssim C_{N,s_0} \mathcal D.
\ee
We justify the spatial estimate directly. For a compactly supported smooth function \(f\), integration by parts gives \(\| \nabla^2 f\|_{L^2} = \| \Delta f\|_{L^2} \). Since \(g^{ab} \) is a small perturbation of \(\delta^{ab} \), the perturbation is absorbed in this identity. On an annulus \(r\simeq R\), we apply it to a cutoff times \(v\) minus its average on the enlarged annulus. Poincar\'e's inequality controls the cutoff terms by \(R^{-1} \| \nabla v\|_{L^2} \). Consequently, the estimate controls \(\|R\nabla^2 v\|_{L^2} \) by \(\|R g^{ab} \del_{ab} v\|_{L^2} \) and \(\| \nabla v\|_{L^2} \) on that enlarged annulus. The time-time term is controlled by the boost identity. The mixed term has the factor \(R|g^{0a}| \lesssim(m+A)R^{1-\lambda} \), and the source is controlled by \eqref{equa:seven-homogeneous-physical-source-weights}. Summing over dyadic annuli is legitimate because the enlarged annuli have bounded overlap. The same argument with the factor \(R^a \) yields the weighted estimate used below; its lower-order term is \(R^{a-1} \nabla v\).

The second-boost identity reads
\be
\label{equa:seven-homogeneous-two-boost-jets}
L_aL_b v=t^2\del_{ab} v+ tx_b\del_{at} v+ tx_a \del_{bt} v+x_ax_b\del_t^2 v+ t\delta_{ab} \del_t v+x_a \del_b v.
\ee
We differentiate this spacetime identity and then restrict to \(t=T_0\). We use \eqref{equa:seven-homogeneous-weighted-second-jets} for the terms of the form \(x_a \del_{bi} v\), and the differentiated first-boost identity for the terms of the form \(x_a \del_{tbi} v\). Contraction in \((a,b) \) gives
\be
\label{equa:seven-homogeneous-two-time-jets}
\| \langle r\rangle^2\del_\alpha \del_t^2 v\|_{L^2} \lesssim C_{N,s_0} \mathcal D.
\ee
We next differentiate the physical equation in time and apply the spatial elliptic estimate to \(\del_t v\), with weight \(\langle r\rangle^{2-\lambda} \). Its time-time and mixed time-space terms are controlled by \eqref{equa:seven-homogeneous-two-time-jets}. The differentiated coefficient terms carry \(\langle r\rangle^{1-2\lambda} \del^2 v\), and the differentiated source is bounded by \eqref{equa:seven-homogeneous-physical-source-weights}. We then apply the same argument to \(\del_a v\). The mixed third derivatives have already been controlled by the first application. The annular lower-order terms carry only \(\langle r\rangle^{1-\lambda} \del^2 v\), which is controlled by the weighted physical norm. We conclude that
\be
\label{equa:seven-homogeneous-weighted-third-jets}
\| \langle r\rangle^{2-\lambda} \del^3Z\psi\|_{L^2} \lesssim C_{N,s_0} \mathcal D, \qquad \ord Z\leq p_{\mathrm w}.
\ee
The same estimate applies to every lower-order commutator in \(F_Z\). Combining it with \eqref{equa:seven-homogeneous-slice-coefficients} gives the required flat-source moments:
\be
\label{equa:seven-homogeneous-flat-source-moments}
\| \langle r\rangle S_Z\|_{L^2} + \| \langle r\rangle^2\del S_Z\|_{L^2} \lesssim C_{N,s_0}(m+A) \mathcal D.
\ee
In the metric terms, the second norm uses \(\langle r\rangle^2|YH| \lesssim(m+A) \langle r\rangle^{2-\lambda} \) and \(\langle r\rangle^2| \del YH| \lesssim(m+A) \langle r\rangle^{1-\lambda} \). The effective-field terms use \(1-\mu<\mu\). Thus no negative power of \(\coeff\) occurs.

\paragraph{Free commuted energy and decay}

The first jets of \(w_Z\) and \(v\) agree on \(t=T_0\), while \(\del_t^2 w_Z-\del_t^2 v=S_Z\). The first-boost derivative correction is \(\del_t L_a w_Z-\del_t L_a v=x_a S_Z\). At the second-boost level, the corrections are
\bse
\be
\label{equa:seven-homogeneous-free-boost-value}
L_aL_b w_Z-L_aL_b v=x_ax_b S_Z.
\ee
\be
\label{equa:seven-homogeneous-free-boost-time}
\del_t L_aL_b w_Z-\del_t L_aL_b v=T_0(x_b\del_a S_Z+x_a \del_b S_Z+ \delta_{ab}S_Z) +x_ax_b\del_t S_Z.
\ee
\ese
Spatial derivatives of \eqref{equa:seven-homogeneous-free-boost-value} give the other corrections. The flat-source moment bound therefore shows that the flat Cauchy energy of \(L^Jw_Z\), for \(|J| \leq2\), is bounded by \(C_{N,s_0} \mathcal D^2\).

Since boosts commute with the flat wave operator, conservation of the flat energy and positivity of the flux on spacelike graphs imply
\be
\label{equa:seven-homogeneous-hyperboloidal-energy}
\sum_{|J| \leq2} \int_{\Hcal_s} \Bigl(\frac{s^2}{t^2}| \del_t L^Jw_Z|^2 + \sum_a| \delsH_aL^Jw_Z|^2\Bigr)dx\lesssim C_{N,s_0} \mathcal D^2, \qquad s\geq T_0.
\ee
We obtain the inequality first on the graph \(t= \min\{\sqrt{s^2 +r^2},T\} \) for compactly supported smooth Cauchy data. Its outer flat portion has nonnegative energy. Passage to arbitrary finite-energy Cauchy data follows by energy approximation, and \(T\to\infty\) follows by monotone convergence of the hyperboloidal portion. For \(F_0[w]= (s/t) \del_t w\) and \(F_a[w]= \delsH_a w\), the commutation identities are
\be
\label{equa:seven-homogeneous-Sobolev-commutation}
L_b F_0[w]=F_0[L_b w]-\frac st F_b[w], \qquad L_b F_a[w]=F_a[L_b w]-\frac{x_a}{t}F_b[w].
\ee
The coefficients and their boost derivatives are bounded on each hyperboloid. Hence the hyperboloidal Sobolev inequality applied to \(F_0[w_Z] \) and \(F_a[w_Z] \) uses only the energies of \(L^Jw_Z\) with \(|J| \leq2\), and gives \eqref{equa:seven-homogeneous-wave-gain}. The conversion uses at most \(p_{\mathrm w} +2\) commuted energy derivatives and \(p_{\mathrm w} +6\) derivatives for the coefficient bounds, both within \(p_{\mathbf{high}} \).

\paragraph{Cauchy data of the matter correction}

The localized correction \(3\coeff Z(a_{\mathrm m}F_\psi) \) in \eqref{equa:seven-matter-localized-equation} changes both Cauchy data. On \(t=T_0\), we set
\be
\label{equa:seven-matter-correction-Cauchy-data}
f_Z\coloneqq3\coeff Z(a_{\mathrm m}F_\psi)|_{t=T_0}, \qquad g_Z\coloneqq3\coeff\del_tZ(a_{\mathrm m}F_\psi)|_{t=T_0}, \qquad \ord Z\leq p_{\mathrm w}.
\ee
The fixed-slice product estimates imply
\be
\label{equa:seven-matter-correction-Cauchy-moments}
\sum_{\ell=1}^3\| \langle r\rangle^2\nabla_x^\ell f_Z\|_{L^2} + \sum_{\ell=0}^2\| \langle r\rangle^2\nabla_x^\ell g_Z\|_{L^2} \lesssim C_{N,T_0, \coeff_0}A^2.
\ee
We first estimate the product before multiplication by \(a_{\mathrm m} \). A low scalar first-derivative factor gives \(A\langle r\rangle^{-1-\mu} \), while the remaining differentiated scalar is measured with the energy weight \(\langle r\rangle^\mu\). The inequality \(1-\mu<\mu\) absorbs the two moments. The largest scalar derivative requires \(p_{\mathrm w} +3\leq p_{\mathbf{low}} \) commuted energy orders. At \(N=19\), the smaller factor has at most five commutations, so its two rotations and radial Sobolev derivative stay within ten energy orders. If the largest derivative falls on a metric coefficient, the two scalar first derivatives leave the weight \(\langle r\rangle^{-2\mu} \), controlled by its Hardy weight \(\langle r\rangle^{\lambda-1} \) since \(\lambda+2\mu>1\). The bounded part of the slice follows from the direct finite-time estimates. The cutoff and its derivatives have bounded support on \(t=T_0\), so the product rule preserves these moment estimates.

We denote by \(w_{Z, \mathrm{corr}} \) the free wave with Cauchy data \((f_Z,g_Z) \). Each first derivative of \(L^Jw_{Z, \mathrm{corr}} \), \(|J| \leq2\), on this slice is a sum of spatial derivatives of \(f_Z\) up to order three and of \(g_Z\) up to order two, with polynomial coefficients in \(x\) of degree at most two. Higher time derivatives are replaced using the free equation. Thus \eqref{equa:seven-matter-correction-Cauchy-moments} controls its two-boost energy, and the preceding hyperboloidal argument gives
\be
\label{equa:seven-matter-correction-homogeneous-gain}
| \del w_{Z, \mathrm{corr}}|+ \frac ts| \delsH w_{Z, \mathrm{corr}}| \lesssim C_{N,T_0, \coeff_0}A^2t^{-1/2}s^{-1} \leq C_{N,T_0, \coeff_0}A^2s^{-3/2}.
\ee
The reconstruction of \(Z\psi\) therefore contains \(+3\coeff Z(a_{\mathrm m}F_\psi)-w_{Z, \mathrm{corr}} \), in addition to the original homogeneous wave and the corrected source volumes. At targets \(Q\geq2\), the local term equals \(3\coeff ZF_\psi\). The Cauchy term keeps the localized initial values even though the cutoff is constant at those targets.

\paragraph{Reconstruction of the massless wave}

\begin{proof}[Proof of~\autoref{lem:seven-massless-core-gain}]
\bse
The direct estimate \eqref{equa:seven-direct-massless-central-reduction} treats \(t\geq s^{1+2b} \). On the remaining targets, \(Q\geq s^{1-2b}/2\), so bounded \(Q\) also gives a bounded spacetime region, covered by the direct estimates. For large \(Q\), we use the commuted localized equation \eqref{equa:seven-matter-localized-equation}. Its contributions are controlled as follows.

\noindent{\bf 1.} \emph{Homogeneous waves from the Cauchy data.} The original homogeneous wave has amplitude \(CA\). The finite initial-time matter contribution and the two corrected Cauchy data have amplitude \(CA^2\), by \eqref{equa:seven-exterior-matter-initial-strip} and \eqref{equa:seven-matter-correction-homogeneous-gain}.

\noindent{\bf 2.} \emph{Estimating the metric source integrals.} The interior metric volume is controlled by \eqref{equa:seven-radial-metric-total-gain} and \eqref{equa:seven-complement-metric-gap}. The exterior metric volume and the kept \(q=1\) trace satisfy \eqref{equa:seven-exterior-metric-target-gain} and \eqref{equa:seven-exterior-metric-retained-trace}.

\noindent{\bf 3.} \emph{Estimating the matter correction.} The matter correction, its exterior source, and the localization strip are estimated in Sections~\ref{section-central-matter-correction} and \ref{section-central-exterior-matter}. Their local reconstruction term satisfies \eqref{equa:seven-matter-local-correction}. Every source contribution has amplitude at most \(C_{N,B_{s_1}, \coeff_0}A^2\), with positive excess decay. The source estimates apply to derivatives of the actual signed representation. The exceptional order at \(N=19\) is covered by \eqref{equa:seven-cone-exceptional-trace}; its loss occurs once and is absorbed by the strict source gaps. More explicitly, even the weaker restrictions \(\delta<1/8\), \(\rho=\delta/64\), and \(c<\rho/23\) give, with \(\varepsilon_N=\delta\),
\be
\label{equa:audit-exceptional-gap-check}
\gamma_{\mathrm{met}}>\frac16,\qquad \gamma_2>\frac1{80},\qquad \frac12-11b-e_{\mathrm{rem}}>\frac18.
\ee
The angular, regular, and additional principal-source gaps are larger than \(1/4\); all these margins are obtained before summing the source integrals. Thus no extra factor \(s^{\varepsilon_N}\) remains in the reconstructed field. The smallness condition gives \(B_{s_1}\leq2\), so all constants are uniform in \(s_1\). Summation, including both Cauchy terms of the localized correction, yields
\be
\label{equa:seven-central-reconstruction-gain}
t^{1/2}s| \del Z\psi|+ t^{3/2}| \delsH Z\psi| \leq C_N(A+A^2)s^{\eta_j}, \qquad \ord Z\leq p_{\mathrm w}, \quad\rank Z\leq j.
\ee
Division by \(t^{1/2}s\), followed by \(t\geq s\), proves \eqref{equa:seven-massless-core-gain}. The boundary-amplitude definition and \(C_{\mathrm{tr}} \geq1\) give
\be
\label{equa:seven-central-amplitude-comparison}
A+A^2\leq\mathcal B_3^{\mathrm{low}}(s) \leq\theta^{-1} \mathcal B_3^{\mathrm{low}}(s),
\ee
which proves \eqref{equa:seven-sufficient-massless-gain}.
\ese
\end{proof}

\paragraph{Estimate in the lightlike region}

\begin{proposition}[High-order matter-source estimate]
\label{lem:seven-massless-two-region-gain}
For \(0\leq j\leq p_{\mathrm w} \), the characteristic and direct Sobolev estimates give, on \(\Mint_{s, \mathrm{opt}} \)
\be
\label{equa:seven-massless-light-like-gain}
| \del\psi|_{p_{\mathrm w},j} \lesssim {C_N\over \theta} \mathcal B_3^{\mathrm{low}}(s) t^{-1}s^{\eta_j}, \qquad | \delsH\psi|_{p_{\mathrm w},j} \lesssim C_NA t^{-3/2}s^{\beta_{j+2}}.
\ee
Under the hypotheses of~\autoref{lem:seven-massless-core-gain}, the commuted matter source satisfies
\be
\label{equa:seven-massless-high-allocation}
\begin{aligned}
\mathcal K_{p_{\mathbf{high}},k}^\psi(s) & \lesssim C_Na_\psi(s)s^{-1} \mathbf P_{p_{\mathbf{high}},k}(s) + C_Na_\psi(s) \sum_{i =1}^{k}s^{-1+ \eta_i} \mathbf P_{p_{\mathbf{high}},k-i}(s) \\
& \quad + C_Na_\psi(s) \sum_{i =0}^{k}s^{-3/2 + \eta_i} \Bigl(U_{p_{\mathbf{high}},k-i}(s) + \mathbf W_{p_{\mathbf{high}},k-i}(s) \Bigr) \\
& \quad + C_N\mathbf R_{\psi;p_{\mathbf{high}},k}^{\mathrm{crit}}(s) + C_NA^2s^{-3/2 + \nu_k},
\end{aligned}
\ee
where the matter coefficient is defined by
\be
\label{equa:seven-massless-product-amplitude}
a_\psi(s) \coloneqq A+ \theta^{-1} \mathcal B_3^{\mathrm{low}}(s).
\ee
All constants are independent of \(0<\coeff\leq\coeff_0\).
\end{proposition}

\begin{proof}
The two lightlike bounds follow from the characteristic estimate and direct Sobolev. For the source estimate, we use the low--high decomposition of \eqref{equa:seven-matter-commutator-L2}; a low factor has at most \(p_{\mathrm w} \) commutations. There are four derivative placements.

\noindent{\bf 1.} \emph{Lower-order derivatives of the matter field.} If \(Z_2\psi\) is the low factor of rank \(i\), the semi-hyperboloidal contraction and \eqref{equa:seven-massless-core-gain} give
\be
\label{equa:seven-massless-global-low-high-product}
\|g(\nabla Z_1\fieldPhi, \nabla Z_2\psi) \|_{L^2(\Mint_s)} \lesssim C_N(A+A^2)s^{-3/2 + \eta_i}U_{p_{\mathbf{high}},k-i}(s).
\ee
Indeed, the mixed terms pair either a good high derivative with \(| \del Z_2\psi| \), or a weighted bad high derivative with \((t/s)| \delsH Z_2\psi| \). The flat bad--bad term has the factor \((s/t)^2\), while \((t/s)|H^{\H00}| \lesssim1\).

\noindent{\bf 2.} \emph{Lower-order derivatives of the effective field.} If the effective-field factor is low, its refined good or weighted bad derivative is placed in \(L^\infty\), and the high matter derivative is measured by \(\mathbf P\). Together with \eqref{equa:seven-low-metric-coefficient}, these terms give the terms involving \(\mathbf P\) in \eqref{equa:seven-massless-high-allocation}.

\noindent{\bf 3.} \emph{Using the wave gauge for translated metric coefficients.} A high metric coefficient containing a translation is controlled by the commuted wave gauge. The estimate \eqref{equa:seven-e-translated-metric-product}, proved below, gives the term with \(\mathbf W\) and an integrable trace remainder.

\noindent{\bf 4.} \emph{Metric coefficients commuted only with boosts.} A high boost-only metric coefficient paired with a low Hessian remains in \(\mathbf R_{\psi;p_{\mathbf{high}},k}^{\mathrm{crit}} \). The terms involving the reference metric and the lower-order products are those separated in \eqref{equa:seven-matter-commutator-L2}. The identity \(\eta_i+ \nu_{k-i} = \nu_k\) preserves the remaining high rank. Since \(A+A^2\lesssim a_\psi(s) \), summation proves \eqref{equa:seven-massless-high-allocation}.
\end{proof}


\section{Metric coefficients and critical commutators} 
\label{section-high-metric}


\subsection{\texorpdfstring{Weighted metric coefficients and traces at the interface}{Weighted metric coefficients and traces at the interface}}
\label{section-high-coefficients}

\paragraph{Weighted control after subtraction of the initial value}

The critical products pair a high metric coefficient with a low Hessian. We separate three estimates: the gauge controls coefficients containing a translation; a weighted identity controls the variation of a boost-only coefficient; and the exterior spacetime integral controls its initial value. Each estimate preserves the commutation rank.

First of all, we separate the high boost-only coefficient into its initial value and its subsequent variation. For a real scalar field $f$, we write $t_s(x)= \sqrt{s^2 + |x|^2}$ and $f_s(x)=f(t_s(x),x)$. The first data at fixed spatial position into the interior slab has parameter $s_{\mathrm{in}}(x)= \max\{s_0, \sqrt{2|x|+1} \}$, and we set
\be
\label{equa:seven-e-decomposition}
f_{\mathrm{in}}(x)=f(t_{s_{\mathrm{in}}(x)}(x),x), \qquad f_0(s,x)=f_s(x)-f_{\mathrm{in}}(x).
\ee
The function $f_{\mathrm{in}}$ is independent of $s$. We have $f_0(s_0,x)=0$ on the initial ball $B_{R_{s_0}}$, and $f_0(s,x)=0$ on the moving sphere $|x|=R_s$. For $0\leq\alpha \leq1$, we use the weight $w_\alpha=s^{-1}(t/s)^\alpha$ and the two norms
\be
\label{equa:seven-e-weighted-norms}
F_\alpha(s)= \|w_\alpha f_0\|_{L^2(B_{R_s})}, \qquad M_\alpha(s)= \|(s/t)^{1-\alpha} \del_tf\|_{L^2(\Mint_s)}.
\ee
At fixed $x$, we have $\del_sf_0= (s/t) \del_tf$ and $\del_s\log w_\alpha=-s^{-1}(1+ \alpha r^2/t^2)$. Differentiation keeps the full favorable term:
\be
\label{equa:seven-e-weighted-identity}
\frac{d}{ds}F_\alpha^2 + \frac2sF_\alpha^2 + \frac{2\alpha}{s} \|(r/t)w_\alpha f_0\|_2^2= \frac2s\int_{B_{R_s}}w_\alpha f_0(s/t)^{1-\alpha} \del_tf\, dx.
\ee
The moving-boundary term vanishes because $f_0=0$ there. Cauchy--Schwarz gives $F_\alpha'+s^{-1}F_\alpha \leq s^{-1}M_\alpha$, first where $F_\alpha>0$ and then by regularization. Since $F_\alpha(s_0)=0$, we obtain
\be
\label{equa:seven-e-weighted-estimate}
F_\alpha(s) \leq s^{-1} \int_{s_0}^sM_\alpha(\tau) \, d\tau.
\ee
The substitution $r=R_\tau$, with $dr= \tau\, d\tau$, gives the identity
\be
\label{equa:seven-e-trace-cost}
\|f_{\mathrm{in}} \|_{L^2(B_{R_s})}^2= \|f_{s_0} \|_{L^2(B_{R_{s_0}})}^2 + \int_{s_0}^s\tau\|f\|_{L^2(\mathbb S_\tau)}^2\, d\tau.
\ee

For the expanding ball of radius $R_b(s)$, the first entry parameter at fixed $x$ is $s_0$ on the initial ball and $R_b^{-1}(|x|)$ outside it. The variation $f_0$ still vanishes on the current sphere, so \eqref{equa:seven-e-weighted-identity} keeps its form. The measure in the last integral above becomes $R_b'(\tau)\,d\tau$; the identity~\eqref{equa:audit-moving-interface-initial-profile} keeps the same bound because $R_b'(\tau)\simeq\tau$.

\paragraph{The semi-hyperboloidal gauge identity}

The tensor $K$ in \eqref{equa:seven-densitized-metric-definition} satisfies $\del_\alpha K^{\alpha \beta} =0$. We write $z_a=x_a/t$ and $\ell_\H=dt-z_a dx^a$, and define
\be
\label{equa:seven-e-gauge-components}
K_\H=K(\ell_\H, \ell_\H), \qquad F^a=K^{a0}-z_bK^{ab}.
\ee
Substitution of the divergence equations and $\delsH_az_b=t^{-1}(\delta_{ab}-z_az_b)$ gives
\be
\label{equa:seven-e-gauge-identity}
\del_tK_\H=-\delsH_aF^a-t^{-1}(\delta_{ab}-z_az_b)K^{ab} +2t^{-1}z_aF^a.
\ee
The same identity holds for iterated tensor Lie derivatives of $K$ by Lorentz boosts: these vector fields are affine and divergence-free, so the commuted tensor still has zero divergence. The frame formula $\mathcal L_{L_c} \ell_\H=-z_c\ell_\H$ gives $L_cK_\H= (\mathcal L_{L_c}K) (\ell_\H, \ell_\H)-2z_cK_\H$. Its iteration therefore yields, without an additional order or rank,
\be
\label{equa:seven-e-gauge-commuted}
| \del_tL^JK_\H| \lesssim C_J\sum_{|J'| \leq|J|} \bigl(| \delsH L^{J'}K|+ t^{-1}|L^{J'}K| \bigr).
\ee
The identity $H^{\H00} =j_g^{-1}K_\H+(1-j_g^{-1}) (s/t)^2$ keeps the lightlike factor under boosts. Thus the derivative norm in \eqref{equa:seven-e-weighted-estimate}, for $f=L^JH^{\H00}$, is reduced to good metric derivatives, Hardy terms with their traces, and the differentiated inverse-metric products. This identity gives the weighted coefficient argument; the critical product still requires the corresponding wave Hessian estimate.

\paragraph{An unnormalized exterior sphere trace}

The merging annulus has width one. At fixed $s$, the function $T_s(r)= \fTime(s,r)$ equals $R_s+1$ at $r=R_s$ and is constant, with value in $[R_s+1,R_s+2]$, for $r\geq R_s+1$. Consequently, with $w_s(r)= \crochet(T_s(r),r)$ and $\lambda>1/2$, we have
\be
\label{equa:seven-e-exterior-trace-kernel}
w_s(r) \geq\tfrac13(1+r-R_s), \qquad R_s^2\int_{R_s}^\infty r^{-2}w_s(r)^{-2\lambda} \, dr\leq\frac{3^{2\lambda}}{2\lambda-1}.
\ee
For a decaying dynamic field, the fundamental theorem of calculus on the fixed hypersurface gives
\be
\label{equa:seven-e-radial-trace-identity}
v(T_s(R_s),R_s\omega)=-\int_{R_s}^\infty\delsME_rv(T_s(r),r\omega) \, dr.
\ee
Cauchy--Schwarz with \eqref{equa:seven-e-exterior-trace-kernel}, followed by integration over the sphere, proves
\be
\label{equa:seven-e-exterior-sphere-trace}
\|v\|_{L^2(\mathbb S_s)} \leq\frac{3^\lambda}{\sqrt{2\lambda-1}} \| \crochet^\lambda \delsME_rv\|_{L^2(\MME_s,dx)}.
\ee
For finite-energy asymptotically Euclidean fields, we first truncate the radial integral and pass along a sequence realizing the zero asymptotic value of the dynamic perturbation. The trace costs neither an angular commutation nor an additional derivative. To compare it with the exterior energy, we set $\ell_s=dt-T_s'(r)dr$ and use the coordinate normal density $N_s=-j_g g^{-1} \ell_s$. The stress-tensor identity is
\be
\label{equa:seven-e-exterior-energy-comparison}
2Q^{\mathrm{wave}}[v](N_s, \del_t)=j_g\bigl(-g^{-1}(\ell_s, \ell_s)| \del_tv|^2 +g^{ab} \delsME_av\delsME_bv\bigr).
\ee
The spacelike-hypersurface condition and positivity of $g^{ab}$ therefore control the weighted radial derivative in \eqref{equa:seven-e-exterior-sphere-trace}. Applying this estimate to Cartesian components of $L^Ju$, then using the inverse-metric identity, gives at the same order and rank
\be
\label{equa:seven-e-metric-sphere-trace}
\mathsf B^H_{2;p,k}(s) \lesssim C_N\bigl(A_0 + \Fenergy^{\ME,p,k}_\lambda(s,u) \bigr), \qquad p\leq p_{\mathbf{high}}.
\ee
The conversion to $H^{\H00}$ is made only on $\mathbb S_s$, where $x/t$ and all its boost derivatives are bounded. In each inverse-metric product, the highest-order dynamic factor is measured by the sphere norm and the other factors by the lower pointwise bounds. The reference coefficient has sphere norm $\Ocal(A_0)$ and is estimated separately from the dynamic weighted energy.


\subsection{\texorpdfstring{Antisymmetric null forms in the metric source}{Antisymmetric null forms in the metric source}}
\label{section:seven-antisymmetric-metric-source}

We write \(K^{\alpha \beta} =j_gg^{\alpha \beta}-(\gMink)^{\alpha \beta} \), as in \eqref{equa:seven-densitized-metric-definition}. The wave gauge gives \(\del_\alpha K^{\alpha \beta} =0\). The quadratic tensor and the antisymmetric null form are
\be
\label{equa:seven-trace-antisymmetric-definition}
C^{\lambda \delta}{\coloneqq} (\gMink)_{\alpha \beta}K^{\alpha \lambda}K^{\beta \delta}, \qquad 
Q_{\alpha \beta}(f,v) \coloneqq\del_\alpha f\, \del_\beta v-\del_\beta f\, \del_\alpha v, \quad 0\leq\alpha, \beta \leq3.
\ee
Direct differentiation, with both divergences kept, gives the identity
\be
\label{equa:seven-trace-double-divergence-null}
\del_\lambda \del_\delta C^{\lambda \delta} = (\gMink)_{\alpha \beta}(\del_\delta K^{\alpha \lambda}) (\del_\lambda K^{\beta \delta})= (\gMink)_{\alpha \beta}Q_{\delta \lambda}(K^{\alpha \lambda},K^{\beta \delta}).
\ee
The terms with two derivatives on the same factor vanish by the gauge, and the product subtracted in the last expression is \((\del_\lambda K^{\alpha \lambda}) (\del_\delta K^{\beta \delta})=0\). This bilinear identity also holds for two distinct divergence-free tensors. Tensor Lie derivatives by boosts preserve the divergence identity, so all ordered boost partitions keep this null structure.

\begin{proposition}[Retarded estimate for the metric double divergence]
\label{lem:seven-trace-antisymmetric-retarded}
Let \(\ell=p_{\mathbf{low}} +2\), as in~\autoref{lem:seven-boost-v-interpolation}, and choose the free exponent \(\delta \) subject to the preceding weight conditions and
\be
\label{equa:seven-trace-interpolation-exponent-choice}
\delta(26\ell+115)<32.
\ee
Under the finite bootstrap, the propagated wave gauge, and~\autoref{lem:seven-first-full-metric-refinement}, the zero-data retarded contribution of \(L^I\del_\lambda \del_\delta C^{\lambda \delta} \), split over the interior and exterior parts of the past cone, is bounded by \(C_NA^2/t\) for \(|I| \leq p_{\mathbf{low}} \). The constants are independent of \(\coeff\) and the terminal time.
\end{proposition}

\begin{proof}
\bse

\noindent{\bf 1.} \emph{The interior source.} The identities \(\del_a=t^{-1}L_a-z_a \del_t\) and \(Q_{0a}(f,v)=t^{-1}(\del_tf\,L_av-L_af\, \del_tv) \) give \(|Q_{\alpha \beta}(f,v)| \lesssim t^{-1}(|Lf|| \del v|+ | \del f||Lv|) \) for every pair of spacetime indices. For a partition \(i+j\leq\ell-2\), we combine~\autoref{lem:seven-boost-v-interpolation} with \(| \del L^jK| \lesssim At^{-1/2}s^{-1+b} \), where \(b=13\delta/64\). For the placement with the boost on the first factor, the exponents are
\be
\label{equa:seven-trace-null-interpolation-powers}
\aligned 
&
t^{-1}|L^{i+1}K| \,| \del L^jK| \lesssim A^2t^{-2-\epsilon_i}s^{-1+g_i},
\quad 
\\
& \epsilon_i\coloneqq\frac{1-a_i}{2}, \; g_i\coloneqq2\delta(1-a_i) +b(a_i+1), \; a_i\coloneqq\frac{i+1}{\ell}.
\endaligned
\ee
Since \(a_i\leq1-1/\ell\), we have \(g_i/\epsilon_i\leq\delta(26\ell+115)/32<1\). The inequality \(s\leq t\) consequently gives
\be
\label{equa:seven-trace-null-interior-source}
|L^I\del_\lambda \del_\delta C^{\lambda \delta}| \leq C_NA^2t^{-2}s^{-1}, \qquad |I| \leq p_{\mathbf{low}}, \quad q\geq1.
\ee

\noindent{\bf 2.} \emph{The positive cone kernel.} For a target \((T,x) \), we put \(R=|x| \), \(P=T+R\), and \(Q=T-R\geq1\). At a source point we use \(p'= \tau+r'\), \(q'= \tau-r'\). The radial Kirchhoff majorant of \eqref{equa:seven-trace-null-interior-source} is bounded by
\be
\label{equa:seven-trace-null-interior-kernel}
\frac{C_NA^2}{R} \int_Q^P(p')^{-3/2} \, dp'\int_1^Q(q')^{-1/2} \, dq'\leq\frac{C_NA^2}{T}.
\ee
For \(R\geq T/2\), the two integrals are bounded by \(2Q^{-1/2} \) and \(2Q^{1/2} \). For \(R\leq T/2\), the first is at most \(CR T^{-3/2} \), and the second is at most \(CT^{1/2} \). The latter argument includes the limit \(R=0\). The actual initial graph only reduces this positive integration domain.

\noindent{\bf 3.} \emph{The exterior source.} The direct exterior metric jets give, with \(w=2 +r'-\tau\),
\be
\label{equa:seven-trace-null-exterior-source}
|L^I\del_\lambda \del_\delta C^{\lambda \delta}| \leq C_NA^2(r')^{-3}w^{1-2\lambda} \tau^\delta, \qquad |I| \leq p_{\mathbf{low}}.
\ee
Near the cone we pair a good derivative \(Ar'^{-2}w^{1-\lambda}s_{\mathrm e}^{\delta} \) with a Cartesian derivative \(Ar'^{-1}w^{-\lambda}s_{\mathrm e}^{\delta} \). In the far region \(w\simeq r'\), the direct Cartesian bounds give the same result. These estimates require at most \(p_{\mathbf{low}} +6\leq N\) metric orders. The explicit reference derivatives are bounded by \(C_NA_0(r')^{-2} \); their cross and quadratic terms are no larger. Since \(r'\simeq p'\) on this cone portion, integration in \(q'\leq1\) gives
\be
\label{equa:seven-trace-null-exterior-kernel}
\frac{C_NA^2}{R} \int_Q^P(p')^{-2\lambda+ \delta} \, dp'\leq\frac{C_NA^2}{T}.
\ee
Here \(2\lambda-\delta>1\); the two target regions are treated as in the interior calculation. Splitting the actual source into its two integration regions creates no derivative of a cutoff and no additional interface term. The Cauchy contribution remains separate from this zero-data estimate.
\ese
\end{proof}


\subsection{\texorpdfstring{Weighted Hessians and coefficient variation}{Weighted Hessians and coefficient variation}}
\label{section-weighted-Hessians}

\paragraph{Weighted low Hessians}

For \(0\leq j\leq3\), the lightlike Hessian estimate gives the following bound after multiplication by \(s/t\), with constants depending on \(N,B_{s_1}, \coeff_0\):
\be
\label{equa:seven-e-weighted-low-Hessian}
\sup_{\Mint_s} \frac st\bigl(| \del^2u|_{3,j} + | \del^2H|_{3,j} + | \del^2\psi|_{3,j} \bigr) \lesssim A s^{-5/2 + \nu_{j+3}} \lesssim A s^{-2 + \eta_j}.
\ee
We use \eqref{equa:seven-cone-light-like-Hessian-reduction}, with its smooth semi-hyperboloidal extension to the spatial origin. The term \(t(| \del L_aw|+ | \del w|)/s^2\) costs at most six lower energy orders. All quadratic derivative sources, including the matter stress, have size \(A^2t^{-1}s^{-2 +24\rho+(j+8)c} \), where \(c=C_{\mathrm{rk}} \theta \). The effective-field potential is bounded by \(A^2t^{-3}s^{24\rho+(j+8)c} \), using \(\fieldvarphi = \coeff^{-1/2} \fieldPhi\). Thus the direct calculation gives
\be
\label{equa:seven-e-direct-Hessian-terms}
| \del^2(u,H, \psi)|_{3,j} \lesssim C_{N,B_{s_1}, \coeff_0} \bigl(A t^{1/2}s^{-3+ \nu_{j+3}} +A^2t s^{-4+24\rho+(j+8)c} +A_0t^{-1}s^{-2} \bigr).
\ee
Principal commutators are estimated by induction on total order. Their bad coefficient contributes \(C_{N,B_{s_1}}(t/s^2)s^{\eta_i} \) times a lower Hessian, and \(\eta_i+ \nu_{j-i+3} = \nu_{j+3} \); the other components have a good derivative. The inverse-metric conversion adds only quadratic first derivatives. Multiplication by \(s/t\), followed by \(t\geq s\), proves \eqref{equa:seven-e-weighted-low-Hessian}, since \(12\rho+3c<1/2\) and \(12\rho+5c<1/2\). These estimates use lower direct energies and the refined bad coefficient through order three.

\paragraph{Metric coefficients containing a translation}

A commuted coefficient containing a translation can be written as \(\del(YH^{\H00}) \), up to frame terms with an additional factor \(t^{-1} \). The commuted gauge gives its time derivative, and \(\del_a= \delsH_a-(x_a/t) \del_t\) then controls every Cartesian derivative. Boosts preserve \(\ell_\H\) up to a bounded scalar factor; translations of \(\ell_\H\) contribute \(t^{-1} \). Their differentiated frame terms are controlled by energy and Hardy with its trace. For \(Y\preccurlyeq(p-1,k) \), the inverse-metric conversion therefore gives
\be
\label{equa:seven-e-translated-metric-gauge}
\| \del(YH^{\H00}) \|_{L^2(\Mint_s)} \lesssim\mathbf Q^u_{p,k} +A\sum_{i =0}^{\min\{k, \lfloor p/2\rfloor\}}s^{\nu_{i+2}} \mathbf Q^u_{p,k-i} +A_0s^{-1/2}.
\ee
The two nonlinear placements are \((t/s)|Z_i u| \) against a weighted high derivative and \(t| \del Z_i u| \) against a Hardy high factor. Each is bounded by \(CA s^{\nu_{i+2}} \), as in the gauge calculation below. When the metric coefficient has order above \(p_{\mathbf{low}} \), the remaining wave operator has at most three commutations. The lightlike Hessian estimate \eqref{equa:seven-cone-light-like-Hessian}, extended to the center, yields \(| \del^2Z\psi|_{3,j} \lesssim A s^{-2 + \nu_{j+3}} \), since \(p=t+r\lesssim s^2\). Consequently, the contribution \(\mathbf R^{\mathrm{tr}}_{\psi;p,k} \) of these translated high coefficients satisfies
\be
\label{equa:seven-e-translated-metric-product}
\mathbf R^{\mathrm{tr}}_{\psi;p,k}(s) \lesssim A\sum_{\ell=0}^k s^{-3/2 + \eta_\ell} \mathbf W_{p,k-\ell}(s) +A^2s^{-3/2 + \nu_k}.
\ee
Before weakening the powers, the two coefficients of \(\mathbf Q^u\) are \(A s^{-2 + \eta_j+12\rho+5c} \) and \(A^2s^{-2 + \eta_{i+j} +24\rho+9c} \). The inequalities \(12\rho+5c<1/2\) and \(24\rho+9c<1/2\) give the decay without losing a rank. The normalized Hardy trace is bounded by \(A s^{-1+ \delta} \); it and the reference term give the last term in \eqref{equa:seven-e-translated-metric-product}.

\paragraph{The varying part of the critical coefficient}

We take \(\alpha=1\) in \eqref{equa:seven-e-weighted-norms} and sum the squared norms over the boost operators \(p_{\mathbf{low}}<|J| \leq\min\{p,k\} \), with \(f=L^JH^{\H00} \); the resulting norms are denoted by \(F_{1;p,k} \) and \(M_{1;p,k} \). We keep the favorable terms of \eqref{equa:seven-e-weighted-identity}. In the gauge conversion, the two nonlinear placements have the pointwise coefficients \((t/s)|Z_i u| \) and \(t| \del Z_i u| \). Direct Sobolev and Hardy, including the normalized sphere trace, bound each by \(CA s^{\nu_{i+2}} \), since \(t^{1/2}/s\lesssim1\). With \(\mathbf Q^u\) from \eqref{equa:seven-metric-Hardy-energy}, the gauge identity therefore gives
\be
\label{equa:seven-e-alpha-one-gauge}
M_{1;p,k} \lesssim C_N\mathbf Q^u_{p,k} +C_NA\sum_{i =0}^{\min\{k, \lfloor p/2\rfloor\}}s^{\nu_{i+2}} \mathbf Q^u_{p,k-i} +C_NA_0s^{-1/2}, \qquad p\leq p_{\mathbf{high}}.
\ee
Each distinguished pointwise factor costs at most \(\lfloor p/2\rfloor+2\leq p_{\mathbf{low}} \) orders. Additional low factors in a differentiated smooth coefficient are uniformly bounded; their unused ranks are included in the cumulative high norm. The reference contribution uses \(\|t^{-2} \|_{L^2(\Mint_s)} \lesssim s^{-1/2} \). The normalized trace in \(\mathbf Q^u\) remains in this estimate.

We denote by \(\mathbf R^{\mathrm{var}}_{v;p,k} \) the part of \(\mathbf R^{\mathrm{crit}}_{v;p,k} \) obtained by replacing each high coefficient \(f\) by \(f_0\), for \(v=u, \psi\). The weighted Hessian gives
\be
\label{equa:seven-e-varying-product}
\mathbf R^{\mathrm{var}}_{v;p,k}(s) \lesssim C_{N,B_{s_1}, \coeff_0}A\sum_{j=0}^{\min\{3,k\}}s^{-3/2 + \nu_{j+3}}F_{1;p,k-j}(s).
\ee
The bound \eqref{equa:seven-e-weighted-estimate} and a change in the order of integration give
\be
\label{equa:seven-e-varying-Fubini}
\int_{s_0}^{s_1} s^{-3/2 + \nu_{j+3}}F_{1;p,k-j}(s) \, ds\leq\frac1{3/2-\nu_{j+3}} \int_{s_0}^{s_1}\tau^{-3/2 + \nu_{j+3}}M_{1;p,k-j}(\tau) \, d\tau.
\ee
The resulting coefficients of \(\mathbf Q^u_{p,k-j} \) and \(\mathbf Q^u_{p,k-i-j} \) are, respectively,
\be
\label{equa:seven-e-varying-rank-kernels}
C_{N,B_{s_1}, \coeff_0}A\tau^{-3/2 + \eta_j+12\rho+5c}, \qquad C_{N,B_{s_1}, \coeff_0}A^2\tau^{-3/2 + \eta_{i+j} +24\rho+9c}.
\ee
Their excess decay is positive because \(24\rho+9c<1/2\). The radial trace estimate \eqref{equa:seven-e-exterior-sphere-trace} and the coarse exterior bootstrap give
\be
\label{equa:seven-e-normalized-trace-bound}
\mathsf T_{2;p,k}[u] \lesssim s^{-1} \Fenergy^{\ME,p,k}_\lambda(s,u) \lesssim A s^{-1+ \delta}.
\ee
These trace contributions are integrable as well. The reference term is bounded after integration by \(C_{N,B_{s_1}, \coeff_0}AA_0\). This estimates the variation at the required rank. The initial value is controlled by the integrated exterior trace in~\autoref{lem:seven-e-good-bulk}.


\subsection{\texorpdfstring{Integrated initial traces}{Integrated initial traces}}
\label{section-e-bulk}

\paragraph{Exterior spacetime control}

The wave gauge controls the integrated initial trace by three complementary estimates:
\begin{itemize}
\item the good exterior spacetime integral controls tangential derivatives;
\item an outgoing Hardy identity controls the undifferentiated dynamic coefficient, with its upper-hypersurface term;
\item a radial gauge identity gives the sphere trace, while the reference coefficients are integrated explicitly.
\end{itemize} We write \(w=2 +r-t\) on \(t-r\leq1\); this weight is comparable to \(\crochet\). For a dynamic metric component, the exterior bootstrap controls
\be
\label{equa:seven-e-good-bulk-input}
\sum_{Z\preccurlyeq(p,p)} \Bigl(\int_{\MME_{[s_0,s_1]}}w^{2\lambda-1}| \delsN Zu|^2\, dx\, dt+ \|w^{\lambda-1}Zu\|_{L^2(\MME_{s_1})}^2\Bigr) \lesssim A^2{s_1}^{2\delta}, \qquad p=p_{\mathbf{high}}.
\ee
Here \(J= \del_s\fTime(s,r) \) is the foliation Jacobian, so \(dx\, dt=J\, dx\, ds\). The second term is the exterior Hardy norm of the dynamic perturbation; the reference metric is estimated separately.

\paragraph{Radial trace of the coefficient}

\begin{proposition}[Integrated boost-only initial trace]
\label{lem:seven-e-good-bulk}
For \(p=p_{\mathbf{high}} \), \(0\leq k\leq p\), the initial values of \(f_J=L^JH^{\H00} \) are measured by
\be
\label{equa:seven-e-integrated-profile}
E_{0;p,k}(s_1)^2\coloneqq{s_1}^{-2} \sum_{p_{\mathbf{low}}<|J| \leq\min\{p,k\}} \|(f_J)_{\mathrm{in}} \|_{L^2(B_{R_{s_1}})}^2.
\ee
Under the exterior bootstrap and the propagated wave gauge, the asymptotically Euclidean normalization gives
\be
\label{equa:seven-e-good-bulk-bound}
E_{0;p,k}(s_1) \lesssim C_N\bigl(A {s_1}^{1-2\lambda+ \delta} +A {s_1}^{-1} +A_0\bigr).
\ee
The terms produced by the fixed reference coefficients have the pointwise majorant \(C_NA_0\langle r\rangle^{-1} \). After separating these terms, the last summand in \eqref{equa:seven-e-good-bulk-bound} is absent. The sphere norms in the proof use Euclidean area; their control preserves the differential order and boost rank.
\end{proposition}

\begin{proof}

The initial estimate follows from three distinct calculations.

\noindent{\bf 1.} \emph{Spacetime Hardy estimate.} We first consider a dynamic component \(v=Zu\). At fixed \(q=t-r\) and sphere direction, we set \(p=t+r\), so that \(\del_p=L/2\). The identity
\bse
\be
\label{equa:seven-e-outgoing-Hardy-identity}
\del_p(r|v|^2)= \tfrac12|v|^2 +2r\operatorname{Re}(\overline v\del_pv)
\ee
is integrated between the initial and upper Euclidean--hyperboloidal hypersurfaces. Its contribution on the initial hypersurface has a favorable sign. Cauchy--Schwarz yields
\be
\label{equa:seven-e-outgoing-Hardy-line}
\int|v|^2\, dp\leq4r_{\mathrm{end}}|v_{\mathrm{end}}|^2 +16\int r^2| \del_pv|^2\, dp.
\ee
On the upper hypersurface, \(|dq|= (1-\del_r\fTime(s_1,r)) \, dr\). Since \(0\leq1-\del_r\fTime\leq1\) and \(w\lesssim r\), its boundary contribution satisfies
\be
\label{equa:seven-e-outgoing-Hardy-top}
\int w^{2\lambda-1}r_{\mathrm{end}}|v_{\mathrm{end}}|^2\,|dq| \, d\omega \lesssim\|w^{\lambda-1}v\|_{L^2(\MME_{s_1})}^2.
\ee
Integration of \eqref{equa:seven-e-outgoing-Hardy-line} in \((q, \omega) \), using \(dt\, dr= \tfrac12dp\, dq\), therefore gives
\be
\label{equa:seven-e-outgoing-Hardy-bulk}
\int_{\MME_{[s_0,s_1]}}w^{2\lambda-1}r^{-2}|Zu|^2\, dx\, dt\lesssim A^2{s_1}^{2\delta}.
\ee
\ese
The upper-hypersurface contribution is the Hardy norm in \eqref{equa:seven-e-good-bulk-input}. Truncation in the incoming variable and passage to the limit are justified by these integrable norms.

\noindent{\bf 2.} \emph{Radial trace and wave gauge.} We take the radial trace on a physical time slice. For a symmetric tensor \(P\), we use
\bse
\be
\label{equa:seven-e-null-gauge-components}
P_N=P^{00}-2\omega_aP^{a0} + \omega_a \omega_bP^{ab}, \quad F^a=P^{a0}-\omega_bP^{ab}, \quad R^\beta= \del_\alpha P^{\alpha \beta}, \quad\Pi_{ab} = \delta_{ab}-\omega_a \omega_b.
\ee
The divergence identity gives
\be
\label{equa:seven-e-null-gauge-radial}
\del_tP_N=-\delsN_aF^a-r^{-1} \Pi_{ab}P^{ab} +R^0-\omega_aR^a, \qquad | \del_rP_N| \lesssim| \delsN P|+r^{-1}|P|+ |R|.
\ee
The second bound uses \(\del_rP_N=LP_N-\del_tP_N\). We apply it to the tensor Lie derivatives of \(K-K^\star\), where \(K^\star\) is the densitized inverse of the prescribed reference metric minus the Minkowski inverse. The full tensor \(K\) has zero divergence, and affine divergence-free boosts preserve this identity. Thus \(R\) contains only differentiated reference coefficients, bounded by \(C_NA_0r^{-2} \). Their radial integrals have the explicit bound \(C_NA_0r^{-1} \).

The dynamic composition terms obey, at the same order,
\be
\label{equa:seven-e-dynamic-density-bulk}
|P| \lesssim\sum_{\ord Z\leq p}|Zu|, \qquad | \delsN P| \lesssim\sum_{\ord Z\leq p}| \delsN Zu|+A r^{-1} \sum_{\ord Z\leq p}|Zu|.
\ee
Indeed, the direct exterior estimates give \(|Zu_{\mathrm{low}}|+r| \delsN Zu_{\mathrm{low}}| \lesssim A\). On \(r\leq2t\), the outgoing and angular identities give \(| \delsN Zu_{\mathrm{low}}| \lesssim A r^{-2}w^{1-\lambda}s_{\mathrm e}^{\delta} \). On \(r\geq2t\), the Cartesian derivative estimate gives the same bound because \(w\simeq r\). The inequalities \(w\lesssim r\) and \(s_{\mathrm e}^2\simeq t\lesssim r\) make its product with \(r\) bounded. Rotations can be written as commutators of two boosts, including in the far exterior. Together with the five-order sphere Sobolev count, the safe lower-factor cost is therefore \(\lfloor p/2\rfloor+7\leq p_{\mathbf{high}} \), equal to \(14\) at \(N=19\). Equations \eqref{equa:seven-e-good-bulk-input} and \eqref{equa:seven-e-outgoing-Hardy-bulk} consequently control the spacetime integral of \(w^{2\lambda-1}(| \delsN P|^2 +r^{-2}|P|^2) \) by \(CA^2{s_1}^{2\delta} \).

For \(t\geq T_0\), the whole exterior part of the physical slice lies above the initial hypersurface. We set \(R=t-1\). The dynamic field tends to zero at spatial infinity. The radial fundamental theorem and Cauchy--Schwarz use the kernel
\be
\label{equa:seven-e-physical-trace-kernel}
R^2\int_R^\infty r^{-2}(1+r-R)^{1-2\lambda} \, dr\lesssim C_\lambda R^{2-2\lambda}, \qquad\tfrac12<\lambda<1.
\ee
\ese
The substitution \(r=R(1+y) \) proves this bound by the integrability of \((1+y)^{-2}y^{1-2\lambda} \). With the reference contribution kept separately, this proves
\bse
\be
\label{equa:seven-e-physical-sphere-trace}
\|P_N(t, \cdot) \|_{L^2(r=t-1)}^2\lesssim t^{2-2\lambda} \int_{r\geq t-1}w^{2\lambda-1}(| \delsN P|^2 +r^{-2}|P|^2) \, dx+C_NA_0^2.
\ee
For \(t\leq t_b(s_1)\coloneqq ({s_1}^2 +1)/2\), the physical exterior slice is below the upper hypersurface \(s_1\). Integration of \eqref{equa:seven-e-physical-sphere-trace} therefore gives
\be
\label{equa:seven-e-physical-trace-integral}
\int_{T_0}^{t_b(s_1)} \|P_N(t, \cdot) \|_{L^2(r=t-1)}^2\, dt\lesssim A^2{s_1}^{4-4\lambda+2\delta} +A_0^2{s_1}^2.
\ee
The bounded initial time interval is controlled separately and contributes \(CA/s_1\) to the initial norm. This is why the physical radial fundamental theorem is started at \(T_0\), rather than at the inner corner of the initial hypersurface.

\noindent{\bf 3.} \emph{Recovering the scalar metric coefficient and its initial norm.} We convert the tensorial trace to the scalar coefficient. With \(\ell_N=dt-dr\), the frame identities are
\be
\label{equa:seven-e-null-boost-conversion}
\mathcal L_{L_c} \ell_N=-\omega_c\ell_N-(q/r) \Pi_{ca} \, dx^a, \qquad\ell_\H= \ell_N+(q/t) \, dr.
\ee
All boost derivatives of \(q/r\) and \(q/t\) keep these factors. At \(q=1\), the resulting frame errors are therefore \(\Ocal(r^{-1}) \) times components of no greater order or rank. The density identity \(H^{\H00} =j_g^{-1}K_\H+(1-j_g^{-1}) (s/t)^2\) has the same property, since \((s/t)^2=\Ocal(r^{-1}) \) there. If the highest derivative falls on the density, the remaining lower metric value is \(\Ocal(A r^{-1}s^\delta) \); if it falls on \(K_\H\), the lower coefficient is bounded. The full component sphere norm is \(\Ocal(A s^\delta) \) by \eqref{equa:seven-e-exterior-sphere-trace}. Thus the dynamic frame and density remainders have sphere norms
\be
\label{equa:seven-e-frame-remainder-trace}
\Ocal(A s^{-2 + \delta} +A^2s^{-2 +2\delta}),
\ee
\ese
whose squared integrals against \(s\, ds\) are bounded. The reference values and reference-divergence integrals keep their pointwise \(C_NA_0\langle r\rangle^{-1} \) majorant. The spacetime Hardy estimate was used only for dynamic fields, never for the reference.

Finally, \(dt_b=s\, ds\) on the interface. Substitution of \eqref{equa:seven-e-physical-trace-integral} and \eqref{equa:seven-e-frame-remainder-trace} into the identity \eqref{equa:seven-e-trace-cost} proves \eqref{equa:seven-e-good-bulk-bound}. All radial integrations follow first for smooth truncated fields and then by the asymptotically Euclidean limit and these weighted norms. \qedhere

\end{proof}

\paragraph{Pointwise metric trace on the interface}

\begin{proposition}[Pointwise metric interface trace]
\label{lem:seven-pointwise-metric-interface-bulk}
Under the hypotheses of~\autoref{lem:seven-e-good-bulk}, the metric interface amplitude satisfies
\be
\label{equa:seven-pointwise-metric-interface-bulk}
\max_{0\leq k\leq p_{\mathbf{low}}} \mathsf B_{00;k}(s_1) \leq C_N(A_0 +A), \qquad s_1\geq s_0.
\ee
The constant is independent of \(s_1\) and \(0<\coeff\leq\coeff_0\). This estimate uses the coarse exterior energy and good spacetime integral, together with the propagated wave gauge.
\end{proposition}

\begin{proof}

The tensor argument in~\autoref{lem:seven-e-good-bulk} applies to mixed translation--boost operators of order at most \(p_{\mathbf{high}} \). Translations, like boosts, are affine divergence-free vector fields, so their tensor Lie derivatives preserve the divergence identity for the full densitized inverse metric. Translation derivatives of \(\ell_N\) and \(\ell_\H\) have coefficients \(\Ocal(r^{-1}) \) on \(q=1\). They therefore belong to the frame remainders already estimated in \eqref{equa:seven-e-frame-remainder-trace}. The differentiated reference divergence still has the bound \(C_NA_0r^{-2} \). The dynamic density products use the same lower-factor count:
\bse
\be
\label{equa:seven-mixed-e-order-count}
\lfloor p/2\rfloor+7\leq p_{\mathbf{high}}, \qquad p\leq p_{\mathbf{high}}, \qquad N\geq19.
\ee
Consequently, the physical trace calculation \eqref{equa:seven-e-physical-trace-integral}, followed by the frame and density conversion, gives
\be
\label{equa:seven-mixed-e-time-integral}
\sum_{Z\preccurlyeq(p,p)} \int_{T_0}^{T} \|ZH^{\H00}(t, \cdot) \|_{L^2(r=t-1)}^2\, dt \leq C_N\bigl(A^2T^{2-2\lambda+ \delta} +A^2 +A_0^2T\bigr), \qquad p\leq p_{\mathbf{high}}.
\ee
\ese
Here \(T_0\) is the initial-time ceiling in \eqref{equa:seven-cone-initial-time-ceiling}, and the sphere norm uses the physical Euclidean area \((t-1)^2d\omega \). To obtain \eqref{equa:seven-mixed-e-time-integral}, we take the upper hypersurface \(s= \sqrt{2T-1} \); all physical slices in the integral lie below that hypersurface. The bounded initial time interval contributes at most \(C_NA^2\). The spacetime Hardy inequality continues to be applied only to the dynamic perturbation, with the reference contribution kept separate.

For a mixed operator \(Z\), we use
\bse
\be
\label{equa:seven-cone-time-trace-function}
f_Z(t, \omega)\coloneqq (t-1)ZH^{\H00}(t,(t-1) \omega), \qquad \|f_Z(t) \|_{L^2(\mathbb S^2)} = \|ZH^{\H00}(t, \cdot) \|_{L^2(r=t-1)}.
\ee
On \(q=1\), the outgoing derivative satisfies the identities
\be
\label{equa:seven-cone-time-boost-identity}
t(\del_t+ \del_r)= \omega^aL_a+ \del_t, \qquad t{d f_Z\over dt} ={t\over t-1}f_Z+(t-1) \omega^aL_a(ZH^{\H00}) +(t-1) \del_t(ZH^{\H00}).
\ee
\ese
On a moving sphere of physical radius $r_b(t)$, the trace is $f_Z(t,\omega)=r_b(t)ZH^{\H00}(t,r_b(t)\omega)$. The derivative $\del_t+r_b'(t)\del_r$ gives the identity~\eqref{equa:audit-moving-interface-time-boost}. Its bounded coefficients still cost one commutation. The surface area is $r_b(t)^2d\omega$, and the time-to-hypersurface conversion is given in \eqref{equa:audit-moving-interface-radius}. These changes preserve the weights and derivative orders below. Thus \(t\, df_Z/dt\) is controlled by physical sphere traces with one additional commutation. For \(T\geq2T_0\), the Hilbert-valued Sobolev inequality on the past interval \([T/2,T] \) gives
\bse
\be
\label{equa:seven-past-time-Sobolev}
\|f_Z(T) \|_{L^2(\mathbb S^2)}^2\leq {C\over T} \int_{T/2}^{T} \Bigl(\|f_Z(t) \|_{L^2(\mathbb S^2)}^2 + \left\|t{d f_Z\over dt} \right\|_{L^2(\mathbb S^2)}^2\Bigr)dt.
\ee
Using \eqref{equa:seven-mixed-e-time-integral} at one higher order therefore yields
\be
\label{equa:seven-pointwise-physical-metric-trace}
\|ZH^{\H00}(T, \cdot) \|_{L^2(r=T-1)} \leq C_N\bigl(A T^{1/2-\lambda+ \delta/2} +A T^{-1/2} +A_0\bigr), \qquad \ord Z\leq p_{\mathbf{high}}-1.
\ee
The exponent \(1-2\lambda+ \delta \) is negative under the prescribed weight gaps. The bounded initial interval is controlled by the same fixed-hypersurface trace estimate and the finite-time energy bounds.

Finally, on the interface the rotation identity
\be
\label{equa:seven-interface-rotation-order-count}
\Omega_{ab} ={x_a \over t}L_b-{x_b\over t}L_a, \qquad |J|+2 +1\leq p_{\mathbf{low}} +3=p_{\mathbf{high}}-1
\ee
\ese
converts two rotations into two boosts with bounded coefficients. This conversion is used only on \(q=1\). One additional commutation pays for \eqref{equa:seven-cone-time-boost-identity}; in particular, the largest order is \(13\) when \(N=19\). Applying \eqref{equa:seven-pointwise-physical-metric-trace} to the resulting boost operators bounds every sphere norm in \eqref{equa:seven-metric-pointwise-traces} by \(C_N(A+A_0) \). Since \(s^2/R_s\) is bounded and \(s^{-\eta_j} \leq1\), this proves \eqref{equa:seven-pointwise-metric-interface-bulk}.

\end{proof}

\paragraph{Products involving the initial metric coefficient}

For the initial contribution, the full wave Hessian is bounded by \(C_{N,B_{s_1}, \coeff_0}A s^{-2 + \nu_{j+3}} \). The integrated initial estimate therefore gives the integrable coefficient
\be
\label{equa:seven-e-dynamic-integrable-kernel}
C_{N,B_{s_1}, \coeff_0}A^2s^{-2\lambda+ \delta+ \nu_{j+3}}, \qquad 0\leq j\leq3, \qquad 2\lambda-\delta-\nu_6>1.
\ee
The exponent inequality follows from \(\lambda>3/4\), \(\delta<1/8\), \(\rho= \delta/64\), and \(c(N+4)<\rho\). The reference contribution is estimated separately. For the reference contribution, \eqref{equa:seven-cone-light-like-Hessian}, with its extension to the center, and \(\| \langle r\rangle^{-1}t^{1/2} \|_{L^2(B_{R_s})} \lesssim s^2\) give
\be
\label{equa:seven-e-reference-light-like-product}
\|f_{\mathrm{ref,in}} \del^2Zv\|_2\lesssim C_{N,B_{s_1}}AA_0s^{-1+ \nu_{j+3}}, \qquad \int_{s_0}^{s_1}\|f_{\mathrm{ref,in}} \del^2Zv\|_2\, ds\lesssim\frac{C_{N,B_{s_1}}AA_0}{\rho}{s_1}^{\nu_k}.
\ee
Here \(v\in\{u, \psi\} \), \(j\leq3\), and the high boost-only coefficient has more than \(p_{\mathbf{low}} \) boosts, so \(\nu_{j+3} \leq\nu_k\). The integrated initial estimate therefore controls both contributions to the wave commutator.


\section{Boost derivatives of the metric coefficients}  
\label{section:seven-boosted-coefficients}


\subsection{\texorpdfstring{Wave Hessians and dilation derivatives}{Wave Hessians and dilation derivatives}}
\label{section:seven-coarse-dilation}

\paragraph{Purpose}

The coupled ray estimates keep the metric amplitudes as unknown quantities. We estimate those amplitudes on the same finite bootstrap slab. The proof has three steps. 
\begin{itemize}
\item The trace of the densitized inverse metric satisfies a scalar wave equation whose quadratic null terms admit an explicit local correction;
\item Contraction with the polynomial covector \(X= (t,-x) \) preserves the boost commutators and removes derivatives of the moving frame from the represented equation;
\item The resulting metric bound and the two ray bounds close together by induction on the commutation rank.
\end{itemize}
We keep the reference metric, the Cauchy contribution, and the nonlinear sources separate throughout this argument. We use \(\ell=p_{\mathbf{low}} +2\), \(b=13\delta/64\), \(\chi =s/t\), and \(z=x/t\). In cone integrals, \((p,q) \) are the source lightlike coordinates and \((P,Q)= (T+R,T-R) \) are the target coordinates, with \(R=|x| \).

\paragraph{Coarse lightlike Hessians at the higher wave order}

\begin{proposition}[Coarse commuted wave Hessians]
\label{lem:seven-coarse-high-light-like-Hessian}
Under the coarse energy, Hardy, and metric spacetime bounds in the interior and exterior regions, the propagated wave gauge, and the prescribed initial and reference bounds, one has, for every translation--boost operator of order at most \(p_{\mathbf{low}} \),
\be
\label{equa:seven-coarse-high-light-like-Hessian}
| \del^2Zv| \lesssim C_NA t^{1/2}s^{-3+2\delta}, \qquad | \del\delsH Zv|+ t^{-1}| \del Zv| \lesssim C_NA t^{-3/2}s^{-1+b}, \qquad b\coloneqq13\delta/64,
\ee
where \(v\) is \(\psi\) or a Cartesian component of \(h=g-\gMink\). The constants are independent of the finite bootstrap interval and of \(0<\coeff\leq\coeff_0\).
\end{proposition}

\begin{proof}
\bse
We write \(\chi =s/t\) and use \(s\leq t\lesssim s^2\).

\noindent{\bf 1.} \emph{Commuted bad coefficients.} The proof of~\autoref{lem:seven-pointwise-metric-interface-bulk}, before insertion of its rank weights, gives \(t|(\mathcal L_ZH)^{\H00}(t,(t-1) \omega)| \leq C_NA\) through order \(p_{\mathbf{low}} \). Its dynamic term is \(CA t^{1/2-\lambda+ \delta/2} \), and the reference term is \(CA_0\). The direct radial gauge estimate therefore gives
\be
\label{equa:seven-coarse-high-bad-coefficient}
\chi^{-2}|(\mathcal L_ZH)^{\H00}| \lesssim C_N\bigl(A t^{-1/2}s^\delta+A^2s^{-1+2\delta} +At/s^2\bigr) \lesssim C_NA.
\ee
This applies to boost-only operators; an operator containing a translation has the stronger direct gauge derivative bound and \(t^{-1} \) frame terms. Reordering preserves at least one translation. The uncommuted inequality permits inversion of the lightlike Hessian identity; nonempty coefficient operators multiply strictly lower-order Hessians. The interface count is \(p_{\mathbf{low}} +3\leq p_{\mathbf{high}} \).

\noindent{\bf 2.} \emph{Nonprincipal wave sources.} Direct first-derivative estimates bound the metric and massless quadratic stress by \(CA^2t^{-1}s^{-2 +2\delta} \). In an effective-field product, the factor with fewer differentiations uses the lower mass energy, whereas its higher partner keeps the factor \(s^{1/2} \) from the high mass energy. Their counts are \(\lfloor p_{\mathbf{low}}/2\rfloor+3\leq p_{\mathbf{low}} \) and \(p_{\mathbf{low}} +3\leq p_{\mathbf{high}} \). Consequently the sufficient bound for either wave source is
\be
\label{equa:seven-coarse-high-wave-source}
|Z F_v| \lesssim C_{N, \coeff_0}A^2\bigl(t^{-1}s^{-2 +2\delta} + t^{-3}s^{1/2 +2b} \bigr).
\ee
For the potential, the equation~\eqref{equa:seven-Ufrak-rescaled} gives \(\coeff\Ufrak_\coeff(\fieldphi)=(\fieldvarphi)^2\bigl(\tfrac12+\fieldPhi\mathcal U_{1,\coeff}(\fieldPhi)\bigr)\), with a coefficient uniformly bounded in the required differentiability norms. For \(g(d\fieldPhi,d\psi) \), the good derivative in the semi-hyperboloidal contraction gives the same bound; its bad--bad coefficient is \(-\chi^2 +H^{\H00} \). The estimate in the first step controls this coefficient also after tensor Lie commutation. Higher derivatives on a smooth coefficient leave two lower scalar factors.

\noindent{\bf 3.} \emph{Induction on differential order.} The algebraic Hessian reduction has the form
\be
\label{equa:seven-coarse-high-Hessian-reduction}
| \del^2Zv| \lesssim \frac{t}{s^2} \bigl(| \del LZv|+ | \del Zv| \bigr) + \frac{t^2}{s^2}| \Box_gZv|.
\ee
Relative to \(At^{1/2}s^{-3+2\delta} \), the two nonprincipal terms in \eqref{equa:seven-coarse-high-wave-source} have respective ratios \(At^{1/2}/s\) and \(A(s/t)^{3/2}s^{2b-2\delta} \), both bounded by \(C_NA\). A bad commuted coefficient is controlled by \eqref{equa:seven-coarse-high-bad-coefficient} and acts on a lower-order Hessian. A good Hessian follows directly from \(\delsH=t^{-1}L\); after lightlike inversion its product with a metric coefficient has ratio at most \(CA t^{-1/2}s^{2b-2\delta} \). Frame terms contain \(t^{-1} \) times a first derivative and satisfy the same estimate. Finite induction proves the first inequality; direct Sobolev proves the second, using at most \(p_{\mathbf{low}} +3\) first-jet orders. The field term in the Hardy estimate and the two further boosts used for the dilation below have maximal count \(p_{\mathbf{low}} +4=p_{\mathbf{high}} \). \qedhere
\ese
\end{proof}

\paragraph{Dilation derivatives of the wave fields}

For \(\mathcal D=t\del_t+x^a \del_a=s\Lcal\), \(a=s^2/t\), and a Cartesian wave component \(v\in\{h_{\alpha \beta}, \psi\} \), the identity is
\be
\label{equa:seven-polynomial-dilation-square}
\mathcal D^2v=a^2\del_t^2v+2az^aL_a \del_tv+z^az^bL_aL_bv+a(\del_t+z^a \del_a)v.
\ee
Proposition~\ref{lem:seven-coarse-high-light-like-Hessian}, together with the direct first-derivative and Hardy estimates, therefore gives \(| \mathcal D^2L^Jv| \lesssim At^{-1/2}s^{2\delta} \). Its largest order is \(|J|+4\leq p_{\mathbf{high}} \). We interpolate on the past dilations \((t,x) \mapsto(e^\epsilon t,e^\epsilon x) \), with \(-h_0\leq\epsilon\leq0\) and \(e^{-h_0} \geq1/2\). These curves satisfy \(q\geq1/2\) and remain below the terminal hypersurface, since \(t-r\fTime_r\geq t-r>0\). For \(t\geq4T_0\) they also remain above the initial hypersurface. The band \(1/2\leq q\leq1\) uses direct exterior jets; its largest count is \(p_{\mathbf{low}} +7\leq N\). The bounded-time region is estimated by the prescribed local norms.

We put \(a_j=j/\ell\) and \(g_j=2\delta(1-a_j) +ba_j\).~\autoref{lem:seven-boost-v-interpolation} and the one-dimensional endpoint interpolation inequality give
\be
\label{equa:seven-polynomial-dilation-interpolation}
| \mathcal D L^Jv| \leq C_NA t^{-3/4+a_j/4}s^{(g_j+2\delta)/2}, \qquad |J|=j\leq p_{\mathbf{low}}.
\ee
All weights are comparable on the fixed dilation interval. The term involving \(L^Jv\) in the interpolation inequality is smaller than its geometric mean term. For two wave factors whose boost orders add to at most \(\ell-2\), it follows that
\be
\label{equa:seven-polynomial-dilation-source}
| \mathcal D L^Iv\, \mathcal D L^Jw| \leq C_NA^2t^{-1-\epsilon_S}, \qquad \epsilon_S\coloneqq \frac{\ell+2}{4\ell}-\frac{\delta(397\ell+230)}{128\ell}>0.
\ee
The same numerical condition on \(\delta \) gives this positive gap, since \(\ell(26\ell-230)>0\). For an interior target, the actual radial Kirchhoff integral satisfies
\be
\label{equa:seven-polynomial-dilation-kernel}
T^{-1}| \mathcal I_B(T,x)| \leq \frac{C_NA^2Q}{RT} \int_Q^P p^{-\epsilon_S} \, dp\leq C_NA^2.
\ee
For \(R\leq T/2\), the average of \(p^{-\epsilon_S} \) is \(\Ocal(T^{-\epsilon_S}) \); for \(R\geq T/2\), it is \(\Ocal(P^{-\epsilon_S}) \). The first case includes \(R=0\) by continuity.


\subsection{\texorpdfstring{The densitized trace and its retarded estimate}{The densitized trace and its retarded estimate}}
\label{section:seven-density-trace}

\paragraph{Trace equation and local correction}

We use \(K=j_gg^{-1}-\gMink^{-1} \), \(h=g-\gMink\), and the two scalar traces \(\mathcal T= \operatorname{tr}_{\gMink}K\) and \(h_{\mathrm{tr}} = \operatorname{tr}_{\gMink}h\). These traces and the contractions in \(Q_0\) and \(C\) use \(\gMink\); the matrix differentials below use \(g\). We write \(Q_0(f,v)= \gMink^{\alpha \beta} \del_\alpha f\del_\beta v\), with contraction of the tensor indices when the arguments are tensors. The determinant expansion in dimension four gives
\be
\label{equa:seven-density-trace-expansion}
K^{\alpha \beta} = \tfrac12h_{\mathrm{tr}} \gMink^{\alpha \beta}-h^{\alpha \beta} +h^\alpha{}_{\gamma}h^{\gamma \beta}-\tfrac12h_{\mathrm{tr}} h^{\alpha \beta} +(\tfrac18h_{\mathrm{tr}}^2-\tfrac14\langle h,h\rangle) \gMink^{\alpha \beta} +\Ocal(h^3), \qquad \mathcal T=h_{\mathrm{tr}} +\Ocal(h^3).
\ee
The quadratic terms have zero Minkowski trace. Contraction of the reduced metric equation and substitution of the divergence-free density give the equation
\be
\label{equa:seven-density-trace-equation}
\Box_g\mathcal T= \tfrac12Q_0(K,K) + \tfrac14Q_0(\mathcal T, \mathcal T) + \del_\lambda \del_\delta C^{\lambda \delta}-3Q_0(\fieldPhi, \fieldPhi)-16\pi Q_0(\psi, \psi)-4\coeff\Ufrak(\fieldphi) +R_3.
\ee
Here \(C\) is defined in \eqref{equa:seven-trace-antisymmetric-definition}. The remainder \(R_3\) consists of the differences between the exact coefficients and their quadratic Taylor terms; each derivative product contains an additional undifferentiated metric or scalar factor. The potential is kept in \eqref{equa:seven-density-trace-equation}, with its sign. To determine the coefficients of this remainder, we regard \(K(g)=j_gg^{-1}-\gMink^{-1}\) as a matrix map. Its differential on a symmetric covariant matrix \(v\) is
\be
\label{equa:audit-density-map-differential}
\mathrm D K_g[v]=j_g\bigl(\tfrac12\operatorname{tr}_g(v)g^{-1}-g^{-1}vg^{-1}\bigr).
\ee
The componentwise chain rule gives the metric source and the scalar contributions separately:
\be
\label{equa:audit-density-exact-source-generator}
\mathcal M_K\coloneqq\mathrm D K_g[\Fbb]+g^{\mu\nu}\mathrm D^2K_g[\del_\mu g,\del_\nu g],\qquad \Box_gK=\mathcal M_K+\mathrm D K_g[\Gbb_\coeff+\Mbb_\coeff].
\ee
Thus the exact remainder in the trace equation is
\be
\label{equa:audit-density-exact-remainder-generator}
\begin{aligned}
R_3&=\operatorname{tr}_{\gMink}\bigl(\mathcal M_K+\mathrm D K_g[\Gbb_\coeff+\Mbb_\coeff]\bigr)-\tfrac12Q_0(K,K)-\tfrac14Q_0(\mathcal T,\mathcal T)\\
&\quad-(\gMink)_{\alpha\beta}(\del_\delta K^{\alpha\lambda})(\del_\lambda K^{\beta\delta})+3Q_0(\fieldPhi,\fieldPhi)+16\pi Q_0(\psi,\psi)+4\coeff\Ufrak(\fieldphi).
\end{aligned}
\ee
All the coefficients follow by differentiating the matrix formula and the reduced source~\eqref{equa:audit-exact-reduced-metric-source}; the ordered product and composition rules count their multiplicities. This expression contains first metric derivatives. The local Hessian terms in~\eqref{equa:seven-density-trace-flat-equation}, below, are estimated separately.

We verify the quadratic metric terms explicitly. In this calculation all indices range over the four spacetime coordinates and are raised with \(\gMink\). We set
\be
\label{equa:audit-trace-quadratic-notation}
D_{\alpha\beta\gamma}\coloneqq\del_\alpha h_{\beta\gamma},\qquad Q_h\coloneqq D_{\alpha\beta\gamma}D^{\alpha\beta\gamma},\qquad B_h\coloneqq D_{\alpha\beta\gamma}D^{\beta\alpha\gamma}.
\ee
The linear Christoffel expression is
\be
\label{equa:audit-trace-linear-Christoffel}
\Gamma^{(1)}_{\alpha\beta\gamma}=\tfrac12\bigl(D_{\beta\alpha\gamma}+D_{\gamma\alpha\beta}-D_{\alpha\beta\gamma}\bigr).
\ee
Direct contraction gives
\be
\label{equa:audit-trace-Christoffel-contractions}
D_\rho{}^{\alpha\beta}\Gamma^{(1)\rho}_{\alpha\beta}=B_h-\tfrac12Q_h,\qquad \gMink^{\alpha\beta}\Gamma^{(1)\rho}_{\beta\lambda}\Gamma^{(1)\lambda}_{\alpha\rho}=\tfrac12B_h-\tfrac14Q_h.
\ee
Hence the Minkowski trace of the quadratic metric source is \(B_h+Q_h/2\). The density gauge gives \(\del^\alpha h_{\alpha\beta}=\del_\beta h_{\mathrm{tr}}/2+\Ocal(h\del h)\), and the expansion~\eqref{equa:seven-density-trace-expansion} yields
\be
\label{equa:audit-trace-density-quadratic-identities}
Q_0(K,K)=Q_h+O\bigl(h(\del h)^2\bigr),\qquad \del_\lambda\del_\delta C^{\lambda\delta}=B_h-\tfrac14Q_0(h_{\mathrm{tr}},h_{\mathrm{tr}})+O\bigl(h(\del h)^2\bigr).
\ee
The second identity uses the formula \(\del_\lambda\del_\delta C^{\lambda\delta}=(\gMink)_{\alpha\beta}(\del_\delta K^{\alpha\lambda})(\del_\lambda K^{\beta\delta})\). These contractions establish the first three terms of \eqref{equa:seven-density-trace-equation} modulo cubic terms. Moreover, \(\mathcal T-h_{\mathrm{tr}}=\Ocal(h^3)\), so its wave operator consists of \(\Ocal(h^2)\Box_g h+\Ocal(h(\del h)^2)\), with smooth coefficients. Substitution of the reduced metric equation controls this contribution without an additional Hessian estimate. The scalar derivative terms keep their quadratic coefficients. The potential \(-4\coeff\Ufrak(\fieldphi)\) remains exact; its remaining metric factors are estimated using \(\coeff\Ufrak=(\fieldvarphi)^2/2+\Ocal(\fieldPhi(\fieldvarphi)^2)\).

We define the local correction by
\be
\label{equa:seven-density-trace-normal-form}
\widehat{\mathcal T}{\coloneqq} \mathcal T+Q_{\mathrm{loc}}, \qquad Q_{\mathrm{loc}}{\coloneqq}-\tfrac14\langle K,K\rangle-\tfrac18\mathcal T^2 + \tfrac32\fieldPhi^2 +8\pi\psi^2.
\ee
The effective-field equation is \(\Box_g\fieldPhi = (3\coeff)^{-1} \fieldPhi+G\), whereas \(\Box_g\psi =F_\psi\). The product rule therefore yields
\be
\label{equa:seven-density-trace-flat-equation}
\Box\widehat{\mathcal T} =-H^{\alpha \beta} \del_\alpha \del_\beta h_{\mathrm{tr}} + \del_\lambda \del_\delta C^{\lambda \delta}-(\fieldvarphi)^2 + \widetilde R_3-H^{\alpha \beta} \del_\alpha \del_\beta(Q_{\mathrm{loc}} + \mathcal T-h_{\mathrm{tr}}).
\ee
The remainder before the last principal correction is
\be
\label{equa:seven-density-trace-remainder}
\begin{aligned}
\widetilde R_3&=R_3+2(\fieldvarphi)^2-4\coeff\Ufrak(\fieldphi)-\tfrac12\langle K, \Box_gK\rangle-\tfrac14\mathcal T\Box_g\mathcal T+3\fieldPhi G+16\pi\psi F_\psi\\
& \quad +H^{\alpha \beta} \bigl(-\tfrac12\langle\del_\alpha K, \del_\beta K\rangle-\tfrac14\del_\alpha \mathcal T\del_\beta \mathcal T+3\del_\alpha \fieldPhi\del_\beta \fieldPhi+16\pi\del_\alpha \psi\del_\beta \psi\bigr).
\end{aligned}
\ee
In particular, the leading potential in \eqref{equa:seven-density-trace-flat-equation} is \(-(\fieldvarphi)^2\): the correction \(3\fieldPhi^2/2\) adds \(\coeff^{-1} \fieldPhi^2= (\fieldvarphi)^2\) to the term \(-2(\fieldvarphi)^2\) in the original trace. The factorization \(\coeff\Ufrak= (\fieldvarphi)^2/2 +\Ocal(\fieldPhi(\fieldvarphi)^2) \) makes every remaining potential coefficient uniform in \(\coeff\).

\paragraph{Finite-slab amplitudes and the principal term}

On a finite bootstrap slab \(s_0\leq s\leq s_1\), we abbreviate the amplitudes defined by the equations~\eqref{equa:seven-pointwise-coefficient-amplitudes}, \eqref{equa:seven-weak-ray-amplitudes}, and \eqref{equa:seven-strong-ray-coupled-amplitude} by
\be
\label{equa:seven-density-trace-amplitudes}
K_i = \mathcal K_i(s_1), \qquad V_i^{(0)} = \mathcal V_i^{(0)}(s_1), \qquad V_i^{(2)} = \mathcal V_i^{(2)}(s_1).
\ee
The operators in \(K_i\) contain only boosts; \(\mathcal L_I\) acts on the tensor indices as well as the coefficients. We set \(\ell=p_{\mathbf{low}} +2\), \(b=13\delta/64\), and keep the numerical condition in \eqref{equa:seven-trace-interpolation-exponent-choice}.~\autoref{lem:seven-coarse-high-light-like-Hessian} gives, for every wave component \(v\in\{h_{\alpha \beta}, \psi\} \),
\be
\label{equa:seven-density-trace-coarse-Hessian}
| \del^2Zv| \leq C_NAt^{1/2}s^{-3+2\delta}, \qquad | \del\delsH Zv|+ t^{-1}| \del Zv| \leq C_NAt^{-3/2}s^{-1+b}, \qquad \ord Z\leq p_{\mathbf{low}}.
\ee
Its largest direct first-jet count is \(p_{\mathbf{low}} +3\leq p_{\mathbf{high}} \). In particular, the exponent \(2\delta \) keeps the higher effective mass factor \(s^{1/2} \); the estimate depends only on the coarse bounds, not on the amplitudes \(K_i\).

At a source point we write \(p=t+r\), \(q=t-r\), and \(\sigma= \sqrt{pq} \). Subtracting the prescribed reference coefficient in the radial gauge gives, with \(\chi_0= \lambda-1/2-\delta/2>0\),
\be
\label{equa:seven-density-trace-dynamic-gauge}
|(\mathcal L_I(H-H_\star))^{\H00}| \leq C_N\bigl(Aqp^{-3/2} \sigma^\delta+A^2p^{-3/2}q^{1/2} \sigma^{2\delta} +A_0qp^{-2} +Ap^{-1-\chi_0} +Ap^{-2 + \delta/2} +A^2p^{-2 + \delta} \bigr).
\ee
The last three terms are the dynamic interface value; the \(A_0qp^{-2} \) term is the differentiated reference. These terms are kept separately. For a coefficient of boost order \(i\), the first term can also be bounded by \(C_N(A+K_i)p^{-1} \sigma^{\eta_i} \), up to those remaining terms. For an interior target \((T,x) \), we put \(R=|x| \), \(P=T+R\), and \(Q=T-R\). The positive radial integration domain lies in \(Q\leq p\leq P\), \(1\leq q\leq Q\). After extracting the factor \(s_{\mathrm{target}}^{\eta_i} \), the minimum of these two bounds, paired with \eqref{equa:seven-density-trace-coarse-Hessian}, satisfies
\be
\label{equa:seven-density-trace-minimum-kernel}
\int_1^Q |r'F_{\min}| \, dq\leq C_NA(A+K_i)p^{-5/4+9\delta/4}, \qquad \frac1R\int_Q^Pp^{-1-\epsilon} \, dp\leq\frac{C_\epsilon}{T}Q^{-\epsilon} \quad(\epsilon>0).
\ee
We split the first integral at \(q= \sqrt p\): below that value we use \(Aqp^{-3/2} \sigma^\delta \), and above it we use \((A+K_i)p^{-1} \). The two powers are integrable because \(\delta<1/9\), which follows from the chosen numerical condition. The quadratic gauge term gives \(p^{-3/2 +4\delta} \), and the initial data gives \(p^{-1-\chi_0 + \delta} \). Both powers are smaller than \(-1\). For a good principal component with \(a_i =i/\ell\) and \(g_i =2\delta(1-a_i) +ba_i\), the corresponding power is
\be
\label{equa:seven-density-trace-good-principal-power}
\int_1^Q |r'F_{\mathrm{good}}| \, dq\leq C_NA^2p^{-3/2 +a_i/2 +g_i+b}, \qquad (1-a_i)/2>g_i+b\quad(i\leq\ell-2).
\ee
Thus every dynamic principal term contributes at most \(C_N(A^2 +A\sum_{i\leq k}K_i)T^{-1}s_{\mathrm{target}}^{\eta_k} \). The reference coefficient \(A_0/p\) is treated by signed integration, rather than included in the lightlike minimum.

\paragraph{Integration of the reference-metric coefficients and estimates for effective-field products}

On the partition \(|q| \leq Q/2\), we apply the two incoming integrations in \eqref{equa:seven-cone-double-integration} to the reference principal coefficient. Its normalized coefficient \(C=r'H_{\star,N}/4\) satisfies \(| \Omega^aC| \leq C_Nm\), \(| \del_q\Omega^aC| \leq C_Nm/p\), and \(| \del_q^2\Omega^aC| \leq C_Nm/p^2\), for \(a \leq2\). The resulting volume coefficients are bounded by \(C_NmQ^{-2} \). The pointwise estimate with two angular derivatives costs \(p_{\mathbf{low}} +4=p_{\mathbf{high}} \) in the interior. The two interface expressions at \(q=1\) are opposite evaluations of the same signed expression and cancel. The near partition misses the initial hypersurface when \(Q>4T_0\); for bounded \(Q\), the original source has a bounded positive integral. The reference trace contribution satisfies
\be
\label{equa:seven-density-trace-reference-bound}
| \mathcal I_{\mathcal T, \star}| \leq C_NmA\bigl(P^{-1/2 +b/2}Q^{-1+b/2} +P^{-1+ \delta}Q^{-1/2 + \delta} +P^{-1+ \delta/2}Q^{-\lambda} \bigr).
\ee
The first term is the near interior volume. The second is the complementary interior volume, for which the Hessian exponent is \(2\delta \), and the third is the exterior volume. The far exterior part uses Cartesian derivatives; its estimates do not divide rotations by the source time.

\paragraph{\texorpdfstring{Exterior Hessian of the metric trace}{Exterior Hessian of the metric trace}}

The exterior reference integral requires the trace equation. On \(q=\tau-r'\leq1\), write \(w=2-q\) and \(p=\tau+r'\). In the region \(r'\simeq\tau\), the required estimate is
\be
\label{equa:seven-audited-exterior-trace-Hessian}
|\del^2Zh_{\mathrm{tr}}|\leq C_NA p^{-1+\delta/2}w^{-1-\lambda},\qquad \ord Z\leq p_{\mathbf{low}}.
\ee
For bounded \(w\), this is the direct kinetic Hessian bound. For \(w\geq4\), the exterior wave reduction multiplies the source by \(p/w\). The quadratic derivative terms of the trace equation are null forms, by the contractions~\eqref{equa:audit-trace-Christoffel-contractions}--\eqref{equa:audit-trace-density-quadratic-identities} and the equation~\eqref{equa:seven-trace-double-divergence-null}. The metric contribution is bounded by \(C_NA^2p^{-3+\delta}w^{1-2\lambda}\). The scalar derivative terms have the corresponding bound with \(\mu\) in place of \(\lambda\), and the massive reconstruction~\eqref{equa:seven-polynomial-exterior-mass-value} bounds the potential by \(C_NA^2p^{-3+3\delta/2}w^{1-2\mu}\). After multiplication by \(p/w\), the ratios to the proposed Hessian bound are at most
\be
\label{equa:audit-trace-Hessian-source-ratios}
C_NA p^{-1+\delta/2}w^{1-\lambda},\qquad C_NA p^{-1+\delta}w^{1+\lambda-2\mu}.
\ee
These ratios are bounded, since \(w\lesssim p\), \(\lambda>\delta/2\), and \(2\mu-\lambda>\delta\). The cubic coefficient corrections have the stronger ratios \(C_NA^2p^{-1+\delta}w^{1/2-2\lambda}\) or \(C_NA^2p^{-1+\delta}w^{1/2-2\mu}\).

The commuted principal terms are treated by induction on the total order. On \(4\leq w\leq4p^{\delta/\lambda}\), the displaced-cone coefficient bound~\eqref{equa:seven-polynomial-exterior-bad-dynamic} gives
\be
\label{equa:audit-trace-Hessian-coefficient-ratio}
\frac pw|(\mathcal L_I(H-H^\star))_N|\lesssim A p^{-\lambda+1/2+\delta/2}w^{-1/2}+A p^{-1+\delta/2}w^{1-\lambda}+A^2p^{-1+\delta}w^{1-2\lambda}\lesssim A+A^2.
\ee
Outside this region, the direct coefficient bound gives \(C_NA p^{\delta/2}w^{-\lambda}\lesssim C_NA\). The reference multiplier is bounded by \(C_Nm/w\). Good Hessians have one additional inverse power of \(p\), and their coefficient products are smaller. The tangent term in the wave reduction is bounded directly by the right-hand side of the equation~\eqref{equa:seven-audited-exterior-trace-Hessian}. Finite induction proves this estimate, using four additional closing operators near the cone. In \(r'\geq2\tau\), Cartesian reduction gives \(C_NA p^{-2-\lambda+\delta/2}\); its largest high-mass source has relative factor \(C_NA p^{1/2+\lambda-2\mu+\delta/2}\), which decreases.

The complementary exterior reference volume is consequently controlled by the radial source \(C_NmA p^{-1+\delta/2}w^{-1-\lambda}\). Integration over \(w\gtrsim Q\) gives \(C_NmA P^{-1+\delta/2}Q^{-\lambda}\). On the near partition, two signed incoming integrations give the same bound from the pointwise estimate \(C_NA p^{-1+\delta/2}w^{1-\lambda}\). This proves the exterior term in the equation~\eqref{equa:seven-density-trace-reference-bound}.

\paragraph{\texorpdfstring{Effective-field products in the trace equation}{Effective-field products in the trace equation}}

In each boost partition of the potential, the factor with fewer derivatives has order at most \(\lfloor p_{\mathbf{low}}/2\rfloor\leq p_{\mathbf{low}}-3\). The two effective-field norms give
\be
\label{equa:seven-density-trace-potential-product}
|Z_1\fieldvarphi\,Z_2\fieldvarphi| \leq V_i^{(2)}V_j^{(0)}t^{-2}s^{-1+ \eta_i+ \eta_j}, \qquad i = \rank Z_1, \quad j= \rank Z_2, \quad \ord Z_1\leq\ord Z_2.
\ee
Direct integration gives a trace contribution bounded by \(C_NV_i^{(2)}V_j^{(0)}T^{-1}Q^{\eta_i+ \eta_j} \). The local wave corrections in \eqref{equa:seven-density-trace-normal-form} satisfy \(|L^JQ_{\mathrm{loc,w}}| \leq C_NA^2t^{-3/2 +2\delta+b} \leq C_NA^2/t\), by the two-factor boost interpolation. The effective correction has the stronger bound \(|L^J(\fieldPhi^2)| \leq C_{N, \coeff_0}A^2t^{-3}s^{1/2 +2b} \). Their differentiated principal terms keep the separate bounds \eqref{equa:seven-polynomial-wave-local-cubic}--\eqref{equa:seven-polynomial-effective-local-cubic}. These estimates, and the low--high decomposition of the additional effective-field factors, give retarded contributions at most \(C_{N, \coeff_0}(A^2 +A^3)/T\). The exterior bounds are obtained with three pointwise orders near \(r'\simeq\tau\), four for the massive reconstruction, and at most seven in the far region; hence all counts are at most \(N\). The mass term on the merging annulus is transported using the mass norm of a translation operator.

For later integration in the polynomial equation, we keep more than the weighted trace bound alone. The regular terms satisfy
\be
\label{equa:seven-density-trace-regular-bound}
|L^J\mathcal T_{\mathrm{reg}}(T,x)| \leq\frac{C_N}{T} \Bigg(a_{\mathrm C} +A^2 + \Bigl(A\sum_{i\leq k}K_i+ \sum_{i+j\leq k, \ i\leq\lfloor p_{\mathbf{low}}/2\rfloor}V_i^{(2)}V_j^{(0)} \Bigr)s_{\mathrm{target}}^{\eta_k} \Bigg), \qquad |J|=k.
\ee
The remainder consists of the reference volumes in \eqref{equa:seven-density-trace-reference-bound} and the free-wave contribution from the reference-metric Cauchy data, bounded by \(C_NA_0T^{-1}[1+ \log(P/Q)] \). The logarithm follows directly from the spherical average of the inverse-square Cauchy jet. The dynamic Cauchy contribution is bounded by \(C_Na_{\mathrm C}T^{-1}Q^{1/2-\lambda} \), and is already included in \eqref{equa:seven-density-trace-regular-bound}. The uniform initial-cap estimate gives \(a_{\mathrm C} \leq C_{N,s_0}A_0\); the contribution of the reference metric is estimated separately from the dynamic Hardy estimate. These separate formulas keep the decay needed when \(2\mathcal T\) is integrated once more.

\paragraph{Local corrections and cubic products}

We write the trace normal form with the flat operator and keep its principal term on the linear trace. Its local quadratic corrections then produce explicit cubic sources, including \(H\del^2R_2(h, \psi) \) and \(H\del^2(\fieldPhi^2) \). Subscripts in the following bounds indicate boost orders. For two wave factors estimated by pointwise interpolation with \(a= (i+j)/\ell\) and \(i+j+n\leq\ell-2\), their product with a lightlike Hessian satisfies
\be
\label{equa:seven-polynomial-wave-local-cubic}
|H_i v_j\del^2v_n| \leq C_NA^3t^{-3/2 +a/2}s^{-3+G_a}, \qquad G_a\coloneqq6\delta-2\delta a+ba.
\ee
Radial integration uses the integrable weight \(q^{-3/2 +G_a/2} \) and leaves \(p^{-2 +a/2 +G_a/2} \), whose exponent is smaller than \(-1\). For a cubic product with one undifferentiated metric factor and two wave derivatives, we instead set \(a=i/\ell\) and use
\be
\label{equa:seven-polynomial-wave-derivative-cubic}
|H_i\del v_j\del w_n| \leq C_NA^3t^{-2 +a/2}s^{-2 +G'_a}, \qquad G'_a\coloneqq2\delta(1-a) +ba+2b.
\ee
The weaker envelope with \(t^{-3/2 +a/2} \) already suffices for both the trace and the polynomial equation. Indeed, after multiplying by \(p^2\) for the latter, its actual radial integral obeys
\be
\label{equa:seven-polynomial-wave-cubic-kernel}
\frac{C_NA^3}{RT} \int_Q^Pp^{1/2 +a/2 +G'_a/2} \, dp\int_1^Qq^{-1+G'_a/2} \, dq\leq C_{N, \delta}A^3P^{-\epsilon_a}, \qquad \epsilon_a\coloneqq \frac{1-a}{2}-G'_a>0.
\ee
At \(a=1-2/\ell\), positivity follows from \(\delta(39\ell+230)<64\), which is implied by the chosen condition on \(\delta \). The trace integral is no larger after its weight \(s^2/T\) is inserted. Smooth coefficient products have the same estimates by the composition form of the boost interpolation.

For \(H\del^2(\fieldPhi^2) \), we use the uniform mass norms of the two effective-field jets. In this undifferentiated representation of the metric coefficient, a scalar Hessian is estimated as \(\del^2L^J\fieldPhi=\sqrt\coeff\,\del^2L^J\fieldvarphi\). The mass norm and its two Sobolev boosts require \(|J|+4\leq p_{\mathbf{low}}+4=p_{\mathbf{high}}\) commutations; at \(N=19\), the largest count is exactly fourteen. If both are at the lower energy level, their bound is stronger. If a scalar jet of Cartesian order at most two reaches the high level, its boost order is at least \(p_{\mathbf{low}}-3\); hence the metric coefficient receives at most three boosts. The largest term satisfies
\be
\label{equa:seven-polynomial-effective-local-cubic}
|H_i\del^aL^j\fieldPhi\, \del^{2-a}L^n\fieldPhi| \leq C_{N, \coeff_0} \coeff A^3t^{-3-\epsilon_\Phi}, \qquad \epsilon_\Phi{\coloneqq} \tfrac12-\frac3{2\ell}-2\delta-2b>0.
\ee
The high mass contributes \(s^{1/2 + \nu} \), and its lower partner contributes \(s^\alpha \); the pointwise interpolation bound for the metric gives the time-exponent gap. The weighted trace integral is bounded by \(CA^3Q^2R^{-1} \int_Q^Pp^{-2-\epsilon_\Phi} \, dp\), which is uniformly bounded. This calculation keeps the positive factor \(\coeff\) and uses no negative power of that parameter. For the wave product in \eqref{equa:seven-polynomial-wave-local-cubic}, the polynomial kernel is bounded by \(C_NA^3P^{-1+a/2 +G_a/2} \): its radial integrand is \(A^3p^{a/2 +G_a/2}q^{-3/2 +G_a/2} \), and \(G_a<1\). Thus the same local correction is controlled in both represented equations.


\subsection{\texorpdfstring{Polynomial projection and quadratic sources}{Polynomial projection and quadratic sources}}
\label{section:seven-polynomial-projection}

\paragraph{Polynomial projection and a direct wave-gauge estimate}

We use the densitized inverse metric \(K=j_gg^{-1}-\gMink^{-1} \), whose divergence vanishes. For the covector \(X= (t,-x) \), we define the scalar polynomial projection and the Minkowski trace by
\be
\label{equa:seven-polynomial-coefficient-definition}
B\coloneqq X_\alpha X_\beta K^{\alpha \beta}, \qquad \mathcal T\coloneqq \operatorname{tr}_{\gMink}K, \qquad \Box B=X_\alpha X_\beta \Box K^{\alpha \beta} +2\mathcal T.
\ee
Indeed, \(\del_\gamma X_\alpha=-(\gMink)_{\alpha \gamma} \), and the mixed product term is \(-4X_\beta \del_\alpha K^{\alpha \beta} =0\). Since \(\mathcal L_{L_a}X=0\), the same identity applies to every tensorial Lie operator. The conversion to the inverse-metric coefficient is
\be
\label{equa:seven-polynomial-coefficient-conversion}
\frac Bt=tj_gH^{\H00}-\frac{s^2}{t}(j_g-1), \qquad j_g-1= \tfrac12\mathcal T+R(K), \qquad R(0)=dR(0)=0.
\ee
The quadratic remainder is controlled by~\autoref{lem:seven-boost-v-interpolation}. Thus the two linear quantities to estimate are \(B/t\) and \((s^2/t) \mathcal T\).

\begin{proposition}[Direct bound for the densitized projection]
\label{lem:seven-densitized-projection-overlap}
Let \(\ell=p_{\mathbf{low}} +2\), \(b=13\delta/64\), and suppose \(\delta(26\ell+115)<32\). Under the coarse metric estimates, the propagated wave gauge, and the established metric interface bounds, define
\be
\label{equa:seven-polynomial-overlap-exponents}
\beta{\coloneqq} \tfrac12 + \frac1{4\ell}, \qquad g\coloneqq \frac{\delta(13\ell+115)}{64\ell}, \qquad \gamma_g\coloneqq \frac1{4\ell}-\Bigl(\tfrac34+ \frac1{8\ell} \Bigr)g>0.
\ee
For \(Q=t-r\), \(P=t+r\), \(Q\leq P^\beta \), and \(|J| \leq p_{\mathbf{low}} \), one has
\be
\label{equa:seven-polynomial-direct-overlap}
t^{-1}|L^JB(t,x)| \leq C_N(A_0 +A) +C_NA P^{-\gamma_g}.
\ee
This estimate uses the coarse interface bound already established in \eqref{equa:seven-pointwise-metric-interface-bulk}. Since \(A_0\lesssim A\), it gives the bound of order \(A\) needed for the coefficient estimate.
\end{proposition}

\begin{proof}
\bse

\noindent{\bf 1.} \emph{Radial identity for the metric contraction.} For a symmetric tensor \(P_J= \mathcal L_JK\), we write \((P_J)_\H=P_J^{00}-2z_aP_J^{0a} +z_az_bP_J^{ab} \), with \(z=x/t\). The contracted divergence equation and \(E_a=t^{-1}L_a \) give the identity
\be
\label{equa:seven-polynomial-densitized-gauge}
\del_t(P_J)_\H=-E_aP_J^{a0} +z_bE_aP_J^{ab} +2t^{-1}z_aP_J^{0a}-2t^{-1}z_az_bP_J^{ab}.
\ee
The last two terms differentiate the coefficients \(z\). The time derivatives of the tensor components are the contraction of \(\del_\alpha P_J^{\alpha \beta} =0\) with \((1,-z) \). Since \(\del_r= \omega^aE_a-(r/t) \del_t\), differentiation of the contraction also gives
\be
\label{equa:seven-polynomial-radial-gauge}
| \del_r(P_J)_\H| \leq C_Nt^{-1}(|LP_J|+ |P_J|).
\ee

\noindent{\bf 2.} \emph{Radial integration from the interface.} We integrate at fixed \(t\) from \(r\) to \(t-1\). The coarse interface estimate gives \(t|(P_J)_\H(t,t-1)| \leq C_N(A_0 +A)\); the difference between the null and semi-hyperboloidal contractions there has an additional factor \(t^{-1} \). The boost interpolation at order \(|J|+1\leq \ell-1\) then yields
\be
\label{equa:seven-polynomial-radial-interpolation}
t|(P_J)_\H(t,r)| \leq C_N(A_0 +A) +C_NAq\,t^{-1+(\ell-1)/(2\ell)}s^g.
\ee
The bounded initial-time part is included in the prescribed initial norms. Substitution of \(s^2=PQ\) and \(Q\leq P^\beta \) proves the assertion. Positivity of \(\gamma_g\) follows from \(26\ell^2-243\ell-115>0\) for \(\ell\geq12\). This calculation uses the divergence-free density directly; the determinant factor is kept in \eqref{equa:seven-polynomial-coefficient-conversion}.
\ese
\end{proof}

\paragraph{Quadratic sources of the polynomial equation}

We describe the metric quadratic source before applying its majorant. If \(F_{\alpha \beta} \) denotes the quadratic reduced Einstein expression, its part with uncontracted derivative indices is
\be
\label{equa:seven-polynomial-metric-nonnull-tensor}
P_{\alpha \beta} =-\tfrac12\langle\del_\alpha h, \del_\beta h\rangle+ \tfrac14\del_\alpha h_{\mathrm{tr}} \del_\beta h_{\mathrm{tr}}, \qquad F_{\alpha \beta}-P_{\alpha \beta} = \text{a sum of classical null forms}.
\ee
Differentiating the quadratic Taylor expansion \eqref{equa:seven-density-trace-expansion} gives the metric source of the densitized tensor:
\be
\label{equa:seven-polynomial-metric-quadratic-tensor}
\begin{aligned}
(Q_K^{(2)})^{\alpha \beta}&= \tfrac12\operatorname{tr}_{\gMink}F\, \gMink^{\alpha \beta}-F^{\alpha \beta} +2Q_0(h^\alpha{}_{\gamma},h^{\gamma \beta})-Q_0(h_{\mathrm{tr}},h^{\alpha \beta}) \\
& \quad + \bigl(\tfrac14Q_0(h_{\mathrm{tr}},h_{\mathrm{tr}})-\tfrac12Q_0(h,h) \bigr) \gMink^{\alpha \beta}.
\end{aligned}
\ee
Since \(\gMink^{-1}(X,X)=-s^2\), contraction of the first two terms belonging to \(P\) gives
\be
\label{equa:seven-polynomial-metric-dilation-source}
X_\alpha X_\beta \bigl(-P^{\alpha \beta} + \tfrac12\gMink^{\alpha \beta} \operatorname{tr}_{\gMink}P\bigr)= \tfrac12\langle\mathcal D h, \mathcal D h\rangle-\tfrac14(\mathcal D h_{\mathrm{tr}})^2 + \tfrac14s^2Q_0(h,h)-\tfrac18s^2Q_0(h_{\mathrm{tr}},h_{\mathrm{tr}}).
\ee
The remaining quadratic terms are bounded by \(Ct^2\) times classical null forms. Their boost partitions keep these identities because \(\mathcal L_LX=0\) and \(\mathcal L_L\gMink=0\). The two types of null form are estimated separately. The antisymmetric ones satisfy the interpolation bound in~\autoref{lem:seven-trace-antisymmetric-retarded}. For the scalar contraction we use the identity
\be
\label{equa:seven-polynomial-null-dilation-identity}
s^2Q_0(f,v)=-(\mathcal D f) (\mathcal D v) +(\chi^2\delta_{ab} +z_az_b) (L_af) (L_bv).
\ee
The two additional boosts in the last product have total order at most \(p_{\mathbf{low}} +2= \ell\). The pointwise interpolation estimate and \eqref{equa:seven-polynomial-dilation-source} therefore give \(t^2|Q_0| \leq C_NA^2t^{1-\epsilon}s^{-2} \), with \(\epsilon= \min\{\epsilon_S,1/2-2\delta-b\}>0\). The two radial kernels are
\be
\label{equa:seven-polynomial-null-kernels}
\frac{| \mathcal I[t^2Q_{ab}]|}{T} \leq C_NA^2\sqrt{Q/P}, \qquad \frac{| \mathcal I[t^2Q_0]|}{T} \leq C_{N, \epsilon}A^2P^{-\epsilon} \log(2Q) \leq C_{N, \epsilon}A^2.
\ee
Indeed, their radial integrands are bounded, respectively, by \(A^2p^{1/2}q^{-1/2} \) and \(A^2p^{1-\epsilon}q^{-1} \), with the factor \((RT)^{-1} \) outside the integrals. The two target-radius regions give the same bounds, including the continuous limit at the center. A metric cubic source has the stronger bound \(|X X R_3| \leq C_NA^3t^{1/2}s^{-2 +e} \), where \(e= \delta+2b<1/2\); its kernel is bounded by \(C_{N,e}A^3P^{-1/2 +e} \).

The effective and matter contributions to the densitized equation are
\be
\label{equa:seven-polynomial-exact-scalar-tensor}
3j_g T_g[d\fieldPhi]^{\alpha \beta} +16\pi j_ge^{-\fieldPhi}T_g[d\psi]^{\alpha \beta}-j_g\coeff\Ufrak(\fieldphi)g^{\alpha \beta}, \qquad T_g[df]^{\alpha \beta}{\coloneqq} \nabla^\alpha f\nabla^\beta f-\tfrac12g^{\alpha \beta}g(df,df).
\ee
Its leading flat stress satisfies \(XXT_{\gMink}[df]= [(\mathcal D f)^2 +(\chi^2\delta+zz) (Lf) (Lf)]/2\). For \(f= \psi\), the preceding interpolation applies. For the effective field, the ray derivative and the additional boost derivative give
\be
\label{equa:seven-polynomial-effective-dilation-jets}
| \mathcal D Z_{\mathrm l} \fieldPhi| \leq C_N(A+V_i^{(2)}) \chi^2s^{-1/2 + \eta_i}, \qquad | \mathcal D Z_{\mathrm h} \fieldPhi| \leq C_N(A+V_j^{(0)})s^{-1/2 + \eta_j}.
\ee
The letters \(\mathrm l, \mathrm h\) refer to the lower and higher differential orders, not to ranks. In the identity \(\mathcal D\fieldPhi = (s^2/t) \del_t\fieldPhi+z^aL_a \fieldPhi\), the second term has the direct lower bound \(C\sqrt\coeff A\chi^{3/2}s^{-3/2 + \alpha} \), or the direct higher bound \(C\sqrt\coeff A\chi^{3/2}s^{-1+ \nu} \). The interior inequality \(\chi\gtrsim s^{-1} \) bounds these by the terms containing \(A\) in \eqref{equa:seven-polynomial-effective-dilation-jets}. The leading stress and potential thus have the source bound \(C_N(A+V_i^{(2)}) (A+V_j^{(0)})t^{-2}s^{1+ \eta_i+ \eta_j} \). Its polynomial kernel satisfies
\be
\label{equa:seven-polynomial-effective-kernel}
T^{-1}| \mathcal I_{B, \mathrm{eff}}| \leq C_Ns_{\mathrm{target}}^{\eta_k} \sum_{i+j\leq k, \ i\leq\lfloor p_{\mathbf{low}}/2\rfloor}(A+V_i^{(2)}) (A+V_j^{(0)}).
\ee
The tangential product in the projected stress uses the direct mass estimates, with at most one higher-level factor:
\be
\label{equa:seven-polynomial-effective-tangential-product}
|LZ_{\mathrm l} \fieldPhi\,LZ_{\mathrm h} \fieldPhi| \leq C_N\coeff A^2t^{-3}s^{1/2 +2b+ \eta_i+ \eta_j} \leq C_{N, \coeff_0}A^2t^{-2}s^{1+ \eta_i+ \eta_j}.
\ee
The lower factor costs at most \(\lfloor p_{\mathbf{low}}/2\rfloor+3\leq p_{\mathbf{low}} \), and the higher one at most \(p_{\mathbf{low}} +3\leq p_{\mathbf{high}} \). It therefore obeys the same polynomial kernel bound. Additional undifferentiated metric or scalar factors in this stress have an integrable kernel. If at least four boosts fall on a coefficient, both mass factors remain at the lower energy level, and the source is bounded by \(C_NA^3t^{-3/2}s^{\delta+2b} \). Otherwise the coefficient receives at most three boosts; interpolation bounds the source by \(C_NA^3t^{-2 +3/(2\ell)}s^{1/2 +2\delta+2b} \), whose time gain is positive. These estimates keep the higher mass factor where it occurs and use no negative power of \(\coeff\).

\paragraph{Terms from the reference metric and the metric trace}

The complementary targets satisfy \(Q\geq P^\beta \). The signed double incoming integration of the reference principal term, with its angular derivatives kept, gives
\be
\label{equa:seven-polynomial-reference-overlap}
\begin{aligned}
T^{-1}| \mathcal I_{B, \star}|& \leq C_NmA\bigl(P^{1/2 +b/2}Q^{-1+b/2} +P^\delta Q^{-1/2 + \delta} \bigr) \\
& \leq C_NmA\bigl(P^{-\gamma_\star} +P^{-\gamma_c} \bigr), \qquad \gamma_\star{\coloneqq} \frac1{4\ell}-\Bigl(\tfrac34+ \frac1{8\ell} \Bigr)b>0, \quad \gamma_c\coloneqq \tfrac\beta2-\delta(1+ \beta)>0.
\end{aligned}
\ee
The largest commutation order in these pointwise estimates is \(p_{\mathbf{low}} +4=p_{\mathbf{high}} \): two angular derivatives and two Sobolev boosts. The interior lower trace and exterior upper trace at \(q'=1\) are opposite evaluations of the same signed boundary expression and cancel before estimation. The partition avoids the initial hypersurface for \(Q>4T_0\); the bounded range belongs to the direct estimate above. The exterior reference volumes have the stronger weights \(mA P^{\delta/2}Q^{-\lambda} \) and \(mA^2P^{1/2 + \delta}Q^{-2\mu} \).

The term \(2\mathcal T\) in the polynomial equation keeps \eqref{equa:seven-density-trace-regular-bound}. A component bounded by \(C\tau^{-1} \sigma^{\eta_k} \) contributes at most \(Cs^{\eta_k}Q/T\). In particular this applies to the dynamic principal term and to the potential, with their separate amplitudes. The near reference trace volume has the actual integral
\be
\label{equa:seven-polynomial-reference-trace-kernel}
\frac{CmA}{RT} \int_Q^Pp^{1/2 +b/2} \, dp\int_1^Qq^{-1+b/2} \, dq\leq C_bmA P^{-1/2 +b}.
\ee
The complementary reference trace contributes at most \(C_NmA P^{-1/2 +2\delta} \). Both terms are bounded. The factor \(1/b\) depends only on the fixed numerical exponents. Finally the free-wave contribution from the reference-metric Cauchy data contributes at most \(C_NA_0(Q/T)[1+ \log(P/Q)] \), since \(\int_1^Q\log(p/q) \, dq\leq Q[1+ \log(p/Q)] \). This quantity is also bounded. These calculations use the separate trace decay, rather than the weaker aggregate bound \(| \mathcal T| \lesssim q^{-1} \).

On the exterior part of the source cone, we estimate \(2\mathcal T\) directly. The radial metric estimate gives \(|L^J\mathcal T_{\mathrm{dyn}}| \leq C_NAp^{-1+ \delta/2}w^{1/2-\lambda} \), whereas the prescribed reference is bounded by \(C_Nm/p\). The range \(q\leq1\), \(\tau\geq T_0\) has \(1\lesssim w\lesssim p\). Consequently, we obtain
\be
\label{equa:seven-polynomial-exterior-trace-kernel}
T^{-1}| \mathcal I_{2\mathcal T, \mathrm{ext}}| \leq\frac{C_N}{RT} \int_Q^P\bigl(Ap^{3/2-\lambda+ \delta/2} +mp\bigr) \, dp\leq C_NAP^{1/2-\lambda+ \delta/2} +C_Nm.
\ee
The dynamic exponent is negative. This contribution uses the metric energy at all required orders and adds only an auxiliary \(C_NA\), independently of the interior trace estimate.


\subsection{\texorpdfstring{Exterior source integrals}{Exterior source integrals}}
\label{section:seven-polynomial-exterior}

We estimate the exterior sources of the trace normal form and of the polynomial projection using the coarse bootstrap. We write \(p= \tau+r'\), \(q= \tau-r'\leq1\), \(w=2-q\), \(a= \delta/\lambda \), and \(\chi_\lambda= \lambda-1/2-\delta/2\). The target variables are \(P=T+R\), \(Q=T-R\geq1\), with \(R=|x| \). All source commutation orders below are at most \(p_{\mathbf{low}} \).

\noindent{\bf 1.} \emph{Near-cone scalar jets.} On the flat part with \(r'\simeq\tau\), radial \(H^1\) and spherical \(H^2\) give
\be
\label{equa:seven-polynomial-exterior-three-word}
r'w^\sigma|f| \lesssim\sum_{b\leq1, \ |J| \leq2} \|w^\sigma \del_r^b\Omega^Jf\|_{L^2}.
\ee
Here each rotation costs one boost, since \(\Omega_{ab} = (x_a/\tau)L_b-(x_b/\tau)L_a \) has bounded coefficients. We transport the outer trace through the merging annulus, of width one, by the hypersurface-tangent radial derivative. For a first jet this derivative is measured by the kinetic energy of a translation operator, because \(1-\fTime_r\leq\zeta \). For a field controlled by the mass term, the mass term of the additional translation controls this derivative. Thus the cost remains three operators throughout this annulus. One additional translation or boost has cost \(p_{\mathbf{low}} +4=p_{\mathbf{high}} \), within the closing exterior energy. In particular, we have
\be
\label{equa:seven-polynomial-exterior-closing-jets}
| \del Z\fieldPhi|+ |Z\fieldvarphi| \lesssim A p^{-1+ \delta/2}w^{-\mu}, \qquad | \del Zh|+ | \del^2Zh| \lesssim A p^{-1+ \delta/2}w^{-\lambda}.
\ee
The identity \(Z\fieldvarphi =3\sqrt\coeff\bigl(Z\Box_g\fieldPhi-ZG_\Phi\bigr) \), with the flat contraction \((w/p) \del^2\fieldPhi+p^{-1}(\del L\fieldPhi+ \del\fieldPhi) \), further gives
\be
\label{equa:seven-polynomial-exterior-mass-value}
|Z\fieldvarphi| \lesssim C_{N, \coeff_0}A p^{-2 + \delta}w^{1-\mu}, \qquad |Z((\fieldvarphi)^2)| \lesssim C_{N, \coeff_0}A^2p^{-3+3\delta/2}w^{1-2\mu}.
\ee
The metric correction is bounded by \(CA^2p^{-2 + \delta}w^{1-\lambda-\mu} \). The potential and matter trace have the stronger weight \(w^{-2\mu} \). These estimates use the kinetic Hessian, at cost four additional operators; they do not require a second mass jet.

\noindent{\bf 2.} \emph{Dynamic principal term.} We separate \(H_{\mathrm d} =H-H^\star\). The radial pointwise estimate and~\autoref{lem:eight-displaced-cone-coefficient} give, on \(w\leq4p^a \),
\be
\label{equa:seven-polynomial-exterior-bad-dynamic}
|(\mathcal L_IH_{\mathrm d})_N| \lesssim A p^{-1-\chi_\lambda}w^{1/2} +A p^{-2 + \delta/2}w^{2-\lambda} +A^2p^{-2 + \delta}w^{2-2\lambda}.
\ee
The direct Hessian in \eqref{equa:seven-polynomial-exterior-closing-jets} bounds its complementary wave factor. The other contractions use a good Hessian \(CAp^{-2 + \delta/2}w^{1-\lambda} \). If \(F_T\) is the corresponding trace source and \(F_B\) its polynomial contraction, integration in \(q\) therefore gives
\be
\label{equa:seven-polynomial-exterior-collar-integrals}
\int|r'F_T| \, dq\lesssim (A^2 +A^3) \sum_{j=1}^4p^{-1-g_j}, \qquad \int|r'F_B| \, dq\lesssim (A^2 +A^3) \sum_{j=1}^4p^{1-g_j},
\ee
where the four exponents are defined by
\be
\label{equa:seven-polynomial-exterior-collar-gaps}
g_1\coloneqq \lambda-\tfrac12-\frac{3\delta}{2\lambda}, \quad g_2\coloneqq1-\delta-a(3-2\lambda), \quad g_3\coloneqq1-\tfrac32\delta-a(3-3\lambda), \quad g_4\coloneqq1-\delta-a(\tfrac52-2\lambda).
\ee
All four numbers are positive. On \(w\geq2p^a \), within \(r'\simeq\tau\), the exterior lightlike Hessian reduction gives
\be
\label{equa:seven-polynomial-exterior-light-like-Hessian}
| \del^2ZK| \lesssim A p^{-1+ \delta/2}w^{-1-\lambda}.
\ee
Its principal perturbation has multiplier \(CAp^{\delta/2}w^{-\lambda} +Cm/w\). The closing effective-field source has ratio \(CAp^{\delta/2}w^{\lambda-2\mu} \), also small. Induction on order consequently uses only the four additional closing operators. Multiplication by \(| \mathcal L_IH_{\mathrm d}| \lesssim Ap^{-1+ \delta/2}w^{1/2-\lambda} \) gives \eqref{equa:seven-polynomial-exterior-collar-integrals} with gap \(g_{\mathrm o} = \delta(1-1/(2\lambda))>\delta/3\). The bounded range preceding the flat region is covered by the fixed local norms.

In the far region \(r'\geq2\tau\), where \(w\simeq p\), the Cartesian reduction uses at most \(p_{\mathbf{low}} +7\leq N\) orders. It gives \(| \del^2ZK| \lesssim Ap^{-2-\lambda+ \delta/2} \). The high effective-field contribution has the decreasing ratio \(Ap^{1/2 + \lambda-2\mu+ \delta/2} \). The resulting gap is \(g_{\mathrm f} =2\lambda-1/2-\delta>0\).

\noindent{\bf 3.} \emph{Quadratic and cubic sources.} The trace metric null source is \(CA^2p^{-3+ \delta}w^{1-2\lambda} \). For the polynomial source we project the non-null metric form before taking absolute values, using \(| \mathcal D Zh| \lesssim Ap^{-1+ \delta/2}w^{1-\lambda} \), where \(\mathcal D= \tau\del_\tau+x'\cdot\nabla \). The trace and polynomial radial integrals then have powers \(p^{-2\lambda+ \delta} \) and \(p^{2-2\lambda+ \delta} \), with positive gap \(2\lambda-1-\delta \).

The trace potential in \eqref{equa:seven-polynomial-exterior-mass-value} has gap \(2\mu-1-3\delta/2\). In the far region the high--low mass product is \(CA^2p^{-3/2-2\mu+ \delta} \), with gap \(2\mu-3/2-\delta>0\). For the polynomial effective-field source we may keep the high mass loss also near the cone: the two dilation jets give \(CA^2p^{-3/2 + \delta}w^{2-2\mu} \), while the potential contraction uses \(| \gMink^{-1}(X,X)| \lesssim pw\). Their radial integrals have power \(p^{5/2-2\mu+ \delta} \) and the same positive gap. The kinetic trace is included in the projected scalar stress.

A cubic metric trace source is bounded by \(CA^3p^{-3+3\delta/2}w^{1/2-3\lambda} \), whose \(q\)-weight is integrable. Multiplication by \(p^2\) gives its polynomial estimate, with the same gap \(1-3\delta/2\). For an effective-field coefficient correction, we may keep one high mass loss: its polynomial source is bounded by \(CA^3p^{-1/2 +3\delta/2} \) times \(w^{1/2-\lambda-2\mu} \) or \(w^{-\mu-2\lambda} \). These weights are integrable and give the positive gap \(1/2-3\delta/2\); the trace source is smaller by \(p^2\). The local normal-form corrections use the closing kinetic Hessian in \(H\del^2(\fieldPhi^2) \), and the integrable weights \(w^{-2\lambda} \), \(w^{1-\lambda-2\mu} \), or \(w^{1-3\mu} \). In the far region the single high mass loss leaves at least the gap \(2\mu-3/2-\delta \). Reference cross products have amplitude \(mA\); their derivative weights remain integrable.

\noindent{\bf 4.} \emph{Target integral.} The actual exterior source domain satisfies \(Q\leq p\leq P\), \(2T_0-p\leq q\leq1\). Each preceding radial bound is \(C(A^2 +A^3+mA)p^{-1-g} \) for the trace, or \(C(A^2 +A^3+mA)p^{1-g} \) for the polynomial, with \(g>0\). The positive Kirchhoff kernel gives
\be
\label{equa:seven-polynomial-exterior-target-kernels}
\frac QR\int_Q^Pp^{-1-g} \, dp\lesssim_g1, \qquad \frac1{RT} \int_Q^Pp^{1-g} \, dp\lesssim_g1.
\ee
We may decrease \(g\) to a number in \((0,1) \). Splitting \(R\leq T/2\) and \(R\geq T/2\) proves both inequalities, with continuous limits at \(R=0\). Consequently, we obtain
\be
\label{equa:seven-polynomial-exterior-source-bound}
Q| \mathcal I_{T, \mathrm{ext}}|+T^{-1}| \mathcal I_{B, \mathrm{ext}}| \lesssim C_N(A^2 +A^3+mA).
\ee
The principal reference source keeps its signed representation. The estimates above depend only on the coarse bootstrap and include the actual finite initial strip.


\subsection{\texorpdfstring{Cauchy data and the rank estimate}{Cauchy data and the rank estimate}}
\label{section:seven-j-coefficient-closure}

\paragraph{Polynomial Cauchy data}

We denote by \(a_{\mathrm C} \) a bound for the actual commuted Cauchy norms on the fixed Cartesian slice used by the representation, with the prescribed reference kept separately. The weighted annular estimates give
\be
\label{equa:seven-polynomial-Cauchy-data}
|f_B(y)| \leq C_Na_{\mathrm C} \langle y\rangle^{3/2-\lambda}, \qquad |g_B(y)|+ | \nabla f_B(y)| \leq C_Na_{\mathrm C} \langle y\rangle^{1/2-\lambda}.
\ee
These data are noncompact. Their free wave is nevertheless controlled by the Cauchy formula: the average of \(\langle x+(T-T_0) \omega \rangle^{1/2-\lambda} \) on the unit sphere is bounded by \(C\langle T\rangle^{1/2-\lambda} \) for \(|x| \leq T-1\). Radial integration proves this estimate also when the sphere passes through the origin. Consequently, we obtain
\be
\label{equa:seven-polynomial-Cauchy-value}
T^{-1}|B_{\mathrm{free}}(T,x)| \leq C_NA_0 +C_Na_{\mathrm C} \langle T\rangle^{1/2-\lambda}.
\ee
For targets with \(s\geq T_0\), the region between the initial graph and \(t=T_0\) lies in the existing slab, as proved before the equation~\eqref{equa:seven-first-wave-matter-source}. The uniform energy estimate on this compact region gives \(a_{\mathrm C}\leq C_{N,s_0}A_0\). Its mass term remains in the positive effective-field energy; the highest pointwise count on this compact region is \(p_{\mathbf{low}}+3\leq p_{\mathbf{high}}\). The noncompact flat part is already on the initial hypersurface and uses at most \(p_{\mathbf{low}}+5\leq N\) prescribed orders. The constant is independent of \(\coeff\) and of the coarse bootstrap multiplier. For \(s<T_0\), the equation~\eqref{equa:audit-bounded-initial-Gronwall} gives the coefficient bounds directly on the available hypersurface. The polynomial estimate remains an auxiliary coefficient bound, distinct from the strict improvement of an energy bootstrap.

\paragraph{Uniform transfer of the initial data}

\begin{lemma}[Transfer to the initial Cartesian slice]
\label{lem:seven-uniform-initial-cap}
Under the prescribed initial energy bounds and the coarse geometric and energy bootstrap, on every available slab with terminal parameter \(s_1\geq T_0\), the dynamic Cauchy amplitude in \eqref{equa:seven-polynomial-Cauchy-data} satisfies
\be
\label{equa:seven-polynomial-initial-amplitude}
a_{\mathrm C} \leq C_{N,s_0, \mathrm{data}}A_0.
\ee
The constant is independent of \(\coeff\), the bootstrap multiplier \(C_1\), and the terminal time. The reference contribution remains separate.
\end{lemma}

\begin{proof}
\bse

\noindent{\bf 1.} \emph{The compact transfer region.} We put \(R_0=\rhoE(s_0)\) and use \(T_0\) from \eqref{equa:seven-cone-initial-time-ceiling}. Since \(s_1\geq T_0\) and the terminal graph has height at least \(s_1\), the entire transfer region belongs to the available development. The initial graph is already the slice \(t=T_0\) for \(r\geq R_0\). Its transfer region is therefore the compact cap \(\fTime(s_0,r)\leq t\leq T_0\), \(r\leq R_0\). For \(m=p_{\mathbf{high}}\), we sum the physical wave energies of \(Zu\) and \(Z\psi\), keeping the matter weight \(e^{-\fieldPhi}\), and the massive energies of \(Z\fieldPhi\). The mass term is \((3\coeff)^{-1}|Z\fieldPhi|^2=|Z\fieldvarphi|^2/3\). Adding the \(L^2\) norm of the commuted dynamic metric gives an energy equivalent to
\be
\label{equa:seven-initial-cap-energy}
\mathcal E_m(t)^2\simeq\sum_{\ord Z\leq m} \int_{\{r\leq R_0:\, \fTime(s_0,r) \leq t\}} \bigl(|Zu|^2 + | \del Zu|^2 + | \del Z\psi|^2 + | \del Z\fieldPhi|^2 + |Z\fieldvarphi|^2\bigr) \, dx.
\ee
The initial flux is bounded by \(CA_0^2\) through \eqref{equa:seven-initial-robust-high-bound} and \eqref{equa:seven-exterior-initial-assumptions}; the commuted metric is controlled by Hardy with its initial trace. The fixed initial graph is uniformly spacelike. Its normal and Cartesian jets are related by \eqref{equa:eight-dynamic-initial-jet-reconstruction}.

\noindent{\bf 2.} \emph{Uniform energy estimate.} Every nonempty principal commutator leaves at most \(m-1\) operators on the field, so its high Hessian is controlled by the order-\(m\) energy. When the metric coefficient has the higher order, the low Hessian costs at most \(\lfloor m/2\rfloor+3\leq p_{\mathbf{low}} \) orders. All these direct bounds are uniform on the fixed cap. The potential factorization controls the remaining effective-field factors by \(\fieldvarphi\); the constant mass is kept on the left and commutes with every operator. The only source without a dynamic factor is the prescribed reference residual, of size \(CA_0\). Thus the current identities give
\be
\label{equa:seven-initial-cap-Gronwall}
\mathcal E_m(t)^2\leq CA_0^2 +C\int_{s_0}^{t} \bigl((1+A) \mathcal E_m(\tau)^2 +A_0^2\bigr) \, d\tau, \qquad \sup_{s_0\leq t\leq T_0} \mathcal E_m(t) \leq CA_0.
\ee
There is no lateral flux: the cap meets \(r=R_0\) only at \(t=T_0\). Gronwall is applied over a fixed interval, with \(A\) below a fixed geometric threshold. Its constant has no dependence on \(C_1\) or negative power of \(\coeff\).

\noindent{\bf 3.} \emph{Cauchy norms.} On the compact portion of \(t=T_0\), Sobolev for the first Cartesian jet costs at most \(p_{\mathbf{low}} +3\leq p_{\mathbf{high}} \) operators; spatial rotations have bounded coefficients there. Smooth inversion and the determinant formula transfer the bound to the dynamic densitized metric. Outside this portion, the weighted annular argument of \eqref{equa:seven-first-full-metric-data}--\eqref{equa:seven-first-full-metric-data-identities} uses only prescribed initial traces. Its two rotations and one additional boost cost at most \(p_{\mathbf{low}} +5\leq N\), within the exterior initial norm. Those traces have amplitude \(CA_0\), so the same bound holds for \(a_{\mathrm C} \). \qedhere
\ese
\end{proof}

\begin{proposition}[Coupled bound for the boost coefficients]
\label{prop:seven-polynomial-coefficient-reduction}
Suppose that the finite bootstrap, the propagated wave gauge, the prescribed reference bounds, and the initial Cauchy norms hold. The numerical exponents and amplitudes are those of \eqref{equa:seven-trace-interpolation-exponent-choice} and \eqref{equa:seven-density-trace-amplitudes}, respectively. For every \(0\leq k\leq p_{\mathbf{low}} \), these finite-slab quantities satisfy
\be
\label{equa:seven-polynomial-coefficient-coupling}
\begin{aligned}
\mathcal K_k(s_1)& \leq C_N(A_0 +a_0 +a_{\mathrm C} +A+A^2 +mA) +C_NA\sum_{i\leq k} \mathcal K_i(s_1) \\
& \quad +C_N\sum_{i+j\leq k, \ i\leq\lfloor p_{\mathbf{low}}/2\rfloor}(A+ \mathcal V_i^{(2)}(s_1)) (A+ \mathcal V_j^{(0)}(s_1)).
\end{aligned}
\ee
The constants depend on the fixed exponents, the admissibility bounds, and \(N\), and are independent of \(s_1\), \(\coeff\), and the bootstrap multiplier. The estimate keeps the weak and strong effective-field norms as unknown amplitudes; they are not additional hypotheses on the development.
\end{proposition}

\begin{proof}
\bse

\noindent{\bf 1.} \emph{Cauchy representation of the metric projection.} We apply the scalar Cauchy representation to \(L^JB\), \(|J|=k\), using the polynomial identity \eqref{equa:seven-polynomial-coefficient-definition}. Its commuted tensor source is the ordered Lie expansion of \(XX(\Box_gK-H^{\alpha \beta} \del_\alpha \del_\beta K) \). In the dynamic principal term, multiplication by \(XX\) adds at most \(Cp^2\) to each source bound in \eqref{equa:seven-density-trace-minimum-kernel} and \eqref{equa:seven-density-trace-good-principal-power}. A radial source with integral \(Cp^{-1-\epsilon} \) then has polynomial contribution
\be
\label{equa:seven-polynomial-principal-kernel}
\frac{C}{RT} \int_Q^Pp^{1-\epsilon} \, dp\leq C_\epsilon, \qquad \epsilon>0.
\ee
This gives \(C_N(A^2 +A\sum_{i\leq k} \mathcal K_i(s_1))s_{\mathrm{target}}^{\eta_k} \). The Taylor difference between a component of \(K\) and its linear part has the local-product estimates already proved. The nonprincipal interior terms are controlled by \eqref{equa:seven-polynomial-null-kernels}, \eqref{equa:seven-polynomial-effective-kernel}, and the cubic kernels. The exterior contributions are bounded by \eqref{equa:seven-polynomial-exterior-source-bound}; their derivative counts are at most \(p_{\mathbf{low}} +7\leq N\). The Cauchy part is \eqref{equa:seven-polynomial-Cauchy-value}.

\noindent{\bf 2.} \emph{Separating the two target regions.} For targets \(Q\geq P^\beta \), the reference principal term has the two distinct bounds in \eqref{equa:seven-polynomial-reference-overlap}, together with the exterior bounds stated there. For \(Q\leq P^\beta \),~\autoref{lem:seven-densitized-projection-overlap} directly controls \(L^JB/t\). Its interface term is \(C_N(A_0+A)\), while its variation is \(C_NAP^{-\gamma_g} \). The signed reference calculation and the positive integration of the other source terms are needed only in the complementary target region. The source \(2\mathcal T\) uses \eqref{equa:seven-density-trace-regular-bound} and the separate reference kernels, as calculated above.

\noindent{\bf 3.} \emph{Recovering the inverse-metric coefficient.} We commute the determinant conversion before taking absolute values. Ordered partitions of a boost operator \(I\) give the identity
\be
\label{equa:seven-polynomial-commuted-conversion}
t j_g(\mathcal L_IH)^{\H00} =t^{-1}L^IB+ \frac{s^2}{t}L^I(j_g-1)-t\sum_{I_1\sqcup I_2=I, \ I_1\ne\varnothing}(L^{I_1}j_g) (\mathcal L_{I_2}H)^{\H00}.
\ee
The sum is over partitions of the ordered boost operator, preserving the order within each factor and counting the Leibniz multiplicities. This formula follows by contracting the Lie derivative of \(K=j_gH+(j_g-1) \gMink^{-1} \) with \(XX\). The determinant expansion gives \(L^I(j_g-1)=L^I\mathcal T/2 +L^IR(K) \). Its quadratic remainder has a weighted pointwise bound by the boost interpolation. Every nonempty derivative of \(j_g\) is bounded by \(C_NA\) from the same interpolation, so the last sum contributes \(C_NA\sum_{i<k} \mathcal K_i(s_1) s_{\mathrm{target}}^{\eta_k} \). The factor \(j_g^{-1} \) is uniformly bounded. Combining these estimates proves \eqref{equa:seven-polynomial-coefficient-coupling}. The isolated \(A\) on its right-hand side is the auxiliary direct variation bound; the initial amplitudes remain those appearing in the Cauchy and interface estimates. \qedhere
\ese
\end{proof}

\paragraph{Induction for the metric coefficient and the rays}

We combine \eqref{equa:seven-polynomial-coefficient-coupling} with the weak and strong ray inequalities. The source amplitudes from the initial slice remain distinct: \(A_0\) controls the reference, \(a_0\) the exterior initial energy, and \(a_{\mathrm C} \) the actual Cauchy norms in the wave representation. On the coarse bootstrap slab they are bounded by \(C_NA\). The terms \(A^2\) and \(mA\) come from nonlinear sources and reference interactions, whereas the direct near-cone variation contributes \(C_NA\).

We suppress the argument \(s_1\), set \(p_s=p_{\mathbf{low}}-3\), and write \(P_i = \mathcal V_i^{(2)} \), \(V_j= \mathcal V_j^{(0)} \), and \(K_k= \mathcal K_k\). For \(0\leq k\leq p_{\mathbf{low}} \), the coefficient estimate has the sufficient form
\be
\label{equa:seven-coupled-coefficient-ray-system}
\begin{aligned}
K_k& \leq C_N(A_0 +a_0 +a_{\mathrm C} +A+A^2 +mA) +C_NA\sum_{i =0}^kK_i +C_N\sum_{\substack{i+j\leq k\; ;\; i\leq p_s}}(A+P_i) (A+V_j), \\
V_k& \leq C_NA\Bigl(1+ \sum_{i =1}^kK_i\Bigr) + \frac{C_N}{\eta_k} \sum_{i =1}^kK_iV_{k-i}, \qquad k\geq1.
\end{aligned}
\ee
The same second inequality holds for \(P_k\) in its range \(k\leq p_s\). In each actual product the strong factor has differential order at most \(\lfloor p_{\mathbf{low}}/2\rfloor\leq p_s\); the enlarged product sum therefore covers its rank. This formulation distinguishes differential order from boost rank. The direct terms \(A\) in each factor keep the tangential contributions in the conversion from ray derivatives to dilation derivatives.

\noindent{\bf 1.} \emph{Rank zero.} \(P_0 +V_0\leq C_NA\) independently of \(K_0\). The first line therefore gives \(K_0\leq C_NA+C_NAK_0\), and smallness of \(A\) gives \(K_0\leq C_NA\). 
\noindent{\bf 2.} \emph{Positive ranks.} Suppose that all amplitudes at ranks below \(k\) have been bounded by \(C_{N,k-1}A\). The current-rank terms are \((A+P_0)V_k\) and \((A+V_0)P_k\), the latter only when \(k\leq p_s\). We put \(\widetilde P_k=P_k\) in that range and \(\widetilde P_k=0\) otherwise. Consequently, we obtain
\be
\label{equa:seven-coupled-current-rank-system}
V_k+ \widetilde P_k\leq C_{N,k}A+C_NA(1+ \eta_k^{-1})K_k, \qquad K_k\leq C_{N,k}A+C_NAK_k+C_NA(V_k+ \widetilde P_k).
\ee
Substitution leaves the current coefficient multiplier \(C_N[A+A^2(1+ \eta_k^{-1})] \). Since \(\eta_k\geq c=C_{\mathrm{rk}} \theta \), a fixed smallness condition on \(A\) and \(A/c\) absorbs this multiplier uniformly over the finite rank range. We obtain \(K_k+V_k\leq C_{N,k}A\), and then recover \(P_k\leq C_{N,k}A\) whenever \(k\leq p_s\). Finite induction proves
\be
\label{equa:seven-coupled-coefficient-ray-bound}
\max_{k\leq p_{\mathbf{low}}}(K_k+V_k) + \max_{k\leq p_s}P_k\leq C_NA.
\ee

This is an auxiliary pointwise estimate on the bootstrap slab. The strict improvement of the energy amplitudes follows from their separate energy inequalities, with the initial energy terms and the quadratic source amplitudes kept.


\section{High-order estimates and continuation}  
\label{section-11}


\subsection{\texorpdfstring{Uniform and fixed-parameter energies}{Uniform and fixed-parameter energies}}
\label{section-10-4}

\paragraph{Auxiliary high-order hierarchy}

The pointwise bounds and critical commutator estimates are combined in the high-order energy hierarchy. The resulting uniform bounds and their rescaling give the interior bootstrap improvement and the fixed-parameter continuation estimate. The initial and interface terms remain explicit throughout.

First of all, we estimate the metric, matter, and rescaled effective-field energies at order \(p_{\mathbf{high}} \). Their uniform bounds control the differentiated nonlinear products. At the same order, rescaling gives the fixed-parameter continuation norm. The uniform high-order energies control the four additional derivatives in \eqref{equa:seven-sharp-boost-only-H00}, together with the radial metric profile. We set
\be
\label{equa:seven-robust-high-energy}
\mathsf E_{\mathrm h,k}^{\mathrm{rob}}(s) \coloneqq \mathbf W_{p_{\mathbf{high}},k}(s) + \mathbf P_{p_{\mathbf{high}},k}(s) +s^{-1/2}U_{p_{\mathbf{high}},k}(s) + \mathbf H_{p_{\mathbf{high}},k}(s).
\ee
We first estimate the three differentiated energies, then recover the radial metric profile. In addition to \(\mathsf E_{\mathrm h,k}^{\mathrm{rob}} \), we use
\be
\label{equa:seven-high-differentiated-energy}
\mathbf E_{\mathrm h,k}(s)\coloneqq \mathbf W_{p_{\mathbf{high}},k}(s) + \mathbf P_{p_{\mathbf{high}},k}(s) +s^{-1/2}U_{p_{\mathbf{high}},k}(s).
\ee

\begin{proposition}[High-order uniform energy estimates]
\label{lem:seven-robust-high-triangular}
Under the interior and exterior bootstrap assumptions, the propagated wave gauge, and the parameter smallness conditions, one has, for \(0\leq k\leq p_{\mathbf{high}} \),
\bse
\be
\label{equa:seven-robust-high-Volterra}
\begin{aligned}
\mathbf E_{\mathrm h,k}(s)& \leq C_N\mathcal I_{\mathrm h,k}^{\mathrm{rob}}(s)s^{\nu_k} + \frac{C_N}{\rho\theta} \bigl(\mathcal B_3^{\mathrm{low}}(s) \bigr)^2s^{\nu_k} +C_N\mathcal B_3^{\mathrm{low}}(s) \int_{s_0}^s\tau^{-1} \mathbf E_{\mathrm h,k}(\tau) \, d\tau \\
& \quad + \frac{C_N}{\theta} \mathcal B_3^{\mathrm{low}}(s) \sum_{j=1}^k\int_{s_0}^s\tau^{-1+ \eta_j} \mathbf E_{\mathrm h,k-j}(\tau) \, d\tau.
\end{aligned}
\ee
The radial profile is recovered with its Hardy and sphere traces. Consequently, one obtains
\be
\label{equa:seven-robust-high-improved-raw}
\mathsf E_{\mathrm h,k}^{\mathrm{rob}}(s) \leq C_N\Bigl(\mathcal I_{\mathrm h,k}^{\mathrm{rob}}(s) + \frac{(\mathcal B_3^{\mathrm{low}}(s))^2}{\rho\theta} \Bigr)s^{\nu_k}.
\ee
\ese
The constants are independent of \(0<\coeff\leq\coeff_0\). The interface quantities remain those of \eqref{equa:seven-robust-high-interface-input}; their bounds from the initial data belong to the argument in Part~III.
\end{proposition}

\begin{proof}

We work on a fixed bootstrap slab ending at \(s_1\), and write \(B= \mathcal B_3^{\mathrm{low}}(s_1) \), \(c=C_{\mathrm{rk}} \theta \), and \(p=p_{\mathbf{high}} \). The smallness condition bounds \(B_{s_1}\), so the constants of the lightlike Hessian estimates are uniform on the slab. The low--high decomposition uses
\bse
\be
\label{equa:seven-high-low-order-count}
\lfloor p_{\mathbf{high}}/2\rfloor\leq p_{\mathbf{low}}-3, \qquad p_{\mathbf{high}}-p_{\mathbf{low}} =4, \qquad N\geq19.
\ee

\noindent{\bf 1.} \emph{The normalized interface flux.} For the high effective-field energy, the factor \(s^{-1} \) is applied to the squared balance law before taking a square root. We set \(F_k(\sigma)= \mathsf F_{\mathrm{eff}}^{\mathrm{unif}, \Lscr,p,k}(\sigma;s_0) \). Since \(F_k(s_0)=0\), integration by parts gives
\be
\label{equa:seven-high-normalized-interface-flux}
\begin{aligned}
\int_{s_0}^{\sigma} \tau^{-1} \, dF_k(\tau)
& = \sigma^{-1}F_k(\sigma) + \int_{s_0}^{\sigma} \tau^{-2}F_k(\tau) \, d\tau,
 \\
\sup_{s_0\leq\sigma \leq s} \left| \int_{s_0}^{\sigma} \tau^{-1} \, dF_k(\tau) \right|
& \leq\left(1+ \frac1{2\nu_k} \right) \bigl(\mathcal I^U_{p,k}(s) \bigr)^2s^{2\nu_k}.
\end{aligned}
\ee
\ese
Thus the flux contribution is bounded by \(C_{\nu_k} \mathcal I^U_{p,k}(s)s^{\nu_k} \), with no sign assumption. The constant is allowed to depend on the fixed growth exponents. The metric and matter fluxes need no additional time normalization.

\noindent{\bf 2.} \emph{Critical forcing terms.} We estimate the critical terms using the coarse bootstrap bounds. For the variation of a wave coefficient, \eqref{equa:seven-e-varying-Fubini}--\eqref{equa:seven-e-varying-rank-kernels} and \(\mathbf W_{p,k-j} \leq A s^{\nu_{k-j}} \) give the two powers
\bse
\be
\label{equa:seven-high-critical-forcing-exponents}
-\tfrac32 +24\rho+(k+7)c, \qquad -\tfrac32 +36\rho+(k+11)c.
\ee
Both are smaller than \(-1\), since \(k\leq N-5\) and \((N+4)c<\rho\). Their amplitudes are \(C_NA^2\) and \(C_NA^3\). The normalized Hardy-trace and reference contributions have stronger decay. The integrated initial estimate \eqref{equa:seven-e-good-bulk-bound} gives a bounded initial norm, with a decaying dynamic part. In its pairing with the unweighted wave Hessian, we use
\be
\label{equa:seven-high-e-integral-rank}
\int_{s_0}^s\tau^{-1+ \eta_{j+1}} \, d\tau\leq\frac{s^{\nu_k}}{\nu_k} \leq\frac{s^{\nu_k}}{12\rho}, \qquad 0\leq j\leq\min\{3,k\}.
\ee
For the effective field, its energy is normalized at the differential-inequality level. Thus its critical term becomes \(C_N\theta^{-1}B\sum_j s^{-1+ \eta_j} \mathbf H_{p,k-j} \). The coarse profile bound \(\mathbf H_{p,k-j} \leq A s^{\nu_{k-j}} \) and \(\eta_j+ \nu_{k-j} = \nu_k\) give the same integral. Since \(A_0\leq A\leq B\), these estimates prove
\be
\label{equa:seven-high-critical-integrated-forcing}
\int_{s_0}^s\Bigl(\mathbf R^{\mathrm{crit}}_{u;p,k} + \mathbf R^{\mathrm{crit}}_{\psi;p,k} + \tau^{-1/2} \mathbf R^{\mathrm{crit}}_{\fieldPhi;p,k} \Bigr) (\tau) \, d\tau\lesssim\frac{C_NB^2}{\rho\theta}s^{\nu_k}.
\ee
\ese

\noindent{\bf 3.} \emph{Coefficients of the differentiated energies.} The remaining terms are estimated directly in the three energies. We keep the sharper rank-zero coefficient in this step. If the effective-field derivative is the low factor in the matter source, the lower direct estimate gives
\bse
\be
\label{equa:seven-high-low-effective-direct-coefficient}
\frac ts\Bigl(\frac st| \del\fieldPhi|_{p_{\mathrm w},i} + | \delsH\fieldPhi|_{p_{\mathrm w},i} \Bigr) \lesssim A t^{-1/2}s^{-1+ \alpha_{i+2}} \leq A s^{-3/2 + \alpha_{i+2}} \leq A s^{-1+ \eta_i}.
\ee
The last inequality uses \(\alpha_{i+2}-\eta_i =3\rho+2c<1/2\). If the matter derivative is low, the central estimate gives \(C_N(A+A^2)s^{-3/2 + \eta_i}U_{p,k-i} =C_N(A+A^2)s^{-1+ \eta_i} \widehat U_{p,k-i} \). The rank-zero metric coefficient has size \(C_NB/s\); positive coefficient ranks have size \(C_N\theta^{-1}B s^{-1+ \eta_i} \). These are the coefficients of \eqref{equa:seven-robust-high-Volterra}, with coefficient \(C_NB\) at rank zero.

The translated high metric coefficient is controlled by \eqref{equa:seven-e-translated-metric-product}, including its normalized trace. For the metric and effective-field equations, \eqref{equa:seven-G-source-structure}--\eqref{equa:seven-M-source-structure} and \eqref{equa:seven-S-exact}--\eqref{equa:seven-W-product} give the same two kernels: the factor with at most half the commutations is estimated pointwise, and the other is measured by its differentiated energy. The reference sources and oriented fluxes remain in \(\mathcal I^{\mathrm{rob}} \).

Some refined effective-field factors, notably in \(\Gbb_\coeff\) and the potential terms, keep the coefficient \(\theta^{-1}B\). We estimate their current-rank contributions by the coarse high bootstrap and include them in the quadratic forcing. More generally, the Leibniz split gives the following estimate whenever such a factor has rank \(i\):
\be
\label{equa:seven-high-refined-coefficient-forcing}
\frac B\theta \int_{s_0}^s\tau^{-1+ \eta_i} \mathbf E_{\mathrm h,k-i}(\tau) \, d\tau\lesssim\frac{BA}{\theta \nu_k}s^{\nu_k} \leq\frac{B^2}{12\rho\theta}s^{\nu_k}, \qquad0\leq i\leq k.
\ee
Here the instantaneous normalization of the high effective-field energy has already been made, and \(\eta_i+ \nu_{k-i} = \nu_k\). Thus every current-rank term with the available coefficient \(\theta^{-1}B\) is covered by the forcing; the terms kept in the current-rank Volterra integral have coefficient \(C_NB\). Positive-rank terms may keep the factor \(\theta^{-1} \) in the lower-rank integrals.

\noindent{\bf 4.} \emph{Rank induction.} Combining these bounds with the energy identities proves \eqref{equa:seven-robust-high-Volterra}. The instantaneous factor \(s^{-1/2} \) is applied to the differential effective-field inequality; its favorable term \((2s)^{-1} \widehat U\) is kept before passing to the weaker integral estimate. Gronwall at rank zero and induction using \(\eta_j+ \nu_{k-j} = \nu_k\) yield
\be
\label{equa:seven-high-three-energy-bound}
\mathbf E_{\mathrm h,k}(s) \lesssim C_N\Bigl(\mathcal I_{\mathrm h,k}^{\mathrm{rob}}(s) + \frac{B^2}{\rho\theta} \Bigr)s^{\nu_k}.
\ee
\ese

\noindent{\bf 5.} \emph{Recovery of the radial metric profile.} The norm \(M_{0;p,k} \) is defined by the same sum of squares over high boost-only operators as \(M_{1;p,k} \), with \(\alpha=0\). The initial decomposition and the favorable identity at \(\alpha=0\) give
\bse
\be
\label{equa:seven-high-profile-damped-recovery}
\begin{aligned}
\mathbf H_{p,k}(s)& \leq E_{0;p,k}(s) + \frac1s\int_{s_0}^sM_{0;p,k}(\tau) \, d\tau, \\
M_{0;p,k}& \lesssim C_N\mathbf Q^u_{p,k} +C_NA\sum_{i =0}^{\min\{k, \lfloor p/2\rfloor\}}s^{-1/2 + \nu_{i+2}} \mathbf Q^u_{p,k-i} +C_NA_0s^{-1/2}.
\end{aligned}
\ee
The two nonlinear placements are the same as in \eqref{equa:seven-e-alpha-one-gauge}, with the additional factor \(s/t\). Their coefficients are bounded by \(CA s^{-1/2 + \nu_{i+2}} \). The wave-energy average has denominator \(1+ \nu_k\), rather than \(\nu_k\). The initial identity gives
\be
\label{equa:seven-high-profile-e-interface}
E_{0;p,k}(s) \leq\frac{s_0}{s} \mathbf H_{p,k}(s_0) + \left\| \tau^{-1/2} \mathsf B^H_{2;p,k}(\tau) \right\|_{L^2((s_0,s);d\tau)}.
\ee
\ese
This bound and the average \(s^{-1} \int_{s_0}^s\mathsf T_{2;p,k}[u] \) are included explicitly in \(\mathcal I_{\mathrm h,k}^{\mathrm{rob}} \). The nonlinear wave averages may use the coarse bootstrap; their amplitude is \(C_NA^2\), with the decaying power \(-1/2 +24\rho+(k+6)c\). The nonlinear trace averages are bounded by the same trace average, since \(-1/2 + \nu_{i+2}<0\). The reference average is bounded by \(C_NA_0\). Substitution in \eqref{equa:seven-high-profile-damped-recovery}, followed by the choice \(s_1=s\), proves \eqref{equa:seven-robust-high-improved-raw}. \qedhere

\end{proof}

\paragraph{Fixed-parameter continuation by rescaling}

The uniform and fixed-parameter high energies contain the same derivatives and the same radial metric profile. Their only difference is the normalization of the effective field. By \(D= \coeff^{-1/2}U\), we have, for \(0\leq k\leq p_{\mathbf{high}} \),
\be
\label{equa:seven-fixed-high-rescaling}
\mathsf E_k^{\mathrm{tame}} = \mathsf E_{\mathrm h,k}^{\mathrm{rob}} +(\coeff^{-1/2}-s^{-1/2})U_{p_{\mathbf{high}},k} \leq(1+ \coeff^{-1/2})s^{1/2} \mathsf E_{\mathrm h,k}^{\mathrm{rob}}.
\ee
We set \(\mathsf E_{\mathrm{top}}^{\mathrm{tame}}{\coloneqq} \mathsf E_{p_{\mathbf{high}}}^{\mathrm{tame}} \). The robust high-order estimate and rescaling yield
\be
\label{equa:seven-tame-high-final}
\mathsf E_{\mathrm{top}}^{\mathrm{tame}}(s) \leq C_N(1+ \coeff^{-1/2}) \Bigl(\mathcal I_{\mathrm h,p_{\mathbf{high}}}^{\mathrm{rob}}(s) + \frac{(\mathcal B_3^{\mathrm{low}}(s))^2}{\rho\theta} \Bigr)s^{1/2 + \nu_{p_{\mathbf{high}}}}.
\ee
This polynomial bound is finite on every finite interval for each fixed \(\coeff>0\). It gives the high-order interior norm in the continuation criterion, with the singular factor \(\coeff^{-1/2} \) explicit. Together with the exterior high-order estimate, it applies to the  Cauchy development.

\paragraph{Uniform lower-order and degenerate estimates}

Combining the three lower-level estimates gives
\bse
\be
\label{equa:seven-uniform-improved-raw}
X_{p_{\mathbf{low}},k}(s) \leq C_N\mathcal J_{p_{\mathbf{low}},k}(s)s^{\nu_k},
\ee
where we define
\be
\label{equa:seven-uniform-input-J}
\mathcal J_{p_{\mathbf{low}},k}(s) = \mathcal U_k^{\mathbf{low}}(s) + \mathcal L_k(s) + \mathcal W_k^{\mathbf{low}}(s).
\ee
The separate estimates \eqref{equa:seven-U-improved-raw} and \eqref{equa:seven-P-low-improved-raw} keep the sharper exponents \(\alpha_k\) and \(\chi_k\), respectively.

We define
\be
\label{equa:seven-robust-source-input}
\mathcal R_{p_{\mathbf{low}},k}^{\mathrm{rob}}(s) = A_0 + \mathcal J_{p_{\mathbf{low}},k}(s) + {A\over \rho} \mathcal J_{p_{\mathbf{low}},k}(s).
\ee
The same rank induction in the degenerate lower inequality, without suppressing the source factor, gives
\be
\label{equa:seven-Ydeg-triangular}
D_{p_{\mathbf{low}},k}(s) \leq C_N\left[ \mathcal I^{\mathrm{sing}}_{p_{\mathbf{low}},k}(s) + \coeff^{-1/2} \mathcal R_{p_{\mathbf{low}},k}^{\mathrm{rob}}(s) \right]s^\delta.
\ee
\ese
The factor \(\coeff^{-1/2} \), the degenerate flux, and the initial degenerate energy are all kept explicitly.  The closed lower estimate gives the dependence on \(\coeff\).


\subsection{\texorpdfstring{Singular bounds and local convergence}{Singular bounds and local convergence}}
\label{section-11-1}

It is useful to spell out what \eqref{equa:seven-Ydeg-triangular} implies for the  original effective field at the uniform lower level.  Suppressing only constants independent of $\coeff$, we have, for \(0\leq k\leq p_{\mathbf{low}} \),
\bse
\be
\label{equa:seven-closed-phi-L2}
\| \,| \fieldphi|_{p_{\mathbf{low}},k} \|_{L^2} \lesssim \left[ \mathcal I^{\mathrm{sing}}_{p_{\mathbf{low}},k}(s) + \coeff^{-1/2} \mathcal R_{p_{\mathbf{low}},k}^{\mathrm{rob}}(s) \right] s^\delta
\ee
and we have
\be
\label{equa:seven-closed-dphi-L2}
\Bigl\|{s\over t}| \del\fieldphi|_{p_{\mathbf{low}},k} \Bigr\|_{L^2} + \| \,| \delsH\fieldphi|_{p_{\mathbf{low}},k} \|_{L^2} \lesssim\left[\coeff^{-1/2} \mathcal I^{\mathrm{sing}}_{p_{\mathbf{low}},k}(s) + \coeff^{-1} \mathcal R_{p_{\mathbf{low}},k}^{\mathrm{rob}}(s) \right]s^\delta.
\ee
Thus the estimate for an unscaled derivative may blow up like \(\coeff^{-1} \) when it is recovered from the robust source hierarchy. This estimate remains part of the theorem for each fixed \(f(R) \) theory.

Conversely, multiplying \eqref{equa:seven-closed-phi-L2} by $\coeff$ gives the positive-power estimate
\be
\label{equa:seven-closed-Phi-vanishing}
\| \,|{\fieldPhi}|_{p_{\mathbf{low}},k} \|_{L^2} \lesssim \left[ \coeff\mathcal I^{\mathrm{sing}}_{p_{\mathbf{low}},k}(s) + \coeff^{1/2} \mathcal R_{p_{\mathbf{low}},k}^{\mathrm{rob}}(s) \right] s^\delta.
\ee
\ese
By \eqref{equa:seven-energy-input-consequence} and \eqref{equa:seven-closure-amplitude-consequences}, this gives the explicit positive-power consequence
\be
\label{equa:seven-closed-Phi-positive-power}
\| \,|{\fieldPhi}|_{p_{\mathbf{low}},k} \|_{L^2} \lesssim C_\star\coeff^{1/2} \bigl(A_0 + \mathcal R_{\mathrm{cl}} \bigr)s^\delta.
\ee
The seminorm in \eqref{equa:seven-closed-Phi-positive-power} includes every translation--boost operator through that order, and first derivatives one order lower. The uniform high hierarchy and the massive equation give the stronger local conclusions below. The weighted comparison in Part~III additionally uses convergence of the metric--matter initial data. Both \eqref{equa:seven-closed-dphi-L2} and \eqref{equa:seven-closed-Phi-vanishing} enter the final convergence argument. In \eqref{equa:seven-energy-inputs}, the singular amplitude \(\mathcal I^{\mathrm{sing}}_{p_{\mathbf{low}},k} \) keeps the initial energy \(D_{p_{\mathbf{low}},k}(s_0) \) and the interface flux \(| \mathsf F_{\mathrm{eff}}^{\mathrm{sing}, \Lscr,p_{\mathbf{low}},k}|^{1/2} \).

At the additional fixed-\(\coeff\) continuation level, the energy comparison \eqref{equa:seven-D-coercivity} gives the algebraic bounds
\be
\label{equa:seven-fixed-high-singular-consequences}
\begin{aligned}
\| \,| \fieldphi|_{p_{\mathbf{high}}} \|_{L^2} & \lesssim \mathsf E_{\mathrm{top}}^{\mathrm{tame}}(s), \qquad \| \,|{\fieldPhi}|_{p_{\mathbf{high}}} \|_{L^2} \lesssim \coeff\, \mathsf E_{\mathrm{top}}^{\mathrm{tame}}(s), \\
\Bigl\|{s\over t}| \del\fieldphi|_{p_{\mathbf{high}}} \Bigr\|_{L^2} + \| \,| \delsH\fieldphi|_{p_{\mathbf{high}}} \|_{L^2} & \lesssim \coeff^{-1/2} \mathsf E_{\mathrm{top}}^{\mathrm{tame}}(s).
\end{aligned}
\ee
The polynomial growth in \eqref{equa:seven-tame-high-final} additionally requires the high-order estimate of~\autoref{lem:seven-robust-high-triangular}.

\begin{proposition}[Local vanishing from the commuted robust bounds]
\label{proposition:seven-local-vanishing-from-robust}
Let \(s_1\geq s_0\) be fixed. Assume the field equations, the uniform potential factorization \eqref{equa:seven-uniform-potential-hypothesis}--\eqref{equa:seven-uniform-potential-CN}, and the improved interior bounds \eqref{equa:seven-final-uniform-improvement}--\eqref{equa:seven-final-robust-high-improvement} on \([s_0,s_1] \), with an amplitude \(A\leq1\) independent of \(\coeff\). For \(p=p_{\mathbf{high}} =N-5\), one has
\be
\label{equa:seven-local-Phi-commuted-vanishing}
\sup_{s_0\leq s\leq {s_1}} \Bigl(\| \,| \fieldPhi|_p\|_{L^2(\Mint_s)} + \| \,| \del\fieldPhi|_{p-1} \|_{L^2(\Mint_s)} \Bigr) \leq C_{s_1}\coeff^{1/2}A.
\ee
The equation also gives
\be
\label{equa:seven-local-varphi-vanishing}
\sup_{s_0\leq s\leq {s_1}} \| \,| \fieldvarphi|_{p-2} \|_{L^2(\Mint_s)} \leq C_{s_1}\Bigl(\coeff A+ \coeff^{1/2}A^2\Bigr)
\ee
and, for the residual \(\Theta= \fieldphi+ \mathcal S\),
\be
\label{equa:seven-local-residual-vanishing}
\sup_{s_0\leq s\leq {s_1}} \Bigl(\| \,| \Theta|_{p-2} \|_{L^2(\Mint_s)} + \coeff^{1/2} \| \,| \del\Theta|_{p-2} \|_{L^2(\Mint_s)} \Bigr) \leq C_{s_1}\coeff^{1/2}A.
\ee
Here \(| \cdot|_j\) contains all Cartesian translations and Lorentz boosts of total order at most \(j\), as in \eqref{equa-pointwise-semi-norm-pk-section3}. The constant may depend on \(s_1\), \(N\), the fixed geometry, and \(\coeff_0\), but not on \(0<\coeff\leq\coeff_0\).
\end{proposition}

These norms are taken on the whole truncated interior hypersurface, which is bounded for \(s\leq s_1\). On a bounded part of the initial hypersurface where \(\Sigma_0\) coincides with \(\Mcal_{s_0}\), the equation~\eqref{equa:seven-local-residual-vanishing} therefore gives preparation of order \(q=N-6\), with amplitude \(\Ocal(\sqrt\coeff A)\). This is a consequence of uniform bounds for the commuted energies, which include all Cartesian translations and hence the time derivatives determined by the equations. These hypotheses constrain the rapid oscillations more strongly than smallness of the undifferentiated fields alone. The global preparation amplitude in the equation~\eqref{equa:commuted-preparation-amplitude} also measures the noncompact exterior region with \(w_0\simeq1+r\); that weighted control remains an additional requirement for the global comparison.

\begin{proof}
\bse

\noindent{\bf 1.} \emph{Mass and kinetic terms.} On the fixed slab, \(t/s\) is bounded by a constant depending on \(s_1\). The high energy comparison \eqref{equa:seven-U-coercivity} and \eqref{equa:seven-final-robust-high-improvement} therefore give
\be
\label{equa:seven-local-mass-kinetic-jets}
\| \,| \fieldvarphi|_p\|_{L^2(\Mint_s)} + \| \,| \del\fieldPhi|_p\|_{L^2(\Mint_s)} \leq C_{s_1} A, \qquad \| \,| \fieldPhi|_p\|_{L^2(\Mint_s)} \leq C_{s_1}\coeff^{1/2}A.
\ee
The last bound uses \(\fieldPhi = \coeff^{1/2} \fieldvarphi\). Every first derivative of an operator of order at most \(p-1\) belongs to the order-\(p\) translation--boost family, after commuting translations through boosts. This proves \eqref{equa:seven-local-Phi-commuted-vanishing} without using the kinetic bound at that lower derivative order.

\noindent{\bf 2.} \emph{The massive equation.} In wave coordinates, \(\Box_g\fieldPhi =g^{\alpha \beta} \del_{\alpha \beta} \fieldPhi\). The metric and matter energies, with the Hardy and pointwise estimates already established, control the coefficients and products on the fixed slab. The trace source \(\mathcal S=-8\pi e^{-\fieldPhi}g(\nabla \psi, \nabla \psi) \) and the potential factorization \(W_\coeff(\fieldphi)= (\fieldvarphi)^2\mathcal W_\coeff(\fieldPhi) \) give
\be
\label{equa:seven-local-equation-product-bounds}
\begin{aligned}
\| \,| \Box_g\fieldPhi|_{p-2} \|_{L^2}& \leq C_{s_1}\coeff^{1/2}A,& \qquad \| \,| \del\Box_g\fieldPhi|_{p-2} \|_{L^2}& \leq C_{s_1} A, \\
\| \,| \mathcal S|_{p-2} \|_{L^2} + \| \,|W_\coeff|_{p-2} \|_{L^2} + \| \,| \del W_\coeff|_{p-2} \|_{L^2}& \leq C_{s_1} A^2.
\end{aligned}
\ee
All norms here are on \(\Mint_s\). These are the low--high product estimates for the sources in \eqref{equa:seven-matter-product-rescaled-sources}. For the first term, the largest effective-field derivative has order \(p\) and is controlled by the mass part. For its derivative, the largest order is \(p+1\), controlled by the kinetic part at commutation order \(p\). The differentiated potential uses effective-field mass jets only through order \(p-1\); the trace source uses matter derivatives only through order \(p-1\). In each product the lower factor has at least two remaining Sobolev orders, since \(p\geq14\).

The trace equation and its algebraic consequence are
\be
\label{equa:seven-local-massive-reconstruction}
3\Box_g\fieldPhi-\fieldphi = \mathcal S+W_\coeff, \qquad \fieldvarphi = \coeff^{1/2} \Bigl(3\Box_g\fieldPhi-\mathcal S-W_\coeff\Bigr).
\ee
Combining \eqref{equa:seven-local-equation-product-bounds} with this identity proves \eqref{equa:seven-local-varphi-vanishing}. Applying the same product inequality with one improved \(\fieldvarphi\) factor then gives
\be
\label{equa:seven-local-potential-vanishing}
\| \,|W_\coeff|_{p-2} \|_{L^2(\Mint_s)} \leq C_{s_1} A\Bigl(\coeff A+ \coeff^{1/2}A^2\Bigr).
\ee

\noindent{\bf 3.} \emph{The trace residual.} The definition of \(\Theta \) and \eqref{equa:seven-local-massive-reconstruction} imply
\be
\label{equa:seven-local-residual-reconstruction}
\Theta=3\Box_g\fieldPhi-W_\coeff, \qquad \coeff^{1/2} \del\Theta=3\coeff^{1/2} \del\Box_g\fieldPhi-\coeff^{1/2} \del W_\coeff.
\ee
Equations \eqref{equa:seven-local-equation-product-bounds} and \eqref{equa:seven-local-potential-vanishing} prove \eqref{equa:seven-local-residual-vanishing}. On a bounded part of the initial hypersurface coinciding with the initial interior hypersurface, the normal derivatives and the tangential derivatives used in the comparison norm are bounded combinations of the Cartesian derivatives, and \(w_0\) is bounded independently of \(\coeff\). Thus the local preparation consequence uses the operators of order \(q-1=N-7=p-2\). \qedhere 
\ese
\end{proof}


\subsection{\texorpdfstring{Initial and interface amplitudes}{Initial and interface amplitudes}}
\label{subsection:seven-initial-interface-amplitudes}

\paragraph{Initial energy amplitudes}

We separate the initial quantities and the interface outputs of the Cauchy problem on the full domain from the estimates proved in the open interior.  We take \(C_{\mathrm{data}} \) and \(C_{\mathrm{bd}} \) to be independent of \(0<\coeff\leq\coeff_0\).  The uniform lower initial energies satisfy
\bse
\be
\label{equa:seven-initial-energy-bound}
\sup_{0\leq k\leq p_{\mathbf{low}}} \Bigl(s_0^{-\alpha_k}U_{p_{\mathbf{low}},k}(s_0) +s_0^{-\beta_k} \mathbf P_{p_{\mathbf{low}},k}(s_0) +s_0^{-\nu_k} \mathbf W_{p_{\mathbf{low}},k}(s_0) \Bigr) + \sup_{0\leq k\leq p_{\mathbf{low}}}s_0^{-\chi_k} \mathbf P_{p_{\mathbf{low}},k}(s_0) \leq C_{\mathrm{data}}A_0.
\ee
The identity \(D= \coeff^{-1/2}U\) then gives
\be
\label{equa:seven-initial-degenerate-bound}
\sup_{0\leq k\leq p_{\mathbf{low}}} s_0^{-\delta}D_{p_{\mathbf{low}},k}(s_0) \leq C_{\mathrm{data}} \coeff^{-1/2}A_0.
\ee
The auxiliary robust high initial data satisfy
\be
\label{equa:seven-initial-robust-high-bound}
\max_{0\leq k\leq p_{\mathbf{high}}} \Bigl( s_0^{-\nu_k} \mathbf W_{p_{\mathbf{high}},k}(s_0) +s_0^{-\nu_k} \mathbf P_{p_{\mathbf{high}},k}(s_0) \quad {} +s_0^{-1/2-\nu_k}U_{p_{\mathbf{high}},k}(s_0) +s_0^{-\nu_k} \mathbf H_{p_{\mathbf{high}},k}(s_0) \Bigr) \leq C_{\mathrm{data}}A_0.
\ee
\ese
It implies the explicitly singular initial estimate \(D_{p_{\mathbf{high}},k}(s_0) \leq C_{\mathrm{data}} \coeff^{-1/2}A_0s_0^{1/2 + \nu_k} \).

\paragraph{Interface bounds and energy amplitudes}

For the lower family, we abbreviate
\be
\mathsf T_{2;p_{\mathbf{low}},k}^{\mathrm{all}} = \mathsf T_{2;p_{\mathbf{low}},k}[u] + \mathsf T_{2;p_{\mathbf{low}},k}[\psi] + \mathsf T_{2;p_{\mathbf{low}},k}[{\fieldvarphi}].
\ee
The information on the interior--exterior interface used in the bootstrap improvement will be established by the exterior analysis and consists of the following quantities:
\bse
\label{equa:seven-interface-data-bounds}
\be
\label{equa:seven-interface-flux-data}
\sup_{\substack{0\leq k\leq p_{\mathbf{low}} \\
s_0\leq\tau\leq s}} \tau^{-\eta_{\Lscr}} \mathcal F_{p_{\mathbf{low}},k}^{\Lscr, \mathrm{unif}}(\tau;s_0) \leq C_{\mathrm{bd}}A_0,
\ee
\be
\label{equa:seven-interface-L2-trace-data}
\sup_{\substack{0\leq k\leq p_{\mathbf{low}} \\
s_0\leq\tau\leq s}} \tau^{-\eta_{\Lscr}} \mathsf T_{2;p_{\mathbf{low}},k}^{\mathrm{all}}(\tau) \leq C_{\mathrm{bd}}A_0,
\ee
\be
\label{equa:seven-interface-pointwise-data}
\max_{0\leq k\leq p_{\mathbf{low}}-3} \mathcal A_{p_{\mathbf{low}}-3,k}^{\mathrm{bd}}(s) + \max_{0\leq k\leq p_{\mathbf{low}}} \mathsf B_{00;k}(s) \leq C_{\mathrm{bd}}(A_0 +A+A^2),
\ee
\be
\label{equa:seven-interface-robust-high-data}
\max_{0\leq k\leq p_{\mathbf{high}}} \mathcal I_{\mathrm h,k}^{\mathrm{rob}}(s_1) \leq C_{\mathrm{bd}}A_0\qquad\hbox{for every }{s_1}\geq s_0,
\ee
\be
\label{equa:seven-interface-fixed-high-data}
\mathcal I_{p_{\mathbf{high}}}^{\mathrm{tame}}(s_1) <+ \infty\qquad\hbox{for every finite }{s_1}\geq s_0.
\ee
\ese
The last line follows from the uniform high interface bound by \eqref{equa:seven-fixed-interface-from-robust}. The other four bounds distinguish integrated fluxes, sphere traces, characteristic initial values, and high-order traces. Proposition~\ref{proposition:eight-c-top-exterior} will control the positive exterior flux and good bulk. Proposition~\ref{lem:eight-rank-local-interface-elimination} will convert these energy bounds into the interface estimates by radial, physical-time, and characteristic arguments.

By \({\fieldvarphi} = \coeff^{1/2} \fieldphi\) and \eqref{equa:seven-flux-rescaling}, the lower unscaled trace and flux obey
\be
\label{equa:seven-interface-degenerate-consequences}
\mathsf T_{2;p_{\mathbf{low}},k}[\fieldphi](s) = \coeff^{-1/2} \mathsf T_{2;p_{\mathbf{low}},k}[{\fieldvarphi}](s), \qquad | \mathsf F_{\mathrm{eff}}^{\mathrm{sing}, \Lscr, p_{\mathbf{low}},k}|^{1/2} = \coeff^{-1/2}| \mathsf F_{\mathrm{eff}}^{\mathrm{unif}, \Lscr, p_{\mathbf{low}},k}|^{1/2}.
\ee
Thus these formulas keep the singular interface information.

Since \(\eta_{\Lscr}<\chi_k<\alpha_k, \beta_k, \nu_k\), \eqref{equa:seven-initial-energy-bound}-- \eqref{equa:seven-interface-L2-trace-data} imply
\be
\label{equa:seven-energy-input-consequence}
\begin{aligned}
& \sup_{0\leq k\leq p_{\mathbf{low}}} \Bigl( \mathcal I^U_{p_{\mathbf{low}},k} + \mathcal I^P_{p_{\mathbf{low}},k} + \mathcal I^W_{p_{\mathbf{low}},k} + \mathcal I^{P, \mathrm l}_k \Bigr) \leq C_\star A_0, \qquad \mathcal I^{\mathrm{sing}}_{p_{\mathbf{low}},k} \leq C_\star \coeff^{-1/2}A_0, \quad \mathcal T^u_{p_{\mathbf{low}},k}(s) \leq C_\star A_0, \\
& \max_{0\leq k\leq p_{\mathbf{high}}} \mathcal I_{\mathrm h,k}^{\mathrm{rob}}(s) \leq C_\star A_0, \qquad C_\star=C_N(1+C_{\mathrm{data}} +C_{\mathrm{bd}}).
\end{aligned}
\ee
Moreover, we have
\be
\label{equa:seven-closure-boundary-amplitude}
\mathcal B_3^{\mathrm{low}}(s) \leq \mathcal B_{\mathrm{cl}}, \quad \mathcal B_{\mathrm{cl}} =C_\star(A_0 +A+A^2),
\ee
and the definition of \(C_{\mathrm{tr}} \), followed by the final smallness choice, gives \(C_{\mathrm{tr}}(s) \leq2\).

\paragraph{Explicit bounds for the energy amplitudes}

We define the amplitudes used below by
\be
\label{equa:seven-closure-explicit-amplitudes}
\begin{aligned}
\mathcal U_{\mathrm{cl}} &=C_N\Bigl(C_\star A_0 + {\mathcal B_{\mathrm{cl}}^2\over \rho\theta} \Bigr), \qquad \mathcal L_{\mathrm{cl}} =C_N\Bigl(C_\star A_0 + {\mathcal B_{\mathrm{cl}}^2 + \mathcal B_{\mathrm{cl}} \mathcal U_{\mathrm{cl}} \over \rho\theta} \Bigr), \\
\mathcal W_{\mathrm{cl}} &= C_N\Bigl( C_\star A_0 + {A_0\over \zeta_\star} +(A+A_0)C_\star A_0 + {\mathcal B_{\mathrm{cl}}^2\over \widetilde\mu_0} + {\mathcal B_{\mathrm{cl}} (\mathcal U_{\mathrm{cl}} + \mathcal L_{\mathrm{cl}}) \over \rho\theta} \Bigr), \\
\mathcal J_{\mathrm{cl}} &= \mathcal U_{\mathrm{cl}} + \mathcal L_{\mathrm{cl}} + \mathcal W_{\mathrm{cl}}, \qquad \mathcal R_{\mathrm{cl}} =A_0 + \Bigl(1+ {A\over \rho} \Bigr) \mathcal J_{\mathrm{cl}}.
\end{aligned}
\ee
Equations \eqref{equa:seven-U-improved-raw}, \eqref{equa:seven-P-low-improved-raw}, and \eqref{equa:seven-W-improved-raw} imply
\be
\label{equa:seven-closure-amplitude-consequences}
\mathcal U_k^{\mathbf{low}} \leq\mathcal U_{\mathrm{cl}}, \quad \mathcal L_k\leq\mathcal L_{\mathrm{cl}}, \quad \mathcal W_k^{\mathbf{low}} \leq\mathcal W_{\mathrm{cl}}, \quad \mathcal J_{p_{\mathbf{low}},k} \leq\mathcal J_{\mathrm{cl}}.
\ee
Consequently, we have \(\mathcal R_{p_{\mathbf{low}},k}^{\mathrm{rob}} \leq\mathcal R_{\mathrm{cl}} \).

We also recover the undifferentiated metric without adding an interior bootstrap norm.  We use the radial semi-hyperboloidal derivative \(\delsH_r\) defined in Section~\ref{section-3-4}.  The Sobolev inequality on \(\mathbb S_s\), followed by integration at fixed \(s\), gives
\be
\label{equa:seven-undifferentiated-metric-recovery}
\begin{aligned}
|u(t,r \omega)| & \leq |u(t(R_s),R_s\omega)| + \int_r^{R_s}| \delsH_ru(t(\rho), \rho\omega)| \, d\rho \\
& \leq C_NR_s^{-1/2} \mathsf T_{2;p_{\mathbf{low}},2}[u](s) + C_NA s^\delta \int_0^{R_s}(s^2 + \rho^2)^{-3/4} \, d\rho \leq C_N(C_{\mathrm{bd}}A_0 +A)s^{-1/2 + \delta}.
\end{aligned}
\ee
Likewise, \eqref{equa:seven-refined-effective-field} and \({\fieldPhi} = \coeff^{1/2}{\fieldvarphi} \) imply
\be
\label{equa:seven-Phi-geometric-pointwise}
\|{\fieldPhi} \|_{L^\infty(\Mint_s)} \leq C_N\coeff^{1/2} \mathcal B_{\mathrm{cl}} \leq C_N\coeff_0^{1/2} \mathcal B_{\mathrm{cl}}.
\ee
The first inequality gives the positive-power estimate, while the second is the weaker consequence used to improve the geometric bootstrap range.


\subsection{Bootstrap improvement and interior continuation}
\label{section-11-2}

\paragraph{Strict improvement of the energy hierarchy}

\begin{proposition}[Interior improvement under the bootstrap]
\label{proposition:seven-interior-improvement}
Let the interior and exterior fields be restrictions of the same smooth Cauchy development on a finite bootstrap interval. Assume that
\begin{itemize}
\item the field equations of Proposition~\ref{propo41} hold with \(V\equiv0\), the potential hypotheses \eqref{equa:seven-uniform-potential-hypothesis}--\eqref{equa:seven-uniform-potential-CN} hold, and the initial constraints and compatibility conditions are satisfied;
\item the two interior energy levels \eqref{equa:seven-uniform-bootstrap}, the high radial bound \eqref{equa:seven-uniform-high-radial-bootstrap}, and the degenerate bound \eqref{equa:seven-degenerate-bootstrap} hold; the exterior hierarchy includes metric and matter energies through \(N\), the effective-field energy through \(p_{\mathbf{high}} \), and the order-\(N\) effective-field energy with its factor \(s^{-1} \), as in \eqref{equa:eight-mobile-coupled-leaf}; these energies and the metric spacetime and Hardy norms in \eqref{equa:seven-e-good-bulk-input} satisfy their coarse bootstrap bounds with amplitude \(A\);
\item the reference metric is \eqref{equa-defineMS-new}, with \(m\leq C_{\mathrm{ref}}A_0\), and the interior and exterior geometric bootstrap ranges hold, including \eqref{equa:seven-adapted-coercivity-hypothesis};
\item the initial estimates \eqref{equa:seven-initial-energy-bound}--\eqref{equa:seven-initial-degenerate-bound}, \eqref{equa:seven-initial-robust-high-bound}, and the prescribed exterior weighted initial bounds hold, together with the interface estimates \eqref{equa:seven-interface-data-bounds}.
\end{itemize}
The coefficient estimates on this slab were established in Proposition~\ref{prop:seven-polynomial-coefficient-reduction} and the equation~\eqref{equa:seven-coupled-coefficient-ray-bound}. After the fixed constants have been chosen, there is an \(\varepsilon_0>0\), independent of \(0<\coeff\leq\coeff_0\), such that \(A=C_1A_0\leq\varepsilon_0\) implies, for \(0\leq k\leq p_{\mathbf{low}} \),
\be
\label{equa:seven-final-uniform-improvement}
U_{p_{\mathbf{low}},k}(s) \leq\frac A2s^{\alpha_k}, \qquad \mathbf P_{p_{\mathbf{low}},k}(s) \leq\frac A2s^{\beta_k}, \qquad \mathbf W_{p_{\mathbf{low}},k}(s) \leq\frac A2s^{\nu_k}.
\ee
The sharper matter estimate \(\mathbf P_{p_{\mathbf{low}},k} \leq C_N\mathcal L_{\mathrm{cl}}s^{\chi_k} \) also holds.

The auxiliary robust high hierarchy is strictly improved as well:
\be
\label{equa:seven-final-robust-high-improvement}
\mathsf E_{\mathrm h,k}^{\mathrm{rob}}(s) \leq {A\over2}s^{\nu_k}, \quad 0\leq k\leq p_{\mathbf{high}}.
\ee
In particular, the rescaling keeps the singular high estimate
\be
\label{equa:seven-final-high-D-from-U}
D_{p_{\mathbf{high}},k}(s) = \coeff^{-1/2}U_{p_{\mathbf{high}},k}(s) \leq {A\over2} \coeff^{-1/2}s^{1/2 + \nu_k}.
\ee
For each fixed \(\coeff>0\), let \(B_\coeff\) satisfy
\be
\label{equa:seven-Bcoeff-final-choice}
B_\coeff \geq 4C_N\coeff^{-1/2} \Bigl(C_\star A_0 + \mathcal R_{\mathrm{cl}} \Bigr).
\ee
Then one has
\be
\label{equa:seven-final-degenerate-improvement}
D_{p_{\mathbf{low}},k}(s) \leq {B_\coeff\over2}s^\delta.
\ee
The separate high fixed-$\coeff$ energy satisfies \eqref{equa:seven-tame-high-final}, and hence remains finite on each finite interval.  The pointwise estimates \eqref{equa:seven-refined-pointwise-system}, \eqref{equa:seven-refined-effective-field-singular}, and the high singular consequences \eqref{equa:seven-fixed-high-singular-consequences} hold on the same interval.

If the geometric bootstrap permits \(|{\fieldPhi}| \leq c_0\) and energy-comparison constants in \([c_\H/2,2C_\H] \), then the rank-zero estimates improve them to
\be
\label{equa:seven-geometric-improvement}
|{\fieldPhi}| \leq {c_0\over2}, \quad {3c_\H\over4}e_\H[q] \leq Q^{\mathrm{wave}}_{\alpha \beta}[q] \,N_s^\alpha X_\H^\beta \leq {3C_\H\over2}e_\H[q].
\ee
The weighted matter and effective-field forms have the same strict improvement. The exterior argument gives the boundary quantities required in this proposition.
\end{proposition}

\begin{proof}

First, we choose
\bse
\be
\label{equa:seven-final-C1-choice}
C_1\geq2^{10}C_NC_\star(1+ \zeta_\star^{-1}), \quad A=C_1A_0.
\ee
After \(\rho, \theta, \widetilde\mu_0, \zeta_\star\), and \(C_1\) have been fixed in that order, we choose \(A_0\) sufficiently small so that
\be
\label{equa:seven-final-smallness-choice}
C_NA\Bigl( 1+ \theta^{-1} +(\rho\theta)^{-1} + \widetilde\mu_0^{-1} + \zeta_\star^{-1} \Bigr) \leq2^{-12} \min\{1, \rho, \theta \}.
\ee
We include the smallness required by the coefficient induction in \eqref{equa:seven-final-smallness-choice}. Its constants are fixed before \(C_1\) and \(A_0\); the resulting auxiliary coefficient \(C_NA\) multiplies an energy or another small field in the energy inequalities. We also reduce the initial amplitude, with all fixed constants unchanged, so that
\be
\label{equa:seven-final-central-smallness}
C_Np_{\mathrm w} \bigl[A+C_\star(A_0 +A+A^2) \bigr] \leq\varepsilon_{\mathrm{cw}} \theta.
\ee
\ese
Since \(\mathcal B_3^{\mathrm{low}} \leq\mathcal B_{\mathrm{cl}} =C_\star(A_0 +A+A^2) \), this verifies \eqref{equa:seven-massless-core-smallness} and gives \(B_{s_1}\leq2\). The choice depends on the prescribed data size, uniformly in the bootstrap interval and in \(\coeff\).

The linear parts of the three amplitudes in \eqref{equa:seven-closure-explicit-amplitudes} are then at most \(A/16\); every other term contains at least two small amplitudes and is also at most \(A/16\).  The three lower estimates therefore give \eqref{equa:seven-final-uniform-improvement}.  Equations \eqref{equa:seven-energy-input-consequence}, \eqref{equa:seven-Ydeg-triangular}, and \eqref{equa:seven-Bcoeff-final-choice} give the degenerate improvement without absorbing \(B_\coeff\) into itself.

Proposition~\ref{lem:seven-robust-high-triangular}, \eqref{equa:seven-interface-robust-high-data}, and the same smallness choice give \eqref{equa:seven-final-robust-high-improvement}.  The identity \(D= \coeff^{-1/2}U\) then gives \eqref{equa:seven-final-high-D-from-U}; its negative power is kept.

Exact rescaling in \eqref{equa:seven-fixed-high-rescaling} gives the polynomial bound \eqref{equa:seven-tame-high-final}, finite on every finite interval at fixed \(\coeff\). To improve the geometric range, we also need the weighted inverse-metric component near the cone. On the graph \(t= (s^2 + |x|^2)^{1/2} \), the definition of \(N_s\) and the wave energy tensor give the identity
\bse
\be
\label{equa:seven-adapted-density-exact}
Q^{\mathrm{wave}}_{\alpha \beta}[q]N_s^\alpha X_\H^\beta = \frac{\sqrt{| \det g|}}2\Bigl(-g^{\H00}| \del_tq|^2 +g^{ab} \delsH_aq\, \delsH_bq\Bigr).
\ee
Consequently, for small Cartesian \(H\), we obtain
\be
\label{equa:seven-adapted-density-error}
\left|Q^{\mathrm{wave}}_{\alpha \beta}[q]N_s^\alpha X_\H^\beta-\frac12e_\H[q] \right| \leq C\Bigl(|H|+(t/s)^2|H^{\H00}| \Bigr)e_\H[q].
\ee
The rank-zero metric estimate and \(t\leq(s^2 +1)/2\) on \(\Mint_s\), with \(s\geq2\), imply
\be
\label{equa:seven-adapted-H00-smallness}
(t/s)^2|H^{\H00}| \leq C\theta^{-1} \mathcal B_{\mathrm{cl}} \frac{t}{s^2} \leq C\theta^{-1} \mathcal B_{\mathrm{cl}}.
\ee
\ese
Together with \eqref{equa:seven-undifferentiated-metric-recovery} and \eqref{equa:seven-Phi-geometric-pointwise}, this proves the strict energy comparison. The factor \(e^{-\fieldPhi} \) and the positive effective-field mass term obey the same comparison after reducing the initial smallness threshold. The massless estimate and the critical-commutator bounds apply on the bootstrap interval; the interface bounds give their amplitudes. This proves the interior improvement.

\end{proof}

\paragraph{Interior continuation}

The global continuation argument also uses the exterior Cauchy data. Their weighted norms, defined in \eqref{equa:eight-exterior-X-definitions}--\eqref{equa:eight-exterior-order-weight}, measure the metric, matter, and rescaled effective field on the exterior initial hypersurface. The required bounds are
\be
\label{equa:seven-exterior-initial-assumptions}
\max_{\substack{p\in\{N,p_{\mathbf{high}},p_{\mathbf{low}} \} \\
0\leq k\leq p}} s_0^{-\delta} \mathsf X^{\mathrm{unif}}_{p,k}(s_0) \leq C_{\mathrm{data}}A_0, \qquad \mathsf X^{\mathrm{fix}}_{N,N}(s_0)< + \infty.
\ee
Future-global continuation uses, in addition to the interior estimates, the improvement of the exterior energies and interface quantities in Part~III.

Proposition~\ref{proposition:seven-interior-improvement} concludes the interior argument: once the five interface estimates are provided, it improves the lower energies, the normalized high energies, and the geometric bounds on the bootstrap interval. The estimates for the unscaled effective field remain explicit in the equations~\eqref{equa:seven-closed-phi-L2}--\eqref{equa:seven-closed-dphi-L2}, and rescaling gives the finite high-order norm~\eqref{equa:seven-tame-high-final} for each fixed \(\coeff>0\).

\paragraph{Completion of the global existence and convergence arguments}

In Part~III, Proposition~\ref{prop:audit-moving-interface-interior-transfer} will transfer these estimates to the expanding interface with their radial and cone traces. Proposition~\ref{prop:mobile-simultaneous-improvement} will then give the interface bounds and improve the interior, exterior, and geometric estimates simultaneously. Its final step proves global continuation, using the order-\(N\) estimate of Lemma~\ref{lem:audit-fixed-parameter-joined} on each finite slab. The initial order-\(N\) norm is finite by smoothness on the compact interior and by the weighted exterior assumption; all smallness choices preceding the fixed-parameter bound are independent of \(\coeff\).

For convergence, Proposition~\ref{proposition:seven-local-vanishing-from-robust} already gives the strong interior consequence of the trace equation on finite slabs. The later estimate on the full domain will also keep the weighted preparation amplitude and cancel the two interface fluxes before estimating its source. These estimates, together with the positive-power bound~\eqref{equa:seven-closed-Phi-positive-power}, enter the comparison with the Einstein development.


\part{Exterior estimates, global stability, and convergence}


\section{Exterior geometry and weighted inequalities}
\label{section-12}


\subsection{Exterior strategy and interface requirements}
\label{section-12-1}

\paragraph{Purpose}

Part~III establishes the exterior energy and interface estimates needed to complete the global stability proof. These estimates also control the spatial tails in the convergence to Einstein gravity. We keep throughout this part the global convention \(s_0\geq2\) fixed at the beginning of Part~II. The Schwarzschild mass \(m\), the reference metric, the foliation, and all cutoff functions are fixed once and for all, independently of \(\coeff\); only the solution and its initial data vary with \(\coeff\).

This section collects standard functional inequalities of Sobolev, Hardy, and Poincar\'e type, together with weighted energy identities for the wave equation. We state the precise weights and interface terms used in the subsequent estimates.

\paragraph{Exterior weight and boundary terms}

Throughout Part~III, we use the affine weight
\be
\label{equa:eight-affine-exterior-current-weight}
w\coloneqq 2+r-t, \qquad \crochet(t,r)=1+\aleph(r-t), \qquad w|_{\Lscr}=1, \qquad \tfrac12\crochet\leq w\leq2\crochet.
\ee
Here \(\aleph\) is the smooth nondecreasing function fixed in~\eqref{equa:global-foliation-exterior-weight-section3}, with \(\aleph(q)=0\) for \(q\leq-1\) and \(\aleph(q)=q+1\) for \(q\geq0\). The comparison of the two weights is uniform. All hypersurface norms use the coordinate measure \(dx\); the densitized normal \(N_s=j_sn_s\) gives the geometric density in the energy currents.

\paragraph{The expanding interface used in the global argument}

The following auxiliary construction uses a nonnegative reference mass \(m\geq0\) and a fixed coarse amplitude \(A\). For a parameter-dependent family, \(A\) is chosen as one uniform upper bound, so the auxiliary interface is independent of \(\coeff\). We put \(t_i\coloneqq(s_0^2+1)/2\). We choose the decay exponent
\be
\label{equa:eight-mobile-interface-exponents}
e_0\coloneqq\frac\delta6.
\ee
We fix a sufficiently large constant \(K\), put \(h(t)\coloneqq K(A+A^2)t^{-1-e_0}\), and define
\be
\label{equa:eight-mobile-interface-definition}
q_b(t)\coloneqq1-\int_{t_i}^t h(\tau)\,d\tau,\qquad r_b(t)\coloneqq t-q_b(t),\qquad \Lscr_b\coloneqq\{r=r_b(t),\ t\geq t_i\}.
\ee
The amplitude restriction \(K(A+A^2)t_i^{-e_0}/e_0\leq1/4\) gives \(3/4\leq q_b\leq1\). Thus \(r_b'(t)=1+h(t)\). The interface starts at the initial sphere \(r=(s_0^2-1)/2\), \(t=t_i\). On its exterior we use
\be
\label{equa:eight-mobile-exterior-weight}
w_b(t,r)\coloneqq1+r-t+q_b(t),\qquad \tfrac34w\leq w_b\leq w,\qquad dw_b=dr-(1+h(t))dt.
\ee
The exterior weight is one on \(\Lscr_b\) and is uniformly comparable to \(\crochet\). For the current on the full domain, we use the weight one on the interior and \(w_b^{2\eta}\) on the exterior. This piecewise weight is continuous and locally Lipschitz; its weak first derivative has no interface measure. The resulting hypersurface norms are equivalent to the norms with the smooth weight \(\crochet\) on the same domains.

The functional and characteristic estimates in Sections~\ref{section-12}--\ref{section-14} are first stated for the prescribed foliation. Section~\ref{section:small-mass-moving-interface} proves their transfer to the domains bounded by \(\Lscr_b\), establishes positivity of the corresponding currents, and closes the interior and exterior estimates simultaneously. The initial hypersurface and the Cauchy data remain unchanged.

There are two different boundary considerations in the exterior argument.
\bei
\item A \emph{signed flux} will arise as the boundary term of an energy identity. We write the interior and exterior identities for the same current and add them. The opposite normals cancel that term exactly.
\item A positive norm for a \emph{boundary trace} arises as a measure of the size of a field on the interface. We estimate it from the exterior hypersurface by radial integration and angular differentiation. This boundary trace is kept explicitly in the pointwise and Hardy estimates, with its precise derivative count and effective-field normalization.
\eei
Both operations concern restrictions of one solution to the Cauchy problem. At a given rank, we first substitute the trace inequalities in the energy estimate and absorb their small same-rank coefficients. After that rank has been estimated, we recover its traces for the next step. This is how the two arguments form a single continuation proof.

The geometry also explains the weights arising in our boundary terms. On the interface, \(r\simeq t\simeq s^2\), so spatial decay \(r^{-1} \) becomes \(s^{-2} \), while \(dt\simeq s\, ds\) changes the power in an integrated flux. On the interface, each angular derivative can be written as a combination of Lorentz boosts with bounded coefficients. Controlling three additional angular derivatives thus requires estimates at boost rank \(k+3\) instead of \(k\), with a correspondingly larger growth exponent. The strict inequalities imposed below on the weight exponents ensure that the radial decay compensates for this additional growth. We also express the wave equation in Schwarzschild lightlike coordinates, using the tortoise coordinate introduced below. Its logarithmic correction to \(r\) is unbounded along the interface and is kept explicitly in that coordinate.

\paragraph{\texorpdfstring{Traces and fluxes at the interface}{Traces and fluxes at the interface}}

We established the sufficient pointwise estimate for the massless field~\eqref{equa:seven-sufficient-massless-gain} in Proposition~\ref{lem:seven-massless-core-gain}, under the interior--exterior energy bootstrap. In Proposition~\ref{lem:seven-e-good-bulk}, we bounded the integrated high-order initial trace of the metric by the coarse exterior spacetime integral of its good derivatives. We now estimate the interface quantities for the same solution to the Cauchy problem, with the normal orientation opposite to that of the interior, the current normalization~\eqref{equa:eight-c-interface-current-normalization}, and the same dependence on \(\coeff\). We organize the argument into five steps. At each rank, we estimate the exterior currents and traces at the interface together with the interior energy. Signed interface fluxes cancel when we add the interior and exterior identities, while radial and angular estimates control the trace norms.

\begin{enumerate}
\item \emph{Oriented fluxes.}  With the exterior and interior orientations, we have
\be
d\Sigma_\alpha^{\Lscr, \mathrm{ext}} =-d\Sigma_\alpha^{\Lscr, \mathrm{int}}.
\ee
The exterior multiplier is \(X_{\ME} = \del_t=X_\H\) on the whole Euclidean--merging slab, and hence in particular on \(\Lscr\). The matter current is
\be
e^{-{\fieldPhi}}Q^{\mathrm{wave} \, \alpha}{}_{\beta}[Z\psi]X_\H^\beta,
\ee
and the two effective-field fluxes satisfy
\be
\mathsf F_{\mathrm{eff}}^{\mathrm{unif}, \Lscr,p,k} = \coeff\, \mathsf F_{\mathrm{eff}}^{\mathrm{sing}, \Lscr,p,k},
\ee
and hence we have
\be
{\bigl| \mathsf F_{\mathrm{eff}}^{\mathrm{sing}, \Lscr,p,k} \bigr|^{1/2} = \coeff^{-1/2} \bigl| \mathsf F_{\mathrm{eff}}^{\mathrm{unif}, \Lscr,p,k} \bigr|^{1/2}.}
\ee

\item \emph{Sphere traces.} We estimate the three traces \(\mathsf T_{2;p_{\mathbf{low}},k}[u] \), \(\mathsf T_{2;p_{\mathbf{low}},k}[\psi] \), and \(\mathsf T_{2;p_{\mathbf{low}},k}[{\fieldvarphi}] \), defined in~\eqref{equa:seven-L2-trace-definition}, separately from the flux. We fix the foliation parameter \(s\) and the direction \(\omega\in\mathbb S^2\), and integrate along the curve \(r\mapsto(\fTime(s,r),r\omega)\), from the interface radius \(R_s\) to spatial infinity. This curve lies in the exterior hypersurface \(\MME_s\); the derivative \(\delsME_r\) differentiates along it with respect to \(r\), at fixed \(s\) and \(\omega\). Integration over the angular directions then gives the radial inequality
\be
R_s^{-1} \|Zv\|_{L^2(\mathbb S_s)}^2 \lesssim \int_{\MME_s} \bigl(| \delsME_rZv|^2 + r^{-2}|Zv|^2\bigr)J\, dx.
\ee
For ${\fieldvarphi}$, we use one additional Cartesian translation in the uniform effective-field hierarchy; its mass term controls \(\del Z{\fieldvarphi} \). Exact rescaling gives \(\mathsf T_2[\fieldphi]= \coeff^{-1/2} \mathsf T_2[{\fieldvarphi}] \).

\item \emph{Pointwise traces at the interface and angular derivatives.} The exterior weighted Sobolev and characteristic estimates control \(\mathsf B_{00;k} \), \(\mathsf B_{\mathrm{char};p,k} \), and \(\mathsf B_{\psi;k} \), defined in~\eqref{equa:seven-metric-pointwise-traces}, as well as \(\mathsf B_{\mathrm{ray};p,k}^{\fieldphi} \), defined in~\eqref{equa:seven-ray-trace-definition}. The small exponents \(\eta_k\) describe how the allowed polynomial growth in \(s\) increases with the number \(k\) of Lorentz boosts. Three additional angular derivatives require three additional boosts, so the growth bound acquires the factor \(s^{\eta_{k+3}-\eta_k}\). On \(\Lscr\), where \(r\simeq t\simeq s^2\), the radial weight gives enough decay to compensate for this additional growth, since
\be
\eta_{k+3}-\eta_k=3C_{\mathrm{rk}} \theta<2\mu-1.
\ee
At this stage we have an integrated \(L^2\) flux estimate, not yet a pointwise estimate.

\item \emph{Exact Schwarzschild characteristics.} In the reference end we use
\be
\mathbf r\coloneqq r+m, \quad D_m(r)\coloneqq{r-m\over r+m} =1-{2m\over\mathbf r}, \quad \mathbf r_*\coloneqq r+2m\log(r-m) +c_*,
\ee
The bold symbol \(\mathbf r\) denotes the Schwarzschild areal radius, while \(r=|x|\) remains the radial coordinate of the wave chart. The constant \(c_*\in\mathbb R\) determines the additive normalization of the tortoise coordinate \(\mathbf r_*\). We also set
\be
u_*\coloneqq t-\mathbf r_*, \qquad L_*\coloneqq \del_t+ \del_{\mathbf r_*}, \qquad \underL_*\coloneqq \del_t-\del_{\mathbf r_*}.
\ee
With \(F_Z\coloneqq \Box_{g^\star}Z\psi\), the wave equation for the commuted massless field \(Z\psi\) on the Schwarzschild reference metric takes the form
\be
L_*\underL_*(\mathbf r Z\psi) -{D_m\over {\mathbf r}^2} \Delta_{\mathbb S^2}(\mathbf r Z\psi) + {2mD_m\over {\mathbf r}^3}(\mathbf r Z\psi) =-D_m\mathbf r F_Z.
\ee
Its characteristic integration follows \(u_*=t-\mathbf r_*= \mathrm{constant} \), starting from the initial exterior hypersurface. The conversion from \(\Box_gZ\psi\) to \(F_Z\) is the metric-difference term in \eqref{equa:eight-Schwarzschild-exact-remainder}. On the interface, we have
\be
\underL_*(\mathbf r Z\psi) = \underL(rZ\psi) +mL(Z\psi) +(1-D_m)Z\psi.
\ee
Along the interface \(\Lscr\), we have \(du_*/ds\sim-4m/s\). Since \(s^{-1}\) is not integrable on \([s_0,\infty)\), this produces a logarithmically unbounded change in the retarded time \(u_*\), even for small \(m>0\). We must therefore keep this contribution explicitly in the characteristic coordinate to obtain estimates uniform in time.

\item \emph{Robust high-order control and fixed-$\coeff$ continuation.} We denote by \(s_1\) the upper value of the foliation parameter on the slab \(s_0\leq s\leq s_1\). At the high differentiation order \(p_{\mathbf{high}}=N-5\) fixed in~\eqref{equa:seven-differentiation-levels}, the exterior argument controls the metric and matter fluxes without normalization, the uniform effective-field flux with the factor \({s_1}^{-1/2} \), and the two distinct high traces collected in \(\mathcal I^{\mathrm{tr}}_{p_{\mathbf{high}},p_{\mathbf{high}}}(s_1) \). The weights are those in \eqref{equa:seven-robust-high-interface-input}; consequently, \(\mathcal I_{\mathrm h,p_{\mathbf{high}}}^{\mathrm{rob}}(s_1) \lesssim A_0\). The gap \(p_{\mathbf{high}}-p_{\mathbf{low}}=4\) provides the two angular derivatives and the two additional derivatives needed for Sobolev embedding, as specified in~\eqref{equa:seven-boost-only-available-orders}. These bounds are used to prove the pointwise estimate~\eqref{equa:seven-sharp-boost-only-H00} for metric coefficients differentiated only by Lorentz boosts. For the separate fixed-parameter hierarchy, the degenerate high effective-field flux and the same two traces make \(\mathcal I_{p_{\mathbf{high}}}^{\mathrm{tame}}(s_1) \) finite on every finite interval.  Its estimate may contain \(\coeff^{-1/2} \) and the corresponding fixed-parameter exponential growth; these singular factors remain explicit.
\end{enumerate}

The weighted energy and Hardy estimates follow the arguments of~\cite[Sections~3.3 and~4.2]{LeFloch-Ma-2024a}, while the commutator estimates follow~\cite[Sections~5 and~6]{LeFloch-Ma-2024a} and~\cite[Appendix~B]{LeFloch-Ma-2024b}. We derive the currents with the matter factor, the Schwarzschild lightlike coordinate, the angular derivative counts, and both effective-field normalizations. Section~\ref{section:small-mass-moving-interface} establishes positivity and simultaneous improvement on the expanding domains. Section~\ref{section:eight-interface-rank-estimates} gives the resulting interface estimates used in Section~\ref{section-16}.


\subsection{\texorpdfstring{Weighted functional inequalities}{Weighted functional inequalities}}
\label{section-12-2}

\paragraph{Functional inequalities on the Euclidean--merging hypersurfaces}

The radial Hardy and Sobolev inequalities are now stated on the foliation under consideration. The current identities also use the metric and field equations. Their constants are independent of the terminal time and, for the rescaled effective field, of \(\coeff\).

\begin{proposition}[Weighted radial Hardy inequality with an interface trace]
\label{lem:eight-weighted-radial-hardy}
Let \(v\) be a smooth field on a Euclidean--merging hypersurface, with finite \(\|\crochet^{\lambda-1}v\|_{L^2(dx)}\) and \(\|\crochet^\lambda\delsME_rv\|_{L^2(dx)}\), and let \(\lambda>1/2\). Then one has
\be
\label{equa:eight-weighted-radial-hardy}
\| \crochet^{\lambda-1}v\|_{L^2(\MME_s)}^2 +R_s^{-1} \| \crochet^{\lambda}v\|_{L^2(\mathbb S_s)}^2 \leq C_\lambda \| \crochet^{\lambda} \delsME_rv\|_{L^2(\MME_s)}^2.
\ee
Here \(R_s=(s^2-1)/2\), and the sphere norm uses its Euclidean area measure.
\end{proposition}

The proof uses the coordinate measure \(dx=r^2\,dr\,d\omega\), rather than the induced geometric volume.

\begin{proof}
\bse
The estimate is a one-dimensional integration on each ray of a fixed hypersurface, followed by integration over the sphere. We keep the inner endpoint term, because it is the trace needed by the interior argument. For a fixed value of \(s\), we first pull the spacetime field back to the hypersurface and set
\be
v_s(x)\coloneqq v\bigl(\fTime(s,|x|),x\bigr).
\ee
The chain rule and the definition of the Euclidean--merging frame give the identities
\be
\label{equa:eight-fixed-leaf-radial-derivative}
\del_{x^a}v_s= \Bigl(\del_a+ {x^a \over r} \del_r\fTime\, \del_t\Bigr)v = \delsME_av, \qquad \del_rv_s={x^a \over r} \del_{x^a}v_s= \delsME_rv.
\ee
We set \(R_E=\rhoE(s)=R_s+1\), \(a(r)=1+r-R_E\), and \(\beta=2\lambda-1>0\). On the flat portion \(r\geq R_E\), the weights \(a\) and \(\crochet\) are uniformly comparable. Integrating \(\del_r(r^2a^\beta|v_s|^2)\) in \(r\) and over the unit sphere gives
\be
\label{equa:eight-Hardy-flat-positive-identity}
\begin{aligned}
R_E^2\|v_s(R_E,\cdot)\|_{L^2(\mathbb S^2)}^2+\int_{r\geq R_E}(\beta a^{\beta-1}+2r^{-1}a^\beta)|v_s|^2\,dx
&=-2\int_{r\geq R_E}a^\beta v_s\del_rv_s\,dx.
\end{aligned}
\ee
For finite weighted norms, we first integrate up to a radius \(R\). Radial averaging of \(a^{2\lambda-2}|v_s|^2\) selects radii tending to infinity for which \(R^2a(R)^\beta\|v_s(R,\cdot)\|_{L^2(\mathbb S^2)}^2\) tends to zero. Cauchy--Schwarz on the right and Young's inequality then give
\be
\label{equa:eight-Hardy-flat-estimate}
R_E^2\|v_s(R_E,\cdot)\|_{L^2(\mathbb S^2)}^2+\|a^{\lambda-1}v_s\|_{L^2(r\geq R_E)}^2\leq C_\lambda\|a^\lambda\del_rv_s\|_{L^2(r\geq R_E)}^2.
\ee
On the merging interval \([R_s,R_E]\), the weights are uniformly comparable to one and the radii are comparable. The fundamental theorem of calculus, followed by Cauchy--Schwarz on this interval of length one, bounds its volume integral and its inner sphere trace by the outer trace and the integral of \(|\del_rv_s|^2\). Since \(R_s\geq1\), the resulting trace bound also controls \(R_s^{-1}\|v\|_{L^2(\mathbb S_s)}^2\). The chain-rule identity above proves the assertion with \(\del_rv_s=\delsME_rv\).
\ese
\end{proof}

\paragraph{Weighted Sobolev estimate}

\begin{lemma}[Exterior weighted Sobolev estimates]
\label{lem:eight-exterior-weighted-sobolev}
Let \(\eta \in\{\lambda, \mu\} \), and let \(v\) be a wave field tending to zero at spatial infinity on each hypersurface, with finite weighted energy and Hardy norm. For \(p+5\leq N\) and \(0\leq k\leq p\), one has on the whole exterior hypersurface
\be
\label{equa:eight-exterior-weighted-sobolev}
r w^\eta \bigl(| \del v|_{p,k} + | \delsME v|_{p,k} \bigr) +r w^{\eta-1/2}|v|_{p,k} \lesssim\Fenergy^{\ME,p+5,k+5}_\eta(s,v).
\ee
The stronger weight on the field itself follows by radial integration from spatial infinity; this part requires only four additional commutations. For the effective field, the uniform estimate is
\be
\label{equa:eight-exterior-weighted-sobolev-mass}
r w^\mu\bigl(| \del\fieldPhi|_{p,k} + | \delsME\fieldPhi|_{p,k} + | \fieldvarphi|_{p,k} \bigr) \lesssim U^{\mathrm{ext}}_{p+5,k+5}(s).
\ee
On \(r\leq5t/4\), the same estimates hold with \((p+5,k+5) \) replaced by \((p+3,k+3) \), provided that these orders are available. All norms use \(dx\), and all constants are independent of \(s\) and \(\coeff\).
\end{lemma}
\begin{proof}
\bse
On the flat part of the hypersurface, the local radial \(H^1\) inequality and sphere \(H^2\) inequality apply to \(f= \del Zv\) on a unit radial interval. As in \eqref{equa:seven-exterior-local-sphere-bound}, they give
\be
\label{equa:eight-exterior-Sobolev-local-norms}
rw^\eta|f(r, \omega)| \lesssim\sum_{|J| \leq2} \Bigl(\|w^\eta \Omega^Jf\|_{L^2(\mathcal A_r)} + \|w^\eta \del_r\Omega^Jf\|_{L^2(\mathcal A_r)} \Bigr), \qquad \mathcal A_r= \{r\leq|x| \leq r+1\}.
\ee
The same flat-end estimate gives the outer trace of the width-one merging interval. On this interval, we integrate the hypersurface-radial derivative. Its expression \(\delsME_a= \delsN_a-(1-\del_r\fTime) \omega_a \del_t\), with \(1-\del_r\fTime\leq\zeta \), is controlled by the kinetic energy of a translation operator. Cauchy--Schwarz transports the outer trace to every point of the interval. This keeps the energy factor \(\zeta \) rather than dividing by it.

Globally, \(\Omega_{ab} = [L_a,L_b] \), so two rotations cost four boost operators. The radial derivative of \(f\) costs one further translation. This proves the derivative estimate with five additional commutations. For the field itself, the finite Hardy norm and integrable radial derivative give zero limit at infinity. On the flat portion, radial integration of \(\Omega^JZv\), followed by Cauchy--Schwarz, uses
\be
\label{equa:eight-exterior-value-radial-kernel}
\int_r^\infty\rho^{-2}(2+\rho-t)^{-2\eta}\,d\rho\leq\frac{r^{-2}w^{1-2\eta}}{2\eta-1},\qquad \eta>\tfrac12.
\ee
Sphere Sobolev with two rotations then gives \(rw^{\eta-1/2}|v|_{p,k}\lesssim\Fenergy_\eta^{\ME,p+4,k+4}(s,v)\). On the merging interval, the width-one radial integration transports this outer trace; its tangential derivative is already controlled by the kinetic energy, with no additional commutation or division by \(\zeta\). For \(Z\fieldvarphi\), both \(Z\fieldvarphi\) and its translation derivatives are controlled by the positive mass terms of the corresponding energies. This proves \eqref{equa:eight-exterior-weighted-sobolev-mass}.

In the near-cone region, the identity
\be
\label{equa:eight-exterior-Sobolev-near-rotations}
\Omega_{ab} = (x_a/t)L_b-(x_b/t)L_a
\ee
has bounded coefficients and bounded commuted coefficients on the local radial intervals in use. Each rotation then costs one boost. The two angular derivatives and the radial derivative require three additional commutations.
\ese
\end{proof}

\paragraph{Weighted product estimate}

\begin{proposition}[Low--high weighted product calculus]
\label{proposition:eight-low-high-products}
Let \(v_{\mathrm l} \) and \(v_{\mathrm h} \) be the low- and high-order factors in a commuted semilinear source, with the low factor of order at most \(N-5\).  Suppose that, for some \(\eta \in\{\lambda, \mu\} \) and \(\gamma_0>0\), the low-order factor satisfies
\be
\label{equa:eight-source-admissible-low-factor}
{\bigl\|J\zeta^{-1} \crochet^{\lambda-\eta}v_{\mathrm l} \bigr\|_{L^\infty(\MME_s)} \leq A s^{-1-\gamma_0}.}
\ee
Then the corresponding Leibniz summand satisfies
\be
\label{equa:eight-low-high-products}
{\bigl\|J\zeta^{-1} \crochet^\lambda v_{\mathrm l}v_{\mathrm h} \bigr\|_{L^2(\MME_s)} \lesssim A s^{-1-\gamma_0} \bigl\| \crochet^\eta v_{\mathrm h} \bigr\|_{L^2(\MME_s)}.}
\ee
If, in addition, \(\|\crochet^\eta v_{\mathrm h}\|_{L^2}\leq As^\delta\) and \(\delta<\gamma_0/2\), the product is bounded by \(CA^2s^{-1-\gamma_0/2}\), which is integrable in \(s\).

\end{proposition}

For a high Cartesian derivative, the kinetic energy controls \(\zeta\del Zv\): the estimate is therefore applied with \(v_{\mathrm h}=\zeta\del Zv\), placing the compensating factor \(\zeta^{-1}\) in \(v_{\mathrm l}\). This factor must be included when checking the hypothesis~\eqref{equa:eight-source-admissible-low-factor}. The principal commutators are estimated separately against the positive spacetime terms.

\begin{proof}
H\"older's inequality with the weights split as in \eqref{equa:eight-source-admissible-low-factor} gives \eqref{equa:eight-low-high-products}.  Under the additional high-factor hypothesis, the weighted norm is bounded by \(As^\delta\), so \(s^{-1-\gamma_0}s^\delta \leq s^{-1-\gamma_0/2} \).  For an effective-field derivative we use \(\del\fieldPhi = \coeff\del\fieldphi\), and for a zero-order effective-field factor we use \(\fieldvarphi = \coeff^{1/2} \fieldphi\).  In the unscaled source, each derivative factor \(\del\fieldphi\) contributes \(\coeff^{-1}\), and each undifferentiated factor \(\fieldphi\) contributes \(\coeff^{-1/2}\). The resulting parameter power depends on the source under consideration.
\end{proof}


\subsection{\texorpdfstring{Weighted wave energies and interface fluxes}{Weighted wave energies and interface fluxes}}
\label{section:eight-weighted-wave-energies}

\paragraph{Weighted currents and energy densities}

In the exterior domain \(q=t-r\leq1\), we use the affine weight \(w\) from~\eqref{equa:eight-affine-exterior-current-weight}. For a wave field satisfying \(\Box_gv=F_v\), the multiplier is \(X_{\ME}=\del_t=X_\H\), and the weighted current is
\be
\label{equa:eight-weighted-current-definition}
\mathcal P^\alpha_{\lambda,g}[v]\coloneqq-2w^{2\lambda}Q^{\mathrm{wave} \, \alpha}{}_{\beta,g}[v]X_{\ME}^\beta.
\ee
We use the following coordinate normalization of the interior and exterior energies and interface fluxes:
\be
\label{equa:eight-c-interface-current-normalization}
\begin{aligned}
\widehat{\calE}_{\mathrm w}^{\mathrm{int}}&\coloneqq2\calE_{\mathrm w}^{\H}, \qquad \widehat{\mathsf F}_{\mathrm w}^{\Lscr, \mathrm{int}}\coloneqq-2\mathsf F_{\mathrm w}^{\Lscr, \mathrm{int}}, \\
\widehat{\calE}_{\mathrm w}^{\mathrm{ext}}(s,v)&\coloneqq2\int_{\MME_s}w^{2\lambda}Q^{\mathrm{wave}}_{\alpha \beta}[v]N_s^\alpha X_{\ME}^\beta \, dx, \quad \widehat{\mathsf F}_{\mathrm w}^{\Lscr, \mathrm{ext}}\coloneqq \int_{\Lscr} \mathcal P^\alpha_{\lambda,g}[v] \, d\Sigma_\alpha^{\Lscr, \mathrm{ext}}.
\end{aligned}
\ee
With \(H_N=H^{\mathcal N00}\), the quadratic form entering the interface flux is
\be
\label{equa:eight-affine-null-flux-density}
D_g[v]\coloneqq g^{\mathcal N ab} \delsN_av\delsN_bv-H_N| \del_tv|^2.
\ee
The corresponding signed bulk density and its decomposition into nonnegative terms are
\be
\label{equa:eight-weighted-current-bulk-density}
\Omega_{\lambda,g}[v]\coloneqq2\lambda w^{2\lambda-1}D_g[v],
\ee
\be
\label{equa:eight-affine-current-bulk-splitting}
\begin{aligned}
\Omega_{\lambda,g}&= \Omega^+_{\lambda,g}-\Omega^-_{\lambda,g}, \\
\Omega^+_{\lambda,g}&\coloneqq2\lambda w^{2\lambda-1} \bigl(g^{\mathcal N ab} \delsN_av\delsN_bv+(-H_N)_+ | \del_tv|^2\bigr), \\
\Omega^-_{\lambda,g}&\coloneqq2\lambda w^{2\lambda-1}(H_N)_+ | \del_tv|^2.
\end{aligned}
\ee
The prescribed reference metric is stationary in the exterior. Writing \(u=g-g^\star\), we define the signed deformation integral by
\be
\label{equa:eight-weighted-current-deformation-definition}
\mathcal D_u(v)\coloneqq-\int_{\MME_{[s_0,s]}}w^{2\lambda}Q^{\mathrm{wave} \, \alpha \beta}[v] \del_tu_{\alpha \beta} \, d\mu_g.
\ee
We write \(\calE^{\ME}_\lambda\coloneqq\widehat{\calE}_{\mathrm w}^{\mathrm{ext}}\), \(\calE^{\Lscr}_\lambda\coloneqq\widehat{\mathsf F}_{\mathrm w}^{\Lscr,\mathrm{ext}}\), \(\mathcal G^{\ME}_\lambda\coloneqq\int_{\MME_{[s_0,s]}}\Omega^+_{\lambda,g}\,d\mu_g\), and \(\mathcal M^{\ME}_\lambda\coloneqq\int_{\MME_{[s_0,s]}}\Omega^-_{\lambda,g}\,d\mu_g\).

\begin{proposition}[Exterior weighted energy identity and positivity]
\label{proposition:eight-weighted-current-identity}
Assume that \(g^{\mathcal N ab}\) is uniformly positive definite in the exterior and that \(H_N\leq0\) on the interface. Let \(v\) be a smooth wave field with finite weighted energy and integrable source and deformation terms on the finite slab. On the noncompact exterior, assume that the outer-cylinder flux vanishes along an exhaustion. Then \(v\) satisfies the exterior energy balance
\be
\label{equa:eight-weighted-current-identity}
\begin{aligned}
\calE^{\ME}_\lambda(s,v) + \calE^{\Lscr}_\lambda(s,v;s_0) + \mathcal G^{\ME}_\lambda(s_0,s;v)
&= \calE^{\ME}_\lambda(s_0,v)-2\int_{\MME_{[s_0,s]}}w^{2\lambda}F_v\del_tv\, d\mu_g\\
&\quad+ \mathcal D_u(v) + \mathcal M^{\ME}_\lambda(s_0,s;v).
\end{aligned}
\ee
The interface flux is nonnegative:
\be
\label{equa:eight-interface-flux-positivity}
\calE^{\Lscr}_\lambda(s,v;s_0)= \int_{\Lscr_{[s_0,s]}}D_g[v]j_g\, dx\geq0,
\ee
The positive bulk controls the good derivatives up to the interface:
\be
\label{equa:eight-weighted-current-positivity}
\Omega^+_{\lambda,g}[v] \geq c_\lambda w^{2\lambda-1} \bigl(| \delsN v|^2 +(-H_N)_+ | \del_tv|^2\bigr) \geq0.
\ee
\end{proposition}

\begin{proof}
On \(0\leq q\leq1\), the weights satisfy \(1\leq w,\crochet\leq2\); on \(q\leq0\), they agree. Since \(dw=dr-dt\), contraction of the stress tensor gives
\be
\label{equa:eight-affine-weight-contraction}
-4\lambda w^{2\lambda-1}(\del_\alpha w)Q^{\mathrm{wave} \, \alpha}{}_{\beta}[v]X_{\ME}^\beta=-\Omega_{\lambda,g}[v].
\ee
The stress-tensor divergence identity and the product rule yield
\be
\label{equa:eight-weighted-current-pointwise-identity}
\nabla_\alpha \mathcal P^\alpha_{\lambda,g}[v]+ \Omega_{\lambda,g}[v] =-2w^{2\lambda}F_v\del_tv-w^{2\lambda}Q^{\mathrm{wave} \, \alpha \beta}[v] \del_tg_{\alpha \beta}.
\ee
Separating the positive and negative bulk contributions gives
\be
\label{equa:eight-affine-current-positive-identity}
\nabla_\alpha \mathcal P^\alpha_{\lambda,g}[v]+ \Omega^+_{\lambda,g}[v] =-2w^{2\lambda}F_v\del_tv-w^{2\lambda}Q^{\mathrm{wave} \, \alpha \beta}[v] \del_tg_{\alpha \beta} + \Omega^-_{\lambda,g}[v].
\ee
Integration proves~\eqref{equa:eight-weighted-current-identity}. The outward exterior surface element is \((dt-dr)j_g\,dx\); the same contraction at \(w=1\), with \(H_N\leq0\) on the interface, gives~\eqref{equa:eight-interface-flux-positivity}. Uniform positive definiteness of \(g^{\mathcal N ab}\) gives~\eqref{equa:eight-weighted-current-positivity}. On a finite radial truncation, we keep the signed outer-cylinder flux. Lemma~\ref{lem:eight-outer-cylinder-tail} justifies passage to infinity; comparability of \(w\) and \(\crochet\) preserves its finite-energy hypotheses.
\end{proof}

\paragraph{Matter and effective-field currents}

The matter current has stress tensor \(e^{-\fieldPhi}Q^{\mathrm{wave}}\) and weight exponent \(\mu\); derivatives of \(e^{-\fieldPhi}\) remain in the signed source term. The effective-field currents have tensors \(Q^{\mathrm{sing}}\) and \(Q^{\mathrm{unif}}\), also with exponent \(\mu\). Their interface densities are
\be
\label{equa:eight-affine-effective-field-interface-density}
3\coeff D_g[Z\fieldphi]+ |Z\fieldphi|^2, \qquad 3D_g[Z\fieldPhi]+ |Z\fieldvarphi|^2,
\ee
The corresponding bulk densities are these expressions multiplied by \(2\mu w^{2\mu-1}\), with the same positive--negative splitting of the bad derivative term. The mass term remains in the positive bulk. The singular and uniform current identities differ by the exact factor \(\coeff\). With the weight \(w^{2\mu}\) in~\eqref{equa:eight-exterior-effective-field-energy-definitions}, their coordinate-current energies are twice the singular and uniform exterior energies, respectively. Estimates with \(\crochet\) have fixed comparison constants. These assertions follow from~\eqref{equa:seven-exact-effective-field-tensors}, \(\fieldPhi=\coeff\fieldphi\), and \(\fieldvarphi=\coeff^{1/2}\fieldphi\).

\paragraph{Interior and exterior energy identities}

The interior wave deformation integral is
\be
\widehat{\mathcal D}^{\mathrm{int}}_g(v)\coloneqq-\int_{\Mint_{[s_0,s]}}Q^{\mathrm{wave} \, \alpha \beta}[v] \pi^{(X_\H)}_{\alpha \beta} \, d\mu_g.
\ee
On a finite slab \([s_0,s_1]\), let \(E^i_\bullet\) and \(E^e_\bullet\) denote the commuted interior and exterior energies in coordinate normalization, where \(\bullet\in\{\mathrm w,\psi,\mathrm c\}\) and \(\mathrm c\) denotes the uniformly rescaled effective field. We write \(F_\bullet\) for the nonnegative exterior interface flux, \(B_\bullet\) for the positive exterior bulk, and \(S^i_\bullet,S^e_\bullet\) for the signed source and deformation integrals. The latter include the exterior \(\Omega^-\) contribution. We define the energy and source combinations
\be
\label{equa:eight-augmented-energy-components}
E_\bullet\coloneqq E^i_\bullet+2E^e_\bullet, \qquad S_\bullet\coloneqq S^i_\bullet+2S^e_\bullet.
\ee
At the high effective-field level, only the interior energy is multiplied by \(s^{-1}\):
\be
\label{equa:eight-augmented-high-effective-field-energy}
E^{\mathrm h}_{\mathrm c}(s)\coloneqq s^{-1}E^i_{\mathrm c}(s) +2E^e_{\mathrm c}(s).
\ee

\begin{proposition}[\texorpdfstring{Energy balance with the interface flux}{Energy balance with the interface flux}]
\label{lem:eight-augmented-interface-energy}
Under the hypotheses of Proposition~\ref{proposition:eight-weighted-current-identity}, let \(v\) be the same smooth solution on the interior and exterior domains, with finite interior energy and integrable interior source and deformation terms. Adding the two wave identities with equal coefficients cancels their signed interface fluxes and gives
\be
\label{equa:eight-c-current-joined-identity}
\begin{aligned}
\widehat{\calE}_{\mathrm w}^{\mathrm{int}}(s,v) + \widehat{\calE}_{\mathrm w}^{\mathrm{ext}}(s,v) + \mathcal G^{\ME}_\lambda(s_0,s;v) &= \widehat{\calE}_{\mathrm w}^{\mathrm{int}}(s_0,v) + \widehat{\calE}_{\mathrm w}^{\mathrm{ext}}(s_0,v) \\
& \quad -2\int_{\Mint_{[s_0,s]}}F_v\del_tv\, d\mu_g-2\int_{\MME_{[s_0,s]}}w^{2\lambda}F_v\del_tv\, d\mu_g\\
& \quad + \widehat{\mathcal D}^{\mathrm{int}}_g(v) + \mathcal D_u(v) + \mathcal M^{\ME}_\lambda(s_0,s;v).
\end{aligned}
\ee
Adding the interior identity to twice the exterior identity instead keeps a positive interface flux. With the preceding energy normalization, each of the three fields satisfies
\be
\label{equa:eight-augmented-interface-identity}
E_\bullet(s) +F_\bullet(s) +2B_\bullet(s)=E_\bullet(s_0) +S_\bullet(s), \qquad F_\bullet(s)=2\bigl| \mathsf F_\bullet^{\Lscr, \mathrm{int},p,k}(s;s_0) \bigr|.
\ee
At the high effective-field level, the balance is
\be
\label{equa:eight-augmented-effective-field-normalization}
E^{\mathrm h}_{\mathrm c}(s) + \int_{s_0}^{s}(2-\tau^{-1}) \, dF_{\mathrm c}(\tau) +2B_{\mathrm c}(s) + \int_{s_0}^{s} \tau^{-2}E^i_{\mathrm c}(\tau) \, d\tau =E^{\mathrm h}_{\mathrm c}(s_0) + \int_{s_0}^{s} \tau^{-1} \, dS^i_{\mathrm c}(\tau) +2S^e_{\mathrm c}(s).
\ee
All four terms on the left of the last identity are nonnegative, since \(s_0\geq2\). For smooth solutions, the Stieltjes integrals are ordinary integrals of the corresponding densities. The equal-coefficient cancellation also holds for the matter and effective-field currents.
\end{proposition}

\begin{proof}
The interior wave identity is
\be
\label{equa:eight-c-current-interior-identity}
\widehat{\calE}_{\mathrm w}^{\mathrm{int}}(s,v) + \widehat{\mathsf F}_{\mathrm w}^{\Lscr, \mathrm{int}}(s,v;s_0) = \widehat{\calE}_{\mathrm w}^{\mathrm{int}}(s_0,v)-2\int_{\Mint_{[s_0,s]}}F_v\del_tv\, d\mu_g -\int_{\Mint_{[s_0,s]}}Q^{\mathrm{wave} \, \alpha \beta}[v] \pi^{(X_\H)}_{\alpha \beta} \, d\mu_g.
\ee
On the interface, the two multipliers are \(\del_t\), the exterior weight equals one, and the induced normals are opposite. The two current traces therefore cancel when this identity is added to~\eqref{equa:eight-weighted-current-identity}, proving~\eqref{equa:eight-c-current-joined-identity}. In the notation for all three fields, the two identities read
\be
\label{equa:eight-augmented-r-identities}
E^i_\bullet(s)-F_\bullet(s)=E^i_\bullet(s_0) +S^i_\bullet(s), \qquad E^e_\bullet(s) +F_\bullet(s) +B_\bullet(s)=E^e_\bullet(s_0) +S^e_\bullet(s).
\ee
The signs and factors follow from \(\widehat{\mathsf F}^{\Lscr,\mathrm{int}}=-\widehat{\mathsf F}^{\Lscr,\mathrm{ext}}=-2\mathsf F^{\Lscr,\mathrm{int}}\) and~\eqref{equa:eight-interface-flux-positivity}. Adding the first identity to twice the second proves~\eqref{equa:eight-augmented-interface-identity}. The singular effective-field identity follows by division of every term of the uniform identity by \(\coeff\). For the high-order normalization, the product rule gives \(s^{-1}dE^i_{\mathrm c}=d(s^{-1}E^i_{\mathrm c})+s^{-2}E^i_{\mathrm c}\,ds\). Multiplication of the interior identity by \(s^{-1}\), followed by addition of twice the exterior identity, proves~\eqref{equa:eight-augmented-effective-field-normalization}. Monotonicity of the positive flux also gives
\be
\label{equa:eight-augmented-effective-field-flux-comparison}
s^{-1}F_{\mathrm c}(s) \leq F_{\mathrm c}(s) \leq\int_{s_0}^{s}(2-\tau^{-1}) \, dF_{\mathrm c}(\tau).
\ee
Thus the high-order identity controls the exterior energy and flux at their original scales, and also the required normalization \(s^{-1/2}|\mathsf F_{\mathrm{eff}}^{\mathrm{unif}}|^{1/2}\). On a finite radial truncation, both identities keeping the interface flux keep twice the signed outer-cylinder flux. The same outer-flux limit justifies passage to infinity.
\end{proof}

\paragraph{Consequences}

At \(p=p_{\mathbf{low}} \), \((\Eenergy^+_{p,k})^2\) is the sum of the three nonnegative left-hand sides of \eqref{equa:eight-augmented-interface-identity}. At \(p=p_{\mathbf{high}} \), it is the sum of the left-hand sides for the unweighted metric and matter identities and for \eqref{equa:eight-augmented-effective-field-normalization}. This definition keeps both the full good exterior bulk and the favorable term \(\int\tau^{-2}E^i_{\mathrm c} \, d\tau\). An estimate furnished by the nonlinear source analysis on the finite slab of the form
\be
\label{equa:eight-augmented-energy-target}
\Eenergy^+_{p,k}(s) \leq M a \,s^{d_k}, \qquad s_0\leq s\leq s_1, \qquad a\coloneqq C_0\eps+A^2, \qquad d_k\coloneqq (k+1) \delta_{\mathrm G} + \eta_k,
\ee
has the following interface consequences:
\bei
\item If \(d_{p_{\mathbf{low}}} \leq\eta_{\Lscr} \), the lower aggregate flux is bounded by \(CMa \) after multiplication by \(s^{-\eta_{\Lscr}} \).
\item If \(d_k\leq\nu_k<1\) for \(k\leq p_{\mathbf{high}} \), the prescribed high initial bounds and the fixed-hypersurface trace estimates imply
\be
\label{equa:eight-augmented-robust-interface-consequence}
\max_{0\leq k\leq p_{\mathbf{high}}} \mathcal I_{\mathrm h,k}^{\mathrm{rob}}(s_1) \leq C_{N, \mathrm{gaps}}(A_0 +Ma).
\ee
The constant includes the fixed factor \(\zeta_\star^{-1} \) and is independent of \(s_1\), \(A\), and \(\coeff\). If \(a \leq C A_0\), this is the required bound by the initial size. Exact rescaling then gives the fixed-\(\coeff\) finiteness statement.
\eei

To verify the second consequence, \eqref{equa:seven-e-exterior-sphere-trace} and \eqref{equa:seven-e-metric-sphere-trace} give
\be
\label{equa:eight-augmented-high-sphere-consequences}
\mathsf T_{2;p,k}[u](s) \leq CM a \,s^{-1+d_k}, \qquad \mathsf B^H_{2;p,k}(s) \leq C(A_0 +Ma \,s^{d_k}), \qquad p=p_{\mathbf{high}}.
\ee
Since \(s_0\geq1\), \(d_k\leq\nu_k\), and \(\nu_k>0\), direct integration yields
\be
\label{equa:eight-augmented-high-trace-integrals}
\begin{aligned}
\sup_{s_0\leq\sigma \leq s_1} \sigma^{-\nu_k} \| \tau^{-1/2} \mathsf B^H_{2;p,k}(\tau) \|_{L^2(s_0, \sigma)} & \leq {C(A_0 +Ma) \over\sqrt{2\nu_k}}, \\
\sup_{s_0\leq\sigma \leq s_1} \sigma^{-\nu_k} \Bigl(\int_{s_0}^{\sigma} \kappa_u^\star(\tau) \mathsf T_{2;p,k}[u](\tau) \, d\tau+ {1\over\sigma} \int_{s_0}^{\sigma} \mathsf T_{2;p,k}[u](\tau) \, d\tau\Bigr) & \leq CM a(1+A+A_0).
\end{aligned}
\ee
The first bound uses \(\tau^{2d_k-1} \leq\tau^{2\nu_k-1} \); the second uses \(d_k<1\) and \eqref{equa:seven-metric-integrable-kernel-L1}. The metric and matter fluxes are direct summands of \((\Eenergy^+_{p,k})^2\); the effective-field flux uses \eqref{equa:eight-augmented-effective-field-flux-comparison}. These estimates prove \eqref{equa:eight-augmented-robust-interface-consequence}.

For the rank exponents in \eqref{equa:eight-augmented-energy-target}, sufficient strict margins are
\be
\label{equa:eight-augmented-interface-exponent-margins}
(p_{\mathbf{low}} +1) \delta_{\mathrm G}<9C_{\mathrm{rk}} \theta, \qquad (p_{\mathbf{high}} +1) \delta_{\mathrm G}<12\rho+2C_{\mathrm{rk}} \theta.
\ee
They can be imposed by reducing the initial amplitude at fixed rank weights. The estimate \eqref{equa:eight-augmented-energy-target} is the remaining nonlinear energy estimate needed for these interface consequences; the identities above keep its signed source and deformation terms explicitly.


\section{Exterior characteristics and traces at the interface}
\label{section:exterior-characteristics}


\subsection{\texorpdfstring{Outgoing metric coefficients and matter decay}{Outgoing metric coefficients and matter decay}}
\label{section:eight-outgoing-characteristics}

\paragraph{Purpose}

We estimate the outgoing metric coefficient and integrate the matter and good-metric characteristic equations. Together with the weighted currents, these estimates control the traces at the interface at each commutation rank.

\paragraph{Outgoing metric coefficient}

We separate the dynamic metric coefficient from the prescribed Schwarzschild contribution before estimating the characteristic equation. The same exterior metric energies and good spacetime integral control the traces on \(q=t-r=q_0\leq1\). We write \(w_0=2-q_0\), and use the exponents
\be
\label{equa:eight-displaced-cone-exponents}
\chi_\lambda\coloneqq \lambda-\tfrac12-\tfrac\delta2, \qquad a_\lambda\coloneqq \frac\delta \lambda, \qquad\gamma_\lambda\coloneqq \chi_\lambda-a_\lambda.
\ee

The reference coefficient is the null contraction
\be
\label{equa:eight-reference-null-coefficient}
H^\star\coloneqq (g^\star)^{-1}-\gMink^{-1}, \qquad H_N^\star\coloneqq H^\star(\ell_N, \ell_N), \qquad\ell_N=dt-dr.
\ee

\begin{lemma}[Metric coefficient on an outgoing cone]
\label{lem:eight-displaced-cone-coefficient}
Under the coarse exterior metric energy and good spacetime bootstrap through order \(N\), and the propagated wave gauge, one has, at \((T,(T-q_0) \omega) \), for \(T\geq2T_0\), \(q_0\leq1\), \(|q_0| \leq T/4\), and \(\ord Z\leq N-5\),
\be
\label{equa:eight-displaced-cone-coefficient}
r|Z(H_N-H_N^\star)| \lesssim C_N\Bigl(Aw_0^{1/2}T^{-\chi_\lambda} +A T^{-1+ \delta/2}w_0^{2-\lambda} +A^2T^{-1+ \delta}w_0^{2-2\lambda} \Bigr).
\ee
The same estimate holds for \((\mathcal L_Z(H-H^\star))_N\), where \(\mathcal L_Z\) is the corresponding operator of tensor Lie derivatives. The constant is independent of the terminal time and of \(0<\coeff\leq\coeff_0\). In particular, one has the following bounds.
\begin{itemize}
\item The prescribed reference satisfies \(r|ZH_N^\star| \leq C_NA_0\).
\item In the collar \(w_0\leq T^{a_\lambda} \), the dynamic coefficient satisfies \(r|Z(H_N-H_N^\star)| \lesssim A T^{-\gamma_\lambda} \), with \(\gamma_\lambda>1/48\).
\item In the remaining near-cone region, where \(w_0\geq T^{a_\lambda}\), the direct estimate gives \(rw_0^{-1}|Z(H_N-H_N^\star)|\lesssim AT^{-\delta/2}\).
\end{itemize}
\end{lemma}

\begin{proof}
\bse
We put \(h=K-K^\star\), where \(K=j_g g^{-1}-\gMink^{-1} \) and \(K^\star=j_\star(g^\star)^{-1}-\gMink^{-1} \). The explicit Schwarzschild coefficients in \eqref{equa:reference-Schwarzschild-on-exterior} give
\be
\label{equa:eight-reference-densitized-components}
j_\star= (1+m/r)^2, \quad (K^\star)^{00} =1-\frac{(r+m)^3}{r^2(r-m)}, \quad (K^\star)^{0a} =0, \quad (K^\star)^{ab} =-\frac{m^2x_ax_b}{r^4}.
\ee
Thus \(\del_\alpha(K^\star)^{\alpha \beta} =0\), and the dynamic tensor \(h\) is divergence-free. Tensor Lie derivatives by translations, boosts and rotations preserve this identity. For \(P_J= \mathcal L_Jh\), the density conversion and outgoing Hardy estimate of Section~\ref{section-e-bulk} yield
\be
\label{equa:eight-displaced-cone-density-bulk}
\sum_{\ord J\leq N} \int_{\MME_{[s_0,s_1]}}w^{2\lambda-1} \bigl(| \delsN P_J|^2 +r^{-2}|P_J|^2\bigr) \, dx\, dt\lesssim A^2{s_1}^{2\delta}.
\ee
Here \(w=2 +r-t\). The differentiated density products have a lower factor of order at most \(\lfloor N/2\rfloor\); their safe pointwise count is \(\lfloor N/2\rfloor+7\leq N\). The metric energy at order \(N\) has growth \(As^\delta \); the extra top-order factor in \eqref{equa:eight-exterior-order-weight} belongs only to the effective-field energy.

We use the past interval \(I_T= [T/2,T] \), put \(R(t)=t-q_0\), and choose the upper hypersurface \(S_T\) through \((T,R(T)) \). It covers every radial slice in the calculation: \(f(t)= \fTime(S_T,t-q_0)-t\) satisfies \(f(T)=0\) and \(f'(t)= \del_r\fTime-1\leq0\), so \(\fTime(S_T,r) \geq t\) when \(t\in I_T\) and \(r\geq R(t) \). Also \(S_T^2\simeq T\). The radial gauge identity \eqref{equa:seven-e-null-gauge-radial} has zero divergence remainder for \(P_J\). Its radial Cauchy--Schwarz kernel satisfies
\be
\label{equa:eight-displaced-cone-kernel}
R^2\int_R^\infty r^{-2}(w_0 +r-R)^{1-2\lambda} \, dr\lesssim C_\lambda R^{2-2\lambda}.
\ee
The substitution \(r=R(1+y) \) bounds this kernel by the integral of \((1+y)^{-2}y^{1-2\lambda} \). With \(\ell_N=dt-dr\) and \(F_J(t, \omega)=R(t)P_J(\ell_N, \ell_N) (t,R(t) \omega) \), integration in \(t\) consequently gives
\be
\label{equa:eight-displaced-cone-time-Ltwo}
\sum_{\ord J\leq N} \int_{I_T} \|F_J(t) \|_{L^2(\mathbb S^2)}^2\, dt\lesssim A^2T^{2-2\lambda+ \delta}.
\ee
\ese

The relation \(tL= \omega^aL_a+q_0\del_t\), together with \eqref{equa:seven-e-null-boost-conversion}, gives
\be
\label{equa:eight-displaced-cone-time-identity}
t\frac{dF_J}{dt} = \Bigl(\frac tr-2\Bigr)F_J+r\omega^a(\mathcal L_{L_a}P_J)_N+q_0r(\mathcal L_{\del_t}P_J)_N.
\ee
The angular frame term cancels because \(\omega^a \Pi_{ab} =0\); the factor \(q_0\del_t\) remains explicit. The one-dimensional trace inequality therefore gives
\be
\label{equa:eight-displaced-cone-time-supremum}
\|F_J(T) \|_2^2\lesssim T^{-1} \|F_J\|_{L^2(I_T;L^2)}^2 + \|F_J\|_{L^2(I_T;L^2)} \|F_J'\|_{L^2(I_T;L^2)} \lesssim A^2w_0T^{-2\chi_\lambda}.
\ee
Rotations preserve \(r\) and \(\ell_N\). Two rotations in the sphere Sobolev inequality cost four boosts, since \(\Omega_{ab} = [L_a,L_b] \), and the trace in time costs one further operator. Thus the spherical supremum uses at most \(\ord Z+5\leq N\) derivatives.

\bse
We finally convert the tensor derivatives to scalar derivatives of the coefficient. The boost conversion~\eqref{equa:seven-e-null-boost-conversion} has two distinct contributions. For any symmetric contravariant two-tensor \(T\), it reads
\be
\label{equa:audit-exterior-scalar-Lie-conversion}
L_a(T_N)=(\mathcal L_{L_a}T)_N-2\omega_aT_N-2\frac{t-r}{r}\Pi_{ab}T(dx^b,\ell_N).
\ee
The parallel term is a lower-order contraction \(T_N\), controlled by the same estimate. Only the transverse term carries \((t-r)/r\); translation differentiation of \(\ell_N\) contributes \(r^{-1}\). Iteration therefore gives lower-order null contractions with bounded coefficients, together with \(C_Nw_0/r\) times full components of no greater order. The direct exterior pointwise bound \(|Zu| \lesssim A r^{-1}w_0^{1-\lambda}T^{\delta/2} \) therefore gives the second term of \eqref{equa:eight-displaced-cone-coefficient}. The identity
\be
\label{equa:eight-displaced-cone-density-conversion}
H_N-H_N^\star=j_g^{-1}h_N+(j_g^{-1}-j_\star^{-1})K_N^\star
\ee
gives its third term when derivatives hit a density factor and a lower-order tensor factor. The lower density factors are bounded, and mixed reference--dynamic products are covered by \(A_0\leq A\). This proves \eqref{equa:eight-displaced-cone-coefficient}.

Tensor Lie derivatives satisfy the same density product rule. Translations and boosts preserve \(\gMink^{-1} \), so contraction after differentiation proves the estimate for \((\mathcal L_Z(H-H^\star))_N\). These tensor coefficients enter the commuted principal operator: for every generating translation or boost, we have
\be
\label{equa:eight-principal-Lie-commutator}
[g^{\alpha \beta} \del_\alpha \del_\beta,Z]v=-(\mathcal L_Zg^{-1})^{\alpha \beta} \del_\alpha \del_\beta v.
\ee
In the collar we may replace \(w_0^{1/2} \) by \(w_0\). The three resulting gains satisfy
\be
\label{equa:eight-displaced-cone-collar-gaps}
\chi_\lambda-a_\lambda>\tfrac1{48}, \qquad 1-\tfrac\delta2-a_\lambda(2-\lambda)>\tfrac{35}{48}, \qquad 1-\delta-a_\lambda(2-2\lambda)>\tfrac{19}{24}.
\ee
These inequalities follow from \(\delta \leq(\mu-\lambda)/2<(1-\lambda)/2\) and \(\lambda>3/4\); also \(\gamma_\lambda<1/2\), so it is smaller than the other two gains. In the complement, the direct bound has the factor \(T^{\delta/2}w_0^{-\lambda} \leq T^{-\delta/2} \) after division by \(w_0\). The explicit reference contribution satisfies \(r|ZH_N^\star| \lesssim A_0\) throughout the region.
\ese
\end{proof}

\paragraph{Low-order matter decay from the coarse exterior bounds}

The outgoing-cone coefficient estimate gives a pointwise matter bound at the lower differentiation level. We use the coarse energies through order \(N\), including the metric good spacetime integral, on the same finite bootstrap interval. We put \(w=2 +r-t\), \(\gamma= (1+ \delta)/2\), and \(c=C_{\mathrm{rk}} \theta \).

\begin{proposition}[Low-order exterior matter estimate]
\label{lem:eight-low-matter-characteristic-Lie}
The coarse bootstrap, the propagated wave gauge, and the massless matter equation are assumed. The geometric bootstrap gives \(|H|<h_0\ll1\) and \(H_N\leq0\) on \(t-1\leq r\leq5t/4\). If \(C_Nm\leq c\) and the prescribed amplitude is sufficiently small, then one has
\be
\label{equa:eight-low-matter-characteristic-Lie}
r w^\gamma| \del\psi|_{p,k} \lesssim C_NA s_{\mathrm e}^{\eta_k}, \qquad p\leq p_{\mathbf{low}} =N-9, \quad0\leq k\leq p.
\ee
The constants may depend on the fixed exponents, \(N\), \(s_0\), and \(\coeff_0\), and are independent of the terminal time and of \(0<\coeff\leq\coeff_0\).
\end{proposition}

\begin{proof}

We first commute the principal operator tensorially. For an ordered operator \(Z=X_1\cdots X_d\) of translations and boosts, \(Z_I\) denotes the composition of operators indexed by \(I\subset\{1, \ldots,d\} \), in its original order, and \(\mathcal L_I\) denotes the corresponding tensor Lie derivatives. With \(F_\psi =g^{-1}(d\fieldPhi,d\psi) \), affine commutation gives
\bse
\be
\label{equa:eight-low-matter-ordered-Lie-commutation}
\Box_gZ\psi =ZF_\psi-\sum_{\varnothing\ne I\subset\{1, \ldots,d\}}(\mathcal L_Ig^{-1})^{\alpha \beta} \del_\alpha \del_\beta Z_{I^c} \psi.
\ee
Indeed, for \(\mathcal D_T=T^{\alpha \beta} \del_\alpha \del_\beta \), we have \([\mathcal D_T,X]=-\mathcal D_{\mathcal L_XT} \) and \(X\mathcal D_T= \mathcal D_TX+ \mathcal D_{\mathcal L_XT} \). These identities prove the formula by induction, preserving the order of the operators. Since translations and boosts preserve the Minkowski tensor, the coefficients are Lie derivatives of \(H^\star+(H-H^\star) \).

We set \(E_a= \delsN_a= \del_a+ \omega_a \del_t\) and \(\Pi_{ab} = \delta_{ab}-\omega_a \omega_b\). The contraction is
\be
\label{equa:eight-low-matter-Lie-frame}
T^{\alpha \beta} \del_\alpha \del_\beta V=T_N\del_t^2V+2(T^{0a}-\omega_bT^{ab}) \del_tE_aV+T^{ab}E_aE_bV-r^{-1} \Pi_{ab}T^{ab} \del_tV.
\ee
The last term follows from \(E_a \omega_b= \Pi_{ab}/r\). All terms other than \(T_N\del_t^2V\) thus contain a good second derivative or an additional factor \(r^{-1} \). The coupling has the ordered-partition formula
\be
\label{equa:eight-low-matter-Lie-coupling}
ZF_\psi = \sum_{I_0\sqcup I_1\sqcup I_2= \{1, \ldots,d\}}(\mathcal L_{I_0}g^{-1}) (dZ_{I_1} \fieldPhi,dZ_{I_2} \psi).
\ee
\ese
Its Minkowski part has one good derivative; each other term contains a metric coefficient.

In \(t-1\leq r\leq5t/4\), the coarse exterior Sobolev estimates and \(tE_a=L_a+q\omega_a \del_t\) give, for operators of order at most \(p\),
\bse
\be
\label{equa:eight-low-matter-direct-jets}
\begin{aligned}
| \del Z\psi|+ | \del^2Z\psi|& \lesssim A t^{-1+ \delta/2}w^{-\mu}, \qquad |Z\psi| \lesssim A t^{-1+ \delta/2}w^{1-\mu}, \\
| \del E Z\psi|& \lesssim A t^{-2 + \delta/2}w^{1-\mu}, \qquad | \del Z\fieldPhi|+ t w^{-1}|E Z\fieldPhi| \lesssim A t^{-1/2 + \delta/2}w^{-\mu}.
\end{aligned}
\ee
The effective-field bound keeps its top-energy factor \(s_{\mathrm e} \simeq t^{1/2} \); its constant may depend on \(\coeff_0\). These bounds cost at most \(p+6\) energy orders. The angular term \(\Delta_{\mathbb S^2}Z\psi\) costs four boost operators and five Sobolev orders, hence \(p+9\leq N\). This is the largest order in the argument. Every factor in the preceding product formulas obeys these bounds separately.

For \(V=Z\psi\), we use \(\mathcal P_{H_N} = \del_t+(4+H_N) (4-H_N)^{-1} \del_r\). The radial equation takes the form
\be
\label{equa:eight-low-matter-weighted-transport}
\Bigl(\mathcal P_{H_N}-\frac{2\gamma H_N}{w(4-H_N)} \Bigr) \bigl(w^\gamma \underL(rV) \bigr)= \frac{4rw^\gamma}{4-H_N} \Bigl(-\Box_gV+r^{-2} \Delta_{\mathbb S^2}V+R_H[V] \Bigr),
\ee
where \(|R_H[V]| \lesssim |H|(| \del EV|+r^{-1}| \del V|+r^{-2}|V|) \). The zeroth-order term has the favorable sign. The angular, good-Hessian, and matter-coupling contributions to the absolute value of the right-hand side are bounded, respectively, by
\be
\label{equa:eight-low-matter-integrable-sources}
C_NA t^{-1-e_{\mathrm a}}, \qquad C_NA^2t^{-1-e_{\mathrm g}}, \qquad C_N(A^2 +A^3)t^{-3/2 +3\delta/2}.
\ee
The positive exponents are
\be
\label{equa:eight-low-matter-source-exponents}
e_{\mathrm a}\coloneqq \mu-\tfrac12-\delta>0, \qquad e_{\mathrm g}\coloneqq \lambda+ \mu-\tfrac32-\tfrac{3\delta}2>0.
\ee
The angular contribution satisfies
\be
\label{equa:eight-low-matter-angular-weight}
r^{-1}w^\gamma| \Delta_{\mathbb S^2}Z\psi| \lesssim At^{-2 + \delta/2}w^{\gamma+1-\mu}.
\ee
\ese
The good-Hessian contribution is bounded by \(A^2t^{-2 + \delta}w^{\gamma+2-\lambda-\mu} \). The flat coupling contributes at most \(A^2t^{-3/2 + \delta}w^{\gamma+1-2\mu} \); its metric correction has one additional metric factor. These bounds use \(w\lesssim t\), \(\mu-\lambda \geq2\delta \), and \(\lambda>3/4\).

We estimate the dynamic bad coefficient in two regions. In \(w\leq4t^{a_\lambda} \),~\autoref{lem:eight-displaced-cone-coefficient} and the direct Hessian give the leading product \(C_NA^2t^{-1-\chi_\lambda+ \delta/2}w^{1+ \gamma-\mu} \), where the weaker factor \(w\) has been used in place of \(w^{1/2} \). Its gain, and the gains of the frame and density terms, satisfy
\bse
\be
\label{equa:eight-low-matter-collar-source-gaps}
\begin{aligned}
\chi_\lambda-\tfrac\delta2-a_\lambda(1+ \gamma-\mu)&>\tfrac1{32}, \\
1-\delta-a_\lambda(\gamma+2-\lambda-\mu)&>\tfrac23, \qquad 1-\tfrac{3\delta}2-a_\lambda(\gamma+2-2\lambda-\mu)>\tfrac34.
\end{aligned}
\ee
In the complement \(w\geq4t^{a_\lambda} \), we have \(-q=w-2\geq w/2\). The exterior Hessian reduction gives
\be
\label{equa:eight-low-matter-light-like-Hessian}
| \del^2Z\psi| \lesssim w^{-1}(| \del LZ\psi|+ | \del Z\psi|) +(t/w)| \Box_gZ\psi| \lesssim A t^{-1+ \delta/2}w^{-1-\mu}.
\ee
The coefficient \((t/w)|H| \lesssim A t^{\delta/2}w^{-\lambda} \lesssim A t^{-\delta/2} \) is absorbed. The commutator terms close by induction on total order; the null effective-field source has the additional factor \(t^{-1/2 + \delta/2}w^{1-\mu} \lesssim1\). The largest order remains \(p+6\). The dynamic principal product is therefore bounded by
\be
\label{equa:eight-low-matter-outer-source-gap}
C_NA^2t^{-1+ \delta}w^{\gamma-\lambda-\mu} \lesssim C_NA^2t^{-1-e_{\mathrm o}}, \qquad e_{\mathrm o}\coloneqq a_\lambda(\lambda+ \mu-\gamma)-\delta= \frac{\delta(\mu-\gamma)} \lambda>0.
\ee
\ese

The reference coefficients in \eqref{equa:eight-low-matter-ordered-Lie-commutation} have a different estimate. If their operator contains a translation, they are \(\Ocal(mr^{-2}) \), including after boosts, and the direct Hessian gives an integrable source. If their operator contains boosts only, their size is \(\Ocal(mr^{-1}) \). In this case \(\del_tZ_{I^c} \psi\) has order at most \(p\) and strictly smaller boost rank, so \(\del_t^2Z_{I^c} \psi\) is estimated by the lower-rank matter gradient. The remaining reference terms belong to the good-Hessian estimate by \eqref{equa:eight-low-matter-Lie-frame}.

On \(r\geq5t/4\), the direct estimate already gives \(rw^\gamma| \del\psi|_{p,k} \lesssim A\), since \(w\simeq r\) and \(\gamma-\mu+ \delta/2<0\). The bounded interval \(t\leq2T_0\) is also controlled directly, with a constant depending on \(T_0\). In the near region, a backward \(\mathcal P_{H_N} \)-curve reaches the initial hypersurface or \(r=5t/4\): its retarded coordinate satisfies \(q'=-2H_N/(4-H_N) \geq0\) forward in time. The initial term is controlled by the data and the latter term by the direct estimate. Both have size \(C_NA\). The curve remains below the terminal hypersurface: along it, the derivative of \(t-\fTime(s_1,r) \) is \(1-(\del_r\fTime) (4+H_N)/(4-H_N)>0\), since \(H_N\leq0\) and \(0\leq\del_r\fTime<1\).

We denote by \(Y_k(t) \) the supremum of \(rw^\gamma| \del\psi|_{p,k} \) over the near region available in the current finite bootstrap slab at time \(t\). The good derivative and the term \(w^\gamma V\) in the conversion from \(\underL(rV) \) to \(r\del V\) are bounded directly by \(C_NA\). Integration of the preceding equation, keeping its favorable zeroth-order term, gives
\be
\label{equa:eight-low-matter-reference-rank-Volterra}
Y_k(t) \leq C_N(A+A^2) +C_Nm\sum_{j<k} \int_{T_0}^t\tau^{-1}Y_j(\tau) \, d\tau.
\ee
The sum is empty at rank zero. For \(k\geq1\), insertion of the lower-rank bounds and \(\int_1^t\tau^{-1+cj/2} \, d\tau\leq2t^{ck/2}/(ck) \) gives a total coefficient at most \(C_Nm/c\). The smallness choice therefore yields \(Y_k(t) \lesssim C_NA t^{ck/2} \). Since \(s_{\mathrm e}^2\simeq t\), this proves the conclusion. All traces of the displaced-cone estimate are taken on past time intervals within the same finite bootstrap slab.

\end{proof}


\subsection{\texorpdfstring{Characteristic estimates for the projected metric}{Characteristic estimates for the projected metric}}
\label{section:eight-good-metric-characteristics}

We use \(\ell=dt-dr\), \(E_a= \del_a+ \omega_a \del_t\), and the family of contractions \(\mathcal A_\omega T= (T(E_a, \del_\beta))_{a, \beta} \). Since \(\ell(E_a)=0\), this family annihilates \(\ell\otimes\ell\). It contains all tensor components having at least one semi-null tangential direction, including the transverse trace-free components.

\begin{proposition}[Characteristic bounds for the projected metric]
\label{lem:eight-good-metric-characteristic}
Under the finite coarse bootstrap and geometric hypotheses of~\autoref{lem:eight-low-matter-characteristic-Lie}, the metric equation and wave gauge imply
\be
\label{equa:eight-good-metric-characteristic}
r w^\gamma| \del\mathcal A_\omega \mathcal L_Zu| \lesssim C_NA s_{\mathrm e}^{\eta_k}, \qquad \mathsf B_{\mathrm{char};p,k}(s_1) \lesssim C_NA, \qquad p\leq N-9, \quad Z\preccurlyeq(p,k).
\ee
Here \(\gamma= (1+ \delta)/2\); the constants may depend on the fixed exponents and \(\coeff_0\), and are independent of the terminal time and \(0<\coeff\leq\coeff_0\). In particular, the trace estimate applies at \(p=p_{\mathbf{low}}-3\).
\end{proposition}

\begin{proof}
\bse

\noindent{\bf 1.} \emph{Quadratic metric terms.} We first calculate the leading quadratic tensor. At a point where \(g= \gMink\), we write a plane-null first jet as \(\del_\alpha g_{\mu\nu} = \ell_\alpha B_{\mu\nu} \), put \(T= \operatorname{tr}_{\gMink}B\), \(I=B:B\), and impose its linear wave constraint. The constraint and Christoffel symbols are
\be
\label{equa:eight-good-metric-plane-null-Christoffel}
\ell^\alpha B_{\alpha \nu} = \tfrac12\ell_\nu T, \qquad \Gamma^\rho_{\mu\nu} = \tfrac12(\ell_\mu B_\nu{}^\rho+ \ell_\nu B_\mu{}^\rho-\ell^\rho B_{\mu\nu}).
\ee
Setting the second metric jet to zero isolates the quadratic polynomial. Differentiation of the inverse metric gives
\be
\label{equa:eight-good-metric-plane-null-contractions}
\begin{aligned}
\del_\rho\Gamma^\rho_{\mu\nu}-\del_\nu\Gamma^\rho_{\mu\rho}&= (I/2-T^2/4) \ell_\mu\ell_\nu, \\
\Gamma^\rho_{\rho a} \Gamma^a_{\mu\nu}-\Gamma^\rho_{\nu a} \Gamma^a_{\mu\rho}&= (T^2/8-I/4) \ell_\mu\ell_\nu, \qquad \del_\mu\Gamma_\nu=-(T^2/4-I/2) \ell_\mu\ell_\nu.
\end{aligned}
\ee
Consequently the reduced quadratic term is \((T^2/4-I/2) \ell\otimes\ell\), and its \(\mathcal A_\omega \)-projection vanishes. For a general first jet, \(\del_\alpha= \ell_\alpha \del_t+E_\alpha \), with \(E_0=0\). The linear constraint has a bounded right inverse on symmetric tensors, while the wave-gauge identity is
\be
\label{equa:eight-good-metric-gauge-defect}
(\gMink)^{\alpha \beta}(\del_\alpha h_{\beta \nu}-\tfrac12\del_\nu h_{\alpha \beta})=-H^{\alpha \beta}(\del_\alpha h_{\beta \nu}-\tfrac12\del_\nu h_{\alpha \beta}).
\ee
Correcting the plane-null jet by the bounded right inverse of the linear constraint and expanding the quadratic polynomial therefore proves
\be
\label{equa:eight-good-metric-projected-quadratic}
| \mathcal A_\omega \mathbb F| \lesssim |E h| \,| \del h|+ |H| \,| \del h|^2.
\ee
The same bound with the finite ordered partitions holds after tensor Lie commutation: translations and boosts preserve \(\gMink\), and the flat contractions are tensorial. The projection commutator obeys
\be
\label{equa:eight-good-metric-projection-error}
|[\Box_g, \mathcal A_\omega]h_Z| \lesssim r^{-1}|E h_Z|+r^{-2}|h_Z|+r^{-1}|H| \,| \del h_Z|.
\ee
This follows because the spatial derivatives of its coefficients are tangent to spheres. The stationary reference terms have an additional radial decay.
\ese

\bse
\noindent{\bf 2.} \emph{Potential and derivative terms of the effective field.} The potential is controlled directly from the massive equation. On \(r\leq5t/4\), the near-cone Sobolev estimate based on the mass term costs three commutations. Thus \(p+3\leq N-5\) gives \(|Z\fieldvarphi| \lesssim At^{-1+ \delta/2}w^{-\mu} \). Writing \(\Box_g\fieldPhi-(3\coeff)^{-1} \fieldPhi =G\), flat commutation yields
\be
\label{equa:eight-good-metric-mass-reconstruction}
Z\fieldvarphi =3\sqrt{\coeff} \bigl(\Box_{\gMink}Z\fieldPhi+Z(H^{\alpha \beta} \del_\alpha \del_\beta \fieldPhi)-ZG\bigr), \qquad |Z\fieldvarphi| \lesssim C_{\coeff_0}At^{-3/2 + \delta}w^{1-\mu}.
\ee
Indeed, the flat contraction has the bound \((w/t)| \del^2Z\fieldPhi|+\allowbreak t^{-1}(| \del LZ\fieldPhi|+ | \del Z\fieldPhi|) \). The direct high effective-field estimate keeps its factor \(s_{\mathrm e} \simeq t^{1/2} \) and costs at most \(p+6\leq N\). Its metric correction and \(G\) are bounded by \(A^2t^{-3/2 + \delta}w^{1-\lambda-\mu} \) and \(A^2t^{-2 + \delta}w^{-2\mu} \), respectively. Thus the projected derivative source and the potential source satisfy
\be
\label{equa:eight-good-metric-effective-field-sources}
| \mathcal A_\omega(d\fieldPhi\otimes d\fieldPhi)|_p\lesssim A^2t^{-5/2 + \delta}w^{1-2\mu}, \qquad |g(\fieldvarphi)^2|_p\lesssim A^2t^{-5/2 +3\delta/2}w^{1-2\mu}.
\ee
In the first product, the good derivative is paired with a factor at the closing differentiation level. In the second, the reconstruction is applied to the higher-order factor. The projected matter derivative source is stronger.
\ese

\noindent{\bf 3.} \emph{Projected Hessians away from the cone.} With \(a_\lambda \) from \eqref{equa:eight-displaced-cone-exponents}, we put \(V_Z= \mathcal A_\omega \mathcal L_Zu\). In a principal commutator the nonempty coefficient operator leaves a field operator of order at most \(p-1\). Since \(p-1+5\leq N-5\), its effective-field source factors belong to the closing level. On \(w\geq4t^{a_\lambda}\), the exterior wave-Hessian reduction gives the following estimates; the unprojected field has order at most \(p-1\)
\be
\label{equa:eight-good-metric-outer-Hessians}
|\del^2Zu|\lesssim At^{-1+\delta/2}w^{-1-\lambda}, \qquad |\del^2V_Z|+\frac{t}{w}|\del E V_Z|\lesssim At^{-1+\delta/2}w^{-1-\lambda}.
\ee
For the projected Hessian the order may equal \(p\): the quadratic, projection, and effective-field sources above replace the unprojected equation. The coefficient \((t/w)|H| \lesssim At^{-\delta/2} \) is absorbed. The effective-field-source ratio to the desired Hessian bound is at most \(A t^{-1/2 + \delta}w^{1+ \lambda-2\mu} \), which decays. The good-Hessian bound follows from \(tE_aV=L_aV+q\omega_a \del_tV\).

\bse
\noindent{\bf 4.} \emph{Characteristic integration and induction on the boost rank.} We apply the weighted characteristic identity \eqref{equa:eight-low-matter-weighted-transport} to each \(V_Z\). The angular term costs at most \(p+9\leq N\). The angular and linear projection terms have an integrable majorant \(C_NAt^{-1-e_a} \), where \(e_a= \lambda-1/2-\delta>0\). The effective-field terms, after multiplication by \(rw^\gamma \), are bounded by \(C_NA^2t^{-3/2 +2\delta} \). For the principal good-Hessian terms we use the direct estimate in \(w\leq4t^{a_\lambda} \) and \eqref{equa:eight-good-metric-outer-Hessians} elsewhere. Their respective positive exponent gaps are
\be
\label{equa:eight-good-metric-good-source-gaps}
1-\delta-a_\lambda(\gamma+2-2\lambda)>\tfrac{67}{96}, \qquad \min\{1-\delta,2\lambda-\gamma-\delta \}>0.
\ee
For a dynamic bad coefficient, the \(w^{1/2} \) term of~\autoref{lem:eight-displaced-cone-coefficient} is kept. With \(\chi_\lambda \) from \eqref{equa:eight-displaced-cone-exponents}, its collar gap and the outer gap are
\be
\label{equa:eight-good-metric-bad-source-gaps}
\chi_\lambda-\tfrac\delta2-a_\lambda(\tfrac12 + \gamma-\lambda)= \lambda-\tfrac12-\frac{\delta(1+ \delta/2)} \lambda>\tfrac7{96}, \qquad a_\lambda(2\lambda-\gamma)-\delta= \frac{\delta(\lambda-\gamma)} \lambda>0.
\ee
The frame and density errors have respectively the positive collar gaps \(1-\delta-a_\lambda(\gamma+2-2\lambda) \) and \(1-3\delta/2-a_\lambda(\gamma+2-3\lambda) \). A translated reference coefficient is \(\Ocal(mr^{-2}) \); a boost-only reference coefficient is \(\Ocal(mr^{-1}) \) and multiplies a strictly lower-rank projected field. Its Hessian is the first derivative of a translation operator of total order at most \(p\). Thus the characteristic supremum satisfies
\be
\label{equa:eight-good-metric-rank-transport}
Y_k(t) \leq C_N(A+A^2) +C_Nm\sum_{j<k} \int_{T_0}^t\tau^{-1}Y_j(\tau) \, d\tau.
\ee
The initial and outer-boundary values are controlled by the direct bounds. The geometric argument and finite-slab restriction are the same as for the matter characteristic. The rank induction, with \(C_Nm\leq c\), proves the first assertion.
\ese

\noindent{\bf 5.} \emph{Traces at the interface.} Finally, on \(q=1\), the frame conversion is
\be
\label{equa:eight-good-metric-interface-frame}
\delH_a=E_a-\frac qt\omega_a \del_t, \qquad \mathcal L_{L_a} \ell=-\omega_a \ell-\frac qr\Pi_{ab} \, dx^b.
\ee
Boost derivatives preserve the factors \(q/t\) and \(q/r\), while their translations are \(\Ocal(t^{-1}) \) there. Thus each scalar-commuted component in \(\mathscr G(u) \) is a bounded linear combination of tensorially commuted good contractions of no higher order and rank, plus \(\Ocal(t^{-1}) \) Cartesian metric components. Cartesian derivatives of the bounded coefficients are \(\Ocal(t^{-1}) \). The characteristic trace of the error, after multiplication by \(r\), is bounded by \(C_NAt^{-1+ \delta/2} \). The term \(-V\) in \(\underL(rV)=r\underL V-V\) is bounded directly. This proves the bound for \(\mathsf B_{\mathrm{char};p,k} \).
\end{proof}


\subsection{\texorpdfstring{Traces at the interface from exterior energy bounds}{Traces at the interface from exterior energy bounds}}
\label{section:eight-interface-rank-estimates}

\paragraph{Exterior energies and traces at the interface}

The following estimates extract traces from positive exterior energy bounds on the prescribed foliation. At this stage the characteristic arguments require the geometric hypothesis \(H_N\leq0\) in \(t-1\leq r\leq5t/4\). Proposition~\ref{prop:eight-mobile-characteristics} will replace this hypothesis in the global argument by the geometry of the expanding interface.

We specify the quantities required by the trace argument. We let \(U^{\mathrm{ext}}_{p,k}\) denote the square root of the exterior integral of \(w^{2\mu}Q^{\mathrm{unif}}[Z\fieldphi](N_s,\del_t)\), summed over \(Z\preccurlyeq(p,k)\). With \(n=N-5\) and \(k'=\min\{k,n\}\), we set
\be
\label{equa:conditional-trace-energy-levels}
e_k(s)\coloneqq\bigl(\Fenergy^{\ME,N,k}_\lambda(s,u)\bigr)^2+\bigl(\Fenergy^{\ME,N,k}_\mu(s,\psi)\bigr)^2+\bigl(U^{\mathrm{ext}}_{n,k'}(s)\bigr)^2+s^{-2}\bigl(U^{\mathrm{ext}}_{N,k}(s)\bigr)^2.
\ee
We let \(\mathcal F_k\) and \(\mathcal G_k\) be the sums of the nonnegative fluxes and positive bulk integrals of these four current identities, including all good-derivative terms. The order-\(N\) effective-field contributions are integrated with weight \(s^{-2}\); the other three currents keep weight one. We use
\be
\label{equa:conditional-positive-trace-energy}
\mathcal Y_k(s_1)^2\coloneqq e_k(s_1)+\mathcal F_k(s_1)+\mathcal G_k(s_1)+4\int_{s_0}^{s_1}s^{-3}\bigl(U^{\mathrm{ext}}_{N,k}(s)\bigr)^2\,ds,\qquad a_0^2\coloneqq e_N(s_0).
\ee
The trace estimate is conditional on \(\mathcal Y_k(s)\leq C_Na_0s^{d_k}\). The adapted positive energy in Section~\ref{section:small-mass-moving-interface} controls the corresponding quantities on the expanding domains. For the present extraction of traces, the exponents have the form
\be
\label{equa:eight-interface-exterior-rank-exponents}
d_k\coloneqq \eta_k+(k+1)D\varepsilon_{\mathrm g}, \qquad \varepsilon_{\mathrm g}>0,
\ee
with \(D\geq2\) fixed. The unnormalized metric, matter, and closing effective-field fluxes, together with radial and characteristic estimates, give the interface bounds used in Part~II. The required exponent margins are
\be
\label{equa:eight-exterior-interface-margins}
d_N<\delta, \qquad (p_{\mathbf{low}} +1)D\varepsilon_{\mathrm g}<9C_{\mathrm{rk}} \theta, \qquad (p_{\mathbf{high}} +1)D\varepsilon_{\mathrm g}<12\rho+2C_{\mathrm{rk}} \theta.
\ee
These inequalities imply \(d_{p_{\mathbf{low}}}<\eta_{\Lscr}\) and \(d_k<\nu_k<1\) for \(k\leq p_{\mathbf{high}}\). In the global application, Proposition~\ref{prop:eight-mobile-coupled-estimate} gives \(\varepsilon_{\mathrm g}=C_NA\), so the margins follow by reducing the initial amplitude at fixed rank weights.

\begin{proposition}[Interface bounds from the exterior energies]
\label{lem:eight-rank-local-interface-elimination}
Assume the finite coarse bootstrap, the geometric sign \(H_N\leq0\) on \(t-1\leq r\leq5t/4\), the positive exterior bound \(\mathcal Y_k(s)\leq C_Na_0s^{d_k}\) with~\eqref{equa:conditional-positive-trace-energy} and~\eqref{equa:eight-interface-exterior-rank-exponents}, the margins~\eqref{equa:eight-exterior-interface-margins}, and the prescribed interior initial bounds. Then the interface estimates~\eqref{equa:seven-interface-data-bounds} hold. In particular, the flux and sphere bounds have amplitude \(C_Na_0\), and
\be
\label{equa:eight-rank-local-interface-ledger}
\max_{k\leq p_{\mathbf{high}}} \mathcal I_{\mathrm h,k}^{\mathrm{rob}}(s_1) \leq C_{N, \mathrm{gaps}}(A_0 +a_0), \qquad \max_{k\leq p_{\mathbf{low}}} \mathsf B_{00;k}(s_1) \leq C_N(A_0 +a_0).
\ee
For \(p\leq p_{\mathbf{low}}-3\), the strong massive ray data satisfies \(\mathsf B_{\mathrm{ray};p,k}^{\fieldphi}(s_1) \leq C_{N,s_0, \coeff_0}a_0\). The constants are independent of \(C_1\), \(s_1\), and \(0<\coeff\leq\coeff_0\).
\end{proposition}

\begin{proof}
\bse

\noindent{\bf 1.} \emph{Fluxes and sphere traces.} For \(p\leq p_{\mathbf{high}} \), the three unnormalized exterior fluxes are nonnegative terms in \(\mathcal Y_k^2\). Their coordinate normalization gives \(\mathcal F_{p,k}^{\Lscr,\mathrm{unif}}(s_1;s_0)\lesssim a_0s_1^{d_k}\), and rescaling multiplies the aggregate fixed-parameter bound by at most \(1+\coeff^{-1/2}\). The fixed-hypersurface radial estimate \eqref{equa:seven-e-exterior-sphere-trace} and \(R_s^{1/2} \simeq s\) also give
\be
\label{equa:eight-rank-local-sphere-traces}
\mathsf T_{2;p,k}[u](s) + \mathsf T_{2;p,k}[\psi](s) \leq C_Na_0s^{-1+d_k}, \qquad \mathsf T_{2;p,k}[\fieldvarphi](s) \leq C_Na_0s^{-1+d_k} \quad(p+1\leq p_{\mathbf{high}}).
\ee
For the second estimate, the radial derivative of \(Z\fieldvarphi\) on the hypersurface is a bounded combination of its four translation derivatives, which are controlled by the mass term at order \(p+1\) and the same boost rank. This covers \(p=p_{\mathbf{low}} \). The inverse-metric conversion keeps the contribution of the reference metric outside the Hardy estimate and gives \(\mathsf B^H_{2;p_{\mathbf{high}},k}(s) \leq C_N(A_0 +a_0s^{d_k}) \). Direct integration therefore yields
\be
\label{equa:eight-exterior-interface-time-traces}
\begin{aligned}
\sup_{s_0\leq\sigma \leq s_1} \sigma^{-\nu_k} \| \tau^{-1/2} \mathsf B^H_{2;p_{\mathbf{high}},k}(\tau) \|_{L^2(s_0, \sigma)}& \leq {C_N(A_0 +a_0) \over\sqrt{2\nu_k}}, \\
\sup_{s_0\leq\sigma \leq s_1} \sigma^{-\nu_k} \Bigl(\int_{s_0}^{\sigma} \kappa_u^\star(\tau) \mathsf T_{2;p_{\mathbf{high}},k}[u](\tau) \, d\tau+ {1\over\sigma} \int_{s_0}^{\sigma} \mathsf T_{2;p_{\mathbf{high}},k}[u](\tau) \, d\tau\Bigr)& \leq C_Na_0(1+A+A_0).
\end{aligned}
\ee
The first line uses \(d_k\leq\nu_k\); the second uses \(\tau^{-1+d_k} \leq1\) and \eqref{equa:seven-metric-integrable-kernel-L1}. The flux bounds, these time traces, and the prescribed initial estimates prove the robust inequality. Exact rescaling gives the fixed-\(\coeff\) conclusion.
\ese

\noindent{\bf 2.} \emph{The metric pointwise trace.} We apply the proof of~\autoref{lem:seven-pointwise-metric-interface-bulk} with the metric energy and good bulk controlled by \(\mathcal Y_N\). Every dynamic density product has a factor measured by this improved norm; its remaining factors are bounded by the coarse small amplitude. The contribution keeps its separate bound by \(A_0\). The physical time trace thus becomes
\be
\label{equa:eight-exterior-improved-metric-time-trace}
\|ZH^{\H00}(T, \cdot) \|_{L^2(r=T-1)} \leq C_N\bigl(A_0 +a_0T^{1/2-\lambda+d_N/2} +a_0T^{-1/2} \bigr), \qquad \ord Z\leq p_{\mathbf{high}}-1.
\ee
On \(q=1\), the identity \(\Omega_{ab} = (x_a/t)L_b-(x_b/t)L_a \) converts the two sphere rotations into two boost operators with bounded coefficients. The time trace adds one operator, so the largest order is \(p_{\mathbf{low}} +3\leq p_{\mathbf{high}} \). Since \(d_N<\delta<2\lambda-1\), the dynamic term in \eqref{equa:eight-exterior-improved-metric-time-trace} decays. The factors \(s^2/R_s\) are bounded and \(s^{-\eta_j} \leq1\), proving the bound for \(\mathsf B_{00;k} \).

\bse
\noindent{\bf 3.} \emph{Initial estimates for the effective field along rays.} We use the exterior closing energies at every operator \(Q\) of order at most \(p+1\), where \(p\leq p_{\mathbf{low}}-3\). Flat commutation of the massive equation gives the identity
\be
\label{equa:eight-exterior-massive-e-reconstruction}
Q\fieldvarphi =3\sqrt\coeff\bigl(\Box_{\gMink}Q\fieldPhi+Q(H^{\alpha \beta} \del_\alpha \del_\beta \fieldPhi)-QG_\Phi\bigr), \qquad G_\Phi = (\mathcal S+W_\coeff)/3.
\ee
At \(q=1\), we have \(w=1\) and \(t\simeq s^2\). The flat contraction is bounded by \(Ct^{-1}(| \del^2Q\fieldPhi|+ | \del LQ\fieldPhi|+ | \del Q\fieldPhi|) \). The exterior scalar first-jet estimate costs five operators and the Hessian costs six, so the largest effective-field order is \(p+7\leq N-5\). The direct closing bounds, the ordered metric products, and the source \(G_\Phi\) consequently give
\be
\label{equa:eight-exterior-massive-e-values}
|Q\fieldvarphi| \leq C_{N, \coeff_0}a_0t^{-2}s^{2\delta} \leq C_{N, \coeff_0}a_0s^{-4+2\delta}, \qquad \ord Q\leq p+1.
\ee
In each nonlinear product we use one improved factor with amplitude \(a_0\) and the remaining small coarse factors. The potential term is quadratic in \(\fieldvarphi\), and the smooth coefficient bounds introduce no negative power of \(\coeff\). Applying \eqref{equa:eight-exterior-massive-e-values} to translations and boosts, with \(\fieldPhi = \sqrt\coeff\, \fieldvarphi\), gives
\be
\label{equa:eight-exterior-massive-e-velocity}
| \fieldPhi|_{p+1,k+1} \leq C_{N, \coeff_0}a_0s^{-4+2\delta}, \qquad | \Lcal Z\fieldPhi| \leq C_{N, \coeff_0}a_0s^{-5+2\delta}, \quad Z\preccurlyeq(p,k).
\ee
Indeed, \(\Lcal= (s/t) \del_t+x^a(st)^{-1}L_a \) has coefficients bounded by \(Cs^{-1} \) on the interface. Substitution in \eqref{equa:seven-strong-massive-ray-e-definition} gives
\be
\label{equa:eight-exterior-strong-ray-en}
s^{-\eta_k} \widehat{\mathcal B}_Z(s) \leq C_{N,s_0, \coeff_0}a_0s^{-1/2 +2\delta-\eta_k} \leq C_{N,s_0, \coeff_0}a_0.
\ee
The last inequality uses \(2\delta<1/2\). The initial-hyperboloid portion is controlled by the prescribed initial norms.

\noindent{\bf 4.} \emph{Initial estimates for the matter and projected metric along characteristics.} We apply the proofs of Propositions~\ref{lem:eight-low-matter-characteristic-Lie} and \ref{lem:eight-good-metric-characteristic} after the  exterior estimate. Every direct hypersurface or good-bulk bound required there, through order \(N\), is then bounded by \(C_Na_0s^{d_N} \leq C_Na_0s^\delta \), with the factor \(s\) kept on the top effective-field energy. The linear angular and initial terms have amplitude \(C_Na_0\). Each nonlinear product uses one improved factor and the remaining coarse factors, giving amplitudes \(C_NAa_0\) or \(C_NA^2a_0\). The displaced dynamic metric coefficient has the same improvement because its radial and time traces use the improved metric hypersurface and good-bulk norms. The reference contribution remains the triangular term \(C_Nm\sum_{j<k} \int\tau^{-1}Y_j(\tau) \, d\tau\). The same finite-rank induction therefore gives characteristic amplitudes \(C_Na_0\). Together with the preceding metric and massive traces, these prove \eqref{equa:seven-interface-pointwise-data}. Finally, \(a_0\leq C_{\mathrm{data}}A_0\) gives the data-amplitude bounds with constants independent of \(C_1\).
\ese
\end{proof}

The simultaneous rank induction is proved once, in Proposition~\ref{prop:eight-mobile-coupled-estimate}, with all lower-rank energy and bulk contributions kept.

For fixed \(\coeff>0\), define \(\Eenergy^{\mathrm{tot,fix}}_{p,k} \) by replacing the rescaled effective-field current with its singular normalization in the same energy and flux sum. Exact rescaling gives
\be
\label{equa:eight-augmented-fixed-scale}
\Eenergy^{\mathrm{tot,fix}}_{p,k} \leq \max\{1, \coeff^{-1/2} \} \Eenergy^+_{p,k}.
\ee
At the high level the interior effective-field energy still has the factor \(s^{-1} \); removing it costs \(s^{1/2} \). This finite-parameter conversion preserves the positive flux and leaves its singular constant explicit.

\paragraph{Orientation and normalization of interface currents}

\begin{proposition}[Oriented interface currents]
\label{proposition:eight-interface-traces-fluxes}
On a finite slab of a smooth Cauchy development, the two restrictions of each geometric current agree on the interface. In the  coordinate normalization, one has
\be
\label{equa:eight-interface-oriented-flux}
{ \begin{aligned}
d\Sigma_\alpha^{\Lscr, \mathrm{ext}} &=-d\Sigma_\alpha^{\Lscr, \mathrm{int}}, \\
\widehat{\mathsf F}_{\bullet}^{\Lscr, \mathrm{ext},p,k} &=-\widehat{\mathsf F}_{\bullet}^{\Lscr, \mathrm{int},p,k}, \qquad \bullet\in\{\mathrm w, \psi,(\mathrm{eff}, \mathrm{sing}), (\mathrm{eff}, \mathrm{unif}) \}, \\
\widehat{\mathsf F}_{\mathrm{eff}}^{\mathrm{unif}, \Lscr, \mathrm{side},p,k} &= \coeff\, \widehat{\mathsf F}_{\mathrm{eff}}^{\mathrm{sing}, \Lscr, \mathrm{side},p,k}, \qquad \mathrm{side} \in\{\mathrm{int}, \mathrm{ext} \}.
\end{aligned} }
\ee
These identities hold at every available commutation order. Once Proposition~\ref{proposition:eight-c-top-exterior} has provided the exterior energy bounds, the conditional estimates proved in Proposition~\ref{lem:eight-rank-local-interface-elimination} will give the quantitative flux and sphere-trace bounds.
\end{proposition}

\begin{proof}
The multipliers agree on \(\Lscr\), and the exterior weight is one there. The field and its derivatives are restrictions of the same smooth solution. The two current integrands therefore agree before orientation is imposed. Their outward directed surface elements have opposite signs, proving the first two identities. The rescaled effective-field stress tensor is exactly \(\coeff\) times the singular tensor, including its zeroth-order term. This proves the last identity. Consequently, the square-root flux norms differ by the factor \(\coeff^{1/2} \). Positivity and quantitative bounds use the kept exterior flux in~\autoref{lem:eight-augmented-interface-energy}.
\end{proof}


\section{\texorpdfstring{Exterior equations and weighted energies}{Exterior equations and weighted energies}}
\label{section-13}


\subsection{\texorpdfstring{Effective-field commutators and currents}{Effective-field commutators and currents}}
\label{section-13-1}

\paragraph{Purpose}

We derive the radial--angular equations and the singular and rescaled effective-field currents. We then specify the weighted exterior norms and obtain their direct decay estimates.

\paragraph{Radial--angular decomposition of the exterior equations}

Whenever a radial second derivative appears, we separate the good null and angular terms before solving for it. This avoids a derivative loss at the top order and keeps the long-range Schwarzschild coefficient on the left-hand side.

\begin{lemma}[Radial--angular decomposition]
\label{lem:eight-radial-angular-decomposition}
Let \(v\) be any commuted wave component.  In the Minkowski null frame \((L, \underL,r^{-1} \Omega_A) \), write \(g^{\alpha \beta} = (g^\star)^{\alpha \beta} + \widehat H^{\alpha \beta} \), where \(g^\star \) is the fixed reference end.  In the Euclidean region one can write
\be
\label{equa:eight-radial-angular-decomposition}
\begin{aligned}
\Box_gv&=-a^{\underL L} \, \underL Lv +a^{LL}L^2v +a^{\underL\underL} \underL^2v+a^{LA}L(r^{-1} \Omega_Av) \\
& \quad +a^{\underL A} \underL(r^{-1} \Omega_Av) +a^{AB}r^{-2} \Omega_A\Omega_Bv+r^{-1}b^\alpha \del_\alpha v+ \mathcal R_g^{(1)}[v].
\end{aligned}
\ee
Here \(a^{\underL L} =a_\star^{\underL L} +\Ocal(\widehat H) \), where \(a_\star^{\underL L} \) is bounded above and below by positive constants. Moreover, the reference coefficients satisfy
\be
\label{equa:eight-reference-null-coefficients}
a_\star^{LL},a_\star^{\underL\underL} =\Ocal(m/r), \qquad a_\star^{LA} =a_\star^{\underL A} =0, \qquad a_\star^{AB}-\gamma^{AB} =\Ocal(m/r),
\ee
and every dynamic principal coefficient differs from its reference value by \(\Ocal(\widehat H) \). The remainder \(\mathcal R_g^{(1)}[v] \) contains only first-order terms and satisfies
\be
\label{equa:eight-radial-angular-first-order-remainder}
| \mathcal R_g^{(1)}[v]| \lesssim \bigl(| \del\widehat H|+r^{-1}| \widehat H| \bigr)| \del v| +r^{-2}| \widehat H| \,|v|,
\ee
together with the explicitly prescribed reference lower-order coefficients. After commutation, the coefficient of the mixed null derivative remains invertible. The square-null and null-angular second derivatives remain explicit and are treated by the null decomposition and the rank induction.

\end{lemma}
\begin{proof}
We decompose in the Minkowski null frame, while keeping \(g^\star\) in the fixed coefficients.  We split the inverse metric into its mixed-null, square-null, null-angular, and angular blocks. The principal part is the sum of the six second-order terms in \eqref{equa:eight-radial-angular-decomposition}; the frame connection coefficients contain only one derivative of \(v\). The angular derivatives are \(r^{-1} \Omega_A\) and therefore give the factors \(r^{-1} \) and \(r^{-2} \) in that formula. By \eqref{equa:inverse-metric-dynamic-identity}, the dynamic parts of all principal coefficients are \(\Ocal(\widehat H) \), while their differentiated contributions are bounded by \(\Ocal(\del\widehat H) +\Ocal(r^{-1} \widehat H) \). The fixed reference coefficients give \eqref{equa:eight-reference-null-coefficients} and the prescribed lower-order terms. We commute with \(Z\) and expand by the Leibniz rule. If a derivative falls on a principal coefficient, the resulting term contains a lower-order coefficient and a second derivative of a field of lower commutation order or lower boost rank. The pointwise smallness of \(\widehat H\) keeps the coefficient of the mixed null derivative bounded away from zero. The remaining square-null and null-angular blocks have the triangular structure in the statement.
\end{proof}

\begin{proposition}[Commutation before radial decomposition]
\label{proposition:eight-angular-commutators}
Let \(Z=X_1\cdots X_d\) be an ordered operator of translations and boosts. For a subset \(I\subset\{1, \ldots,d\} \), let \(Z_I\) and \(\mathcal L_I\) be the corresponding ordered operators of scalar and tensor Lie derivatives. In wave coordinates, one has
\be
\label{equa:eight-angular-commutator-bound}
[\Box_g,Z]v=-\sum_{\varnothing\ne I\subset\{1, \ldots,d\}}(\mathcal L_Ig^{-1})^{\alpha \beta} \del_\alpha \del_\beta Z_{I^c}v.
\ee
Every coefficient on the right is a derivative of \(H=g^{-1}-\gMink^{-1} \). The radial--angular decomposition is applied to this identity after the full operator has been commuted. On the interface, each sphere rotation costs one boost operator; on the whole exterior hypersurface it costs at most two.
\end{proposition}
\begin{proof}
\bse
For an affine vector field \(X\) and a contravariant tensor \(T\), direct differentiation gives
\be
\label{equa:eight-affine-tensor-commutator}
[T^{\alpha \beta} \del_\alpha \del_\beta,X]v=-(\mathcal L_XT)^{\alpha \beta} \del_\alpha \del_\beta v.
\ee
Induction on the ordered operator gives \eqref{equa:eight-angular-commutator-bound}. Translations and boosts are Minkowski Killing fields, so every nonempty Lie operator annihilates \(\gMink^{-1} \). The separate angular and radial parts of the flat wave operator satisfy
\be
\label{equa:eight-flat-radial-angular-cancellation}
[Z,r^{-2} \Delta_{\mathbb S^2}]=-[Z,-\del_t^2 + \del_r^2 +2r^{-1} \del_r].
\ee
Their sum vanishes, as required by \([Z, \Box_{\gMink}]=0\). For the angular derivative count, we use \eqref{equa:eight-exterior-Sobolev-near-rotations} on \(r=t-1\), and \(\Omega_{ab} = [L_a,L_b] \) globally.~\autoref{lem:eight-exterior-weighted-sobolev} includes the additional radial differentiation.
\ese
\end{proof}

\paragraph{Singular and rescaled effective-field identities}

The effective-field current has a fixed-parameter and a uniform normalization. Multiplication of its entire balance by \(\coeff\) rescales the energy, source, flux, and bulk together. The nonlinear estimates keep the source pairing and the signed deformation terms.

\begin{proposition}[\texorpdfstring{Effective-field current normalizations}{Effective-field current normalizations}]
\label{proposition:eight-effective-field-normalized-energy}
Suppose that \(3\coeff \, \Box_gZ\fieldphi-Z\fieldphi =F_Z^{\fieldphi} \). The two rescaled sources are
\be
\label{equa:eight-effective-field-source-normalizations}
F_Z^{\fieldPhi}\coloneqq \coeff F_Z^{\fieldphi}, \qquad F_Z^{\fieldvarphi}\coloneqq \sqrt\coeff F_Z^{\fieldphi}.
\ee
The three scalar equations are exactly equivalent:
\be
\label{equa:eight-effective-field-three-normalized-equations}
3\coeff \, \Box_gZ\fieldphi-Z\fieldphi =F_Z^{\fieldphi}, \qquad 3\coeff \, \Box_gZ\fieldPhi-Z\fieldPhi =F_Z^{\fieldPhi}, \qquad 3\coeff \, \Box_gZ\fieldvarphi-Z\fieldvarphi =F_Z^{\fieldvarphi}.
\ee
The coordinate-current energies and all their signed source, deformation, flux, and bulk contributions obey the same rescaling. In particular,
\be
\label{equa:eight-effective-field-source-normalization-energy}
\mathcal E_{\mathrm{unif}} = \coeff\mathcal E_{\mathrm{sing}}, \qquad \mathcal F_{\mathrm{unif}} = \coeff\mathcal F_{\mathrm{sing}}, \qquad \mathcal G^\pm_{\mathrm{unif}} = \coeff\mathcal G^\pm_{\mathrm{sing}}.
\ee
On an exterior slab, the singular balance is
\be
\label{equa:eight-effective-field-singular-energy}
\mathcal E_{\mathrm{sing}}(s) + \mathcal F_{\mathrm{sing}}(s) + \mathcal G^+_{\mathrm{sing}}(s)= \mathcal E_{\mathrm{sing}}(s_0)-2\int w^{2\mu}F_Z^{\fieldphi} \del_tZ\fieldphi\, dV_g+ \mathcal D_{\mathrm{sing}}(s) + \mathcal G^-_{\mathrm{sing}}(s).
\ee
Its uniform counterpart is
\be
\label{equa:eight-effective-field-uniform-energy}
\mathcal E_{\mathrm{unif}}(s) + \mathcal F_{\mathrm{unif}}(s) + \mathcal G^+_{\mathrm{unif}}(s)= \mathcal E_{\mathrm{unif}}(s_0)-2\int w^{2\mu}F_Z^{\fieldphi} \del_tZ\fieldPhi\, dV_g+ \mathcal D_{\mathrm{unif}}(s) + \mathcal G^-_{\mathrm{unif}}(s).
\ee
Here the slab integrals, outward interface flux, and positive--negative bulk splitting are those of Proposition~\ref{proposition:eight-weighted-current-identity}. The source norm occurring in the uniform pairing has the alternative representations
\be
\label{equa:eight-effective-field-uniform-source-equivalence}
\|J\zeta^{-1}w^\mu F_Z^{\fieldphi} \|_2= \coeff^{-1} \|J\zeta^{-1}w^\mu F_Z^{\fieldPhi} \|_2= \coeff^{-1/2} \|J\zeta^{-1}w^\mu F_Z^{\fieldvarphi} \|_2.
\ee
Cauchy--Schwarz in the singular pairing introduces \(\coeff^{-1/2} \), whereas the uniform pairing introduces no negative power of \(\coeff\).
\end{proposition}
\begin{proof}
The singular stress tensor has derivative coefficient \(3\coeff\) and mass coefficient one. Its divergence is \(F_Z^{\fieldphi} \, dZ\fieldphi\). Multiplication by the current factor \(-2w^{2\mu} \del_t\) gives \eqref{equa:eight-effective-field-singular-energy}. The uniform stress tensor is exactly \(\coeff\) times the singular tensor, including its mass term. Multiplying the entire balance by \(\coeff\) gives \eqref{equa:eight-effective-field-uniform-energy}; the source pairs with \(\del_tZ\fieldPhi\). The kinetic energy controls \(\sqrt\coeff\, \zeta \del Z\fieldphi\) or \(\zeta \del Z\fieldPhi\), respectively. Together with \(dV_g=j_gJ\, dx\, ds\), this proves the two source coefficients. The source, deformation, and adverse bulk are subsequently estimated together with the other field identities in Proposition~\ref{proposition:eight-c-top-exterior}.
\end{proof}

\begin{lemma}[Source in the commuted effective-field equation]
\label{lem:eight-effective-field-source-allocation}
With the convention for ordered compositions of Proposition~\ref{proposition:eight-angular-commutators}, the source in the unscaled commuted equation is
\be
\label{equa:eight-effective-field-source-allocation}
F_Z^{\fieldphi} =Z\mathcal S+ZW_\coeff-3\sum_{\varnothing\ne I\subset\{1, \ldots, \ord Z\}}(\mathcal L_Ig^{-1})^{\alpha \beta} \del_\alpha \del_\beta Z_{I^c} \fieldPhi.
\ee
The nonlinearities are \(\mathcal S=-8\pi e^{-\fieldPhi}g^{-1}(d\psi,d\psi) \) and \(W_\coeff= (\fieldvarphi)^2\mathcal W_\coeff(\fieldPhi) \). Under the  exterior bootstrap, the potential source satisfies, for the closing and auxiliary top orders,
\be
\label{equa:eight-effective-field-source-allocation-bound}
\sum_{Z\preccurlyeq(p,k)} \|J\zeta^{-1}w^\mu ZW_\coeff\|_2\lesssim C_{N, \coeff_0}A s^{-1}U^{\mathrm{ext}}_{p,k}(s), \qquad p\in\{N-5,N\}.
\ee
The principal source and the matter trace are estimated in their energy pairings, with the independently assigned good-bulk fractions of Section~\ref{section:exterior-coupled-energies}. Their sum enters the integrated rank estimate of Proposition~\ref{proposition:eight-c-top-exterior}.
\end{lemma}
\begin{proof}
We commute \(3\coeff \, \Box_g\fieldphi-\fieldphi = \mathcal S+W_\coeff\), apply \eqref{equa:eight-angular-commutator-bound}, and use \(\coeff\del^2\fieldphi = \del^2\fieldPhi\). This proves the source formula. In the potential, one differentiated effective-field factor is placed in the weighted mass norm and a lower factor satisfies \(|Z_{\mathrm l} \fieldvarphi| \lesssim C_{N, \coeff_0}A/(rw^{(1+ \delta)/2}) \), as proved in \eqref{equa:eight-low-effective-stress-coefficient}. The coefficient obeys \(J\zeta^{-1}/r\lesssim s^{-1} \). A derivative on \(\mathcal W_\coeff(\fieldPhi) \) is controlled by its uniformly bounded rescaled derivatives and the mass norm of \(\fieldPhi = \sqrt\coeff\, \fieldvarphi\); the extra lower factors are small. Summing the finite Leibniz terms proves \eqref{equa:eight-effective-field-source-allocation-bound}. The matter trace keeps its flat null contraction and metric corrections. Its high good derivative is treated by a matter good-bulk fraction, while the principal effective-field commutators use Lemmas~\ref{lem:eight-massive-principal-commutator} and \ref{lem:eight-massive-top-principal}.
\end{proof}


\subsection{\texorpdfstring{Matter currents and traces at the interface}{Matter currents and traces at the interface}}
\label{section-13-2}

\paragraph{\texorpdfstring{Matter current and trace at the interface}{Matter current and trace at the interface}}

The matter equation determines the factor \(e^{-\fieldPhi}\) in its energy current. At order zero this factor gives the exchange with the effective field; differentiated coupling terms remain at positive orders.

\begin{lemma}[\texorpdfstring{Weighted matter current}{Weighted matter current}]
\label{lem:eight-exact-matter-current}
Let \(\mathscr L_{\fieldPhi}\coloneqq \Box_g-g(\nabla{\fieldPhi}, \nabla \, \cdot\,) \). For the commuted equation \(\mathscr L_{\fieldPhi}Z\psi = \mathcal K_Z^\psi\) and every exterior multiplier \(X\), one has
\be
\label{equa:eight-exact-matter-current}
{ \begin{aligned}
\nabla_\alpha \!\Bigl(e^{-{\fieldPhi}} Q^\alpha{}_{\beta}[Z\psi]X^\beta \Bigr) &=e^{-{\fieldPhi}} \mathcal K_Z^\psi X(Z\psi) {+ {1\over2}e^{-{\fieldPhi}}Q^{\alpha \beta}[Z\psi]{}^{(X)} \!\pi_{\alpha \beta}} \\
& \quad + {1\over2}e^{-{\fieldPhi}}(X{\fieldPhi}) g(\nabla Z\psi, \nabla Z\psi).
\end{aligned} }
\ee
At order zero, this last term cancels, after the effective-field and matter identities are added with the physical normalization, the corresponding effective-field--matter exchange term. At positive order, \(\mathcal K_Z^\psi = [\mathscr L_{\fieldPhi},Z] \psi\) contains the ordered principal commutators and differentiated scalar coupling in \eqref{equa:eight-low-matter-ordered-Lie-commutation}--\eqref{equa:eight-low-matter-Lie-coupling}. Their energy pairings are estimated separately in Proposition~\ref{proposition:eight-c-top-exterior}.

\end{lemma}
\begin{proof}
We apply the product rule to \(e^{-\fieldPhi}Q^\alpha{}_{\beta}[Z\psi]X^\beta\) and substitute the commuted matter equation. The drift term cancels the contraction of \(d\fieldPhi\) with the differentiated wave tensor, leaving the final term in the equation~\eqref{equa:eight-exact-matter-current}. For \(Z=1\), this term cancels the trace source \(-8\pi e^{-\fieldPhi}g(\nabla\psi,\nabla\psi)\) after adding the uniform effective-field identity and \(16\pi\) times the matter identity. At positive order, the differentiated coupling and metric commutators are estimated in Section~\ref{section:exterior-coupled-energies}. The factor \(e^{-\fieldPhi}\) is positive and uniformly comparable to one.
\end{proof}

\begin{proposition}[Schwarzschild lightlike interface identity]
\label{proposition:eight-schwarzschild-light-like-endpoint}
In the reference end, let \(L_*= \del_t+ \del_{\mathbf r_*} \) and \(\underL_*= \del_t-\del_{\mathbf r_*} \) be the Schwarzschild lightlike derivatives introduced in Section~\ref{section-12-1}.  If \(W= \mathbf r Z\psi\), then
\be
\label{equa:eight-schwarzschild-light-like-equation}
{ L_*\underL_*W-{D_m\over {\mathbf r}^2} \Delta_{\mathbb S^2}W + {2mD_m\over {\mathbf r}^3}W =-D_m\mathbf r F_Z^\psi+ \mathcal R_Z^\psi. }
\ee
Here \(F_Z^\psi = \Box_gZ\psi\), \(\widehat H=g^{-1}-(g^\star)^{-1} \), and
\be
\label{equa:eight-Schwarzschild-exact-remainder}
\mathcal R_Z^\psi =D_m\mathbf r\, \widehat H^{\alpha \beta} \del_\alpha \del_\beta Z\psi.
\ee
At a point of \(\Lscr\), the incoming lightlike derivative is given by
\be
\label{equa:eight-schwarzschild-interface-endpoint}
{ \underL_*(\mathbf r Z\psi) = \underL(rZ\psi) +mL(Z\psi) +(1-D_m)Z\psi. }
\ee
The logarithmic factor in \(\mathbf r_*\) is kept in the trace estimate; it is not absorbed into the perturbative remainder. Along \(\Lscr\), the lightlike label is
\be
u_*(s)=1-2m\log(r(s)-m)-c_*,
\ee
where \(r(s)= (s^2-1)/2\), and hence one has
\be
\label{equa:eight-Schwarzschild-interface-light-like-derivative}
{{du_*\over ds} =-{2ms\over r(s)-m} =-{4m\over s} +O_m(s^{-3}).}
\ee

\end{proposition}
\begin{proof}
For the reference metric, we write the radial part of the wave operator in the coordinates \((t,\mathbf r_*) \).  The identities \(d\mathbf r_*/dr=D_m^{-1} \), \(\del_{\mathbf r_*} =D_m\del_r\), and \(\del_{\mathbf r_*} \mathbf r=D_m\) show that conjugation by \(\mathbf r\) removes the first-order radial coefficient and leaves the potential \(2mD_m{\mathbf r}^{-3} \). The angular Laplacian contributes \(D_m{\mathbf r}^{-2} \Delta_{\mathbb S^2} \), giving the equation. Both metrics are in the harmonic coordinates used here, so \(\Box_{g^\star}Z\psi =F_Z^\psi-\widehat H^{\alpha \beta} \del_\alpha \del_\beta Z\psi\). This gives \eqref{equa:eight-Schwarzschild-exact-remainder}; all commutators already belong to \(F_Z^\psi\). This proves \eqref{equa:eight-schwarzschild-light-like-equation}. The angular Laplacian is estimated as a source term, without assigning it a pointwise sign. The quantitative characteristic bounds for the perturbed metric use the operator \(\mathcal P_{H_N}\) in~\eqref{equa:eight-low-matter-weighted-transport}, whose angular source was bounded in~\eqref{equa:eight-low-matter-angular-weight}. On the interface \(\Lscr\), we differentiate \(\mathbf r=r+m\) and use \(\underL_*r=-D_m\). Expanding \(\underL_*(\mathbf r Z\psi) \) in the pair \(L= \del_t+ \del_r\), \(\underL= \del_t-\del_r\) gives \eqref{equa:eight-schwarzschild-interface-endpoint}. Finally, substituting \(r(s)= (s^2-1)/2\) into \(u_*=1-2m\log(r(s)-m)-c_*\) gives \eqref{equa:eight-Schwarzschild-interface-light-like-derivative}.  Hence the logarithmic correction is part of the principal interface geometry and cannot be classified as an integrable perturbative error.
\end{proof}

\paragraph{Quantitative interface and continuation estimates}

The interior--exterior interface is not a physical boundary and carries no independent data: the interior and exterior traces are restrictions of the same Cauchy development.   The calculation keeps the Jacobian and the rank weights explicitly. This is also the step which transfers the exterior rank-weight margin to the interior center estimate.

\begin{lemma}[Direct pointwise estimates and traces on spheres]
\label{lem:eight-quantitative-cone-trace}
Under the coarse exterior bootstrap, let \(Z\) have order \(p\leq N-5\) and rank \(k\). On \(\Lscr_s\), one has
\be
\label{equa:eight-quantitative-cone-trace}
t\bigl(| \del Zu|+ | \delsN Zu|+ | \del Z\psi| \bigr) \lesssim As^\delta, \qquad t\bigl(| \del Z\fieldPhi|+ |Z\fieldvarphi| \bigr) \lesssim As^{1+ \delta}.
\ee
The sphere traces satisfy
\be
\label{equa:eight-quantitative-cone-flux}
\int_{s_0}^{s_1} \tau^{-1} \Bigl(\mathsf T_{2;p,k}[u]^2 + \mathsf T_{2;p,k}[\psi]^2 + \mathsf T_{2;p,k}[\fieldvarphi]^2\Bigr) (\tau) \, d\tau\lesssim C_\delta A^2{s_1}^{2\delta}.
\ee
These are unsigned trace bounds. The sharper rank-resolved interface estimates are those of~\autoref{lem:eight-rank-local-interface-elimination}.
\end{lemma}

The estimates above give the finite-slab flux and trace bounds. Their use in the global argument is justified in Proposition~\ref{prop:mobile-simultaneous-improvement}, after the positive currents and the interior estimates have been transferred to the expanding interface.

\begin{proof}
On the interface, \(r\simeq t\simeq s^2\) and \(w=1\). The near-cone version of~\autoref{lem:eight-exterior-weighted-sobolev} uses \(p+3\leq N-2\) commutations. The metric and matter energies have growth \(As^\delta \), while the effective-field energy at these auxiliary orders is bounded by \(As^{1+ \delta} \). This proves \eqref{equa:eight-quantitative-cone-trace}. On the other hand, the fixed-hypersurface radial estimate gives
\be
\label{equa:eight-direct-sphere-trace-growth}
\mathsf T_{2;p,k}[u](s) + \mathsf T_{2;p,k}[\psi](s) \lesssim As^{-1+ \delta}, \qquad \mathsf T_{2;p,k}[\fieldvarphi](s) \lesssim As^\delta.
\ee
For the last estimate, the hypersurface-radial derivative of \(Z\fieldvarphi\) uses one additional translation derivative, hence the mass energy at order \(p+1\leq N-4\), bounded by the auxiliary order-\(N\) energy. The factor \(R_s^{-1/2} \simeq s^{-1} \) cancels its extra growth \(s\). Inserting these bounds and integrating \(\tau^{-1+2\delta} \) proves \eqref{equa:eight-quantitative-cone-flux}. The oriented current is not used in this radial trace argument.
\end{proof}


\subsection{Weighted exterior norms and direct decay estimates}
\label{section-13-3}

\paragraph{Exterior weights and bootstrap scales}

We turn to the Euclidean--merging domain $\MME_{[s_0,s_1]}$.  In this domain the hypersurfaces interpolate between hyperboloids and asymptotically Euclidean slices.  The weight $\zeta$ in the energy density is comparable to one in the far Euclidean region and to $s/t$ at the interior--exterior interface.  We use the fixed smooth, one-sided weight \(\crochet=1+ \aleph(r-t) \) defined in \eqref{equa:global-foliation-exterior-weight-section3}; it is constant on the interior side of the cone, grows linearly on the outgoing Euclidean side, and satisfies \(| \delsN\crochet| \lesssim1\). The exponents are fixed so that
\be
\label{equa:eight-exterior-weight-gaps}
{3\over4}<\lambda<\mu<1, \quad 0<\delta \ll\mu-\lambda, \quad 3C_{\mathrm{rk}} \theta<2\lambda-1<2\mu-1.
\ee
Throughout Sections~\ref{section-12}--\ref{section-15}, \(\lambda \) is a fixed exterior weight exponent and is distinct from the singular parameter \(\coeff\).  The metric is measured with the \(\lambda \)-weighted wave energy, whereas the  effective field and the weighted matter wave use the weight \(\mu\).  This slight separation makes the quadratic effective-field sources integrable near spacelike infinity.  The assumption \(V\equiv0\) remains in force: \(\psi\) is never assigned a Klein--Gordon mass.

For \(\eta \in\{\lambda, \mu\} \), we define the commuted exterior hypersurface norm by
\be
\label{equa:eight-exterior-wave-norm-definition}
\bigl(\Fenergy^{\ME,p,k}_\eta(s,v) \bigr)^2\coloneqq \sum_{Z\preccurlyeq(p,k)} \calE^{\ME}_\eta(s,Zv).
\ee
For a metric field the sum also extends over its components.  For \(v= \psi\), the energy is generated by the weighted current \(e^{-\fieldPhi}Q[Z\psi] \); thus the notation never replaces the matter current by an unweighted wave current.

For each order \(p\) and rank \(k\), the norms \(D^{\mathrm{ext}}_{p,k}\) and \(U^{\mathrm{ext}}_{p,k}\) measure the singular and rescaled effective-field energies. Their kinetic terms control \((\zeta\del_t Z\fieldphi,\delsN Z\fieldphi)\) with factor \(\sqrt\coeff\), or \((\zeta\del_t Z\fieldPhi,\delsN Z\fieldPhi)\), respectively; their mass terms control \(Z\fieldphi\) or \(Z\fieldvarphi\). We define these norms by
\bse
\be
\label{equa:eight-exterior-effective-field-energy-definitions}
\bigl(D^{\mathrm{ext}}_{p,k}(s) \bigr)^2\coloneqq \sum_{Z\preccurlyeq(p,k)} \int_{\MME_s}w^{2\mu}Q^{\mathrm{sing}}_{\alpha \beta}[Z\fieldphi]N_s^\alpha X_{\ME}^\beta \, dx, \qquad U^{\mathrm{ext}}_{p,k}(s)\coloneqq \coeff^{1/2}D^{\mathrm{ext}}_{p,k}(s),
\ee
\be
\label{equa:eight-exterior-X-definitions}
{\begin{aligned}
\mathsf X^{\mathrm{fix}}_{p,k}(s)&\coloneqq \Fenergy^{\ME,p,k}_\lambda(s,u) + \Fenergy^{\ME,p,k}_{\mu}(s, \psi) + \omega_p^{\mathrm{ext}}(s)^{-1}D^{\mathrm{ext}}_{p,k}(s), \\
\mathsf X^{\mathrm{unif}}_{p,k}(s)&\coloneqq \Fenergy^{\ME,p,k}_\lambda(s,u) + \Fenergy^{\ME,p,k}_{\mu}(s, \psi) + \omega_p^{\mathrm{ext}}(s)^{-1}U^{\mathrm{ext}}_{p,k}(s).
\end{aligned}}
\ee
\be
\label{equa:eight-exterior-order-weight}
{\omega_p^{\mathrm{ext}}(s)\coloneqq \begin{cases}
s,&p=N, \\
1,&p\leq N-5.
\end{cases} \qquad \mathsf X_{p,k}\coloneqq \mathsf X^{\mathrm{unif}}_{p,k}.}
\ee
\ese
The norms \(\Bulk^{\ME, \mathrm{fix}}_{p,k}(s_0,s) \) and \(\Bulk^{\ME, \mathrm{unif}}_{p,k}(s_0,s) \) are the square roots of the summed positive bulk integrals for the metric, matter, and effective-field currents. Their effective-field contributions use the singular and uniform normalizations, respectively. The adverse integrals \(\mathcal M^{\ME} \) remain on the source side of the identities. We set \(\Bulk^{\ME}_{p,k}\coloneqq \Bulk^{\ME, \mathrm{unif}}_{p,k} \). Both \(\mathsf X\) and \(\Bulk^{\ME} \) are norms; their squares are the hypersurface energy and spacetime integral. The factor \((\omega_p^{\mathrm{ext}})^{-1} \) permits the top-order effective-field growth and is absent for \(p\leq N-5\). The instantaneous bulk norm is defined by
\be
\label{equa:eight-exterior-instantaneous-bulk-definition}
\bigl(\Bulk_{p,k}^{\ME, \mathrm{unif}}(s_0,s) \bigr)^2 = \int_{s_0}^{s} \bigl(\bulkdens_{p,k}^{\ME, \mathrm{unif}}(\tau) \bigr)^2\, d\tau, \qquad \bulkdens_{p,k}^{\ME}\coloneqq \bulkdens_{p,k}^{\ME, \mathrm{unif}}.
\ee
The effective-field summand of the instantaneous squared bulk is multiplied by \(\omega_p^{\mathrm{ext}}(s)^{-2} \). Thus its order-\(N\) bulk and nonnegative flux are integrated with the time weight \(s^{-2} \); the metric and matter terms keep weight one. All closing-order terms have weight one. Multiplying the order-\(N\) differential energy identity by \(s^{-2} \) also produces the positive integral \(4\int_{s_0}^{s_1} s^{-3}(U^{\mathrm{ext}}_{N,k}(s))^2\, ds\), kept in \eqref{equa:eight-mobile-coupled-positive-energy}.

Table~\ref{table:exterior-parameter-scales} distinguishes the exterior energy normalizations and their interface controls.
\begin{table}[htbp]
\centering
{\renewcommand{\arraystretch}{1.2}
\begin{tabular}{>{\raggedright\arraybackslash}p{.29\textwidth}>{\raggedright\arraybackslash}p{.36\textwidth}>{\raggedright\arraybackslash}p{.27\textwidth}}
\hline
Field & Defining equations and estimates & Dependence on \(\coeff\) \\
\hline
Metric \(u\) & \eqref{equa:eight-exterior-wave-norm-definition}, \eqref{equa:eight-interface-flux-positivity} & Uniform in \(\coeff\) \\
Matter \(\psi\) & \eqref{equa:eight-exterior-wave-norm-definition}, \eqref{equa:eight-exact-matter-current} & Uniform in \(\coeff\) \\
Effective field \(\fieldphi\) & \eqref{equa:eight-exterior-effective-field-energy-definitions}, \eqref{equa:eight-affine-effective-field-interface-density} & Constants may depend on \(\coeff\) \\
Rescaled fields \(\fieldPhi,\fieldvarphi\) & \eqref{equa:eight-exterior-effective-field-energy-definitions}, \eqref{equa:eight-affine-effective-field-interface-density} & Uniform in \(\coeff\) \\
Conformal field \(\fieldPhi\) & \eqref{equa:eight-uniform-radial-mass-value} on fixed finite slabs & Vanishing as \(\coeff\to0\) \\
\hline
\end{tabular}
}
\caption{Exterior controls at the three parameter scales}
\label{table:exterior-parameter-scales}
\end{table}

We assume on the bootstrap interval that
\be
\mathsf X_{N,N}(s) \le As^\delta, \quad \mathsf X_{N-5,N-5}(s) \le As^\delta, \quad A\coloneqq C_1(\epss+ C_0\eps).
\ee
For the interface argument we keep the sharper lower, rank-resolved subcollection
\be
\label{equa:eight-robust-rank-bootstrap}
\mathsf E_{\mathrm{ext};p_{\mathbf{low}},k}^{\mathrm{rob}}(s) \leq As^{\eta_{\Lscr}}, \quad 0\leq k\leq p_{\mathbf{low}},
\ee
where \(\mathsf E_{\mathrm{ext}}^{\mathrm{rob}} \) is the sum of the weighted metric wave energy, the exact \(e^{-{\fieldPhi}} \)-matter wave energy, the uniform effective-field energy in \((\del{\fieldPhi},{\fieldvarphi}) \), their Hardy terms, and the positive tangential bulk norms, all commuted with the standard family \(Z= \del^IL^J\).  No unscaled effective-field derivative occurs in \eqref{equa:eight-robust-rank-bootstrap}.  The coarse \(s^\delta \) bound above is used for harmless products; the smaller exponent \(\eta_{\Lscr} \) is propagated for the actual interface currents, and we also assume that the spacetime bulk term generated by the weighted wave energy obeys
\be
\sum_{\ord Z\le N} \int_{s_0}^{s} \int_{\MME_\tau} \crochet^{2\lambda-1}| \delsN Zu|^2J\, dx\, d\tau \le A^2s^{2\delta}.
\ee
On the prescribed foliation, the light-bending sign guarantees positivity of this bulk term near the interface. The simultaneous improvement with amplitude \(A/2\) will be proved on the adapted foliation in Section~\ref{section:mobile-simultaneous-improvement}, where the moving current gives positivity under the coarse estimates.

\paragraph{Positive exterior currents}

The two-sided Euclidean--hyperboloidal energy comparisons give
\be
\| \crochet^\lambda \zeta| \del u|_p\|_{L^2(\MME_s)} + \| \crochet^\lambda| \delsN u|_p\|_{L^2(\MME_s)} \lesssim As^\delta, \quad p\le N,
\ee
and the Hardy--Poincar\'e inequality yields
\be
\| \crochet^{-1+ \lambda}|u|_{p,k} \|_{L^2(\MME_s)} \lesssim \delta^{-1}As^\delta.
\ee
For the  effective field, the uniform quantities satisfy
\be
\begin{aligned}
\| \crochet^\mu\zeta \coeff| \del\fieldphi|_p\|_{L^2} + \| \crochet^\mu\coeff| \delsN\fieldphi|_p\|_{L^2} + \| \crochet^\mu\coeff^{1/2}| \fieldphi|_p\|_{L^2} & \lesssim A\begin{cases}
s^{1+ \delta},&p=N, \\
s^\delta,&p=N-5.
\end{cases}
\end{aligned}
\ee
These effective-field estimates are written only in rescaled combinations; hence their constants do not deteriorate as $\coeff\to0$.  The massless matter estimates are instead
\be
\| \crochet^\mu\zeta| \del\psi|_p\|_{L^2} + \| \crochet^\mu| \delsN\psi|_p\|_{L^2} + \| \crochet^{-1+ \mu}| \psi|_{p,k} \|_{L^2} \lesssim As^\delta,
\ee
where the last term follows from the exterior Hardy inequality.

\paragraph{Direct exterior decay}

The global five-order estimate gives the following bounds throughout \(\MME_s\):
\be
\label{equa:eight-safe-global-wave-decay}
r w^\lambda \bigl(| \del u|_{N-5} + | \delsN u|_{N-5} \bigr) +r w^{\lambda-1}|u|_{N-5} \lesssim As^\delta.
\ee
For \(p\in\{N-5,N\} \), the two scalar fields satisfy
\be
\label{equa:eight-safe-global-scalar-decay}
r w^\mu\bigl(| \del\fieldPhi|_{p-5} + | \delsN\fieldPhi|_{p-5} + | \fieldvarphi|_{p-5} \bigr) \lesssim A\omega_p^{\mathrm{ext}}(s)s^\delta, \quad r w^\mu| \del\psi|_{p-5} +r w^{\mu-1}| \psi|_{p-5} \lesssim As^\delta.
\ee
In \(r\leq5t/4\), the subscripts \(N-5\) and \(p-5\) in these estimates may be replaced by \(N-3\) and \(p-3\), respectively. The two effective-field levels remain distinct: \(\omega_{N-5}^{\mathrm{ext}} =1\), whereas \(\omega_N^{\mathrm{ext}} =s\). The mass reconstruction in \eqref{equa:eight-good-metric-mass-reconstruction} provides the additional near-cone gain used below, at its specified commutation orders.


\section{Coupled exterior energy estimates}
\label{section:exterior-coupled-energies}


\subsection{Metric and matter source estimates}
\label{section-13-4}

\paragraph{Purpose}

We estimate the nonlinear sources and principal commutators against the positive exterior currents. After assigning separate fractions of the positive bulk terms, we obtain the coupled energy estimate at the closing and auxiliary orders.

\paragraph{Matter stress terms in the metric energy estimate}

We estimate the matter stress tensor directly by the metric and matter hypersurface energies. We use \(w=2+r-t\) and the near-cone region \(\mathcal C_s\coloneqq\{t-1\leq r\leq5t/4\}\cap\MME_s\). For \(0\leq j\leq k\leq p\), we abbreviate the square roots of the energies by
\be
\label{equa:audit-stress-leaf-energies}
E_u(s)\coloneqq\Fenergy^{\ME,p,k}_\lambda(s,u),\qquad E_{\psi,j}(s)\coloneqq\Fenergy^{\ME,p,k-j}_\mu(s,\psi).
\ee

\begin{proposition}[Weighted estimate for the matter stress tensor]
\label{lem:eight-weighted-matter-stress}
For \(p\in\{N-5,N\}\), \(0\leq k\leq p\), and \(d=\lfloor p/2\rfloor\), the hypotheses of Proposition~\ref{lem:eight-low-matter-characteristic-Lie} give
\be
\label{equa:eight-stress-pointwise-condition}
| \del\psi|_{d,j} \lesssim C_NA r^{-1}w^{-(1+ \delta)/2}s^{\eta_j} \quad\hbox{on } \mathcal C_s, \qquad 0\leq j\leq d.
\ee
The contributions of \(-16\pi Z(e^{-\fieldPhi} \del\psi\del\psi) \) to the metric squared-energy identity satisfy the following estimates, with every Leibniz term included. The symbol \(\mathcal S^{\mathrm{near}}_{\psi;p,k} \) denotes the sum of the near-cone pairings with undifferentiated \(e^{-\fieldPhi} \).
\begin{itemize}
\item Near the cone, the terms with undifferentiated \(e^{-\fieldPhi}\) satisfy
\be
\label{equa:audit-stress-direct-energy}
|\mathcal S_{\psi;p,k}^{\mathrm{near}}|\leq C_NA\int_{s_0}^{s_1}s^{-1}E_u(s)^2\,ds+C_NA\sum_{j=0}^{\min\{k,d\}}\int_{s_0}^{s_1}s^{-1+2\eta_j}E_{\psi,j}(s)^2\,ds.
\ee
\item Outside \(\mathcal C_s\), their contribution is bounded by \(C_NA\int_{s_0}^{s_1}s^{-1-\gamma_{\mathrm f}}(\mathsf X^{\mathrm{unif}}_{p,k}(s))^2\, ds\), where \(\gamma_{\mathrm f} =4\mu-2\lambda-\delta>0\).
\item The terms in which a derivative falls on \(e^{-\fieldPhi} \) are bounded by \(C_{N, \coeff_0}A^2\int_{s_0}^{s_1}s^{-2 +2\delta}(\mathsf X^{\mathrm{unif}}_{p,k}(s))^2\, ds\).
\end{itemize}
The near-cone estimate has coefficient \(C_NA/s\) at the current rank, with the lower-rank terms. Its constants are independent of the reference mass.
\end{proposition}

\begin{proof}
\bse
We consider a Leibniz term with \(U=Zu\), \(V=Z_{\mathrm h} \psi\), and a lower factor \(\psi_{\mathrm l} \) of rank \(j\). The high factor has order at most \(p\) and rank at most \(k-j\). The coordinate volume is \(dV_g=j_gJ\, dx\, ds\), and the foliation and matter-current comparisons give
\be
\label{equa:eight-stress-Jacobian-energy}
J\zeta^{-2}r^{-1} \lesssim s^{-1}, \qquad \int_{\MME_s}j_ge^{-\fieldPhi}w^{2\mu} \zeta^2| \del V|^2\, dx\lesssim\bigl(\Fenergy^{\ME,p,k-j}_\mu(s, \psi) \bigr)^2.
\ee
Indeed, in the merging interval, \(b=s/\sqrt{s^2 +r^2} \simeq s^{-1} \), \(\zeta^2\gtrsim b^2 +1-\xi\), and \(J\lesssim b+s(1-\xi) \lesssim s\zeta^2\); on the flat exterior, \(\zeta=1\) and \(J\lesssim s\). In both regions \(r\gtrsim s^2\). The second estimate keeps the factor \(\zeta \) on every high Cartesian derivative.

The source integrand is \(j_gJw^{2\lambda}e^{-\fieldPhi}|\del\psi_{\mathrm l}||\del V||\del_tU|\), up to a fixed constant. After the two high derivatives are put in their weighted kinetic energies, the remaining coefficient satisfies
\be
\label{equa:audit-stress-direct-coefficient}
J\zeta^{-2}w^{\lambda-\mu}|\del\psi_{\mathrm l}|\lesssim A s^{-1+\eta_j}w^{\lambda-\mu-(1+\delta)/2}\leq A s^{-1+\eta_j}.
\ee
Here \(\mu>\lambda\), \(w\geq1\), and \(e^{-\fieldPhi}\) is uniformly bounded. Weighted Cauchy--Schwarz therefore bounds the pairing on each hypersurface by \(C_NAs^{-1+\eta_j}E_uE_{\psi,j}\). Applying \(2ab\leq a^2+b^2\) with \(a=E_u\), \(b=s^{\eta_j}E_{\psi,j}\), and summing proves~\eqref{equa:audit-stress-direct-energy}. A lower operator of rank \(j\) leaves rank at most \(k-j\) on the high factor, and \(d\leq p_{\mathbf{low}}\) for \(N\geq19\), so every lower factor is within the stated pointwise range.

Outside \(\mathcal C_s\), we have \(w\simeq r\), and the direct derivative estimate gives
\be
\label{equa:eight-stress-far-coefficient}
J\zeta^{-2}w^{\lambda-\mu}| \del\psi_{\mathrm l}| \lesssim A s^{1+ \delta}r^{-1+ \lambda-2\mu} \lesssim A s^{-1-\gamma_{\mathrm f}}.
\ee
This proves the second assertion by the two hypersurface energies. For a differentiated scalar coefficient, a lower \(Z\fieldPhi\) contributes the additional pointwise bound \(C_{\coeff_0}A r^{-1}w^{-\mu}s^\delta \). If the highest derivative instead falls on \(\fieldPhi\), we use its weighted mass norm and put both matter derivatives in the direct supremum norm. The coefficient is bounded by \(C_{\coeff_0}A^2s^{-3+2\delta} \). The top effective-field normalization at \(p=N\) loses one factor \(s\), giving the third assertion. The lower Sobolev count is \(\lfloor N/2\rfloor+5\leq N-5\).
\ese
\end{proof}

\paragraph{Low effective-field factors in the metric stress}

The massive reconstruction also bounds the low factors in the metric effective-field stress on \(\mathcal C_s\), where \(r\simeq t\). If \(d\leq\lfloor N/2\rfloor\), then \(d+1\leq N-9\). The closing direct estimate and \eqref{equa:eight-good-metric-mass-reconstruction}, applied to the translation operator for the derivative term, give
\be
\label{equa:eight-low-effective-stress-two-bounds}
| \del\fieldPhi|_d+ | \fieldvarphi|_d\lesssim C_{N, \coeff_0}A\min\{t^{-1+ \delta/2}w^{-\mu}, \ t^{-3/2 + \delta}w^{1-\mu} \}.
\ee
The relation \(\fieldPhi = \sqrt{\coeff} \, \fieldvarphi\) introduces only a constant depending on \(\coeff_0\). In \(w\geq t^{2\delta} \), the first bound has the normalized exponent \(\delta/2-2\delta(\mu-\gamma)<0\). In \(w\leq t^{2\delta} \), the second has exponent at most \(-1/2 +5\delta/2<0\). Thus
\be
\label{equa:eight-low-effective-stress-coefficient}
rw^\gamma \bigl(| \del\fieldPhi|_d+ | \fieldvarphi|_d\bigr) \lesssim C_{N, \coeff_0}A, \qquad \gamma= (1+ \delta)/2.
\ee
At the closing orders \(p\leq N-5\), the direct energy estimate also applies to the derivative stress and the potential. The high derivative is measured by \(w^\mu\zeta\del Z\fieldPhi\), and the potential by \(w^\mu Z\fieldvarphi\). Their coefficients contain \(J\zeta^{-2}\) and \(J\zeta^{-1}\), respectively; the latter is smaller since \(0<\zeta\leq1\). Thus the undifferentiated-coefficient pairings satisfy
\be
\label{equa:eight-low-effective-stress-Young}
|\mathcal S_{\mathrm c;p,k}^{\mathrm{near}}|\leq C_{N,\coeff_0}A\int_{s_0}^{s_1}s^{-1}\bigl(E_u(s)^2+(U^{\mathrm{ext}}_{p,k}(s))^2\bigr)\,ds,\qquad 0\leq k\leq p\leq N-5.
\ee
The uniform exterior effective-field energy keeps its original scale in this formula. Differentiated smooth coefficients are estimated separately by their pointwise bounds and the high weighted Hardy or mass norm.

\paragraph{Metric quadratic terms in the energy identity}

For symmetric covariant tensors \(B,C\), the leading reduced-Einstein quadratic form is \(P(B,C)= \tfrac14\operatorname{tr}_{\gMink}B\operatorname{tr}_{\gMink}C-\tfrac12B:C\). In a null frame \((L, \underL,e_1,e_2) \), we put \(b_B=B_{L\underL} \), \(T_B= \sum_AB_{AA} \), and \(\widehat B_{AB} =B_{AB}-\tfrac12T_B\delta_{AB} \). Direct contraction gives
\be
\label{equa:eight-quasi-null-shear-decomposition}
P(B,C)=-\tfrac12\widehat B:\widehat C-\tfrac14(T_Bb_C+b_BT_C)-\tfrac18(B_{LL}C_{\underL\underL} +B_{\underL\underL}C_{LL}) + \tfrac12\sum_A(B_{LA}C_{\underL A} +B_{\underL A}C_{LA}).
\ee
For \(B= \del_th_Z\), the wave-gauge identity controls \(B_{LL} \), \(B_{LA} \), and \(T_B\) by good derivatives and the finite products \(H\del h\). The only remaining quadratic product has two transverse trace-free factors.~\autoref{lem:eight-good-metric-characteristic} controls its lower factor by \(C_NAr^{-1}w^{-(1+ \delta)/2}s^{\eta_j} \). Weighted Cauchy--Schwarz, with both energy weights equal to \(\lambda\), gives
\be
\label{equa:eight-quasi-null-shear-pairing}
\begin{aligned}
|\mathcal S_{\mathrm{shear};p,k}^{\mathrm{near}}|&\leq C_NA\int_{s_0}^{s_1}s^{-1}E_u(s)^2\,ds\\
&\quad+C_NA\sum_{j=0}^{\min\{k,\lfloor p/2\rfloor\}}\int_{s_0}^{s_1}s^{-1+2\eta_j}\bigl(\Fenergy^{\ME,p,k-j}_\lambda(s,u)\bigr)^2\,ds.
\end{aligned}
\ee
The remaining weight is \(w^{-(1+\delta)/2}\leq1\). For a good high derivative, Young's inequality spends an independently chosen fraction \(\epsilon_g\) of the metric good bulk and leaves the integrable coefficient \(C_{N, \epsilon_g}A^2s^{-3+2\delta} \). A good low derivative leaves \(C_NAs^{-1-2\lambda+ \delta} \). The cubic gauge terms use the same coefficient \(C_NA^2s^{-3+2\delta} \), with the high undifferentiated metric factor in its weighted Hardy norm. Outside \(r\leq5t/4\), the direct low derivative gives the integrable coefficient \(C_NAs^{-1-2\lambda+ \delta} \). The good-bulk fractions are assigned separately from those used for the other sources; the adverse density \(\Omega^-\) remains a separate source.

\paragraph{Products of densitized metric coefficients and wave Hessians}

The dynamic densitized tensor is \(K-K^\star\), where \(K=j_gg^{-1}-\gMink^{-1} \). On the exact harmonic Schwarzschild exterior, \(P= \mathcal L_I(K-K^\star) \) is divergence free. We fix a finite exterior slab with terminal parameter \(s_1\). For \(t\geq T_0\), we let \(I_{s_1}(t)\) be its available radial interval on the physical time slice. The monotonicity of the terminal graph implies that the outward ray from any available point remains in this interval. The radial gauge identity and decay at infinity give, for \(R\simeq t\),
\be
\label{equa:eight-high-density-radial-pairing}
|P_N(t,R\omega)|\lesssim R^{-\lambda}B(t,\omega),\qquad B(t,\omega)^2\coloneqq\int_{I_{s_1}(t)}w^{2\lambda-1}(|EP|^2+r^{-2}|P|^2)r^2\,dr.
\ee
The radial identity bounds \(|\del_rP_N|\) by \(C(|EP|+r^{-1}|P|)\), without an angular derivative of \(P\). The Cauchy--Schwarz kernel is \(\int_R^\infty r^{-2}(2+r-t)^{1-2\lambda}\,dr\lesssim R^{-2\lambda}\): after scaling by \(R\), its only possible endpoint singularity has exponent \(1-2\lambda>-1\). The finite dynamic Hardy norm excludes a nonzero radial limit. We use \(B\) only in \(L^2(dt\,d\omega)\), not in angular \(L^\infty\). In the bounded initial range,~\eqref{equa:audit-bounded-initial-Gronwall} controls the low factors on the available hypersurfaces. We place the high coefficient directly in its actual Hardy norm and the high target in its actual energy. Their pairing has coefficient \(C_NA\) on this fixed interval and uses no pointwise derivative of the high coefficient.

With \(a_\lambda= \delta/\lambda \) from \eqref{equa:eight-displaced-cone-exponents}, we let \(f= \del^2Zv\), where \(v\in\{u, \psi\} \), \(\operatorname{ord}Z\leq\lfloor N/2\rfloor\), and \(\sigma= \lambda \) or \(\mu\), respectively. The direct and exterior lightlike wave field estimates give
\be
\label{equa:eight-high-density-low-Hessian}
|f| \lesssim At^{-1+ \delta/2}w^{-\sigma} \quad(w\leq4t^{a_\lambda}), \qquad |f| \lesssim At^{-1+ \delta/2}w^{-1-\sigma} \quad(w\geq4t^{a_\lambda}).
\ee
These estimates use at most \(\operatorname{ord}Z+6\leq N\) energy orders. The effective-field factors in their wave sources satisfy \(\operatorname{ord}Z+5\leq N-5\), so they belong to the closing level. The lightlike reduction absorbs \((t/w)|H| \lesssim At^{-\delta/2} \).

For a high wave \(U\), we set \(\mathcal C(t)\coloneqq I_{s_1}(t)\cap[t-1,5t/4]\) and \(\mathcal C\coloneqq\{(t,r\omega):t\geq T_0,\ r\in\mathcal C(t),\ \omega\in\mathbb S^2\}\). Cauchy--Schwarz on this available radial interval gives
\be
\label{equa:eight-high-density-radial-Cauchy}
\Bigl(\int_{\mathcal C(t)}R^{2-\lambda}w^{2\sigma}|f| \,| \del_tU| \, dR\Bigr)^2\lesssim A^2t^{-2\lambda+ \delta+a_\lambda} \int_{\mathcal C(t)}R^2w^{2\sigma}| \del_tU|^2\, dR.
\ee
The first Cauchy factor is bounded by \(A^2t^{-2\lambda+ \delta}(\int_1^{4t^{a_\lambda}}dw+ \int_{4t^{a_\lambda}}^{Ct}w^{-2}dw) \). Young's inequality, followed by \(J\zeta^{-2} \lesssim s\), therefore proves
\be
\label{equa:eight-high-density-wave-pairing}
\begin{aligned}
\int_{\mathcal C}w^{2\sigma}|P_Nf\del_tU| \, dx\, dt& \leq\epsilon\int B^2\, dt\, d\omega+C_{N, \epsilon}A^2\int_{s_0}^{s_1} s^{-1-\gamma_0} \|w^\sigma \zeta \del_tU\|_2^2\, ds, \\
\gamma_0&\coloneqq4\lambda-2-2\delta-2\delta/\lambda>\tfrac5{12}.
\end{aligned}
\ee
The dynamic composition estimate \(|EP| \lesssim\sum|EZu|+Ar^{-1} \sum|Zu| \), followed by outgoing Hardy, keeps the terminal trace:
\be
\label{equa:eight-high-density-bulk-Hardy}
\int B^2\, dt\, d\omega \lesssim C_N\sum_{\operatorname{ord}Z\leq|I|} \Bigl(\int w^{2\lambda-1}|EZu|^2\, dx\, dt+ \|w^{\lambda-1}Zu\|_{L^2(\MME_{s_1})}^2\Bigr).
\ee
Thus \eqref{equa:eight-high-density-wave-pairing} spends a specified fraction of the good metric bulk and terminal Hardy norm. This treats the bad null contraction. To estimate the full tensor contraction, we also need the good Hessians and the spatial end. For an actually high coefficient, \(|I|\geq N-4\), the complementary field \(V=Zv\) has at most four operators. From \(tE_a=L_a+q\omega_a\del_t\) and~\eqref{equa:eight-high-density-low-Hessian}, we obtain, in \(t-1\leq r\leq5t/4\),
\bse
\be
\label{equa:audit-high-density-good-Hessian}
|E\del V|+r^{-1}|\del V|\lesssim At^{-2+\delta/2}\begin{cases}w^{1-\sigma},&w\leq4t^{a_\lambda},\\w^{-\sigma},&w\geq4t^{a_\lambda}.\end{cases}
\ee
In the spatial end, the identity~\eqref{equa:seven-exterior-spacelike-full-Hessian} and the exterior wave equation give
\be
\label{equa:audit-high-density-spatial-Hessian}
|\del^2V|\lesssim A s^\delta r^{-2}w^{-\sigma},\qquad r\geq5t/4.
\ee
\ese
Here \(1-t^2/r^2\geq9/25\), so the spatial Hessian identity has bounded coefficients. We use the mixed \(H^2\) Sobolev inequality on a unit radial interval times \(\mathbb S^2\), whose derivative sum consists of \(\del_r^j\Omega^KF\) with \(j+|K|\leq2\). These intervals lie in the flat portion of the hypersurface. Each rotation is a commutator of two boosts, so this sum costs at most four commutations, not separate radial and angular losses. Together with the rotation in the spatial Hessian identity, the total is at most \(\ord Z+6\). The effective-field factors in the wave sources cost at most \(\ord Z+5\) closing orders. In particular, for \(\ord Z\leq4\), both counts lie below the available orders at \(N=19\). The source terms are estimated by the same lower-order induction and absorption of \(|H|\) used in the Hessian reduction.

We place the high coefficient in its dynamic weighted Hardy norm. The factor multiplying that norm and the actual target kinetic energy satisfies
\bse
\be
\label{equa:audit-high-density-good-pairing}
J\zeta^{-1}w^{\sigma+1-\lambda}(|E\del V|+r^{-1}|\del V|)\lesssim A\begin{cases}s^{-3+\delta+2a_\lambda(2-\lambda)},&w\leq4t^{a_\lambda},\\s^{-1-2\lambda+\delta},&w\geq4t^{a_\lambda},\ r\leq5t/4.\end{cases}
\ee
In \(r\geq5t/4\), we estimate the whole contraction instead, using \(w\simeq r\), \(r\gtrsim s^2\), and~\eqref{equa:audit-high-density-spatial-Hessian}:
\be
\label{equa:audit-high-density-spatial-pairing}
J\zeta^{-1}w^{\sigma+1-\lambda}|\del^2V|\lesssim A s^{1+\delta}r^{-1-\lambda}\lesssim A s^{-1-2\lambda+\delta}.
\ee
\ese
Both coefficients are integrable in \(s\): the collar exponent is less than \(-59/24\), and \(-1-2\lambda+\delta<-1\). These terms multiply the actual metric Hardy norm and the actual target energy and spend no additional good bulk. No angular derivative is applied to the highest coefficient. The contribution of the reference metric is estimated separately from this Hardy inequality. Its translated principal coefficients are \(\Ocal(mr^{-2}) \); its boost-only coefficients are \(\Ocal(mr^{-1}) \) and multiply a translated field operator of strictly lower boost rank and no higher total order.

Finally, \(g^{-1} =j_g^{-1}(\gMink^{-1} +K) \), \(\mathcal L_X\gMink^{-1} =0\), and \(\gMink^{-1}(\ell, \ell)=0\) separate the principal coefficient into \(j_g^{-1}P_N\) and the finite density products. For \(Z=X_1\cdots X_d\), the ordered product rule gives the conversion
\be
\label{equa:audit-exterior-density-all-orders}
(\mathcal L_Zg^{-1})_N=\sum_{A\subset\{1,\ldots,d\}}(Z_Aj_g^{-1})(\mathcal L_{A^c}K)_N.
\ee
The Minkowski tensor disappears in this null contraction, including the term with every derivative on \(j_g^{-1}\); its other components are estimated by the good-Hessian bounds above. Subtracting the corresponding reference identity gives the dynamic products. When the highest derivative falls on the density, the low \(K_N\) factor is bounded directly. Relative to the high metric Hardy norm its collar multiplier is
\be
\label{equa:eight-high-density-conversion-error}
J\zeta^{-1}w^{\sigma+1-\lambda}|K_{N, \mathrm{low}}f| \lesssim A^2s^{-3+2\delta+2a_\lambda(2-2\lambda)}, \qquad -3+2\delta+2a_\lambda(2-2\lambda)<-\tfrac{31}{12}.
\ee
Outside the collar the additional \(w^{-1} \) in \eqref{equa:eight-high-density-low-Hessian} leaves the stronger coefficient \(A^2s^{-3+2\delta} \). These terms pair the high wave energy with the high metric Hardy norm and spend no additional bulk. The remaining principal terms with a low coefficient and a high Hessian, as well as the massive commutators, keep their distinct estimates.

\paragraph{Sign of the outgoing metric coefficient at rank zero}

The sign of the outgoing metric coefficient controls the fixed-interface current. For the global argument, Proposition~\ref{prop:eight-mobile-positive-current} will instead prove positivity on the expanding interface from the coarse coefficient bounds and a small reference mass. The characteristic estimate of Proposition~\ref{lem:eight-good-metric-characteristic} has the separate order range \(p\leq N-9\).


\subsection{Principal effective-field commutators}
\label{section:exterior-effective-commutators}

\paragraph{Commutator estimate through order \(N-5\)}

A radial pointwise estimate on the flat part of each exterior hypersurface controls the principal effective-field commutator without inverting the massive equation. The prescribed reference is separated from the inverse metric throughout this estimate.

\begin{lemma}[Uniform bounds for the exterior principal commutators]
\label{lem:eight-massive-principal-commutator}
Under the coarse metric energy and good spacetime bounds through order \(N\), the propagated wave gauge, and the fixed exterior reference metric, let \(p\leq N-5\), \(0\leq k\leq p\). Denote by \(\mathcal S^{\Phi}_{\mathrm{comm};p,k} \) the principal-commutator pairing in the uniform effective-field identity, and by \(\mathcal G^{\Phi}_{\mathrm{good};p,k} \) its positive good-derivative bulk. For every \(\epsilon>0\), one has
\be
\label{equa:eight-massive-principal-pairing}
\begin{aligned}
| \mathcal S^{\Phi}_{\mathrm{comm};p,k}|& \leq\epsilon\mathcal G^{\Phi}_{\mathrm{good};p,k} +C_{N, \epsilon}(A+A^2) \int_{s_0}^{s_1} s^{-1-\beta} \bigl(U^{\mathrm{ext}}_{p,k}(s) \bigr)^2\, ds\\
& \quad +C_Nm\int_{s_0}^{s_1} s^{-1}U^{\mathrm{ext}}_{p,k}(s) \sum_{j<k}U^{\mathrm{ext}}_{p,j}(s) \, ds+C_Nm\int_{s_0}^{s_1} s^{-3} \bigl(U^{\mathrm{ext}}_{p,k}(s) \bigr)^2\, ds,
\end{aligned}
\ee
where \(\beta= \min\{\delta/3,1/24\}>0\). The mass and bad-derivative parts of the positive effective-field bulk remain available. All exterior energies keep their unnormalized scale.
\end{lemma}

\begin{proof}
\bse

\noindent{\bf 1.} \emph{Radial coefficient estimate.} We first estimate \(D_I= \mathcal L_I(H-H^\star) \). In \(w\geq4\), the hypersurface is flat and its outward radial ray has constant physical time. For \(|I| \leq N-5\), the coarse metric energy and smooth inverse-metric composition give
\be
\label{equa:eight-massive-coefficient-radial-energy}
\sum_{|K| \leq2} \int_r^\infty\rho^2(2 + \rho-t)^{2\lambda} \| \del_\rho\Omega^KD_I(t, \rho\, \cdot) \|_{L^2(\mathbb S^2)}^2\, d\rho\lesssim C_NA^2s^{2\delta}.
\ee
Two rotations cost four boost operators. The radial derivative is already measured by the kinetic energy, so the required index is \(|I|+4\leq N-1\). The metric bound is \(As^\delta \) at every order through \(N\); the extra top-order factor \(s\) belongs only to the effective field. All additional angular ranks are taken in this coarse family. Radial Cauchy--Schwarz uses the kernel
\be
\label{equa:eight-massive-coefficient-radial-kernel}
\int_r^\infty\rho^{-2}(2 + \rho-t)^{-2\lambda} \, d\rho\leq\frac{r^{-2}w^{1-2\lambda}}{2\lambda-1}.
\ee
It gives an integrable radial derivative and hence an \(L^2(\mathbb S^2) \) limit at infinity. The dynamic weighted Hardy norm forces that limit to vanish. The fundamental theorem of calculus and sphere Sobolev therefore yield
\be
\label{equa:eight-massive-coefficient-flat-value}
|D_I(t,r\omega)| \lesssim C_NA r^{-1}w^{1/2-\lambda}s^\delta.
\ee
This argument applies to the dynamic tensor, not to the reference metric. Reference derivatives appearing in the composition multiply \(u\) and are controlled by its dynamic Hardy norm.
\ese

\bse
\noindent{\bf 2.} \emph{Near-cone and outer regions.} We use a wider collar, defined by
\be
\label{equa:eight-massive-collar-definition}
b\coloneqq \frac{4\delta}{3(2\lambda-1)}, \qquad \mathcal C_b\coloneqq \{w\leq4t^b\}, \qquad \mathcal O_b\coloneqq \{w>4t^b\}.
\ee
Since \(b<1/3\), the collar lies in \(|q| \leq t/4\) once \(t\) exceeds a fixed threshold. The bounded preceding interval is controlled directly. On \(\mathcal O_b\), the Jacobian comparison and \(t\simeq s^2\) give
\be
\label{equa:eight-massive-coefficient-outer-multiplier}
J\zeta^{-2}|D_I| \lesssim C_NA s^{-1+ \delta}w^{1/2-\lambda} \lesssim C_NA s^{-1-\delta/3}.
\ee
On \(\mathcal C_b\),~\autoref{lem:eight-displaced-cone-coefficient} controls \((D_I)_N\). With \(\chi_\lambda= \lambda-1/2-\delta/2\) from \eqref{equa:eight-displaced-cone-exponents}, its three Jacobian-weighted terms have the positive gaps
\be
\label{equa:eight-massive-collar-bad-gaps}
2\chi_\lambda-b>\tfrac1{24}, \qquad 2-\delta-2b(2-\lambda)>\tfrac{25}{24}, \qquad 2-2\delta-2b(2-2\lambda)>\tfrac{17}{12}.
\ee
These inequalities follow from \(\lambda>3/4\), \(\delta<1/8\); their limiting values occur at \(\lambda=3/4\), \(\delta=1/8\). Thus \(J\zeta^{-2}|(D_I)_N| \lesssim C_N(A+A^2)s^{-1-\beta} \) in the collar.
\ese

\bse
\noindent{\bf 3.} \emph{Principal commutator terms and the positive spacetime integral.} The mass term commutes with translations and boosts. The principal part of the commuted equation is
\be
\label{equa:eight-massive-ordered-principal-commutator}
3\Box_gZ\fieldPhi-\coeff^{-1}Z\fieldPhi =Z(\mathcal S+W_\coeff)-3\sum_{\varnothing\ne I\subset\{1, \ldots, \ord Z\}}(\mathcal L_Ig^{-1})^{\alpha \beta} \del_\alpha \del_\beta Z_{I^c} \fieldPhi.
\ee
We put \(V=Z_{I^c} \fieldPhi\), \(U=Z\fieldPhi\). Every \(\del_\alpha V\) has commutation order at most \(p\) and boost rank at most \(k-\operatorname{rank}I\). Its first derivative, the Hessian of \(V\), is therefore measured by \(U^{\mathrm{ext}}_{p,k} \). This uses an additional translation, not a boost. Estimate~\eqref{equa:eight-massive-coefficient-outer-multiplier} controls all principal terms on \(\mathcal O_b\) directly.

In the collar, the semi-null contraction \eqref{equa:eight-low-matter-Lie-frame} separates \((D_I)_N\del_t^2V\) from good second derivatives and \(r^{-1} \del V\). The bad term uses \eqref{equa:eight-massive-collar-bad-gaps}. Each good second derivative is \(E\del_\alpha V\), up to \(r^{-1} \del V\); it belongs to the good bulk at the same order and rank. Young's inequality uses
\be
\label{equa:eight-massive-good-Young-factors}
X= \sqrt{j_gJ} \,w^{\mu-1/2}|E\del_\alpha V|, \qquad Y= \sqrt{j_gJ} \,w^{\mu+1/2}|D_I| \,| \del_tU|.
\ee
After division by the target kinetic density, its second factor has coefficient
\be
\label{equa:eight-massive-good-Young-remainder}
J\zeta^{-2}w|D_I|^2\lesssim C_NA^2s^{-3+2\delta}w^{3-2\lambda} \lesssim C_NA^2s^{-1-\beta_g}, \qquad \beta_g\coloneqq2-2\delta-2b(3-2\lambda)>\tfrac34.
\ee
The positive matrix \(g^{\mathcal N ab} \), the kinetic factor \(3\), and the finite number of commutators are included when choosing fractions whose sum is at most \(\epsilon\). The \(r^{-1} \) remainder has the integrable coefficient \(C_NAs^{-1-2\lambda+ \delta} \). This proves the dynamic part of \eqref{equa:eight-massive-principal-pairing}.

\noindent{\bf 4.} \emph{Reference coefficients and parameter dependence.} Applying an operator containing a translation to a coefficient of the reference metric gives \(\Ocal(mr^{-2}) \), giving the last integral in the conclusion. A coefficient of the reference metric differentiated by a nonempty boost-only operator is \(\Ocal(mr^{-1}) \), and its translated field operator has strictly smaller boost rank. This gives the penultimate integral. Finally, multiplication of the singular current identity by \(\coeff\) gives the uniform identity and its source pairing. Dividing the proved inequality by \(\coeff\) recovers the singular estimate with its singular energies. The commutator proof introduces no further negative power of \(\coeff\).
\ese
\end{proof}

\paragraph{Principal effective-field commutators at order \(N\)}

The auxiliary top-order estimate permits \(U^{\mathrm{ext}}_{N,N} \leq As^{1+ \delta} \). The metric and matter energies keep the smaller bound \(As^\delta \), as specified in \eqref{equa:eight-exterior-order-weight}. We use this distinction when a high metric coefficient multiplies a low effective-field Hessian.

\begin{proposition}[Principal commutators at the highest order in the effective-field equation]
\label{lem:eight-massive-top-principal}
Under the coarse bootstrap, propagated wave gauge, and prescribed exterior reference metric, one writes \(U_N=U^{\mathrm{ext}}_{N,N} \) and \(\alpha=2-2\lambda+2\delta \). For every \(\epsilon>0\), the principal-commutator pairing satisfies
\be
\label{equa:eight-massive-top-principal-pairing}
\begin{aligned}
| \mathcal S^{\Phi}_{\mathrm{comm};N,N}|& \leq\epsilon\mathcal G^{\Phi}_{\mathrm{good};N,N} +C_{N, \epsilon}(A+A^2) \int_{s_0}^{s_1} s^{-1-\beta}U_N(s)^2\, ds\\
& \quad +C_Nm\int_{s_0}^{s_1} s^{-1}U_N(s)^2\, ds+C_NA^2\int_{s_0}^{s_1} s^{\alpha-1}U_N(s) \, ds,
\end{aligned}
\ee
where \(\beta= \min\{\delta/3,1/24\} \). The positive bad-derivative and mass bulk terms remain available. The constants are independent of \(\coeff\) and of the bootstrap constant \(C_1\).
\end{proposition}

\begin{proof}
\bse

\noindent{\bf 1.} \emph{Coefficients below order \(N-4\).} We use the ordered commutator \eqref{equa:eight-massive-ordered-principal-commutator}. For \(|I| \leq N-5\), the radial and displaced-cone estimates in the proof of~\autoref{lem:eight-massive-principal-commutator} apply to \(D_I= \mathcal L_I(H-H^\star) \). Their restriction concerns the coefficient order, so the same proof applies with target energy \(U_N\). Each translated field operator \(\del_\alpha Z_{I^c} \fieldPhi\) has order at most \(N\); its good derivative is measured by \(\mathcal G^{\Phi}_{\mathrm{good};N,N} \). The assigned fractions sum to at most \(\epsilon\).

\noindent{\bf 2.} \emph{Metric coefficients at high differential order and wave Hessians at low orders.} If \(|I| \geq N-4\), then \(\ord Z_{I^c} \leq4\). The direct exterior estimate, extended across the merging strip by its hypersurface-radial derivative and flat-end trace, uses only the closing effective-field energy and gives
\be
\label{equa:eight-massive-top-low-Hessian}
| \del^2Z_{I^c} \fieldPhi| \lesssim A r^{-1}w^{-\mu}s^\delta.
\ee
Its safe differential cost is \(\ord Z_{I^c} +6\leq10\leq N-5\): one translation forms the first-derivative field, and the sphere and radial estimate uses five further commutations. Smooth inverse-metric composition and the dynamic Hardy estimate also give \(\|w^{\lambda-1}D_I\|_{L^2} \lesssim C_NM_N\), where \(M_N\) denotes the order-\(N\) metric energy and dynamic Hardy norm. The coarse bootstrap bounds \(M_N\) by \(C_NAs^\delta \). The contribution of the reference metric remains separate. Since \(J\zeta^{-1} \lesssim s\), \(w\lesssim r\), and \(r\gtrsim s^2\), we obtain
\be
\label{equa:eight-massive-top-high-coefficient}
\|J\zeta^{-1}w^\mu D_I^{\alpha \beta} \del_\alpha \del_\beta Z_{I^c} \fieldPhi\|_{L^2} \lesssim C_NA s^{1-2\lambda+ \delta}M_N\lesssim C_NA^2s^{\alpha-1}.
\ee
Pairing this source with \(U_N\) uses no positive bulk. Applying an operator containing a translation to a coefficient of the reference metric gives \(\Ocal(mr^{-2}) \), while a coefficient of the reference metric differentiated by a nonempty boost-only operator is \(\Ocal(mr^{-1}) \) and multiplies a translated field operator of strictly lower boost rank. Summing these terms gives the reference integral in \eqref{equa:eight-massive-top-principal-pairing}. At this auxiliary order, their sum is bounded by the small coefficient \(C_Nm/s\), without requiring a finer rank growth.

\noindent{\bf 3.} \emph{Growth and parameter dependence.} The forcing exponent leaves a strict margin in the allowed top-order estimate:
\be
\label{equa:eight-massive-top-growth-gap}
0<\alpha<\tfrac34, \qquad 1+ \delta-\alpha=2\lambda-1-\delta>\tfrac38.
\ee
The mass commutes, and every effective-field derivative above is a derivative of \(\fieldPhi\). Dividing the uniform current identity and its estimate by \(\coeff\) recovers the singular normalization, including all flux and bulk terms.
\ese
\end{proof}

\paragraph{Auxiliary top-order effective-field energy}

We next estimate the other terms of the uniform effective-field identity at order \(N\). The metric and matter energies are used with their coarse bound \(As^\delta \), while the effective field is allowed the growth \(As^{1+ \delta} \). The interface term has the sign given by \eqref{equa:eight-affine-effective-field-interface-density}.

\begin{proposition}[Energy bound for the effective field at the highest differential order]
\label{proposition:eight-massive-top-energy}
Under the finite coarse bootstrap, the potential hypotheses, and the propagated geometric conditions, assume on the prescribed foliation that \(H_N\leq0\) in \(t-1\leq r\leq5t/4\). One writes \(U_N=U^{\mathrm{ext}}_{N,N} \), \(\alpha=2-2\lambda+2\delta \), and \(\beta= \min\{\delta/3,1/24\} \). Denote the nonnegative effective-field interface flux by \(\mathcal F_N^\Phi\), and its positive good-derivative, bad-derivative, and mass bulk terms by \(\mathcal G^\Phi_{\mathrm{good};N,N} \), \(\mathcal G^\Phi_{\mathrm{bad};N,N} \), and \(\mathcal G^\Phi_{\mathrm{mass};N,N} \). For sufficiently small \(A\), one has
\be
\label{equa:eight-massive-top-full-energy}
\begin{aligned}
&2U_N(s_1)^2 + \mathcal F_N^\Phi+ \tfrac12\mathcal G^\Phi_{\mathrm{good};N,N} + \mathcal G^\Phi_{\mathrm{bad};N,N} + \mathcal G^\Phi_{\mathrm{mass};N,N} \\
& \leq2U_N(s_0)^2 +C_NA\int_{s_0}^{s_1} s^{-1}U_N(s)^2\, ds\\
& \quad +C_N(A+A^2) \int_{s_0}^{s_1} s^{-1-\beta}U_N(s)^2\, ds+C_NA^2\int_{s_0}^{s_1} s^{\alpha-1}U_N(s) \, ds.
\end{aligned}
\ee
Consequently, with constants independent of \(\coeff\), \(C_1\), and the terminal time, one has
\be
\label{equa:eight-massive-top-bound}
U_N(s_1) \leq C_N\bigl(U_N(s_0) +A^2\bigr){s_1}^{\alpha+C_NA}.
\ee
The choice \(C_NA<(2\lambda-1-\delta)/2\) leaves a strict margin below the permitted exponent \(1+ \delta \). With the initial-data bound and sufficiently large fixed \(C_1\), followed by a sufficiently small initial amplitude, this improves \(U_N\leq As^{1+ \delta} \) to \(U_N\leq(A/2)s^{1+ \delta} \).
\end{proposition}

For the expanding interface, Proposition~\ref{prop:eight-mobile-coupled-estimate} uses the same source pairings with the positive current of Proposition~\ref{prop:eight-mobile-positive-current}. Its positivity follows from the moving-boundary term in the equation~\eqref{equa:eight-mobile-absorption}.

\begin{proof}
\bse

\noindent{\bf 1.} \emph{Principal commutators and the trace of the matter stress tensor.} We use \eqref{equa:eight-exterior-effective-field-energy-definitions}; the coordinate-current energy is exactly \(2U_N^2\).~\autoref{lem:eight-massive-top-principal} estimates the principal commutators with at most one quarter of the good bulk. The mass of the reference metric satisfies \(m\lesssim A\). We write \(P_N\) and \(M_N\) for the order-\(N\) matter energy and the metric energy with its dynamic Hardy norm, respectively; both are bounded by \(C_NAs^\delta \).

In the trace source \(\mathcal S=-8\pi e^{-\fieldPhi}g(\del\psi, \del\psi) \), the flat null form contains a good matter derivative. A high good derivative is controlled by the hypersurface norm without \(\zeta \), whereas a low good derivative has the additional radial decay. The respective multipliers are
\be
\label{equa:eight-massive-top-null-source-factors}
J\zeta^{-1}| \del\psi_{\mathrm l}| \lesssim A s^{-1+ \delta}, \qquad J\zeta^{-2}|E\psi_{\mathrm l}| \lesssim A s^{-1-2\mu+ \delta}.
\ee
These terms contribute the source norm \(C_NA^2s^{-1+2\delta} \). A high dynamic metric coefficient and two low matter derivatives are estimated by Hardy; their multiplier is \(J\zeta^{-1}w^{\mu+1-\lambda}| \del\psi_{\mathrm l}|^2\lesssim A^2s^{-3+2\delta} \). A low metric coefficient and a high matter derivative satisfy the same bound. A high \(\fieldPhi\) in the exponential uses \(\fieldPhi = \sqrt{\coeff} \, \fieldvarphi\) and the positive mass norm. These corrections therefore have source norm at most \(C_{N, \coeff_0}A^2s^{-3+2\delta}(M_N+P_N+U_N) \). The matter good bulk has not been used.

\noindent{\bf 2.} \emph{Potential.} The potential factorization \(W_\coeff= (\fieldvarphi)^2\mathcal W_\coeff(\fieldPhi) \) has uniformly bounded rescaled coefficients. Estimate~\eqref{equa:eight-low-effective-stress-coefficient} gives \(| \fieldvarphi_{\mathrm l}| \lesssim A/(rw^\gamma) \), where \(\gamma= (1+ \delta)/2\). In the far region, the direct estimate implies the same bound since \(s^\delta w^{\gamma-\mu} \lesssim1\). With the high factor in the mass norm, the source multiplier is \(J\zeta^{-1}| \fieldvarphi_{\mathrm l}| \lesssim A/s\). A high differentiated smooth coefficient has two low effective-field factors and the stronger integrable multiplier \(C_{N, \coeff_0}A^2s^{-3+2\delta} \). All lower factors have order at most \(\lfloor N/2\rfloor\); their direct first derivatives use at most \(\lfloor N/2\rfloor+5\leq N-5\) commutations. The additional good matter derivative may use the order-\(N\) matter energy, which keeps the growth \(As^\delta \).

\noindent{\bf 3.} \emph{Deformation and signed weight terms.} The deformation is estimated from its algebraic expression, with \(j_g= \sqrt{| \det g|} \):
\be
\label{equa:eight-massive-top-deformation-algebra}
-j_gQ^{\mathrm{unif} \, \alpha \beta}[Z\fieldphi] \del_tg_{\alpha \beta} =3\del_t(j_gg^{\alpha \beta}) \del_\alpha Z\fieldPhi\del_\beta Z\fieldPhi+(\del_tj_g)|Z\fieldvarphi|^2.
\ee
The densitized wave gauge bounds the bad kinetic coefficient by good metric derivatives and \(r^{-1} \) metric components. In inverse-metric variables, the additional product is \(H\del g\). The Jacobian-weighted bounds for these contributions are \(C_NAs^{-1-2\lambda+ \delta} \) and \(C_NA^2s^{-3+2\delta} \). Mixed good and bad field derivatives use Young's inequality with a further assigned fraction of the good bulk; its remainder has coefficient \(C_NA^2s^{-3+2\delta} \). The good--good terms are absorbed since \(w| \del g| \lesssim C_NA\). We choose these fractions so that the total spent, including the principal commutators, is at most one half.

The mass coefficient is \(\del_tj_g= (j_g/2)g^{\alpha \beta} \del_tg_{\alpha \beta} \). Its flat trace is the spherical trace minus the \(L\underL\) contraction, so~\autoref{lem:eight-good-metric-characteristic} at rank zero and the quadratic inverse-metric correction give
\be
\label{equa:eight-massive-top-mass-deformation}
J| \del_tj_g| \lesssim C_NAs^{-1} +C_NA^2s^{-3+2\delta}.
\ee
The multiplier in this formula is \(J\), with no loss of \(\zeta \), because the mass is measured directly in the positive hypersurface energy. The reference metric is stationary. The adverse bulk \(\Omega^-\) vanishes in the region where light bending gives \(H_N\leq0\); in the remaining far region \(w\simeq r\), its energy coefficient is bounded by \(C_NAs^{-1-2\lambda+ \delta} \). Its signed origin remains the splitting \eqref{equa:eight-affine-current-bulk-splitting}.

\noindent{\bf 4.} \emph{Scalar energy inequality.} Since \(\alpha>2\delta \), the trace-source forcing is bounded by \(C_NA^2s^{\alpha-1} \). All integrable same-energy coefficients above are bounded by \(C_N(A+A^2)s^{-1-\beta} \). The positive interface, bad-derivative bulk, and mass bulk have not been spent. Substitution into the exterior identity proves \eqref{equa:eight-massive-top-full-energy}. For its scalar consequence, we take the positive square root of the monotone right-hand-side envelope, differentiate this envelope, and apply the linear Gronwall inequality. A positive regularization handles vanishing initial energy. The integrable coefficient contributes a bounded factor, while \(C_NA/s\) contributes \({s_1}^{C_NA} \); integration of the forcing gives \(C_NA^2{s_1}^\alpha \). This proves \eqref{equa:eight-massive-top-bound} and the improvement.
\ese
\end{proof}


\subsection{\texorpdfstring{Normalization and nonlinear terms at the highest order}{Normalization and nonlinear terms at the highest order}}
\label{section-13-5}

\paragraph{Coupled exterior energy estimate at two differential orders}

The metric and matter top energies can be estimated together with the two effective-field levels. The closing effective energy keeps its uniform scale, whereas the order-\(N\) identity receives the time weight \(s^{-2} \). This weight is applied to the hypersurface balance before integration. If \(\mathcal E_{\mathrm c;N,k} =2(U^{\mathrm{ext}}_{N,k})^2\), the identity becomes
\be
\label{equa:eight-top-normalized-positive-identity}
\del_s(s^{-2} \mathcal E_{\mathrm c;N,k}) +s^{-2} \del_s\mathcal F_{\mathrm c;N,k} +s^{-2} \del_s\mathcal G_{\mathrm c;N,k} +2s^{-3} \mathcal E_{\mathrm c;N,k} =s^{-2} \del_s\mathcal S_{\mathrm c;N,k}.
\ee
Here \(\mathcal S\) contains the signed sources, deformation, and adverse bulk. Thus the normalized identity keeps its nonnegative interface and bulk terms and adds the positive integral \(4\int s^{-3}(U^{\mathrm{ext}}_{N,k})^2\, ds\).

\paragraph{Effective-field stress at the highest differential order}
When the high effective factor has order greater than \(N-5\), the complementary factor has order at most four. At these orders the massive reconstruction uses only closing effective-field energies. In \(r\simeq t\), it gives
\be
\label{equa:eight-top-strong-low-effective-factor}
| \del Z\fieldPhi|+ |Z\fieldvarphi| \lesssim C_{N, \coeff_0}A t^{-2 + \delta}w^{1-\mu}, \qquad \ord Z\leq4.
\ee
We apply the mass reconstruction \eqref{equa:eight-good-metric-mass-reconstruction} at the closing order. The flat contraction is bounded by the sum of \((w/t)| \del^2Z\fieldPhi|\) and \(t^{-1}(| \del LZ\fieldPhi|+ | \del Z\fieldPhi|)\). Its direct closing bounds give \(At^{-2 + \delta/2}w^{1-\mu} \); the metric and source products are bounded by \(A^2t^{-2 + \delta}w^{1-\lambda-\mu} \) and \(A^2t^{-2 + \delta}w^{-2\mu} \). Applying the formula to a translation operator proves the derivative estimate, with largest safe count \(4+7=11\leq N-5\). Thus the stress multiplier relative to \(U^{\mathrm{ext}}_{N,k}/s\) and the metric energy satisfies
\be
\label{equa:eight-top-effective-stress-gap}
sJ\zeta^{-2}w^{\lambda-\mu}| \del\fieldPhi_{\mathrm l}| \lesssim A s^{-2 +2\delta}w^{1+ \lambda-2\mu} \lesssim A s^{-1-\epsilon_\Phi}, \qquad \epsilon_\Phi>\tfrac12.
\ee
If \(1+ \lambda-2\mu\geq0\), we use \(w\lesssim s^2\) and \(2\lambda-4\mu+2\delta<-3/2\); otherwise we use \(w\geq1\) and \(-2 +2\delta<-7/4\). The far region uses the direct bound and gives the stronger exponent \(2\lambda-4\mu+ \delta<-3/2\). The potential contribution is at least as strong. At the closing order, the direct hypersurface-energy estimate~\eqref{equa:audit-stress-direct-energy} applies instead; it leaves a coefficient \(C_NA/s\) and requires no division by the reference mass.

\paragraph{Matter coupling and the normalized good bulk} A high effective derivative of order at most \(N-5\) uses the closing effective good bulk and the direct lower matter factor. For a higher effective derivative, the lower matter operator has at most four commutations.~\autoref{lem:eight-low-matter-characteristic-Lie} gives its characteristic bound at boost rank \(j\) as \(CA r^{-1}w^{-\gamma}s^{\eta_j} \), where \(\gamma= (1+ \delta)/2\). If \(g_{k-j}(s)^2\) denotes the order-\(N\) effective good-bulk density multiplied by \(s^{-2} \), the term with a high good effective derivative is bounded by
\be
\label{equa:eight-top-matter-effective-good-pairing}
C_NA\int_{s_0}^{s_1} s^{-1/2 + \eta_j}g_{k-j}(s) \Fenergy^{\ME,N,k}_\mu(s, \psi) \, ds.
\ee
The multiplier is \(s\sqrt J\, \zeta^{-1}w^{1/2}| \del\psi_{\mathrm l}| \lesssim As^{-1/2 + \eta_j}w^{-\delta/2} \). At \(j=0\), Young's inequality spends an assigned fraction of the normalized effective good bulk and leaves \(C_NA^2s^{-1} \) times the matter energy. At \(j>0\), it gives a small current-rank coefficient and the explicit lower-rank bulk term
\be
\label{equa:eight-top-lower-rank-good-bulk}
C_NA\int_{s_0}^{s_1} s^{-1} \bigl(\Fenergy^{\ME,N,k}_\mu(s, \psi) \bigr)^2\, ds+C_NA\int_{s_0}^{s_1} s^{2\eta_j}g_{k-j}(s)^2\, ds.
\ee
The second integral is at most \(C_NA s_1^{2\eta_j}\mathcal Y_{k-j}(s_1)^2\), where \(\mathcal Y_{k-j}^2\) denotes any positive energy quantity keeping this lower-rank good bulk. In the global argument it will be the adapted energy defined in~\eqref{equa:eight-mobile-coupled-positive-energy}. A high bad effective derivative paired with a low good matter derivative has the integrable coefficient \(As^{-2\mu+ \delta} \) after normalization. Metric corrections, high smooth coefficients, and the matter-current remainder have integrable coefficients or use independently assigned good-bulk fractions.

These estimates are combined with the principal-commutator bounds in Proposition~\ref{prop:eight-mobile-coupled-estimate}. Its rank induction keeps the lower-rank bulk term in~\eqref{equa:eight-top-lower-rank-good-bulk} and uses only the small coefficient \(C_NA\).


\section{\texorpdfstring{Initial norms and high-order continuation}{Initial norms and high-order continuation}}
\label{section-14}


\subsection{\texorpdfstring{Initial norms and effective-field decay}{Initial norms and effective-field decay}}
\label{section-14-1}
\label{section-14-2}
\label{section-14-3}
\label{section-14-4}

The initial metric norm and the pointwise estimates from the effective-field mass term will be used in the global continuation and convergence arguments. We first isolate the dynamic metric contribution:
\be
\label{equa:eight-light-bending-dynamic-data}
a_{\mathrm{dyn}}\coloneqq\Fenergy^{\ME,p_{\mathbf{high}},p_{\mathbf{high}}}_\lambda(s_0,u).
\ee
\paragraph{Initial dynamic metric amplitude}

The amplitude \(a_{\mathrm{dyn}} \) can be expressed entirely in terms of the prescribed reduced Cauchy data. On the fixed initial graph \(\Sigma_0= \Mcal_{s_0} \), the traces of each commuted metric component are
\be
\label{equa:eight-dynamic-initial-traces}
f_Z(x)\coloneqq (Zu) (\fTime(s_0,|x|),x), \qquad b_Z(x)\coloneqq (n_0Zu) (\fTime(s_0,|x|),x), \qquad \ord Z\leq p_{\mathbf{high}}.
\ee
All components and operators are summed as in \eqref{equa:eight-exterior-wave-norm-definition}. The traces concern all ten reduced metric components. Their normal jets are determined recursively by the reduced equations and the chosen initial wave-coordinate completion; their largest total order is \(p_{\mathbf{high}} +1=N-4\). The comparison below applies to these commuted Cauchy traces already measured by the initial energy hypotheses.

\begin{lemma}[Initial metric norm comparison]
\label{lem:eight-dynamic-initial-norm-comparison}
Under the prescribed initial metric comparison, the following equivalence holds with constants independent of \(\coeff\), the bootstrap constant, and the terminal time:
\be
\label{equa:eight-dynamic-initial-norm-comparison}
a_{\mathrm{dyn}}^2\simeq \sum_{\ord Z\leq p_{\mathbf{high}}} \int_{\MME_{s_0}}(1+r)^{2\lambda} \Bigl(| \nabla_xf_Z|^2 + |b_Z|^2\Bigr) \, dx.
\ee
\end{lemma}

\begin{proof}
\bse
We write \(T_a= \del_a \fTime(s_0,|x|) \) and \(n_0=n^0\del_t+n^a \del_a \). The tangent vectors are \(\del_a+T_a \del_t\), so the reconstruction is
\be
\label{equa:eight-dynamic-initial-jet-reconstruction}
(\del_tZu)|_{\Sigma_0} = \frac{b_Z-n^a \del_af_Z}{n^0-n^aT_a}, \qquad (\del_aZu)|_{\Sigma_0} = \del_af_Z-T_a(\del_tZu)|_{\Sigma_0}.
\ee
The denominator is the positive normal length of \(dt-d\fTime(s_0,|x|) \) and is bounded below on the fixed initial graph. The positive wave-current comparison therefore gives
\be
\label{equa:eight-dynamic-initial-energy-comparison}
\calE^{\ME}_\lambda(s_0,Zu) \simeq \int_{\MME_{s_0}}w^{2\lambda} \Bigl(| \nabla_xf_Z|^2 + |b_Z|^2\Bigr) \, dx.
\ee
Here \(\zeta(s_0,r) \) has a fixed positive lower bound and \(w=2 +r-\fTime(s_0,r) \simeq1+r\). Summation proves the equivalence. Neither this reconstruction nor the current comparison divides by \(\coeff\).
\ese
\end{proof}

The aggregate initial bound \eqref{equa:seven-exterior-initial-assumptions} gives
\be
\label{equa:eight-dynamic-data-aggregate-bound}
a_{\mathrm{dyn}} \leq\mathsf X^{\mathrm{unif}}_{p_{\mathbf{high}},p_{\mathbf{high}}}(s_0) \leq C_{\mathrm{data}}s_0^\delta A_0.
\ee

\paragraph{Mass and the initial constraints}
The constraints also determine the scale of the reference mass. We consider smooth small constrained data on the fixed complete initial graph \(\Sigma_0\simeq\mathbb R^3\), without inner boundary, and take \(V=0\). We let \((\gamma_{\mathrm M},k_{\mathrm M})\) be the geometry induced by Minkowski spacetime on this graph, and set \(\delta\gamma\coloneqq g_0-\gamma_{\mathrm M}\) and \(\delta k\coloneqq k_0-k_{\mathrm M}\). The Minkowski time translation decomposes as \(T=\nu_{\mathrm M}n_{\mathrm M}+Y_{\mathrm M}\). Outside a fixed compact set, \(\nu_{\mathrm M}=1\), \(Y_{\mathrm M}=0\), \(\gamma_{\mathrm M}\) is Euclidean, and \(k_{\mathrm M}=0\). Pairing the linearized Hamiltonian constraint with \(\nu_{\mathrm M}\) and the linearized momentum constraint with \(-2Y_{\mathrm M}\) cancels their interior linear terms, since \(T\) is Killing. In particular, the identity \(\mathcal L_{Y_{\mathrm M}}\gamma_{\mathrm M}=2\nu_{\mathrm M}k_{\mathrm M}\) gives this cancellation for the variations of the second fundamental form, with the convention \(k=-\frac12\mathcal L_n\gamma\). We write \(\mu_0\coloneqq\mu_0^{\mathrm{eff}}+\mu_0^{\mathrm{mat}}\), \(J_0\coloneqq J_0^{\mathrm{eff}}+J_0^{\mathrm{mat}}\) for the sources in~\eqref{consttransHan}, and \(j_{\mathrm M}\coloneqq\sqrt{\det\gamma_{\mathrm M}}\). Integration of the exact constraints gives
\be
\label{equa:audit-constraint-charge}
16\pi m=\int_{\Sigma_0}\bigl(\nu_{\mathrm M}\mu_0-2Y_{\mathrm M}^aJ_{0a}\bigr)j_{\mathrm M}\,dx+\mathcal R_{\mathrm{quad}}.
\ee
Here \(\mathcal R_{\mathrm{quad}}\) is the nonlinear remainder after linearization. With \(D\) the connection of \(\gamma_{\mathrm M}\) and \(B_0\) a fixed ball containing the non-Euclidean part of the graph, integration by parts of \(\delta\gamma D^2\delta\gamma\) yields
\be
\label{equa:audit-constraint-quadratic-remainder}
|\mathcal R_{\mathrm{quad}}|\leq C_{s_0}\bigl(\|D\delta\gamma\|_2^2+\|\delta k\|_2^2+\|\delta\gamma\|_{L^2(B_0)}^2\bigr)\leq C_{s_0}(m^2+A_0^2).
\ee
These norms use the coordinate measure \(dx\) and the fixed metric \(\gamma_{\mathrm M}\). The compact terms involving \(k_{\mathrm M}\) and \(Y_{\mathrm M}\) satisfy the same bound. To justify the boundary limit, the initial weighted energies and Hardy inequality give \(r^\lambda Du,r^\lambda D^2u,r^{\lambda-1}u\in L^2\) on the Euclidean end. Since \(\lambda>1/2\), the terms \(uD^2u\) and \((m/r)D^2u\) are integrable. The weighted first-derivative bound gives radii \(R_j\to\infty\) such that \(R_j^{1+2\lambda}\int_{|x|=R_j}|Du|^2R_j^2\,d\omega\to0\); by Cauchy--Schwarz, the dynamic ADM flux is then \(o(R_j^{1/2-\lambda})\). The quadratic boundary terms vanish along this exhaustion. The exact constraints and the integrable remainder give the existence of the full limit of the linear ADM flux integral, which therefore equals \(16\pi m\). The source integral is bounded by \(C_{s_0}A_0^2\), uniformly in \(\coeff\): its terms are quadratic in \(\del\fieldPhi\), \(\fieldvarphi\), and \(\del\psi\), with \(\coeff\Ufrak(\fieldphi)\lesssim\fieldvarphi^2\). Absorption of \(C_{s_0}m^2\) for sufficiently small \(m\) gives
\be
\label{equa:audit-constraint-mass-bound}
m\leq C_{s_0}A_0^2.
\ee
Thus the constraints yield a quadratic mass scale, consistent with the families constructed in Proposition~\ref{prop:nontrivial-small-mass-prepared-data}. The reference mass can be much smaller than the perturbation amplitude; the expanding-interface estimates below allow this regime.

\paragraph{Initial data with an exact Schwarzschild exterior}
For a special class of data, domain of dependence determines the exterior solution directly. Suppose that, on \(\MME_{s_0}\), the conformal initial metric and second fundamental form are induced by Schwarzschild on the prescribed graph, the scalar Cauchy data vanish, and the lapse, shift, and reduced initial jets agree with its harmonic coordinates. On the curved merging portion, the second fundamental form is the induced Schwarzschild form. For \(s_0\geq2\), this entire exterior lies in the exact Schwarzschild part of the reference metric. A future causal curve satisfies
\be
\label{equa:audit-Schwarzschild-causal-speed}
\left|\frac{dr}{dt}\right|\leq D_m=\frac{r-m}{r+m}\leq1,\qquad \frac{d(t-r)}{dt}\geq1-D_m\geq0.
\ee
Consequently, the causal past of a point with \(t-r\leq1\), down to the initial graph, remains in the exterior. The graph is spacelike there, since \(D_m^{-1}-D_m(\del_r\fTime(s_0,r))^2>0\). Uniqueness and domain of dependence for the reduced system give, throughout the exterior of any existing development,
\be
\label{equa:audit-Schwarzschild-exact-exterior}
g=g^\star=g_{\mathrm{Sch}},\qquad u=\psi=\fieldPhi=0\quad\hbox{on }\MME.
\ee
The dynamic and scalar traces at the interface vanish, while the reference-metric traces keep their Schwarzschild values. The expanding construction below also treats general small exterior perturbations.

\paragraph{Pointwise estimates from the mass term}

To estimate the field pointwise on an exterior hypersurface, we use the mass terms of five additional commuted energies. Two sphere rotations cost at most four boost operators, and a radial derivative costs one translation operator. The mass term at these orders gives
\be
\label{equa:eight-radial-degenerate-effective-field-bound}
r w^{\mu}| \fieldphi|_{p,k} \lesssim D^{\mathrm{ext}}_{p+5,k+5}(s) = \coeff^{-1/2}U^{\mathrm{ext}}_{p+5,k+5}(s), \qquad p+5\leq N.
\ee
Indeed, for the restriction \(q_s(r, \omega)=Z\fieldphi(\fTime(s,r),r\omega) \), we have the identity
\be
\label{equa:eight-concluding-radial-mass-derivative}
\del_rq_s= (\del_r Z\fieldphi+ \del_r\fTime(s,r) \, \del_tZ\fieldphi)|_{\MME_s}.
\ee
Both translation derivatives are measured by the mass part of the next commuted energy. Local radial \(H^1\) and sphere \(H^2\) Sobolev, with the coordinate measure \(dx\), prove the estimate because \(0\leq\del_r\fTime<1\) and \(w\) varies by a bounded factor on unit radial intervals. The same calculation applied to \(\fieldvarphi\) gives the uniform pointwise estimate from the mass term. More explicitly,
\be
\label{equa:eight-uniform-radial-mass-value}
\begin{aligned}
r w^\mu| \fieldvarphi|_{p,k}& \lesssim U^{\mathrm{ext}}_{p+5,k+5}(s),&&p+5\leq N, \\
r w^\mu| \del\fieldPhi|_{p,k}& \lesssim\sqrt\coeff\,U^{\mathrm{ext}}_{p+6,k+5}(s),&&p+6\leq N.
\end{aligned}
\ee
The second line applies the first to a translation operator, using \(\fieldPhi = \sqrt\coeff\, \fieldvarphi\). All constants are uniform in \(s\) and \(\coeff\). The mass part controls these translation derivatives directly, with no loss from the hypersurface degeneracy. These pointwise bounds are distinct from the fixed-sphere \(L^2\) traces provided by the interface lemma.


\subsection{\texorpdfstring{High regularity at fixed parameter}{High regularity at fixed parameter}}
\label{section-14-5}

For this finite-time argument, we keep the unnormalized currents on the entire hypersurface. We set $W_\eta\coloneqq1$ on $\Mint$ and $W_\eta\coloneqq\crochet^\eta$ on $\MME$, for $\eta\in\{\lambda,\mu\}$. These weights are continuous across $\Lscr$, where both traces equal one. With the tensors from the equations~\eqref{equa:seven-exact-wave-energy} and~\eqref{equa:seven-exact-effective-field-tensors}, we define the squared energy
\be
\label{equa:audit-fixed-parameter-joined-energy}
\mathcal E^{\mathrm{fix}}_p(s)\coloneqq2\sum_{Z\preccurlyeq(p,p)}\int_{\Mcal_s}\Bigl(W_\lambda^2Q^{\mathrm{wave}}[Zu]+W_\mu^2e^{-\fieldPhi}Q^{\mathrm{wave}}[Z\psi]+W_\mu^2Q^{\mathrm{unif}}[Z\fieldphi]\Bigr)(N_s,\del_t)\,dx.
\ee
The metric contribution is summed over its Cartesian components. The normal density $N_s$ and the multiplier $\del_t$ have the same normalization on both parts of the hypersurface.

\begin{lemma}[High regularity on finite slabs]
\label{lem:audit-fixed-parameter-joined}
Suppose $N\geq19$, $\coeff>0$, and $s_1<+\infty$, and set $n_{\mathrm{low}}\coloneqq N-5$. Consider a smooth solution on $s_0\leq s<s_*\leq s_1$ whose metric, spacelike hypersurfaces, and timelike multiplier remain uniformly nondegenerate. Suppose the fields remain in a fixed interval on which the coefficients and potentials are smooth and
\be
\label{equa:audit-fixed-parameter-low-hypothesis}
\sup_{s_0\leq s<s_*}\mathcal E^{\mathrm{fix}}_{n_{\mathrm{low}}}(s)\leq M_{s_1}^2,\qquad \mathcal E^{\mathrm{fix}}_N(s_0)<+\infty.
\ee
Then one has
\be
\label{equa:audit-fixed-parameter-high-bound}
\sup_{s_0\leq s<s_*}\mathcal E^{\mathrm{fix}}_N(s)\leq C_{\coeff,s_1}\Bigl(\mathcal E^{\mathrm{fix}}_N(s_0)+\int_{s_0}^{s_1} R_N(\tau)^2\,d\tau\Bigr),
\ee
where $R_N$ depends only on the prescribed reference metric. The constant also depends on $N$, $M_{s_1}$, the coefficient interval, and the geometric nondegeneracy bounds, but is independent of $s_*$ and of the high energy on the evolving slab.
\end{lemma}

\begin{proof}
\bse
On a finite slab the current energy is equivalent to the weighted $L^2$ norm of $\del Zu$, $\del Z\psi$, $\del Z\fieldPhi$, and $\coeff^{-1/2}Z\fieldPhi$. The weights $W_\eta$ are comparable to $(1+r)^\eta$, with constants depending on $s_1$. The radial Hardy inequality, including the interior trace, also gives
\be
\label{equa:audit-fixed-parameter-Hardy}
\|W_\lambda(1+r)^{-1}Zu\|_2^2\leq C_{s_1}\mathcal E^{\mathrm{fix}}_N(s),\qquad Z\preccurlyeq(N,N).
\ee
We use the asymptotically Euclidean normalization to eliminate the constant at infinity. Adding the interior and exterior current identities cancels the interface flux exactly, as in the equation~\eqref{equa:eight-c-current-joined-identity}; this cancellation uses equality of the two weights and not a sign of the flux. The derivatives of the weights and the deformation of the multiplier contribute at most $C_{\coeff,s_1}\mathcal E^{\mathrm{fix}}_N$. All finite-slab comparisons may depend on $s_1$.

We verify the derivative count in the principal commutators. Since each generator is affine, their terms have the form $C_i\del^2Z_jv$, where $i\geq1$, $i+j\leq N$, and $v\in\{u,\psi,\fieldPhi\}$. Here $C_i$ is a metric coefficient commuted $i$ times. We set $a\coloneqq\lfloor(N+1)/2\rfloor$ and distinguish two placements.

\noindent{\bf 1.} \emph{Pointwise control of the metric coefficient.} If $i\leq a$, we estimate $C_i$ pointwise. Outside a fixed ball, two angular derivatives and the radial primitive give $|Z_i u|\leq C M_{s_1} r^{-\lambda-1/2}$. Two rotations cost at most four boosts, and the radial derivative is already measured by the energy; thus this estimate requires at most $i+4\leq n_{\mathrm{low}}$ commutations. On the bounded part of the hypersurface the ordinary Sobolev inequality applies. The complementary Hessian has at most $j+2\leq N+1$ derivatives and is measured by $\mathcal E^{\mathrm{fix}}_N$.

\noindent{\bf 2.} \emph{Hardy control of the metric coefficient.} If $i\geq a+1$, we place $C_i$ in the Hardy norm. The complementary Hessian satisfies $|\del^2Z_jv|\leq C M_{s_1}(1+r)^{-1}W_\eta^{-1}$, with $\eta=\lambda$ for $u$ and $\eta=\mu$ for the scalar fields. Its pointwise estimate costs at most $j+6\leq n_{\mathrm{low}}$ commutations. The weight comparison follows from
\be
\label{equa:audit-fixed-parameter-weight-placement}
W_\eta|C_i\del^2Z_jv|=\bigl(W_\lambda(1+r)^{-1}|C_i|\bigr)\bigl((1+r)W_\eta W_\lambda^{-1}|\del^2Z_jv|\bigr).
\ee
The second factor is bounded by $CM_{s_1}$, since $W_\lambda\geq1$. The inequalities $a+4\leq N-5$ and $N-a-1+6\leq N-5$ hold for every $N\geq19$. For quadratic products of first derivatives, the pointwise factor costs at most $\lfloor N/2\rfloor+5\leq N-5$ orders. Differentiation of the smooth coefficients has the same property: only the factor with the largest order is measured by the high energy. The remaining first derivatives give the decay required by the Hardy factor. The fixed mass $(3\coeff)^{-1}$ commutes with every generator; the derivatives of the nonlinear potentials are bounded with constants depending on $\coeff$. Consequently, if $F_Z$ denotes a commuted source after the principal operator has been kept on the left, its appropriate weight satisfies
\be
\label{equa:audit-fixed-parameter-source-bound}
\sum_{Z\preccurlyeq(N,N)}\|W_{\mathrm{target}}F_Z\|_2\leq C_{\coeff,s_1}\sqrt{\mathcal E^{\mathrm{fix}}_N(s)}+R_N(s).
\ee
Here $W_{\mathrm{target}}$ is $W_\lambda$ for the metric equation and $W_\mu$ for either scalar equation. The Schwarzschild coefficients remain separate: their explicit decay gives pointwise multiplier bounds, while the dynamic Hardy norm applies only to $u$. The prescribed source $\mathcal R^\star$ vanishes in the exact Schwarzschild region; only its localized error and derivatives enter $R_N$, which is square-integrable in time on every finite slab. The current identities, followed by Cauchy--Schwarz, therefore give
\be
\label{equa:audit-fixed-parameter-Gronwall}
\mathcal E^{\mathrm{fix}}_N(s)\leq C_{s_1}\mathcal E^{\mathrm{fix}}_N(s_0)+C_{\coeff,s_1}\int_{s_0}^s\mathcal E^{\mathrm{fix}}_N(\tau)\,d\tau+C_{s_1}\int_{s_0}^sR_N(\tau)^2\,d\tau.
\ee
We first integrate on radially truncated domains and remove the outer boundary along a sequence using the finite weighted energy. Gronwall's inequality proves the conclusion.
\ese
\end{proof}


\section{\texorpdfstring{An outgoing interface for a small reference mass}{An outgoing interface for a small reference mass}}
\label{section:small-mass-moving-interface}


\subsection{\texorpdfstring{The moving foliation and radial estimates}{The moving foliation and radial estimates}}
\label{section:mobile-foliation}
\paragraph{Purpose}
We now construct the foliation adapted to the interface introduced in Section~\ref{section-12-1}. We prove positivity of its currents, transfer the interior estimates to the adapted domains, and close the global energy argument. The initial hypersurface and its data are unchanged.

\paragraph{An expanding interface}
We use the interface $q_b$, $r_b$, and $\Lscr_b$ from the equation~\eqref{equa:eight-mobile-interface-definition}, and its function $h$ and exponent $e_0$. We additionally require $K(A+A^2)t_i^{-e_0}\leq1/8$.

We distinguish the adapted geometric objects from the fixed-interface objects. We set $d_b(t)\coloneqq q_b(t)+r_b(t)q_b'(t)$. The function $t^2-r_b(t)^2$ is strictly increasing since its derivative is $2d_b\geq5/4$. Its inverse and the radial cutoff define
\be
\label{equa:audit-moving-interface-foliation-definition}
t_b(s)^2-r_b(t_b(s))^2=s^2,\qquad R_b(s)\coloneqq r_b(t_b(s)),\qquad \xi_b(s,r)\coloneqq1-\chi_{\mathrm{fol}}(r-R_b(s)).
\ee
The adapted time, Jacobian, and energy coefficient are defined by
\be
\label{equa:audit-moving-interface-foliation-coefficients}
\del_rT_b=\frac{r\xi_b}{\sqrt{s^2+r^2}},\quad T_b(s,0)=s,\quad J_b\coloneqq\del_sT_b,\quad \zeta_b^2\coloneqq1-\frac{r^2\xi_b^2}{s^2+r^2}.
\ee
The adapted regions and sphere are
\be
\label{equa:audit-moving-interface-domains}
\begin{aligned}
\Mcal_s^b&\coloneqq\{(T_b(s,|x|),x)\},&{\Mint}_s^b&\coloneqq\Mcal_s^b\cap\{r\leq R_b(s)\},\\
{\MME}_s^b&\coloneqq\Mcal_s^b\cap\{r\geq R_b(s)\},&\mathbb S_s^b&\coloneqq{\Mint}_s^b\cap\Lscr_b.
\end{aligned}
\ee
For any of the wave, matter, or effective-field stress tensors $Q$ used in the current identities, we define its adapted energy and the coordinate normal by
\be
\label{equa:audit-moving-interface-energy-definition}
\mathcal E_b^Q(s)\coloneqq2\int_{r\leq R_b(s)}Q(N_b,\del_t)\,dx+2\int_{r\geq R_b(s)}w_b^{2\eta}Q(N_b,\del_t)\,dx,\qquad N_b\coloneqq-j_g g^{-1}(dt-\del_rT_b\,dr).
\ee
The exterior exponent is $\eta=\lambda$ for the metric and $\eta=\mu$ for either scalar field. The commuted energies are the sums of these expressions over the original ordered operators; their singular or uniform effective-field normalization is kept. The two integrals also define the separate interior and exterior energies. Every conditional estimate below refers to these adapted domains, weights, and energies.

\begin{lemma}[Geometry of the expanding interface]
\label{lem:audit-moving-interface-geometry}
The interface determines an increasing radius $R_b(s)$ on each hyperboloid and a Euclidean--hyperboloidal foliation with the following properties.
\begin{itemize}
\item The initial hypersurface is unchanged. The new interior is the ball $r\leq R_b(s)$ on $t^2-r^2=s^2$, and $R_b(s)\simeq s^2$, $R_b'(s)\simeq s$.
\item The Jacobian, the energy coefficient, and the exterior radial trace inequalities satisfy their previous bounds, with constants independent of the parameter $\coeff$ and of the terminal time. The exterior weight equals one on the interface.
\item The ray estimates keep their differential orders and weights. The moving-boundary identities are given below, including the full sphere traces.
\end{itemize}
\end{lemma}
\begin{proof}
\bse
Differentiation of the defining equation for $t_b$ gives
\be
\label{equa:audit-moving-interface-radius}
\frac{d}{dt}\bigl(t^2-r_b(t)^2\bigr)=2d_b(t)\geq\frac54,\qquad t_b'(s)=\frac{s}{d_b(t_b(s))},\qquad R_b'(s)=\frac{s(1-q_b'(t_b(s)))}{d_b(t_b(s))}.
\ee
In particular, we obtain
\be
\label{equa:audit-moving-interface-times}
t_b(s)=\frac{s^2+q_b(t_b(s))^2}{2q_b(t_b(s))},\qquad R_b(s)=\frac{s^2-q_b(t_b(s))^2}{2q_b(t_b(s))},\qquad R_b(s_0)=\frac{s_0^2-1}{2}.
\ee
The definitions of the adapted cutoff and time agree with the original ones at $s=s_0$. On the interval $R_b\leq r\leq R_b+1$, differentiation, with $a(s,r)\coloneqq r/\sqrt{s^2+r^2}$, yields the formula
\be
\label{equa:audit-moving-interface-Jacobian}
J_b=\frac{s}{\sqrt{s^2+r^2}}-\int_{R_b}^r\del_sa(s,\rho)\chi_{\mathrm{fol}}(\rho-R_b)\,d\rho+R_b'\int_{R_b}^r a(s,\rho)\chi_{\mathrm{fol}}'(\rho-R_b)\,d\rho.
\ee
Since $\del_sa\leq0$, $a\simeq1$, and $R_b'\simeq s$ on this interval, we obtain
\be
\label{equa:audit-moving-interface-comparisons}
J_b\simeq\xi_b\frac{s}{\sqrt{s^2+r^2}}+(1-\xi_b)s,\qquad \zeta_b^2=1-\frac{r^2\xi_b^2}{s^2+r^2},\qquad J_b\zeta_b^{-2}\lesssim s\quad\hbox{in }{\MME}^b.
\ee
The time $T_b$ equals $t_b$ at $r=R_b$ and is constant with value in $[t_b,t_b+1]$ for $r\geq R_b+1$. We keep the exterior weight $w_b$ of the equation~\eqref{equa:eight-mobile-exterior-weight}, which equals one on the interface. In the merging interval, $1-\del_rT_b\gtrsim t^{-1}$, whereas $th(t)\leq K(A+A^2)t_i^{-e_0}$. After decreasing the amplitude, the radial derivative satisfies
\be
\label{equa:audit-moving-interface-radial-weight}
\frac{d}{dr}w_b(T_b(s,r),r)=1-(1+h(T_b(s,r)))\del_rT_b(s,r)\geq c(1-\del_rT_b(s,r))>0.
\ee
In the flat region this derivative is one. Thus the exterior of the adapted hypersurface is exactly its portion with $t-r\leq q_b(t)$. The weight is comparable to $1+r-R_b(s)$ on this hypersurface. The radial Hardy proof and its sphere term therefore apply, with a fixed change of constants; the initial weighted norms remain uniformly equivalent. The new exterior is contained in $r\geq t-1$, where the prescribed metric is exactly Schwarzschild.
\ese

\bse
For a ray through $(t,x)$, we keep its parametrization $(\lambda t/s,\lambda x/s)$. If $s_0r/s\leq R_b(s_0)$, we set $\lambda_{\mathrm{in}}=s_0$. Otherwise $s_0t/s>t_i$, and the function $\lambda(t-r)/s-q_b(\lambda t/s)$ is defined and strictly increasing for $\lambda\geq s_0$; its unique zero is the entry parameter. Consequently, we have
\be
\label{equa:audit-moving-interface-ray-entry}
\begin{aligned}
&\max\left\{s_0,\frac{3s}{4(t-r)}\right\}\leq\lambda_{\mathrm{in}}\leq\max\left\{s_0,\frac{s}{t-r}\right\},\\
&\lambda_{\mathrm{in}}\frac{t-r}{s}=q_b\left(\frac{\lambda_{\mathrm{in}}t}{s}\right)\qquad\hbox{at an interface entry}.
\end{aligned}
\ee
The ray field and its reduced equation are unchanged. At an interface entry, $t_b\simeq\lambda_{\mathrm{in}}^2$, so all interface powers in the oscillator estimates remain valid. The boosts are tangent to the same hyperboloids; at fixed $s$ their one-sided Sobolev and interpolation arguments take place in the ball $B_{R_b(s)}$, with the same uniform rescaled geometry.

We finally keep the moving-boundary contributions. For $E_v(s)\coloneqq s^{-1}\|v\|_{L^2(B_{R_b(s)})}$, differentiation gives
\be
\label{equa:audit-moving-interface-radial-profile}
\frac{dE_v^2}{ds}+\frac2sE_v^2=\frac2{s^2}\int_{B_{R_b(s)}}v\frac{s}{t}\del_tv\,dx+\frac{R_b'(s)}{s^2}\|v\|_{L^2(\mathbb S_s^b)}^2.
\ee
The trace integral in the radial profile estimate is therefore $\int_{s_0}^s\tau^{-2}R_b'(\tau)\|v\|_{L^2(\mathbb S_\tau^b)}^2\,d\tau$. For the weighted variation of a coefficient, we define $s_{\mathrm{in}}(x)=s_0$ on the initial ball and $s_{\mathrm{in}}(x)=R_b^{-1}(|x|)$ outside it. Then $f_0=f_s-f_{\mathrm{in}}$ vanishes on the moving boundary, and the identity~\eqref{equa:seven-e-weighted-identity} is unchanged, including its damping. Its initial-value term instead satisfies the identity
\be
\label{equa:audit-moving-interface-initial-profile}
\|f_{\mathrm{in}}\|_{L^2(B_{R_b(s)})}^2=\|f_{s_0}\|_{L^2(B_{R_b(s_0)})}^2+\int_{s_0}^sR_b'(\tau)\|f\|_{L^2(\mathbb S_\tau^b)}^2\,d\tau.
\ee
The comparison $R_b'(\tau)\simeq\tau$ preserves every resulting exponent and commutation rank.
\ese
\end{proof}

\paragraph{Outgoing Hardy estimate on the moving exterior}

We write ${\MME}^b_s$ for the exterior part of the adapted hypersurface, $\mathbb S^b_s$ for its intersection with $\Lscr_b$, and $t_b(s)$ for the time of that intersection. The space-time domain between two such hypersurfaces is denoted by ${\MME}^b_{[s_0,s_1]}$. In the following radial calculation we use $w=2+r-t$, which is constant on an outgoing Minkowski ray and is uniformly comparable to $w_b$. The exterior energy norm $\Fenergy^{b,p,k}_\lambda(s,u)$ is the norm $\Fenergy^{\ME,p,k}_\lambda(s,u)$ restricted to ${\MME}^b_s$, with weight $w_b^\lambda$.

\begin{lemma}[Outgoing Hardy estimate with the moving boundary]
\label{lem:eight-mobile-outgoing-Hardy}
Under the exterior energy and spacelike-hypersurface comparisons, with $\lambda>1/2$ and $h(t)=K(A+A^2)t^{-1-e_0}$, one has, with every sum taken over $Z\preccurlyeq(p,k)$,
\be
\label{equa:eight-mobile-outgoing-Hardy-bound}
\begin{aligned}
&\sum_Z\int_{{\MME}^b_{[s_0,s_1]}}w_b^{2\lambda-1}r^{-2}|Zu|^2\,dx\,dt\\
&\lesssim\sum_Z\left(\int_{{\MME}^b_{[s_0,s_1]}}w_b^{2\lambda-1}|\delsN Zu|^2\,dx\,dt+\|w_b^{\lambda-1}Zu\|_{L^2({\MME}^b_{s_1})}^2\right)\\
&\quad+CK(A+A^2)\int_{s_0}^{s_1} s^{-3-2e_0}\bigl(\Fenergy^{b,p,k}_\lambda(s,u)\bigr)^2\,ds.
\end{aligned}
\ee
This estimate preserves the order and rank. The constant $C$ depends only on the fixed exponent and foliation comparisons, independently of $K$.
\end{lemma}

\begin{proof}
\bse
For $v=Zu$, we put $p=t+r$, $q=t-r$, and $a\coloneqq2\lambda-1$. Since $\del_p=L/2$ and $\del_pw=0$, we have the identity
\be
\label{equa:eight-mobile-Hardy-exact-line}
\del_p(w^a r|v|^2)=\tfrac12w^a|v|^2+2w^a r\operatorname{Re}(\overline v\del_pv).
\ee
A segment at fixed $(q,\omega)$ starts on the initial hypersurface and ends either on the upper hypersurface or on the moving boundary. The initial term has a favorable sign. If $B_{\mathrm{in}}$, $B_{\mathrm{top}}$, and $B_{\mathrm{mov}}$ denote the corresponding integrals of $w^ar|v|^2$ in $(q,\omega)$, integration gives
\be
\label{equa:eight-mobile-Hardy-exact-integral}
\tfrac12\int w^a|v|^2\,dp\,dq\,d\omega+B_{\mathrm{in}}
=B_{\mathrm{top}}+B_{\mathrm{mov}}-2\int w^ar\operatorname{Re}(\overline v\del_pv)\,dp\,dq\,d\omega.
\ee
The upper-hypersurface term is bounded by the terminal Hardy norm as in the equation~\eqref{equa:seven-e-outgoing-Hardy-top}. On the moving part, $-dq=h(t)\,dt$, and the additional term is
\be
\label{equa:eight-mobile-Hardy-boundary}
B_{\mathrm{mov}}=\int_{t_i}^{t_b(s_1)}(2-q_b(t))^a\frac{h(t)}{r_b(t)}\|v(t,\cdot)\|_{L^2(r=r_b(t))}^2\,dt.
\ee
The radial trace argument of the equation~\eqref{equa:seven-e-exterior-sphere-trace} applies on each adapted hypersurface. Its kernel remains bounded because the merging interval has width one, $w_b=1$ at its inner boundary, and the radial hypersurface derivative is controlled by the positive energy. Consequently, we obtain
\be
\label{equa:eight-mobile-Hardy-boundary-bound}
\sum_{Z\preccurlyeq(p,k)}B_{\mathrm{mov}}[Zu]\lesssim CK(A+A^2)\int_{s_0}^{s_1} s^{-3-2e_0}\bigl(\Fenergy^{b,p,k}_\lambda(s,u)\bigr)^2\,ds.
\ee
We used $r_b(t)\simeq t\simeq s^2$ and $dt_b/ds\lesssim s$. Cauchy--Schwarz in the last term of the equation~\eqref{equa:eight-mobile-Hardy-exact-integral}, followed by $dx\,dt=\tfrac12r^2dp\,dq\,d\omega$, proves the result. The two weights are comparable throughout the domain. The finite-energy passage from radial truncations is the same as in the equation~\eqref{equa:seven-e-outgoing-Hardy-bulk}.
\ese
\end{proof}


\subsection{\texorpdfstring{Positive energy currents on the moving exterior domain}{Positive energy currents on the moving exterior domain}}
\label{section:mobile-currents}
\paragraph{Decay of the outgoing metric coefficient}
The coefficient estimate is uniform over the interfaces satisfying the two amplitude restrictions in Section~\ref{section:mobile-foliation}. In particular, its constant is independent of $K$. To verify this dependence, we use the coarse metric bound $\Fenergy^{b,N,N}_\lambda(s,u)\leq As^\delta$ in the additional Hardy term from the equation~\eqref{equa:eight-mobile-Hardy-boundary-bound}. Since $K(A+A^2)t_i^{-e_0}\leq1/8$, we obtain
\be
\label{equa:mobile-Hardy-uniform-interface-constant}
K(A+A^2)\int_{s_0}^{s_1} s^{-3-2e_0}\bigl(\Fenergy^{b,N,N}_\lambda(s,u)\bigr)^2\,ds\leq\frac{A^2t_i^{e_0}s_0^{-2-2e_0+2\delta}}{8(2+2e_0-2\delta)}\leq C_{s_0,\delta}A^2.
\ee
The foliation comparisons are also uniform, since $3/4\leq q_b\leq1$ and $th(t)\leq1/8$. Thus the outgoing Hardy estimate gives the density bound used in the equation~\eqref{equa:eight-displaced-cone-density-bulk}, with a constant independent of $K$. Under the coarse metric energy and good spacetime bounds, the propagated wave gauge, and the dynamic decay at spatial infinity, we consequently obtain at every available exterior point
\be
\label{equa:eight-mobile-global-null-coefficient}
|H_N-H_N^\star|(t,x)\leq C_N(A+A^2)t^{-1-e_0}\qquad(r\geq r_b(t),\ t\geq t_i).
\ee
Indeed, for \(w\geq4t^b\), the radial estimate \eqref{equa:eight-massive-coefficient-flat-value} and \(s\simeq t^{1/2}\) give
\be
\label{equa:eight-mobile-outer-null-coefficient}
|H-H^\star|\leq C_NA r^{-1}t^{\delta/2}w^{1/2-\lambda}\leq C_NA r^{-1}t^{-\delta/6}.
\ee
For \(w\leq4t^b\), the three terms of \eqref{equa:eight-displaced-cone-coefficient} have gains
\be
\label{equa:eight-mobile-collar-null-gains}
\chi_\lambda-b/2>1/48,\qquad1-\delta/2-b(2-\lambda)>25/48,\qquad1-\delta-b(2-2\lambda)>17/24.
\ee
They are all greater than \(e_0\), since \(\delta<1/8\). In the far region \(r\geq5t/4\), the radial estimate has the stronger decay \(C_NA t^{-1-\chi_\lambda}\). A bounded initial time interval is included by increasing the constant, depending on \(s_0\) and the fixed exponents. The estimates require only the orders already used for the radial and outgoing-cone coefficient bounds.

\begin{proposition}[Positive outgoing current on the moving interface]
\label{prop:eight-mobile-positive-current}
Assume the coarse near-Minkowski range \(-g^{00}\geq1/2\), \((g^{ab})\geq\tfrac12 I\), and \((g^{\mathcal N ab})\geq\tfrac12 I\). The constant \(K\) is increased, if necessary, so that \(h\geq4|H_N-H_N^\star|\). For \(\Box_gv=F_v\), the wave current is \(\mathcal J^\alpha=-2Q^{\mathrm{wave}\,\alpha}{}_0[v]\). The following quantities are defined by
\be
\label{equa:eight-mobile-wave-flux-density}
K_{b,g}[v]\coloneqq D_g[v]+h\mathcal J^0[v],\qquad\mathcal J^0[v]=-g^{00}|\del_tv|^2+g^{ab}\del_av\del_bv.
\ee
Then one has
\be
\label{equa:eight-mobile-wave-flux-lower-bound}
K_{b,g}[v]\geq\tfrac12|\delsN v|^2-H_N^\star|\del_tv|^2+\tfrac h2\mathcal J^0[v]\geq0.
\ee
On the finite interface segment $\Lscr_b^{s_0,s_1}\coloneqq\Lscr_b\cap\{s_0\leq s\leq s_1\}$, the outward flux has the coordinate normalization
\be
\label{equa:eight-mobile-wave-flux-measure}
F_b[v](s_1)\coloneqq\int_{\Lscr_b^{s_0,s_1}}j_gK_{b,g}[v]r_b(t)^2\,dt\,d\omega=\int_{\Lscr_b^{s_0,s_1}}\frac{j_gK_{b,g}[v]}{1+h(t)}\,dx.
\ee
For the uniform effective-field current, its density is
\be
\label{equa:eight-mobile-effective-flux-density}
K_{b,g}^{\mathrm c}[v,z]\coloneqq3D_g[v]+z^2+h\mathcal J_{\mathrm c}^0[v,z],\qquad\mathcal J_{\mathrm c}^0[v,z]=3\mathcal J^0[v]+z^2,
\ee
where \(v=Z\fieldPhi\), \(z=Z\fieldvarphi\). It satisfies
\be
\label{equa:eight-mobile-effective-flux-lower-bound}
K_{b,g}^{\mathrm c}[v,z]\geq\tfrac32|\delsN v|^2-3H_N^\star|\del_tv|^2+z^2+\tfrac h2\mathcal J_{\mathrm c}^0[v,z]\geq0.
\ee
The weighted wave current \(\mathcal P^\alpha=w_b^{2\eta}\mathcal J^\alpha\) has the identity
\be
\label{equa:eight-mobile-current-identity}
\nabla_\alpha\mathcal P^\alpha+2\eta w_b^{2\eta-1}K_{b,g}[v]=-2w_b^{2\eta}F_v\del_tv-w_b^{2\eta}Q^{\mathrm{wave}\,\alpha\beta}[v]\del_tg_{\alpha\beta}.
\ee
The same formula holds for the effective-field current, with \(K_{b,g}^{\mathrm c}\), its source, and its stress tensor. The matter current has the positive multiplier \(e^{-\fieldPhi}\); derivatives of this factor stay in its usual signed source. All constants are independent of \(\coeff\).
\end{proposition}

\begin{proof}
The moving boundary is \(q-q_b(t)=0\). Its exterior directed covector is proportional to \(dq+hdt\), so its contraction with \(\mathcal J\) is \(D_g+h\mathcal J^0\). In Cartesian components the mixed terms cancel in \(\mathcal J^0\). We write \(H_N^{\mathrm d}\coloneqq H_N-H_N^\star\). The reference metric satisfies \(H_N^\star\leq0\), while
\be
\label{equa:eight-mobile-absorption}
\tfrac h2\mathcal J^0\geq\tfrac h4|\del_tv|^2\geq|H_N^{\mathrm d}|\,|\del_tv|^2.
\ee
This proves the wave lower bound. The same calculation with the factor three and the positive mass density proves the effective-field lower bound. Finally, \(dw_b(\mathcal J)=-K_{b,g}\), and the product rule proves the current identity. There is no derivative of \(h\) in this calculation. The singular current is obtained by dividing the uniform current by \(\coeff\), as before.

For a target point in the moving exterior, its outgoing cone has parameter \(q_0\leq q_b(T)\). At earlier times \(t\leq T\), monotonicity gives \(q_0\leq q_b(t)\). Hence the fixed outgoing-cone segment and every outward radial slice used in the coefficient estimate remain in the moving exterior. We let \(S_T\) be the adapted hypersurface through \((T,T-q_0)\). Since \(0\leq\del_rT_b\leq1\), the function \(T_b(S_T,t-q_0)-t\) is nonincreasing and vanishes at \(T\). Thus these past radial slices lie below that terminal hypersurface. For \(T\geq2T_0\), the interval \([T/2,T]\) also lies above the entire initial hypersurface, so every integral in the outgoing-cone proof is available in the finite slab. On a bounded preceding time interval the coefficient follows from the initial norms and the coarse estimates, with a fixed constant. In \(w\geq4t^b\), all outward radial slices lie in the flat part of the adapted foliation; the same radial kernel applies. The constants in \(s\simeq t^{1/2}\) and \(w_b\simeq w\) are uniform.

The outgoing Hardy estimate in Lemma~\ref{lem:eight-mobile-outgoing-Hardy} includes every terminal contribution on the moving boundary. It gives the density integral required in the outgoing-cone estimate with the same orders and polynomial growth.
\end{proof}

\paragraph{Positive energy on the adapted hypersurfaces}
We set $a\coloneqq\del_rT_b$ and $n_b\coloneqq dt-a\,dr$. In the merging interval, the inverse metric has the contraction
\be
\label{equa:eight-mobile-leaf-normal}
g^{-1}(n_b,n_b)=-\zeta_b^2+H_N+2(1-a)(H^{0r}-H^{rr})+(1-a)^2H^{rr}.
\ee
Here $H^{0r}=\omega_aH^{0a}$ and $H^{rr}=\omega_a\omega_bH^{ab}$. Since $1-a\leq\zeta_b^2$ and $\zeta_b^2\gtrsim t^{-1}$ in this interval, the last two terms are bounded by $C|H|\zeta_b^2$, while the equation~\eqref{equa:eight-mobile-global-null-coefficient} bounds the positive part of $H_N$ by $C_N(A+A^2)t^{-1-e_0}$. The Schwarzschild term is nonpositive and has size $\Ocal(m/t)$. For small $A$ and $m$, we therefore have $-g^{-1}(n_b,n_b)\simeq\zeta_b^2$. On the flat exterior, $a=0$ and $-g^{00}\simeq1$. The adapted exterior energies thus control the same weighted first derivatives and scalar mass terms as the original hypersurface energies.

\paragraph{Normalization of the moving interface flux}

For a coordinate current $\mathcal J^\alpha=-2Q^\alpha{}_0$, the outward exterior directed surface element is
\be
\label{equa:eight-mobile-directed-surface}
d\Sigma_\alpha^{b,\mathrm e}=j_gr_b(t)^2(1+h(t),-\omega)_\alpha\,dt\,d\omega.
\ee
The weight $w_b$ equals one on the boundary. Thus, with $K_{b,g}$ from the equation~\eqref{equa:eight-mobile-wave-flux-density}, the coordinate flux is exactly
\be
\label{equa:eight-mobile-coordinate-flux-normalization}
F_{\mathrm w}^b(s_1)=\int_{t_i}^{t_b(s_1)}\int_{\mathbb S^2}j_gK_{b,g}[v]r_b(t)^2\,d\omega\,dt
=\int_{\Lscr_b\cap\{s_0\leq s\leq s_1\}}\frac{j_gK_{b,g}[v]}{1+h}\,dx.
\ee
The second formula uses $dr_b=(1+h)dt$. The matter current inserts $e^{-\fieldPhi}$; the effective-field current uses $K_{b,g}^{\mathrm c}[Z\fieldPhi,Z\fieldvarphi]$. Since the interior orientation is opposite and its geometric current is $-\mathcal J/2$, its flux satisfies
\be
\label{equa:eight-mobile-flux-opposite-normalizations}
F_\bullet^b=2\mathsf F_\bullet^{\Lscr_b,\mathrm{int}}\geq0,\qquad
\mathsf F_{\mathrm{eff}}^{\mathrm{unif},\Lscr_b}=\coeff\mathsf F_{\mathrm{eff}}^{\mathrm{sing},\Lscr_b}.
\ee
Here $\bullet\in\{\mathrm w,\psi,\mathrm c\}$, with the same component and operator sums as in the equation~\eqref{equa:seven-exact-lateral-fluxes}. In coordinate normalization the two balance laws therefore read
\be
\label{equa:eight-mobile-opposite-energy-identities}
E^i_\bullet-F_\bullet^b=E^i_\bullet(s_0)+S^i_\bullet,\qquad
E^e_\bullet+F_\bullet^b+B_\bullet^b=E^e_\bullet(s_0)+S^e_\bullet.
\ee
The sources keep their signs. At the high effective-field level, multiplication of the interior differential identity by $s^{-1}$ gives
\be
\label{equa:eight-mobile-effective-normalization}
\begin{aligned}
&s^{-1}E^i_{\mathrm c}(s)+2E^e_{\mathrm c}(s)+\int_{s_0}^s(2-\tau^{-1})\,dF^b_{\mathrm c}(\tau)+2B^b_{\mathrm c}(s)+\int_{s_0}^s\tau^{-2}E^i_{\mathrm c}(\tau)\,d\tau\\
&=s_0^{-1}E^i_{\mathrm c}(s_0)+2E^e_{\mathrm c}(s_0)+\int_{s_0}^s\tau^{-1}\,dS^i_{\mathrm c}(\tau)+2S^e_{\mathrm c}(s).
\end{aligned}
\ee
All flux and damping terms on the left are nonnegative. The extra term $h\mathcal J^0_{\mathrm c}$ belongs to both oriented currents and hence to the same flux $F^b_{\mathrm c}$. In particular, the original normalization of the high interior interface term remains controlled:
\be
\label{equa:eight-mobile-effective-flux-comparison}
s^{-1}F^b_{\mathrm c}(s)\leq F^b_{\mathrm c}(s)\leq\int_{s_0}^s(2-\tau^{-1})\,dF^b_{\mathrm c}(\tau),\qquad s\geq s_0\geq2.
\ee
The lower bound in the equation~\eqref{equa:eight-mobile-effective-flux-lower-bound} also controls the good derivatives and the mass trace with their original unweighted scales. Thus the additional Cartesian energy in the moving flux preserves every flux comparison used in Part~II.

\paragraph{Characteristics in the moving exterior}

We use the optical weight \(w=2-q\) in the characteristic equation; it is equivalent to \(w_b\) on the moving exterior. We recall the characteristic operator and weight exponent:
\be
\label{equa:eight-mobile-characteristic-definition}
\mathcal P_{H_N}=\del_t+\frac{4+H_N}{4-H_N}\del_r,\qquad\gamma=\frac{1+\delta}2.
\ee

\begin{proposition}[Characteristic estimates for the matter field and projected metric components]
\label{prop:eight-mobile-characteristics}
Under the coarse metric and scalar bounds on the adapted foliation, the wave gauge, the potential hypotheses, and \(C_Nm\leq c=C_{\mathrm{rk}}\theta\), the conclusions of \eqref{equa:eight-low-matter-characteristic-Lie} and \eqref{equa:eight-good-metric-characteristic} hold in the moving exterior, for \(p\leq N-9\), without a sign assumption on \(H_N\). Their constants are independent of the terminal time and of \(\coeff\). After improvement of the exterior hypersurface and good spacetime norms to amplitude \(C_Na_0\), the characteristic amplitude and its normalized trace on \(\Lscr_b\) can be bounded by \(C_N(a_0+A_0+A^2)\).
\end{proposition}

\begin{proof}
\bse

\noindent{\bf 1.} \emph{Availability of a backward characteristic.} Along an integral curve parameterized by \(t\), we have
\be
\label{equa:eight-mobile-characteristic-boundary}
\frac{dq}{dt}=-\frac{2H_N}{4-H_N},\qquad\frac{d}{dt}(q-q_b)=h-\frac{2H_N}{4-H_N}\geq0.
\ee
The last inequality follows from \(|H_N|\ll1\), \(H_N^\star\leq0\), and \(h\geq4|H_N-H_N^\star|\). Thus a backward curve remains in \(q\leq q_b(t)\). On the adapted terminal hypersurface, the hyperboloidal and merging portions satisfy \(1-\del_rT_b\geq c_0/t\); on the flat portion \(\del_rT_b=0\). After reducing the amplitude, \eqref{equa:eight-mobile-global-null-coefficient} gives
\be
\label{equa:eight-mobile-characteristic-terminal-leaf}
\frac d{dt}\bigl(t-T_b(s_1,r(t))\bigr)=1-\del_rT_b(s_1,r(t))\frac{4+H_N}{4-H_N}\geq\frac{c_0}{2t}>0
\ee
in the near-cone part. The derivative is one in the flat part. The backward curve consequently stays below the terminal hypersurface until it reaches the initial hypersurface or the outer boundary \(r=5t/4\). It requires no incoming data on \(\Lscr_b\).
\ese

\bse
\noindent{\bf 2.} \emph{The coefficient of the transported quantity.} For \(V=Z\psi\), or \(V=\mathcal A_\omega\mathcal L_Zu\), put \(Y_V=w^\gamma\underL(rV)\). The weighted characteristic equation has the form
\be
\label{equa:eight-mobile-characteristic-equation}
\mathcal P_{H_N}Y_V=\frac{2\gamma H_N}{w(4-H_N)}Y_V+\frac{4rw^\gamma}{4-H_N}\bigl(-\Box_gV+r^{-2}\Delta_{\mathbb S^2}V+R_H[V]\bigr).
\ee
Its adverse zeroth-order coefficient is integrable:
\be
\label{equa:eight-mobile-characteristic-integrable-coefficient}
\left(\frac{2\gamma H_N}{w(4-H_N)}\right)_+\leq C_N(A+A^2)t^{-1-e_0},\qquad\int_{t_i}^{\infty}t^{-1-e_0}\,dt=\frac{t_i^{-e_0}}{e_0}.
\ee
The factor in the resulting Gronwall estimate is bounded independently of the terminal time.
\ese

\bse
\noindent{\bf 3.} \emph{Derivative counts and decay rates in the source terms.} The ordered Lie commutators and the outgoing-frame contraction \eqref{equa:eight-low-matter-Lie-frame} are unchanged. Their pointwise estimates use \(w_b\simeq w\), \(s^2\simeq t\), and the same radial coefficient bounds. For matter, the angular and good-Hessian powers are \(t^{-1-e_{\mathrm a}}\) and \(t^{-1-e_{\mathrm g}}\), with the strictly positive exponents in \eqref{equa:eight-low-matter-source-exponents}. The effective-field coupling is bounded by \(C_N(A^2+A^3)t^{-3/2+3\delta/2}\). The dynamic bad-coefficient terms keep the positive collar and outer gaps in \eqref{equa:eight-low-matter-collar-source-gaps} and \eqref{equa:eight-low-matter-outer-source-gap}. For the projected metric, the projection removes the plane-null quadratic term, and the remaining sources keep the positive gaps in \eqref{equa:eight-good-metric-good-source-gaps} and \eqref{equa:eight-good-metric-bad-source-gaps}. The angular term uses \(p+9\leq N\) operators; direct Hessians use \(p+6\leq N\). No derivative of \(q_b\) enters these equations, since their commutators and weight remain the original spacetime operators and \(w\).

\noindent{\bf 4.} \emph{Growth with boost rank and bounds for traces at the interface.} A translated reference coefficient contributes an integrable source, while a boost-only reference coefficient is \(\Ocal(m/r)\) and multiplies a strictly lower boost rank. We let \(Y_k(t)\) be the near-cone supremum of \(rw^\gamma|\del V|\) through order \(p\) and rank \(k\). The initial and outer values and the preceding source bounds give
\be
\label{equa:eight-mobile-characteristic-rank-inequality}
Y_k(t)\leq C_NL+C_N(A+A^2)\int_{t_i}^t\tau^{-1-e_0}Y_k(\tau)\,d\tau+C_Nm\sum_{j<k}\int_{t_i}^t\tau^{-1}Y_j(\tau)\,d\tau,
\ee
where \(L=A+A^2\) under the coarse bounds. After improvement of the exterior energies, every linear angular, initial, and outer term has amplitude \(C_N(a_0+A_0)\), while each nonlinear term may keep the coarse amplitude \(C_NA^2\); hence we may take \(L=a_0+A_0+A^2\). Gronwall first removes the integrable current-rank coefficient. Induction on \(k\), using \(C_Nm\leq c\), then gives
\be
\label{equa:eight-mobile-characteristic-rank-bound}
Y_k(t)\leq C_NL t^{ck/2}\leq C_NL s^{\eta_k}.
\ee
Finally, on \(q=q_b(t)\), the frame identities \eqref{equa:eight-good-metric-interface-frame} still have coefficients \(q/t=\Ocal(t^{-1})\) and \(q/r=\Ocal(t^{-1})\), because \(3/4\leq q_b\leq1\). The scalar- and tensor-commuted good components therefore have the same trace bounds and rank weights. This proves the normalized characteristic estimate.
\ese
\end{proof}
\paragraph{The coupled exterior estimate with a small reference mass}

All exterior energies and integrals in this paragraph are taken on the moving domain with weight $w_b$. The coordinate squared energies are denoted by $E^b_{u;p,k}$, $E^b_{\psi;p,k}$, and $E^b_{\mathrm c;p,k}$, with $E^b_{\mathrm c;p,k}=2(U^b_{p,k})^2$. We use the same derivative levels as before, $n\coloneqq N-5$ and $k'\coloneqq\min\{k,n\}$, and define
\be
\label{equa:eight-mobile-coupled-leaf}
e^b_k(s)\coloneqq E^b_{u;N,k}(s)+E^b_{\psi;N,k}(s)+E^b_{\mathrm c;n,k'}(s)+s^{-2}E^b_{\mathrm c;N,k}(s),\qquad a_{0,b}^2\coloneqq e^b_N(s_0).
\ee
The aggregate flux $\mathcal F^b_k$ is the sum of the first three unnormalized fluxes and $\int_{s_0}^{s_1} s^{-2}\,dF^b_{\mathrm c;N,k}(s)$. We denote the aggregate bulk with the same normalization by $\mathcal B^b_k$. For $f\in\{u,\psi,\fieldPhi\}$, we set $(\eta_u,\eta_\psi,\eta_{\fieldPhi})\coloneqq(\lambda,\mu,\mu)$ and $(b_u,b_\psi,b_{\fieldPhi})\coloneqq(1,e^{-\fieldPhi},3)$, and define its good-derivative contribution by
\be
\label{equa:eight-mobile-good-bulk-components}
\mathcal G^{b,f}_{p,k}(s_1)\coloneqq\sum_{Z\preccurlyeq(p,k)}\int_{{\MME}^b_{[s_0,s_1]}}2\eta_f w_b^{2\eta_f-1}b_f g^{\mathcal N ab}\delsN_a Zf\delsN_b Zf\,j_g\,dx\,dt.
\ee
The metric components are summed separately. The positive current identity gives
\be
\label{equa:eight-mobile-good-rest-bulk}
\mathcal G^b_{\mathrm{good};k}\coloneqq\mathcal G^{b,u}_{N,k}+\mathcal G^{b,\psi}_{N,k}+\mathcal G^{b,\fieldPhi}_{n,k'}+\int_{s_0}^{s_1} s^{-2}\,d\mathcal G^{b,\fieldPhi}_{N,k}(s),\qquad
\mathcal G^b_{\mathrm{rest};k}\coloneqq\mathcal B^b_k-\mathcal G^b_{\mathrm{good};k}\geq0.
\ee
Indeed, the remaining wave density is $-H_N|\del_tv|^2+h\mathcal J^0[v]\geq-H_N^\star|\del_tv|^2+(h/2)\mathcal J^0[v]$; the effective-field density also keeps its positive mass contribution. We define
\be
\label{equa:eight-mobile-coupled-positive-energy}
\mathcal Y^b_k(s_1)^2\coloneqq\tfrac12e^b_k(s_1)+\mathcal F^b_k(s_1)+\tfrac12\mathcal G^b_{\mathrm{good};k}(s_1)+\mathcal G^b_{\mathrm{rest};k}(s_1)+4\int_{s_0}^{s_1} s^{-3}(U^b_{N,k}(s))^2\,ds.
\ee

\begin{proposition}[Coupled estimate on the moving exterior]
\label{prop:eight-mobile-coupled-estimate}
\label{proposition:eight-c-top-exterior}
Suppose the coarse exterior bounds, the propagated wave gauge, the potential hypotheses, and the adapted-foliation comparisons hold. The moving interface and its currents are those of Proposition~\ref{prop:eight-mobile-positive-current}. Assume $0\leq m\leq C_{\mathrm{ref}}A_0$ and $C_Nm\leq C_{\mathrm{rk}}\theta$. Then, for a fixed sufficiently large $D$ and sufficiently small $A$, one has
\be
\label{equa:eight-mobile-coupled-rank-bound}
\mathcal Y^b_k(s_1)\leq C_Na_{0,b}{s_1}^{d_k},\qquad d_k\coloneqq\eta_k+(k+1)D\epsilon_{\mathrm g},\qquad\epsilon_{\mathrm g}\coloneqq C_NA,\qquad0\leq k\leq N.
\ee
The constants are independent of $\coeff$, $C_1$, and the terminal time. The original initial bounds give $a_{0,b}\leq C_{\mathrm{data}}A_0$. Consequently, $d_N<\delta$ and $C_Na_{0,b}\leq A/2$ improve the four exterior energy bounds at orders $N$ and $N-5$.
\end{proposition}

\begin{proof}
\bse

\noindent{\bf 1.} \emph{Energy currents and normalization of the effective-field energy.} The identity in Proposition~\ref{prop:eight-mobile-positive-current} keeps a nonnegative boundary flux and a nonnegative bulk throughout the exterior. We multiply the order-$N$ effective-field identity by $s^{-2}$ before integration. The resulting damping is $4\int s^{-3}(U^b_{N,k})^2\,ds$. The other three identities keep weight one. These are the normalizations in the equation~\eqref{equa:eight-mobile-coupled-positive-energy}.

\noindent{\bf 2.} \emph{Stress tensors estimated by hypersurface energies.} Proposition~\ref{prop:eight-mobile-characteristics} gives the lower matter and transverse metric bounds $CA r^{-1}w^{-\gamma}s^{\eta_j}$, with $\gamma=(1+\delta)/2$. The geometric comparison $J_b\zeta_b^{-2}/r\lesssim s^{-1}$ gives
\be
\label{equa:eight-mobile-stress-leaf-multiplier}
J_b\zeta_b^{-2}w_b^{\lambda-\mu}|\del\psi_{\mathrm l,j}|\lesssim As^{-1+\eta_j},\qquad
J_b\zeta_b^{-2}|\del u_{\mathrm{shear},\mathrm l,j}|\lesssim As^{-1+\eta_j}.
\ee
Weighted Cauchy--Schwarz therefore bounds each pairing by a current-rank term $CA s^{-1}$ times the metric squared energy and a term $CA s^{-1+2\eta_j}$ times the complementary squared energy at rank $k-j$. The estimate~\eqref{equa:eight-low-effective-stress-coefficient} gives the identical conclusion for the effective-field derivative and potential stress when the higher factor belongs to the closing energy. The potential uses $J_b\zeta_b^{-1}\leq J_b\zeta_b^{-2}$. For an effective-field factor of order above $N-5$, the complementary order is at most four, and the stronger estimate~\eqref{equa:eight-top-strong-low-effective-factor} gives the integrable coefficient in~\eqref{equa:eight-top-effective-stress-gap} relative to $U^b_{N,k}/s$. Differentiated scalar coefficients and the far-region stress keep their integrable estimates. These steps use the hypersurface energies and preserve every fraction of the good-derivative bulk assigned to the remaining sources.

\noindent{\bf 3.} \emph{Principal terms and the remaining sources.} The principal estimates~\eqref{equa:eight-massive-principal-pairing} and~\eqref{equa:eight-massive-top-principal-pairing} use the displaced-cone coefficient, the radial coefficient estimate, and specified fractions of the good-derivative bulk. Their proofs require no lower bound for $m$. For the wave fields, the radial and displaced-cone argument applies to coefficient operators of order at most \(N-5\), since the mass enters only the energy and the complementary translated field has order at most \(N\). A coefficient operator of order at least \(N-4\) leaves at most four operators on the wave Hessian, so~\eqref{equa:eight-high-density-wave-pairing} applies, with the terminal metric Hardy norm. The remaining tensor components and the spatial end are controlled by~\eqref{equa:audit-high-density-good-pairing}--\eqref{equa:audit-high-density-spatial-pairing}. Its radial \(L^2(\mathbb S^2)\) bound adds no angular derivative to the highest-order coefficient; the coefficient compositions require at most \(\lfloor N/2\rfloor+7\leq N\) coarse orders.

To apply this argument on the moving domain, let \(I^b_{s_1}(t)\) be the intersection of the available exterior slab with the ray at fixed \(t\geq T_0\) and angle. It is empty or an outward interval by radial monotonicity of \(T_b(s_1,r)\). With \(P=\mathcal L_I(K-K^\star)\), the actual radial norm is
\be
\label{equa:audit-mobile-high-density-radial-norm}
B^b(t,\omega)^2\coloneqq\int_{I^b_{s_1}(t)}w_b^{2\lambda-1}(|EP|^2+r^{-2}|P|^2)r^2\,dr,\qquad |P_N(t,R\omega)|\lesssim R^{-\lambda}B^b(t,\omega).
\ee
Thus radial integration uses no point above the terminal hypersurface. The dynamic composition estimate, followed by Lemma~\ref{lem:eight-mobile-outgoing-Hardy}, gives
\be
\label{equa:audit-mobile-high-density-Hardy}
\begin{aligned}
\int(B^b)^2\,dt\,d\omega&\leq C_N\sum_{Z\preccurlyeq(|I|,\rank I)}\Bigl(\int_{{\MME}^b_{[s_0,s_1]}}w_b^{2\lambda-1}|EZu|^2\,dx\,dt+\|w_b^{\lambda-1}Zu\|_{L^2({\MME}^b_{s_1})}^2\Bigr)\\
&\quad+C_NK(A+A^2)\int_{s_0}^{s_1}s^{-3-2e_0}\bigl(\Fenergy^{b,|I|,\rank I}_\lambda(s,u)\bigr)^2\,ds.
\end{aligned}
\ee
The terminal Hardy term is kept and compared with the terminal metric energy. The last coefficient comes from \(h/r_b\lesssim K(A+A^2)t^{-2-e_0}\) and \(dt/ds\lesssim s\); it is integrable and multiplies an actual squared energy. The bounded initial range is controlled by~\eqref{equa:audit-bounded-initial-Gronwall}. Since \(w_b\simeq w\), the Hessian bounds and their exponents remain unchanged. These estimates use no inverse power of the reference mass. The density deformation identities and the matter current use the same good-derivative terms and integrable coefficients. The higher effective-field contribution to the matter equation keeps~\eqref{equa:eight-top-lower-rank-good-bulk}; its rank-$l$ integral is bounded by $CA {s_1}^{2\eta_{k-l}}\mathcal Y^b_l(s_1)^2$. The reference terms have coefficient $C_Nm/s$ at lower boost rank, or $C_Nm/s^3$ when a translation occurs. Since $m\lesssim A$, all current-rank coefficients are bounded by $\epsilon_{\mathrm g}/s$.

For the highest-order sources, we keep an actual energy before using the coarse bootstrap on the other factor. We denote by \(M^b_{N,k}\) the metric energy together with its dynamic Hardy norm, and by \(P^b_{N,k}\) the matter energy. The proof of~\eqref{equa:eight-massive-top-high-coefficient} gives the source norm \(C_NA s^{1-2\lambda+\delta}M^b_{N,k}\); its coefficient has rank at most \(k\), and the lower Hessian requires at most ten orders. The matter trace source in~\eqref{equa:eight-massive-top-null-source-factors} similarly has norm \(C_NA s^{-1+\delta}P^b_{N,k}\). After the time normalization, their pairings satisfy
\be
\label{equa:mobile-top-sources-homogeneous}
A\left(s^{-2\lambda+\delta}M^b_{N,k}+s^{-2+\delta}P^b_{N,k}\right)\frac{U^b_{N,k}}s\leq C_NA s^{-1-\beta_H}(\mathcal Y^b_k)^2,\qquad\beta_H\coloneqq2\lambda-1-\delta>0.
\ee
Here \(M^b_{N,k}+P^b_{N,k}+U^b_{N,k}/s\leq C_N\mathcal Y^b_k\), and \(\beta_H<1-\delta\). The differentiated metric and exponential coefficients leave the integrable term \(C_NA^2s^{-3+2\delta}(\mathcal Y^b_k)^2\). At order \(n\), the high good matter derivative in the trace source is paired against its positive bulk, with multiplier \(C_NA s^{-3/2+\delta}\); Young's inequality leaves a chosen bulk fraction and \(C_{N,\epsilon}A^2s^{-3+2\delta}(U^b_{n,k'})^2\). When the good derivative is on the lower factor, the coefficient is \(C_NA s^{-1-2\mu+\delta}\). Thus these source terms are homogeneous in the energies being estimated. The exterior reference is an exact stationary vacuum metric, so it contributes only coefficient terms, not an independent forcing.

The term \(h\mathcal J^0\) in the moving current absorbs the dynamic contribution that was included in \(\Omega^-\) for the fixed-interface identity: the coefficient estimate gives \(-H_N|\del_tv|^2+h\mathcal J^0\geq-H_N^\star|\del_tv|^2+(h/2)\mathcal J^0\geq0\). Thus this step requires no sign assumption on the full coefficient \(H_N\). The lower factors use at most $\lfloor N/2\rfloor\leq N-9$ characteristic orders. The low effective-field reconstruction has the same maximum count $4+7\leq N-5$. The highest metric coefficient is measured in $L^2$ and adds no angular commutation. Thus all derivatives are among the original exterior orders $N$ and $N-5$.

\noindent{\bf 4.} \emph{Rank induction.} We choose the finitely many Young fractions after summing the component and commutator contributions. These sums include all labelled-subset multiplicities in~\eqref{equa-curved-wave-commutator-coeff}, the composition multiplicities in~\eqref{equa:audit-ordered-scalar-composition}, and the density products in~\eqref{equa:audit-exterior-density-all-orders}. The matter-current remainder and the matter trace source use separate fractions from the principal commutators. If $C_j\epsilon_j$ is the coefficient of the terminal metric energy provided by its Hardy comparison, we require $\sum_jC_j\epsilon_j\leq1/4$. We also require that the sum of the good-bulk fractions is at most $1/2$. The constants are fixed before the amplitude is reduced; these choices leave at least $e^b_k/2$ on the left. For some fixed $\beta_*>0$, the remaining positive quantity satisfies
\be
\label{equa:eight-mobile-coupled-rank-inequality}
\begin{aligned}
\mathcal Y^b_k(s_1)^2&\leq C_Na_{0,b}^2+C_N(A+A^2)\int_{s_0}^{s_1} s^{-1-\beta_*}\mathcal Y^b_k(s)^2\,ds+\epsilon_{\mathrm g}\int_{s_0}^{s_1} s^{-1}\mathcal Y^b_k(s)^2\,ds\\
&\quad+\epsilon_{\mathrm g}\sum_{l<k}\int_{s_0}^{s_1} s^{-1+2\eta_{k-l}}\mathcal Y^b_l(s)^2\,ds+C_NA\sum_{l<k}{s_1}^{2\eta_{k-l}}\mathcal Y^b_l(s_1)^2.
\end{aligned}
\ee
At rank zero, Gronwall applies directly. If the preceding ranks have been estimated with exponent $d_l$, the lower-rank terms have exponent $2\eta_k+2(l+1)D\epsilon_{\mathrm g}<2d_k$. Their Volterra denominator is at least $(2D-1)\epsilon_{\mathrm g}$, so its coefficient stays bounded as $A$ tends to zero. The terminal good-bulk contribution has the same lower exponent. Finite induction proves the conclusion. Finally, $w_b$ is uniformly comparable to the original initial weight and the initial domain is unchanged, proving $a_{0,b}\leq C_{\mathrm{data}}A_0$.
\ese
\end{proof}


\subsection{\texorpdfstring{Traces and retarded integrals in the interior}{Traces and retarded integrals in the interior}}
\label{section:mobile-interior-transfer}
\paragraph{Bounds for traces at fixed physical time}
For the physical-time trace, we set $f_Z(t,\omega)\coloneqq r_b(t)ZH^{\H00}(t,r_b(t)\omega)$. The conversion is
\be
\label{equa:audit-moving-interface-time-boost}
t\frac{df_Z}{dt}=\frac{t r_b'}{r_b}f_Z+r_br_b'\omega^aL_a(ZH^{\H00})+r_bd_b\del_t(ZH^{\H00}).
\ee
The coefficients $r_b'$ and $d_b$ are bounded. The time trace still costs one commutation and the two sphere derivatives cost two boosts. The upper hypersurface for the integrated trace is $S_b(T)\coloneqq\sqrt{2Tq_b(T)-q_b(T)^2}$. Its parameter is comparable to $\sqrt T$ and it covers every earlier interface sphere.

\paragraph{The lower boundary of a retarded cone}
In source coordinates $p=t+r$ and $q=t-r$, we define $a_b(p)$ by $a_b=q_b((p+a_b)/2)$. The boundary of the interior cone integral is $q\geq a_b(p)$, and its first two derivatives satisfy
\be
\label{equa:audit-moving-interface-cone-boundary}
a_b'=\frac{q_b'}{2-q_b'},\qquad a_b''=\frac{2q_b''}{(2-q_b')^3},\qquad p|a_b'|+p^2|a_b''|\lesssim K(A+A^2)p^{-e_0}.
\ee
The arguments of $q_b'$ and $q_b''$ are $(p+a_b(p))/2$. The lower trace in an incoming integration is evaluated at $\max\{a_b(p),s_0^2/p\}$. The volume derivative formulas remain valid after the following boundary contribution is kept: for a physical integrand $W$, extended by zero outside the interior, we have
\be
\label{equa:audit-moving-interface-cone-distribution}
\del_p\bigl(\boldsymbol1_{q\geq a_b(p)}W\bigr)=\boldsymbol1_{q\geq a_b(p)}\del_pW-a_b'(p)\delta_{q=a_b(p)}W.
\ee
For example, the additional term in $\del_Q\mathcal I_W$ is
\be
\label{equa:audit-moving-interface-cone-target-trace}
-2\int_0^1\int_0^{2\pi}\chi(a_b/Q)\frac{(1-\xi)(p-a_b)}{Q-a_b}a_b'W^\sharp(p,a_b,\vartheta)\,d\vartheta\,d\xi.
\ee
The $Q_+$ derivative has coefficient $\xi(p-a_b)/(Q_+-a_b)$, and the tangential derivative has coefficient $2\sqrt{\xi(1-\xi)}\cos\vartheta(p-a_b)/\sqrt{(Q-a_b)(Q_+-a_b)}$. Under the source bound~\eqref{equa:seven-target-cone-source-condition}, these additional kernels have the extra factor $K(A+A^2)Q^{a-1-e_0}$, where $a<1$. They preserve the strict gains in the differentiated cone estimates. The explicit target-dependent angular coefficients keep their separate derivatives.

For the differentiated target formulas, we assume $Q\geq8T_0$, where $T_0$ bounds the initial time as in the equation~\eqref{equa:seven-exterior-metric-retained-trace}; the bounded complementary range keeps the initial-time terms. We differentiate the signed trace on its actual boundary. We write $D\coloneqq Q_+-Q$, $p=Q+D\xi$, $q=a_b(p)$, $A_-\coloneqq Q-q$, $A_+\coloneqq Q_+-q$, and $\beta\coloneqq\xi(1-\xi)$. For a physical boundary integrand $B(p,q,\omega)$, we set $V(p,\omega)\coloneqq B(p,a_b(p),\omega)$. Its outgoing derivative and the correction to the polar-angle derivative are
\be
\label{equa:audit-moving-interface-trace-derivative-definition}
V_p=(\del_p+a_b'\del_q)B,\qquad F_{b,\mathrm c}\coloneqq1+\frac{a_b'D^2\beta}{A_-A_+},\qquad \Lambda_\xi\coloneqq\lambda_\xi+Da_b'\lambda_q=-\frac{\sin\lambda}{2\beta}F_{b,\mathrm c}.
\ee
Here $\Lambda_\xi$ is the total derivative of the polar angle on $q=a_b(p)$. Since $D^2\beta=(p-Q)(Q_+-p)\leq pQ_+$ and $A_-A_+\geq QQ_+/4$, we have
\be
\label{equa:audit-moving-interface-trace-angle-denominator}
|F_{b,\mathrm c}-1|\leq C K(A+A^2)Q^{-1-e_0},\qquad \frac12\leq F_{b,\mathrm c}\leq\frac32.
\ee
The chain rule and elimination of $V_\lambda$ give
\be
\label{equa:audit-moving-interface-trace-target-derivatives}
\del_QV^\sharp=\frac{(1-\xi)(p-q)}{A_-F_{b,\mathrm c}}V_p^\sharp+B_-\del_\xi V^\sharp,\qquad \del_{Q_+}V^\sharp=\frac{\xi(p-q)}{A_+F_{b,\mathrm c}}V_p^\sharp+B_+\del_\xi V^\sharp,
\ee
where the two coefficients are
\be
\label{equa:audit-moving-interface-trace-target-coefficients}
B_-\coloneqq-\frac{\beta}{A_-F_{b,\mathrm c}}\left(1-\frac{a_b'D(1-\xi)}{A_+}\right),\qquad B_+\coloneqq\frac{\beta}{A_+F_{b,\mathrm c}}\left(1+\frac{a_b'D\xi}{A_-}\right).
\ee
Both coefficients vanish at $\xi=0,1$. Their first $\xi$ derivatives have bounds $C/Q$ and $C/Q_+$, respectively; their differences from the fixed-boundary coefficients have the extra factor $C K(A+A^2)Q^{-1-e_0}$. This follows directly from $p|a_b'|+p^2|a_b''|\lesssim K(A+A^2)p^{-e_0}$, using $D\xi\leq p$ and $D(1-\xi)\leq Q_+$. Consequently the integrations by parts in $\xi$ keep their original powers and have no polar-axis trace.

For the target angular derivative, we set $b_{\mathrm c}\coloneqq(p-q)\sqrt\beta/\sqrt{A_-A_+}$. The source rotation identity gives the same azimuthal integral as before, while the elimination of its polar derivative is
\be
\label{equa:audit-moving-interface-trace-angular-elimination}
V_\lambda^\sharp=-\frac{b_{\mathrm c}}{F_{b,\mathrm c}}\del_\xi V^\sharp+D\frac{b_{\mathrm c}}{F_{b,\mathrm c}}V_p^\sharp.
\ee
After integration in $\xi$, the coefficient of $V$ is $\del_\xi(b_{\mathrm c}/F_{b,\mathrm c})+\cot\lambda$. Its difference from the previous coefficient, with the factor $F_{b,\mathrm c}$ kept, has the form
\be
\label{equa:audit-moving-interface-trace-angular-correction}
\del_\xi\left(\frac{b_{\mathrm c}}{F_{b,\mathrm c}}\right)+\cot\lambda-\frac{2D\sqrt\beta}{F_{b,\mathrm c}\sqrt{A_-A_+}}=\frac{F_{b,\mathrm c}-1}{F_{b,\mathrm c}}\cot\lambda+\frac{Da_b'}{F_{b,\mathrm c}}\del_q b_{\mathrm c}-\frac{b_{\mathrm c}}{F_{b,\mathrm c}^2}\del_\xi F_{b,\mathrm c}.
\ee
Each term on the right contains a factor $D$. After division by the target radius $D/2$, their absolute sum is at most $C K(A+A^2)Q^{-1-e_0}\sqrt\beta/\sqrt{A_-A_+}$. Indeed, the formulas
\be
\label{equa:audit-moving-interface-trace-angular-coefficients}
\cot\lambda=\frac{(A_-+A_+)\xi-A_-}{2\sqrt\beta\sqrt{A_-A_+}},\qquad \del_q b_{\mathrm c}=\frac{\sqrt\beta}{\sqrt{A_-A_+}}\left(-1+\frac{(p-q)(A_-+A_+)}{2A_-A_+}\right)
\ee
give this bound together with~\eqref{equa:audit-moving-interface-cone-boundary}. To verify the cancellation at the axes and at the target center explicitly, we set \(S_{\mathrm c}\coloneqq A_-A_+\), \(M_{\mathrm c}\coloneqq p-q\), \(N_{\mathrm c}\coloneqq(A_-+A_+)\xi-A_-\), and \(a_j\coloneqq a_b^{(j)}(p)\), \(j=1,2\). Writing \(F=F_{b,\mathrm c}\) in this calculation, we have
\bse
\be
\label{equa:audit-moving-angular-closed-remainder}
\begin{aligned}
E_{\mathrm{ang}}&\coloneqq\del_\xi(b_{\mathrm c}/F)+\cot\lambda-\frac{2D\sqrt\beta}{F\sqrt{S_{\mathrm c}}}\\
&=\frac{D^2\sqrt\beta}{S_{\mathrm c}^{3/2}F^2}\Bigl(a_1\bigl(N_{\mathrm c}-M_{\mathrm c}(1-2\xi)\bigr)-D\beta\bigl(M_{\mathrm c}a_2+2a_1^2\bigr)\Bigr).
\end{aligned}
\ee
Indeed, \(\del_qb_{\mathrm c}=D\beta\cot\lambda/S_{\mathrm c}\), while \(\del_\xi F=D^3a_2\beta/S_{\mathrm c}+a_1D^2(1-2\xi)/S_{\mathrm c}+a_1^2D^3\beta(A_-+A_+)/S_{\mathrm c}^2\). Substitution and \(DN_{\mathrm c}-M_{\mathrm c}(A_-+A_+)=-2S_{\mathrm c}\) give the identity. Since \(S_{\mathrm c}\simeq QQ_+\), \(M_{\mathrm c}\leq p\), \(|N_{\mathrm c}|\leq M_{\mathrm c}\), and \(D^2\beta\leq pQ_+\), the bounds for \(a_b'\) and \(a_b''\) imply
\be
\label{equa:audit-moving-angular-uniform-remainder}
\frac{2|E_{\mathrm{ang}}|}{D}\leq C K(A+A^2)Q^{-1-e_0}\frac{\sqrt\beta}{\sqrt{A_-A_+}}.
\ee
\ese
This estimate is uniform in \(Q_+/Q\). The factor \(\sqrt\beta\) removes the polar-axis terms. At fixed \(Q\), the remainder is \(\Ocal(D^2)\), so its contribution after division by the target radius \(D/2\) tends to zero. Only two derivatives of the boundary function occur, and no additional derivative of the fields is required. Thus the angular derivative of the signed trace uses only $|B|+p|V_p|$, with the original kernel and a uniformly bounded factor.

For the signed metric trace, we recall $v=Z_2\psi$ and the coefficients $P_N$, $a_P$ from the radial decomposition~\eqref{equa:seven-radial-metric-full-decomposition}. We have $B=C\del_qv$, where $C=r'(P_N+a_P)$ and $r'=(p-q)/2$ is the source radius. The exterior estimates at $q=a_b(p)$ give
\be
\label{equa:audit-moving-interface-trace-jets}
|C|+|\del_qC|+p|\del_pC|\lesssim A p^{\delta/2},\qquad |\del_qv|+|\del_q^2v|+p|\del_p\del_qv|\lesssim A p^{-1+\delta/2}.
\ee
The bad contraction in $C$ uses the displaced-cone bound; the other coefficient has the explicit factor $q/p$. A Cartesian derivative of $C$ uses the direct first-derivative estimate. Since $L_r=p\del_p-q\del_q$, the outgoing scalar Hessian satisfies
\be
\label{equa:audit-moving-interface-outgoing-scalar-Hessian}
p\del_p\del_qv=\del_q(L_rv)+\del_qv+q\del_q^2v.
\ee
The Hessian and $\del_q(L_rv)$ each cost one additional commutation. With the five exterior Sobolev orders, the total remains $p_{\mathrm w}+6$. The product rule consequently gives
\be
\label{equa:audit-moving-interface-trace-source-bound}
|B|+p|(\del_p+a_b'\del_q)B|\lesssim A^2p^{-1+\delta},\qquad p|a_b'\del_qB|\lesssim K(A+A^2)A^2p^{-1+\delta-e_0}.
\ee
The massive correction cutoff, still zero for $q\leq1$, vanishes on the new interface. Its support and differential orders remain unchanged. Its complementary source on $a_b(p)\leq q\leq1$ is estimated by the direct interior scalar bounds on this bounded-width strip; here $q\simeq1$ and $\sigma\simeq\sqrt p$. These bounds are the same ones used on $1\leq q\leq2$ and use the same orders. The initial-time contribution is unchanged because the initial hypersurface is unchanged. The reference expressions on the two sides of the interface are opposite evaluations of the same signed trace and cancel at $q=a_b(p)$.

Finally, the Schwarzschild interface label and its derivative are
\be
\label{equa:audit-moving-interface-Schwarzschild-label}
u_*(s)=q_b(t_b(s))-2m\log(R_b(s)-m)-c_*,\qquad u_*'(s)=q_b'(t_b)t_b'-\frac{2mR_b'}{R_b-m}.
\ee
In particular, $u_*'=-4m/s+\Ocal(K(A+A^2)(1+m)s^{-1-2e_0}+ms^{-3})$. The Schwarzschild logarithm is kept, while the displacement of the interface contributes an integrable correction.

\begin{proposition}[Interior estimates on the adapted foliation]
\label{prop:audit-moving-interface-interior-transfer}
Assume the field equations, the propagated wave gauge, the potential and prescribed-reference hypotheses, and the coarse energy bounds of Part~II on the adapted foliation. The Cauchy norms are those on the unchanged initial hypersurface, and the interface fluxes and traces are measured on $\mathbb S_s^b$ with their original ranks and powers of $s$. Then the conditional interior estimates of Part~II remain valid with the same commutation orders and time exponents, after replacement of their domains and traces by these adapted ones. In particular, one keeps:
\begin{itemize}
\item the Sobolev and Hardy estimates, the high radial coefficient profile, and the damped variation estimate, with the trace factors $R_b'(s)$ in~\eqref{equa:audit-moving-interface-radial-profile} and~\eqref{equa:audit-moving-interface-initial-profile};
\item the ray estimates with the actual lower parameter in~\eqref{equa:audit-moving-interface-ray-entry}, and the interior characteristic estimates stopped at $\Lscr_b$;
\item the differentiated central wave estimate~\eqref{equa:seven-sufficient-massless-gain}, including the signed metric trace, the massive product correction, and the Cauchy terms;
\item the conditional commuted interior energy inequalities with the adapted oriented interface currents.
\end{itemize}
The constants are independent of the terminal time and of $0<\coeff\leq\coeff_0$. The assertion preserves the other, nongeometric hypotheses of each individual estimate.
\end{proposition}
\begin{proof}
The Sobolev and Hardy arguments take place on the uniformly regular balls $B_{R_b(s)}$; their radial and variation identities were proved above. The ray equations are pointwise identities on the same hyperboloids. Their lower parameters remain comparable to the previous ones, and the time at their first intersection with the interface is comparable to the square of the ray parameter. The interior characteristic identities for $dq/dt$ and $ds/dt$ are unchanged; their estimates require only the uniform positive lower bound $q\geq3/4$ instead of $q\geq1$. Their central stopping surface $r/t=3/5$ is unchanged.

The past dilations used in the equation~\eqref{equa:seven-polynomial-dilation-interpolation} satisfy $q\geq3/8$, because their dilation factor lies in $[1/2,1]$ and the target has $q\geq3/4$. Along these curves, $t-r\del_rT_b\geq t-r>0$, so they remain below the terminal hypersurface. For $t\geq4T_0$ they remain above the initial hypersurface. Their exterior portions lie in the bounded strip $3/8\leq q\leq q_b(t)$, where the direct exterior jets give the same bounds for $\mathcal D^2L^Jv$ and the same comparison of weights as in the original dilation argument. The largest exterior count remains $p_{\mathbf{low}}+7\leq N$. The bounded-time portion is controlled by the initial norms through~\eqref{equa:audit-bounded-initial-Gronwall}. Hence the dilation interpolation and its positive source exponent remain valid. In the cone-source comparisons, replacing \(q\geq1\) by \(q\geq3/4\) only replaces a bound by one with a fixed multiplicative constant. It leaves every exponent, including the single exceptional loss at \(N=19\), unchanged.

In the central representation, the coefficient and source inequalities are local and keep their powers for $q\geq3/4$. The positive cone kernels were integrated over $q>0$, so the new domain is still an admissible subset. The volume derivative contributes exactly~\eqref{equa:audit-moving-interface-cone-distribution}; its added trace has the strictly smaller kernel calculated above. The signed boundary contribution uses~\eqref{equa:audit-moving-interface-trace-target-derivatives}--\eqref{equa:audit-moving-interface-trace-angular-coefficients} and the jets~\eqref{equa:audit-moving-interface-trace-source-bound}. These computations preserve the original strict gaps and introduce no additional derivative. The massive correction, the bounded initial-time contribution, and the reference cancellation were treated above. This proves the differentiated central estimate on the adapted domain.

Finally, the commuted equations and all interior product decompositions are unchanged. The interior energy identities use the coordinate normal $N_b$ and their actual moving-boundary currents; the weighted exterior and interior currents have identical weights on $\Lscr_b$ and cancel with opposite orientations when the interior and exterior identities are added. The already transferred pointwise bounds, radial profiles, and sphere traces therefore give the same conditional interior energy inequalities. Every effective-field factor keeps its original singular or uniform normalization, and none of the geometric conversions contains a power of $\coeff$.
\end{proof}



\subsection{\texorpdfstring{Simultaneous improvement and global continuation}{Simultaneous improvement and global continuation}}
\label{section:mobile-simultaneous-improvement}
\paragraph{Energy and trace norms on the adapted foliation}
We use the adapted domains and currents of Section~\ref{section:small-mass-moving-interface}. To distinguish the two energy levels, we write $l\coloneqq N-9$ and $n\coloneqq N-5$. A superscript $b$ on an interior energy, interface flux, or trace means that its defining expression is evaluated on the adapted domain, with the same commuted operators and effective-field normalization. The high trace in the equation~\eqref{equa:seven-robust-high-interface-input} uses the radial factor from the equation~\eqref{equa:audit-moving-interface-radial-profile}: its first term is
\be
\label{equa:mobile-high-trace-factor}
\left\|\tau^{-1}\sqrt{R_b'(\tau)}\,\mathsf B^{H,b}_{2;p,k}(\tau)\right\|_{L^2((s_0,\sigma);d\tau)}.
\ee
The other two terms keep their original expressions with adapted sphere traces. We denote the resulting quantities by $\mathcal I^{\mathrm{tr},b}_{p,k}$ and $\mathcal I^{\mathrm{rob},b}_{\mathrm h,k}$.

\begin{proposition}[Simultaneous improvement with a small reference mass]
\label{prop:mobile-simultaneous-improvement}
Let $N\geq19$, and fix the exponents in the equations~\eqref{equa:seven-exponents}--\eqref{equa:seven-global-exponent-restrictions}. Suppose the initial data satisfy the Hamiltonian, momentum, and wave-gauge constraints, the geometric reconstruction in Proposition~\ref{pro:40S}, the potential hypotheses, and the initial estimates~\eqref{equa:seven-initial-energy-bound}--\eqref{equa:seven-initial-robust-high-bound} and~\eqref{equa:seven-exterior-initial-assumptions}. Assume the prescribed reference metric has mass $0\leq m\leq C_{\mathrm{ref}}A_0$. The data are smooth, with finite order-$N$ energy at each fixed $\coeff>0$.

There are constants $K,C_1$, and $\varepsilon_0>0$, independent of $0<\coeff\leq\coeff_0$, such that $A_0\leq\varepsilon_0$ gives a future-global development on the adapted foliation defined by $A=C_1A_0$. On every finite available bootstrap interval, the following estimates strictly improve their coarse bounds:
\begin{itemize}
\item The exterior quantity satisfies $\mathcal Y^b_k(s)\leq C_Na_{0,b}s^{d_k}$, with the exponents and positive flux and bulk terms of Proposition~\ref{prop:eight-mobile-coupled-estimate}.
\item For $0\leq k\leq l$, the interior energies satisfy $U^b_{l,k}\leq(A/2)s^{\alpha_k}$, $\mathbf P^b_{l,k}\leq(A/2)s^{\beta_k}$, and $\mathbf W^b_{l,k}\leq(A/2)s^{\nu_k}$. At the high level, $\mathsf E^{\mathrm{rob},b}_{\mathrm h,k}\leq(A/2)s^{\nu_k}$ for $0\leq k\leq n$.
\item The five interface bounds~\eqref{equa:seven-interface-data-bounds} hold with adapted traces and currents. Their constants are independent of $C_1$, the terminal time, and $\coeff$.
\item The metric and the rescaled effective field remain strictly inside the geometric and potential ranges. The unscaled effective-field energies and the order-$N$ energy remain finite on every finite slab for each fixed $\coeff>0$.
\end{itemize}
The smallness threshold is independent of the interior order-$N$ initial norm and can be chosen uniformly for the reference masses in the stated range. Bounds for that highest norm may depend on its finite initial value and on $\coeff$.
\end{proposition}

The amplitude \(A_0=\epss+C_0\eps\) includes the size \(\epss\) of the reference metric as well as the perturbation. The condition \(m\leq C_{\mathrm{ref}}A_0\) expresses this convention; it does not bound the perturbation by the mass. The estimates below are uniform as \(m\) tends to zero, with the reference cutoffs and all exponent gaps fixed. In particular, they impose no smallness condition on \(A^2/m\).

\begin{proof}
\bse
\noindent{\bf 1.} \emph{Fixed constants and the exterior improvement.} We first fix $N,s_0,\lambda,\mu,\delta$, the rank constant $C_{\mathrm{rk}}$, and then $\theta,\zeta_\star$, the potential bounds, and the data and reference constants. Thus $c\coloneqq C_{\mathrm{rk}}\theta$ and $\rho=\delta/64$ are fixed, with $c(N+4)<\rho$. We fix the Young fractions in the coupled current estimates, choose $K$ larger than four times the constant in the equation~\eqref{equa:eight-mobile-global-null-coefficient}, and fix the rank-induction constant $D$. All these choices precede $C_1$ and $A_0$.

We include in the finite bootstrap the metric good-derivative integral through order \(N\), with the normalization of~\eqref{equa:eight-mobile-good-bulk-components}:
\be
\label{equa:mobile-order-N-good-bulk-bootstrap}
\mathcal G^{b,u}_{N,N}(s)\leq A^2s^{2\delta},\qquad s_0\leq s\leq s_1.
\ee
This integral vanishes at \(s=s_0\), so local existence and continuity start its bootstrap without an additional initial-data condition. Together with the coarse metric energy, it gives the order-\(N\) density estimate~\eqref{equa:eight-displaced-cone-density-bulk} through the moving Hardy inequality. The coefficient estimate and $H_N^\star\leq0$ then give $h\geq4|H_N-H_N^\star|$. Proposition~\ref{prop:eight-mobile-positive-current} therefore gives the positive flux and bulk used by Proposition~\ref{prop:eight-mobile-coupled-estimate}. In particular, we obtain
\be
\label{equa:mobile-simultaneous-exterior}
\mathcal Y^b_k(s)\leq C_Na_{0,b}s^{d_k},\qquad a_{0,b}\leq C_{\mathrm{data}}A_0,\qquad d_k=ck+(k+1)D\epsilon_{\mathrm g},\qquad\epsilon_{\mathrm g}=C_NA.
\ee
Since \(\mathcal Y_N^b\) keeps one half of \(\mathcal G^{b,u}_{N,N}\), the same estimate gives \(\mathcal G^{b,u}_{N,N}(s)\leq C_Na_{0,b}^2s^{2d_N}\). The choices \(d_N<\delta\) and sufficiently large fixed \(C_1\) improve~\eqref{equa:mobile-order-N-good-bulk-bootstrap} to \((A^2/4)s^{2\delta}\). The coefficient estimate has used only the coarse version of this bound. The initial hypersurface is unchanged, and its two weights are uniformly comparable. The good-derivative term kept in $\mathcal Y^b_N$ and the radial Hardy inequality on its upper hypersurface also give the exterior estimates needed for the high interior metric profile:
\be
\label{equa:mobile-simultaneous-good-bulk-Hardy}
\sum_{Z\preccurlyeq(n,n)}\left(\int_{{\MME}^b_{[s_0,s_1]}}w_b^{2\lambda-1}|\delsN Zu|^2\,dx\,dt+\|w_b^{\lambda-1}Zu\|_{L^2({\MME}^b_{s_1})}^2\right)\leq C_Na_{0,b}^2{s_1}^{2d_N}.
\ee
The Hardy term concerns the dynamic perturbation; the reference contribution is bounded separately by $C_NA_0$. The outgoing integrations use Lemma~\ref{lem:eight-mobile-outgoing-Hardy}, including its terminal-hypersurface term. This gives the spacetime and terminal-hypersurface bounds in~\eqref{equa:seven-e-good-bulk-input} on the adapted domain.

We require the following strict exponent margins, all attainable by decreasing $A$:
\be
\label{equa:mobile-simultaneous-exponent-margins}
(N+1)D\epsilon_{\mathrm g}\leq\frac{\delta-cN}{2},\qquad(l+1)D\epsilon_{\mathrm g}\leq\frac{9c}{2},\qquad(n+1)D\epsilon_{\mathrm g}\leq\frac{12\rho+2c}{2}.
\ee
They imply $d_N<\delta$, $d_l\leq\eta_{\Lscr}-9c/2$, and $d_k\leq\nu_k-(6\rho+c)$ for $k\leq n$. The constants in the interface estimates below consequently depend only on fixed positive exponent gaps.
\ese

\noindent{\bf 2.} \emph{The five interface estimates.} We verify each type of interface estimate used by Part~II.

\noindent{\bf 2.1.} \emph{Integrated fluxes.} The metric, matter, and order-$n$ effective-field fluxes are unnormalized nonnegative terms of $(\mathcal Y^b_k)^2$. Their restriction to order $l$ gives
\be
\label{equa:mobile-simultaneous-low-flux}
\sup_{s_0\leq\tau\leq s_1}\tau^{-\eta_{\Lscr}}\mathcal F^{\Lscr_b,\mathrm{unif}}_{l,k}(\tau;s_0)\leq C_Na_{0,b},\qquad0\leq k\leq l.
\ee
Here the bound uses $d_k\leq d_l<\eta_{\Lscr}$. The interior current is the negative of the exterior current on the same hypersurface. In Cartesian radial coordinates the flux density is $j_gK_{b,g}/(1+h)$; the factor $1/(1+h)$ is kept when passing between the current normalization and the interface norms. It is bounded above and below uniformly. Exact multiplication of the effective-field current by $\coeff$ gives its uniform normalization.

\bse
\noindent{\bf 2.2.} \emph{Traces on spheres and time-integrated traces at high differential order.} The radial trace estimate on each adapted hypersurface and $R_b^{1/2}\simeq s$ imply
\be
\label{equa:mobile-simultaneous-sphere-traces}
\mathsf T^b_{2;p,k}[u]+\mathsf T^b_{2;p,k}[\psi]\leq C_Na_{0,b}s^{-1+d_k},\qquad\mathsf T^b_{2;p,k}[\fieldvarphi]\leq C_Na_{0,b}s^{-1+d_k}\quad(p+1\leq n).
\ee
The first estimate holds for $p\leq n$. For the effective field, the hypersurface-radial derivative is a bounded combination of translation derivatives, which are measured by the mass term at order $p+1$. In particular, these bounds give all three traces at order $l$. For the inverse metric we keep the reference contribution separate and obtain $\mathsf B^{H,b}_{2;n,k}\leq C_N(A_0+a_{0,b}s^{d_k})$. Since $R_b'(s)\leq C s$, direct integration gives
\be
\label{equa:mobile-simultaneous-high-time-traces}
\sup_{s_0\leq\sigma\leq s_1}\sigma^{-\nu_k}\mathcal I^{\mathrm{tr},b}_{n,k}(\sigma)\leq C_{N,\mathrm{gaps}}(A_0+a_{0,b}).
\ee
Indeed, the square of the first term is at most $C(A_0+a_{0,b})^2\int_{s_0}^{\sigma}\tau^{-1+2\nu_k}\,d\tau$, and hence is bounded by $C(A_0+a_{0,b})^2\sigma^{2\nu_k}/(2\nu_k)$. The other two terms use $\tau^{-1+d_k}\leq1$ and the integrable metric kernel~\eqref{equa:seven-metric-integrable-kernel-L1}.
\ese

\bse
\noindent{\bf 2.3.} \emph{Pointwise and ray traces.} The proof of Proposition~\ref{lem:seven-pointwise-metric-interface-bulk}, with the physical-time conversion~\eqref{equa:audit-moving-interface-time-boost} and the improved exterior metric energy and good bulk, gives
\be
\label{equa:mobile-simultaneous-metric-time-trace}
\|ZH^{\H00}(T,\cdot)\|_{L^2(r=r_b(T))}\leq C_N\bigl(A_0+a_{0,b}T^{1/2-\lambda+d_N/2}+a_{0,b}T^{-1/2}\bigr),\qquad\ord Z\leq n-1.
\ee
Two sphere derivatives and one time-trace derivative use at most $l+3\leq n$ commutations. Since $d_N<\delta<2\lambda-1$ and $s^2/R_b$ is bounded, we obtain $\mathsf B^b_{00;k}\leq C_N(A_0+a_{0,b})$ for $k\leq l$.

For the strong effective-field ray data, the reconstruction~\eqref{equa:eight-exterior-massive-e-reconstruction} is evaluated at $q=q_b(t)\simeq1$, where $t\simeq s^2$. If $p\leq l-3$ and $\ord Q\leq p+1$, its flat contraction and ordered nonlinear products give
\be
\label{equa:mobile-simultaneous-massive-ray}
|Q\fieldvarphi|\leq C_{N,\coeff_0}a_{0,b}s^{-4+2\delta},\qquad|\Lcal Z\fieldPhi|\leq C_{N,\coeff_0}a_{0,b}s^{-5+2\delta},\qquad Z\preccurlyeq(p,k).
\ee
The largest required order is $p+7\leq n$; every nonlinear product uses one improved factor and bounded coarse factors. The normalized ray quantity is at most $C_{N,s_0,\coeff_0}a_{0,b}s^{-1/2+2\delta-\eta_k}$ and is therefore bounded. Its initial-hypersurface portion is bounded by $C_{\mathrm{data}}A_0$. Proposition~\ref{prop:eight-mobile-characteristics} gives the remaining matter and good-metric traces with amplitude $C_N(a_{0,b}+A_0+A^2)$. The pointwise interface quantity also includes the coarse interior initial and central values; thus
\be
\label{equa:mobile-simultaneous-pointwise-interface}
\max_{k\leq l-3}\mathcal A^{\mathrm{bd},b}_{l-3,k}(s_1)+\max_{k\leq l}\mathsf B^b_{00;k}(s_1)\leq C_N(A_0+A+A^2).
\ee
\ese

\bse

\noindent{\bf 2.4.} \emph{Uniformly controlled traces at high differential order.} At order $n$, the three flux terms entering $\mathcal I^U$, $\mathcal I^P$, and $\mathcal I^W$ are bounded by $C_Na_{0,b}s^{d_k}$. The effective-field term additionally has the factor $s^{-1/2}$. The inequality $d_k<\nu_k$, the high trace bound~\eqref{equa:mobile-simultaneous-high-time-traces}, and the prescribed initial values give
\be
\label{equa:mobile-simultaneous-high-interface}
\max_{k\leq n}\mathcal I^{\mathrm{rob},b}_{\mathrm h,k}(s_1)\leq C_{N,\mathrm{gaps}}(A_0+a_{0,b})\leq C_{\mathrm{bd}}A_0.
\ee
The contribution $A_0/\zeta_\star$ is kept in this inequality. The constant $C_{\mathrm{bd}}$ is fixed independently of $C_1$ and $A_0$.

\noindent{\bf 2.5.} \emph{Traces at high differential order for a fixed parameter.} The current rescaling gives
\be
\label{equa:mobile-simultaneous-fixed-interface}
\mathcal I^{\mathrm{tame},b}_k(s_1)\leq C_N(1+\coeff^{-1/2}){s_1}^{1/2+\nu_k}\mathcal I^{\mathrm{rob},b}_{\mathrm h,k}(s_1),\qquad k\leq n.
\ee
This quantity is finite for every fixed $\coeff>0$ and finite $s_1$. Thus all five interface estimates have been obtained from the original Cauchy norms and the exterior estimate.

\noindent{\bf 3.} \emph{The interior improvement and the smallness choice.} We set $C_\star\coloneqq C_N(1+C_{\mathrm{data}}+C_{\mathrm{bd}})$ and choose $C_1$ so large that the equation~\eqref{equa:seven-final-C1-choice} holds and $\sqrt2 C_NC_{\mathrm{data}}\leq C_1/4$ in~\eqref{equa:mobile-simultaneous-exterior}. The exterior hypersurface energies and their positive spacetime terms then improve with a fixed margin. We next decrease $A_0$, with every preceding constant fixed, so that the exponent margins~\eqref{equa:mobile-simultaneous-exponent-margins}, $C_NC_{\mathrm{ref}}A_0\leq c$, and the interface conditions
\be
\label{equa:mobile-simultaneous-interface-smallness}
\frac{K(A+A^2)t_i^{-e_0}}{e_0}\leq\frac14,\qquad K(A+A^2)t_i^{-e_0}\leq\frac18
\ee
hold. The reference coefficients and their required derivatives are smooth functions of \(m/r\) on the exterior, where \(r\geq r_b(t_i)>0\). Choosing a fixed upper bound \(m_0<r_b(t_i)/2\) keeps their denominators uniformly away from zero. Their contributions have coefficients \(C_Nm/s\) or \(C_Nm/s^3\), while the dynamic source estimates use \(C_NA/s\) and integrable coefficients. Thus the constants fixed above and the resulting threshold for \(A_0\) are uniform for \(0\leq m\leq m_0\). We also impose the finite smallness conditions~\eqref{equa:seven-final-smallness-choice} and~\eqref{equa:seven-final-central-smallness}, with constants enlarged by the fixed geometric comparisons of Section~\ref{section:mobile-interior-transfer}.

We include in this same choice the geometric smallness bounds in Step~4 and the inequality \(CA<1/2\) used for causal completeness in Lemma~\ref{lem:final-direct-causal-completeness}, where \(C\) is the fixed coefficient of \(A\log T\) in the geodesic energy estimate. This coefficient depends only on the fixed geometry and exponent gaps. Thus all these conditions are finitely many upper bounds on \(A_0\), after \(C_1\) and the other constants have been fixed.

Proposition~\ref{prop:audit-moving-interface-interior-transfer} gives the conditional estimates on the actual adapted domains, including the differentiated cone terms and the radial profile. Substitution of the five interface bounds above into those estimates gives the same three amplitudes $\mathcal U_{\mathrm{cl}},\mathcal L_{\mathrm{cl}},\mathcal W_{\mathrm{cl}}$ in the equation~\eqref{equa:seven-closure-explicit-amplitudes}, with $\mathcal B_{\mathrm{cl}}=C_\star(A_0+A+A^2)$. In particular, the lower energy amplitudes are bounded by these expressions and the high one by
\be
\label{equa:mobile-simultaneous-interior-raw}
\sup_{s_0\leq s\leq s_1}s^{-\nu_k}\mathsf E^{\mathrm{rob},b}_{\mathrm h,k}(s)\leq C_N\left(C_\star A_0+\frac{\mathcal B_{\mathrm{cl}}^2}{\rho\theta}\right),\qquad k\leq n.
\ee
Each expression has a linear part bounded by a fixed multiple of $C_\star A_0(1+\zeta_\star^{-1})$; every other term has at least two factors from $A_0,A,\mathcal B_{\mathrm{cl}}$. The choice of $C_1$ makes the linear parts at most $A/16$. A further reduction of $A_0$ makes each nonlinear part at most $A/16$ and verifies the coefficient and central-wave smallness conditions. The lower estimates and~\eqref{equa:mobile-simultaneous-interior-raw} therefore give the strict interior bounds in the statement. At fixed $\coeff$, we choose $B_\coeff$ by~\eqref{equa:seven-Bcoeff-final-choice}; the identity $D^b=\coeff^{-1/2}U^b$ and the singular energy inequality give the corresponding strict degenerate bound. This last choice does not enter any of the uniform smallness conditions.
\ese

\noindent{\bf 4.} \emph{Geometric estimates.} The radial metric recovery on $B_{R_b(s)}$, the rank-zero coefficient estimate, and $t\leq C s^2$ on the adapted interior give
\be
\label{equa:mobile-simultaneous-interior-geometry}
|u|\leq C_N(C_{\mathrm{bd}}A_0+A)s^{-1/2+\delta},\qquad(t/s)^2|H^{\H00}|\leq C_N\theta^{-1}\mathcal B_{\mathrm{cl}},\qquad|\fieldPhi|\leq C_N\sqrt{\coeff_0}\,\mathcal B_{\mathrm{cl}}.
\ee
The hyperboloidal energy identity~\eqref{equa:seven-adapted-density-exact} thus gives the strict interior energy comparison after a further uniform reduction of $A_0$.

For the exterior normal covector $n_b=dt-a\,dr$, where $a=\del_rT_b$, we use the quadratic-form conversion~\eqref{equa:eight-mobile-leaf-normal}. On the merging interval, $\zeta_b^2\gtrsim t^{-1}$. The Schwarzschild term has size $\Ocal(m/t)$, while the dynamic term is bounded by $C_N(A+A^2)t^{-1-e_0}$. The remaining terms have size at most $C|H|\zeta_b^2$. Thus $|H(n_b,n_b)|\leq C_N(m+A+A^2)\zeta_b^2$. On the flat exterior, $a=0$. These bounds strictly preserve the spacelike hypersurfaces and the positive wave energy forms. The factors $e^{-\fieldPhi}$ and the positive effective-field mass density preserve their comparison constants as well. This completes the improvement of every geometric bootstrap bound.

\noindent{\bf 5.} \emph{High regularity and continuation.} For fixed $\coeff>0$ and finite $s_1$, the improved interior order-$n$ energies and exterior order-$N$ energies bound the adapted energy $\mathcal E^{\mathrm{fix},b}_n$ on $[s_0,s_1]$. In this energy the weight is one in the interior and $w_b^\eta$ in the exterior. It is continuous on $\Lscr_b$, and its two directed interface currents cancel exactly. The weights are comparable to $(1+r)^\eta$ on a finite slab. Their derivatives and the adapted geometric coefficients are bounded on that slab.

We can therefore apply the proof of Lemma~\ref{lem:audit-fixed-parameter-joined} with these weights and domains. Its derivative count remains explicit: for a principal product $C_i\del^2Z_jv$, $i+j\leq N$, we put $a_N\coloneqq\lfloor(N+1)/2\rfloor$. If $i\leq a_N$, the coefficient estimate costs $i+4\leq n$ orders. If $i>a_N$, the lower Hessian costs $j+6\leq N-a_N-1+6\leq n$ orders and the high coefficient is measured by the Hardy norm on the full hypersurface. A lower first-derivative factor costs at most $\lfloor N/2\rfloor+5\leq n$. The reference metric is estimated separately, so the Hardy norm applies only to its dynamic perturbation. The current identity and Gronwall consequently give
\be
\label{equa:mobile-simultaneous-high-continuation}
\sup_{s_0\leq s<s_*}\mathcal E^{\mathrm{fix},b}_N(s)\leq C_{\coeff,s_1}\left(\mathcal E^{\mathrm{fix},b}_N(s_0)+\int_{s_0}^{s_1} R_N(\tau)^2\,d\tau\right),\qquad s_*\leq s_1<+\infty.
\ee
The constant depends on the finite lower-order slab bound and the fixed coefficient range, but not on $s_*$ or the evolving order-$N$ energy. The initial order-$N$ quantity on the right is only required to be finite; its size need not be uniform in $\coeff$.

Local existence and the initial constraints start a nonempty bootstrap interval and propagate the wave gauge. At a hypothetical finite maximal time, the strict energy and geometric improvements exclude exit through a coarse bound. The estimate~\eqref{equa:mobile-simultaneous-high-continuation} preserves the regularity needed for the local continuation criterion. The development therefore extends past that time. All choices preceding $B_\coeff$ were independent of $\coeff$, which proves the uniform bounds. This proves future-global existence for every fixed $\coeff>0$, and hence the existence assertion of Theorem~\ref{theorem:final-global-stability-and-limit}.
\end{proof}

\paragraph{Foliation used for the limiting family}
For a family of Cauchy data, we use the constants $K,C_1$ chosen above and a single upper bound $A_0\leq\varepsilon_0$ for the initial amplitudes. The resulting foliation is independent of $\coeff$. In the subsequent convergence argument, the symbols for hypersurfaces, domains, weights, normal densities, energies, and traces at the interface refer to these adapted objects. In particular, $a_0$ means $a_{0,b}$ and the exterior rank exponents are those of~\eqref{equa:mobile-simultaneous-exterior}. The coordinates, the differential operators, the reference metric, and the Cauchy hypersurface remain unchanged. Estimates stated on the original hypersurfaces are used through Proposition~\ref{prop:audit-moving-interface-interior-transfer} and the moving exterior estimates, rather than through an identification of integrals on different hypersurfaces. Since the initial hypersurface is identical and its weights are uniformly equivalent, the original initial smallness and preparation conditions are preserved. The uniform time-tail estimates use $t\leq C s^2$ in the adapted interior and $t\simeq s^2$ in the near-cone exterior; their positive exponents and their uniformity in $\coeff$ are unchanged.


\section{\texorpdfstring{Constraint propagation and estimates for the Einstein limit}{Constraint propagation and estimates for the Einstein limit}}
\label{section-15}


\subsection{\texorpdfstring{Estimates at the interface and spatial infinity}{Estimates at the interface and spatial infinity}}
\label{section-15-1}

\paragraph{Purpose}

The global estimates established in Section~\ref{section:mobile-simultaneous-improvement} combine the interior analysis of Section~\ref{section-11} with the exterior current estimates. We now justify the identities on the full exterior domain, propagate the geometric constraints, and estimate the prepared trace residual on the full hypersurface for the convergence argument in Section~\ref{section-16}.

In this section and in Section~\ref{section-16}, all energy and trace estimates refer to the adapted foliation just constructed. We omit the superscript or subscript \(b\) on its hypersurfaces, domains, interface, weight, Jacobian, and energy quantities. Thus \(\Lscr\), \(w\), and \(\mathcal Y_k\) here denote \(\Lscr_b\), \(w_b\), and \(\mathcal Y_k^b\), respectively. Their definitions keep the same differential orders and effective-field normalizations. The initial hypersurface is unchanged, and the estimates in the physical coordinates \((t,x)\) use the uniform comparisons proved in Section~\ref{section:mobile-foliation}.

\paragraph{\texorpdfstring{Bounds at the interface for the interior estimates}{Bounds at the interface for the interior estimates}}

The five estimates required by the interior argument have already been obtained during the simultaneous improvement of the interior and exterior bounds. Their differential orders are given here for later use.

\begin{proposition}[\texorpdfstring{Bounds at the interface from the exterior evolution}{Bounds at the interface from the exterior evolution}]
\label{proposition:eight-exterior-interface-outputs}
Under the hypotheses of Proposition~\ref{prop:mobile-simultaneous-improvement}, the restriction of the Cauchy development to \(\Lscr\) satisfies the five estimates~\eqref{equa:seven-interface-data-bounds} on every finite slab. The closing fluxes are controlled through order \(N-5\); an effective-field sphere trace requires one additional translation derivative; the characteristic entries are used at order \(p_{\mathbf{low}}-3=N-12\). The uniform constants are independent of \(s_1\) and \(\coeff\), with \(a_0\leq C_{\mathrm{data}}A_0\). The unscaled effective-field estimates keep their parameter factors and are finite for each fixed \(\coeff>0\).
\end{proposition}

\begin{proof}
These are the interface conclusions of Proposition~\ref{prop:mobile-simultaneous-improvement}. Its sphere estimates control the extra translation derivative by the effective-field mass term. Its characteristic estimates use Proposition~\ref{prop:eight-mobile-characteristics}, which transfers the matter and projected-metric estimates to the expanding domain without a sign assumption on \(H_N\). The rescaling~\eqref{equa:seven-fixed-interface-from-robust} gives the fixed-parameter conclusion.
\end{proof}

\paragraph{Truncating before removing the outer boundary}

We first apply the divergence theorem on finite radial truncations. For a fixed interval \([s_0,s_1]\), these are
\be
\MME_{R;[s_0,s_1]}\coloneqq \MME_{[s_0,s_1]} \cap\{r \leq R\}, \quad R\gg s_1^2,
\ee
whose outer boundary is the timelike cylinder \(\mathcal C_R\coloneqq\{r=R\}\cap\MME_{[s_0,s_1]}\). We keep its signed flux and obtain estimates independent of \(R\) before passing to infinity.

\begin{lemma}[Finite-radius weighted identity]
\label{lem:eight-finite-radius-flux}
Let \(Z= \del^IL^J\) and let \(X\) be the Euclidean multiplier used in the weighted current of Section~\ref{section:eight-weighted-wave-energies}. For \(v=Zu\), \(v=Z\psi\), or \(v=Z\fieldphi\), denote by \(\mathcal E_R(s,v) \) the corresponding flux on the truncated hypersurface and by \(\mathcal F_R(s_0,s,v) \) the flux through the interface \(\Lscr\). Then the finite-radius balance law is
\be
\label{equa:eight-finite-radius-flux}
{ \mathcal E_R(s,v) + \mathcal F_R(s_0,s,v) + \mathcal B_R(s_0,s,v) = \mathcal E_R(s_0,v) + \int_{\MME_{R;[s_0,s]}} \mathcal S_Z[v;X] \,J\, dx\, d\tau. }
\ee
Here \(\mathcal B_R\) is the oriented flux through \(\mathcal C_R\), and \(\mathcal S_Z\) contains the commuted source, the deformation tensor, and the derivative of the weight.  For the  effective field the identity holds at both scales: the fixed-parameter version uses \(\mathcal E_R^{\mathrm{sing}} \), whereas multiplication of the whole identity by \(\coeff\) gives \(\mathcal E_R^{\mathrm{unif}} = \coeff\mathcal E_R^{\mathrm{sing}} \) and the corresponding relation for \(\mathcal F_R\) and \(\mathcal B_R\).

\end{lemma}
\begin{proof}
We apply the divergence theorem to the current \(\mathcal P^\alpha_{\lambda,g}[v] \) in \eqref{equa:eight-weighted-current-definition} on \(\MME_{R;[s_0,s]} \).  The two hypersurface contributions have the usual signs.  The normal to the interface is the opposite of the normal used for the interior region, so its contribution is the oriented interface flux.  The remaining boundary is \(\mathcal C_R\), which gives \(\mathcal B_R\).  The deformation identity decomposes the bulk divergence into the positive quadratic form, the weight derivative, and the commuted source; these are the three pieces collected in \(\mathcal S_Z\).  For the matter field the current includes \(e^{-\fieldPhi}\). At order zero, its exchange term cancels the effective-field trace source after multiplication of the matter identity by \(16\pi\). At positive order, the differentiated coupling and the metric commutators remain in \(\mathcal S_Z\) and are estimated as in~\autoref{lem:eight-exact-matter-current}.

For \(v=Z\fieldphi\), the stress tensor has derivative part multiplied by \(3\coeff\) and zero-order part multiplied by one.  Multiplying the complete identity by \(\coeff\) changes the variables to \(Z\fieldPhi = \coeff Z\fieldphi\) and \(Z\fieldvarphi = \coeff^{1/2}Z\fieldphi\).  This operation keeps every term, and the fixed-parameter identity remains available for each prescribed \(\coeff>0\).  This proves the formula.
\end{proof}

\paragraph{Vanishing of the outer-cylinder flux}

\begin{lemma}[Vanishing of the outer-cylinder contribution]
\label{lem:eight-outer-cylinder-tail}
Assume the weighted bootstrap bounds of Section~\ref{section-13-3}.  Let \(\mathscr C\) be the finite collection of all metric, matter, and uniformly normalized effective-field currents occurring at the commutation orders of the proof.  For every finite \(s_1\), there is one sequence \(R_j\to\infty\) for which all oriented outer-cylinder contributions in this collection satisfy
\be
\label{equa:eight-outer-cylinder-tail}
{ \lim_{j\to\infty} \max_{P\in\mathscr C} \ \sup_{s_0\leq\sigma \leq s_1} \bigl| \mathcal B_{R_j}[P](s_0, \sigma) \bigr|=0. }
\ee
For each fixed \(\coeff>0\), the finitely many unscaled effective-field currents may be added to \(\mathscr C\); their bounds are allowed to depend on \(\coeff\).  On any finite interval on which \(\mathcal E_{\Theta,q}^{\mathrm{sing}} \) is finite, the finitely many trace-residual currents \(Q^{\mathrm{sing}}[Z\Theta] \), \(\ord Z\leq q-1\), may also be added to \(\mathscr C\).  For these currents the co-area sequence may depend on the fixed value of \(\coeff\), which does not affect an estimate whose constants are uniform in \(\coeff\).  The flux through a timelike cylinder may have either sign.

\end{lemma}
\begin{proof}
We use radial averaging to select one sequence of outer cylinders for all the finitely many commuted currents. This suffices for passing to infinity in their identities, since the other terms have uniquely determined limits. For a current \(P\in\mathscr C\), let \(\mathcal J_P^r\) denote the radial component of its flux density.  The co-area formula on \(R\leq r\leq2R\) gives
\be
\label{equa:eight-outer-cylinder-co-area-bound}
\int_R^{2R} \sup_{s_0\leq\sigma \leq s_1} \bigl| \mathcal B_\rho[P](s_0, \sigma) \bigr| \, d\rho \leq \int_{s_0}^{s_1} \!\int_{\{R\leq r\leq2R\}} | \mathcal J_P^r|J\, dx\, d\tau.
\ee
After summing this inequality over the finite collection \(\mathscr C\), we choose \(\rho_R\in[R,2R] \).  The maximum cylinder flux at \(\rho_R\) is bounded by \(R^{-1} \) times the sum of the annular tails.  These tails tend to zero by continuity of the weighted energy and Hardy integrals.  Taking \(R=2^j\) gives a sequence \(R_j\coloneqq \rho_{2^j} \). The bounds for the current densities have constants independent of \(\coeff\). The selected radii may depend on \(\coeff\); a uniform bound on the energies alone would not give uniform decay of their spatial tails.  For the unscaled field it is performed after \(\coeff>0\) has been fixed; its finite bound is allowed to contain the negative powers provided by the singular effective-field energy.  This proves \eqref{equa:eight-outer-cylinder-tail} under the finite-energy hypothesis, without a radiation condition at infinity.
\end{proof}

\paragraph{Removal of the artificial outer boundary}

\begin{proposition}[Removal of the artificial outer boundary]
\label{proposition:eight-passage-to-infinity}
Along the co-area sequence of~\autoref{lem:eight-outer-cylinder-tail}, the finite-radius identities of~\autoref{lem:eight-finite-radius-flux} converge as \(R\to\infty\) to the weighted identities on \(\MME_{[s_0,s]} \).  The convergence is valid simultaneously for the metric, the massless matter wave, the uniform effective-field variables, and, for each fixed \(\coeff>0\), the unscaled effective-field variable.

\end{proposition}
\begin{proof}
The positive hypersurface and spacetime terms converge monotonically. The source and deformation terms are absolutely integrable by the estimates in Proposition~\ref{prop:eight-mobile-coupled-estimate}; the interface lies inside every sufficiently large truncation. The outer flux tends to zero along the sequence in~\autoref{lem:eight-outer-cylinder-tail}. These facts give the identity on the full exterior. The same argument applies to the unscaled effective field for each fixed \(\coeff>0\), keeping its parameter-dependent bound.
\end{proof}


\subsection{\texorpdfstring{High-order commutators and gauge propagation}{High-order commutators and gauge propagation}}
\label{section-15-2}

\paragraph{Affine commutators and the current}

The principal commutator contains the full Hessian, including its angular components. We first keep its tensor identity and then use the signed source estimates of the exterior current.

\begin{lemma}[\texorpdfstring{Affine commutator identity}{Affine commutator identity}]
\label{lem:eight-null-top-order}
Let \(Z=X_1\cdots X_n\), where every \(X_i\) is a Cartesian translation or a Lorentz boost. For a subset \(I\subset\{1, \ldots,n\} \), let \(Z_I\) and \(\mathcal L_I\) denote the scalar and tensorial compositions in their original order, with the identity operator for the empty subset. Then one has
\be
\label{equa:eight-null-decomposition}
[Z,H^{\alpha \beta} \del_{\alpha \beta}]v = \sum_{\varnothing\ne I\subset\{1, \ldots,n\}} (\mathcal L_I H)^{\alpha \beta} \del_{\alpha \beta}(Z_{I^c}v).
\ee
For a single affine generator, the coefficient is
\be
\label{equa:eight-affine-Lie-principal-coefficient}
(\mathcal L_XH)^{\alpha \beta} =XH^{\alpha \beta}-H^{\gamma \beta} \del_\gamma X^\alpha-H^{\alpha \gamma} \del_\gamma X^\beta.
\ee
\end{lemma}

\begin{proof}
Since \(\del^2X=0\), the product rule gives \([X,H^{\alpha \beta} \del_{\alpha \beta}]v= (\mathcal L_XH)^{\alpha \beta} \del_{\alpha \beta}v\). Repeated application of this identity gives the ordered-subset sum. The term with \(I= \varnothing\) is \(H^{\alpha \beta} \del_{\alpha \beta}Zv\) and cancels in the commutator. The identity is valid in Cartesian components; any subsequent change to the semi-null frame keeps every transformed Hessian and its frame-derivative term.
\end{proof}

The current argument estimates these contractions after pairing them with the multiplier derivative. Its two derivative distributions have distinct estimates.
\begin{itemize}
\item A coefficient operator of order at most \(N-5\) is treated by the radial wave-gauge estimates in~\autoref{lem:eight-massive-principal-commutator}.
\item A higher coefficient operator leaves at most four operators on the wave Hessian. The estimate \eqref{equa:eight-high-density-wave-pairing} keeps the good metric bulk and its terminal Hardy contribution. The other tensor components and the spatial end use~\eqref{equa:audit-high-density-good-pairing}--\eqref{equa:audit-high-density-spatial-pairing}; their coefficients are integrable.
\end{itemize}
The semilinear metric, matter, and effective-field sources are estimated in the same summed current. The fractions of positive bulk spent in these pairings are specified in Proposition~\ref{prop:eight-mobile-coupled-estimate}; its remaining positive quantity is \(\mathcal Y_k\).

Proposition~\ref{proposition:eight-exterior-package} below gives the resulting energy bounds together with the sphere estimates and their distinct differential orders.

\paragraph{The reduced system implies the geometric system}

We recover the geometric equations by uniqueness for the homogeneous subsidiary equation of the wave gauge. This is a finite-time argument: smoothness of the reduced solution gives locally bounded coefficients. Once the gauge vector vanishes, all its commuted derivatives vanish as well. The commuted identity below makes explicit how the derivative levels are coupled.

\begin{lemma}[Commutation of the gauge subsidiary system]
\label{lem:eight-gauge-constraint-source}
Let \(\Gamma^\alpha=g^{\mu\nu} \Gamma^\alpha_{\mu\nu} \) be the wave-gauge constraint. On every smooth finite slab, the reduced equations imply the homogeneous system \eqref{equa:seven-subsidiary-wave}. Denoting its lower-order operator by \((\mathcal A\Gamma)^\alpha\coloneqq A^\alpha{}_{\beta}{}^\mu\nabla_\mu\Gamma^\beta+B^\alpha{}_{\beta} \Gamma^\beta \), one has, for every coordinate-component commutator \(Z\) through the available regularity order,
\be
\label{equa:eight-gauge-constraint-source}
\Box_g(Z\Gamma)= [\Box_g,Z] \Gamma-Z(\mathcal A\Gamma).
\ee
Here \(\Box_g\) is the covariant wave operator on vectors; its first-order connection terms are included in the smooth subsidiary coefficients. The right-hand side is homogeneous and linear in \(\Gamma \) and its derivatives.
\end{lemma}

\begin{proof}
The total source tensor \eqref{equa:seven-total-source-subsidiary} satisfies the divergence identity \eqref{equa:seven-total-source-gauge-divergence}. Together with the contracted Bianchi identity and the reduced metric equation, this gives \eqref{equa:seven-subsidiary-wave}. Changing from the component wave operator to the covariant vector wave operator changes only the smooth coefficients \(A,B\). Applying \(Z\) to \(\Box_g\Gamma=-\mathcal A\Gamma \) gives the commuted identity. The product rule preserves homogeneity in the gauge vector and its derivatives.
\end{proof}

\begin{proposition}[Propagation of the gauge constraint and cancellation of the interface fluxes]
\label{proposition:eight-gauge-compatibility}
Assume the initial constraints and their compatibility conditions vanish. On each finite domain of dependence in a slab \([s_0,s_1] \), a positive wave energy \(\mathcal E_\Gamma \) for the uncommuted subsidiary system, including the local \(L^2\) norm of \(\Gamma\), satisfies
\be
\label{equa:eight-gauge-constraint-energy}
{ \mathcal E_\Gamma(s)^{1/2} \lesssim \mathcal E_\Gamma(s_0)^{1/2} + \int_{s_0}^{s}C_{s_1, \coeff}(\tau) \mathcal E_\Gamma(\tau)^{1/2} \, d\tau, }
\ee
with \(C_{s_1, \coeff} \in L^1([s_0,s_1]) \), also depending on the chosen domain. The two contributions of the interior--exterior interface cancel with opposite orientations.  Consequently, \(\Gamma\equiv0\), and the reduced solution satisfies the geometric equations in conformal variables.  Proposition~\ref{pro:40S} then propagates the curvature compatibility relation and identifies the solution with a solution of the original \(f(R) \) equations for every fixed \(\coeff>0\).

\end{proposition}
\begin{proof}
We first apply the wave-energy estimate to the uncommuted subsidiary equation on a finite domain of dependence. The lower-order subsidiary terms are controlled by the energy including the local \(L^2\) norm of \(\Gamma\); the smooth metric coefficients give the integrable coefficient in \eqref{equa:eight-gauge-constraint-energy}. The initial constraints give zero Cauchy data for \(\Gamma \). Gronwall's inequality therefore gives \(\Gamma=0\) on this domain. Exhaustion by finite domains of dependence proves the assertion throughout the smooth development, for every fixed \(\coeff>0\).

If the energy identity is split into interior and exterior pieces, the interface values agree and the induced normals are opposite, so the signed fluxes cancel. Thus the  split gives the same uniqueness argument. Differentiating \(\Gamma=0\) now gives \(Z\Gamma=0\) through the available regularity order. Substitution in the reduced metric equation recovers the geometric metric equation, and Proposition~\ref{pro:40S} propagates the curvature compatibility relation.
\end{proof}


\subsection{Exterior energy bounds and estimates on distant annuli}
\label{section-15-3}

\paragraph{Fixed-parameter and rescaled exterior estimates}

The exterior energy estimate and the sphere-trace estimate have different derivative ranges, which are specified separately.

\begin{proposition}[Exterior energy bounds and traces on spheres]
\label{proposition:eight-exterior-package}
\label{proposition:eight-top-order-absorption}
Under the hypotheses of Proposition~\ref{prop:eight-mobile-coupled-estimate} and the exponent margins \eqref{equa:mobile-simultaneous-exponent-margins}, let \(a_0\) be the actual initial amplitude. On every finite available bootstrap slab, for \(p\in\{N-5,N\} \) and \(0\leq k\leq p\), one has
\be
\label{equa:eight-top-order-absorption}
\mathsf X_{p,k}^{\mathrm{unif}}(s)+\Bulk^{\ME,\mathrm{unif}}_{p,k}(s_0,s)\leq C_N\mathcal Y_k(s)\leq C_Na_0s^{d_k}.
\ee
The corresponding fixed-parameter estimate is
\be
\label{equa:eight-exterior-package-summary}
\mathsf X_{p,k}^{\mathrm{fix}}(s)+\Bulk^{\ME,\mathrm{fix}}_{p,k}(s_0,s)\leq C_N(1+\coeff^{-1/2})a_0s^{d_k}.
\ee
The effective-field sphere traces satisfy, for \(p+1\leq N-5\) and \(0\leq k\leq p\),
\be
\label{equa:eight-exterior-package-mass-traces}
\begin{aligned}
\coeff^{1/2} \mathsf T_{2;p,k}[\fieldphi](s) = \mathsf T_{2;p,k}[\fieldvarphi](s)& \leq C_Na_0s^{-1+d_k}, \\
\mathsf T_{2;p,k}[\fieldPhi](s) = \coeff^{1/2} \mathsf T_{2;p,k}[\fieldvarphi](s)& \leq C_N\coeff^{1/2}a_0s^{-1+d_k}.
\end{aligned}
\ee
All constants are independent of the terminal time and of \(0<\coeff\leq\coeff_0\). At order \(N\), the effective-field hypersurface norm is divided by \(s\), and its flux and bulk integrals carry the time weight \(\tau^{-2}\); at order \(N-5\), their weight is one, as specified after~\eqref{equa:eight-exterior-instantaneous-bulk-definition}.
\end{proposition}

\begin{proof}
The current comparisons and positive fractions kept in the energy in~\eqref{equa:eight-mobile-coupled-positive-energy} bound the hypersurface energies and bulk integrals by \(C_N\mathcal Y_k\). Proposition~\ref{prop:eight-mobile-coupled-estimate} gives the uniform bound; its rank induction keeps the coefficient \(C_NA\) and the terminal lower-rank good-bulk contribution in~\eqref{equa:eight-mobile-coupled-rank-inequality}. The relation \(D^{\mathrm{ext}}_{p,k}=\coeff^{-1/2}U^{\mathrm{ext}}_{p,k}\) gives the fixed-parameter estimate. For a sphere trace, the radial derivative of \(Z\fieldvarphi\) is a bounded combination of translation derivatives of order \(p+1\), controlled by the mass term in the closing energy. The radial estimate used in Proposition~\ref{prop:mobile-simultaneous-improvement} gives \(s^{-1+d_k}\); the two rescaling identities then give the stated effective-field traces.
\end{proof}

\paragraph{Positive bulk estimates on distant annuli}

The annular bounds needed at spacelike infinity follow directly from the positive bulk kept in the energy. On a fixed slab, the first radius is chosen beyond the entire merging region so that the exterior weight is comparable to the radius.

\begin{lemma}[Spacetime integral estimates on distant annuli]
\label{lem:eight-dyadic-annular-trace}
Let \(s_1\geq s_0\), \(p\leq N-5\), and \(0\leq k\leq p\). Suppose that \(R_0\geq C_{\mathrm{fol}}(1+{s_1}^2) \), where the fixed foliation constant is sufficiently large. For \(R_j=2^jR_0\) and \(\mathcal A_j(s)\coloneqq \{R_j\leq r\leq2R_j\} \cap\MME_s\), one has
\be
\label{equa:eight-dyadic-annular-trace}
\begin{aligned}
\sum_{j\geq0} \int_{s_0}^{s_1}\!\int_{\mathcal A_j(s)} \Bigl(R_j^{2\lambda-1}| \delsN u|_{p,k}^2 +R_j^{2\mu-1} \bigl(| \delsN\psi|_{p,k}^2 + | \delsN\fieldPhi|_{p,k}^2 + | \fieldvarphi|_{p,k}^2\bigr) \Bigr)J\, dx\, ds \leq C_Na_0^2{s_1}^{2d_k}.
\end{aligned}
\ee
The constant is independent of \(R_0\), \(s_1\), and \(\coeff\).
\end{lemma}

\begin{proof}
On every annulus, \(w\simeq r\simeq R_j\), with fixed comparison constants. The metric good-derivative bulk has weight \(w^{2\lambda-1} \); the matter good-derivative bulk, the effective good-derivative bulk, and the effective mass bulk have weight \(w^{2\mu-1} \). The density \(j_g\) is uniformly comparable to one. At the orders in the statement these four positive terms occur without top-order time normalization in \(\mathcal Y_k(s_1)^2\). The annuli have disjoint interiors, so summation and Proposition~\ref{prop:eight-mobile-coupled-estimate} give the estimate.
\end{proof}


\subsection{\texorpdfstring{Propagation of preparation for the Einstein limit}{Propagation of preparation for the Einstein limit}}
\label{section-15-4}

\paragraph{Energy of the trace residual and cancellation of the interface fluxes}

For later use, we define the singular energy of the trace residual on the full hypersurface. We use the adapted hypersurfaces and normal density \(N_b\), together with the exterior weight \(w_b\) in the equation~\eqref{equa:eight-mobile-exterior-weight}. We use the Cartesian time multiplier \(X_{\ME} = \del_t=X_\H\) fixed in Proposition~\ref{proposition:eight-weighted-current-identity}. With the singular effective-field tensor \(Q^{\mathrm{sing}}\) from \eqref{equa:seven-exact-effective-field-tensors}, we set
\be
\label{equa:eight-joined-trace-residual-energy}
\mathcal E_{\Theta,q}^{\mathrm{sing}}(s)\coloneqq2\sum_{\ord Z\leq q-1} \Bigl(\calE_{\mathrm{eff}}^{\mathrm{sing}, \H}(s,Z\Theta) + \int_{{\MME}_s^b} w_b^{2\mu}Q^{\mathrm{sing}}_{\alpha \beta}[Z\Theta] \,N_b^\alpha X_{\ME}^\beta \, dx\Bigr), \qquad q\geq1.
\ee
On every fixed slab \(s_0\leq s\leq s_1\), the energy comparisons in Sections~\ref{section-8-2} and~\ref{section-13-3} give the following equivalence, with constants independent of \(\coeff\):
\be
\label{equa:eight-joined-trace-residual-coercivity}
\mathcal E_{\Theta,q}^{\mathrm{sing}}(s) \simeq_{s_1}\sum_{\ord Z\leq q-1} \int_{\Mcal_s^b}w_s^2\bigl(|Z\Theta|^2 + \coeff| \Ds Z\Theta|^2\bigr) \, dx, \qquad w_s\coloneqq \mathbf 1_{{\Mint}_s^b} + w_b^\mu\mathbf 1_{{\MME}_s^b}.
\ee
The finite-slab dependence comes from expressing the Cartesian time-current energy in the unit-normal frame. In the interior, the uniformly controlled derivatives are \((s/t) \del_t\) and the semi-hyperboloidal derivatives, as in \eqref{equa:seven-D-coercivity}.

The factor \(2\) in \eqref{equa:eight-joined-trace-residual-energy} is the \(\mathcal P=-2QX\) coordinate-current normalization. The interior energy is taken on \({\Mint}_s^b\). On \(\Lscr_b\), the exterior weight satisfies \(w_b=1\), both multipliers are \(\del_t\), and the two induced orientations are opposite. Their signed interface fluxes therefore cancel exactly. Formula \eqref{equa:eight-joined-trace-residual-coercivity} also specifies the topology of every strong trace-residual convergence statement below: the residual itself converges without rescaling, whereas its differentiated component carries the factor \(\coeff^{1/2}\).

\paragraph{Initial-data control of the trace residual}

The trace residual is \(\Theta=\fieldphi+\mathcal S\), where \(\mathcal S\) is the matter trace source in the equation~\eqref{equa:seven-Theta-equation}. Its singular energy measures preparation. We estimate its evolution on each fixed slab with constants independent of \(\coeff\).

\begin{lemma}[Equation for the commuted trace residual]
\label{lem:eight-trace-residual-source}
Let \(\Theta \) denote the trace residual appearing in \eqref{equa:seven-Theta-equation}.  For every commutator \(Z\) in the convergence range, the commuted equation is
\be
\label{equa:eight-trace-residual-source}
3\coeff\, \Box_g Z\Theta-Z\Theta = \coeff\, \mathcal R_Z^{\Theta},
\ee
where the source \(\mathcal R_Z^\Theta \) is defined by
\be
\label{equa:eight-trace-residual-source-definition}
\mathcal R_Z^{\Theta} \coloneqq \coeff^{-1}Z W_\coeff(\fieldphi) +3Z\bigl(\Box_g\mathcal S\bigr) +3[\Box_g,Z] \Theta.
\ee
Under the uniform bounds through order \(p_{\mathbf{high}}=N-5\), for every fixed \(s_1\geq s_0\) and \(1\leq q\leq p_{\mathbf{high}}-1\), one has the following estimate on \([s_0,s_1] \). The source norm is
\be
\label{equa:eight-trace-residual-source-norm}
\|F\|_{L^2_{\mathrm{tot}, \eta}(s)}\coloneqq \|F\|_{L^2({\Mint}_s^b)} + \|J_b\zeta_b^{-1}w_b^\eta F\|_{L^2({\MME}_s^b)}, \qquad\eta \in\{\lambda, \mu\}.
\ee
The commuted sources satisfy
\be
\label{equa:eight-trace-residual-source-bound}
\coeff^{1/2} \sum_{\ord Z\leq q-1} \| \mathcal R_Z^\Theta \|_{L^2_{\mathrm{tot}, \mu}(s)} \leq C_{s_1,q}A\mathcal E_{\Theta,q}^{\mathrm{sing}}(s)^{1/2} +C_{s_1,q} \coeff^{1/2}A^2.
\ee
The constants depend on the fixed slab and robust bounds, and are independent of \(\coeff\).

\end{lemma}
\begin{proof}
\bse
We check the parameter scale using the potential factorization, and the differential order using a Bochner identity and the robust kinetic energies.

\noindent{\bf 1.} \emph{Commuted residual equation.} Commuting \eqref{equa:seven-Theta-equation} gives
\be
3\coeff \, \Box_gZ\Theta-Z\Theta =ZW_\coeff(\fieldphi) +3\coeff Z(\Box_g\mathcal S) +3\coeff[\Box_g,Z] \Theta,
\ee
which is \eqref{equa:eight-trace-residual-source}-- \eqref{equa:eight-trace-residual-source-definition}.  The uniformly smooth factorization \(W_\coeff(\fieldphi)= (\fieldvarphi)^2\mathcal W_\coeff(\fieldPhi) \), together with \(\fieldvarphi = \coeff^{1/2}(\Theta-\mathcal S) \), gives
\be
\label{equa:eight-trace-potential-linearization}
\coeff^{-1/2}W_\coeff(\fieldphi) = \bigl((\fieldvarphi-\coeff^{1/2} \mathcal S) \Theta+ \coeff^{1/2} \mathcal S^2\bigr) \mathcal W_\coeff(\fieldPhi).
\ee
On the fixed slab, we apply spatial Sobolev with two derivatives to the Cartesian scalar jets multiplied by the smooth weight \(\crochet^\mu\). This weight is comparable to the weight that is one in the interior and \(w_b^\mu\) in the exterior. The graph coefficients and the first two spatial derivatives of \(\log\crochet\) are bounded by constants depending on \(s_1\), and each spatial differentiation along the graph is a bounded combination of Cartesian translations. The robust energies therefore control \(w_b^\mu\del Z\psi\), \(w_b^\mu\del Z\fieldPhi\), and \(w_b^\mu Z\fieldvarphi\) pointwise for \(\ord Z\leq N-7\); the interior bounds have weight one. Metric coefficients and differentiated connection coefficients are estimated in unweighted \(L^\infty\), using the metric Hardy estimate for undifferentiated factors. At \(N=19\), twelve commutations on \(\mathcal S\) contain a thirteenth derivative of \(\psi\). Its two Sobolev derivatives reach order fifteen, measured by the kinetic energy at commutation order fourteen. The ordered product rule now gives
\be
\label{equa:audit-trace-fixed-slab-coefficient}
\sum_{\ord Z\leq N-7}\|w_b^{2\mu}Z\mathcal S\|_{L^\infty({\MME}_s^b)}\leq C_{s_1,N}A^2.
\ee
Thus a high commutation on \(\mathcal S\) can be paired with a lower commutation on \(\Theta\) in \(L^2(w_b^{2\mu}dx)\). Every commuted coefficient multiplying \(\Theta \) in this identity is bounded by \(C_{s_1}A\), while the last term is bounded in the source norm by \(C_{s_1}\coeff^{1/2}A^4\). Since \(A\leq1\), this is no larger than the forcing in \eqref{equa:eight-trace-residual-source-bound}.

\noindent{\bf 2.} \emph{Second derivatives of the matter trace.} We next verify that \(\Box_g\mathcal S\) causes no hidden loss of derivatives.  We set \(Q_\psi\coloneqq g(\nabla \psi, \nabla \psi) \), so that \(\mathcal S=-8\pi e^{-\fieldPhi}Q_\psi\).  The matter equation \(\Box_g\psi =g(\nabla \fieldPhi, \nabla \psi) \) and the Bochner identity give the formula
\be
\label{equa:eight-trace-source-Bochner}
{\begin{aligned}
\Box_g\mathcal S&=-8\pi e^{-\fieldPhi} \bigl(2\nabla_{\alpha \beta} \psi\nabla^{\alpha \beta} \psi +2R_{\alpha \beta}(g) \nabla^\alpha \psi\nabla^\beta \psi +2\nabla_{\alpha \beta} \fieldPhi\nabla^\alpha \psi\nabla^\beta \psi\\
& \quad -2\nabla_{\alpha \beta} \psi\nabla^\alpha \fieldPhi\nabla^\beta \psi + \bigl(-\Box_g\fieldPhi+g(\nabla \fieldPhi, \nabla \fieldPhi) \bigr)Q_\psi\bigr).
\end{aligned}}
\ee
Moreover, the effective-field equation and the definition of \(\Theta \) imply the cancellation
\be
\label{equa:eight-Box-Phi-trace-residual}
{3\Box_g\fieldPhi = \fieldphi+ \mathcal S+W_\coeff(\fieldphi)= \Theta+W_\coeff(\fieldphi).}
\ee
In the Bochner formula, we estimate \(\Box_g\fieldPhi\) directly as a contraction of its covariant Hessian, using the robust kinetic energies. This step is independent of the smallness of \(\Theta\); the identity~\eqref{equa:eight-Box-Phi-trace-residual} identifies the trace relation but is not needed for this source bound. For \(d\coloneqq\ord Z\leq q-1\leq p_{\mathbf{high}}-2\), each differentiated Hessian has at most \(d+2\) derivatives and is controlled by the kinetic energy through commutation order \(d+1\leq N-6\). The translations include \(\del_t\), so the same count applies to the time derivatives. On a fixed slab, \(s/t\) and \(\zeta_b\) have positive lower bounds, independently of \(\coeff\). We express the Ricci tensor through the conformal field equation; its factors are then metric coefficients, scalar first derivatives, and the rescaled potential. We put the metric and connection coefficients in unweighted \(L^\infty\) by the fixed-slab Sobolev argument above. The highest scalar derivative is placed in \(L^2\) with weight \(w_b^\mu\); the other scalar factors have the two Sobolev orders needed for \(L^\infty\). This uses the Hardy weight \(w_b^{\lambda-1}\) for an undifferentiated metric factor only in its pointwise estimate, rather than assigning it the scalar \(L^2\) weight. Thus \(\|Z\Box_g\mathcal S\|_{L^2_{\mathrm{tot},\mu}}\leq C_{s_1,q}A^2\), and multiplication by \(\coeff^{1/2}\) gives the required vanishing forcing.

\noindent{\bf 3.} \emph{Metric commutators.} The affine commutators commute with the flat wave operator. Their tensor Lie derivatives therefore express \([\Box_g,Z] \Theta \) as a sum of differentiated metric coefficients times Hessians of lower operators in \(\Theta \), with at most \(q\) derivatives of \(\Theta \). The coefficient is bounded by \(C_{s_1}A\) in \(L^\infty\), using the two remaining Sobolev orders and the metric Hardy estimate. The singular energy controls \(\coeff^{1/2} \del Z'\Theta \) for \(\ord Z'\leq q-1\). The matrices of first derivatives of the affine fields are constant, so no additional unbounded coordinate factor is introduced. These estimates prove \eqref{equa:eight-trace-residual-source-bound}.
\ese
\end{proof}

\begin{proposition}[Uniform trace-residual estimate under preparation]
\label{proposition:eight-uniform-trace-residual}
Let the initial family satisfy the robust assumptions and the preparation hypothesis of Definition~\ref{definition:Einstein-compatible-commuted-preparation} at the top order \(q_\star=N-6\).  On the Cauchy development, for every finite \(s_1\geq s_0\) and every \(1\leq q\leq N-6\), one has
\be
\label{equa:eight-preparation-modulus-estimate}
\sup_{s_0\leq s\leq s_1} \mathcal E_{\Theta,q}^{\mathrm{sing}}(s)^{1/2} \lesssim_{s_1}\mathrm{Prep}^{\mathrm{com}}_{\coeff,q}[\Theta]+ \coeff^{1/2} \varepsilon^2.
\ee
Consequently, an Einstein-compatible family that is well prepared in the commuted topology has strong convergence of the trace residual on each hypersurface.  If the family has preparation rate \(\sigma>0\), then, with \(\sigma_\star\coloneqq \min\{\sigma,1/2\} \), one has
\be
\label{equa:eight-positive-power-trace-residual}
\sup_{s_0\leq s\leq s_1} \mathcal E_{\Theta,q}^{\mathrm{sing}}(s)^{1/2} \lesssim_{s_1}\coeff^{\sigma_\star} \varepsilon.
\ee
The implicit constants are independent of \(\coeff\), although they may depend on the fixed terminal time \(s_1\). Since the commutations include all Cartesian translations, the energy controls derivatives of \(\Theta\) through order \(q-1\) without parameter weights on each fixed finite slab; its order-\(q\) derivative component keeps the factor \(\sqrt\coeff\). Unweighted convergence at this last order does not follow from this estimate.

\end{proposition}
\begin{proof}
\bse
We fix \(s_1<+\infty\) and \(1\leq q\leq N-6\). The constants below may depend on \(s_1,q\), but are independent of \(\coeff\). We write \(E_q(s)\coloneqq\mathcal E_{\Theta,q}^{\mathrm{sing}}(s)^{1/2}\).

\noindent{\bf 1.} \emph{Finite energy before the uniform estimate.} For each fixed \(\coeff>0\), the closed effective-field hierarchy gives finite unscaled energies on \([s_0,s_1]\). Since \(\Theta=\fieldphi+\mathcal S\), the product estimates for the matter trace give \(E_q(s)<+\infty\) on this interval. The largest derivative of \(\psi\) needed here has order \(q+1\leq N-5\). This preliminary finiteness may depend on \(\coeff\); the estimate below removes that dependence.

\noindent{\bf 2.} \emph{Energy identity and passage to infinity.} We apply the singular current identity to \(Z\Theta\), \(\ord Z\leq q-1\), on the full domain truncated at \(r=R\). We denote its energy by \(E_{q;R}^2\). The interior and exterior multipliers are both \(\del_t\), and \(w_b=1\) on \(\Lscr_b\), so the signed interface fluxes cancel. With \(j_g=\sqrt{|\det g|}\) and \(w_s\) from the equation~\eqref{equa:eight-joined-trace-residual-coercivity}, the source pairing satisfies
\be
\label{equa:eight-trace-source-pairing}
2\coeff\left|\int_{\Mcal_{s;R}^b}J_bj_gw_s^2\mathcal R_Z^\Theta\del_tZ\Theta\,dx\right|\lesssim\sqrt\coeff\,\|\mathcal R_Z^\Theta\|_{L^2_{\mathrm{tot},\mu;R}(s)}E_{q;R}(s).
\ee
In the interior, this follows from \(J_b=s/t\); in the exterior, the factor \(J_b\zeta_b^{-1}\) is part of the source norm. The derivatives of \(w_b\), including the term \(-h(t)dt\) in its differential, contribute at most \(C_{s_1,q}E_q^2\) on the fixed slab. The metric deformation terms have the same bound. To remove the outer boundary, we sum the absolute cylinder fluxes of the finitely many residual currents. Radial averaging gives
\be
\label{equa:eight-trace-outer-cylinder-co-area}
\int_R^{2R}\sup_{s_0\leq s\leq s_1}\mathcal B_{\Theta,q;\rho}^{\mathrm{abs}}(s_0,s)\,d\rho
{\lesssim\sum_{\ord Z\leq q-1}\int_{s_0}^{s_1}\int_{{\MME}_s^b\cap\{R\leq r\leq2R\}}w_b^{2\mu}(|Z\Theta|^2+\coeff|\del Z\Theta|^2)J_bj_g\,dx\,ds.}
\ee
The right-hand side has a vanishing annular tail for each fixed \(\coeff\), by the preliminary finite-energy bound. Thus Lemma~\ref{lem:eight-outer-cylinder-tail} gives a sequence along which the entire supremum of the cylinder error tends to zero. Positive terms pass to the limit by Fatou's lemma; source and deformation terms pass by dominated convergence, using the source estimate in the equation~\eqref{equa:eight-trace-residual-source-bound}.

\noindent{\bf 3.} \emph{Uniform estimate and preparation rate.} The initial-energy comparison and the source estimate give
\be
\label{equa:eight-trace-bootstrap-Volterra}
E_q(s)\leq C_{s_1,q}\mathrm{Prep}^{\mathrm{com}}_{\coeff,q}[\Theta]+C_{s_1,q}(1+A)\int_{s_0}^sE_q(\tau)\,d\tau+C_{s_1,q}\sqrt\coeff\,A^2(s-s_0).
\ee
We apply Gronwall's inequality and use \(A\leq C_A\varepsilon\), with \(C_A\) independent of \(\coeff\). Absorbing the fixed interval length and the exponential factor into the constant gives
\be
\label{equa:eight-trace-bootstrap-lower-orders}
\sup_{s_0\leq s\leq s_1}E_q(s)\leq C_{s_1,q}\bigl(\mathrm{Prep}^{\mathrm{com}}_{\coeff,q}[\Theta]+\sqrt\coeff\,\varepsilon^2\bigr).
\ee
This proves the preparation-modulus estimate at every stated order. A preparation rate \(\sigma>0\) gives the rate \(\coeff^{\min\{\sigma,1/2\}}\varepsilon\). The energy comparison in the equation~\eqref{equa:eight-joined-trace-residual-coercivity} then gives strong convergence of \(\Theta\) on each hypersurface and energy convergence of \(\sqrt\coeff\,\nabla\Theta\).
\ese
\end{proof}

\paragraph{Scope of the trace-residual estimate}

Proposition~\ref{proposition:eight-uniform-trace-residual} gives the high-order residual estimate. Its initial term is the preparation amplitude \(\mathrm{Prep}^{\mathrm{com}}_{\coeff,q}[\Theta] \), and its right-hand side tends to zero for well-prepared families. Proposition~\ref{proposition:seven-local-vanishing-from-robust} gives the corresponding local interior consequence of the uniform commuted bounds. On each fixed finite slab, this estimate and the effective-field rescaling make the additional source terms tend to zero in the local metric--matter comparison of Section~\ref{section-16-2}. The passage from compact regions to the whole future spacetime is concluded in Section~\ref{section:final-global-uniform-convergence}, using decay uniform in \(\coeff\).

\label{section-15-5}

\paragraph{Normalization of the source terms}

The three normalized effective-field sources in \eqref{equa:eight-effective-field-source-normalizations} satisfy
\be
\label{equa:eight-commuted-source-singular-recovery}
\|w^\mu F_Z^{\fieldphi} \|_{L^2} = \coeff^{-1} \|w^\mu F_Z^{\fieldPhi} \|_{L^2} = \coeff^{-1/2} \|w^\mu F_Z^{\fieldvarphi} \|_{L^2}.
\ee
In each energy identity, the source is paired with the multiplier derivative at its own scale; the good-derivative, bad-derivative, and mass contributions are all kept.

\paragraph{Pointwise bounds with the angular derivative count}

We put \(\eta_u= \lambda \) and \(\eta_\psi = \mu\). For \(v\in\{u, \psi\} \), the weighted Sobolev and Hardy estimates give
\be
\label{equa:eight-local-energy-pointwise}
\sup_{\MME_s} \Bigl(rw^{\eta_v}| \del v|_{p,k} +rw^{\eta_v}| \delsME v|_{p,k} +rw^{\eta_v-1}|v|_{p,k} \Bigr) \lesssim\Fenergy^{\ME,p+5,k+5}_{\eta_v}(s,v), \qquad p+5\leq N.
\ee
The corresponding uniform effective-field estimate is
\be
\label{equa:eight-local-energy-pointwise-effective-field}
\sup_{\MME_s}rw^\mu\bigl(| \del\fieldPhi|_{p,k} + | \fieldvarphi|_{p,k} \bigr) \lesssim U^{\mathrm{ext}}_{p+5,k+5}(s), \qquad p+5\leq N-5.
\ee
Near the cone, on \(r\leq5t/4\), the angular rotations are bounded combinations of boosts, so five can be replaced by three in both formulas. Globally, the identity expressing a rotation as a commutator of two boosts gives the five-order count. In particular, the energy estimate bounds the global pointwise norms by \(C_Na_0s^{d_{k+5}} \) in the closing ranges of these estimates, and the near-cone norms by \(C_Na_0s^{d_{k+3}} \). The fixed-parameter effective-field estimates follow from \(\del\fieldphi = \coeff^{-1} \del\fieldPhi\) and \(\fieldphi = \coeff^{-1/2} \fieldvarphi\). These bounds provide the pointwise regularity needed for continuation at every fixed \(\coeff>0\).

\paragraph{Averaged Hardy estimate at high differential order}

At \(p=p_{\mathbf{high}} =N-5\), the energy estimate and \eqref{equa:seven-e-exterior-sphere-trace} give
\be
\label{equa:eight-improved-high-Hardy-trace}
\mathsf T_{2;p,k}[u](s) \lesssim s^{-1} \Fenergy^{\ME,p,k}_\lambda(s,u) \lesssim a_0s^{-1+d_k}, \qquad0\leq k\leq p.
\ee
Since \(d_k<\delta<1\), \(\nu_k\geq0\), and \(s_0\geq1\), we obtain
\be
\label{equa:eight-improved-high-Hardy-average}
\sup_{s_0\leq\sigma \leq s_1} \sigma^{-\nu_k} \frac1\sigma \int_{s_0}^{\sigma} \mathsf T_{2;p,k}[u](s) \, ds \lesssim a_0\leq C_{\mathrm{data}}A_0.
\ee
Indeed, \(s^{-1+d_k} \leq1\) on the interval of integration. The constant is independent of the bootstrap multiplier \(C_1\) and of the terminal time. The reference contribution \(A_0/\zeta_\star\) has the same property, with dependence only on the fixed exponent \(\zeta_\star\). Thus the averaged Hardy term in \eqref{equa:seven-robust-high-interface-input} has the required initial-data amplitude.


\section{\texorpdfstring{Global stability and the Einstein limit}{Global stability and the Einstein limit}}
\label{section-16}


\subsection{\texorpdfstring{Initial data and conclusions}{Initial data and conclusions}}
\label{section-16-1}

\paragraph{Purpose}

The global developments and uniform estimates follow from Proposition~\ref{prop:mobile-simultaneous-improvement}. We use its foliation, chosen independently of \(\coeff\), throughout this section. The convergence argument has three stages.
\bei
\item \emph{Global existence at fixed parameter.} The improved low-order estimates preserve the hyperbolic and small-field regimes. The high-order norm stays finite up to every finite time. Local continuation therefore gives a global solution, and the propagated constraints identify it with the original geometric equations.
\item \emph{Uniform control of a family.} The metric couples to \(\del\fieldPhi\) and \(\fieldvarphi\), whose energy bounds are uniform in \(\coeff\). The highest-order unscaled effective-field energy bounds keep their parameter dependence. Uniform pointwise decay controls the large-time and large-distance regions.
\item \emph{Identification of the Einstein limit.} Preparation controls the initial position and mass-normalized velocity of the trace residual. The residual equation propagates that control. Its estimate makes the effective-field source small in the metric--matter difference equations, to which we apply a local wave-energy estimate on each compact spacetime region.
\eei
The local rate is the sum of the initial metric--matter discrepancy, the commuted preparation amplitude, and the \(\coeff^{1/2} \)-weighted source error. Uniform decay extends convergence of the fields and their Cartesian derivatives from compact regions to the whole future spacetime. The local rate and the global convergence modulus are given separately below.

\paragraph{Initial norms and convergence topologies}

The fixed-parameter and uniform estimates keep the separate orders \(p_{\mathbf{high}}=N-5\) in the interior and \(N\) in the exterior. The exterior quantity \(\mathcal Y_N\), defined in the equation~\eqref{equa:eight-mobile-coupled-positive-energy}, includes the positive flux and bulk terms with their time weights. The exterior energy comparisons and the equation~\eqref{equa:mobile-simultaneous-exterior} already give
\be
\label{equa:final-rank-five-exterior-control}
\Fenergy_\lambda^{\ME,N,5}(s,u_\coeff)+\Fenergy_\mu^{\ME,N,5}(s,\psi_\coeff)+U^{\mathrm{ext}}_{N-5,5}(s)\leq C\mathcal Y_5(s)\leq C\varepsilon s^{d_5}.
\ee
The index five in \(\mathcal Y_5\) is the boost rank, independently of the total number of translations. This quantity also contains the time-normalized effective-field energy of order \(N\). The closing effective-field energy on the left has no such normalization. The exponent \(0<d_5<\delta\) is fixed independently of \(\coeff\). For the lower fixed-parameter estimate, we use
\be
\label{equa:final-lower-fixed-norm}
\mathsf E_{\mathrm{low}, \coeff}^{\mathrm{fix}}(s)\coloneqq \mathbf W_{p_{\mathbf{low}},p_{\mathbf{low}}}(s)^2 + \mathbf P_{p_{\mathbf{low}},p_{\mathbf{low}}}(s)^2 + D_{p_{\mathbf{low}},p_{\mathbf{low}}}(s)^2 + \bigl(\mathsf X_{p_{\mathbf{low}},p_{\mathbf{low}}}^{\mathrm{fix}}(s) \bigr)^2.
\ee

If \(K\) is compact, we denote by \(I_K\) the compact interval of foliation parameters met by \(K\) and set \(K_s\coloneqq K\cap\Mcal_s\).  All local Sobolev norms below are calculated in the fixed asymptotic coordinate atlas and with the coordinate measure \(dx\); consequently, their definitions are independent of \(\coeff\).  We write
\be
\label{equa:final-local-energy-topology}
\begin{aligned}
\|v\|_{\mathcal X^q(K)}&\coloneqq \sup_{s\in I_K} \Bigl(\|v\|_{H^q(K_s)} + \| \nabla_{t,x}v\|_{H^{q-1}(K_s)} \Bigr), \\
\|(\fieldPhi, \fieldvarphi) \|_{\mathcal Y^q(K)}&\coloneqq \sup_{s\in I_K} \Bigl(\| \fieldPhi\|_{H^q(K_s)} + \| \nabla_{t,x} \fieldPhi\|_{H^{q-1}(K_s)} + \| \fieldvarphi\|_{H^q(K_s)} \Bigr).
\end{aligned}
\ee
Einstein compatibility and well-preparedness in the commuted topology are those of Definition~\ref{definition:Einstein-compatible-commuted-preparation}; interface fluxes and traces are conclusions of the Cauchy evolution and are never prescribed as data.

We distinguish the decay exponent for the field values from that for their Cartesian derivatives:
\be
\label{equa:final-spacetime-decay-exponents}
\beta_\infty\coloneqq1-2\delta,\qquad \gamma_{\mathrm d}\coloneqq\min\{1-\nu_2,1-d_5/2\},\qquad 0<\beta_\infty<\gamma_{\mathrm d}<1.
\ee
Here \(\nu_2\) is the interior rank-two growth exponent defined in the equation~\eqref{equa:seven-wave-rank-exponents}. For a Cartesian spacetime multi-index \(I\), the operator \(\del^I\) includes time as well as spatial derivatives. With \((g_0,\psi_0)\) denoting the Einstein development of the limiting data, we define the derivative discrepancy, for integers \(j\geq1\), by
\be
\label{equa:final-derivative-discrepancy}
\begin{aligned}
d_{\coeff,j}&\coloneqq\sum_{1\leq|I|\leq j}\bigl(|\del^I(g_\coeff-g_0)|+|\del^I(\widetilde g_\coeff-g_0)|+|\del^I(\psi_\coeff-\psi_0)|
+|\del^I\fieldPhi_\coeff|+|\del^I\fieldvarphi_\coeff|\bigr).
\end{aligned}
\ee

\paragraph{Global stability and convergence theorem}

The theorem combines global continuation on the adapted foliation with the estimate for the discrepancy from the Einstein trace relation and the metric--matter comparison.

\begin{theorem}[Global stability and convergence to Einstein gravity]
\label{theorem:final-global-stability-and-limit}

Let \(N\geq19\) be the fixed regularity index of the energy hierarchy, let \(s_0\geq2\), and let \(0<\coeff\leq\coeff_0\).  Let \(f_\coeff\) be an admissible family of order \(M\geq N+3\) in the sense of Definition~\ref{def:admissible-f-family}. Its conformal potentials then satisfy \eqref{equa:seven-uniform-potential-hypothesis}--\eqref{equa:seven-uniform-potential-CN}, by~\autoref{lem:seven-admissible-potential-factorization}. Let the smooth asymptotically Euclidean initial data be
\be
\mathbf d_\coeff= \bigl(g_{0, \coeff},k_{0, \coeff}, \fieldphi_{0, \coeff}, \fieldphi_{1, \coeff}, \psi_{0, \coeff}, \psi_{1, \coeff} \bigr)
\ee
satisfying the Hamiltonian, momentum, and wave-gauge constraints. The physical Cauchy data are reconstructed by \eqref{equa:physical-initial-geometry}--\eqref{equa:physical-initial-normal-data}. Assume that the matter potential vanishes identically. Assume also the robust initial estimates \eqref{equa:seven-initial-energy-bound} and \eqref{equa:seven-initial-robust-high-bound}, and the weighted exterior estimates~\eqref{equa:seven-exterior-initial-assumptions}, with
\be
{A_0\coloneqq \epss+C_0\eps\leq\varepsilon.}
\ee
Suppose also that the initial metric has the asymptotically Schwarzschild decomposition \eqref{equa-decomp-data-3} relative to the reference metric \eqref{equa-defineMS-new}, whose mass \(m\) is fixed independently of \(\coeff\) and satisfies
\be
\label{equa:final-light-bending-data-condition}
0\leq m\leq m_0,\qquad m\leq C_{\mathrm{ref}}\epss.
\ee
The mass conditions express the smallness of the reference geometry. Since \(\epss\leq A_0\), they imply the bound \(m\leq C_{\mathrm{ref}}A_0\) used in the global estimates; \(\epss\) includes the prescribed reference mass, rather than measuring the dynamic perturbation. The auxiliary foliation and its energy comparisons are those of Section~\ref{section:small-mass-moving-interface}. For each fixed \(\coeff>0\), assume that the order-\(N\) energy of all fields is finite; its interior part is not required to be uniformly small. Suppose, in addition, that the family is Einstein-compatible relative to an Einstein--massless-scalar initial data set \(\mathbf d_0\) and well prepared at order \(q_\star=N-6\), in the sense of Definition~\ref{definition:Einstein-compatible-commuted-preparation}. The same initial lapse and shift are chosen throughout the family and for the limiting data. There exists \(\varepsilon_0>0\), independent of \(0<\coeff\leq\coeff_0\), such that, if \(0<\varepsilon\leq\varepsilon_0\), the following conclusions hold.

\begin{enumerate}
\item The reduced conformal system admits a unique future global Cauchy development
\be
(g_\coeff, \fieldphi_\coeff, \psi_\coeff)
\ee
in wave gauge.  The gauge constraint and the curvature compatibility relation are propagated throughout the development.  In particular, one has
\be
\widetilde g_\coeff=e^{-\fieldPhi_\coeff}g_\coeff, \quad \fieldPhi_\coeff= \coeff\fieldphi_\coeff,
\ee
and \((\widetilde g_\coeff, \psi_\coeff) \) solves the original \(f_\coeff(R) \) field equations.  The physical spacetime \((\widetilde g_\coeff, \psi_\coeff) \) is future causally geodesically complete.

\item For every fixed \(\coeff>0\), the  unscaled effective field obeys two complementary fixed-parameter energy estimates. If \(\mathsf E_{\mathrm{low}, \coeff}^{\mathrm{fix}} \) denotes the lower-order part of the fixed-parameter hierarchy, then \bel{equa:final-fixed-coeff-low-estimate}
\sup_{s\geq s_0}s^{-2\delta} \mathsf E_{\mathrm{low}, \coeff}^{\mathrm{fix}}(s) \leq C_\coeff\varepsilon^2.
\ee
At the high orders used for continuation, every finite \(s_1\geq s_0\) satisfies \bel{equa:final-fixed-coeff-top-estimate}
\sup_{s_0\leq s\leq s_1} \Bigl(\bigl(\mathsf E_{\mathrm{top}}^{\mathrm{tame}}(s) \bigr)^2 + \bigl(\mathsf X_{N,N}^{\mathrm{fix}}(s) \bigr)^2\Bigr) \leq C_{\coeff,s_1} \varepsilon^2<+ \infty.
\ee
The interior norm in this estimate has order \(p_{\mathbf{high}}=N-5\); the exterior norm has order \(N\), with its effective-field normalization. Their polynomial bounds follow from \eqref{equa:seven-tame-high-final} and Proposition~\ref{prop:eight-mobile-coupled-estimate} by rescaling. The full order-\(N\) energy stays finite by \eqref{equa:mobile-simultaneous-high-continuation}; this latter bound may depend on its finite initial value. The fixed-parameter constants keep their negative powers of \(\coeff\).

\item The uniformly controlled   quantities satisfy
\be
\label{equa:final-robust-estimate}
\sup_{s\geq s_0}s^{-2\delta} \Bigl(\sum_{F\in\{U, \mathbf P, \mathbf W\}}F_{p_{\mathbf{low}},p_{\mathbf{low}}}(s)^2 + \bigl(\mathsf E_{\mathrm h,p_{\mathbf{high}}}^{\mathrm{rob}}(s) \bigr)^2 + \mathcal Y_N(s)^2\Bigr) \leq C\varepsilon^2.
\ee
The effective-field terms of order \(N\) in \(\mathcal Y_N^2\) carry the factor \(s^{-2} \) on the hypersurface and inside the flux and bulk integrals. The metric, matter, and closing effective-field terms keep weight one.~\autoref{prop:mobile-simultaneous-improvement} gives the sphere traces and characteristic entries at their respective derivative orders.

\item Let \((g_0, \psi_0) \) be the global Einstein development of \(\mathbf d_0\). For \(1\leq q\leq N-7\), one sets
\be
\label{equa:final-prepared-convergence-modulus}
a_{\coeff,q}\coloneqq \omega_\coeff\varepsilon+ \mathrm{Prep}^{\mathrm{com}}_{\coeff,q+1}[\Theta]+ \coeff^{1/2} \varepsilon^2.
\ee
Then, for every compact region \(K\), including regions meeting the initial hypersurface, one has
\be
\label{equa:final-prepared-Einstein-estimate}
\|g_\coeff-g_0\|_{\mathcal X^q(K)} + \| \widetilde g_\coeff-g_0\|_{\mathcal X^q(K)} + \| \psi_\coeff-\psi_0\|_{\mathcal X^q(K)} + \|(\fieldPhi_\coeff, \fieldvarphi_\coeff) \|_{\mathcal Y^q(K)} \leq C_Ka_{\coeff,q}.
\ee
Moreover, for every \(2\leq q\leq N-6\), one has
\be
\label{equa:final-preparation-modulus-estimate}
\sup_{s\in I_K} \Bigl(\| \Theta_\coeff\|_{H^{q-1}(K_s)} + \coeff^{1/2} \| \nabla_{t,x} \Theta_\coeff\|_{H^{q-2}(K_s)} \Bigr) \leq C_K\Bigl(\mathrm{Prep}^{\mathrm{com}}_{\coeff,q}[\Theta]+ \coeff^{1/2} \varepsilon^2\Bigr).
\ee
If the initial data sets have preparation rate \(\sigma>0\), then, with \(\sigma_\star\coloneqq \min\{\sigma,1/2\} \), the residual estimate becomes \bel{equa:final-positive-power-estimate}
\sup_{s\in I_K} \Bigl(\| \Theta_\coeff\|_{H^{q-1}(K_s)} + \coeff^{1/2} \| \nabla_{t,x} \Theta_\coeff\|_{H^{q-2}(K_s)} \Bigr) \leq C_K\coeff^{\sigma_\star} \varepsilon.
\ee
If, in addition, \(\omega_\coeff\leq C\coeff^\sigma \), then \eqref{equa:final-prepared-Einstein-estimate} holds with right-hand side \(C_K\coeff^{\sigma_\star} \varepsilon\).  The family converges without extraction on every compact spacetime region to the Cauchy development \((g_0, \psi_0) \) of the Einstein--massless-scalar system:
\be
G[g_0]_{\alpha \beta} =8\pi T[\psi_0,g_0]_{\alpha \beta}, \qquad\Box_{g_0} \psi_0=0.
\ee
In particular, the limiting effective-field equation is the Einstein trace constraint, and the limiting metric is simultaneously the limit of the conformal and physical \(f(R) \) metrics. The fixed-\(\coeff\) estimates \eqref{equa:final-fixed-coeff-low-estimate}--\eqref{equa:final-fixed-coeff-top-estimate} remain available throughout this comparison.

\item Convergence of the field values is uniform on the entire future domain \(\mathcal D^+ \), including its initial face. With componentwise absolute values in the wave coordinates, one defines
\be
\label{equa:final-global-v-discrepancy}
d_\coeff(t,x)\coloneqq|g_\coeff-g_0|+ | \widetilde g_\coeff-g_0|+ | \psi_\coeff-\psi_0|+ | \fieldPhi_\coeff|+ | \fieldvarphi_\coeff|.
\ee
For every \(0\leq b<\beta_\infty\), one has
\be
\label{equa:final-global-uniform-Einstein-limit}
\sup_{(t,x) \in\mathcal D^+}(1+ t+ |x|)^b d_\coeff(t,x) \longrightarrow0\qquad(\coeff\to0).
\ee
More quantitatively, there are constants \(C_R\), independent of \(\coeff\), such that, for all sufficiently large \(R\),
\be
\label{equa:final-global-convergence-modulus}
\sup_{\mathcal D^+}(1+ t+ |x|)^b d_\coeff\leq\inf_{R\geq R_0} \Bigl(C_Ra_{\coeff,3} +C\varepsilon R^{b-\beta_\infty} \Bigr).
\ee
The local rate in Item~4 and the global modulus in this estimate describe distinct topologies.
\item The Cartesian derivatives also converge uniformly on the entire future spacetime. For every \(1\leq j\leq N-10\) and \(0\leq b<\gamma_{\mathrm d}\), one has
\be
\label{equa:final-global-uniform-derivative-limit}
\sup_{\mathcal D^+}(1+t+|x|)^b d_{\coeff,j}\longrightarrow0\qquad(\coeff\to0).
\ee
More quantitatively, with constants independent of \(\coeff\), one has
\be
\label{equa:final-global-derivative-modulus}
\sup_{\mathcal D^+}(1+t+|x|)^b d_{\coeff,j}\leq\inf_{R\geq R_0}\bigl(C_{R,j}a_{\coeff,j+3}+C_j\varepsilon R^{b-\gamma_{\mathrm d}}\bigr).
\ee
For the two rescaled effective fields, the same convergence holds with the zeroth derivatives included.
\item The unscaled effective field converges to the Einstein scalar curvature, with two fewer derivatives. For \(0\leq j\leq N-12\) and \(0\leq b<\gamma_{\mathrm d}\), one has
\be
\label{equa:final-global-unscaled-effective-limit}
\sup_{\mathcal D^+}(1+t+|x|)^b\sum_{|I|\leq j}|\del^I(\fieldphi_\coeff-R[g_0])|\longrightarrow0\qquad(\coeff\to0).
\ee
\end{enumerate}
\end{theorem}

\paragraph{Convergent quantities and the scope of the estimates}

The proof of Item~4, including the passage \(\coeff\to0\), is given in Section~\ref{section-16-2}. Section~\ref{section:final-global-uniform-convergence} then combines local convergence with decay uniform in \(\coeff\) to prove Item~5. The conclusions have the following distinct meanings.
\begin{itemize}
\item The physical and conformal metrics and the matter field converge uniformly on the entire future spacetime, together with their Cartesian derivatives through order \(N-10\), including the initial hypersurface. The algebraic rates in \(\coeff\) are local; the global conclusions have the moduli in Items~5--6.
\item The rescaled effective fields \(\fieldPhi_\coeff\) and \(\fieldvarphi_\coeff\) vanish. The unscaled field instead recovers the Einstein scalar curvature:
\be
\label{equa:final-unscaled-effective-field-limit}
\fieldphi_\coeff\longrightarrow R[g_0]=8\pi g_0^{\alpha\beta}\del_\alpha\psi_0\del_\beta\psi_0\quad\text{uniformly on }\mathcal D^+\text{ as }\coeff\to0.
\ee
This limit may be nonzero. Item~7 gives its weighted convergence through order \(N-12\); the loss of two derivatives comes from the wave operator in the reconstruction of \(\Theta\). The effective-field evolution becomes the algebraic Einstein trace relation.
\item At the maximal preparation order \(q_\star=N-6\), the trace-residual energy controls Cartesian derivatives through order \(q_\star-1\) without parameter weights on each fixed finite slab, and the next derivative with the factor \(\sqrt\coeff\). Unweighted convergence at this last derivative order is not a conclusion of that estimate. The higher fixed-parameter bounds keep their explicit dependence on \(\coeff\).
\item The global convergence modulus in the equation~\eqref{equa:final-global-convergence-modulus} uses constants \(C_R\) on growing compact regions. The algebraic local rate therefore does not by itself give an algebraic rate for the supremum on the entire future spacetime.
\end{itemize}


\subsection{\texorpdfstring{Global developments and causal completeness}{Global developments and causal completeness}}
\label{section:global-developments-completeness}

\paragraph{Global existence for each fixed \(\coeff>0\)}

The global-existence assertion of Theorem~\ref{theorem:final-global-stability-and-limit} follows from Step~5 of Proposition~\ref{prop:mobile-simultaneous-improvement}, where the continuation argument is concluded. For each fixed \(\coeff>0\), the conformal Cauchy development is therefore future-global, with the uniform bounds and the fixed-parameter estimates established there. The estimate~\eqref{equa:mobile-simultaneous-high-continuation} gives the order-\(N\) regularity needed for continuation.

The initial Hamiltonian, momentum and wave-gauge constraints propagate by Lemma~\ref{lem:seven-local-wave-gauge-propagation}, and Proposition~\ref{pro:40S} recovers the original geometric equations. Lemma~\ref{lem:final-direct-causal-completeness} below will give future causal geodesic completeness of both metrics from the global rank-zero bounds. For well-prepared data, Proposition~\ref{proposition:eight-uniform-trace-residual} already gives the strong estimate for the discrepancy from the Einstein trace relation; a prescribed rate \(\sigma>0\) gives the positive-power estimate.

\paragraph{Causal geodesics of the physical metric}

The global rank-zero decay bounds control affine parameters through the geodesic energy.

\begin{lemma}[Future causal completeness]
\label{lem:final-direct-causal-completeness}
Suppose that the  development is future global and satisfies the rank-zero metric and effective-field estimates established above, with sufficiently small amplitude \(A\). Then both \(g\) and \(\widetilde g=e^{-\fieldPhi}g\) are future causally geodesically complete.
\end{lemma}

\begin{proof}
\bse

\noindent{\bf 1.} \emph{Metric estimates near and away from the cone.} We write \(h=g-\gMink\), \(q=t-r\), and \(\nu= \nu_2<13\delta/64\). We set \(a_{\mathrm{geo}}\coloneqq2\delta\). The first metric estimate and the exterior radial estimate give the following bound for \(h\) in the interior and in the collar \(|q|\leq2t^{a_{\mathrm{geo}}}\). The derivative bounds hold in the indicated interior and exterior regions for sufficiently large \(t\):
\be
\label{equa:final-causal-direct-metric}
|h| \leq C A t^{-1+2\delta},\qquad | \del h| \leq \begin{cases}
C A t^{-1/2}s^{-1+ \nu},&q\geq q_b(t), \\
C A r^{-1}w^{-\lambda}t^{\delta/2},&q\leq q_b(t),
\end{cases} \qquad w\simeq2-q.
\ee
On the exterior part of this collar, \(r\simeq t\), \(w\lesssim t^{2\delta}\), and \(s\lesssim t^{1/2}\). Thus the radial bound gives \(|h|\lesssim A t^{-1+\delta/2+2\delta(1-\lambda)}+m/t\lesssim A t^{-1+2\delta}\). For the fixed contribution \(h^\star=g^\star-\gMink\), the equation~\eqref{equa:seven-reference-pointwise-bounds} gives \(|\del h^\star|\lesssim m(1+t+r)^{-2}\), so its Cartesian derivatives are integrable in time. Outside the collar \(|q| \leq t^{a_{\mathrm{geo}}} \), the last two derivative bounds are integrable in physical time, since
\be
\label{equa:final-causal-exponent-gaps}
\epsilon_{\mathrm{int}}\coloneqq \frac{a_{\mathrm{geo}}-(1+a_{\mathrm{geo}}) \nu}{2}>0, \qquad \epsilon_{\mathrm{ext}}\coloneqq a_{\mathrm{geo}} \lambda-\frac\delta2>0.
\ee
In the collar, \(r\simeq t\) and \(| \del h| \leq CAt^{-1+ \delta/2} \). We use \(E_a^N= \del_a+ \omega_a \del_t\), \(G_{0a} =h(E_a^N, \del_t) \), and \(G_{ab} =h(E_a^N,E_b^N) \). The rank-zero good-component estimates imply
\be
\label{equa:final-causal-good-metric}
| \del_tG_{0a}|+ | \del_tG_{ab}| \leq C A t^{-1} +b_0(t), \qquad \int_{T_1}^{\infty}b_0(t) \, dt< + \infty.
\ee
Indeed, on the interior collar, \(E_a^N= \delH_a+(q/t) \omega_a \del_t\). The conversion error is bounded by \(CA t^{-2 +a_{\mathrm{geo}} + \delta/2}\) plus \(CA t^{-2 +2\delta}\), both integrable in time. In the exterior, Proposition~\ref{prop:eight-mobile-characteristics} at rank zero gives \(CA/(rw^\gamma) \leq CA/t\). The bounded initial time interval is included in the constants.
\ese

\bse

\noindent{\bf 2.} \emph{The conformal factor.} The rank-zero strong effective-field bound gives \(| \del\fieldPhi| \leq CAt^{-1}s^{-1/2} \leq CAt^{-5/4} \) in the interior. In the exterior, \eqref{equa:eight-uniform-radial-mass-value} and the stronger reconstructed bound \eqref{equa:eight-top-strong-low-effective-factor} give
\be
\label{equa:final-causal-conformal-derivative}
| \del\fieldPhi| \leq C_{\coeff_0}A\min\{t^{-1+ \delta/2}w^{-\mu}, \ t^{-3/2 + \delta}w^{1-\mu} \}.
\ee
These estimates require at most seven effective-field orders, within \(p_{\mathbf{high}} =N-5\). The reconstruction is needed only for \(w\leq t^{(1-\delta)/2} \), where \(r\simeq t\); the direct mass estimate is used elsewhere. The two weights meet at \(w=t^{(1-\delta)/2} \), and therefore
\be
\label{equa:final-causal-conformal-integrability}
\epsilon_\Phi\coloneqq \frac\mu2-\frac{(1+ \mu) \delta}{2}>\frac14, \qquad \sup_x| \del\fieldPhi(t,x)| \leq C_{\coeff_0}A t^{-5/4}, \qquad | \fieldPhi| \leq CA.
\ee

\ese

\bse
\noindent{\bf 3.} \emph{Integration through the collar.} For a future causal curve parametrized by \(t\), we write \(v= (1,x'(t)) \) and \(d=1-\omega \cdot x'(t)=q'(t) \). Smallness of \(h\) gives \(|x'| \leq2\), \(d\geq-C|h| \), and \(d^2\leq C d_++C|h|^2\). Since \(\del_t\omega=0\) at fixed \(x\), we obtain the tensor decomposition
\be
\label{equa:final-causal-metric-contraction}
(\del_t h) (v,v)= (\del_t h_{00})d^2 +2dv^a \del_tG_{0a} +v^av^b\del_tG_{ab}.
\ee
We choose \(0\leq\xi\leq1\) smooth, equal to one on \([-1,1] \) and zero outside \([-2,2] \), and set \(P(z)= \int_0^z\xi(y) \, dy\). Both \(P\) and \(z\xi(z) \) are bounded. For \(z=q/t^{a_{\mathrm{geo}}} \), we have
\be
\label{equa:final-causal-collar-primitive}
\frac d{dt} \bigl(t^{a_{\mathrm{geo}}}P(z) \bigr)= \xi(z)q'+a_{\mathrm{geo}}t^{a_{\mathrm{geo}}-1} \bigl(P(z)-z\xi(z) \bigr).
\ee
Integration by parts with \(t^{-1+ \delta/2} \) bounds the signed integral uniformly in its upper limit: the boundary power is \(t^{a_{\mathrm{geo}}-1+ \delta/2} \), and the remainder is integrable because \(a_{\mathrm{geo}} + \delta/2<1\). The negative variation satisfies \((q')_-\leq C|h| \). Consequently,
\be
\label{equa:final-causal-collar-integral}
\int_{T_1}^{T}t^{-1+ \delta/2} \xi(q/t^{a_{\mathrm{geo}}})d_+ \, dt\leq C+CA\int_{T_1}^{T}t^{-2 +5\delta/2} \, dt\leq C.
\ee
This estimate includes repeated passages through the collar. We combine it with \eqref{equa:final-causal-metric-contraction}, the good-component estimate, and the integrable derivatives outside the collar to obtain
\be
\label{equa:final-causal-integrated-deformation}
\int_{T_1}^{T}|(\del_tg) (v,v)| \, dt\leq CA\log(T/T_1) +C_A.
\ee
\ese

\bse

\noindent{\bf 4.} \emph{Affine parameters.} We take \(\widehat g\) equal to \(g\) or \(e^{-\fieldPhi}g\), and let \(V\) be the affine tangent to a future causal \(\widehat g\)-geodesic. The energy \(E=-\widehat g(\del_t,V) \) is comparable to \(V^0>0\). With \(v=V/V^0\), its derivative gives
\be
\label{equa:final-causal-geodesic-energy}
\frac{dE}{d\tau} =-\frac12(\del_t\widehat g) (V,V), \qquad \left| \frac d{dt} \log E\right| \leq C|(\del_t\widehat g) (v,v)|.
\ee
The physical metric has the same causal cone, and its time derivative satisfies
\be
\label{equa:final-causal-conformal-deformation}
(\del_t\widetilde g) (v,v)=e^{-\fieldPhi} \bigl((\del_tg) (v,v)-(\del_t\fieldPhi)g(v,v) \bigr).
\ee
The additional term is integrable by \eqref{equa:final-causal-conformal-integrability} and boundedness of \(v\). Hence \(V^0(t) \leq C_{\mathrm{geo}}t^{CA} \). A finite upper coordinate time cannot be a maximal endpoint: \(|dx/dt| \leq2\) keeps the curve in a bounded spatial set, and the smooth metric and energy equation bound its tangent there. The geodesic equation therefore extends it. Thus an inextendible future geodesic has \(t\to\infty\), and, for \(CA<1\),
\be
\label{equa:final-causal-affine-length}
\tau(T)-\tau(T_1)= \int_{T_1}^{T} \frac{dt}{V^0(t)} \geq C_{\mathrm{geo}}^{-1} \int_{T_1}^{T}t^{-CA} \, dt\longrightarrow+ \infty.
\ee
This proves both the null and timelike completeness assertions.
\ese
\end{proof}


\subsection{\texorpdfstring{Local convergence as \(\coeff\to0\)}{Local convergence as kappa tends to zero}}
\label{section-16-2}

\paragraph{Preparation and vanishing effective-field terms}

We now pass to the limit \(\coeff\to0\) \emph{first} on fixed compact spacetime regions, using the global developments and uniform estimates already established. Einstein compatibility gives convergence of the metric--matter data and the rescaled effective-field data, whereas the preparation amplitude~\eqref{equa:commuted-preparation-amplitude} controls every commuted initial trace-residual component required by the singular energy estimate.

\bse
Proposition~\ref{proposition:eight-uniform-trace-residual} and the exact matching of the interior and exterior trace-residual fluxes give
\be
\label{equa:fifteen-prepared-trace-residual-energy}
\sup_{s_0\leq s\leq s_1} \mathcal E_{\Theta,q}^{\mathrm{sing}}(s)^{1/2} \leq C_{s_1}\Bigl(\mathrm{Prep}^{\mathrm{com}}_{\coeff,q}[\Theta]+ \coeff^{1/2} \varepsilon^2\Bigr).
\ee
The energy comparison~\eqref{equa:eight-joined-trace-residual-coercivity} and local Sobolev inequalities give the estimate in Item~4, for \(2\leq q\leq N-6\):
\be
\label{equa:final-preparation-modulus-estimate-rep}
\sup_{s\in I_K} \Bigl(\| \Theta_\coeff\|_{H^{q-1}(K_s)} + \coeff^{1/2} \| \nabla_{t,x} \Theta_\coeff\|_{H^{q-2}(K_s)} \Bigr) \leq C_K\Bigl(\mathrm{Prep}^{\mathrm{com}}_{\coeff,q}[\Theta]+ \coeff^{1/2} \varepsilon^2\Bigr).
\ee
 If the preparation rate is \(\sigma>0\), the same estimate gives \eqref{equa:final-positive-power-estimate} with \(\sigma_\star= \min\{\sigma,1/2\} \).

The identity \(\fieldphi_\coeff= \Theta_\coeff-\mathcal S_\coeff\) yields the following.  The robust metric--matter hierarchy controls \(\mathcal S_\coeff\) and its commuted derivatives, while \eqref{equa:fifteen-prepared-trace-residual-energy} controls \(\Theta_\coeff\) and \(\coeff^{1/2} \nabla \Theta_\coeff\).  Since \(\fieldPhi_\coeff= \coeff\fieldphi_\coeff\) and \(\fieldvarphi_\coeff= \coeff^{1/2} \fieldphi_\coeff\), we obtain, for \(q\leq N-7\) and every compact slab ending at \(s_1\),
\be
\label{equa:fifteen-prepared-rescaled-control}
\|(\fieldPhi_\coeff, \fieldvarphi_\coeff) \|_{\mathcal Y^q(K)} \leq C_K\Bigl(\mathrm{Prep}^{\mathrm{com}}_{\coeff,q+1}[\Theta]+ \coeff^{1/2} \varepsilon^2\Bigr).
\ee
Preparation at order \(q+1\) controls all components of \(\mathcal Y^q\) in this estimate.
\ese

\paragraph{Global construction of the limiting development}

We construct the Einstein development on the whole future domain before applying the difference estimate. The foliation map \(\iota_b(s,x)=(T_b(s,|x|),x) \) is fixed independently of \(\coeff\). Its images of the sets \([s_0,s_0 +j] \times\{|x| \leq j\} \), \(j\geq1\), exhaust the future domain. We use a slightly larger compact set for each local Sobolev estimate.

\noindent{\bf 1.} \emph{Compactness of the metric and matter fields.} On a slightly larger fixed cylinder in foliation coordinates, the closed robust hierarchy controls the spatial \(L^2\) norms of the Cartesian derivatives through order five. The commutation family includes \(\del_t\); the energy weights have positive lower bounds on this compact cylinder. The field values are controlled by the local Hardy estimates. Spatial Sobolev embedding with two additional derivatives therefore bounds all spacetime derivatives through order three, uniformly in \(\coeff\). These five derivatives are within \(p_{\mathbf{high}}=N-5\). The Arzel\`a--Ascoli theorem gives precompactness in \(C^2\), including the initial face, and a diagonal subsequence converges in \(C^2_{\mathrm{loc}}\) on the future domain. The higher uniform Sobolev bounds also pass weakly to this limit; in particular, they give the derivatives through order \(q+1\leq N-6\) used in the difference estimate below. The metric bounds preserve nondegeneracy and spacelikeness of the foliation hypersurfaces.

\noindent{\bf 2.} \emph{Disappearance of the effective-field contributions.} Proposition~\ref{proposition:eight-uniform-trace-residual} and \eqref{equa:fifteen-prepared-rescaled-control} imply \(\fieldPhi_\coeff\to0\), \(\fieldvarphi_\coeff\to0\), and \(\del\fieldPhi_\coeff\to0\) uniformly on compact sets. The identity \(\coeff^2d\fieldphi_\coeff\otimes d\fieldphi_\coeff=d\fieldPhi_\coeff\otimes d\fieldPhi_\coeff\) and the potential factorization \eqref{equa:seven-Ufrak-rescaled} make the effective-field stress tend to zero. The conformal matter factor tends to one and its derivative coupling tends to zero. Passing to the limit in the reduced metric equation, the matter equation, and the wave-gauge identities gives the Einstein--massless-scalar equations. The metric and matter current bounds pass to the limit by lower semicontinuity.

\noindent{\bf 3.} \emph{Initial data, uniqueness, and regularity.} We choose the same initial lapse and shift throughout the family. Einstein compatibility and the wave-gauge reconstruction of the first metric jets identify the limiting Cauchy data with \(\mathbf d_0\). Local uniqueness identifies the \(C^2\) limit with the Einstein Cauchy solution on every domain of dependence. Local well-posedness and persistence of higher regularity give this development the regularity of its initial data; compactness itself was used only to construct the \(C^2\) solution. The exhaustion covers every finite future slab. Moreover, each such slab has bounded Cartesian time, and the  metric bounds give a uniform coordinate speed bound for causal curves. A causal curve cannot escape to spatial infinity while its foliation parameter remains bounded. The spacelike hypersurfaces are therefore Cauchy hypersurfaces for the resulting global development \((g_0, \psi_0) \). The difference estimate below then proves convergence of the entire family and the announced quantitative rate.

\begin{lemma}[Difference equations and source bounds for well-prepared data]
\label{lem:fifteen-prepared-difference-system}
Define the metric and matter differences by
\be
h_\coeff\coloneqq g_\coeff-g_0, \qquad \chi_\coeff\coloneqq \psi_\coeff-\psi_0,
\ee
where the two developments are written in the same wave coordinates. Set
\be
\label{equa:fifteen-gauge-difference-source}
\mathcal Q^{\mathrm{gauge}}_{\coeff, \alpha \beta}[h_\coeff] \coloneqq \Fbb_{\alpha \beta}(g_\coeff,g_\coeff;\del g_\coeff, \del g_\coeff) - \Fbb_{\alpha \beta}(g_0,g_0;\del g_0, \del g_0)
\ee
and
\be
\label{equa:fifteen-matter-difference-source}
\Delta \Mbb_{\coeff, \alpha \beta} \coloneqq-16\pi\Bigl(e^{-\fieldPhi_\coeff} \del_\alpha \psi_\coeff\del_\beta \psi_\coeff -\del_\alpha \psi_0\del_\beta \psi_0\Bigr).
\ee
Then the difference equations are
\bse
\be
\label{equa:fifteen-exact-metric-difference}
\BoxChapeaud_{g_\coeff}h_{\coeff, \alpha \beta} = \mathcal Q^{\mathrm{gauge}}_{\coeff, \alpha \beta}[h_\coeff] + \bigl(g_0^{\mu\nu}-g_\coeff^{\mu\nu} \bigr) \del_\mu\del_\nu g_{0, \alpha \beta} + \Delta \Mbb_{\coeff, \alpha \beta} + \Gbb_\coeff[\fieldphi_\coeff,g_\coeff]_{\alpha \beta},
\ee
\be
\label{equa:fifteen-exact-matter-difference}
\Box_{g_\coeff} \chi_\coeff = \bigl(g_0^{\mu\nu}-g_\coeff^{\mu\nu} \bigr) \del_\mu\del_\nu\psi_0 +g_\coeff^{\mu\nu} \del_\mu\fieldPhi_\coeff \del_\nu\psi_\coeff.
\ee
\ese
Let \(K\) be a compact spacetime region, \(S_K\coloneqq \sup I_K\), and \(1\leq q\leq N-7\). The equations are pulled back by the fixed foliation map \(\iota(s,x)= (T(s,|x|),x) \). Thus \(e_a\coloneqq \del_{x^a}|_s\) is tangent to the hypersurfaces. Choose \(R_K\) and \(v_K>0\), independent of \(\coeff\), so that the shrinking coordinate balls
\be
\label{equa:fifteen-local-causal-domains}
D_s\coloneqq \{x:|x|<R_K+v_K(S_K-s) \}, \qquad s_0\leq s\leq S_K,
\ee
contain \(\iota^{-1}(K) \) and have inward speed strictly larger than the characteristic speeds of both wave operators on the fixed slab. The finite-slab metric bounds give such a  choice. The lateral boundary therefore has nonnegative outgoing energy flux. Only spatial operators \(e^I\), \(|I| \leq q-1\), are used, with \(\widehat v\coloneqq v\circ\iota \). For each tensor component of \(h_\coeff\) and the scalar \(\chi_\coeff\), the squared \(L^2\) norm is added to the positive local wave energy of \(e^I\widehat v\). Its square-root sum, denoted by \(E_{\coeff,q}^{\mathrm{diff}} \), satisfies
\be
\label{equa:fifteen-local-difference-energy-comparison}
\bigl(E_{\coeff,q}^{\mathrm{diff}}(s) \bigr)^2\simeq_K\sum_{v\in\{h_\coeff, \chi_\coeff\}} \Bigl(\| \widehat v\|_{H^q(D_s)}^2 + \| \del_s\widehat v\|_{H^{q-1}(D_s)}^2\Bigr).
\ee
For \(v\in\{h_\coeff, \chi_\coeff\} \), the pulled-back equation has a first-order wave formulation with \(V^v\coloneqq (\widehat v, \del_s\widehat v, \nabla_x\widehat v) \). Its bounded coefficient matrices are denoted by \(A_\coeff^a,B_\coeff\). The commuted inhomogeneous terms, including the coefficient commutators and foliation terms, are defined by
\be
\label{equa:fifteen-commuted-difference-sources}
\del_s(e^IV^v)=A_\coeff^a e_a(e^IV^v) +B_\coeff e^IV^v+ \mathcal R^v_{\coeff,I}, \qquad |I| \leq q-1.
\ee
The sources are indexed by \(g\) for \(v=h_\coeff\) and by \(\psi\) for \(v= \chi_\coeff\). At the initial hypersurface, the difference energy satisfies
\be
\label{equa:fifteen-initial-difference-energy}
E_{\coeff,q}^{\mathrm{diff}}(s_0) \leq C_K\omega_\coeff\varepsilon.
\ee
There exists a nonnegative \(r_{\coeff,q,K} \in L^1([s_0,S_K]) \) such that
\bse
\be
\label{equa:fifteen-commuted-difference-source-bound}
\sum_{|I| \leq q-1} \Bigl(\| \mathcal R^g_{\coeff,I} \|_{L^2(D_s)} + \| \mathcal R^\psi_{\coeff,I} \|_{L^2(D_s)} \Bigr) \leq C_KE_{\coeff,q}^{\mathrm{diff}}(s) +r_{\coeff,q,K}(s),
\ee
\be
\label{equa:fifteen-integrated-effective-field-difference-source}
\int_{s_0}^{S_K}r_{\coeff,q,K}(s) \, ds\leq C_K\Bigl(\mathrm{Prep}^{\mathrm{com}}_{\coeff,q+1}[\Theta]+ \coeff^{1/2} \varepsilon^2\Bigr).
\ee
\ese
All constants are independent of \(\coeff\).

\end{lemma}

\begin{proof}[Proof]
\bse
We separate the terms containing a difference field from the effective-field terms that remain when the metric and matter differences vanish. The first group enters Gronwall's inequality; preparation makes the time integral of the second group tend to zero.

\medskip\noindent\emph{Difference equations.} We subtract the reduced Einstein--massless-scalar system for \((g_0, \psi_0) \) from \eqref{MainPDE}, with \(V\equiv0\).  Since both metrics satisfy the wave-coordinate condition, the geometric and reduced wave operators agree on scalar functions.  Hence, we have
\be
\Box_{g_\coeff} \psi_0 = \bigl(g_\coeff^{\mu\nu}-g_0^{\mu\nu} \bigr) \del_\mu\del_\nu\psi_0,
\ee
which gives \eqref{equa:fifteen-exact-matter-difference}.  The same calculation componentwise for the metric gives the principal coefficient term in \eqref{equa:fifteen-exact-metric-difference}.  The identities
\be
\begin{aligned}
g_0^{-1}-g_\coeff^{-1} &=g_\coeff^{-1}h_\coeff g_0^{-1}, \\
e^{-\fieldPhi_\coeff} \del\psi_\coeff\del\psi_\coeff- \del\psi_0\del\psi_0 &=e^{-\fieldPhi_\coeff} \bigl(\del\chi_\coeff\del\psi_\coeff + \del\psi_0\del\chi_\coeff\bigr) + \bigl(e^{-\fieldPhi_\coeff}-1\bigr) \del\psi_0\del\psi_0
\end{aligned}
\ee
then give \eqref{equa:fifteen-matter-difference-source} and the remaining terms in the metric equation. The gauge source satisfies \(\mathcal Q^{\mathrm{gauge}}_\coeff[0]=0\).  More explicitly, with \(g_\vartheta\coloneqq g_0 + \vartheta h_\coeff\), the fundamental theorem of calculus gives the representation
\be
\label{equa:fifteen-gauge-difference-linearization}
\mathcal Q^{\mathrm{gauge}}_\coeff[h_\coeff] = \int_0^1 D\Fbb(g_\vartheta, \del g_\vartheta) [h_\coeff, \del h_\coeff] \, d\vartheta.
\ee
Thus every term in the difference of the reduced Einstein nonlinearities contains at least one factor \(h_\coeff\) or \(\del h_\coeff\), while its null and quasi-null decomposition is the one used in the metric estimates of Sections~\ref{section-8} and~\ref{section-14}.

\noindent{\bf 1.} \emph{Local difference equations commuted with spatial derivatives.} We express the two wave equations in \((s,x) \) coordinates and use their first-order wave systems. Spatial differentiation through order \(q-1\) produces the principal energy at order \(q\). All foliation coefficients and their derivatives are bounded on the fixed domain. The robust hierarchy and the limiting Einstein solution bound the metric and matter coefficients there, independently of \(\coeff\). The inverse-metric identity and \eqref{equa:fifteen-gauge-difference-linearization} show that every term vanishing when \(h_\coeff= \chi_\coeff=0\) is bounded by \(C_KE_{\coeff,q}^{\mathrm{diff}} \). The terms containing a differentiated Hessian of \(g_0\) or \(\psi_0\) require at most \(q+1\leq N-6\) derivatives. The first-order formulation introduces no second normal derivative into the initial energy.

\noindent{\bf 2.} \emph{Vanishing effective-field terms.} The remaining source is bounded by the following nonnegative function; all fields in its formula are pulled back to the foliation coordinates, and the contractions are those of the difference equations:
\be
\label{equa:fifteen-effective-field-difference-remainder}
\begin{aligned}
r_{\coeff,q,K}(s)&\coloneqq C_K\sum_{|I| \leq q-1} \bigl(\|e^I((e^{-\fieldPhi_\coeff}-1) \del\psi_0\del\psi_0) \|_{L^2(D_s)} + \|e^I(\del\fieldPhi_\coeff\del\fieldPhi_\coeff) \|_{L^2(D_s)} \\
& \quad + \|e^I(\coeff\Ufrak_\coeff(\fieldphi_\coeff)g_\coeff) \|_{L^2(D_s)} + \|e^I(g_\coeff^{-1}(\del\fieldPhi_\coeff, \del\psi_\coeff)) \|_{L^2(D_s)} \bigr).
\end{aligned}
\ee
We use \(\fieldPhi_\coeff= \coeff(\Theta_\coeff-\mathcal S_\coeff) \), \(\fieldvarphi_\coeff= \coeff^{1/2}(\Theta_\coeff-\mathcal S_\coeff) \), and the potential factorization \eqref{equa:seven-Ufrak-rescaled}. The trace-residual estimate at order \(q+1\) controls the vanishing factor, while the robust bounds control the other factors. The local product estimates on the enlarged compact slab therefore give \eqref{equa:fifteen-integrated-effective-field-difference-source}. These estimates also apply to the foliation derivatives because their coefficients are bounded there.

\noindent{\bf 3.} \emph{Initial energy and lateral flux.} We use the fixed initial graph \(\Sigma_0= \Mcal_{s_0} \), as in~\autoref{lem:eight-dynamic-initial-norm-comparison}. Spatial derivatives through order \(q\) and one normal derivative through spatial order \(q-1\) are controlled by Einstein compatibility in \eqref{equa:commuted-Einstein-compatibility}. We use the same fixed initial lapse and shift for both developments. The wave-gauge equations determine the remaining first metric jets smoothly from the geometric Cauchy data, using only their first spatial derivatives. The fixed coordinate transformation and this  initial gauge completion give \eqref{equa:fifteen-initial-difference-energy}. The added \(L^2\) norms are controlled by the same hypothesis; their energy derivatives are bounded by \(C_KE_{\coeff,q}^{\mathrm{diff}}{}^2\). We integrate over the shrinking balls \(D_s\) and keep their nonnegative outgoing boundary flux. If the integral is split at the interior--exterior interface, the two signed interface fluxes cancel because the coefficients, multiplier, and commuted fields agree. This proves the local energy estimate and the source bound.
\ese
\end{proof}

\paragraph{Strong convergence of the entire family and the local rate}

\begin{proof}[Completion of the proof of Item~4 of Theorem~\ref{theorem:final-global-stability-and-limit}]
We apply the local wave energy identity on the domains \(D_s\) to the spatially commuted equations in~\autoref{lem:fifteen-prepared-difference-system}.  Equations \eqref{equa:fifteen-initial-difference-energy}-- \eqref{equa:fifteen-integrated-effective-field-difference-source} give
\be
\label{equa:fifteen-prepared-difference-energy}
E_{\coeff,q}^{\mathrm{diff}}(s) \leq C_K\omega_\coeff\varepsilon +C_K\int_{s_0}^{s}E_{\coeff,q}^{\mathrm{diff}}(\tau) \, d\tau +C_K\Bigl(\mathrm{Prep}^{\mathrm{com}}_{\coeff,q+1}[\Theta]+ \coeff^{1/2} \varepsilon^2\Bigr).
\ee
On \(K\), the coordinate and unit-normal frames are uniformly equivalent. The comparison \eqref{equa:fifteen-local-difference-energy-comparison} controls the values in \(H^q\) and one normal derivative in \(H^{q-1} \), hence the topology \(\mathcal X^q(K) \) of \eqref{equa:final-local-energy-topology}. Gronwall's inequality proves the estimates for \(g_\coeff-g_0\) and \(\psi_\coeff-\psi_0\) in \eqref{equa:final-prepared-Einstein-estimate}.  The physical-metric estimate follows from \(\widetilde g_\coeff=e^{-\fieldPhi_\coeff}g_\coeff\), and \eqref{equa:fifteen-prepared-rescaled-control} gives the remaining terms.

On each fixed \(K\), the three terms \(\omega_\coeff\varepsilon\), \(\mathrm{Prep}^{\mathrm{com}}_{\coeff,q+1}[\Theta]\), and \(\sqrt\coeff\,\varepsilon^2\) tend to zero as \(\coeff\to0\). The entire family therefore converges in the norms of Item~4. The reduced equations yield the Einstein--massless-scalar system, and the equation~\eqref{equa:fifteen-prepared-trace-residual-energy} gives \(\Theta_\coeff\to0\), hence the Einstein trace constraint. Taking \(q=3\) and using spatial Sobolev embedding also gives the local convergence of the unscaled field; Item~7 will extend it to the entire future spacetime. If compatibility and preparation have rate \(\coeff^\sigma\), the same estimates give the local rate in the theorem. This completes Item~4, including the initial hypersurface and the vacuum specialization. The convergence on the entire future spacetime is proved next.
\end{proof}


\subsection{\texorpdfstring{Global uniform convergence as \(\coeff\to0\)}{Global uniform convergence as kappa tends to zero}}
\label{section:final-global-uniform-convergence}

\paragraph{Uniform decay at large time and distance}

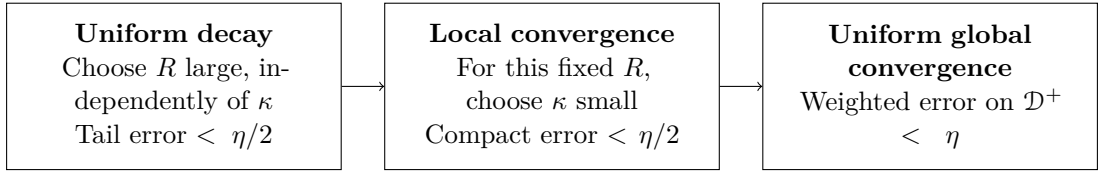
\begin{figure}[htbp]
\centering
\begin{tikzpicture}[every node/.style={draw=black,text=black,align=center,text width=.26\textwidth,inner sep=5pt,minimum height=2.2cm,font=\normalsize}]
\node (tail) at (0,0) {\textbf{Uniform decay}\\Choose \(R\) large, independently of \(\coeff\)\\Tail error \(<\eta/2\)};
\node (local) at (5,0) {\textbf{Local convergence}\\For this fixed \(R\), choose \(\coeff\) small\\Compact error \(<\eta/2\)};
\node (global) at (10,0) {\textbf{Uniform global convergence}\\Weighted error on \(\mathcal D^+\)\\\(<\eta\)};
\draw[->,black] (tail.east) -- (local.west);
\draw[->,black] (local.east) -- (global.west);
\end{tikzpicture}
\caption{From local to uniform convergence. For fixed \(0\leq b<\beta_\infty\) and tolerance \(\eta>0\), the errors are the weighted suprema over \(\mathcal D^+\setminus K_R\), \(K_R\), and \(\mathcal D^+\), respectively. The radius is chosen before the parameter. The same choices apply to derivatives with \(b<\gamma_{\mathrm d}\).}
\label{figure:local-to-global-Einstein-limit}
\end{figure}

The passage from local to global convergence uses the decay of the fields, with \(r=|x|\). We regard \(\mathcal D^+=\{(t,x):t\geq T_b(s_0,|x|)\}\) as including its initial hypersurface. The graph and foliation are fixed independently of \(\coeff\). Since the initial graph is continuous, \(K_R\coloneqq\mathcal D^+\cap\{1+t+r\leq R\}\) is closed and bounded in Cartesian coordinates, hence compact. Figure~\ref{figure:local-to-global-Einstein-limit} shows the order of choices in the proof. We keep the distinct interior and exterior decay rates. The exterior estimate uses the rank-five bound~\eqref{equa:final-rank-five-exterior-control} with \(d_5<\delta\). All constants below are independent of \(\coeff\), but may depend on \(\coeff_0\) and the fixed exponents.

\begin{itemize}
\item In the interior, \(r\leq r_b(t)=t-q_b(t)\), so \(1+t+r\simeq t\) and \(s\leq t\). Proposition~\ref{prop:audit-moving-interface-interior-transfer} applies the interior estimates on these adapted domains. The wave estimate~\eqref{equa:seven-first-full-metric-refinement} and the rank-zero effective-field estimate~\eqref{equa:seven-refined-effective-field}, with \(\eta_0=0\) and the closed interface amplitudes, then give
\be
\label{equa:final-interior-field-decay}
|g_\coeff-\gMink|+|\psi_\coeff|\leq C\varepsilon t^{-1}s^{2\delta},\qquad |\fieldvarphi_\coeff|\leq C\varepsilon(s/t)^2s^{-3/2}.
\ee
These bounds imply the respective Cartesian-time rates \(t^{-1+2\delta}\) and \(t^{-3/2}\). The identity \(\fieldPhi_\coeff=\sqrt\coeff\,\fieldvarphi_\coeff\) keeps an additional factor \(\sqrt\coeff\) for the conformal effective field.

\item In the Euclidean--merging region, \(r\geq r_b(t)\) and \(3/4\leq q_b(t)\leq1\), so \(1+t+r\simeq r\) and \(s^2\lesssim t\lesssim r\). The adapted weight satisfies \(w\simeq2+r-t\) and \(1\leq w\lesssim r\). The weighted Sobolev estimate~\eqref{equa:eight-exterior-weighted-sobolev} and the mass-term estimate~\eqref{equa:eight-uniform-radial-mass-value}, combined with the rank-five bound~\eqref{equa:final-rank-five-exterior-control}, give
\be
\label{equa:final-exterior-field-decay}
|u_\coeff|\leq C\varepsilon r^{-1}w^{1/2-\lambda}s^{d_5},\quad |\psi_\coeff|\leq C\varepsilon r^{-1}w^{1/2-\mu}s^{d_5},\quad |\fieldvarphi_\coeff|\leq C\varepsilon r^{-1}w^{-\mu}s^{d_5}.
\ee
The contribution \(h^\star\) of the reference metric is bounded by \(Cm/r\), with \(m\lesssim\varepsilon\). Since \(\lambda,\mu>1/2\), the powers of \(w\) in all three estimates are nonpositive. The comparison \(s^{d_5}\lesssim r^{d_5/2}\) therefore gives the same rate \((1+t+r)^{-1+d_5/2}\) for the metric, matter, and rescaled effective field. The conformal effective field keeps the additional factor \(\sqrt\coeff\). For the wave-field values, the two angular derivatives require four additional Lorentz boosts. The five commutations in the mass estimate stay within the closing effective-field energy, without the extra top-order time normalization. The bounded initial portion is included by enlarging the constant.
\end{itemize}

Since \(d_5<\delta\), the exterior decay exponent \(1-d_5/2\) exceeds the interior exponent \(1-2\delta\). The interior estimate thus determines the field-value decay valid on the whole future domain:
\be
\label{equa:final-global-field-decay}
|g_\coeff-\gMink|+|\psi_\coeff|+|\fieldPhi_\coeff|+|\fieldvarphi_\coeff|\leq C\varepsilon(1+t+r)^{-\beta_\infty}.
\ee
Local convergence passes this bound to the Einstein fields pointwise, giving
\be
\label{equa:final-Einstein-field-decay}
|g_0-\gMink|+|\psi_0|\leq C\varepsilon(1+t+r)^{-\beta_\infty}.
\ee
The conformal relation and the  metric bound also give \(| \widetilde g_\coeff-g_\coeff| \leq C| \fieldPhi_\coeff| \). Thus the discrepancy in \eqref{equa:final-global-v-discrepancy} satisfies
\be
\label{equa:final-uniform-discrepancy-tail}
d_\coeff(t,x) \leq C\varepsilon(1+ t+r)^{-\beta_\infty}.
\ee

\paragraph{From convergence on compact sets to uniform convergence on the future domain}

\begin{proof}[Proof of Item~5 of Theorem~\ref{theorem:final-global-stability-and-limit}]
\bse
We fix \(0\leq b<\beta_\infty\) and \(R\geq\max\{R_0,s_0+1\}\), independently of \(\coeff\). Since \(s\leq t\), the compact \(K_R\) lies in the fixed foliated cylinder \(K_R^+\coloneqq\iota_b([s_0,R]\times\{|x|\leq2R\})\). We apply the local estimate~\eqref{equa:final-prepared-Einstein-estimate} with \(q=3\) on this larger cylinder. Its spatial sections are regular balls, so spatial Sobolev embedding gives a pointwise bound on \(K_R\), uniformly in \(s\). This argument includes \(s=s_0\) and uses Cartesian coordinates at \(x=0\). Consequently,
\be
\label{equa:final-uniform-compact-v-comparison}
\sup_{1+ t+r\leq R}(1+ t+r)^b d_\coeff\leq C_Ra_{\coeff,3}.
\ee
The constants depend on \(R\), \(b\), and the fixed geometric and analytic parameters, and are independent of \(\coeff\). The tail bound \eqref{equa:final-uniform-discrepancy-tail} gives
\be
\label{equa:final-uniform-exterior-v-comparison}
\sup_{1+ t+r>R}(1+ t+r)^b d_\coeff\leq C\varepsilon R^{b-\beta_\infty}.
\ee
Adding the two bounds and taking the infimum over \(R\geq R_0\), with \(R_0\) enlarged if necessary, proves the equation~\eqref{equa:final-global-convergence-modulus}. For each fixed \(R\), the constant \(C_R\) is finite and independent of \(\coeff\). The modulus \(a_{\coeff,3}=\omega_\coeff\varepsilon+\mathrm{Prep}^{\mathrm{com}}_{\coeff,4}[\Theta]+\sqrt\coeff\,\varepsilon^2\) tends to zero; preparation of order four is available since \(N-6\geq13\). We therefore obtain
\be
\label{equa:final-uniform-two-limit-argument}
\limsup_{\coeff\to0} \sup_{\mathcal D^+}(1+ t+r)^b d_\coeff\leq C\varepsilon R^{b-\beta_\infty} \longrightarrow0\qquad(R\to\infty).
\ee
More explicitly, for any tolerance we first choose \(R\) so that the tail bound is small, using \(b-\beta_\infty<0\), and then choose \(\coeff\) so that \(C_Ra_{\coeff,3}\) is small. This proves Item~5 of Theorem~\ref{theorem:final-global-stability-and-limit} on the entire future spacetime, including its initial hypersurface.
\ese
\end{proof}

\paragraph{Uniform convergence of Cartesian derivatives}

\begin{proof}[Proof of Item~6 of Theorem~\ref{theorem:final-global-stability-and-limit}]
\bse
We fix \(1\leq j\leq N-10\). All multi-indices below are Cartesian spacetime multi-indices; translations have boost rank zero, regardless of their number.

\noindent{\bf 1.} \emph{Interior decay.} Proposition~\ref{prop:audit-moving-interface-interior-transfer} transfers the rank-zero bound~\eqref{equa:seven-direct-pointwise-robust-high} to the adapted interior. The two Sobolev boosts require rank two. The wave estimate uses at most \(j+1\) commutations, and the effective-field mass estimate at most \(j+2\), both within \(N-5\). Together with the reference-metric estimate~\eqref{equa:seven-reference-pointwise-bounds}, they give
\be
\label{equa:final-interior-derivative-decay}
\sum_{1\leq|I|\leq j}(|\del^Ig_\coeff|+|\del^I\psi_\coeff|)\leq C_j\varepsilon t^{-1/2}s^{-1+\nu_2},\qquad \sum_{|I|\leq j}|\del^I\fieldvarphi_\coeff|\leq C_j\varepsilon t^{-3/2}s^{1/2+\nu_2}.
\ee
Since \(t-r\geq3/4\), we have \(\sqrt t\lesssim s\leq t\). The two right-hand sides are therefore bounded by \(C_j\varepsilon t^{-1+\nu_2/2}\) and \(C_j\varepsilon t^{-1+\nu_2}\), respectively. The identity \(\fieldPhi_\coeff=\sqrt\coeff\,\fieldvarphi_\coeff\) gives the same derivative bounds for \(\fieldPhi_\coeff\), with an additional factor \(\sqrt\coeff\). These estimates are in Cartesian coordinates and include the center.

\noindent{\bf 2.} \emph{Exterior decay and the physical metric.} We apply the equation~\eqref{equa:eight-exterior-weighted-sobolev} with \(p=j-1,k=0\), and its mass counterpart~\eqref{equa:eight-exterior-weighted-sobolev-mass} with \(p=j,k=0\). The five additional commutations have rank at most five. Their largest effective-field order is \(j+5\leq N-5\), so the rank-five bound~\eqref{equa:final-rank-five-exterior-control} applies without the top-order time normalization. Since \(w\geq1\), these inequalities imply
\be
\label{equa:final-exterior-derivative-decay}
\sum_{1\leq|I|\leq j}(|\del^Ig_\coeff|+|\del^I\psi_\coeff|)+\sum_{|I|\leq j}|\del^I\fieldvarphi_\coeff|\leq C_j\varepsilon r^{-1}s^{d_5}.
\ee
Here the reference derivatives contribute \(\Ocal(mr^{-2})\) or faster. The Sobolev estimate on the merging interval transports the outer trace along the hypersurface, using a further translation in the energy to control the tangential derivative without a factor \(\zeta^{-1}\). Thus the estimate includes this interval. The comparisons \(s^2\lesssim r\) and \(r\simeq1+t+r\) give decay \((1+t+r)^{-1+d_5/2}\). Differentiating \(\widetilde g_\coeff=e^{-\fieldPhi_\coeff}g_\coeff\) adds products of bounded coefficients and already controlled derivatives, so the physical metric has the same decay. With \(\gamma_{\mathrm d}\) defined in the equation~\eqref{equa:final-spacetime-decay-exponents}, the interior and exterior estimates give
\be
\label{equa:final-global-derivative-tail}
\begin{aligned}
\sum_{1\leq|I|\leq j}(|\del^Ig_\coeff|+|\del^I\widetilde g_\coeff|+|\del^I\psi_\coeff|)+\sum_{|I|\leq j}|\del^I\fieldvarphi_\coeff|&\leq C_j\varepsilon(1+t+r)^{-\gamma_{\mathrm d}},\\
\sum_{|I|\leq j}|\del^I\fieldPhi_\coeff|&\leq C_j\sqrt\coeff\,\varepsilon(1+t+r)^{-\gamma_{\mathrm d}}.
\end{aligned}
\ee
\ese

\bse

\noindent{\bf 3.} \emph{Local convergence of all Cartesian derivatives.} We take \(q=j+3\) in the equation~\eqref{equa:final-prepared-Einstein-estimate}, on a slightly larger compact cylinder. The trace-residual energy estimate~\eqref{equa:eight-preparation-modulus-estimate} is used at order \(j+4\leq N-6\). Its commutations include every Cartesian translation; spatial Sobolev embedding with two additional derivatives controls the Cartesian derivatives of \(\Theta_\coeff\) through order \(j\), including the time derivatives. The identities \(\fieldvarphi_\coeff=\sqrt\coeff(\Theta_\coeff-\mathcal S_\coeff)\) and \(\fieldPhi_\coeff=\sqrt\coeff\,\fieldvarphi_\coeff\) therefore give the vanishing effective-field derivatives directly from preparation and the robust bounds for \(\mathcal S_\coeff\).

For the metric and matter differences, we pull \(y_\coeff\coloneqq(g_\coeff-g_0,\psi_\coeff-\psi_0)\) back to the fixed coordinates \((s,x)\). The local comparison with \(q=j+3\) gives, on slightly enlarged spatial sections,
\be
\label{equa:final-local-time-induction-base}
\sup_s\bigl(\|y_\coeff\|_{H^{j+3}}+\|\del_sy_\coeff\|_{H^{j+2}}\bigr)\leq C_{K,j}a_{\coeff,j+3}.
\ee
Spatial Sobolev embedding \(H^2\hookrightarrow L^\infty\), applied at this stage, controls all jets of total order at most \(j\) with zero or one \(s\)-derivative. The remaining time derivatives follow pointwise from the wave equations~\eqref{equa:fifteen-exact-metric-difference}--\eqref{equa:fifteen-exact-matter-difference}. Their principal coefficient satisfies
\be
\label{equa:final-local-noncharacteristic-time-coefficient}
A_\coeff^{ss}\coloneqq g_\coeff^{-1}(ds,ds),\qquad -A_\coeff^{ss}\geq c_K>0.
\ee
We first solve each equation for \(\del_s^2y_\coeff\), and only then apply \(\del_s^{m-2}\del_x^\alpha\), with \(2\leq m\) and \(m+|\alpha|\leq j\). The principal derivatives on the right are of types \(\del_s\del_a y_\coeff\) and \(\del_a\del_b y_\coeff\); after differentiation, their total orders are at most \(j\), and their numbers of \(s\)-derivatives are at most \(m-1\). The lower-order difference terms have the same property. The coefficient derivatives are uniformly bounded, and the effective-field source derivatives are already controlled by preparation. Pointwise induction on \(m\), starting with the Sobolev bounds just obtained, proves
\be
\label{equa:final-local-Cartesian-derivative-comparison}
\sup_K d_{\coeff,j}\leq C_{K,j}a_{\coeff,j+3}.
\ee
The coordinate transformation is fixed and smooth on the cylinder, so the same estimate holds for Cartesian time derivatives. The conformal relation gives the physical-metric comparison. This argument includes the initial face and uses no inverse power of \(\coeff\).

\ese

\bse
\noindent{\bf 4.} \emph{Uniform convergence on the future domain.} Local convergence passes the derivative bounds to \((g_0,\psi_0)\), hence \(d_{\coeff,j}\leq C_j\varepsilon(1+t+r)^{-\gamma_{\mathrm d}}\). For \(0\leq b<\gamma_{\mathrm d}\), we use the same compact sets \(K_R\) and larger cylinders as in the proof of Item~5. The local estimate and this tail bound yield
\be
\label{equa:final-derivative-compact-tail-estimate}
\sup_{\mathcal D^+}(1+t+r)^b d_{\coeff,j}\leq C_{R,j}a_{\coeff,j+3}+C_j\varepsilon R^{b-\gamma_{\mathrm d}}.
\ee
Taking the infimum gives the equation~\eqref{equa:final-global-derivative-modulus}. We first choose \(R\) large and then \(\coeff\) small. This proves Item~6. The zeroth-order mass bound in the equation~\eqref{equa:final-global-derivative-tail} permits the identical argument for the rescaled fields with their zeroth derivatives included:
\be
\label{equa:final-global-rescaled-jet-limit}
\sup_{\mathcal D^+}(1+t+r)^b\sum_{|I|\leq j}(|\del^I\fieldPhi_\coeff|+|\del^I\fieldvarphi_\coeff|)\longrightarrow0\qquad(\coeff\to0). \qedhere
\ee
\ese
\end{proof}

\paragraph{Global convergence of the unscaled effective field}

\begin{proof}[Proof of Item~7 of Theorem~\ref{theorem:final-global-stability-and-limit}]
\bse
We fix \(0\leq j\leq N-12\). The trace identity~\eqref{equa:eight-Box-Phi-trace-residual} and the potential factorization~\eqref{equa:seven-W-exact} give
\be
\label{equa:final-global-effective-reconstruction}
\Theta_\coeff=3\Box_{g_\coeff}\fieldPhi_\coeff-(\fieldvarphi_\coeff)^2\mathcal W_\coeff(\fieldPhi_\coeff),\qquad \fieldphi_\coeff=\Theta_\coeff-\mathcal S_\coeff.
\ee
In wave coordinates, the wave operator is \(g_\coeff^{\alpha\beta}\del_{\alpha\beta}\). Its derivatives through order \(j\) use at most \(j+2\leq N-10\) derivatives of \(\fieldPhi_\coeff\). The matter trace uses at most \(j+1\) derivatives of \(\psi_\coeff\). All coefficient derivatives are bounded uniformly. The equation~\eqref{equa:final-global-derivative-tail} and the factorization in rescaled variables first bound \(\fieldphi_\coeff\) by \(C_j(\sqrt\coeff\,\varepsilon+\varepsilon^2)(1+t+r)^{-\gamma_{\mathrm d}}\), together with its derivatives through order \(j\). We can then use the other factorization \(W_\coeff=\coeff\fieldphi_\coeff^2\mathcal W_\coeff(\fieldPhi_\coeff)\). Its derivatives have bound \(C_j\coeff\varepsilon^2(1+t+r)^{-2\gamma_{\mathrm d}}\). Substitution into the trace identity gives
\be
\label{equa:final-global-trace-positive-power}
\sum_{|I|\leq j}|\del^I\Theta_\coeff|\leq C_j\sqrt\coeff\,\varepsilon(1+t+r)^{-\gamma_{\mathrm d}}.
\ee
The remaining term is \(\mathcal S_\coeff=-8\pi e^{-\fieldPhi_\coeff}g_\coeff^{-1}(d\psi_\coeff,d\psi_\coeff)\). Its derivatives through order \(j\) converge to those of \(-R[g_0]\), uniformly with weight \((1+t+r)^b\) for \(b<\gamma_{\mathrm d}\). Indeed, in the difference of metric or exponential coefficients we use unweighted uniform convergence; the matter derivatives give the decay. In every other difference term, we place the weight on a derivative of \(\psi_\coeff-\psi_0\) and use Item~6. Leibniz's rule covers all derivatives through order \(j\). The reconstruction~\eqref{equa:final-global-effective-reconstruction} and the bound~\eqref{equa:final-global-trace-positive-power} prove Item~7. In particular, the limiting effective-field relation holds uniformly on the entire future spacetime, not only on compact regions.
\ese
\end{proof}

\paragraph*{Acknowledgments}

PLF was supported by the project ANR-23-CE40-0010-02 funded by the Agence Nationale de la Recherche and by the European Union through the MSCA Staff Exchanges project Einstein--Waves, grant agreement No.~101131233. YM was supported by the Shaanxi Provincial Basic Science Research Fund (Mathematics and Physics) under the grant numbers~22JSZ003 and 23JSY032. 

\paragraph*{Note on AI use}
After we had written the first version, we used ChatGPT 6 to help reorganize it and simplify some arguments before making a human-only check of the paper.

\end{document}